\documentclass{svmono}

\usepackage{amsfonts,amssymb,amsbsy,amsmath,bm,subeqnarray,hyperref}
\usepackage[sort,square]{natbib}
\usepackage[a4paper, twoside, body={13.5cm,17.5cm}, headsep=0.8cm,
nofoot]{geometry}

\usepackage{makeidx} 

\usepackage{epsfig,wrapfig}
\usepackage{graphicx,psfrag} 
\usepackage{multicol} 
\usepackage[mathscr]{euscript}
\usepackage[ps, dvips, all, color]{xy}
\usepackage{multirow}
\usepackage{rotating}
\usepackage{mathabx}
\usepackage{url}

\makeindex 

\newcommand{\proofend}{\hfill$\blacksquare$}
\newcommand{\ppFrac}[2]{\genfrac{[}{]}{0pt}{}{#1}{#2}}
\newcommand{\grule}[5]{#1 \overset{#4}{\longleftarrow} #2
  \overset{#5}{\longrightarrow} #3}  \newcommand{\abb}[3]{#1 \colon #2
  \rightarrow #3} \newcommand{\morfism}[3]{(#1 \overset{#3}
  \rightarrow #2)} \renewcommand{\baselinestretch}{1.3}

\begin{document}
\pagenumbering{roman}
\frontmatter

\setcounter{page}{1}
\pagestyle{empty}
\thispagestyle{empty}
\hbox{\ }

\small\normalsize
\begin{center}
\Large{{MATRIX GRAPH GRAMMARS}}
\ \\
\ \\
\large{by} \\
\ \\
\large{Pedro Pablo P\'erez Velasco}
\ \\
\ \\
Version 1.2
\normalsize
\end{center}
 \cleardoublepage
\thispagestyle{empty}
\hbox{\ }
\vspace{1in}
\renewcommand{\baselinestretch}{1.5}

\begin{center}
\large{\copyright \hbox{ }Copyright by\\
Pedro Pablo P\'erez Velasco \\
2007, 2008, 2009}
\end{center}
\ \\
\ \\
\ \\
\ \\
\ \\
\ \\
\ \\
\ \\
\ \\
\ \\
\ \\
\ \\
\ \\
\ \\
\ \\
\ \\
\ \\
\begin{figure}[htbp]
  \centering
  \includegraphics[scale =  1.3]{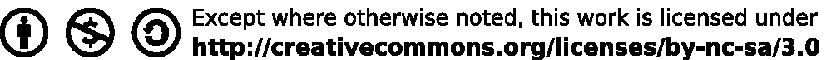}
\end{figure}
\cleardoublepage
\thispagestyle{empty}
\vspace*{3.5cm}
\begin{flushright}

{\large \emph{To my family}}

\end{flushright}
\cleardoublepage
\pagestyle{headings}

\hbox{\ } \vspace{0.75in}
\begin{center}
  \LARGE{{ACKNOWLEDGEMENTS}} \\[0.5in]
\end{center}

These lines are particularly pleasant to write. After all those years,
I have a quite long list of people that have contributed to this
book in one way or another. Unfortunately, I will not be able
to include them all. Apologizes for the absences.

First of all my family. Gema, with neverending patience and love, always
supports me in every single project that I undertake. My unbounded
love and gratitude. Hard to return, though I'll try. My two daughters,
Sof\'ia and Diana, make every single moment worthy. I'm absolutely
grateful for their existence.

My brothers \'Alex and Nina, now living in Switzerland, with whom I
shared so many moments and that I miss so much. My parents, always
supporting also with patience and love, worried if this boy would become a
man (am I?).

Juan, my thesis supervisor, whose advice and interest is invaluable. He has
been actively involved in this project despite his many
responsibilities. Also, I would like to thank the people at the series
of seminars on complexity theory at U.A.M., headed by Roberto
Moriy\'on, for their interest on Matrix Graph Grammars.

Many friends have stoically stood some chats on this topic
\emph{affecting} interest. Thank you very much for your
friendship. KikeSim, GinHz, \'Alvaro Iglesias, Jaime Guerrero, ... All
those who have passed by are not forgotten: People at ELCO (David,
Fabrizio, Juanjo, Juli\'an, Lola, ...), at EADS/SIC (Javier, Sergio,
Roberto, ...), at Isban, at Banco Santander. Almost uncountable.

I am also grateful to those that have worked on the tools used in this
book: Emacs and microEmacs, MikTeX, TeTeX, TeXnicCenter, OpenOffice
and Ubuntu. I would like to highlight the very good surveys available
on different topics on mathematics at the web, in particular at
websites \url{http://mathworld.wolfram.com} and
\url{http://en.wikipedia.org}, and the anonymous people behind them.

Last few years have been particularly intense. A mixture of hard work
and very good luck. I feel that I have received much more than I'm
giving. In humble return, I will try to administer
\url{http://www.mat2gra.info}, with freely available information on
Matrix Graph Grammars such as articles, seminars, presentations,
posters, one e-book (this one you are about to read) and whatever you
may want to contribute with.
\pagestyle{empty}
\cleardoublepage
\pagestyle{headings}
\renewcommand{\baselinestretch}{1.3}

\tableofcontents
\listoffigures \listoftables

\pagenumbering{arabic}
\mainmatter
\chapter{Introduction}
\label{ch:introduction}

This book is one of the subproducts of my dissertation. If its aim had
to be summarized in a single sentence, it could be
\emph{algebraization of graph grammars} or, more accurately,
\emph{study of graph dynamics}.

From the point of view of a computer scientist, graph grammars are a
natural generalization of Chomsky grammars for which a purely
algebraic approach does not exist up to now.  A Chomsky (or string)
grammar is, roughly speaking, a precise description of a formal
language (which in essence is a set of strings).  On a more discrete
mathematical style, it can be said that graph grammars -- Matrix Graph
Grammars in particular -- study dynamics of graphs.  Ideally, this
algebraization would enforce our understanding of grammars in general,
providing new analysis techniques and generalizations of concepts,
problems and results known so far.

In this book we fully develop such theory over the field $GF(2)$ --
the field with two elements -- which covers all graph cases, from
simple graphs (more attractive for a mathematician) to multidigraphs
(more interesting for an applied computer scientist).  The theory is
presented and its basic properties demonstrated in a first stage,
moving to increasingly difficult problems and establishing relations
among them:
\begin{itemize}
\item \textbf{Applicability}, for which two equivalent
  characterizations (necessary and sufficient conditions) are
  provided.
\item \textbf{Independence}. Sequential and parallel independence in
  particular, generalizing previously known results for two elements.
\item \textbf{Restrictions}. The theory developed so far for graph
  constraints and application conditions is significantly generalized.
\item \textbf{Reachability}. The \emph{state equation} for Petri nets
  and related techniques are extended to general Matrix Graph
  Grammars. Also, Matrix Graph Grammars techniques are applied to
  Petri nets.
\end{itemize}

Throughout the book many new concepts are introduced such as
compatibility, coherence, initial and negative graph sets, etc.
Some of them project interesting insights about a given grammar, while
others are used to study previously mentioned problems.

Matrix Graph Grammars have several advantages.  First, many branches
of mathematics are at our disposal.  It is based on Boolean algebra,
so first and second order logics can be applied almost directly.  They
admit a functional representation so many ideas from functional
analysis can be utilized.  On the more algebraic side it is possible
to use group theory and tensor algebra.  Finally, category theory
constructions such as pushouts are available as well.  Second, as it
splits the static definition from the dynamics of the system, it is
possible to study to some extent many properties of the grammar
without the need of an initial state.  Third, although it is a
theoretical tool, Matrix Graph Grammars are quite close to
implementation, being possible to develop tools based on this theory.

This introductory chapter aims to provide some perspective on graph
grammars in general and on Matrix Graph Grammars in particular.  In
Sec.~\ref{sec:historicalOverview} we present a (partial) historical
overview of graph grammars and graph transformation systems taken from
several sources but mainly from~\cite{Kahl} and~\cite{Fundamentals}.
Section~\ref{sec:motivation} introduces those open problems that have
guided our research.  Finally, in Sec.~\ref{sec:dissertationOutline}
we brush over the book and see how \emph{applicability},
\emph{sequential independence} and \emph{reachability} articulate it.

\section{Historical Overview}
\label{sec:historicalOverview}

Research in graph grammars started in the late 60's
\cite{PfR69}\cite{Sch70}, strongly motivated by practical problems in
computer science and since then it has become a very active area.
Currently there is a wide range of applications in different branches
of computer science such as formal language theory, software
engineering, pattern recognition and generation, implementation of
term rewriting, logical and functional programming, compiler
construction, database design and theory, visual programming and
modeling languages and many more (see~\cite{handbook} for references
on these and other topics).

There are different approaches to graph grammars and graph
transformation systems.\footnote{The only difference between a grammar
  and a transformation system is that a grammar considers an initial
  state while a transformation system does not.} Among them, the most
prominent are the algebraic, logical, relational and set-theoretical.

\begin{figure}[htbp]
  \begin{displaymath}
    \xymatrix{
      L \ar[dd]_{m} \ar[rrrr]^{p} &&&& R \ar[dd]^{m^*} \\
      \\
      G \ar@/_17pt/@[blue][rrrr]_{p^*} \ar[rr] && G_1 \ar[rr] && H
    }
  \end{displaymath}
  \caption{Main Steps in a Grammar Rule Application}
  \label{fig:mainSteps}
\end{figure}
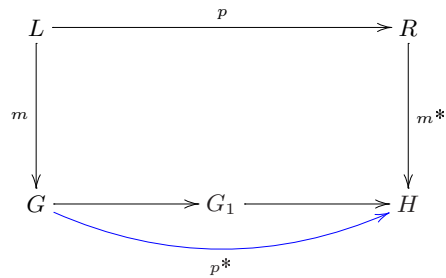

The main steps -- some of which are summarized in Fig.
\ref{fig:mainSteps} -- in all approaches for the application of a
grammar rule $p:L \rightarrow R$ to a host graph $G$ (also known as
\emph{initial state}) to eventually obtain a \emph{final state} $H$
are almost the same:\label{scheme:generalScheme}
\begin{enumerate}
\item Select the grammar rule to be applied ($p:L \rightarrow R$ in
  this case).  In general this step is non-deterministic.
\item Find an occurrence of $L$ in $G$.  In general this step is also
  non-deterministic because there may be several occurrences of $L$ in
  $G$.
\item Check any application condition of the production.
\item Remove elements that appear in $L$ but not in $R$.  There are
  two possibilities for so-called \emph{dangling
    edges}\index{dangling!edge}:\footnote{A dangling edge is one not
    appearing in the rule specification which is incident to one node
    to be eliminated.}
  \begin{enumerate}
  \item Production is not applied.
  \item Dangling edges are deleted too.
  \end{enumerate}
  If the production is to be applied, the system state changes from
  $G$ to $G_1$ (see Fig.~\ref{fig:mainSteps}).
\item Glue $R$ with $G_1$.  The system state changes from $G_1$ to $H$
  (see Fig.~\ref{fig:mainSteps}).
\end{enumerate}

Now we shall briefly review previously mentioned families of
approaches.  The so-called \emph{algebraic approach} to graph grammars
(graph transformation systems) is characterized by relying almost
exclusively on category theory and using gluing of graphs to perform
operations.  It can be divided into at least three main sub
approaches, depending on the categorical construction under use:
\emph{DPO} (Double PushOut, see Sec.~\ref{sec:DPO}), \emph{SPO}
(Single PushOut, see Sec.~\ref{sec:otherCategoricalApproaches}),
\emph{pullback} and \emph{double pullback} (also summarized in Sec.~
\ref{sec:otherCategoricalApproaches}).  We will not comment on others,
like \emph{sesquipushout} for example (see~\cite{CHHK06}).

DPO was initiated by Ehrig, Pfender and Schneider in the early 70's
\cite{EhrigPfenderSchneider} as a generalization of Chomsky grammars
in order to consider graphs instead of strings.  It seems that the
term \emph{algebraic} was appended because graphs might be considered
as a special kind of algebras and because the pushout construction was
perceived more as a concept from universal algebra than from category
theory.  Nowadays it is the more prominent approach to graph
rewriting, with a vast body of theoretical results and several tools
for their implementation.\footnote{For example AGG -- see~\cite{agg}
  or visit \url{http://tfs.cs.tu-berlin.de/agg/} -- and AToM${}^3$ --
  see~\cite{atom3} or visit \url{http://atom3.cs.mcgill.ca/} --.}

By mid and late 80's Raoult~\cite{Raoult}, Kennaway
\cite{Kennaway87}\cite{Kennaway91} and L\"owe~\cite{Lowe} developed
SPO approach probably motivated by some ``restrictions'' of DPO, e.g.
the usage of total instead of partial morphisms.  Raoult and Kennaway
were focused on term graph rewriting while L\"owe took a more general
approach.

In the late 90's a new approach -- although less prominent for now --
emerged by reverting all arrows (using pullbacks instead of pushouts),
proposed by Bauderon~\cite{Bau97}.  It seems that, in contrast to the
pushout construction, pullbacks can handle deletion and duplication
more easily.

DPO has been generalized recently through \emph{adhesive HLR
  categories}, which is summarized in Sec.
\ref{sec:otherCategoricalApproaches} (we are not aware of a similar
initiative for SPO or pullback).  For a detailed account see
\cite{Fundamentals}.  Instead of just considering graphs, all main
ideas in DPO can be extended to higher level structures like labeled
graphs, typed graphs, Petri nets, etc.  This is firstly
accomplished in~\cite{EHKPB91a} and~\cite{EHKPB91b}, starting the
theory of \emph{HLR systems} (High Level Replacement systems).
Independently, Lack and Soboci\'nski in~\cite{LackSobo} introduced the
concept of \emph{adhesive category} and in~\cite{EHPP04} both were
merged to get adhesive HLR categories.

In this book we shall refer to these approaches as
\emph{categorical}, to distinguish from ours which is more algebraic
in nature.

The so-called \emph{set-theoretic approach} (sometimes also known as
\emph{algorithmic} approach) substitutes one structure by another
structure, either nodes or edges.  There are two subfamilies,
\emph{node replacement} and \emph{edge replacement} (also
\emph{hyperedge} replacement), depending on the type of elements to be
replaced.  Node replacement (edNCE) was introduced in
\cite{Nag76}\cite{Nag79} and further investigated in many papers.  It
is based on connecting instead of gluing for embedding one graph into
another.  Many extensions and particular cases have been studied so
far, and many others, such as C-edNCE when considering confluence,
NCE, NLC, dNLC, edNLC and edNCE (see Sec.~\ref{sec:nodeReplacement}
for the meaning of acronyms) are currently on going.  Hyperedge
replacement was introduced in the early 70s by Feder~\cite{Fed71} and
Pavlidis~\cite{Pav72} and has been intensively investigated since
then.  Contrary to the node replacement approach, it is based on
gluing.  Please, see Secs.~\ref{sec:nodeReplacement} and
\ref{sec:hyperedgeReplacement} for a quick introduction.

It is possible to use logics to express graphs and to encode graph
transformation. In Sec.~\ref{sec:msolApproach} this approach with
monadic second order logic is reviewed presenting its foundations and
main results.\footnote{Monadic Second Order Logics, MSOL, lie in
  between first and second order logics.}

The \emph{relational approach} (also \emph{algebraic-relational
  approach}) is based on relational methods to specifying graph
rewriting (in fact it could be applied to more general structures than
graphs).  Once a graph is characterized as a relational structure it
is possible to apply all relational machinery, substituting categories
by allegories and Dedekind categories.  Probably, the main advantage
is that it is possible to give local characterization of concepts.
The roots of this approach seem to date back to the early 1970's with
the papers of Kawahara~\cite{Kaw73a}\cite{Kaw73b}\cite{Kaw73c}
establishing a relational calculus inside topos theory.  An overview
can be found in Sec.~\ref{sec:relationAlgebraicApproach}.

Our approach has been influenced by these approaches to a different
extent, heavily depending on the topic.  The basics of Matrix Graph
Grammars are most influenced by the categorical approach, mainly by
SPO in the shape of productions and to some extent of direct
derivations.  For application conditions and graph constraints, our
inspiration comes almost exclusively from MSOL.  Concerning the
relational approach, our basic structure has a natural representation
in relational terms but the development in both cases is very
different.  The influence of hyperedge replacement and node
replacement, if any, is much more fuzzy.

\section{Motivation}
\label{sec:motivation}

The dissertation that gave rise to this book started as a project to
study simulation protocols (conservative, optimistic, etc.) under
graph transformation systems.  In the first few weeks we missed a
\emph{real} algebraic approach to graph grammars.  ``Real'' in the
sense that there are algebraic representations of graphs very close to
basic algebraic structures such as vector spaces (incidence or
adjacency matrices for example) but the theories available so far do
not make use of them.  As commented above, the main objective of this
book is to give an algebraization of graph grammars.

One advantage foreseen from the very beginning was the fact that nice
interpretations in terms of functional analysis and physics could be
used to move forward, despite the fact that the underlying structure
is binary so, if necessary, it was possible to bring in easily logics
and its powerful methods.

Our schedule included several increasingly difficult problems to be
treated by our approach with the hope of getting better insight and
understanding, trying to generalize whenever possible and, most
importantly, providing a unified body of results in which all concepts
and ideas would fit naturally.

First things first, so we begin with the name of the book:
Matrix Graph Grammars.  It has been chosen to emphasize the algebraic
part of the approach -- although there are also logics, tensors,
operators -- and to recall matrix mechanics as introduced by Born,
Heisenberg and Jordan in the first half of the twentieth
century.\footnote{An alternative was YAGGA, which stands for Yet
  Another Graph Grammar Approach (in the style of the famous ``Yet
  Another...'' series).}  You are kindly invited to visit
\texttt{http://www.mat2gra.info} for further research, a web page
dedicated to this topic that I (hopefully) intend to maintain.

\index{MGG, Matrix Graph Grammar}Section~\ref{sec:historicalOverview}
points out that motivations of some graph grammar approaches have been
quite close to practice, in contrast with Matrix Graph Grammars (MGG)
which is more theoretically driven.  Nonetheless, there is an on-going
project to implement a graph grammar tool based on $\textrm{AToM}^3$
(see~\cite{atom3} or visit \url{http://atom3.cs.mcgill.ca/}) using
algorithms derived from this book (the analysis algorithms are
expected to have a good performance).  We will briefly touch on this
topic in Sec.~\ref{sec:initialDigraphSet}. Appendix
\ref{app:caseStudy} illustrates all the theory with a more or less
realistic case study.

This ``basis for theoretical studies'' intends to provide us with the
capability of solving theoretical problems as those commented below,
which are the backbone of the book.

Informally, a grammar is a set of productions plus an initial graph
which we can safely think of as a collection of functions plus an
initial set.  A sequence of productions would then be a sequence of
functions, applied in order.  Together with the function we specify
the elements that must be found in the initial set (in its domain), so
in order to apply a function we must first find the domain of the
function in the initial set (this process is known as
\emph{matching}).  As productions are applied, the system moves on
transforming the initial set in a sequence of intermediate sets to
eventually arrive to a final state (final set).\footnote{The natural
  interpretation is that functions modify sets, so some dynamics
  arise.}  Actually, we will deal neither with sets nor with functions
but with directed graphs and morphisms.

We will speak of graphs, digraphs or simple digraphs meaning in
all cases simple digraphs.  See Sec.~\ref{sec:graphTheory} for its
definition and main properties.

Once grammar rules have been defined and its main properties
established, the first problem we will address is the characterization
of \emph{applicability}, i.e. give necessary and sufficient conditions
to guarantee that a sequence can be applied to an initial state (also
known as \emph{host graph}) to output a final state (a graph again).
Formally stated for further reference:

\newtheorem{applicability}{Problem}
\begin{applicability}[Applicability]\label{prob:applicability}
  \index{applicability}For a sequence $s_n$ made up of rules in a
  grammar $\mathfrak{G}$ and a simple\footnote{Defined in
    Sec.~\ref{sec:graphTheory}.} digraph $G$, is it possible to apply
  $s_n$ to the host graph $G$?
\end{applicability}

No restriction is set on the output of the sequence except that it is
a simple digraph.  There is a basic problem when deleting nodes known
as \emph{dangling condition}: Are all incident edges eliminated too?
Otherwise the output would not be a digraph.

When we have a production and a matching (for that production) we will
speak of a \emph{direct derivation}.\index{direct derivation} A
sequence of direct derivations is called a
\emph{derivation}.\index{derivation}

A quite natural progression in the study of grammars is the following
question, that we call \emph{independence
  problem}:\footnote{\emph{Independence} from the point of view of the 
  grammar:  It does not matter which path the grammar follows because
  in both cases it finishes in the same state.}

\newtheorem{indepProblem}[applicability]{Problem}
\begin{indepProblem}[Independence]\label{prob:independence}
  \index{independence}For two given derivations $d_n$ and $d'_n$
  applicable to host graph $G$, do they reach the same state?, i.e. is
  $d_n(G) = d'_n(G)$?
\end{indepProblem}

Mind the similarities with \emph{confluence} and \emph{local
  confluence} (see below).  However, independence is a very general
problem and we will be interested in a reduced version of it, known as
\emph{sequential independence}, which is widely addressed in the graph
grammar literature and also in other branches of computer science.  As
far as we know, in the literature~\cite{handbook, Fundamentals} this
problem is addressed for sequences of two direct derivations, being
longer sequences studied pairwise.

\newtheorem{seqIndepProblem}[applicability]{Problem}
\begin{seqIndepProblem}[Sequential
  Independence]\label{prob:sequentialIndependence}
  \index{sequential independence}For two derivations $d_n$ and $d'_n =
  \sigma (d_n)$ applicable to host graph $G$, with $\sigma$ a
  permutation, do they reach the same state?
\end{seqIndepProblem}

Of course, problems~\ref{prob:independence} and
\ref{prob:sequentialIndependence} can be extended easily to consider
any finite number of derivations and, in both cases, there is a
dependence relationship with respect to problem
\ref{prob:applicability}.

Our next step will be to generalize some theory from Petri nets
\cite{Murata}, which can be seen as a particular case of Matrix Graph
Grammars.  In particular, our interest is focused on
\emph{reachability}:

\newtheorem{reachProblem}[applicability]{Problem}
\begin{reachProblem}[Reachability]\label{prob:reachability}
  \index{reachability}For two given states (initial $S_0$ and final
  $S_T$), is there any sequence made up of productions in $G$ that
  transforms $S_0$ into $S_T$?
\end{reachProblem}

In the theory developed so far for Petri nets, reachability is
addressed using the \emph{state equation} (linear system) which is a
necessary condition for the existence of such a sequence (see Chap.
\ref{ch:reachability}).

Problem~\ref{prob:reachability} directly relies on problem
\ref{prob:applicability}.  More interestingly, it is also related to
problems~\ref{prob:independence} and
\ref{prob:sequentialIndependence}: As every solution provided by the
state equation specifies the set of productions to be applied but not
the order (see Sec.~\ref{sec:crashCourseInPetriNets}), sequences
associated to different solutions of the state equation can be
independent but not sequential independent (this is because different
sets of solutions apply each production a different number of times).
So, in particular, reachability can be useful to \emph{split}
independence and sequential independence.

\begin{figure}[htbp]
  \centering
  \includegraphics[scale =
  0.5]{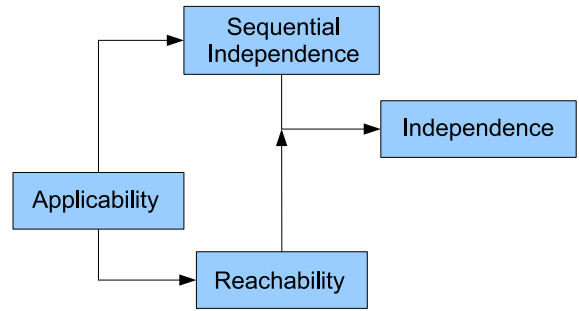}
  \caption{Partial Diagram of Problem Dependencies}
  \label{fig:problemsDepDiagramInitial}
\end{figure}

All these problems with their corresponding dependencies are
summarized in Fig.~\ref{fig:problemsDepDiagramInitial}.  Compare with
the complete diagram that includes mid-term and long-term research in
Fig.~\ref{fig:problemsDepDiagram} on p.
\pageref{fig:problemsDepDiagram}.

Although we will not study confluence in this book (except
some ideas in Chap.~\ref{ch:conclusionsAndFurtherResearch}), just to
make a complete account two further related problems are introduced.
We will briefly review them in the last chapter.

\newtheorem{confluenceProblem}[applicability]{Problem}
\begin{confluenceProblem}[Confluence]\label{prob:confluence}
  \index{confluence}For two given states $S_1$ and $S_2$, do there
  exist two derivations $d_1$ and $d_2$ such that $d_1(S_1) \cong
  d_2(S_2)$?.
\end{confluenceProblem}

\begin{figure}[htbp]
  \centering
  \includegraphics[scale = 0.6]{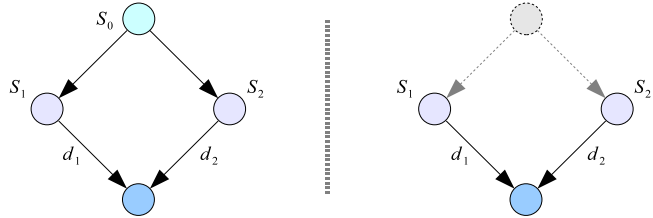}
  \caption{Confluence}
  \label{fig:confluence}
\end{figure}

Strictly speaking this is not confluence as defined in the literature
\cite{Terese}. To the left of Fig.~\ref{fig:confluence} you can find
confluence: For the initial state $S_0$ that independently evolves to
$S_1$ and $S_2$, is it possible to find derivations that close the
diamond?\footnote{The difference between \emph{local confluence} and
  confluence is that in the former to move from $S_0$ to $S_1$ or
  $S_2$ it is mandatory to use a direct derivation and not a
  derivation.} To the right of the same figure we have represented
problem~\ref{prob:confluence}. The difference is that a common initial
state is not assumed.

In mathematics, \emph{existence} and \emph{uniqueness} theorems are
central to any of its branches.  As it is, the analogous terms in
computer science are \emph{termination} and \emph{confluence},
respectively.

In some sense we may think of reachability as opening or broadening
the state space of a given grammar while confluence, as introduced
here, closes or bounds it.

Problem~\ref{prob:confluence} deals with \emph{confluency} of
confluence.  The other part (how to actually get to the states $S_1$
and $S_2$) is more related to reachability.  Note that if one of the
derivations is the identity then problem~\ref{prob:confluence} becomes
problem~\ref{prob:reachability} (reachability).

If we limit to permutation of sequences, as in the derivation of
problem~\ref{prob:sequentialIndependence} out of problem
\ref{prob:independence}, we can pose:

\newtheorem{seqConfluenceProblem}[applicability]{Problem}
\begin{seqConfluenceProblem}[Sequential
  Confluence]\label{prob:sequentialConfluence}
  \index{sequential confluence}For two given initial states, do there
  exist two derivations (one permutation of the other) with isomorphic
  final states?.
\end{seqConfluenceProblem}

Again, it is not difficult to make them consider any finite set of
derivations instead of just two.  Once we know if a grammar is
confluent, the next step is to know how much it takes to get to its
final state.  This is very close to \emph{complexity}. Complexity
theory is not addressed in this book.

To the best of our knowledge, applicability (problem
\ref{prob:applicability}) has not been addressed up to now.
Independence and sequential independence (problems
\ref{prob:independence} and~\ref{prob:sequentialIndependence}) are
very popular.\footnote{Actually, it is sequential independence the one
  normally addressed in the literature. We have introduced
  independence for its potential link with confluence.} See for
example Chaps. 3 and 4 in~\cite{handbook}. Reachability is a key
concept and has been studied and partially characterized in many
papers, mainly in Petri nets theory. See~\cite{Murata}. Confluence is
a concept of fundamental importance to grammar theory. For term
rewriting systems see~\cite{confluence}.

\section{Book Outline}
\label{sec:dissertationOutline}

Based on the problems commented in previous section, the book
is organized in nine chapters plus one appendix.  The First three
chapters, including this one, are introductory.  Chapter~\ref{ch:backgroundAndTheory} provides a short overview of needed
mathematical machinery which includes some basic results from logics
(first and monadic second order), category theory, tensor algebra,
graph theory, functional analysis (notation and some basic results)
and group theory.  We have not used advanced results on any of these
disciplines so probably a quick review should suffice, mainly for
fixing notation.

Graph grammars approaches are discussed in Chap.
\ref{ch:graphGrammarsApproaches}, which essentially expands the
overview in Sec.~\ref{sec:historicalOverview}.  Sections~\ref{sec:DPO}
and~\ref{sec:otherCategoricalApproaches} cover \emph{algebraic}
approaches, for which we prefer the term \emph{categorical}, as
commented above.  Set-theoretic approaches (node and hyperedge
replacement) are covered in Secs.~\ref{sec:nodeReplacement} and
\ref{sec:hyperedgeReplacement}.  Term rewriting through monadic second
order logics is the MSOL approach, to which
Sec.~\ref{sec:msolApproach} is devoted.  The chapter ends with the
relational approach in Sec.~\ref{sec:relationAlgebraicApproach}.  The
objective of this chapter is to get an idea of each approach (and not
to provide a detailed study) in order to, among other things, ease
comparison with Matrix Graph Grammars.

Chapter~\ref{ch:mggFundamentals1} introduces the basics of our proposal
(Sec.~\ref{sec:characterizationAndBasicConcepts}) and prepares to
attack problem~\ref{prob:applicability} by introducing concepts such
as \emph{completion} (Sec.~\ref{sec:completion}), \emph{coherence},
sequences (Sec.~\ref{sec:sequencesAndCoherence}) and the \emph{nihilation
  matrix} (Sec.~\ref{sec:coherenceRevisited}).

Standing on Chapter~\ref{ch:mggFundamentals1},
Chapter~\ref{ch:mggFundamentals2} studies \emph{minimal} and
\emph{negative initial digraphs}
(Secs.~\ref{sec:MID}~and~\ref{sec:NID}), subsequently generalized to
\emph{initial digraph set} in Sec.  \ref{sec:initialDigraphSet}),
\emph{composition} and \emph{compatibility}
(Sec.~\ref{sec:compositionAndCompatibility}) and theorems related to
their properties and characterizations.

Chapter~\ref{ch:matching} covers an essential part of production
applicability: Matching the left hand side (LHS) of a production
inside the host graph.  Dangling edges are covered, dealing with them
with what we call \emph{$\varepsilon$-productions} in Sec.
\ref{sec:matchAndExtendedMatch} and further studied and classified in
Sec.~\ref{sec:internalAndExternalProductions}.  We deal with
\emph{marking} in Sec.~\ref{sec:marking}, which can help in case it is
necessary to guarantee that several productions have to be applied in
the same place.  Minimal and negative initial digraphs are generalized
to the \emph{initial digraph set} in Sec.~\ref{sec:initialDigraphSet}.
In Sec.~\ref{sec:summaryAndConclusions4} we give two characterizations
for applicability (problem~\ref{prob:applicability}).

We will cope with sequential independence (problem
\ref{prob:sequentialIndependence}) for quite general families of
permutations in Chap.~\ref{ch:sequentializationAndParallelism}.
Sameness of minimal initial digraph (called $G$\emph{-congruence}) for
two sequences is addressed in Sec.
\ref{sec:sequentializationGrammarRules}; the case of two derivations
is seen in Sec.~\ref{sec:sequentialIndependenceDerivations}.  Explicit
parallelism is studied in Sec.~\ref{sec:explicitParallelism} through
composition and $G$\emph{-congruence}, which is related to
\emph{initial digraph sets}.

In Chap.~\ref{ch:restrictionsOnRules} graph constraints and
application conditions (preconditions and postconditions) are studied
for Matrix Graph Grammars.  They are introduced in Sec.
\ref{sec:graphConstraintsAndApplicationConditions} where a short
overview of related concepts in other graph grammars approaches is
carried out.  The notion of direct derivation is extended to cope with
application conditions in Matrix Graph Grammars in a very natural
manner in Sec.~\ref{sec:extendingDerivations} and functionally
represented in Sec.~\ref{sec:functionalRepresentation}, where they are
sequentialized.

Chapter~\ref{ch:transformationOfRestrictions} continues with graph
constraints and application conditions. First, some properties such as
consistency are defined and characterized
(Sec.~\ref{sec:consistencyAndCompatibility}). In
Sec.~\ref{sec:movingConditions} we show how it is possible to
transform postconditions into preconditions and vice versa.  Both of
theoretical and of practical importance is the use of variable nodes
because, among other things, it allows us to automatically extend the
theory to include multidigraphs without any change of the theory of
Matrix Graph Grammars in
Sec.~\ref{sec:fromSimpleDigraphsToMultidigraphs}.

In Chap.~\ref{ch:reachability} problem~\ref{prob:reachability}
(reachability) is tackled, extending results from Petri nets to more
general grammars.  Section~\ref{sec:crashCourseInPetriNets} quickly
introduces this theory and summarizes some basic results.  Section
\ref{sec:mggTechniquesForPetriNets} applies some Matrix Graph Grammars
results from previous chapters to Petri nets.  The rest of the chapter
is devoted to extending Petri nets results for reachability to Matrix
Graph Grammars, in particular Sec.~\ref{sec:dPOLikeMGG} covers graph
grammars without dangling edges while Sec.~\ref{sec:sPOLikeMGG} deals
with the general case.

The book ends in Chap.~\ref{ch:conclusionsAndFurtherResearch}
with the conclusions and further research.  A summary of what we think
are our most important contributions can be found there.

Finally, in Appendix A a fully worked case study is presented in which
all main theorems are applied together with detailed explanations and
implementation remarks and advices.

Most of the material presented in this book has been published
\cite{JuanPP_1},~\cite{JuanPP_6},~\cite{JuanPP_2},~\cite{JuanPP_3},
\cite{JuanPP_4} and~\cite{JuanPP_5} and presented in international
congresses: ICM'2006 (International Congress of Mathematicians,
awarded with the second prize of the poster competition in Section 15,
\emph{Mathematical Aspects of Computer Science}), ICGT'2006
(International Conference on Graph Transformations), PNGT'2006 (Petri
Nets and Graph Transformations), PROLE'2007 (VII Jornadas sobre
Programaci\'on y Lenguajes) and GT-VC'2007 (Graph Transformation for
Verification and Concurrency, in CONCUR'2007).

Some further research is now available in
\url{http://www.mat2gra.info} and in the arXiv
(\url{http://arxiv.org}, just look for ``Matrix Graph Grammars'' in
their search engine). Besides, a slight generalization using
\emph{Boolean complexes} have appeared in~\cite{MGG_Combinatorics}.
\chapter{Background and Theory}
\label{ch:backgroundAndTheory}

The Matrix Graph Grammar approach uses many mathematical theories
which might seem distant one from the others.  Nevertheless, there are
some interesting ideas connecting them which we seize to contribute
whenever possible.  Matrix Graph Grammars do not depend on any novel
theorem that opens a new field of research, but aims to put ``old''
problems in a new perspective.

There are excellent books available covering every subject of this
topic.  There are also excellent resources on the web.  We think that
this fast introduction should suffice.  It is intended as a reference
chapter.  All concepts are highlighted in bold to ease their location.

\section{Logics}
\label{sec:logics}

Logics are of fundamental importance to Matrix Graph Grammars for two
reasons.  First, graphs are represented by their adjacency matrices.
As we will be most concerned with simple digraphs, they can be
represented by Boolean matrices (we will come back to this in Sec.
\ref{sec:graphTheory}).\footnote{Multidigraphs are also addressed
  using Boolean matrices. Refer to Sec.~\ref{sec:fromSimpleDigraphsToMultidigraphs}.}  Second, Chap.
\ref{ch:restrictionsOnRules} generalizes graph constraints and
application conditions using monadic second order logics.  Good
references on mathematical logics are~\cite{logic} and~\cite{Smul}.

\index{FOL!first order logic}First-order predicate calculus (more
briefly, first order logic, FOL) generalizes propositional logic,
which deals with propositions: A statement that is either
\textbf{true} or \textbf{false}.\index{propositional logic}

\index{FOL!constant}\index{FOL!variable}\index{FOL!symbol}\index{FOL!function}\index{FOL!connective}\index{FOL!quantifier}FOL
formulas are constructed from \emph{individual constants} ($a$, $b$,
$c$, etc., typically lower-case letters from the beginning of the
alphabet), \emph{individual variables} (x, y, z, etc., typically
lower-case letters from the end of the alphabet), \emph{predicate
  symbols} (P, Q, R, etc., typically upper-case letters),
\emph{function symbols} (f, g, h, etc., typically lower-case letters
from the middle of the alphabet), \emph{propositional connectives}
($\neg$, $\wedge$, $\vee$, $\Rightarrow$, $\Leftrightarrow$) and
\emph{quantifiers} ($\forall$, $\exists$).  Set $\mathcal{C}$ will be
that of individual constants, set $\mathcal{F}$ will be function
symbols and set $\mathcal{P}$ will contain predicate symbols.  Besides
these elements, punctuation symbols are permitted such as parenthesis
and commas.

A formula in which every variable is quantified is a \textbf{closed
  formula} (\emph{open formula} otherwise).\index{closed formula} A
term (formula) that contains no variable is called \textbf{ground}
term (ground formula).\index{ground formula} The \textbf{arity} of any
predicate function $f$ is its number of arguments, normally written as
an upper index, $f^n$, if needed.\index{arity}

The rules for constructing terms and formulas are recursive: Every
element in $\mathcal{C}$ is a term, as it is any individual variable
and also $f^n(t_1, \ldots , t_n)$, where $f^n \in \mathcal{F}$ and
$t_i$ are terms.  Also, $P \in \mathcal{P}$ is a formula\footnote{It
  is called \emph{atomic formula}.} and the application of any
propositional connective or quantifier (or both) to two or more
predicates is also a formula.

In fact, constants are formulas of arity zero so it would be
convenient to omit them and allow formulas of any arity.  Nevertheless
we will follow the traditional exposition and use the term function
when arity is at least 1.

\noindent \textbf{Example}.$\square$As an example of FOL formula, one of the
inference rules of predicate calculus is written:
\begin{equation}
  \exists x P(x) \wedge \forall x Q(x) \Rightarrow \exists x \left[ P(x) \wedge Q(x)\right]. \nonumber
\end{equation}
It reads as if there exists $x$ for which $P$ and for all $x$ $Q$,
then there exists $x$ for which $P$ and $Q$.  For another example,
let's consider the language of ordered Abelian groups.  It has one
constant $0$, one unary function $-$, one binary function $+$ and one
binary relation $\leq$.
\begin{itemize}
\item $0$, $x$, $y$ are atomic terms.
\item $+(x, y)$, $+(x, +(y, -(z)))$ are terms, usually written in
  infix notation as $x + y$, $x + (y + (-z))$.
\item $=(+(x, y), 0)$, $\leq (+(x, +(y, -(z))), +(x, y))$ are atomic
  formulas, usually written in infix notation as $x + y = 0$, $x + y -
  z \leq x + y$.
\item $\left(\forall x \exists y \leq ( +(x, y), z) \right) \wedge
  \left( \exists x =(+(x, y), 0) \right)$ is a formula, more readable
  if written as $\left( \forall x \exists y \; x + y \leq z\right)
  \wedge \left( \exists x \; x + y = 0\right)$.\proofend
\end{itemize}

\index{domain of discourse}The semantics of our language depend on the
\textbf{domain of discourse} ($D$) and on the \textbf{interpretation
  function} $I$.\index{interpretation function} The domain of
discourse (also known as universe of discourse) is the set of objects
we use the FOL to talk about and must be fixed in advance.  In the
example above, for a fixed Abelian group, the domain of discourse are
the elements of the group.

For a given domain of discourse $D$ it is necessary to define an
interpretation function $I$ which assigns meanings to the non-logical
vocabulary, i.e. maps symbols in our language onto the domain:
\begin{itemize}
\item Constants are mapped onto objects in the domain.
\item 0-ary predicates are mapped onto \textbf{true} or
  \textbf{false}, i.e. whether they are true or false in this
  interpretation.
\item N-ary predicates are mapped onto sets of n-ary ordered tuples of
  elements of the domain, i.e. those tuples of members for which the
  predicate holds (for example, a 1-ary predicate is mapped onto a
  subset of $D$).
\end{itemize}

The interpretation of a formula $f$ in our language is then given by
this morphism $I$ together with an assignment of values to any free
variables in $f$.  If $S$ is a variable assignment on $I$ then we can
write $(I,S) \models f$ to mean that $I$ satisfies $f$ under the
assignment $S$ ($f$ is true under interpretation $I$ and assignment
$S$).  Our interpretation function assigns denotations to constants in
the language, while $S$ assigns denotations to free variables.

First-order predicate logic allows variables to range over atomic
symbols in the domain but it does not allow variables to be bound to
predicate symbols, however.  A \textbf{second order logic} (such as
second order predicate logic,~\cite{logic}) does allow this, and
sentences such as $\forall P [P(2)]$ (all predicates apply to number
2) can be written.\index{second order logic, SOL}

\noindent \textbf{Example}.$\square$Starting out with formula:
\begin{equation}
  \beta(X)=\forall x, y, z \left[ \left( P(x,y) \wedge P(x,z)
      \Rightarrow y = z \right) \wedge \left( P(x,z) \wedge P(y,z)
      \Rightarrow x = y \right) \right] \nonumber
\end{equation}
which expresses injectiveness of a binary relation $P$ on its domain,
it is possible to give a characterization of bijection ($X$) between
two sets ($Y_1$, $Y_2$):
\begin{equation}
  \exists X \left[ \beta(X) \wedge \forall x \left( Y_1(x)
      \Leftrightarrow \exists y X(x,y) \right) \wedge \left(
      Y_2(x) \Leftrightarrow \exists y X(y,x) \right) \right].\nonumber
\end{equation}
The bijection $X$ is a binary relation and the sets $Y_1$ and $Y_2$
are unary relations. Hence, $Y_1(x)$ is the same as $x \in Y_1$. See
\cite{handbook}, pp. 319-320 for more details.

Another example is the least upper bound (lub) property for sets of
real numbers (every bounded, nonempty set of real numbers has a
supremum):
\begin{equation}
  \forall A \left[ \left(\exists w (w \in A) \wedge \exists z \forall
      w (w \in A \Rightarrow w \leq z) \right) \Rightarrow \exists x
    \forall y \left( \forall w \in A, (w \leq y) \Leftrightarrow x
      \leq y \right) \right]. \nonumber
\end{equation} \proofend

Second order logic (SOL) is more expressive than FOL under standard
semantics: Quantifiers range over all sets or functions of the
appropriate sort (thus, once the domain of the first order variables
is established, the meaning of the remaining quantifiers is fixed).
It is still possible to increase the order of the logic, for example
by allowing predicates to accept arguments which are themselves
predicates.

\index{monadic second order logic, MSOL}Chapter
\ref{ch:restrictionsOnRules} makes use of \textbf{monadic second order
  logic}, MSOL for short,\footnote{In the literature there are several
  equivalent contractions such as MS, MSO and M2L.} which lies in
between first order and second order logics.  Instead of allowing
quantification over n-ary predicates, MSOL quantifies 0-ary and 1-ary
predicates, i.e. individuals and subsets.  There is no restriction on
the arity of predicates.

A theorem by B\"uchi and Elgot~\cite{Buc60}\cite{Elg61} (see also
\cite{Thomas90}) states that string languages generated by MSOL
formulas correspond to regular languages (see also Sec.
\ref{sec:msolApproach}), so we have an alternative to the use of
regular expressions, appropriate to express patterns (this is one of
the reasons to make use of them in Chap.
\ref{ch:restrictionsOnRules}).\footnote{See~\cite{monadicLogic} for an
  introduction to monadic second order logic. See~\cite{GJJ95} for an
  implementation of a translator of MSOL formula into finite-state
  automaton.}  Another reason is that properties as general as
3-colorability of a graph (see~\cite{handbook}, Chap. 5 and also Sec.
\ref{sec:graphConstraintsAndApplicationConditions}) can be encoded
using MSOL so, for many purposes, it seems to be expressive enough.

\section{Category Theory}
\label{sec:categoryTheory}

Category theory was first introduced by S. Eilenberg and S. Mac Lane
in the early 1940s in connection with their studies in homology
theory (algebraic topology). See~\cite{EML45}.  The reference book in
category theory is~\cite{Mac98}. There are also several very good
surveys on this topic on the web such as
\url{http://www.cs.utwente.nl/~fokkinga/mmf92b.pdf}.

\index{category}A \textbf{category} $\mathcal{C}$ is made up of a
class\footnote{A \emph{class} is a collection of sets or other
  mathematical objects.  A class that is not a set is called a
  \textbf{proper class} and has the properties that it can not be an
  element of a set or a class and is not subject to the
  Zermelo-Fraenkel axioms, thereby avoiding some paradoxes from naive
  set theory.} of objects, a class of morphisms and a binary operation
called \emph{composition} of morphisms, $\left( Obj(\mathcal{C}),
  Hom(\mathcal{C}), \circ \right)$.\index{class} Each morphism $f$ has
a unique source object and a unique target object, $f: A \rightarrow
B$.  There are two axioms for categories:
\begin{enumerate}
\item if $f:A \rightarrow B$, $g:B \rightarrow C$ and $h:C \rightarrow
  D$ then $h \circ (g \circ f) = (h \circ g) \circ f$ (associativity).
\item $\forall X \; \exists 1_X:X \rightarrow X$ such that $\forall
  f:A \rightarrow B$ it is true that $1_B \circ f = f = f \circ 1_A$
  (existence of the identity morphism).
\end{enumerate}

\index{initial object}\index{terminal object}An object $A$ is
\textbf{initial} if and only if $\forall B \, \exists ! f:A
\rightarrow B$, and \textbf{terminal} if $\forall B \, \exists ! g:B
\rightarrow A$.  Not all categories have initial or terminal objects,
although if they exist then they are unique up to a unique
isomorphism.

\index{category!\textbf{Set}}\noindent\textbf{Example}$\square$One first
example is the category \textbf{Set}, where objects are sets and
morphisms are total functions.  Doing set theory in the categorical
language forces to express everything with function composition only
(no explicit arguments, membership, etc).

Notice that morphisms need not be functions.  For example, any
directed graph determines a category in which each node is one object
and each directed edge is a morphism. Composition is concatenation of
paths and the identity is the empty path. This category is at times
called \textbf{Path} category.

Similarly, any preordered set $(A, \leq)$ can be thought of as a
category.  Objects are in this case the elements of $A$ ($a, b \in
A$), and there is a morphism between two given elements whenever $a
\leq b$.  The identity is $a \leq a$.\footnote{These three examples
  can be found in~\cite{Gentle}.}

The empty set $\emptyset$ is the only initial object and every
singleton object (one-element set) is terminal in category
\textbf{Set}.  If as before $(A, \leq)$ is a preordered set, $A$ has
an initial object if and only if it has a smallest element, and a
terminal object if and only if $A$ has a largest element.  In the
category of graphs (to be defined soon) the null graph -- the graph
without nodes and edges -- is an initial object.  The graph with a
single node and a single edge is terminal, except in the category of
simple graphs without loops which does not have a terminal
object. \proofend

\index{multigraph}\noindent\textbf{Example}.$\square$A multigraph $G = \left(
  V, E, s, t \right)$ consists of a set $V$ of vertexes and a set $E$
of edges.  \index{source}\index{target}Functions \textbf{source} and
\textbf{target} $s,t:E \rightarrow V$ respectively return the initial
node and the final node of an edge.

A graph morphism $f:G_1 \rightarrow G_2$, with $f = \left( f_V, f_E
\right)$, consists of two functions $f_V:V_1 \rightarrow V_2$ and
$f_E: E_1 \rightarrow E_2$ such that $f_V \circ s_1 = s_2 \circ f_E$
and $f_V \circ t_1 = t_2 \circ f_E$.  Composition is defined
component-wise, i.e. given $f_1:G_1 \rightarrow G_2$ and $f_2:G_2
\rightarrow G_3$ then $f_2 \circ f_1 = \left( f_{2,V} \circ f_{1,V},
  f_{2,E} \circ f_{1,E} \right):G_1 \rightarrow G_3$.

\index{category!\textbf{Graph}}\index{category!\textbf{Graph}$^\textbf{P}$}The
category of graphs with total morphisms will be denoted \textbf{Graph}
and \textbf{Graph}$^\mathbf{P}$ if morphisms are allowed to be
partial.  \textbf{Graph}$^\mathbf{P}$ will be more interesting for
us. \proofend

\index{functor}Let $\mathcal{C}$ and $\mathcal{D}$ be two categories.
A \textbf{functor} $F: \mathcal{C} \rightarrow \mathcal{D}$ is a
mapping\footnote{Functors can be seen as morphisms between
  categories.} that associates objects in $\mathcal{C}$ with objects
in $\mathcal{D}$ (for some $X \in \mathcal{C}$, $F(X) \in
\mathcal{D}$) and morphisms in $\mathcal{C}$ with morphisms in
$\mathcal{D}$:
\begin{equation}
  f: X \rightarrow Y, f \in \mathcal{C}, F(f): F(X) \rightarrow F(Y), F(f) \in \mathcal{D}.
\end{equation}

Any functor has to keep the category structure (identities and
composition), i.e. it must satisfy the following two properties:
\begin{enumerate}
\item $\forall X \in \mathcal{C}$, $F(1_X) =
  1_{F(X)}$.
\item $\forall f:X \rightarrow Y, g:Y \rightarrow Z$ we have that $F(g
  \circ f) = F(g) \circ F(f)$.
\end{enumerate}

\noindent\textbf{Example}.$\square$The \emph{constant functor} between
categories $\mathcal{C}$ and $\mathcal{D}$ sends every object in
$\mathcal{C}$ to a fixed object in $\mathcal{D}$.  The \emph{diagonal
  functor} is defined between categories $\mathcal{C}$ and
$\mathcal{C}^\mathcal{D}$ and sends each object in $\mathcal{C}$ to
the constant functor in that
object.\footnote{$\mathcal{C}^\mathcal{D}$ is the class of all
  morphisms from $\mathcal{D}$ to $\mathcal{C}$} Let $\mathcal{C}$
denote the category of vector spaces over a fixed field, then the
\emph{tensor product} $V \otimes W$ defines a functor $\mathcal{C}
\times \mathcal{C} \rightarrow \mathcal{C}$. \proofend

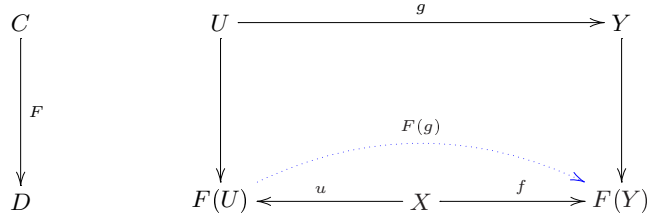
\begin{figure}[htbp]
  \begin{displaymath}
    \xymatrix{
      C \ar[dd]^F && U \ar[dd] \ar[rrrr]^g &&&& Y \ar[dd] \\
      \\
      D && F(U) \ar@{.>}@/^22pt/@[blue][rrrr]^{F(g)} && X \ar[ll]_u \ar[rr]^f && F(Y)
    }
  \end{displaymath}
  \caption{Universal Property}
  \label{fig:universalProperty}
\end{figure}

\index{universal property}All constructions that follow can be
characterized by some abstract property that demands, under some
conditions, the existence of a unique morphism, known as
\textbf{universal properties}.

One concept constantly used is that of \textbf{universal morphism},
which can be easily recognized in the rest of the section: Let $F:
\mathcal{C} \rightarrow \mathcal{D}$ be a functor and let $X \in
\mathcal{D}$, a \emph{universal morphism} from $X$ to $F$ -- where $U
\in \mathcal{C}$ and $u: X \rightarrow F(U)$ -- is the pair $(U, u)$
such that $\forall \, Y \!\! \in \mathcal{C}$ and $\forall f\!:X
\rightarrow F(Y)$, $\exists ! g: U \rightarrow Y$
satisfying:\footnote{In fact, this is a universal property for
  universal morphisms.}
\begin{equation}
  f = F(g) \circ u. \nonumber
\end{equation}
See Fig.~\ref{fig:universalProperty} where blue dotted arrows delimit
the commutative triangle $\left( u, f, F(g) \right)$.

\begin{figure}[htbp]
  \begin{displaymath}
    \xymatrix{
      & P' \ar[ddl]_{\Pi'_X} \ar[ddr]^{\Pi'_Y} \ar@{.>}[d]^u &&& N \ar[ddl]_{\gamma_X} \ar[ddr]^{\gamma_Y} &&& N \ar[ddl]_{\gamma_X} \ar[ddr]^{\gamma_Y} \ar@{.>}[d]^u \\
      & P \ar[dl]^{\Pi_X} \ar[dr]_{\Pi_Y} &&&&&& L \ar[dl]^{\delta_X} \ar[dr]_{\delta_Y} \\
      X && Y & F(X) \ar[rr]_{F(f)} && F(Y) & F(X) \ar[rr]_{F(f)} && F(Y)
    }
  \end{displaymath}
  \caption{Product, Cone and Universal Cone}
  \label{fig:catDiagrams1}
\end{figure}
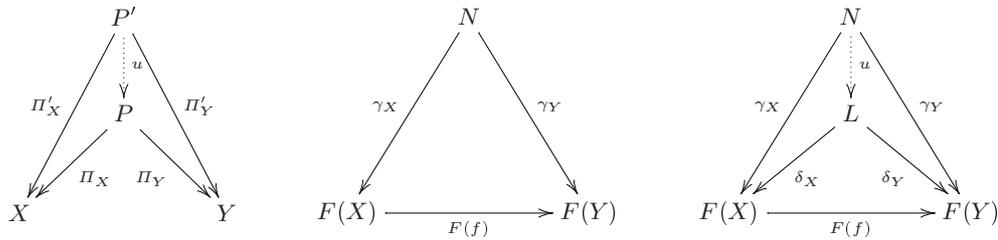

\index{categorical product}The \textbf{product} of objects $X$ and $Y$
is an object $P$ and two morphisms $\Pi_X:P \rightarrow X$ and
$\Pi_Y:P \rightarrow Y$ such that $P$ is terminal.  This definition
can be extended easily to an arbitrary collection of objects.

\index{cone}A \textbf{cone} from $N \in \mathcal{D}$ to functor $F:
\mathcal{C} \rightarrow \mathcal{D}$ is the family of morphisms
$\gamma_{X}:N \rightarrow F(X)$ such that $\forall f: X \rightarrow
Y$, $f \in \mathcal{C}$ we have $F(f) \circ \gamma_X = \gamma_Y$.

\index{limit}A \textbf{limit} is a universal cone, i.e. a cone through
which all other cones factor: A cone $(L,\delta_X)$ of a functor $F:
\mathcal{C} \rightarrow \mathcal{D}$ is a limit of that functor if and
only if for any cone $(N,\gamma_X)$ of $F$, $\exists ! u: N
\rightarrow L$ such that $\gamma_X = \delta_X \circ u$ ($L$ is
terminal).  See Fig.~\ref{fig:catDiagrams1}.

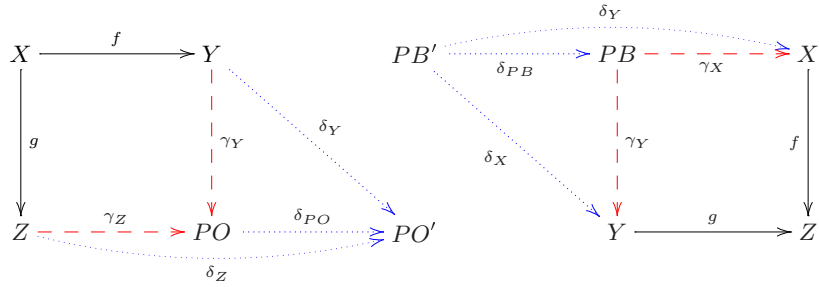
\begin{figure}[htbp]
  \centering
  \begin{displaymath}
    \xymatrix{
      &	X \ar[dd]^g \ar[rr]^f && Y \ar@{-->}@[red][dd]^{\gamma_Y} \ar@{.>}@[blue][ddrr]^{\delta_Y} && PB' \ar@{.>}@[blue][ddrr]_{\delta_X} \ar@{.>}@[blue][rr]_{\delta_{PB}} \ar@/^10pt/@{.>}@[blue][rrrr]^{\delta_Y} && PB \ar@{-->}@[red][dd]^{\gamma_Y} \ar@{-->}@[red][rr]_{\gamma_{X}} && X \ar[dd]_f \\
      \\
      &	Z \ar@{-->}@[red][rr]^{\gamma_Z} \ar@/_10pt/@{.>}@[blue][rrrr]_{\delta_Z} && PO \ar@{.>}@[blue][rr]^{\delta_{PO}} && PO' && Y \ar[rr]^g && Z 
    }
  \end{displaymath}
  \caption{Pushout and Pullback}
  \label{fig:catDiagrams2}
\end{figure}

\index{pullback}A \textbf{pullback}\footnote{Also known as
  \textbf{fibered product} or \textbf{Cartesian square}.} is the limit
of a diagram\footnote{Informally, the diagram is what appears to the
  left of Fig.~\ref{fig:catDiagrams2}. Formally, a diagram of type $I$
  -- the \emph{index} or \emph{scheme} category -- in category $C$ is
  a functor $D:I \rightarrow C$. What objects and morphisms are in $I$
  is irrelevant. Only the way in which they are related is of
  importance.} consisting of two morphisms $f : X \rightarrow Z$ and
$g : Y \rightarrow Z$ with a common codomain.

\index{coproduct}\index{cocone}\index{colimit}\index{pushout}By
reverting all arrows in previous definitions\footnote{Reverting arrows
  is at times called \emph{duality}.} we get the dual concepts:
\textbf{Coproduct}, \textbf{cocone}, \textbf{colimit} and
\textbf{pushout}.  A pushout\footnote{Also known as \textbf{fibered
    coproducts} or \textbf{fibered sums}.} is the colimit of a diagram
consisting of two morphisms $f : X \rightarrow Y$ and $g : X
\rightarrow Z$ with a common domain and can be informally interpreted
as closing the square depicted to the left of
Fig.~\ref{fig:catDiagrams2} by defining the red dashed morphisms
$\gamma_Z$ and $\gamma_Y$.  Fine blue dotted morphisms ($\delta_Y$,
$\delta_Z$ and $\delta_{PO}$) illustrate the universal property of PO
of being the initial object.  We will see in Secs.~\ref{sec:DPO} and
\ref{sec:otherCategoricalApproaches} that the basic pillars of
categorical approaches to graph transformation are the pushout and
pullback diagrams depicted in Fig.~\ref{fig:catDiagrams2}.

\begin{figure}[htbp]
  \centering
  \includegraphics[scale = 0.6]{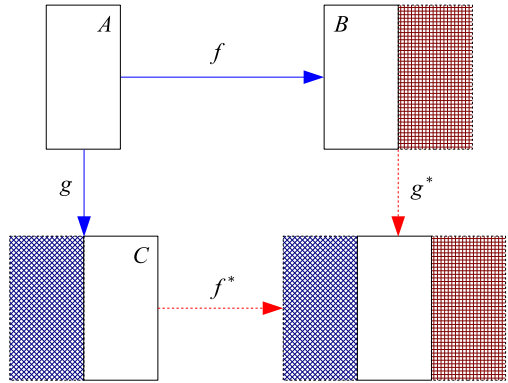}
  \caption{Pushout as Gluing of Sets}
  \label{fig:poIntuition}
\end{figure}

Pushout constructions are very important to graph transformation
systems, in particular to SPO and DPO approaches, but also used to
some extent by most of the rest of the categorical approaches.  The
intuition of a pushout between sets $A$, $B$ and $C$ as in Fig.
\ref{fig:poIntuition} is to glue sets $B$ and $C$ through set $A$ or,
in other words, put $C$ where $A$ is in $B$.

\index{pushout!complement}A \textbf{pushout complement} is a
categorical construction very similar to PO and PB.  In this case,
following the notation on the left of Fig.~\ref{fig:catDiagrams2}, $f$
and $\gamma_Y$ would be given and $g$, $\gamma_Z$ and $Z$ need to be
defined.

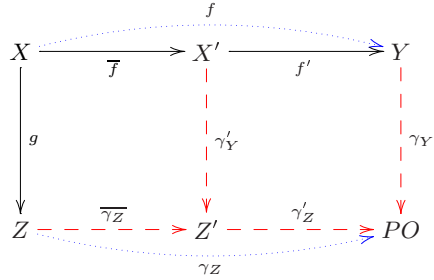
\begin{figure}[htbp]
  \begin{displaymath}
    \xymatrix{
      X \ar[dd]^g \ar[rr]_{\overline{f}} \ar@/^10pt/@{.>}@[blue][rrrr]^{f} && X' \ar@{-->}@[red][dd]^{\gamma'_Y} \ar[rr]_{f'} && Y \ar@{-->}@[red][dd]^{\gamma_Y}\\
      \\ 
      Z \ar@{-->}@[red][rr]^{\overline{\gamma_Z}} \ar@/_10pt/@{.>}@[blue][rrrr]_{\gamma_Z} && Z' \ar@{-->}@[red][rr]^{\gamma'_Z} && PO 
    }
  \end{displaymath}
  \caption{Initial Pushout}
  \label{fig:initialPO}
\end{figure}

\index{pushout!initial}Roughly speaking, an \textbf{initial pushout}
is an initial object in the ``category of pushouts''.\footnote{Initial
  pushouts are needed for the gluing condition and to define HLR
  categories. See below and also
  Sec.~\ref{subsec:adhesiveHLRCategories}.}  Suppose we have a pushout
as depicted to the left of Fig.~\ref{fig:catDiagrams2}, then it is
said to be initial over $\gamma_Y$ if for every pushout $f':X'
\rightarrow Y$ and $\gamma'_Z: Z \rightarrow PO$ (refer to
Fig.~\ref{fig:initialPO}) there exist unique morphisms $\overline{f}:X
\rightarrow X'$ and $\overline{\gamma_Z}:Z \rightarrow Z'$ such that:
\begin{enumerate}
\item $f = f' \circ \overline{f}$ and $\gamma_Z = \gamma'_Z \circ
  \overline{\gamma_Z}$.
\item The square defined by overlined morphisms $\left( f, g,
    \gamma_Y, \gamma_Z \right)$ is a pushout.
\end{enumerate}

\index{category!adhesive HLR}Now we will introduce \textbf{adhesive
  HLR categories}\footnote{HLR stands for \emph{High Level
    Replacement}.} which are very important for a general study of
graph grammars and graph transformation systems.  See
Sec.~\ref{subsec:adhesiveHLRCategories} for an introduction or refer
to~\cite{Fundamentals} for a detailed account.

\index{Van Kampen square}\textbf{Van Kampen squares} are pushout
diagrams closed in some sense under pullbacks.  Given the pushout
diagram $(p,m,p^*,m^*)$ on the floor of the cube in Fig.
\ref{fig:vanKampenSquare} and the two pullbacks $(m,g',m',l')$ and
$(p,r',p',l')$ of the back faces (depicted in dotted red) then the
front faces $(p^*, h',p'^*,g')$ and $(m^*, h',m'^*,r')$ (depicted in
dashed blue) are pullbacks if and only if the top square
$(p',m',p'^*,m'^*)$ is a pushout.  Even in category \textbf{Set} not
all pushouts are van Kampen squares, unless the pushout is defined
along a monomorphism (an injective morphism).  We say that
$(p,m,p^*,m^*)$ is defined along a monomorphism if $p$ is injective
(symmetrically, if $m$ is injective).  A category has pushouts along
monomorphisms if at least one of the given morphism is a monomorphism.

We will be interested in so-called adhesive categories.  A category
$\mathcal{C}$ is called \textbf{adhesive} if it fulfills the following
properties:
\begin{enumerate}
\item $\mathcal{C}$ has pushouts along monomorphisms.
\item $\mathcal{C}$ has pullbacks.
\item Pushouts along monomorphisms are van Kampen squares.
\end{enumerate}

There are important categories that turn out to be adhesive categories
but others are not.
\index{category!\textbf{Top}}\index{category!\textbf{Poset}}For
example, \textbf{Set} and \textbf{Graph} are adhesive categories but
\textbf{Poset} (the category of partial ordered sets) and \textbf{Top}
(topological spaces and continuous functions) are not.

\begin{figure}[htbp]
  \begin{displaymath}
    \xymatrix{
      &&&& L' \ar@{.>}@[red][dd]^{l'} |!{[rrd];[ddll]}\hole \ar@{.>}@[red][drr]^{p'} \ar@{.>}@[red][dllll]^{m'} \\
      G' \ar@{-->}@[blue][dd]^{g'} \ar@{-->}@[blue][drr]^{p'^*} &&&&&& R' \ar@{-->}@[blue][dllll]^(.3){m'^*} \ar@{-->}@[blue][dd]^{r'} \\
      && H' \ar@{-->}[dd]^{h'} && L \ar[dllll]^(.3){m} |!{[ll];[ddll]}\hole \ar[drr]^p \\
      G \ar[drr]^{p^*} &&&&&& R \ar[dllll]^(.3){m^*} \\
      && H
    }
  \end{displaymath}
  \caption{Van Kampen Square}
  \label{fig:vanKampenSquare}
\end{figure}

Axioms of adhesive categories have to be weakened because there are
important categories for graph transformation that do not fulfill them
as e.g. typed attributed graphs.  The main difference between adhesive
categories and adhesive HLR categories is that adhesive properties are
demanded for some subclass $\mathcal{M}$ of monomorphisms and not for
every monomorphism.  A category $\mathcal{C}$ with a set of morphisms
$\mathcal{M}$ is an \textbf{adhesive HLR category} if:
\begin{enumerate}
\item $\mathcal{M}$ is closed under isomorphism composition and
  decomposition ($g \circ f \in \mathcal{M}, g \in \mathcal{M}
  \Rightarrow f \in \mathcal{M}$).
\item $\mathcal{C}$ has pushouts and pullbacks along
  $\mathcal{M}$-morphisms and $\mathcal{M}$-morphisms are closed under
  pushouts and pullbacks.
\item Pushouts in $\mathcal{C}$ along $\mathcal{M}$-morphisms are van
  Kampen squares.
\end{enumerate}

Symmetrically to previous use of the term ``along'', a pushout
along an $\mathcal{M}$-morphism is a pushout where at least one of the
given morphisms is in $\mathcal{M}$.

\index{category!\textbf{PTNets}}\index{category!weak adhesive
  HLR}Among others, category \textbf{PTNets} (place/transition nets)
fails to be an adhesive HLR category so it would be nice to still
consider wider sets of graph grammars by further relaxing the
axiomatic of adhesive HLR categories.  In particular the third axiom
can be weakened if only some cubes in Fig.~\ref{fig:vanKampenSquare}
are considered for the van Kampen property.  In this case we will
speak of \textbf{weak adhesive HLR categories}:

\begin{enumerate}
\item[3'.] Pushouts in $\mathcal{C}$ along $\mathcal{M}$-morphisms are
  weak van Kampen squares, i.e. the van Kampen square property holds
  for all commutative cubes with $p \in \mathcal{M}$ and $m \in
  \mathcal{M}$ or $p \in \mathcal{M}$ and $l', r', g' \in
  \mathcal{M}$.
\end{enumerate}

Adhesive HLR categories enjoy many nice properties concerning pushout
and pullback constructions, allowing us to move forward and backward
easily inside diagrams.  Assuming all involved morphisms to be in
$\mathcal{M}$:
\begin{enumerate}
\item Pushouts along $\mathcal{M}$-morphisms are pullbacks.
\item If a pushout is the composition of two squares in which the
  second is a pullback, then in fact both squares are pushouts and
  pullbacks.
\item The symmetrical van Kampen property for pullbacks also holds
  (see Fig.~\ref{fig:vanKampenSquare}): If the top square $(G', H',
  R', L')$ is a pullback and the front squares $(G', G, H, H')$ and
  $(H', H, R, R')$ are pushouts, then the bottom $(G, H, R, L)$ is a
  pullback if and only if the back faces $(G', G, L, L')$ and $(L', L,
  R, R')$ are pushouts.
\item Pushout complements are unique up to isomorphisms.
\end{enumerate}

It is necessary to be cautious when porting concepts to (weak)
adhesive categories as morphisms involved in the definitions and
theorems have to belong to the set of morphisms $\mathcal{M}$.

\section{Graph Theory}
\label{sec:graphTheory}

In this section simple digraphs are defined, which can be represented
as Boolean matrices. Besides, basic operations on these matrices are
introduced. They will be used in later sections to characterize graph
transformation rules.  Also, compatibility for a graph\footnote{See
  Definition~\ref{def:compatibilityDefinition}.} -- an adjacency
matrix and a vector of nodes -- is defined and studied. This paves the
way to the notion of compatibility of grammar rules\footnote{See
  Definition~\ref{def:prodCompatibility}.} and of
sequence\footnote{See Sec.~\ref{sec:compositionAndCompatibility}.} of
productions.

Graph theory is considered to start with Euler's paper on the seven
bridges of K\"onisberg in 1736.  Since then, there has been an intense
research in the field by, among others, Cayley, Silvester, Tait,
Ramsey, Erd\"os, Szemer\'edy and many more.  Nowadays graph theory is
applied to a wide range of areas in different disciplines in both
science and engineering, such as computer science, chemistry, physics,
topology, and many more.  Among its main branches we can cite extremal
graph theory, geometric graph theory, algebraic graph theory,
probabilistic (also known as random) graph theory and topological
graph theory.  We will just use some basic facts from algebraic graph
theory.

\begin{figure}[htbp]
  \centering
  \includegraphics[scale = 0.6]{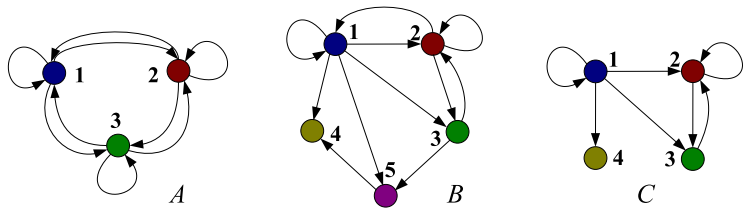}
  \caption{Three, Four and Five Nodes Simple Digraphs}
  \label{fig:CompleteThreeNodesSimpleDigraph}
\end{figure}

\index{simple!digraph}The category of graphs has been introduced in
Sec.~\ref{sec:categoryTheory}.  An easy way to define a \textbf{simple
  digraph} $G=\left(V,E \right)$ is as the structure that consists of
two sets, one of nodes $V=\{V_i \, \vert \,\, i \in I\}$ and one of
edges $E=\{\left(V_i, V_j\right) \in V \times V\}$ (think of arrows as
connecting nodes).\footnote{Mind the difference between this and
  having functions $s$ and $t$, see for example~\cite{Fundamentals}.}
The prefix ``di'' means that edges are directed and the term
``simple'' that at most one arrow is allowed between the same two
nodes.  For example, the complete simple digraph with three vertexes
and two examples of four and five vertexes can be found in Fig.
\ref{fig:CompleteThreeNodesSimpleDigraph}.

\index{adjacency matrix}Any simple digraph $G$ is uniquely determined
through one of its associated matrices, known as \textbf{adjacency
  matrix} \begin{math}A_G\end{math}, whose element
\begin{math}a_{ij}\end{math} is defined to be one if there exists an
arrow joining vertex \emph{i} with vertex \emph{j} and zero otherwise.
This is not the only possible characterization of graphs using
matrices.

\index{line graph}\index{incidence matrix}The \textbf{incidence
  matrix} is an $m \times n$ matrix $I^m_n$, where $m$ is the number
of nodes and $n$ the number of edges,\footnote{The tensor notation is
  explained in Sec.~\ref{sec:tensorAlgebra}.} such that $I^i_j = -1$
if edge $e_j$ leaves the node and $I^i_j = 1$ if edge $e_j$ enters the
node ($I^i_j = 0$ otherwise).  As it is possible to relate the
adjacency and incidence matrices through line graphs, we will mainly
characterize graphs through their adjacency matrices.\footnote{The
  line graph $L(G)$ of graph $G$ is a graph in which each vertex of
  $L(G)$ represents an edge of $G$ and two nodes in $L(G)$ are
  incident if the corresponding edges share an endpoint.  Incidence
  and adjacency matrices are related through the equation:
  \begin{equation}
    A(L(G)) = B(G)^t B(G) - 2 I \nonumber
  \end{equation}
  where $A(L(G))$ is the adjacency matrix of $L(G)$, $B(G)$ its
  incidence matrix and $I$ the identity matrix.}

\index{node!vector}In addition, a vector that we call \textbf{node
  vector} $V_G$ is associated to our digraph $G$, with its elements
equal to one if the corresponding node is in $G$ and zero otherwise.
$V_G$ will be necessary because we will study sequences of
productions, which probably apply to different graphs. Their adjacency
matrices will then refer to different sets of nodes. In order to
operate algebraically we will complete all matrices (refer to
Sec.~\ref{sec:completion} for completion). Node vectors are used to
distinguish which nodes belong to the graph and which ones have been
added for algebraic operation consistency. Next example illustrates
this point.

\noindent\textbf{Example}.$\square$The adjacency matrices $A^E$ and $C^E$ for
first and third graphs of Fig.
\ref{fig:CompleteThreeNodesSimpleDigraph} are:
\begin{displaymath}
  A^E = \left[
    \begin{array}{ccccc}
      \vspace{-6pt}
      1 & 1 & 1 & 0 & \vert \; 1 \\
      \vspace{-6pt}
      1 & 1 & 1 & 0 & \vert \; 2 \\
      \vspace{-6pt}
      1 & 1 & 1 & 0 & \vert \; 3 \\
      \vspace{-6pt}
      0 & 0 & 0 & 0 & \vert \; 4 \\
      \vspace{-10pt}
    \end{array} \right]
  A^N = \left[
    \begin{array}{cc}
      \vspace{-6pt}
      1 & \vert \; 1 \\
      \vspace{-6pt}
      1 & \vert \; 2 \\
      \vspace{-6pt}
      1 & \vert \; 3 \\
      \vspace{-6pt}
      0 & \vert \; 4 \\
      \vspace{-10pt}
    \end{array} \right]
  C^E = \left[
    \begin{array}{ccccc}
      \vspace{-6pt}
      1 & 1 & 1 & 1 & \vert \; 1 \\
      \vspace{-6pt}
      0 & 1 & 1 & 0 & \vert \; 2 \\
      \vspace{-6pt}
      0 & 1 & 0 & 0 & \vert \; 3 \\
      \vspace{-6pt}
      0 & 0 & 0 & 0 & \vert \; 4 \\
      \vspace{-10pt}
    \end{array} \right]
  C^N = \left[
    \begin{array}{cc}
      \vspace{-6pt}
      1 & \vert \; 1 \\
      \vspace{-6pt}
      1 & \vert \; 2 \\
      \vspace{-6pt}
      1 & \vert \; 3 \\
      \vspace{-6pt}
      1 & \vert \; 4 \\
      \vspace{-10pt}
    \end{array} \right]
\end{displaymath}
where $A^N$ and $C^N$ are the corresponding node vectors. A vertically
separated column indicates node ordering, which applies both to rows
and columns. Note that edges incident to node $4$ are considered in
matrix $A^E$.  As there is no node $4$ in $A$, corresponding elements
in the adjacency matrix are zero.  To clearly state that this node
does not belong to graph $A$ we have a zero in the fourth position of
$A^N$. \proofend

Note that simple graphs (without orientation on edges) can be studied
if we limit to the subspace of symmetric adjacency matrices.  In Sec.
\ref{sec:fromSimpleDigraphsToMultidigraphs} we study how to extend
Matrix Graph Grammars approach to consider multigraphs and
multidigraphs. The difference between a simple digraph and a
multidigraph is that simple graphs allow a maximum of one edge
connecting two nodes in each direction, while a multidigraph allows a
finite number of them.

In the literature, depending mainly on the book, there is some
confusion with terminology.  At times, the term graph applies to
multigraphs while other times graph refers to simple graphs (also
known as \emph{relational graphs}).  Whenever found in this book, and
unless otherwise stated, the term \emph{graph} should be understood as
simple digraph.

The basic Boolean operations on graphs are defined component-wise on
their adjacency matrices.  Let $G$ and $H$ be two graphs with
adjacency matrices $\left( g^i_j \right)$ and $\left( h^i_j \right)$,
$i,j \in \{ 1, \ldots n\}$, then:
\begin{equation}
  G \vee H = \left( g^i_j \vee h^i_j \right) \quad G \wedge H = \left(
    g^i_j \wedge h^i_j \right) \quad \overline{G} = \left(
    \overline{g^i_j} \right). \nonumber
\end{equation}

Similarly to ordinary matrix product based on addition and
multiplication by scalars, there is a natural definition for a Boolean
product with the same structure but using Boolean operations
\textbf{and} and \textbf{or}.

\newtheorem{matrixproduct}{Definition}[section]
\begin{matrixproduct}[Boolean Matrix Product]
  \index{Boolean matrix product}For digraphs $G$ and $H$, let $M_G =
  \left( g^i_j\right)_{i,j \in \{1,\ldots, n\}}$ and $M_H = \left(
    h^i_j\right)_{i,j \in \{1,\ldots, n\}}$ be their respective
  adjacency matrices.  The Boolean product is an adjacency matrix
  again whose elements are defined by:
  \begin{equation}
    \left( M_G \odot M_H \right)^i_j = \bigvee_{k=1}^n \left( g^i_k
      \wedge h^k_j \right).
  \end{equation}
\end{matrixproduct}

Element $(i,j)$ in the Boolean product matrix is one if there exists
an edge joining node $i$ in digraph $G$ with some node $k$ in the same
digraph and another edge in digraph $H$ starting in $k$ and ending in
$j$.  The value will be zero otherwise.

If for example we want to check whether node $j$ is reachable starting
in node $i$ in $n$ steps or less, we may calculate $\bigvee_{k=1}^n
A^{(k)}$, where $A^{(k)} = A \odot \stackrel{k)}{\cdots} \odot A$, and
see if element $(i,j)$ is one.\footnote{In order to distinguish when
  we are using the standard or Boolean product, in the latter
  exponents will be enclosed between brackets.}  We will consider
square matrices only as every node can be either initial or terminal
for any edge.

\index{tensor!product!for graphs}Another useful product operation that
can be defined for two simple digraphs $G_1$ and $G_2$ is its
\textbf{tensor product} (defined in Sec.~\ref{sec:tensorAlgebra}) $G =
G_1 \otimes G_2$:
\begin{enumerate}
\item The nodes set is the Cartesian product $V(G) = V(G_1) \times
  V(G_2)$.
\item Two vertices's $u_1 \otimes u_2$ and $v_1 \otimes v_2$ are
  adjacent if and only if $u_1$ is adjacent to $v_1$ in $G_1$ and
  $u_2$ is adjacent to $v_2$ in $G_2$.
\end{enumerate}

In Sec.~\ref{sec:tensorAlgebra} we will see that the adjacency matrix
of $G$ coincide with the tensor product of the adjacency matrices of
$G_1$ and $G_2$.

Definition~\ref{def:vectorNorm},
Proposition~\ref{prop:compatibilityUsingNorm} and the introduction
above of the nodes vector is not standard in graph theory (in fact, as
far as we know, we are introducing them).  The decision of including
them in this introductory section is because they are simple results
very close with what one understands as ``basics'' of a theory.

Given an adjacency matrix and a vector of nodes, a natural question is
whether they define a simple digraph or not.

\newtheorem{CompDef}[matrixproduct]{Definition}
\begin{CompDef}[Compatibility]\label{def:compatibilityDefinition}
  \index{compatibility!graph}A Boolean matrix $M$ and a vector of
  nodes $N$ are \emph{compatible} if they define a simple digraph: No
  edge is incident to any node that does not belong to the digraph.
\end{CompDef}

\index{dangling!edge}An edge incident to some node which does not
belong to the graph (has a zero in the corresponding position of the
nodes vector) is called a \textbf{dangling edge}.

\index{dangling!condition}In the DPO/SPO approaches, this condition is
checked when building a direct derivation, known as \emph{dangling
  condition}.  The idea behind it is to obtain a closed set of
entities, i.e. deletion of nodes outputs a digraph again (every edge
is incident to some node).  Proposition~\ref{prop:compatibilityUsingNorm} below provides a criteria for
testing compatibility for simple
digraphs.

\newtheorem{vectornorm}[matrixproduct]{Definition}
\begin{vectornorm}[Norm of a Boolean Vector]\label{def:vectorNorm}
  \index{norm!of Boolean vector}Let $N = \left( v_1,\ldots,v_n\right)$
  be a Boolean vector.  Its norm $\left\| \, \cdot \, \right\| _1$ is
  given by:
  \begin{equation}
    \left\| N \right\|_1 = \bigvee_{i=1}^n v_i.
  \end{equation}
\end{vectornorm}

\newtheorem{compatibility}[matrixproduct]{Proposition}
\begin{compatibility}\label{prop:compatibilityUsingNorm}
  A pair $\left(M,N\right)$, where $M$ is an adjacency matrix and $N$
  a vector of nodes, is compatible if and only if
  \begin{equation}\label{eq:compeq}
    \left\| \left( M \vee M^t \right) \odot \overline{N}\right\|_1 = 0
  \end{equation}
  where \emph{t} denotes transposition.
\end{compatibility}

\noindent \emph{Proof}\\*
$\square$In an adjacency matrix, row $i$ represents outgoing edges
from vertex $i$, while column $j$ are incoming edges to vertex $j$.
Moreover, $ \left( M \right)_{ik} \wedge \left( \overline{N} \right)
_{k} = 1$ if and only if $ \left( M \right)_{ik} = 1$ and $\left( N
\right) _{k} = 0$, and thus the $i$-th element of vector $M \odot
\overline N$ is one if and only if there is a dangling edge in row
number $i$.  We have just considered outgoing edges; for incoming ones
we have a very similar term: $M^t \odot \overline N$.  To finish the
sufficient part of the proof -- necessity is almost straightforward --
we \textbf{or} both terms and take norms to detect if there is a
$1$.\proofend

\noindent\textbf{Remark}.$\square$We have used in the proof of Proposition~\ref{prop:compatibilityUsingNorm} distribution of $\odot$ and $\vee$,
$\left( M_1 \vee M_2 \right) \odot M_3 = \left( M_1 \odot M_3 \right)
\vee \left( M_2 \odot M_3 \right)$.  In addition, we also have the
distributive law on the left, i.e. $M_3 \odot \left( M_1 \vee M_2
\right) = \left( M_3 \odot M_1 \right) \vee \left( M_3 \odot M_2
\right)$.  Besides, it will be stated without proof that $\left\| \,
  \omega_1 \vee \omega_2 \right\|_1 = \left\| \, \omega_1 \right\|_1
\vee \left\| \, \omega_2 \right\|_1$. \proofend

In Chap.~\ref{ch:matching} we will deal with matching, i.e. finding
the left hand side of a graph grammar rule in the initial state (host
graph). A matching algorithm is not proposed; our approach assumes
that such algorithm is given.  This is closely related to the well
known graph-subgraph isomorphism problem (\textbf{SI}) which is an
\textbf{NP}-complete decision problem if the number of nodes in the
subgraph is strictly smaller than the number of nodes in the graph. We
will brush over complexity theory in
Chap.~\ref{sec:longTermResearchProgram}.

\section{Tensor Algebra}
\label{sec:tensorAlgebra}

Throughout the book, quantities that can be represented by a letter
with subscripts or superscripts attached\footnote{$A^i_{jk}$ for
  example.} will be used, together with some algebraic structure
(tensorial structure).  This section is devoted to a quick
introduction to this topic.  Two very good references are
\cite{Heinbockel} (with relations to physics) and the classic
book~\cite{Sok51}.

\index{tensor}A \textbf{tensor} is a multilinear application between
vector spaces.  It is at times interesting to stay at a more abstract
level and think of a tensor as a system that fulfills certain
notational properties.  Systems can be heterogeneous when there are
different types of elements, but we will only consider homogeneous
systems.  Therefore we will speak of systems or tensors, it does not
matter which.

\index{order}\index{rank}\index{valence}The \textbf{rank}\footnote{The
  terms \textbf{order} and \textbf{valence} are commonly used as
  synonyms.} of a system (tensor) is the number of indexes it has,
taking into account whether they are superscripts or subscripts.  For
example, $A^{i}_{jk}$ is $\ppFrac{1}{2}$-valent or of rank (1,2).
Subscripts or superscripts are referred to as indexes or suffixes.

Algebraic operations of addition and subtraction apply to systems of
the same type and rank.  They are defined component-wise, e.g.
$C^i_{jk} = A^i_{jk} + B^i_{jk}$, provided that some additive
structure is defined on elements of the system.  We do not follow the
Einstein summation convention, which states that when an index appears
twice, one in an upper and one in a lower position, then they are
summed up over all its possible values.

\index{tensor!product}\index{outer product}The product is obtained
multiplying each component of the first system with each component of
the second system, e.g. $C^{imnl}_j = A^i_j \otimes B^{mnl}$.  Such a
product is called \textbf{outer product} or \textbf{tensor product}.
The rank of the result is the sum of the ranks of the factors and
inherits all the indexes of its factors.  All linear relations are
satisfied, i.e. for $v_1, v_2 \in V$, $w \in W$ and $v \otimes w \in V
\otimes W$ the following identities are fulfilled:

\begin{enumerate}
\item $(v_1 + v_2)\otimes w = v_1 \otimes w + v_2\otimes w$.
\item $c v \otimes w = v \otimes c w = c(v \otimes w)$.
\end{enumerate}

To categorically characterize tensor products note that there is a
natural isomorphism between all bilinear maps from $E \times F$ to $G$
and all linear maps from $E \otimes F$ to $G$.  $E \otimes F$ has all
and only the relations that are necessary to ensure that a
homomorphism from $E\otimes F$ to $G$ will be linear (this is a
universal property).  For vector spaces this is quite straightforward,
but in the case of $R$-modules (modules over a ring $R$) this is
normally accomplished by taking the quotient with respect to
appropriate submodules.

\index{Kronecker product}\noindent\textbf{Example}.$\square$The
\textbf{Kronecker product} is a special case of tensor product that we
will use in Chap.~\ref{ch:reachability}.  Given matrices $A = \left(
  a^{i_1}_{j_1} \right)_{m \times n}$ and $B = \left( b^{i_2}_{j_2}
\right)_{p \times q}$, it is defined to be $C = A \otimes B =
(c^i_j)_{mp \times nq}$ where
\begin{equation}
  c^i_j = a^{i_1}_{j_1} \cdot b^{i_2}_{j_2}
\end{equation}
being $i = (i_1 - 1)n + i_2$ and $j = (j_1 - 1)m + j_2$.  The notation
$A = \left( a^i_j \right)_{m \times n}$ denotes a matrix with $m$ rows
and $n$ columns, i.e. $i \in \{ 1, \ldots, m\}$ and $j \in \{ 1,
\ldots, n\}$.  As an example:
\begin{displaymath}
  A = \left[
    \begin{array}{cc}
      a^1_1 & a^1_2
    \end{array} \right]_{1 \times 2}
  B = \left[
    \begin{array}{cc}
      b^1_1 & b^1_2 \\
      b^2_1 & b^2_2
    \end{array} \right]_{2\times 2}
  C = A \otimes B = \left[
    \begin{array}{cccc}
      a^1_1 b^1_1 & a^1_1 b^1_2 & a^1_2 b^1_1 & a^1_2 b^1_2 \\
      a^1_1 b^2_1 & a^1_1 b^2_2 & a^1_2 b^2_1 & a^1_2 b^2_2 
    \end{array} \right]_{2\times 4}
\end{displaymath}

Note that the Kronecker product of the adjacency matrices of two
graphs is the adjacency matrix of the tensor product graph (see Sec.
\ref{sec:graphTheory} for its definition). \proofend

\index{contraction}The operation of \textbf{contraction} happens when
an upper and a lower indexes are set equal and summed up, e.g.
$C^{imnl}_j \longmapsto C^{mnl} = \sum_{j=1}^N C^{jmnl}_j = \sum_{i=j}
C^{imnl}_j$.  For example, the standard multiplication of a vector by
a matrix is a contraction: Consider matrix $A^i_j$ and vector $v^k$
with $i, j, k \in \{1,\ldots , n\}$, then matrix multiplication can be
performed by making $j$ and $k$ equal and summing up, $u^i =
\sum_{j=1}^n A^i_j v^j$.

\index{inner product}The \textbf{inner product} is represented by
$\left\langle \, \cdot \, , \cdot \, \right\rangle$ and is obtained in
two steps:
\begin{enumerate}
\item Take the outer product of the tensors.
\item Perform a contraction on two of its indexes.
\end{enumerate}
In Sec.~\ref{sec:functionalAnalysis} we will extend this notation to
cope with graph grammar rules representation.

\index{contravariance}\index{covariance}Upper indexes are called
\textbf{contravariant} and lower indexes \textbf{covariant}.
Contravariance is associated to the tangent bundle (tangent space) of
a variety and corresponds, so to speak, to columns.  Covariance is the
dual notion and is associated to the cotangent bundle (normal space)
and rows.  As an example, if we have a vector $V$ in a three
dimensional space with basis $\left\lbrace E_1, E_2, E_3
\right\rbrace$ then it can be represented in the form $A = a^1 E_1 +
a^2 E_2 + a^3 E_3$.  Components $a^i$ can be calculated via $a^i =
\left\langle A, E^i \right\rangle$ with $\left\langle E^i, E_j
\right\rangle = \delta^i_j$, where the Kronecker delta function is 1
if $i=j$ and zero if $i \neq j$.  Basis $\left\lbrace E^i
\right\rbrace$ and $\left\lbrace E_i \right\rbrace$ are called
\textbf{reciprocal} or \textbf{dual}.

\index{Kronecker delta}We will not enter the representation of
$\delta$ in integral form or the relation with the Dirac delta
function, of fundamental importance in distribution theory, functional
analysis (see Sec.~\ref{sec:functionalAnalysis}) and quantum
mechanics.  The Kronecker delta can be generalized to an
$\ppFrac{n}{n}$-valent tensor:
\begin{equation}
  \delta^{j_1, \ldots, j_n}_{i_1, \ldots, i_n} = \prod_{k=1}^n \delta_{i_k j_k}.
\end{equation}

\index{Levi-Civita symbol}\index{metric tensor}Besides the Kronecker
delta, there are other very useful tensors such as the \textbf{metric
  tensor}, which can be informally introduced by $g^{ij} E_i = E^j$
and $g_{ij} E^j = E_i$.  Note that $g$ raises or lowers indexes, thus
moving from covariance to contravariance and vice versa.  Related to
$\delta$ and to group theory is the important \textbf{Levi-Civita
  symbol}:
\begin{equation}
  \varepsilon_{\sigma} = \left\{ \begin{array}{ll}
      +1 & \qquad\textrm{if $\sigma$ is an even permutation.} \\
      -1 & \qquad\textrm{if $\sigma$ is an odd permutation.} \\
      0 & \qquad\textrm{otherwise.}\\
    \end{array}\right.
\end{equation}
where $\sigma = (i_1 \; \ldots \; i_n)$ is a permutation of $(1 \;
\ldots \; n)$.  See Sec.~\ref{sec:groupTheory} for definitions and
further results.  Symbols $\delta$ and $\varepsilon$ can be related
through matrix $A = (a_{kl}) = \delta_{i_k j_l}$ and:
\begin{equation}
  \varepsilon_{i_1 \ldots} \varepsilon_{j_1 \ldots} = det(A).
\end{equation}

\section{Functional Analysis}
\label{sec:functionalAnalysis}

Functional analysis is a branch of mathematics focused on the study of
functions -- operators -- in infinite dimensional spaces (although its
results also apply to finite dimensional spaces).  Besides the
algebraic structure (normally a vector space but at times groups) some
other ingredients are normally added such as an inner product (Hilbert
spaces), a norm (Banach spaces) a metric (metric spaces) or just a
topology (topological vector spaces).

\index{operator}An \textbf{operator} is just a function, but the term
is normally employed to call attention to some special aspect.
Examples of operators in mathematics are differential and integral
operators, linear operators (linear transformations), Fourier
transform, etc.

In this book we will call operators to functions that act on functions
with image a function.  Operators will be used, e.g. in Chap.
\ref{ch:matching} to modify productions in order to get a production
or a sequence of productions.

We will need to change productions as commented above and our
inspiration comes from operator theory and functional analysis, but we
would like to put it forward in a quantum mechanics style.  So,
although it will not be used as it is, we will give a very brief
introduction to Hilbert and Banach spaces, bra-ket notation and
duality.

\index{Hilbert space}\index{inner product}\index{scalar product}A
\textbf{Hilbert space} $\mathcal{H}$ is a vector space, complete with
respect to Cauchy sequences over a field $K$ (every Cauchy sequence
has a limit in $\mathcal{H}$), plus a scalar (or inner)
product.\footnote{Inner product $\left\langle \cdot , \cdot
  \right\rangle : \mathcal{H} \times \mathcal{H} \rightarrow K$ axioms
  are:
  \begin{enumerate}
  \item $\forall x, y \in \mathcal{H}, \left\langle x, y \right\rangle
    = \overline{\left\langle y, x \right\rangle}$.
  \item $\forall a,b \in K$, $\forall x , y \in \mathcal{H}$,
    $\left\langle a x, b y \right\rangle = a \left\langle x, y
    \right\rangle + \overline{b}\left\langle x, y \right\rangle$.
  \item $\forall x \in \mathcal{H}$, $\left\langle x, x \right\rangle
    \geq 0$ and $\left\langle x, x \right\rangle = 0$ if and only if
    $x = 0$.
  \end{enumerate}} Completeness ensures that the limit of a convergent
sequence is in the space, facilitating several definitions from
analysis (note that a Hilbert space can be infinite-dimensional).  The
inner product -- $\left\langle u, v\right\rangle$, $u, v \in
\mathcal{H}$ -- equips the structure with the notions of distance and
angle (in particular perpendicularity).  From a geometric point of
view, the scalar product can be interpreted as a projection whereas
analytically it can be seen as an integral.

\index{norm}The inner product gives raise to a
\textbf{norm}\footnote{Norm $\| \cdot \| : \mathcal{B} \rightarrow K$
  axioms are:
  \begin{enumerate}
  \item $\forall x, y \in \mathcal{B}, \| x + y \| \leq \|x\|+\|y\|$.
  \item $\forall a \in K$, $\forall x \in \mathcal{B}$, $\| a x \| =
    |a| \cdot \|x\|$.
  \item $\forall x \in \mathcal{B}$, $\| x \| \geq 0$ and $\|x\| = 0$
    if and only if $x = 0$.
  \end{enumerate}} $\left\| \, \cdot \right\|$ via $\left\| x
\right\|^2 = \left\langle x, x \right\rangle$, $\forall x \in
\mathcal{H}$.  Any norm can be interpreted as a measure of the size of
elements in the vector space.  Every inner product defines a norm but,
in general, the opposite is not true, i.e. norm is a weaker concept
than scalar product.

\index{dual space}The relationship between row and column vectors can
be generalized from an abstract point of view through \textbf{dual
  spaces}.  The dual space $\mathcal{H}^*$ of a Hilbert space
$\mathcal{H}$ over the field $K$ has as elements $x^* \in
\mathcal{H}^*$, linear applications with domain (initial set)
$\mathcal{H}$ and codomain (image) the underlying field $K$, $x^*:
\mathcal{H} \rightarrow K$.

The dual space becomes a vector space defining the addition $\forall
x_1^*, x_2^* \in \mathcal{H}^*$, $x \in \mathcal{H}$ by $(x_1^* +
x_2^*)(x) = x_1^*(x) + x_2^*(x)$ and the scalar product $\forall k \in
K$ by $k x^*(x) = x^*(kx)$.  Using tensor algebra terminology (see
Sec.~\ref{sec:tensorAlgebra}) elements of $\mathcal{H}$ are called
covariant and elements of $\mathcal{H}^*$ contravariant.  Note how in
$\left\langle x, y \right\rangle$ it is possible to think of $x$ as an
element of the vector space and $y$ as an element of the dual space.

Any Hilbert space is isomorphic (or anti-isomorphic) to its dual
space, $\mathcal{H} \cong \mathcal{H}^*$, which is the content of the
\emph{Riesz representation theorem}.  This is particularly relevant to
us because it is a justification of the Dirac bra-ket notation that we
will also use.

\index{Riesz representation theorem}The Riesz representation theorem
can be stated in the following terms: Let $\mathcal{H}$ be a Hilbert
space, $\mathcal{H}^*$ its dual and define $\phi_x (y)= \left\langle
  x, y \right\rangle$, $\phi \in \mathcal{H}^*$.  Then, the mapping
$\Phi:\mathcal{H} \rightarrow \mathcal{H}^*$ such that $x \mapsto
\phi_x$ is an isometric isomorphism.  This means that $\Phi$ is a
bijection and that $\left\| x \right\| = \left\| \phi_x \right\|$.

We will very briefly introduce Banach spaces to illustrate how notions
and ideas from Hilbert spaces, specially notation, is extended in a
more or less natural way.

\index{Banach space}A complete\footnote{Complete in the same sense as
  for Hilbert spaces.} vector space plus a norm is known as a
\textbf{Banach space}, $\mathcal{B}$.  Associated to any Banach space
there exists its dual space, $\mathcal{B}^*$, defined as before.
Contrary to Hilbert spaces, a Banach space is not isometrically
isomorphic to its dual space.

\index{metric}\index{distance}It is possible to define a
\textbf{distance} (also called \textbf{metric}) out of a norm: $d(x,
y) = \left\| x- y \right\|$.  Even though there is no such geometrical
intuition of projection nor angles, it is still possible to use the
notation we are interested in.  Given $x \in \mathcal{B}, x^* \in
\mathcal{B}^*$, instead of writing $x^* \left( x \right)$ (the result
is an element of $\mathcal{K}$) at times $\left\langle x, x^*
\right\rangle$ is preferred.  Although the space and its dual
\emph{live at different levels}, we would like to recover this
geometrical intuition of \emph{projection}.  In some (very nice)
sense, the result of $x^* \left( x \right)$ is the projection of $x$
over $x^*$.

The same applies for an operator $T$ acting on a Banach space
$\mathcal{B}$, $T: \mathcal{B} \rightarrow \mathcal{B}$.  Suppose $f,
g \in \mathcal{B}$, then $g = T \left( f \right) \equiv \left\langle
  f, T \right\rangle$.  This is closer to our situation, so the
application of a production\footnote{See Sec.~\ref{sec:characterizationAndBasicConcepts} for definitions.} can be
written
\begin{equation}
  R=\left\langle L, p \right\rangle.
\end{equation}
The left part is sometimes called \emph{bra} and the right part
\emph{ket}: $\left\langle bra, ket \right\rangle$.

\index{adjoint operator}Besides dual elements, the adjoint of an
operator is also represented using asterisks.  In our case, the
adjoint operator of $T$, represented by $T^*$, is formally defined by
the identity:
\begin{equation}
  \left\langle \, L, T p \, \right\rangle = \left\langle \, T^* L, p \, \right\rangle.
  \label{eq:adjointOperatorDef}
\end{equation}

Roughly speaking, $T$ is an operator (a function) that modifies a
production, being its output a production again, so the left hand side
in (\ref{eq:adjointOperatorDef}) is equivalent to $T \left( p \right)
\left( L \right),$ and the right hand side is just $p \left( T^* L
\right)$.  Note that $T \left( p \right)$ is a production and $T^* L$
is a simple digraph.

In quantum mechanics the possible states of a quantum mechanical
system are represented by unit vectors -- \emph{state vectors} -- in a
Hilbert space $\mathcal{H}$ or \emph{state space} (equivalently,
points in a projective Hilbert space).  Each observable -- property of
the system -- is defined by a linear operator acting on the elements
of the state space.  Each eigenstate of an observable corresponds to
an eigenvector of the operator and the eigenvalue to the value of the
observable in that eigenstate.  An interpretation of $\left\langle
  \psi | \, \phi \right\rangle$ is the probability amplitude for the
state $\psi$ to collapse into the state $\phi$, i.e. the projection of
$\psi$ over $\phi$.  In this case, the notation can be generalized to
metric spaces, topological vector spaces and even vector spaces
without any topology (close to our case as we will deal with graphs
without introducing notions such as metrics, scalar products,
etc).  Two recommended references are~\cite{Kauffman} and
\cite{Penrose}.

This digression on quantum mechanics is justified because along the
present contribution we would like to think in graph grammars as
having a static definition which provokes a dynamic behaviour and the
duality between state and observable.  Besides, the use of the
notation, we would like to keep some ``physical'' (mechanics)
intuition whenever possible.

\section{Group Theory}
\label{sec:groupTheory}

One way to introduce group theory is to define it as the part of
mathematics that study those structures for which the equation $a
\cdot x = b$ has a unique solution.  There is a very nice definition
due to James Newman~\cite{Newman} that I'd like to quote:
\begin{quote}
  The theory of groups is a branch of mathematics in which one does
  something to something and then compares the results with the result
  of doing the same thing to something else, or something else to the
  same thing.
\end{quote}

\index{group}We will be interested in groups, mainly in its notation
and basic results, when dealing with sequentialization in
Chaps.~\ref{ch:mggFundamentals1}~and~\ref{ch:sequentializationAndParallelism}.
A \textbf{group} $G$ is a set together with an operation $\left( G,
  \cdot \right)$ that satisfies the following axioms:

\begin{enumerate}
\item \emph{Closure:} $\forall a, b \in G$, $a \cdot b \in G$.
\item \emph{Associativity:} $\forall a, b, c \in G$, $a \cdot ( b
  \cdot c ) = ( a \cdot b ) \cdot c$.
\item \emph{Identity element:} $\exists e \in G$ such that $a \cdot e
  = e \cdot a = a$.
\item \emph{Inverse element:} $\forall a \in G$ $\exists b \in G$ such
  that $a \cdot b = e = b \cdot a$.
\end{enumerate}

\index{abelian group}Actually, the third and fourth axioms can be
weakened as only one identity per axiom should suffice, but we think
it is worth stressing the fact that if they exist then they work on
both sides.  Normally, the inverse element of $a$ is written $a^{-1}$.
At times the identity element is represented by $1_G$ or $0_G$,
depending on the notation (Abelian or non-Abelian).  A group is called
\textbf{Abelian} or \textbf{commutative} if $\forall a, b \in G$, $a
\cdot b = b \cdot a$.

\index{subgroup}A group $S$ inside a group $G$ is called a
\textbf{subgroup}.  If this is the case, we need $S$ to be closed
under the group operation, it also must have the identity element $e$
and every element in $S$ must have an inverse in $S$.  If $S \subset
G$ and $\forall a, b \in S$ we have that $a \cdot b^{-1} \in S$ then
$S$ is a subgroup.  \emph{Lagrange's theorem} states that the order of
a subgroup (number of elements) necessarily divides the order of the
group.

\index{permutation}We are almost exclusively interested in groups of
permutations: For a given sorted set, a change of order is called a
\textbf{permutation}.  This does not reduce the scope because, by
\emph{Cayley's theorem}, every group is isomorphic to some group of
permutations.

\index{transposition}\index{transposition!even}\index{transposition!odd}\index{parity}A
\textbf{transposition} is a permutation that exchanges the position of
two elements whilst leaving all other objects unmoved.  It is known
that any permutation is equivalent to a product of transpositions.
Furthermore, if a permutation can result from an odd number of
transpositions then it can not result from and even number of
permutations, and vice versa.  A permutation is \textbf{even} if it can
be produced by an even number of exchanges and \textbf{odd} in the
other case.  This is called \textbf{parity}.

\index{signature}The \textbf{signature} of a permutation $\sigma$,
$sgn(\sigma)$, is $+1$ if the permutation is even and $-1$ if it is
odd.  This is the \emph{Levi-Civita symbol} as introduced in Sec.
\ref{sec:tensorAlgebra} if it is extended for non-injective maps with
value zero.

\index{cycle}Any permutation can be decomposed into cycles.  A
\textbf{cycle} is a closed chain inside a permutation (so it is a
permutation itself) which enjoys some nice properties among which we
highlight:

\begin{itemize}
\item Cycles inside a permutation can be chosen to be disjoint.
\item Disjoint cycles commute.
\end{itemize}

Any permutation can be written as a two row matrix where the first row
represents the original ordering of elements and the second the order
once the permutation is applied.

\noindent\textbf{Example}.$\square$The permutation $\sigma$ can be decomposed
into the product of three cycles:
\begin{displaymath}
  \sigma = 
  \left[ \begin{array}{cccccccc}
      1 & 2 & 3 & 4 & 5 & 6 & 7 & 8 \\
      3 & 5 & 7 & 8 & 2 & 4 & 1 & 6
    \end{array} \right] = (1 \; 3 \; 7)(2 \; 5)(4 \; 8 \; 6).
\end{displaymath}

Note that this decomposition is not unique because any decomposition
into transpositions would do (and there are infinitely
many). \proofend

If the permutation turns out to be a cycle, then a clearer notation
can be used: Write in a row, in order, the following element in the
permutation.  In the example above we begin with 1 and note that 1
goes to 3, which goes to 7, which goes back to 1 and hence it is
written $(1 \; 3 \; 7)$.

A cycle with an even number of elements is an odd permutation and a
cycle with an odd number of elements is an even permutation.  In
practice, in order to determine whether a given permutation is even or
odd, one writes the permutation as a product of disjoint cycles: The
permutation is odd if and only if this factorization contains an odd
number of even-length cycles.

\section{Summary and Conclusions}
\label{sec:summaryAndConclusions1}

In this chapter we have quickly reviewed some basic facts of
mathematics that will be used throughout the rest of the book: The
basics of first order, second order and monadic second order logics,
some constructions of category theory such as pushouts and pullbacks
together with the introduction of some categories, graph theory basic
definitions and compatibility, tensor algebra and functional analysis
notations and some basic group theory, paying some attention to
permutations.

Internet is full of very good web pages introducing these branches of
mathematics with deeper explanations and plenty of examples.  It is
not possible to give an exhaustive list of all web pages visited to
make this chapter.  Nevertheless, I would like to highlight the very
good job being performed by the community at
\url{http://planetmath.org/} and \url{http://www.wikipedia.org/}.

Next chapter summarizes current approaches to graph grammars and graph
transformation systems, so it is still introductory. We will put our
hands on Matrix Graph Grammars in Chap.~\ref{ch:mggFundamentals1}.
\chapter{Graph Grammars Approaches}
\label{ch:graphGrammarsApproaches}

Before moving to Matrix Graph Grammars it is necessary to take a look
at other approaches to graph transformation to ``get the taste'',
which is the aim of this chapter.  We will see the basic foundations
leaving comparisons of more advanced topics (like application
conditions) to sporadic remarks in future chapters.

Sections~\ref{sec:DPO} and~\ref{sec:otherCategoricalApproaches} are
devoted to categorical approaches, probably the most developed
formalizations of graph grammars.
On the theoretical side, very nice ideas have put at our disposal the
possibility of using category theory and its generalization power to
study graph grammars, but even more so, a big effort has been
undertaken in order to fill the gap between category theory and
practice with tools such as AGG (see~\cite{Fundamentals}).  Please,
refer to~\cite{Agrawal} for a detailed discussion and comparison of
tools.

In Secs.~\ref{sec:nodeReplacement} and~\ref{sec:hyperedgeReplacement}
two completely different formalisms to the categorical approach are
summarized, at times called \emph{set-theoretic} or even
\emph{algorithmic} approaches.  They are in some sense closer to
implementation than those using category theory. There has been a lot
of research in these two essential approaches so unfortunately we will
just scratch the surface.

Interestingly, it is possible to study graph transformation using
logics, providing us with all powerful methods from this branch of
mathematics, monadic second order logics in particular.  We will brush
over this brilliant approach in Sec.~\ref{sec:msolApproach}.

To finish this review we will briefly touch on the very interesting
\emph{relation-algebraic} approach in
Sec.~\ref{sec:relationAlgebraicApproach}, which has not attracted as
much attention as one should expect.  Finally, the chapter is closed
with a summary in Sec.~\ref{sec:summaryAndConclusions2}.

In this chapter we abuse of bold letters with the intention of
facilitating the search of some definition or result.  It is assumed
that this chapter as well as Chap.~\ref{ch:backgroundAndTheory} will
be mainly used for reference.

\section{Double PushOut (DPO)}
\label{sec:DPO}

\subsection{Basics}
\label{subsec:basics}

\index{double pushout (DPO)}In the DPO approach to graph rewriting, a
direct derivation is represented by a double pushout in category
\textbf{Graph} (multigraphs and total graph morphisms).  Productions
can be defined as three graph components, separating the elements that
should be preserved from the left and right hand sides of the rule.

\index{production!DPO}A \textbf{production} $p :
(\grule{L}{K}{R}{l}{r})$ consists of a production name $p$ and a pair
of injective graph morphisms $\abb{l}{K}{L}$ and $\abb{r}{K}{R}$.
Graphs $L$, $R$ and $K$ are respectively called the left-hand side
(LHS), right-hand side (RHS) and the interface of $p$.  Morphisms $l$
and $r$ are usually injective and can be taken to be inclusions
without loss of generality.

\begin{figure}[htbp]
  \centering
  \includegraphics[scale = 0.59]{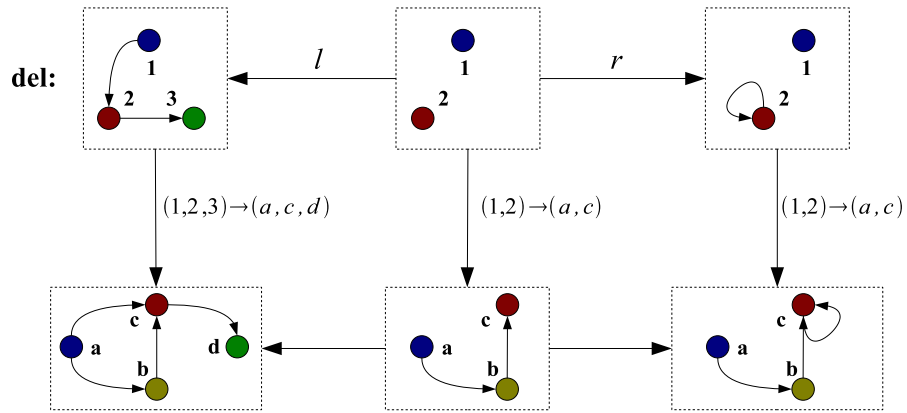}
  \caption{Example of Simple DPO Production}
  \label{fig:DPOruleAppEx}
\end{figure}

\index{interface}The \textbf{interface} $K$ of a production consists
of the elements that should be preserved by the production
application, while elements in $L-K$ are deleted and elements of $R-K$
are added.  Figure~\ref{fig:DPOruleAppEx} shows a simple DPO
production named \emph{del}, that can be applied if a path of three
nodes is found.  If so, the production eliminates the last node and
all edges and creates a loop edge in the second node.

\begin{figure}[htb]
  \centering \makebox{ \xymatrix{
      L \ar[dd]_{m} && K \ar[ll]_{l} \ar[rr]^{r} \ar@{.>}@[red][dd]_{d} && R \ar@{.>}@[red][dd]_{m^*} \\
      \\
      G && D \ar@{.>}@[red][ll]_{l^*} \ar@{.>}@[red][rr]^{r^*} && H\\
    } }
  \caption{Direct Derivation as DPO Construction}
  \label{fig:ruleapp}
\end{figure}
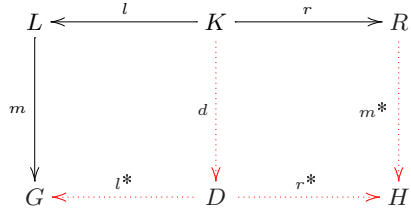

\index{direct derivation!DPO}\index{match!DPO}A \textbf{direct
  derivation} can be defined as an application of a production to a
graph through a match by constructing two pushouts.  A \textbf{match}
is a total morphism from the left hand side of the production onto the
host graph, i.e. it is the operation of finding the LHS of the grammar
rule in the host graph.  Thus, given a graph $G$, a production
$p:(\grule{L}{K}{R}{l}{r})$ and a match $\abb{m}{L}{G}$, a direct 
derivation from $G$ to $H$ using $p$ (based on $m$) exists if and only
if the diagram in Fig.~\ref{fig:ruleapp} can be constructed, where
both squares are required to be pushouts in category \textbf{Graph}.

In Fig.~\ref{fig:ruleapp}, red dotted arrows represent the morphisms
that must be defined in order to close the diagram, i.e. to construct
the pushouts.  $D$ is called the \textbf{context graph}.\index{context
  graph} In particular, if the context graph can not be constructed
then the rule can not be applied.

A direct derivation is written $G \overset{p, m} \Longrightarrow H$ or
simply $G \Longrightarrow H$ if the production and the matching are
known from context.

For example, figure~\ref{fig:DPOruleAppEx} shows the application of
rule \emph{del} to a graph.  Morphisms $m$, $d$ and $m^*$ are depicted
by showing the correspondence of the vertexes in the production and
the graph.

In order to apply a production to a graph $G$, a pushout complement
has to be calculated to obtain graph $D$.  The existence of this
pushout complement is guaranteed if the so-called \textbf{dangling}
and \textbf{identification} conditions are
satisfied.\index{dangling!condition}\index{identification condition}
The first one establishes that a node in $G$ cannot be deleted if this
causes dangling edges.  The second condition states that two different
nodes or edges in $L$ cannot be identified (by means of a
non-injective match) as a single element in $G$ if one of the elements
is deleted and the other is preserved.  Moreover, the injectivity of
$\abb{l}{K}{L}$ guarantees the uniqueness of the pushout complement.
The identification condition plus the dangling condition is at times
known as \textbf{gluing condition}.\index{gluing condition}

In the example in Fig.~\ref{fig:DPOruleAppEx} the match
$(1,2,3)\mapsto (a,b,c)$ does not fulfill the dangling condition, as
the deletion of node $d$ would make edges $(a,c)$ and $(c,d)$ become
dangling, so the production cannot be applied at this match.  One
example (for SPO, but it can be easily translated into DPO) in which
the identification condition fails is depicted to the right of
Fig.~\ref{fig:SPOruleAppEx} on p.~\pageref{fig:SPOruleAppEx}. 

\subsection{Sequentialization and Parallelism}
\label{subsec:sequentializationAndParallelism}

A graph grammar can be defined as $\mathcal{G}=\langle (p:
\grule{L}{K}{R}{l}{r})_{p \in P}, G_0 \rangle$ (see
\cite{DPO:handbook}, Chap. 3), where $(p : \grule{L}{K}{R}{l}{r})_{p
  \in P}$ is a family of productions indexed by their names and $G_0$
is the starting graph of the grammar.  The semantics of the grammar
are all reachable graphs that can be obtained by successively applying
the rules in $\mathcal{G}$.
Events changing a system state can thus be modeled using graph
transformation rules.

In real systems, parallel actions can take place.  Two main approaches
can be followed in order to describe and analyze parallel
computations.  In the first one, parallel actions are sequentialized,
giving rise to different \emph{interleavings} (for example a single
CPU simulating multitasking).  In the second approach, called
\emph{explicit parallelism}, actions are really simultaneous (for
example more than one CPU performing several tasks).

\begin{figure}[htb]
  \centering \makebox{ \xymatrix{
      R_1 \ar[d]_{m^*_1} & K_1 \ar[l]_{r_1} \ar[d]^{k_1} \ar[r]^{l_1}
      & L_1 \ar[dr]_(.7){m_1} \ar@{-->}@[red][drrr]^(.25){i} && L_2
      \ar[dl]^(0.7){m_2} \ar@{-->}@[red][dlll]_(.25){j} & K_2
      \ar[l]_{l_2} \ar[r]^{r_2} \ar[d]_{k_2} & R_2  \ar[d]_{m^*_2} \\
      H_1 & D_1 \ar[l]^{r^*_1} \ar[rr]_{l^*_1} && G && D_2
      \ar[ll]^{l^*_2} \ar[r]_{r^*_2} & H_2 \\
    } }
  \caption{Parallel Independence}
  \label{fig:dpoParallelIndependence}
\end{figure}
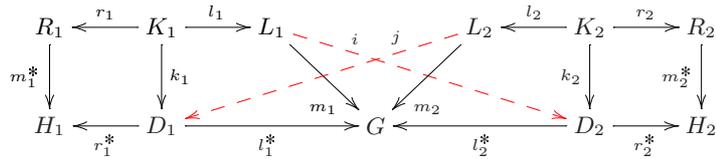

In the interleaving approach, two actions (rule applications) are
considered to be parallel if they can be performed in any order
yielding the same result.  This can be understood in two different
ways.

\index{parallel!independence}The first interpretation is called
\textbf{parallel independence} and states that two alternative direct
derivations $H_1 \overset{p_1} \Longleftarrow G \overset{p_2}
\Longrightarrow H_2$ are independent if there are direct derivations
such that $H_1 \overset{p_2} \Longrightarrow X \overset{p_1}
\Longleftarrow H_2$ (see Fig.~\ref{fig:dpoParallelIndependence}).
That is, both derivations are not in conflict, but one can be
postponed after the other.  It can be characterized using morphisms in
a categorical style saying that two direct derivations (as those
depicted in Fig.~\ref{fig:dpoParallelIndependence}) are parallel
independent if and only if
\begin{equation}\label{eq:dpoParallelIndependence}
  \exists i: L_1 \rightarrow D_2, j: L_2 \rightarrow D_1 \; \vert
  \;l^*_2 \circ i = m_1, l^*_1 \circ j = m_2.
\end{equation}

\index{weak parallel independence}If one element is preserved by one
derivation, but deleted by the other, then the latter is said to be
\textbf{weakly parallel independent} of the first (it is characterized
in equation~\ref{eq:spoParallelIndependence}).  Thus, parallel
independence can be defined as mutual weak parallel independence.

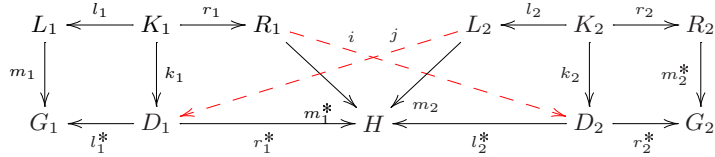
\begin{figure}[htb]
  \centering \makebox{ \xymatrix{
      L_1 \ar[d]_{m_1} & K_1 \ar[l]_{l_1} \ar[d]^{k_1} \ar[r]^{r_1} &
      R_1 \ar[dr]_(.7){m^*_1} \ar@{-->}@[red][drrr]^(.25){i} && L_2
      \ar[dl]^(0.7){m_2} \ar@{-->}@[red][dlll]_(.25){j} & K_2
      \ar[l]_{l_2} \ar[r]^{r_2} \ar[d]_{k_2} & R_2  \ar[d]_{m^*_2} \\
      G_1 & D_1 \ar[l]^{l^*_1} \ar[rr]_{r^*_1} && H && D_2
      \ar[ll]^{l^*_2} \ar[r]_{r^*_2} & G_2 \\
    } }
  \caption{Sequential Independence}
  \label{fig:dpoSequentialIndependence}
\end{figure}

\index{sequential independence}On the other hand (the second
interpretation), two direct derivations are called \textbf{sequential
  independent} if they can be performed in different order with no
changes in the result.  That is, both $G \overset{p_1} \Longrightarrow
H_1 \overset{p_2} \Longrightarrow X$ and $G \overset{p_2}
\Longrightarrow H_2 \overset{p_1} \Longrightarrow X$ yield the same
result (see Fig.~\ref{fig:dpoSequentialIndependence}).  Again,
categorically we say that two derivations are sequential independent
if and only if
\begin{equation}\label{eq:dpoSequentialIndependence}
  \exists i: R_1 \rightarrow D_2, j: L_2 \rightarrow D_1 \; \vert
  \;l^*_2 \circ i = m^*_1, r^*_1 \circ j = m_2.
\end{equation}
Mind the similarities with confluence (problem~\ref{prob:confluence})
and local confluence.

The conditions for sequential and parallel independence are given in
the \textbf{Local Church-Rosser} Theorem~\cite{DPO:handbook}, Chaps. 3
and 4.  It says that two alternative parallel derivations are parallel
independent if their matches only overlap in items that are preserved.
Two consecutive direct derivations are sequential independent if the
match of the second does not depend on elements generated by the
first, and the second derivation does not delete an item that has been
accessed by the first.  Moreover, if two direct alternative
derivations are parallel independent, their concatenation is
sequential independent and vice versa.

\index{parallel!production}The \textbf{explicit parallelism} view
\cite{DPO:handbook,concurrency:handbook3} abstracts from any
application order (no intermediate states are produced).  In this
approach, a derivation is modeled by a single production, called
\textbf{parallel production}.  Given two productions, $p_1$ and $p_2$,
the parallel production $p_1 + p_2$ is the disjoint union of both.
The application of such production is denoted as $G \overset{p_1+p_2}
\Longrightarrow X$.

Two problems arise here: \index{analysis of a
  derivation}\index{synthesis of a derivation}The sequentialization of
a parallel production (\emph{analysis}), and the parallelization of a
derivation (\emph{synthesis}).  In DPO, the \textbf{parallelism
  theorem} states that a parallel derivation $G \overset{p_1+p_2}
\Longrightarrow X$ can be sequentialized into two derivations ($G
\overset{p_1} \Longrightarrow H_1 \overset{p_2} \Longrightarrow X$ and
$G \overset{p_2} \Longrightarrow H_2 \overset{p_1} \Longrightarrow X$)
that are sequential independent.  Conversely, two derivations can be
put in parallel if they are sequentially independent.

\index{amalgamation}This is a limiting case of \textbf{amalgamation},
which specifies that if there are two productions $p_1$ and $p_2$,
then the amalgamated production $p_1 \oplus_{p_0} p_2$ is defined such
that the production $p_1$ and $p_2$ can be applied in parallel and the
amalgamated production $p_0$ (that represents common parts of both)
should be applied only once.

The \textbf{concurrency theorem}\footnote{The concurrency theorem
  appeared in~\cite{Fundamentals} for the first time, to the best of
  our knowledge.  A somehow related concept -- more general, though --
  was introduced simultaneously for Matrix Graph Grammars
  in~\cite{JuanPP_1}.  We will review it in
  Sec.~\ref{sec:explicitParallelism}.} deals with the concurrent
execution of productions that need not be sequentially independent.
Hence, according to previous results, it is not possible to apply them
in parallel.  Anyway, they can be applied concurrently using a
so-called $E$\emph{-concurrent graph production}, $p_1 *_E p_2$.  We
will omit the details, which can be consulted in~\cite{Fundamentals}.

Let the sequence $G \stackrel{p_1,m_1}{\Longrightarrow} H_1
\stackrel{p_2,m_2}{\Longrightarrow} H_2$ be given.  It is possible to
construct a direct derivation $G \stackrel{p_1 *_E
  p_2}{\Longrightarrow} H_2$.  The basic idea is to relate both
productions through an overlapping graph $E$, which is a subgraph of
$H_1$, $E = m^*_1(R_1) \cup m_2(L_2)$.  The corresponding restrictions
$\overline{m^*_1}:R_1 \rightarrow E$ and $\overline{m_2}:L_2
\rightarrow E$ of $m^*_1$ and $m_2$, respectively, must be jointly
surjective.  Also, any direct derivation $G \stackrel{p_1 *_E
  p_2}{\Longrightarrow} H_2$ can be sequentialized.

\subsection{Application Conditions}
\label{subsec:applicationConditions}

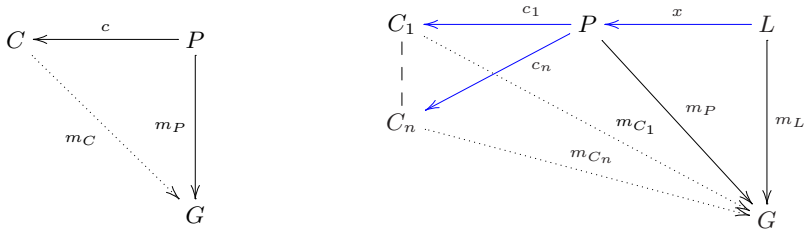
\begin{figure}[htbp]
  \setlength{\unitlength}{1cm}
  \begin{picture}(3.5, 2.5) \put(0.6, 2.3){ \xymatrix{
        C \ar @{.>}[drdr]_{m_C} && P \ar[ll]_c \ar[dd]_{m_P} \\
        \\
        && G } } \put(5.6, 2.5){ \xymatrix{
        C_1 \ar @{--}[d] \ar @{.>}[ddrrrr]^(0.6){m_{C_1}} && P \ar@[blue][ll]_(0.3){c_1} \ar@[blue][dll]^(0.3){c_n} \ar[ddrr]^{m_P} && L \ar@[blue][ll]_{x} \ar[dd]^{m_L}\\
        C_n \ar @{.>}[drrrr]^{m_{C_n}} \\
        &&&& G } }
  \end{picture}
  \caption{Generic Application Condition Diagram}
  \label{fig:ConstrAndAppConds}
\end{figure}

We will make a brief overview of graph constraints and application
conditions.  In~\cite{EEHKP}, graph constraints and application
conditions were developed for the Double Pushout (DPO) approach to
graph transformation and generalized to adhesive HLR categories in
\cite{Fundamentals}.  Atomic constraints were defined to be either
positive or negative.  \index{positive!graph constraint!atomic}A
\textbf{positive atomic graph constraint} $PC \left( c \right)$ (where
$c$ is an arbitrary morphism $\abb{c}{P}{C}$) is satisfied by graph
$G$ if $\forall \, \abb{m_P}{P}{G}$ injective morphism there exists
some $\abb{m_C}{C}{P}$ injective morphism such that $m_P = m_C \circ
c$, mathematically written $G \models PC \left( c \right)$ (see left
part of Fig.~\ref{fig:ConstrAndAppConds}).  It can be interpreted as
\emph{graph $C$ must exist in $G$ if graph $P$ is found in $G$}.

\index{positive!application condition!atomic}Graph morphism $m_L : L
\rightarrow G$ satisfies the \textbf{positive atomic application
  condition} $P \left( c, \bigvee_1^n c_i \right)$ (with
$\abb{c}{L}{P}$ and $\abb{c_i}{P}{C_i}$) if assuming $G \models PC
\left( c \right)$, for all associated morphisms $\abb{m_P}{P}{G},
\exists \, \abb{m_{C_i}}{C_i}{G}$ such that $G \models PC \left( c_i
\right)$.  The notation used is $m_L \models P \left( c, \bigvee_1^n
  c_i \right)$, having also a similar interpretation to that of graph
constraints: Suppose $L$ is found in $G$, if $P$ is also in $G$ then
there must be some $C_i$ in $G$.  Refer to the diagram on the right
side of Fig.~\ref{fig:ConstrAndAppConds}.  \index{positive!graph
  constraint}A \textbf{positive graph constraint} is a Boolean formula
over positive atomic graph constraints.  \index{positive!application
  condition} \index{negative!application condition}
\index{negative!graph constraint}Positive application conditions,
negative application conditions and negative graph constraints are
defined similarly.

\index{application condition} \index{precondition}
\index{postcondition} Finally, an \textbf{application condition} $AC
\left( p \right) = \left( A_L, A_R \right)$ for a production $p: L
\rightarrow R$ consists of a left application condition $A_L$ over $L$
(also known as \textbf{precondition}) and a right application condition
or \textbf{postcondition} $A_R$ over $R$.  A graph transformation
satisfies the application condition if the match satisfies $A_L$ and
the comatch satisfies $A_R$.  In~\cite{EEHKP, HeW95} it is shown that
graph constraints can be transformed into postconditions which
eventually can be translated into preconditions.  In this way, it is
possible to ensure that starting with a host graph that meets certain
restrictions, the application of the production will output a graph
that still satisfies the same restrictions.

DPO approach has been embedded in the weak adhesive HLR categorical
approach, which we will shortly review in the following subsection.


\subsection{Adhesive HLR Categories}
\label{subsec:adhesiveHLRCategories}

This section finishes with a celebrated generalization of
DPO. 
It was during 2004 that \textbf{adhesive HLR categories} were defined
by merging two striking ideas: Adhesive categories~\cite{LackSobo} and
high level replacement systems~\cite{EHKPB91a, EHKPB91b}. See
Sec.~\ref{sec:categoryTheory} for a quick overview of category
theory.

Basic definitions are extended almost immediately to adhesive HLR
systems $(\mathcal{C}, \mathcal{M})$.  A production
$p:(\grule{L}{K}{R}{l}{r})$ consists of three objects $L$, $K$ and
$R$, the left hand side, the gluing object and the right hand side,
respectively, and morphisms $l: K \rightarrow L$ and $r: K \rightarrow
R$ with $l, r \in \mathcal{M}$.  \index{transformation (HLR
  systems)}\index{direct transformation}There is a slight change in
notation and the term derivation is substituted by
\textbf{transformation}, and direct derivation by \textbf{direct
  transformations}.  Adhesive HLR grammars and languages are defined
in the usual way.

\begin{figure}[htb]
  \centering \makebox{ \xymatrix{
      X \ar[dd] \ar[rr]_f \ar@/^10pt/@{.>}@[blue][rrrr]^{\overline{f}} && L \ar@[blue][dd]_{m} && K \ar[ll]^{r} \ar[rr]_{l} && R \\
      \\
      Z \ar[rr]^{\gamma_Z} && G \\
    } }
  \caption{Gluing Condition}
  \label{fig:gluingCondition}
\end{figure}
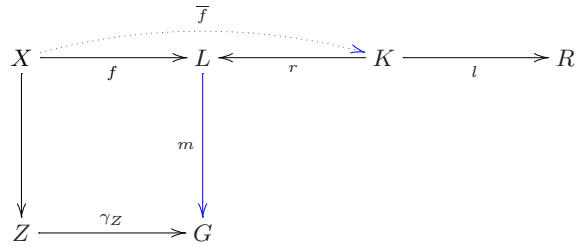

In order to apply a production we have to construct the pushout
complement and a necessary and sufficient condition for it is the
gluing condition.  For adhesive HLR systems this is possible if we can
construct initial pushouts, which is an additional requirement (it
does not follow from the axioms of adhesive HLR categories): A match
$m:L \rightarrow G$ satisfies the gluing condition with respect to a
production $p:(\grule{L}{K}{R}{l}{r})$ if for the initial pushout over
$m$ in Fig.~\ref{fig:gluingCondition} there is a morphism
$\overline{f}:X \rightarrow K$ such that $r \circ \overline{f} = f$.

Parallel and sequential independence are defined analogously to what
has been presented in Sec.~\ref{sec:DPO} and the local Church-Rosser
and the parallelism theorems remain valid.


\section{Other Categorical Approaches}
\label{sec:otherCategoricalApproaches}

This section presents other categorical approaches such as single
pushout (SPO) and pullback and compares them with DPO
(Sec.~\ref{sec:DPO}).

\index{single!pushout (SPO)}In the single pushout approach (SPO) to
graph transformation, rules are modeled with two component graphs ($L$
and $R$) and direct derivations are built with one pushout (which
performs the gluing and the deletion).  SPO relies on category
\textbf{Graph$^\mathbf{P}$} of graphs and partial graph morphisms.

\index{production!SPO}A SPO production $p$ can be defined as
$p:\morfism{L}{R}{r}$, where $r$ is an injective partial graph
morphism.  Those elements for which there is no image defined are
deleted, those for which there is image are preserved and those that
do not have a preimage are added.

\index{match!SPO}A match for a production $p$ in a graph $G$ is a
total morphism $\abb{m}{L}{G}$.  \index{direct derivation!SPO}Given a
production $p$ and a match $m$ for $p$ in $G$, the direct derivation
from $G$ is the pushout of $p$ and $m$ in \textbf{Graph$^\mathbf{P}$}.
As in DPO, a derivation is just a sequence of direct derivations.

\begin{figure}[htbp]
  \centering
  \includegraphics[scale = 0.55]{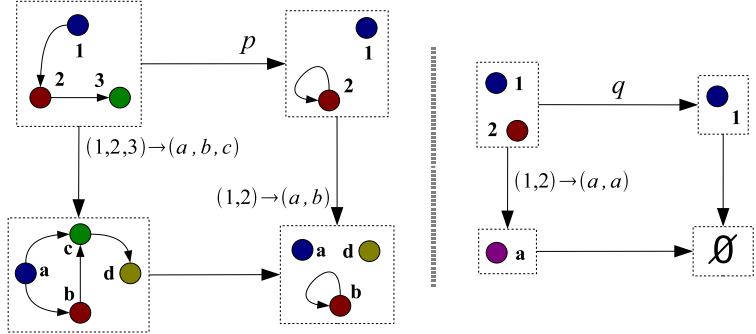}
  \caption{SPO Direct Derivation}
  \label{fig:SPOruleAppEx}
\end{figure}

The left part of Fig.~\ref{fig:SPOruleAppEx} shows an example of the rule
in Fig.~\ref{fig:DPOruleAppEx}, but expressed in the SPO approach.
The production is applied to the same graph $G$ as in
Fig.~\ref{fig:ruleapp} but at a different match.

An important difference with respect to DPO is that in SPO there is no
dangling condition: Any dangling edge is deleted (so rules may have
side effects). In this example, node $c$ and edges $(a, c)$ and $(c,
d)$ are deleted. In addition, in case of a conflict with the
identification condition due to a non-injective matching, the
conflicting elements are deleted.

Due to the way in which SPO has been defined, even though the matching
from the LHS into the host graph is a total morphism, the RHS matching
can be a partial morphism (see the example to the right of
Fig.~\ref{fig:SPOruleAppEx}).

\index{conflict-free condition}In order to guarantee that all
matchings are total it is necessary to ask for the
\textbf{conflict-free condition}: A total morphism $m:L \rightarrow G$
is conflict free for a production $p:L \rightarrow R$ if and only if
\begin{equation}
  m(x) = m(y) \implies \left[ x, y \in dom(p)\textrm{ or }x,y \not \in dom(p) \right].
\end{equation}

Results for explicit parallelism are slightly different in SPO.  In
this approach, a parallel direct derivation $G \overset{p_1+p_2}
\Longrightarrow X$ can be sequentialized into $G \overset{p_1}
\Longrightarrow H_1 \overset{p_2} \Longrightarrow X$ if $G
\overset{p_2} \Longrightarrow H_2$ is weakly parallel independent of
$G \overset{p_1} \Longrightarrow H_1$ (and similarly for the other
sequentialization).  So as this condition may not hold, there are
parallel direct derivations that do not have an equivalent
interleaving sequence.

\begin{figure}[htb]
  \centering \makebox{ \xymatrix{
      R_1 \ar[d]_{m^*_1} && L_1 \ar@[blue][dr]_{m_1} \ar@[blue][ll]_{p_1} && L_2 \ar@[blue][dl]^{m_2} \ar@[blue][rr]^{p_2} && R_2  \ar[d]_{m^*_2} \\
      H_1 &&& G \ar[lll]_{p^*_1} \ar[rrr]^{p^*_2} &&& H_2 \\
    } }
  \caption{SPO Weak Parallel Independence}
  \label{fig:spoParallelIndependence}
\end{figure}
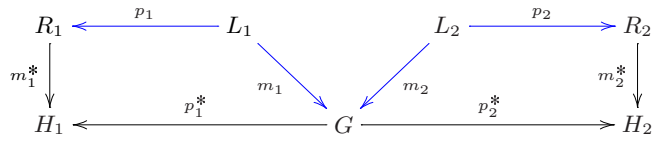

These conditions will be written explicitly because we will make a
comparison in Sec.~\ref{sec:gCongruence}.  Derivation $d_1$ is weakly
parallel independent of derivation $d_2$ (see
Fig.~\ref{fig:spoParallelIndependence}) if
\begin{equation}\label{eq:spoParallelIndependence}
  m(L_2) \cap m_1 \left( m_1 \backslash dom(p_1)\right) = \emptyset.
\end{equation}

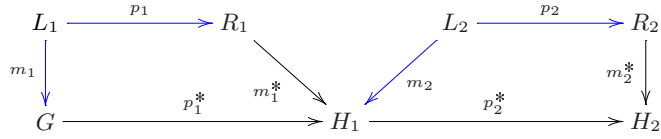
\begin{figure}[htb]
  \centering \makebox{ \xymatrix{
      L_1 \ar@[blue][d]_{m_1} \ar@[blue][rr]^{p_1} && R_1
      \ar[dr]_{m^*_1} && L_2 \ar@[blue][dl]^{m_2} \ar@[blue][rr]^{p_2}
      && R_2  \ar[d]_{m^*_2} \\
      G \ar[rrr]^{p^*_1} &&& H_1 \ar[rrr]^{p^*_2} &&& H_2 \\
    } }
  \caption{SPO Weak Sequential Independence}
  \label{fig:spoSequentialIndependence}
\end{figure}

\index{sequential independence!weak}There is an analogous concept,
similarly defined, known as \textbf{weak sequential independence}.  Let
$d_1$ and $d_2$ be as defined in Fig.~\ref{fig:spoSequentialIndependence}, then $d_2$ is weakly sequentially
independent of $d_1$ if
\begin{equation}\label{eq:spoSequentialIndependence1}
  m_2 \left( L_2 \right) \cap m^*_1 \left( R_1 \backslash p_1 (L_1)
  \right) = \emptyset.
\end{equation}
If additionally
\begin{equation}\label{eq:spoSequentialIndependence2}
  m^*_1 \left( R_1 \right) \cap m_2 \left( L_2 \backslash dom(p_2)
  \right) = \emptyset
\end{equation}
then $d_2$ is \textbf{sequentially independent} of $d_1$.

\begin{figure}[htbp]
  \centering \makebox{ \xymatrix{
      &&& R_2 \ar@{.>}[dl]^{m_{R_2}^*} \ar[ddr]^{m_{R_2}'^*} & \\
      & L_2 \ar@[blue][dl]_{m_{L_2}} \ar[ddr]^(0.7){m_{L_2}'} \ar@[blue][urr]^{p_2} & H_2 \ar@{.>}[rrd]_{p_1'^*}&& \\
      G \ar[drr]^{p_1^*} \ar@{.>}[urr]_(0.3){p_2^*} &&&& X \\
      & L_1 \ar@[blue][drr]_{p_1} \ar@[blue][ul]^{m_{L_1}} \ar@{.>}[uur]_(0.7){m'_{L_1}} & H_1 \ar[urr]^{p_2'^*}  \\
      &&& R_1 \ar[ul]_{m_{R_1}^*} \ar@{.>}[uur]_{m_{R_1}'^*}& \\
    } }
  \caption{Sequential and Parallel Independence.}
  \label{fig:Seq_ParInd}
\end{figure}
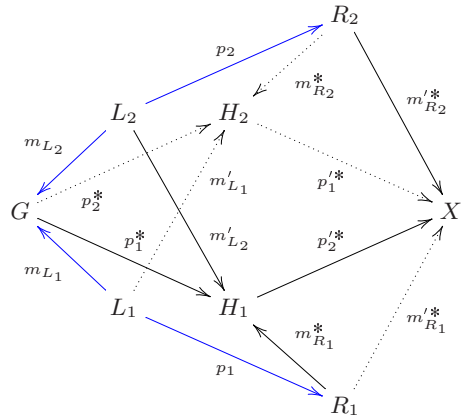

It is possible to synthesize both concepts (weak sequential
independence and parallel independence) in a single diagram.  See
Fig.~\ref{fig:Seq_ParInd}.

Due to the fact that approaches based on the pushout construction can
not replicate substructures naturally, Bauderon and others have
proposed a different setting by using pullbacks instead of pushouts
\cite{Bau95, Bau97, BaJ01}.  \index{single!pullback
  (SPB)}\index{double pullback (DPB)}We will call them SPB and DPB
approaches, depending on the number of pullbacks, similarly to SPO and
DPO.

Note that pullbacks are sub-objects of products (see Sec.~\ref{sec:categoryTheory}) and that products are (in some sense) a
natural replication mechanism.  It has been shown that pullback
approaches are strictly more expressive than those using pushouts, but
they have some drawbacks as well:
\begin{enumerate}
\item The existence condition for pullback complements is much more
  complicated than with pushouts (gluing condition).
\item In general, this condition can not be treated with computers~\cite{Kahl}.
\item There is a loss in comprehensibility and intuitiveness.
\end{enumerate}

\begin{figure}[htbp]
  \centering
  \includegraphics[scale = 0.55]{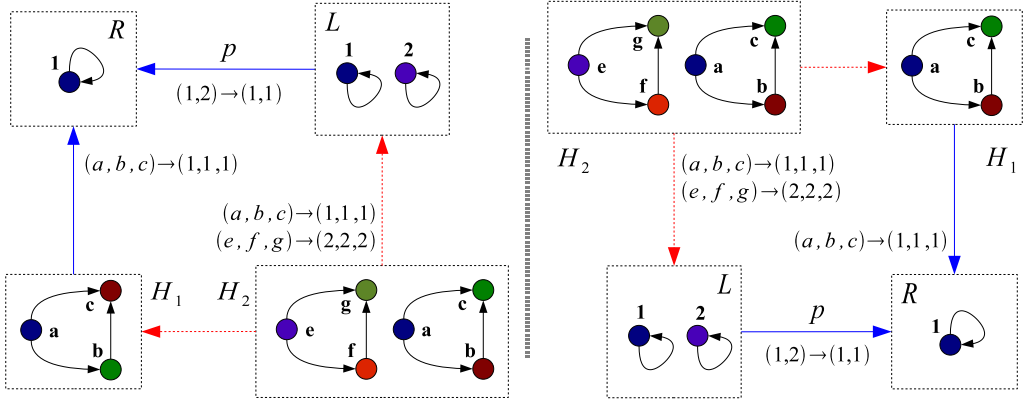}
  \caption{SPB Replication Example}
  \label{fig:spbReplicationExample}
\end{figure}

In Fig.~\ref{fig:spbReplicationExample} what we understand by a
replication that can be handled easily with SPB but not with SPO is
illustrated.  The pullback construction is depicted in dashed red
color on the same production, which is drawn twice.  To the left, the
production on top with the morphism back to front (its LHS on the
right and vice versa) and the system evolves from left to right (as in
SPO or DPO), i.e. the initial state is $H_1$ and the final state is
$H_2$.

To the right of the same figure the production is represented more
naturally for us (the left hand side on the left and the right hand
side on the right) but on the bottom of the figure.  The system
evolves on top from right to left (it should be more intuitive if it
evolved from left to right).  Besides, we notice that what we
understand as the initial state is now given by the RHS of the
production while the final state is given by the left hand
side.\footnote{Anyway, this is not misleading with some practice.}

\section{Node Replacement}
\label{sec:nodeReplacement}


\index{mother graph}\index{daughter graph}Node Replacement grammars
\cite{handbook} (Chap. 1) are a class of graph grammars based on the
replacement of nodes in a graph.  The scheme is similar to the one
described in Sec.~\ref{sec:historicalOverview}, on p.
\pageref{scheme:generalScheme} but with some peculiarities and
notational changes.  There is a \textbf{mother graph} (LHS, normally
it consists of a single node) and a \textbf{daughter graph} (RHS)
together with a gluing construction that defines how the daughter
graph fits in the host graph once the substitution is carried out.
Nodes of the mother graph play a similar role to non-terminals in
Chomsky grammars.  The differences among different node replacement
grammars reside in the way the gluing is performed.

\index{NLC}We will start with \textbf{NLC} grammars (Node Label
Controlled,~\cite{handbook}, Chap. 1) which are defined as the 5-tuple
\begin{equation}
  G = (\Sigma, \Delta, P, C, S)
\end{equation}
where $\Sigma$ are all node labels (alphabet set), $\Delta$ are node
labels ($\Delta \subseteq \Sigma$) that do not appear on the LHS of
any production (alphabet set of terminals, so non-terminals are
$\Sigma - \Delta$), $P$ is the set of productions, $C$ are the gluing
conditions (connection constructions) and $S$ is the initial graph.

\begin{figure}[htbp]
  \centering
  \includegraphics[scale = 0.61]{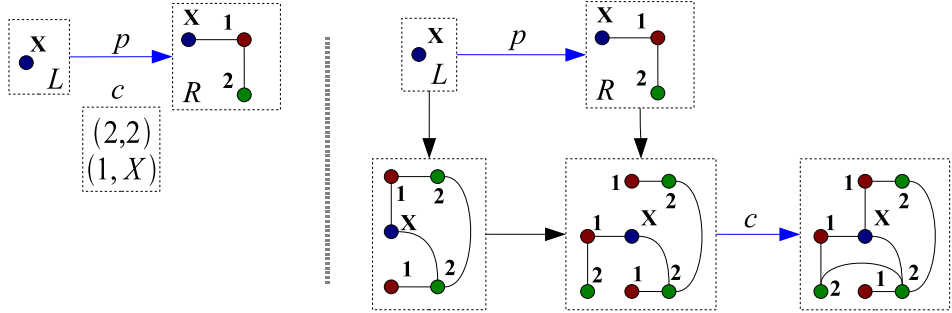}
  \caption{Example of NLC Production}
  \label{fig:nlcExample}
\end{figure}

Here only node labels matter.  Each production is defined as a
non-terminal node producing a graph with terminals and non-terminals
along with a set of connection instructions.  For example, in Fig.~\ref{fig:nlcExample} we see a production $p$ with $X$ in its LHS and a
subgraph in its RHS along with a connection relation $c$ in the box.

Production application (its semantics, also in Fig.~\ref{fig:nlcExample}) consists of deleting the LHS from the host
graph, add the RHS and finally connect the daughter graph with the
start graph.  There are no application conditions.

The linking part is performed according to a connection relation,
which is a pair of node labels of the form $(x, y)$: If the left hand
side node was adjacent to a node labeled $x$ then all nodes in the RHS
with label $y$ will be adjacent to it.

NLC is a class of context-free graph grammars, in particular
recursively defined properties can be described. Also, they are
completely local and have no application conditions which allows to
model derivations by derivation trees.  However, the yield of a
derivation tree is dependent on the order in which productions are
applied.  This property is known as confluence (see problem~\ref{prob:confluence}) and the subclass of NLC grammars that are
confluent is called \textbf{C-NLC}.

\index{NCE}At times it is desirable to refer to a concrete node
instead of to a whole family in the gluing instructions.  This variation
is known as \textbf{NCE} grammar (Neighborhood Controlled Embedding)
and is formally defined to be the tuple
\begin{equation}
  G = (\Sigma, \Delta, P, S)
\end{equation}
where $\Sigma$, $\Delta$ and $S$ are defined as above but productions
in set $P$ are different.

The grammar rule $p: X \rightarrow (D,C)$ contains the production $p:
X \rightarrow D$ and the connection $C$.  The connection is of the
form $(u,x)$ where $u$ is a label and $x$ is a particular node in the
daughter graph.  Note that NCE graph grammars are still NLC-like
grammars, at least concerning replacement.

NCE can be extended in several ways but the most popular one is adding
labels and a direction to edges, giving rise to \textbf{edNCE}
grammars.  Productions in edNCE are equal to those in NCE but
connections differ a little bit, being of the form
\begin{equation}\label{eq:edNCEconnection}
  (\mu, p/q, x, d),
\end{equation}
where $\mu$ is a node label, $p$ and $q$ are edge labels, $x$ is a
node of $D$ and $d \in \{in, out\}$ (which specifies the direction of
the edge).  For example, if $d = in$ the connection in eq.~\eqref{eq:edNCEconnection}, it specifies that the embedding process
should establish an edge with label $q$ to node $x$ of $D$ from each
$\mu$-labeled $p$-neighbor of $m \in M$ (the mother graph) that is an
in-neighbor of $m$.

The expressive power of edNCE is not increased neither if grammar rules change
directions of edges nor if connection instructions make use of
multiple edges.

The graphical representation differs a little from that of DPO and
SPO.  The daughter graph $D$ is included in a box and the area
surrounding it represents its environment.  Non-terminal symbols are
represented by capital letters inside a small box (the large box
itself can be viewed as a non-terminal symbol).  Connection
instructions are directed lines that connect nodes inside (new labels)
with nodes outside (old labels).

\begin{figure}[htbp]
  \centering
  \includegraphics[scale = 0.7]{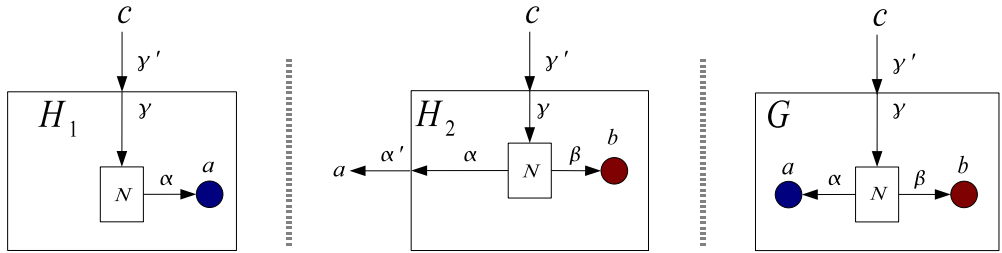}
  \caption{edNCE Node Replacement Example}
  \label{fig:edNCEexample}
\end{figure}

\noindent\textbf{Example}.$\square$The notation $G=H_2[n/H_1]$ is employed for
a derivation, meaning that graph $G$ is obtained by making the
substitution $n \mapsto H_1$ in $H_2$, i.e. by replacing node $n$ in
$H_2$ with graph $H_1$.  In the example of Fig.~\ref{fig:edNCEexample}
(with non-terminal node $N$) we have substituted the non-terminal node
in $H_1$ by $H_2$ attaching nodes according to labels in arrows
($\alpha$) to get $G$. \proofend

Associativity -- reviewed in the next section -- is a natural property
to be demanded on any context-free rewriting framework and is enjoyed
by edNCE grammars.  Some edNCE grammars are context-dependent because
they do not need to be confluent, i.e. the result of a derivation may
depend on the order of application of its productions.  The class of
confluent edNCE grammars is represented by \textbf{C-edNCE}.

C-edNCE grammars fulfill some nice properties such as being closed
under node or edge relabeling.  It is possible to define the notion of
derivation tree as in the case of context-free string grammars (see
\cite{handbook}, Chap. 1).

Many subclasses of edNCE grammars have been -- and are being
-- studied.  Just to mention some, apart from C-edNCE,
\textbf{B-edNCE} (Boundary, in which non-terminal nodes are not
connected),\footnote{The daughter does not have edges between
  non-terminal nodes and in no connection instruction $(\sigma, \beta
  / \gamma, x, d)$ $\sigma$ is non-terminal or, in other words, every
  non-terminal has a boundary of terminal neighbors.}
\textbf{B$_{nd}$-edNCE} (non-terminal neighbor deterministic B-edNCE
grammar),\footnote{The idea behind this extension is that every
  neighbour of a non-terminal is uniquely determined by their labels
  and the direction of the edge joining them.  Therefore, when
  rewriting the non-terminal, it is possible to distinguish between
  neighbours.} \textbf{A-edNCE} (in every connection instruction
$(\sigma, \beta / \gamma, x, d)$ $\sigma$ and $x$ are terminal) and
\textbf{LIN-edNCE} (linear, if every production has at most one
non-terminal node).

\section{Hyperedge Replacement}
\label{sec:hyperedgeReplacement}

The basic idea is similar to node replacement but acting on edges
instead of nodes, i.e. edges are substituted by graphs, playing the
role of non-terminals in Chomsky grammars~\cite{handbook}.

Hyperedge replacement systems are adhesive HLR categories that can be
rewritten as DPO graph transformation systems.

\begin{figure}[htbp]
  \centering
  \includegraphics[scale = 0.69]{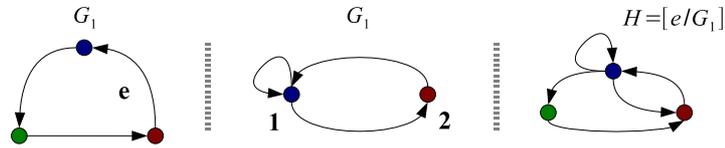}
  \caption{Edge Replacement}
  \label{fig:edgeReplacementExample}
\end{figure}

We will illustrate the ideas with an edge replacement example (instead
of hyperedge replacement, to be defined below) in a very simple case.
Suppose we have a graph as the one depicted to the left of
Fig.~\ref{fig:edgeReplacementExample}, with a labeled edge $e$ to be
substituted by the graph depicted to the center of
Fig.~\ref{fig:edgeReplacementExample}, in which the special nodes 
(\textbf{1} and \textbf{2}) are used as anchor points.  The result is
displayed to the right of Fig.~\ref{fig:edgeReplacementExample}.

A production in essence is what we have done, with a LHS made up of
labels and a graph as RHS.  The notation $H=[e/G_1]$, also $G
\Rightarrow [e/G_1]$, is standard to mean that graph (hypergraph) $H$
is obtained by deleting edge $e$ and plugging in graph $G_1$.

\index{hyperedge}A \textbf{hyperedge} is defined in~\cite{handbook}
(Chap. 2) as an atomic item with a label and an ordered set of
tentacles.  Informally, a \emph{hypergraph} is a set of nodes with a
collection of hyperedges such that each tentacle is attached to one
node.  Note that directed graphs are a special case of hypergraphs.
Normally it is established that the label of a hyperedge is the number
of its tentacles.

\index{string}\index{string!length}Let's provide a formal definition
of hypergraph. For a given string $w$, the length of the string is
denoted by $|w|$.  For a set $A$, $A^*$ is the set of all strings over
$A$.  The free symbolwise extension $f^*:A^* \rightarrow B^*$ of a
mapping $f:A \rightarrow B$ is defined by
\begin{equation}
  f^*(a_1 \cdots a_k) = f(a_1) \cdots f(a_k),
\end{equation}
$\forall k \in \mathbb{N}$ and $a_i \in A$, $i \in \{ 1, \ldots ,
k\}$. Let $\mathcal{C}$ be a set of labels and let $t: \mathcal{C}
\rightarrow \mathbb{N}$ be a typing function.  \index{hypergraph}A
\textbf{hypergraph} $H$ over $\mathcal{C}$ is the tuple
\begin{equation}
  \left( V, E, att, lab, ext \right)
\end{equation}
where $V$ is the set of nodes, $E$ the set of hyperedges, $att:E
\rightarrow V^*$ a mapping that assigns a sequence of pairwise
distinct attachment nodes $att(e)$ to each $e \in E$, $lab:E
\rightarrow \mathcal{C}$ a mapping that labels each hyperedge such
that $t(lab(e)) = | att(e)|$ and $ext \in V^*$ are pairwise distinct
external nodes.  The type of a hyperedge is its number of tentacles
and the type of a hypergraph is its number of external nodes. The set
of hypergraphs will be denoted $\mathcal{H}$, or $\mathcal{H}_C$ if we
need to explicitly refer to the set of types.

\index{hypergraph!isomorphism}Two hypergraphs $H$ and $H'$ are
isomorphic if there exist $i = (i_V, i_E)$, $i_V: H_V \rightarrow
H'_V$ and $i_E:H_E \rightarrow H'_E$ such that:
\begin{enumerate}
\item $i^*_V(att_H(e)) = att_{H'}(i_E(e))$.
\item $\forall e \in E_H$, $lab_H(e) = lab_{H'}(i_E(e))$.
\item $i^*_V(ext_H) = ext_{H'}$.
\end{enumerate}

As it usually happens in algebra, equality is defined up to
isomorphism. If $R = \{e_1, \ldots, e_n\} \subseteq E_H$ is the set of
hyperedges to be replaced and there is a preserving type function $r:
R \rightarrow \mathcal{H}$ ($\forall e \in R$, $t(r(e)) = t(e)$) such
that $r(e_i) = R_i$, then we write it both as $H[e_1/R_1, \ldots,
e_n/R_n]$ or as $H[r]$.

Hyperedge replacement belongs to the gluing approaches and follows the
high level scheme introduced in Sec.~\ref{sec:historicalOverview}: The
replacement of $R$ in $H$ according to $r$ is performed by first
removing $R$ from $E_H$, then $\forall e \in R$ the nodes and
hyperedges of $r(e)$ are disjointly added and the $i$-th external node
of $r(e)$ is fused with the $i$-th attachment node of $e$.

If a hyperedge is replaced its context is not affected.  Therefore,
hyperedge replacement provides a context-free type of rewriting as
long as no additional application conditions are employed.

There are three nice properties fulfilled by hyperedge replacement
grammars that we will briefly comment and that can be compared with
the problems introduced in Sec.~\ref{sec:dissertationOutline}, in
particular problems~\ref{prob:independence},~\ref{prob:sequentialIndependence}~and~\ref{prob:confluence},~\ref{prob:sequentialConfluence}.  Let's assume the hypothesis on
hyperedges necessary so the following formulas make sense:
\begin{itemize}
\item \emph{Sequentialization and Parallelism}: Assuming pairwise
  distinct hyperedges,
  \begin{equation}
    H[e_1/H_1, \ldots, e_n/H_n] = H[e_1/H_1] \cdots H[e_n/H_n].
  \end{equation}
\item \emph{Confluence}: Let $e_1$ and $e_2$ be distinct hyperedges,
  \begin{equation}
    H[e_1/H_1][e_2/H_2] = H[e_2/H_2][e_1/H_1].
  \end{equation}
\item \emph{Associativity}:
  \begin{equation}
    H[e_1/H_1][e_2/H_2] = H\left[e_2/H_2[e_1/H_1]\right].
  \end{equation}
\end{itemize}

Note however that in hyperedge replacement grammars, confluence is a
consequence of the first property which holds due to disjointness of
application of grammar rules.

A production $p$ over the set of non-terminals $N \subseteq
\mathcal{C}$ is an ordered pair $p = (A, R)$ with $A \in N$, $R \in
\mathcal{H}$ and $t(A) = t(R)$.  A direct derivation is the
application of a production, i.e. the replacement of a hyperedge by a
hypergraph.  If $H \in \mathcal{H}$, $e \in E_H$ and $(lab_H(e), R)$
is a production then $H' = H[e/R]$ is a direct derivation and is
represented by $H \Rightarrow H'$.  As always, a derivation is a
sequence of direct derivations.

Formally, a hyperedge replacement grammar is a system $H\!RG = \left(
  N, T, P, S \right)$ where $N$ is the set of non-terminals, $T$ is
the set of terminals, $P$ is the set of productions and $S \in N$ is
the start symbol.

We will finish this section with a simple example that generates the
string-graph language\footnote{This example is adapted (simplified)
  from one that appears in~\cite{handbook}, Chap. 2.} $L \left( A^nB^n
\right) = \{ \left(a^n b^n \right) \vert n \geq 1\}$.  This is the
graph-theoretic counterpart of the Chomsky language that consists of
strings of the form $(a^n b^n)$, $n \geq 1$, i.e. that has any string
with an arbitrary finite number of a's followed by the same number of
b's, e.g. $aabb$, $aaabbb$, etc.

\begin{figure}[htbp]
  \centering
  \includegraphics[scale = 0.6]{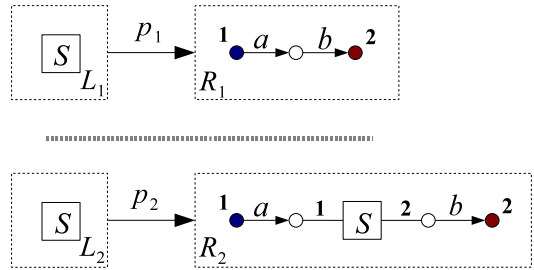}
  \caption{String Grammar Example}
  \label{fig:stringGrammar}
\end{figure}

A black filled circle $\bullet$ represents an external node while
non-filled circles $\circ$ are internal nodes.  A box represents a
hyperedge with attachments with the label inscribed in the box.  A
2-edge is represented by an arrow joining the first node to the
second.

\begin{figure}[htbp]
  \centering
  \includegraphics[scale = 0.56]{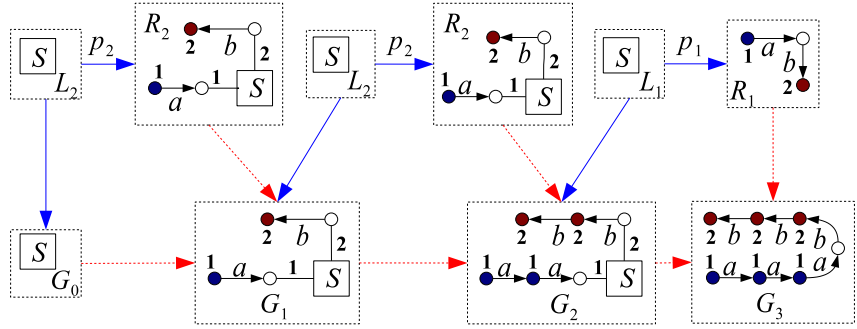}
  \caption{String Grammar Derivation}
  \label{fig:stringGrammarEvolution}
\end{figure}

The grammar is defined as $A^n B^n = \left( \{ S \}, \{ a, b \}, P, S
\right)$, where the set of productions $P = \{ p_1, p_2\}$ is depicted
in Fig.~\ref{fig:stringGrammar}. Production $p_1$ is necessary to get
the graph-string $ab$ and to stop rule application.  The start graph
and an evolution of the grammar -- derivation\footnote{Productions
  inside sequences in this book are applied from right to left, as in
  the composition of functions.} $p_1;p_2;p_2$ -- can be found in
Fig.~\ref{fig:stringGrammarEvolution}.

\section{MSOL Approach}
\label{sec:msolApproach}

It is possible to represent graphs as logical structures, expressing
their properties by logical formulas or, in other words, use logical
formulas to characterize classes of graphs and to establish their
properties out of their logical description.  In this section we will
give a brief introduction to monadic second order logics (MSOL) for
graph transformation.  Refer to Chap. 5 of~\cite{handbook} and
references therein cited.

Currently it is not possible to define graph transformation in terms
of automaton (recall that in language theory it is essential to have
transformations that produce outputs while traversing words or trees).
Quoting B. Courcelle (Chap. 5 of~\cite{handbook}):
\begin{quotation}
  \noindent The deep reason why MSOL logic is so crucial is that it
  replaces for graphs $(\ldots)$ the notion of a finite automaton
  $(\ldots)$
\end{quotation}
The key point here is that these transformations can be defined in
terms of MSOL formulas (called \emph{definable transductions}).

Graph operations will allow us to define context-free sets of graphs
as components of least solutions to systems of equations (without
using any graph rewriting rule) and recognizable sets of graphs
(without using any notion of graph automaton).

Graphs and graph properties are represented using logical structures
and relations.  \index{binary relation}A \textbf{binary relation} $R
\subseteq A \times B$ is a multivalued\footnote{One element may have
  several images.} partial mapping that we will call
\textbf{transduction}.\index{transduction} Recall from Sec.~\ref{sec:logics} that an interpretation in logics in essence defines
semantically a structure in terms of another one, for which MSOL
formulas will be used.

Let $\mathcal{R}$ be a finite set of relation symbols and let
$\rho(R)$ be the arity of $R \in \mathcal{R}$.
\index{R@$\mathcal{R}$-structure}An $\mathcal{R}$\textbf{-structure}
is the tuple $S = \left( D, (R)_{R \in \mathcal{R}} \right)$ such that
$D$ is the (possibly infinite) domain of $S$ and each $R$ is a
$\rho(R)$-ary relation on $D$, this is, a subset of $D^{\rho(R)}$.
The class of $\mathcal{R}$-structures is denoted by
$ST\!R(\mathcal{R})$.

As an example of structure, for a simple digraph $G$ made up of nodes
in $V$ we have the associated $\mathcal{R}$-structure $\left| G
\right|_1 = \left( V, edg \right)$, where $(x, y) \in edg$ if and only
if there is an edge starting in $x$ and ending in $y$.  Note that this
structure represents simple digraphs.

The set of monadic second order formulas over $\mathcal{R}$ with free
variables in $\mathcal{Y}$ is represented by $\textrm{MS}
\left(\mathcal{R}, \mathcal{Y} \right)$.  As commented in Sec.~\ref{sec:logics}, languages defined by MSOL formulas are regular
languages.

Let $\mathcal{Q}$ and $\mathcal{R}$ be two finite ranked sets of
relation symbols and $\mathcal{W}$ a finite set of set variables (the
set of parameters).  \index{definition scheme}A $(\mathcal{R},
\mathcal{Q})$\textbf{-definition scheme} is a tuple of formulas of the
form:
\begin{equation}
  \Delta = \left( \phi, \psi_1, \ldots , \psi_k , \left( \theta_w \right)_{w \in Q^*k} \right).
\end{equation}

The aim of these formulas is to define a structure $T$ in
$ST\!R(\mathcal{Q})$ out of a structure $S$ in $STR(\mathcal{R})$.
The notation needs some comments:
\begin{itemize}
\item $\phi \in \textrm{MS}(\mathcal{R}, \mathcal{W})$ defines the
  domain of the corresponding transduction, i.e. $T$ is defined if
  $\phi$ is \textbf{true} for some assignment in $\mathcal{S}$ of
  values assigned to the parameters.
\item $\psi_i \in \textrm{MS}(\mathcal{R}, \mathcal{W} \cup \{x_i\})$
  defines the domain of $T$ as the disjoint union of elements in the
  domain of $\mathcal{S}$ that satisfy $\psi_i$ for the considered
  assignment.
\item $\theta_w \in \textrm{MS}(\mathcal{R}, \mathcal{W} \cup \{x_1,
  \ldots, x_{\rho(q)}\})$ for $w = \left( q, j \right) \in
  \mathcal{Q}^*k$, where we define $\mathcal{Q}^*k = \left\{ w \;
    \vert \; q \in \mathcal{Q}, j \in [k]^{\rho (q)} \right\}$ and
  $[k]=\{1, \ldots k\}$, $k \in \mathbb{N}$.  Formulas $\theta_w$
  define the relation $q_T$.
\end{itemize}

For a more rigorous definition with some examples, please refer to~\cite{handbook}, Chap. 5.  The important fact of transductions is that
they keep monadic second order properties, i.e. monadic second order
properties of $\mathcal{S}$ can be expressed as monadic second order
properties in $\mathcal{T}$.  Furthermore, the inverse image of a
MS-definable class of structures under a definable transduction is
definable (not so for the image), as well as the composition and the
intersection of a definable structure with the Cartesian product of
two definable structures.  However, there are some ``negative''
results apart from that of the image, e.g. the inverse of a definable
transduction is not definable neither is the intersection of two
definable transductions.

The theory goes far beyond, for example by defining context free sets
of graphs by systems of recursive equations, generalizing in some
sense the concatenation of words in string grammars.  No attention
will be paid to rigorous details and definitions (again, see Chap. 5
in~\cite{handbook}) but a simple classical example of context free
grammars will be reviewed: Let $A=\{a_1, \ldots, a_n\}$ be a finite
alphabet, $\varepsilon$ the empty word and $A^*$ the set of words over
$A$.  Let's consider the context-free grammar $G=\{u \rightarrow auuv,
u \rightarrow avb, v \rightarrow avb, v \rightarrow ab\}$.  The
corresponding system of recursive equations would be:
\begin{equation}
  S = \left\langle u = a.(u.(u.v)) + a.(v.b), v=a.(v.b)+a.b \right\rangle \nonumber
\end{equation}
where ``$.$'' is the concatenation.  It is possible, although we will
not see it, to express node replacement and hyperedge replacement in
terms of systems of recursive equations.

Analogously to the way in which the equational set extends
context-freeness, recognizable sets extend regular languages.  For
example, it is possible to show that every set of finite graphs or
hypergraphs defined by a formula of an appropriate monadic second
order language is recognizable with respect to an appropriate set of
operations (the converse also holds in many cases).

%

\section{Relation-Algebraic Approach}
\label{sec:relationAlgebraicApproach}

We will mainly follow~\cite{mizoguchi} and~\cite{Kahl} in this
section, paying special attention to the justification that the
category \textbf{Graph}$^\mathbf{P}$ has pushouts, which will be used
in Chap.~\ref{ch:matching} for one of the definitions of direct
derivation in Matrix Graph Grammars.

We will deviate from standard relational methods\footnote{Visit the
  RelMiCS initiative at
  \url{http://www2.cs.unibw.de/Proj/relmics/html/}.} notation in favor
of other which is probably more immediate for mathematicians not
acquainted with it and, besides, we think eases comparison with the
rest of the approaches in this chapter.

\index{relation}A \textbf{relation} $r_1$ from $S_1$ to $S_2$ is a
subset of the Cartesian product $S_1 \times S_2$, denoted by $r_1: S_1
\rightharpoondown S_2$.  Its inverse $r^{-1}: S_2 \rightharpoondown
S_1$ is such that $(s_2,s_1) \in r_1^{-1} \Leftrightarrow (s_1,s_2)
\in r_1$.  If $r_2: S_2 \rightharpoondown S_3$ is a relation, the
composition $r_2 r_1 \equiv r_2 \circ r_1: S_1 \rightharpoondown S_3$
is again a relation such that
\begin{equation}
  (s_1, s_3) \in r_2 \circ r_1 \Leftrightarrow \left[ \exists s_2 \in S_2 \; | \; (s_1, s_2) \in r_1, (s_2, s_3)\in S_2\right].
\end{equation}

As relations are sets, naive set operations are available such as
inclusion ($\subseteq$), intersection ($\cap$), union ($\cup$) and
difference ($-$).  \index{category!\textbf{Rel}}It is possible to form
the category \textbf{Rel} of sets and relations (the identity relation
$1_S = S \rightharpoondown S$ is the diagonal set of $S \times S$),
which besides fulfills the following properties:
\begin{itemize}\label{it:relProp}
\item $\left(r^{-1}\right)^{-1} = r$.
\item $(r_2 \circ r_1)^{-1} = r_1^{-1} \circ r_2^{-1}$.
\item Distributive law: $r_2 \circ \left( \bigcup_{\alpha \in A}
    (r_\alpha) \right) \circ r_1 = \bigcup_{\alpha \in A} \left( r_2
    \circ r_\alpha \circ r_1 \right)$.
\end{itemize}

\index{function!partial}\index{function!total}A relation $f:S_1
\rightharpoondown S_2$ such that $f \circ f^{-1} \subseteq 1_{S_2}$ is
called a \textbf{partial function} and it is represented with an arrow
instead of a harpoon, $f:S_1 \rightarrow S_2$.  If $1_{S_1} \subseteq
f^{-1} \circ f$ also, then it is called a \textbf{total function}.
Note that these are the standard set-theoretic definitions of partial
function and total function.  The function $f$ is injective if $f^{-1}
\circ f = 1_{S_1}$ and surjective if $f \circ f^{-1} = 1_{S_2}$.

\index{category!\textbf{Set}$^{\mathbf{P}}$}The category of sets and
partial functions is represented by \textbf{Set}$^{\mathbf{P}}$.  It
can be proved that \textbf{Set}$^\mathbf{P}$ has small limits and
colimits, so in particular it has pushouts.

\index{domain}For a relation $r: S \rightharpoondown T$ its
\textbf{domain} is also a relation $d:S \rightharpoondown S$ and is
given by the formula $d(r) = \left( r^{-1} \circ r \right) \cap 1_S$.

In order to define graph rewriting using relations we need a
relational representation of graphs.  A \textbf{graph} $\left\langle
  S, r \right\rangle$ is a set $S$ plus a relation $r: S
\rightharpoondown S$.
\index{morphism!partial}\index{morphism!partial}A \textbf{partial
  morphism} between graph $\left\langle S_1, r_1 \right\rangle$ and
$\left\langle S_2, r_2 \right\rangle$, $p: S_1 \rightarrow S_2$, is a
partial function $p$ such that:
\begin{equation}
  p \circ r_1 \circ d(p) \subseteq r_2 \circ p.
\end{equation}

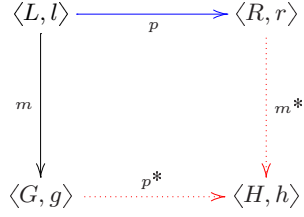
\begin{figure}[htb]
  \centering \makebox{ \xymatrix{
      \left\langle L,l \right\rangle \ar@[blue][rr]_p \ar[dd]_m && \left\langle R,r \right\rangle \ar@{.>}@[red][dd]^{m^*}\\
      \\
      \left\langle G,g \right\rangle \ar@{.>}@[red][rr]^{p^*} && \left\langle H,h \right\rangle \\
    } }
  \caption{Pushout for Simple Graphs (Relational) and Direct
    Derivation}
  \label{fig:relPushout}
\end{figure}

It is not difficult to see that the composition of two partial
morphisms of graphs is again a partial morphism of graphs.  It is a
bit more difficult (although still easy to understand) to show that
the category \textbf{Graph}$^\mathbf{P}$ of simple graphs and partial
morphisms has pushouts (Theorem 3.2 in~\cite{mizoguchi}).  The square
depicted in Fig.~\ref{fig:relPushout} is a pushout in
\textbf{Set}${}^\mathbf{P}$ if the formula for the relation $h$ is
given by:
\begin{equation}\label{eq:relPO}
  h = \left( m^* \circ r \circ m^{*{-1}} \right) \cup \left(p^* \circ g \circ p^{*{-1}}\right).
\end{equation}

A production is defined similarly to the SPO case, as a triple of two
graphs $\left\langle L,l \right\rangle$, $\left\langle R,r
\right\rangle$ and a partial morphism $p: L \rightarrow R$.  A match
for $p$ is a morphism of graphs $M:\left\langle L,l \right\rangle
\rightarrow \left\langle G,g \right\rangle$.  A production plus a
match is a direct derivation.  As always, a derivation is a finite
sequence of direct derivations.

Equation~\eqref{eq:relPO} defines a pushout in category
\textbf{Set}$^\mathbf{P}$ which is different than a rewriting square
(a direct derivation).  If we want the rewriting rule to be a pushout,
the relation in $\left\langle H, h \right\rangle$ must be defined by
the equation:
\begin{equation}
  h = \left( m^* \circ r \circ m^{*{-1}} \right) \cup \left[ p^* \circ \left( g - m^{-1} \circ l \circ m \right) \circ p^{*{-1}}\right].
\end{equation}

The relation-algebraic approach is based almost completely in
relational methods.  To illustrate the main differences with respect
to categorical approaches an example taken from~\cite{Kahl} follows
that deals with categorical products.

\noindent\textbf{Example}.$\square$In order to define the categorical product
-- see Sec.~\ref{sec:categoryTheory} -- it is necessary to check the
universal property of being a terminal object, which is a global
condition (it should be checked against the rest of candidate
elements, in principle all elements in the category).  In contrast, in
relation algebras, the direct product of two objects $X$ and $Y$ is a
triple $\left( P, \Pi_X, \Pi_Y \right)$ satisfying the following
properties:
\begin{itemize}
\item $\Pi_X \circ \Pi^{-1}_X = 1_X$ and $\Pi_Y \circ \Pi^{-1}_Y =
  1_Y$.
\item $\Pi_Y \circ \Pi_X^{-1} = \mathbb{U}$.
\item $\left(\Pi_X^{-1} \circ \Pi_X \right) \cap \left(\Pi_Y^{-1}
    \circ \Pi_Y \right) = 1_P$.
\end{itemize}
where $\mathbb{U}$ is the universal relation (to be defined below).
Note that this is a local condition, in the sense that it only
involves functions without quantification (in Category theory this
sort of characterizations are more like \emph{for all objects in the
  class there exists a unique morphism such that...}). \proofend

\index{allegory}The relational approach is based on the notion of
\textbf{allegory} which is a category $\mathcal{C}$ as defined in
Sec.~\ref{sec:categoryTheory} -- the underlying category -- plus two
operations (${}^{-1}$ and $\cap$) with the following
properties:\footnote{Compare with those on p.~\pageref{it:relProp}.}
\begin{itemize}
\item $\left(r^{-1}\right)^{-1} = r$; $\left( r \circ s \right)^{-1} =
  s^{-1} \circ r^{-1}$; $\left( r_1 \cap r_2 \right)^{-1} = r_1^{-1}
  \cap r_2^{-1}$.
\item $r_1 \circ \left( r_2 \cap r_3 \right) \subseteq \left( r_1
    \circ r_2 \right) \cap \left( r_1 \circ r_3 \right)$.
\item \emph{Modal rul}e: $\left( r_1 \cap r_2 \right) \circ r_3
  \subseteq r_1 \circ \left[ r_3 \cap \left( r_2 \circ r_1^{-1}
    \right) \right]$.
\end{itemize}

\index{relation!universal}The \textbf{universal relation} $\mathbb{U}$
for two objects $X$ and $Y$ in an allegory is the maximal element in
the set of morphisms from $X$ to $Y$, if it exists.
\index{relation!zero}If there is a least element, then it is called an
empty relation or a \textbf{zero relation}.

It is possible to obtain the other modal rule starting with the axioms
of allegories:
\begin{equation}
  \left( r_1 \circ r_2 \right) \circ r_3 \subseteq \left[ r_3 \cap \left( r_2 \circ r_3^{-1} \right) \right] \circ r_2,
\end{equation}
which can be synthesized in the so-called \emph{Dedekind formula}:
\begin{equation}
  \left( r_1 \circ r_2 \right) \circ r_3 \subseteq \left[ r_3 \cap \left( r_2 \circ r_3^{-1} \right) \right] \circ \left[ r_3 \cap \left( r_2 \circ r_1^{-1} \right) \right].
\end{equation}

\index{category!Dedekind}A locally complete distributive allegory is
called a \textbf{Dedekind category}.  \index{allegory!distributive}A
\textbf{distributive allegory} is an allegory with joins and zero
element; \emph{locally completeness} refer to distributivity of
composition with respect to joins.

By using Dedekind categories~\cite{Kahl} provides a variation of the
DPO approach in which graph variables and replication is possible.  We
will not introduce it here because it would take too long, due mainly
to notation and formal definitions, and it is not used in our
approach. 

\index{pullout}As a final remark,~\cite{Kahl} proceeds by defining
pushouts, pullbacks, complements and an amalgamation of pushouts and
pullbacks (called \textbf{pullouts}) over Dedekind categories to
define \textbf{pullout rewriting}.

\section{Summary and Conclusions}
\label{sec:summaryAndConclusions2}


The intention of this quick summary is to make an up-to-date
review of the main approaches to graph grammars and graph
transformation systems: Categorical, relational, set-theoretical and
logical.  The theory developed so far for any of these approaches goes
far beyond what has been exposed here.  The reader is referenced to
cites spread across the chapter for further study.

Throughout the rest of the book we will see that their
influence in Matrix Graph Grammars varies considerably depending on
the topic. For example, our basic diagram for graph rewriting is
similar to that of SPO\footnote{Chapter~\ref{ch:matching} defines what
  a derivation is in Matrix Graph Grammars. Two different but
  equivalent definitions of derivations are provided, one using a
  pushout construction plus an operator defined on productions and
  another with no need of categorical constructions.} but the way to
deal with restrictions on rules (application conditions) is much more
``logical'', so to speak.

We are now in the position to introduce the basics of our proposal for
graph grammars.  This will be carried out in the next chapter, Chap.~\ref{ch:mggFundamentals1}, with the peculiarity that (to some extent)
there is no need for a match of the rule's left hand side, i.e. we
have productions and not direct derivations. This is further studied
in Chapter~\ref{ch:mggFundamentals2} with the notion of \emph{initial
  digraph} and composition.
\chapter{Matrix Graph Grammars Fundamentals}
\label{ch:mggFundamentals1}

In this chapter and the next one, ideas outlined in
Chap.~\ref{ch:introduction} will be soundly based, assuming a
background knowledge on the material of
Secs.~\ref{sec:logics},~\ref{sec:graphTheory}
and~\ref{sec:groupTheory}.  No matching to any host graph is assumed, 
although identification of elements (in essence, nodes) of the same
type will be specified through \emph{completion}.

Analysis techniques developed in this chapter include compatibility of
productions and sequences as well as coherence of sequences. These
concepts will be used to tackle applicability
(problem~\ref{prob:applicability}), sequential independence
(problem~\ref{prob:sequentialIndependence}) and reachability
(problem~\ref{prob:reachability}).

In Sec.~\ref{sec:characterizationAndBasicConcepts} the dynamic nature
of a single grammar rule is developed together with some basic facts.
The operation of \emph{completion} is studied in
Sec.~\ref{sec:completion}, which basically permits algebraic
operations to be performed as one would like.
Section~\ref{sec:sequencesAndCoherence} deals with sequences, i.e.
ordered sets of grammar rules applied one after the other.\footnote{At
  times we will use the term \emph{concatenation} as a synonym. A
  derivation is a concatenation of direct derivations, and not just of
  productions.}  To this end we will introduce the concept of
\emph{coherence}.  Due to their importance, sequences will be studied
in deep detail in Chap.~\ref{ch:sequentializationAndParallelism}.

\section{Productions and Compatibility}
\label{sec:characterizationAndBasicConcepts}

A production (also known as \emph{grammar rule}) is defined as an
application which transforms a simple digraph into another simple
digraph, $p:L \rightarrow R$.  We can describe a production $p$ with
two matrices (those with an $E$ superindex) and two vectors (those
with an $N$ superindex), $p = (L^E, R^E, L^N, R^N)$, where the
components are respectively the left hand side edges matrix $\left(
  L^E \right)$ and nodes vector $\left( L^N \right)$, and the right
hand side edges matrix $\left( R^E \right)$ and nodes vector $\left(
  R^N \right)$.

$L^E$ and $R^E$ are the adjacency matrices and $L^N$ and $R^N$ are the
nodes vector as studied in Sec.~\ref{sec:graphTheory}.  A formal
definition is given for further reference:

\newtheorem{prodDef}[matrixproduct]{Definition}
\begin{prodDef}[Production - Static Formulation]\label{def:prodDef}
  \index{production!static formulation}A \emph{grammar rule} or
  \emph{production} $p$ is a partial morphism\footnote{``Partial
    morphisms'' since some elements in $L$ may not have an image in
    $R$.} between two simple digraphs $L$ and $R$, and can be
  specified by the tuple
  \begin{equation}\label{eq:prodDefinition}
    p = \left( L^E, R^E, L^N, R^N \right),
  \end{equation}
  where \emph{E} stands for \emph{edge} and \emph{N} for \emph{node}.
  \emph{L} is the left hand side and \emph{R} is the right hand side.
\end{prodDef}

It might seem redundant to specify nodes as they are already in the
adjacency matrix. The reason is that they can be added or deleted
during rewriting. Nodes and edges are considered separately, although
it could be possible to synthesize them in a single structure using
tensor algebra. See the construction of the incidence tensor --
Def.~\ref{def:IncidenceMatrixNodes} -- in Sec.~\ref{sec:dPOLikeMGG}.

It is more interesting to characterize the dynamic behaviour of rules
for which matrices will be used, describing the basic actions that can
be performed by a production: Deletion and addition of nodes and
edges.  Our immediate target is to get a dynamic formulation.

In this book $p$ will be injective unless otherwise stated.  A
production models deletion and addition actions on both edges and
nodes, carried out in the order just mentioned, i.e. first deletion
and then addition.  Appropriate matrices are introduced to represent
them.

\newtheorem{ErasingRepositionEdges}[matrixproduct]{Definition}
\begin{ErasingRepositionEdges}[Deletion and Addition of Edges]
  \index{edge!deletion}\index{edge!addition}Matrices for deletion and
  addition of edges are defined elementwise by the formulas
  \begin{equation}
    e^E = \left( e \right) _{ij} = 
    \left\{ \begin{array}{ll}
        1 & \qquad\textrm{if edge $(i,j)$ is to be erased} \\
        0 & \qquad\textrm{otherwise}\\
      \end{array}\right.
  \end{equation}
  \begin{equation}
    r^E = \left( r \right) _{ij} = 
    \left\{ \begin{array}{ll}
        1 & \qquad\textrm{if edge $(i,j)$ is to be added} \\
        0 & \qquad\textrm{otherwise}\\
      \end{array}\right.
  \end{equation}
\end{ErasingRepositionEdges}

For a given production $p$ as above, both matrices can be calculated
through identities:
\begin{eqnarray}
  e^E & \! = & \! L^E \wedge \overline{(L^E \wedge R^E)} = L^E \wedge \left( \overline{L^E} \vee \overline{R^E} \right) = L^E \wedge \overline{R^E}\\
  r^E & \! = & \! R^E \wedge \overline{(L^E \wedge R^E)} = R^E \wedge \left( \overline{R^E} \vee \overline{L^E} \right) = R^E \wedge \overline{L^E}
\end{eqnarray}

\noindent where $L^E \wedge R^E$ are the elements that are preserved
by the rule application (similar to the $K$ component in DPO rules,
see Sec.~\ref{sec:DPO}).  Thus, using previous construction, the
following two conditions hold and will be frequently used: Edges can
be added if they do not currently exist and may be deleted only if
they are present in the left hand side (LHS) of the
production.\index{LHS, Left Hand Side}
\begin{eqnarray}
  r^E \wedge \overline{L^E} & = & R^E \wedge \overline{L^E} \wedge \overline{L^E} = r^E \\
  e^E \wedge L^E & = & L^E \wedge \overline{R^E} \wedge L^E = e^E.
\end{eqnarray}

In a similar way, vectors for the deletion and addition of nodes can
be defined:

\newtheorem{ErasingRepositionNodes}[matrixproduct]{Definition}
\begin{ErasingRepositionNodes}[Deletion and Addition of
  Nodes]\index{node!deletion}\index{node!addition}
  \begin{equation}
    e^N = \left( e \right) _i = 
    \left\{ \begin{array}{ll}
        1 & \qquad\textrm{if node \emph{i} is to be erased} \\
        0 & \qquad\textrm{otherwise}\\
      \end{array}\right.
  \end{equation}
  \begin{equation}
    r^N = \left( r \right) _i = 
    \left\{ \begin{array}{ll}
        1 & \qquad\textrm{if node \emph{i} is to be added} \\
        0 & \qquad\textrm{otherwise}\\
      \end{array}\right.
  \end{equation}
\end{ErasingRepositionNodes}

\begin{figure}[htbp]
  \centering
  \includegraphics[scale = 0.57]{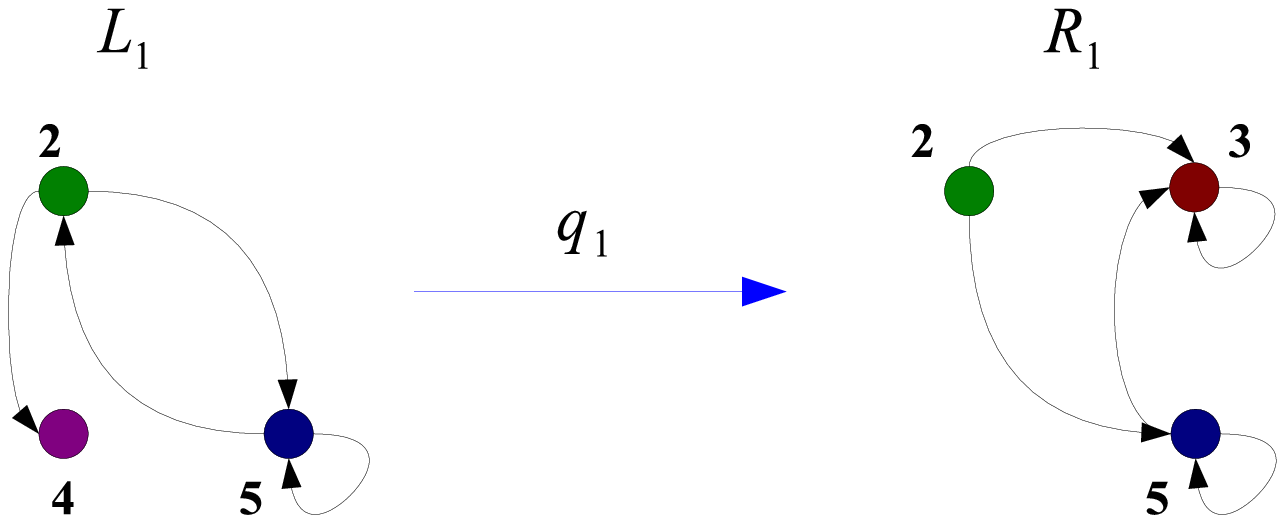}
  \caption{Example of Production}
  \label{fig:FirstProduction1}
\end{figure}

\noindent\textbf{Example}.$\square$An example of production is graphically
depicted in Fig.~\ref{fig:FirstProduction1}. Its associated matrices
are:
\begin{displaymath}
  L_1^E = \left[
    \begin{array}{cccc}
      \vspace{-6pt}
      0 & 1 & 1 & \vert\; 2 \\
      \vspace{-6pt}
      0 & 0 & 0 & \vert\; 4 \\
      \vspace{-6pt}
      1 & 0 & 1 & \vert\; 5 \\
      \vspace{-10pt}
    \end{array} \right]
  \;\;
  L_1^N = \left[
    \begin{array}{cc}
      \vspace{-6pt}
      1 & \vert\; 2 \\
      \vspace{-6pt}
      1 & \vert\; 4 \\
      \vspace{-6pt}
      1 & \vert\; 5 \\
      \vspace{-10pt}
    \end{array} \right]
  \;\;
  R_1^E = \left[
    \begin{array}{cccc}
      \vspace{-6pt}
      0 & 1 & 1 & \vert\; 2 \\
      \vspace{-6pt}
      0 & 1 & 0 & \vert\; 3 \\
      \vspace{-6pt}
      0 & 1 & 1 & \vert\; 5 \\
      \vspace{-10pt}
    \end{array} \right]
  \;\;
  R_1^N = \left[
    \begin{array}{cc}
      \vspace{-6pt}
      1 & \vert\; 2 \\
      \vspace{-6pt}
      1 & \vert\; 3 \\
      \vspace{-6pt}
      1 & \vert\; 5 \\
      \vspace{-10pt}
    \end{array} \right]
\end{displaymath}

\begin{displaymath}
  e_1^E = \left[
    \begin{array}{cccc}
      \vspace{-6pt}
      0 & 1 & 0 & \vert\; 2 \\
      \vspace{-6pt}
      0 & 0 & 0 & \vert\; 4 \\
      \vspace{-6pt}
      1 & 0 & 0 & \vert\; 5 \\
      \vspace{-10pt}
    \end{array} \right]
  \quad
  e_1^N = \left[
    \begin{array}{cc}
      \vspace{-6pt}
      0 & \vert\; 2 \\
      \vspace{-6pt}
      1 & \vert\; 4 \\
      \vspace{-6pt}
      0 & \vert\; 5 \\
      \vspace{-10pt}
    \end{array} \right]
  \quad
  r_1^E = \left[
    \begin{array}{cccc}
      \vspace{-6pt}
      0 & 1 & 0 & \vert\; 2 \\
      \vspace{-6pt}
      0 & 1 & 0 & \vert\; 3 \\
      \vspace{-6pt}
      0 & 1 & 0 & \vert\; 5 \\
      \vspace{-10pt}
    \end{array} \right]
  \quad
  r_1^N = \left[
    \begin{array}{cc}
      \vspace{-6pt}
      0 & \vert\; 2 \\
      \vspace{-6pt}
      1 & \vert\; 3 \\
      \vspace{-6pt}
      0 & \vert\; 5 \\
      \vspace{-10pt}
    \end{array} \right]
\end{displaymath}

The last column of the matrices specify node ordering, which is
assumed to be equal by rows and by columns. The characterization of
productions through matrices will be completed by introducing the
nihilation matrix (Sec.~\ref{sec:coherenceRevisited}) and the negative
initial digraph (Sec.~\ref{sec:NID}). They keep track of all elements
that can not be present in the graph (dangling edges and those to be
added by the production). For an example of production with all its
matrices, please see the one on page~\pageref{ex:first}. \proofend

Now we state some basic properties that relate the adjacency matrices
and $e$ and $r$.

\newtheorem{SimpleEqualities}[matrixproduct]{Proposition}
\begin{SimpleEqualities}[Rewriting
  Identities]\label{prop:simpleEqualities}
  Let $p: L \rightarrow R$ be a production. The following identities
  are fulfilled:
  \begin{equation}\label{eq:r_And_e}
    r^E \wedge \overline{e^E} = r^E \qquad r^N \wedge \overline{e^N} = r^N
  \end{equation}
  \begin{equation}\label{eq:Not_e_And_Not_r}
    e^E \wedge \overline{r^E} = e^E \qquad e^N \wedge \overline{r^N} = e^N
  \end{equation}
  \begin{equation}
    R^E \wedge \overline{e^E} = R^E \qquad R^N \wedge \overline{e^N} = R^N
  \end{equation}
  \begin{equation}\label{eq:L_And_Not_e}
    L^E \wedge \overline{r^E} = L^E \qquad L^N \wedge \overline{r^N} = L^N
  \end{equation}
\end{SimpleEqualities}
\emph{Proof}\\*
$\square$It is straightforward to prove these results using basic
Boolean identities.  Only the first one is included:
\begin{eqnarray}
  r^E \wedge \overline{e^E} & = & \left( \overline{L^E} \wedge R
  \right) \wedge \left( \overline{L^E \wedge \overline{R^E} }
  \right) = \nonumber \\
  & = & \left( \overline{L^E} \wedge R \wedge \overline{L^E}
  \right) \vee \left( \overline{L^E} \wedge R^E \wedge R^E
  \right) = \nonumber \\
  & = & \left( \overline{L^E} \wedge R^E \right) \vee \left(
    \overline{L^E} \wedge R^E \right) = r^E \vee r^E = r^E.
\end{eqnarray}
The rest of the identities follow easily by direct substitution of
definitions.\proofend

First two equations say that edges or nodes cannot be rewritten --
erased and created or vice versa -- by a rule application (a
consequence of the way in which matrices $e$ and $r$ are calculated).
This is because, as we will see in formulas~\eqref{eq:charNodes}
and~\eqref{eq:charEdges}, elements to be deleted are those specified
by $e$ and those to be added are those in $r$, so common elements are:
\begin{equation}
  e \wedge r = e \wedge \overline{r} \wedge r \wedge \overline{e} = 0.
\end{equation}

This contrasts with the DPO approach, in which edges and nodes can be
rewritten in a single rule.\footnote{It might be useful for example to
  forbid a rule application if the dangling condition is violated.
  This is addressed in Matrix Graph Grammars through
  $\varepsilon$-productions, see Chap.~\ref{ch:matching}.}  The
remaining two conditions state that if a node or edge is in the right
hand side (RHS), then it can not be deleted, and that if a node or
edge is in the LHS, then it can not be created.\index{RHS, Right Hand
  Side}

Finally we are ready to characterize a production $p:L \rightarrow R$
using deletion and addition matrices, starting from its LHS:
\begin{equation}\label{eq:charNodes}
  R^N = r^N \vee \left( \overline{e^N} \wedge L^N \right)
\end{equation}
\begin{equation}\label{eq:charEdges}
  R^E = r^E \vee \left( \overline{e^E} \wedge L^E \right).
\end{equation}

The resulting graph $R$ is calculated by first deleting the elements
in the initial graph -- $\overline{e} \wedge L$ -- and then adding the
new elements -- $r \vee \left( \overline{e} \wedge L \right)$ --.  It
can be proved using Proposition~\ref{prop:simpleEqualities} that, in
fact, it doesn't matter whether deletion is carried out first and
addition afterwards or vice versa.\footnote{The order in which actions
  are performed does matter if instead of a single production we
  consider a sequence. See comments after the proof of
  Corollary~\ref{cor:CompLemma}.}

\noindent\textbf{Remark}.$\square$In the rest of the book we will omit $\wedge$
if possible, and avoid unnecessary parenthesis bearing in mind that
$\wedge$ has precedence over $\vee$.  So,
e.g. formula~\eqref{eq:charEdges} will be written
\begin{equation}
  R^E = r^E \vee \overline{e^E} L^E.
\end{equation}

Besides, if there is no possible confusion due to context or a formula
applies to both edges and nodes, superscripts can be omitted.  For
example, the same formula would read $R = r \vee \overline{e}
L$.\proofend

There are two ways to characterize a production so far, either using
its initial and final \emph{states} (see Definition~\ref{def:prodDef})
or the operations it specifies:
\begin{equation}
  p = \left( e^E, r^E, e^N, r^N \right).
\end{equation}

As a matter of fact, they are not completely equivalent.  Using $L$
and $R$ gives more information because those elements which are
present in both of them are mandatory if the production is to be
applied to a host graph, but they do not appear in the \emph {e-r}
characterization.\footnote{This usage of elements whose presence is
  demanded but are not used is a sort of \emph{positive application
    condition}. See Chap.~\ref{ch:restrictionsOnRules}.}  An alternate
and complete definition to~(\ref{eq:prodDefinition}) is
\begin{equation}
  p = \left( L^E, e^E, r^E, L^N, e^N, r^N \right).
\end{equation}

A \emph{dynamic} definition of grammar rule is postponed until
Sec.~\ref{sec:NID}, Definition~\ref{def:dynamicProduction} because
there is a useful matrix (the nihilation matrix) that has not been
introduced yet.

Some conditions have to be imposed on matrices and vectors of nodes
and edges in order to keep compatibility when a rule is applied, that
is, to avoid dangling edges once the rule is applied.  It is not
difficult to extend the definition of compatibility from adjacency
matrices (see Def.~\ref{def:compatibilityDefinition}) to productions:

\newtheorem{prodCompatibility}[matrixproduct]{Definition}
\begin{prodCompatibility}[Compatibility]\label{def:prodCompatibility}
  \index{compatibility!production}A production $p:L \rightarrow R$ is
  compatible if $R = p(L)$ is a simple digraph.
\end{prodCompatibility}

From a conceptual point of view the idea is the same as that of the
dangling condition in DPO.  Also, what is demanded here is
completeness of the underlying space $\textbf{Graph}^\textbf{P}$ with
respect to the operations defined.

Next we enumerate the implications for Matrix Graph Grammars of
compatibility.  Recall that ${}^t$ denotes transposition:

\begin{enumerate}
\item An incoming edge cannot be added $\left( r^E \right)$ to a node
  that is going to be deleted $\left( e^N \right)$:
  \begin{equation}\label{FirstCondCompProd}
    \left\| r^E \odot e^N \right\|_1 = 0.
  \end{equation}
  Similarly, for outgoing edges $\left( r^E \right)^t$, the condition
  is:
  \begin{equation}
    \left\| \left( r^E \right)^t \odot e^N \right\|_1 = 0.
  \end{equation}
\item Another forbidden situation is deleting a node with some
  incoming edge, if that edge is not deleted as well:
  \begin{equation}
    \left\| \overline{e^E} \, L^E \odot e^N \right\|_1= 0.
  \end{equation}
  Similarly for outgoing edges:
  \begin{equation}
    \left\| \left( \overline{e^E} \, L^E \right)^t \odot e^N \right\|_1 = 0.
  \end{equation}
  Note that $\overline{e}^E L^E$ are elements preserved (used but not
  deleted) by production $p$.
\item It is not possible to add an incoming edge $\left( r^E \right)$
  to a node which is neither present in the LHS $\left( \overline{L}^N
  \right)$ nor added $\left( \overline{r}^N \right)$ by the
  production:
  \begin{equation}
    \left\| r^E \odot \left( \overline{r^N} \, \overline{L^N} \right) \right\|_1 = 0.
  \end{equation}
  Similarly, for edges starting in a given node:
  \begin{equation}
    \left\| \left( r^E \right)^t \odot \left( \overline{r^N} \, \overline{L^N} \right) \right\|_1 = 0.
  \end{equation}
\item Finally, our last conditions state that it is not possible that
  an edge reaches a node which does not belong to the LHS and which is
  not going to be added:
  \begin{equation}
    \left\| \left( \overline{e^E} L^E \right) \odot \left( \overline{r^N} \, \overline{L^N} \right) \right\|_1= 0.
  \end{equation}
  And again, for outgoing edges:
  \begin{equation}\label{LastCondCompProd}
    \left\| \left( \overline{e^E} L^E \right)^t \odot \left( \overline{r^N} \, \overline{L^N} \right) \right\|_1 = 0.
  \end{equation}
\end{enumerate}

\noindent Thus we arrive naturally at the next proposition:

\newtheorem{ProdCompatibility}[matrixproduct]{Proposition}
\begin{ProdCompatibility}\label{prop:prodCompatibility}
  Let $p: L \rightarrow R$ be a production. If
  conditions~(\ref{FirstCondCompProd})~--~(\ref{LastCondCompProd}) are
  fulfilled then $R = p(L)$ is compatible.\footnote{$p(L)$ is given
    by~(\ref{eq:charNodes}) and~(\ref{eq:charEdges}).}
\end{ProdCompatibility}

\noindent \emph{Proof}\\*
$\square$ We have to check $\left\| \left( M_E \vee M_E^t \right)
  \odot \overline{M_N} \right\|_1 = 0$, with $M_E = r^E \vee
\overline{e^E} L^E$ and $\overline{M}_N = \overline{r^N} \left( e^N
  \vee \overline{L^N} \right)$.  Applying~(\ref{eq:Not_e_And_Not_r})
in the second equality we have
\begin{eqnarray}\label{eq:OneProdCompatibility}
  \left( M_E \right. & \vee & \left. M_E^t \right) \odot
  \overline{M}_N = \left[ \left( r^E \vee \overline{e^E} L^E
    \right) \vee \left( r^E \vee \overline{e^E} L^E \right)^t
  \right] \odot \left[ \overline{r^N} \left( e^N \vee
      \overline{L^N} \right) \right] = \nonumber \\
  & = & \left[ r^E \vee \overline{e^E} L^E \vee \left( r^E
    \right)^t \vee \left( \overline{e^E} L^E \right)^t \right]
  \odot \left( e^N \vee \overline{r^N} \, \overline{L^N}
  \right).
\end{eqnarray}
Synthesizing conditions~(\ref{FirstCondCompProd})~
--~(\ref{LastCondCompProd}) or
expanding eq.~(\ref{eq:OneProdCompatibility}) the proof is
completed. \proofend

A full example is worked out in the next section, together with
further explanations on node identification across productions and
types.

\section{Types and Completion}
\label{sec:completion}

Besides characterization (with compatibility), in practice we will
need to endorse graphs with some ``semantics'' (types). These types
will impose some restrictions on the way algebraic operations can be
carried out (completion). This section is somewhat informal. For a
more formal exposition, please refer
to~\cite{MGG_computation}~and~\cite{MGG_Combinatorics}, Sec.~2.

Grammars in essence rely on the possibility to apply several morphisms
(productions) in sequence, generating languages.  At grammar design
time we do not know in general which actual initial state is to be
studied but probably we do know which elements make up the system
under consideration and what properties we are going to study. For
example, in a local area network we know that there are messages,
clients, servers, routers, hubs, switches and cables. We also know
that we are interested in dependency, deadlock and failure recovery
although we probably do not know which actual net we want to study.

\index{node!type}It seems natural to introduce \emph{types}, which are
simply a level of abstraction in the set of elements under
consideration.  For example, in previous paragraph, messages, clients,
servers, etc would be types.  So there is a ground level in which
\emph{real} things are (one actual hub) and another a little bit more
abstract level in which \emph{families} of elements live.

\begin{figure}[htbp]
  \centering
  \includegraphics[scale = 0.57]{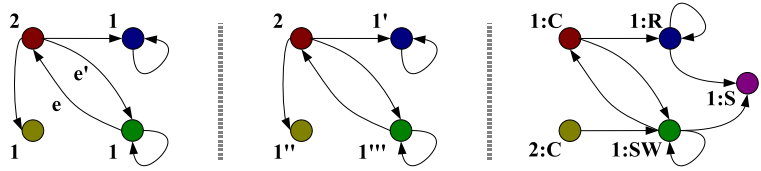}
  \caption{Examples of Types}
  \label{fig:exTypes}
\end{figure}

\noindent\textbf{Example}.\label{ex:typing}$\square$Along this book we will use
two ways of typing productions.  The first manner will be to use
natural numbers $\mathbb{N}>0$ and primes to distinguish between
elements.  To the left side of Fig.~\ref{fig:exTypes} there is a
typical simple digraph with three nodes $1$ (they are of type $1$).
This is correct as long as we do not need to operate with them.
During ``runtime'', i.e. if some algebraic operation is to be carried
out, it is mandatory to distinguish between different elements, so
primes are appended as depicted to the center of the same figure.

For the second way of typing productions, check out a small network to
the left of Fig.~\ref{fig:exTypes} where there are two clients --
(1:C) and (2:C) -- one switch -- (1:SW) -- one router -- (1:R) -- and
one server -- (1:S) --. Types are $C$, $SW$, $R$ and $S$ and instead
of primes we use natural numbers to distinguish among elements of the
same type.

Their adjacency matrices are:
\begin{displaymath}
  \left[
    \begin{array}{ccccl}
      \vspace{-6pt}
      1 & 0 & 0 & 0 & \vert\; 1' \\
      \vspace{-6pt}
      0 & 0 & 0 & 0 & \vert\; 1'' \\
      \vspace{-6pt}
      0 & 0 & 1 & 1 & \vert\; 1''' \\
      \vspace{-6pt}
      1 & 1 & 1 & 0 & \vert\; 2 \\
      \vspace{-10pt}
    \end{array} \right]
  \qquad
  \left[
    \begin{array}{cccccl}
      \vspace{-6pt}
      0 & 0 & 1 & 0 & 1 & \vert\; 1:C \\
      \vspace{-6pt}
      0 & 0 & 0 & 0 & 1 & \vert\; 2:C \\
      \vspace{-6pt}
      0 & 0 & 1 & 1 & 0 & \vert\; 1:R \\
      \vspace{-6pt}
      0 & 0 & 0 & 0 & 0 & \vert\; 1:S \\
      \vspace{-6pt}
      1 & 0 & 0 & 1 & 1 & \vert\; 1:SW \\
      \vspace{-10pt}
    \end{array} \right]
\end{displaymath}
\proofend

Nodes of the same type can be identified across productions or when
performing any kind of operation, while nodes of different types must
remain unrelated.  A production can not change the type of any node.
In some sense, nodes in the left and right hand sides of productions
specify their types. Matching (refer to Chap.~\ref{ch:matching})
transforms them in ``actual'' elements.

\index{edge!type}Types of edges are given by the type of its initial
and terminal nodes.  In the example of Fig.~\ref{fig:exTypes}, the
type of edge $e$ is $(1,2)$ and the type of edge $e'$ is $(2,1)$.  For
edges, types $(1,2)$ and $(2,1)$ are different. See~\cite{Corradini}.

\index{type}A type is just an element of a predefined set
$\mathcal{T}$ and the assignment of types to nodes of a given graph
$G$ is just a (possibly non-injective) total function from the graph
under consideration to the set of types, $t_G: G \rightarrow
\mathcal{T}$, such that it defines an equivalence relation $\sim$ in
$G$.\footnote{A reflexive ($\forall g \in G, g \sim g$), symmetric
  ($\forall g_1, g_2 \in G, \left[ g_1 \sim g_2 \Leftrightarrow g_2
    \sim g_1 \right]$) and transitive ($\forall g_1, g_2, g_3 \in G,
  \left[ g_1 \sim g_2 \right.$, $\left. g_2 \sim g_3 \Rightarrow g_1
    \sim g_3 \right] $) relation.}  It is important to have disjoint
types (something for granted if the relation is an equivalence
relation) so one element does not have two
types.\index{relation!equivalence} In previous example, the first way
of typing nodes would be $\mathcal{T}_1 = \mathbb{N} > 0$ and the
second $\mathcal{T}_2 = \left\{ (\alpha:\beta) \vert \alpha \in
  \mathbb{N}>0, \beta \in \{C,S,R,SW\}\right\}$.

The notion of type is associated to the underlying algebraic structure
and normally will be specified using an extra column on matrices and
vectors.  Conditions and restrictions on types and the way they relate
to each other can be specified using \emph{restrictions} (see
Chap.~\ref{ch:restrictionsOnRules}).

Next we introduce the concept of \emph{completion}.  In previous
sections we have assumed that when operating with matrices and vectors
these had the same size, but in general matrices and vectors represent
graphs with different sets of nodes or edges, although probably there
will be common subsets.

\index{completion}Completion modifies matrices (and vectors) to allow
some specified operation.  Two problems may occur:
\begin{enumerate}
\item Matrices may not fully coincide with respect to the nodes under
  consideration.
\item Even if they are the same, they may well not be ordered as
  needed.
\end{enumerate}

To address the first problem matrices and vectors are enlarged, adding
the missing vertexes to the edge matrix and setting their values to
zero.  To declare that these elements do not belong to the graph under
consideration, the corresponding node vector is also enlarged setting
to zero the newly added vertexes.

If for example an \textbf{and} is specified between two matrices, say
$A \wedge B$, the first thing to do is to reorder elements so it makes
sense to \textbf{and} element by element, i.e. elements representing
the same node are operated.  If we are defining a grammar on a
computer, the tool or environment will automatically do it but some
procedure has to be followed.  For the sake of an example, the
following is proposed:

\begin{enumerate}
\item Find the set $C$ of common elements.
\item Move elements of $C$ upwards by rows in $A$ and $B$, maintaining
  the order.  A similar operation must be done moving corresponding
  elements to the left by columns.
\item Sort common elements in $B$ to obtain the same ordering as in
  $A$.
\item Add remaining elements in $A$ to $B$ sorted as in $A$,
  immediately after the elements accessed in previous step.
\item Add remaining elements in $B$ to $A$ sorted as in $B$.
\end{enumerate}

Addition of elements and reordering (the operations needed for
completion) extend and modify productions syntactically but not from a
semantical point of view.

\begin{figure}[htbp]
  \centering
  \includegraphics[scale = 0.57]{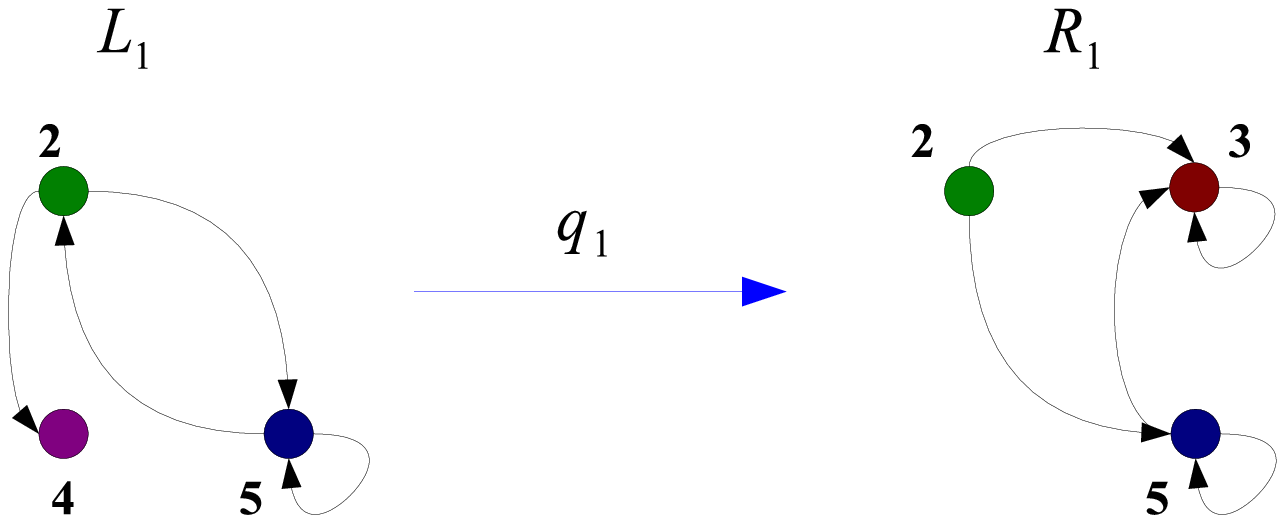}
  \caption{Example of Production (Rep.)}
  \label{fig:FirstProduction}
\end{figure}

\noindent\textbf{Example.}\label{ex:first}$\square$Consider the production
depicted in Fig.~\ref{fig:FirstProduction}. Its associated matrices
are represented below. As already commented above, the notation for
matrices will be extended a little bit in order to specify node and
edges types. It is assumed for the adjacency matrix that it is equally
ordered by rows so we do not add any row. If it is clear from context
or there is a problem with space, this labeling column will not
appear, making it explicit in words if needed.

\begin{displaymath}
  L_1^E = \left[
    \begin{array}{cccc}
      \vspace{-6pt}
      0 & 1 & 1 & \vert\; 2 \\
      \vspace{-6pt}
      0 & 0 & 0 & \vert\; 4 \\
      \vspace{-6pt}
      1 & 0 & 1 & \vert\; 5 \\
      \vspace{-10pt}
    \end{array} \right]
  \;\;
  L_1^N = \left[
    \begin{array}{cc}
      \vspace{-6pt}
      1 & \vert\; 2 \\
      \vspace{-6pt}
      1 & \vert\; 4 \\
      \vspace{-6pt}
      1 & \vert\; 5 \\
      \vspace{-10pt}
    \end{array} \right]
  \;\;
  R_1^E = \left[
    \begin{array}{cccc}
      \vspace{-6pt}
      0 & 1 & 1 & \vert\; 2 \\
      \vspace{-6pt}
      0 & 1 & 0 & \vert\; 3 \\
      \vspace{-6pt}
      0 & 1 & 1 & \vert\; 5 \\
      \vspace{-10pt}
    \end{array} \right]
  \;\;
  R_1^N = \left[
    \begin{array}{cc}
      \vspace{-6pt}
      1 & \vert\; 2 \\
      \vspace{-6pt}
      1 & \vert\; 3 \\
      \vspace{-6pt}
      1 & \vert\; 5 \\
      \vspace{-10pt}
    \end{array} \right]
\end{displaymath}

\begin{displaymath}
  e_1^E = \left[
    \begin{array}{cccc}
      \vspace{-6pt}
      0 & 1 & 0 & \vert\; 2 \\
      \vspace{-6pt}
      0 & 0 & 0 & \vert\; 4 \\
      \vspace{-6pt}
      1 & 0 & 0 & \vert\; 5 \\
      \vspace{-10pt}
    \end{array} \right]
  \quad
  e_1^N = \left[
    \begin{array}{cc}
      \vspace{-6pt}
      0 & \vert\; 2 \\
      \vspace{-6pt}
      1 & \vert\; 4 \\
      \vspace{-6pt}
      0 & \vert\; 5 \\
      \vspace{-10pt}
    \end{array} \right]
  \quad
  r_1^E = \left[
    \begin{array}{cccc}
      \vspace{-6pt}
      0 & 1 & 0 & \vert\; 2 \\
      \vspace{-6pt}
      0 & 1 & 0 & \vert\; 3 \\
      \vspace{-6pt}
      0 & 1 & 0 & \vert\; 5 \\
      \vspace{-10pt}
    \end{array} \right]
  \quad
  r_1^N = \left[
    \begin{array}{cc}
      \vspace{-6pt}
      0 & \vert\; 2 \\
      \vspace{-6pt}
      1 & \vert\; 3 \\
      \vspace{-6pt}
      0 & \vert\; 5 \\
      \vspace{-10pt}
    \end{array} \right]
\end{displaymath}

For example, if the operation $\overline{e_1^E} \, r_1^E$ was to be
performed, then both matrices must be completed.  Following the steps
described above we obtain:

\begin{displaymath}
  e_1^E = \left[
    \begin{array}{ccccc}
      \vspace{-6pt}
      0 & 1 & 0 & 0 & \vert\; 2 \\
      \vspace{-6pt}
      0 & 0 & 0 & 0 & \vert\; 4 \\
      \vspace{-6pt}
      1 & 0 & 0 & 0 & \vert\; 5 \\
      \vspace{-6pt}
      0 & 0 & 0 & 0 & \vert\; 3 \\
      \vspace{-10pt}
    \end{array} \right]
  \,
  r_1^E = \left[
    \begin{array}{ccccc}
      \vspace{-6pt}
      0 & 0 & 0 & 1 & \vert\; 2 \\
      \vspace{-6pt}
      0 & 0 & 0 & 0 & \vert\; 4 \\
      \vspace{-6pt}
      0 & 0 & 0 & 1 & \vert\; 5 \\
      \vspace{-6pt}
      0 & 0 & 0 & 1 & \vert\; 3 \\
      \vspace{-10pt}
    \end{array} \right]
  \,
  L_1^N = \left[
    \begin{array}{cc}
      \vspace{-6pt}
      1 & \vert\; 2 \\
      \vspace{-6pt}
      1 & \vert\; 4 \\
      \vspace{-6pt}
      1 & \vert\; 5 \\
      \vspace{-6pt}
      0 & \vert\; 3 \\
      \vspace{-10pt}
    \end{array} \right]
  \,
  R_1^N = \left[
    \begin{array}{cc}
      \vspace{-6pt}
      1 & \vert\; 2 \\
      \vspace{-6pt}
      0 & \vert\; 4 \\
      \vspace{-6pt}
      1 & \vert\; 5 \\
      \vspace{-6pt}
      1 & \vert\; 3 \\
      \vspace{-10pt}
    \end{array} \right]
\end{displaymath}
where, besides the erasing and addition matrices, the completion of
the nodes vectors for both left and right hand sides are displayed.

Now we check whether $r^N_1 \vee \overline{e^N_1} \, L^N_1$ and $r^E_1
\vee \overline{e^E_1} \, L^E_1$ are compatible, i.e. $R^E_1$ and
$R^N_1$ define a simple digraph.
Proposition~\ref{prop:compatibilityUsingNorm} and
equation~(\ref{eq:compeq}) are used, so we need to compute
eq.~(\ref{eq:OneProdCompatibility}) and, as

\begin{equation}
  r^E_1 \vee \overline{e^E_1} L^E_1 = \left[
    \begin{array}{ccccc}
      \vspace{-6pt}
      0 & 0 & 1 & 1 & \vert\; 2 \\
      \vspace{-6pt}
      0 & 0 & 0 & 0 & \vert\; 4 \\
      \vspace{-6pt}
      0 & 0 & 1 & 1 & \vert\; 5 \\
      \vspace{-6pt}
      0 & 0 & 0 & 1 & \vert\; 3 \\
      \vspace{-10pt}
    \end{array} \right] \qquad 
  \overline{r^N_1} \left( e^N_1 \vee \overline{L^N_1} \right) = \left[
    \begin{array}{cc}
      \vspace{-6pt}
      0 & \vert\; 2 \\
      \vspace{-6pt}
      1 & \vert\; 4 \\
      \vspace{-6pt}
      0 & \vert\; 5 \\
      \vspace{-6pt}
      0 & \vert\; 3 \\
      \vspace{-10pt}
    \end{array} \right] \nonumber
\end{equation}

\noindent substituting we finally arrive at

\begin{equation}
 (\ref{eq:OneProdCompatibility}) = \left(
    \left[
      \begin{array}{ccccc}
        \vspace{-6pt}
        0 & 0 & 1 & 1 & \vert\; 2 \\
        \vspace{-6pt}
        0 & 0 & 0 & 0 & \vert\; 4 \\
        \vspace{-6pt}
        0 & 0 & 1 & 1 & \vert\; 5 \\
        \vspace{-6pt}
        0 & 0 & 0 & 1 & \vert\; 3 \\
        \vspace{-10pt}
      \end{array} \right]
    \vee
    \left[
      \begin{array}{ccccc}
        \vspace{-6pt}
        0 & 0 & 0 & 0 & \vert\; 2 \\
        \vspace{-6pt}
        0 & 0 & 0 & 0 & \vert\; 4 \\
        \vspace{-6pt}
        1 & 0 & 1 & 0 & \vert\; 5 \\
        \vspace{-6pt}
        1 & 0 & 1 & 1 & \vert\; 3 \\
        \vspace{-10pt}
      \end{array} \right] \right)
  \odot
  \left[
    \begin{array}{cc}
      \vspace{-6pt}
      0 & \vert\; 2 \\
      \vspace{-6pt}
      1 & \vert\; 4 \\
      \vspace{-6pt}
      0 & \vert\; 5 \\
      \vspace{-6pt}
      0 & \vert\; 3 \\
      \vspace{-10pt}
    \end{array} \right] 
  = 
  \left[
    \begin{array}{cc}
      \vspace{-6pt}
      0 & \vert\; 2 \\
      \vspace{-6pt}
      0 & \vert\; 4 \\
      \vspace{-6pt}
      0 & \vert\; 5 \\
      \vspace{-6pt}
      0 & \vert\; 3 \\
      \vspace{-10pt}
    \end{array} \right] \nonumber
\end{equation}
as desired.\proofend

It is not possible, once the process of completion has finished, to
have two nodes with the same \emph{number} inside the same
production\footnote{For example, if there are two nodes of type 8,
  after completion there should be one with a $8$ and the other with
  an $8'$.} because from an operational point of view it is mandatory
to know all relations between nodes.  If completion is applied to a
sequence then we will speak of a \emph{completed sequence}.

Note that up to this point only the production itself has been taken
into account, with no reference to the state of the system (host
graph).  Although this is half truth -- as you will promptly see -- we
may say that we are starting the analysis of grammar rules without the
need of any matching, i.e. we will analyze productions and not
necessarily direct derivations, with the advantage of gathering
information at a grammar definition stage. Of course this is a
desirable property as long as results of this analysis can be used
for derivations (during runtime).

In some sense completion and matching are complementary operations:
Inside a sequence of productions, matchings -- as side effect --
differentiate or relate nodes (and hence, edges) of productions.
Completion imposes some restrictions to possible matchings.  If we
have the image of the evolution of a system by the application of a
derivation as depicted in Fig.~\ref{fig:sequence} on
p.~\pageref{fig:sequence}, then matchings can be viewed as
\emph{vertical} identifications, while completions can be seen as
\emph{horizontal} identifications.

The way completion has been introduced, there is a deterministic part
limited to adding dummy elements and a non-deterministic one deciding
on identifications.\footnote{Non-determinism in MGG is not addressed
  in this book. Refer to~\cite{MGG_computation}.}  It should be
possible to define it as an operator whose output would be all
possible relations among elements (of the same type), i.e. completion
of two matrices would not be two matrices anymore, but the set of
matrices in which all possible combinations would be considered (or a
subset if some of them can be discarded).  This is related to the
definition of \emph{initial digraph set} in
Sec.~\ref{sec:initialDigraphSet} and the structure therein studied.

\section{Sequences and Coherence}
\label{sec:sequencesAndCoherence}

Once we are able to characterize a single production, we can proceed
with the study of finite collections of them.\footnote{The term
  \emph{set} instead of collection is avoided because repetition of
  productions is permitted.}  Two main operations, \emph{composition}
and \emph{concatenation},\footnote{Also known as
  \emph{sequentialization}.} which are in fact closely related, are
introduced in this and next sections, along with notions that make it
possible to speak of ``potential definability'': \emph{Coherence} and
\emph{compatibility}.

In order to ease exposition, in this section we shall prove partial
results concerning coherence: we shall consider productions that do
not generate dangling edges. Coherence characterization taking into
account dangling edges can be found in
Sec.~\ref{sec:coherenceRevisited} or somewhat generalized
in~\cite{MGG_Combinatorics}.

\newtheorem{concatenation}[matrixproduct]{Definition}
\begin{concatenation}[Concatenation]
  \index{sequence}\index{concatenation}Let $\mathfrak{G}$ be a
  grammar.  Given a collection of productions $ \{ p_1 , \ldots , p_n
  \} \subset \mathfrak{G}$, the notation $s_n = p_n;p_{n-1};\ldots
  ;p_1$ defines a sequence (concatenation) of productions establishing
  an order in their application, starting with $p_1$ and ending with
  $p_n$.
\end{concatenation}

\noindent\textbf{Remark}.$\square$In the literature of graph transformation,
the concatenation operator is defined back to front, this is, in the
sequence $p_2;p_1$, production $p_2$ would be applied first and $p_1$
right afterwards~\cite{DPO:handbook}.  The ordering already introduced
is preferred because it follows the mathematical way in which
composition is defined and represented. This issue will be raised
again in Sec.~\ref{sec:crashCourseInPetriNets}. \proofend

It is worth stressing that there exists a total order in a sequence,
one production being applied after the previous has finished, and thus
intermediate states are generated.  These intermediate states are
indeed the difference between concatenation and composition of
productions (see Sec.~\ref{sec:compositionAndCompatibility}).  The
study of concatenation is related to the interleaving approach to
concurrency, while composition is related to the explicit parallelism
approach (see Sec.~\ref{sec:DPO}).

A production is \emph{moved forward}, \emph{moved to the front} or
\emph{advanced} if it is shifted one or more positions to the right
inside a sequence of productions, either in a composition or a
concatenation (it is to be applied earlier), e.g. $p_4;p_3;p_2;p_1
\mapsto p_3;p_2;p_1;p_4$.  On the contrary, \emph{move backwards} or
\emph{delay} means shifting the production to the left, which implies
delaying its application, e.g. $p_4;p_3;p_2;p_1 \mapsto
p_1;p_4;p_3;p_2$.

\newtheorem{coherence}[matrixproduct]{Definition}
\begin{coherence}[Coherence]\label{def:CoherenceDefinition}
  \index{coherence}Given the set of productions $\{ p_1, \ldots , p_n
  \}$, the completed sequence $s_n = p_n ; p_{n-1} ; \ldots ; p_1$ is
  called coherent if actions of any production do not prevent actions
  of the productions that follow it, taking into account the effects
  of intermediate productions.
\end{coherence}

Coherence is a concept that deals with potential applicability to a
host graph of a sequence $s_n$ of productions.  It does not guarantee
that the application of $s_n$ and a coherent reordering of $s_n$,
$\sigma \left( s_n \right)$, lead to the same result.  This latter
case is a sort of generalization\footnote{Generalization in the sense
  that, a priori, we are considering any kind of permutation.} of
sequential independence applied to sequences, which will be studied in
Chap.~\ref{ch:sequentializationAndParallelism}.

\begin{figure}[htbp]
  \centering
  \includegraphics[scale = 0.57]{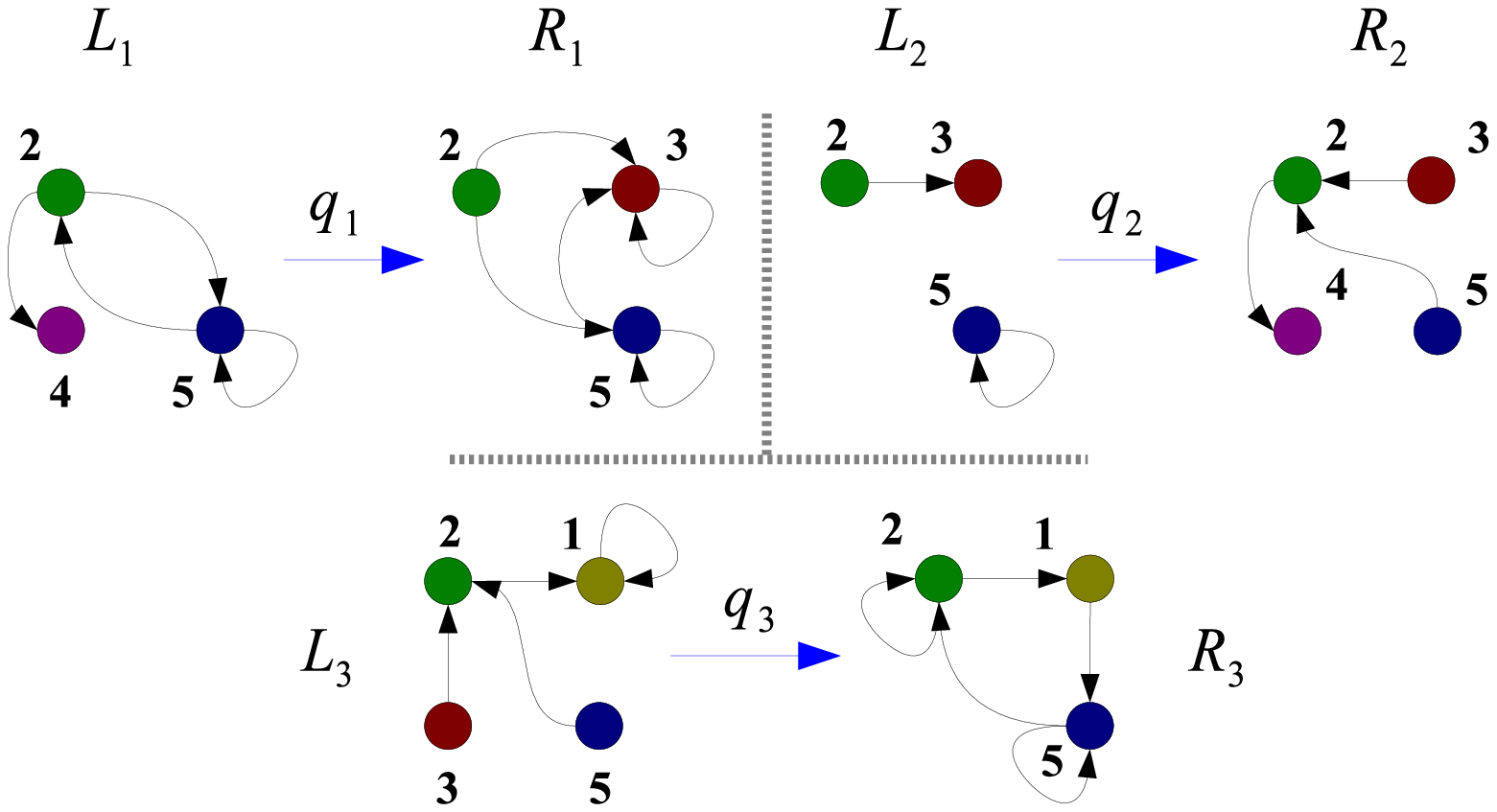}
  \caption{Productions $q_1$, $q_2$ and $q_3$}
  \label{fig:SecThirdProd}
\end{figure}

\noindent\textbf{Example.}\label{ex:second}$\square$We extend previous example
(see Fig.~\ref{fig:FirstProduction} on
p.~\pageref{fig:FirstProduction}) with two more productions.  Recall
that our first production $q_1$ deletes edge $(5,2)$, which starts in
vertex 5 and ends in vertex 2.  As depicted in
Fig.~\ref{fig:SecThirdProd}, production $q_2$ adds this edge and $q_3$
preserves it ($q_3$ used but does not delete this edge). Sequence
$s_3=q_3;q_2;q_1$ would be coherent if only this vertex was
considered.\proofend

Now we study the conditions that have to be satisfied by the matrices
associated with a coherent and dangling-free sequence of productions.
Instead of stating a result concerning conditions on coherence and
proving it immediately afterwards, we begin by discussing the case of
two productions in full detail, we continue with three and we finally
set a theorem -- Theorem~\ref{th:SeqCoherenceTheorem} -- for a finite
number of them.

Let us consider the concatenation $s_2 = p_2 ; p_1$.  In order to
decide whether the application of $p_1$ does not exclude $p_2$, we
impose three conditions on edges:\footnote{Note the similarities and
  differences with \emph{weak sequential independence}. See
  Sec. ~\ref{sec:otherCategoricalApproaches}.}
\begin{enumerate}
\item The first production -- $p_1$ -- does not delete any edge
  ($e^E_1$) used by the second production ($L^E_2$):
  \begin{equation} \label{eq:FirstCondTwoProd} e^E_1 L^E_2 = 0.
  \end{equation}
\item $p_2$ does not add ($r^E_2$) any edge preserved (used but not
  deleted, $\overline{e^E_1} L^E_1$) by $p_1$:
  \begin{equation} \label{eq:SecondCondTwoProd} r^E_2 L^E_1
    \overline{e^E_1} = 0.
  \end{equation}
\item No common edges are added by both productions:
  \begin{equation} \label{eq:ThirdCondTwoProd} r^E_1 r^E_2 = 0.
  \end{equation}
\end{enumerate}

The first condition is needed because if $p_1$ deletes an edge used by
$p_2$, then $p_2$ would not be applicable.  The last two conditions
are mandatory in order to obtain a simple digraph (with at most one
edge in each direction between two nodes).

Conditions~(\ref{eq:SecondCondTwoProd})~and~(\ref{eq:ThirdCondTwoProd})
are equivalent to $r^E_2 R^E_1 = 0$ because, as both are equal to
zero, we can do
\begin{center}
  $0 = r^E_2 L^E_1 \overline{e^E_1} \vee r^E_2 r^E_1 = r^E_2 \left(
    r^E_1 \vee \overline{e^E_1} L^E_1 \right) = r^E_2 R^E_1$
\end{center}
which may be read ``$p_2$ does not add any edge that comes out from
$p_1$'s application''.  All conditions can be synthesized in the
following identity:
\begin{equation}\label{eq:TwoProductions}
  r^E_2 R^E_1 \vee e^E_1 L^E_2 = 0.
\end{equation}

Our immediate target is to obtain a closed formula to represent these
conditions for the case of an arbitrary finite number of productions.
Applying~(\ref{eq:r_And_e}) and~(\ref{eq:Not_e_And_Not_r}),
equation~(\ref{eq:TwoProductions}) can be transformed to get:

\begin{equation}\label{eq:TwoProdTransformedEdges}
  R^E_1 \overline{e^E_2} r^E_2 \vee L^E_2 e^E_1 \, \overline{r^E_1} = 0.
\end{equation}
A similar reasoning gives the corresponding formula for nodes:

\begin{equation}\label{eq:TwoProdTransformedNodes}
  R^N_1 \overline{e^N_2} r^N_2 \vee L^N_2 e^N_1 \, \overline{r^N_1} = 0.
\end{equation}

\noindent\textbf{Remark}.$\square$Note that
conditions~(\ref{eq:SecondCondTwoProd})~and~(\ref{eq:ThirdCondTwoProd})
do not really apply to nodes as apply to edges. For example, if a node
of type 1 is to be added and nodes 1 and $1'$ have already been
appended, then by completion node $1''$ would be added. It is not
possible to add a node that already exists.

However, coherence looks for conditions that guarantee that the
operations specified by the productions of a sequence do not interfere
one with each other. Suppose the same example but this time, for some
unknown reason, the node to be added is completed as $1'$ -- this one
has just been added --. If conditions of the kind
of~(\ref{eq:SecondCondTwoProd})~and~(\ref{eq:ThirdCondTwoProd}) are 
removed, then we would not detect that there is a potential problem if
this sequence is applied.\proofend

\begin{figure}[htbp]
  \centering
  \includegraphics[scale = 0.65]{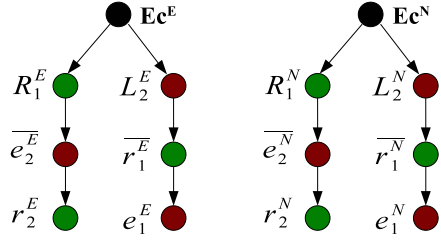}
  \caption{Coherence for Two Productions}
  \label{fig:CoherenceTwoProds}
\end{figure}

Next we introduce a graphical notation for Boolean equations: A
vertical arrow means \textbf{and} while a fork stands for \textbf{or}.
We use these diagrams because formulas grow very fast with the number
of nodes.  As an example, the representation of
equations~(\ref{eq:TwoProdTransformedEdges})~and~(\ref{eq:TwoProdTransformedNodes})
is shown in Fig.~\ref{fig:CoherenceTwoProds}.

\newtheorem{CoherenceProp}[matrixproduct]{Lemma}
\begin{CoherenceProp} \label{SeqCoherenceProp} Let $s_2 = p_2;p_1$ be
  a sequence of productions without dangling edges.  If equations
  ~(\ref{eq:TwoProdTransformedEdges})~and~(\ref{eq:TwoProdTransformedNodes})
  hold, then $s_2$ is coherent.
\end{CoherenceProp}

\noindent \emph{Proof}\\*
$\square$Only edges are considered because a symmetrical reasoning
sets the result for nodes.  Call \textbf{D} the action of deleting an
edge, \textbf{A} its addition and \textbf{P} its preservation, i.e.
the edge appears in both LHS and RHS.
Table~\ref{tab:actionsForTwoProductions} comprises all nine
possibilities for two productions.

\begin{table*}[htbp]
  \centering
  \begin{tabular}{||c|c||c|c||c|c||}
    \hline
    $D_2;D_1$ & \eqref{eq:FirstCondTwoProd} & $D_2;P_1$ & $\surd$ &
    $D_2;A_1$ & $\surd$\\
    \hline
    $P_2;D_1$ & \eqref{eq:FirstCondTwoProd} & $P_2;P_1$ & $\surd$ &
    $P_2;A_1$ & $\surd$\\
    \hline
    $A_2;D_1$ & $\surd$ & $A_2;P_1$ & \eqref{eq:SecondCondTwoProd} &
    $A_2;A_1$ & \eqref{eq:ThirdCondTwoProd}\\
    \hline
  \end{tabular}
  \caption{Possible Actions for Two Productions}
  \label{tab:actionsForTwoProductions}
\end{table*}

A tick means that the action is allowed, while a number refers to the
condition that prohibits the action.  For example, $P_2;D_1$ means
that first production $p_1$ deletes the edge and second $p_2$
preserves it (in this order). If the table is looked up we find that
this is forbidden by equation~(\ref{eq:FirstCondTwoProd}). \proofend

Now we proceed with three productions.  We must check that $p_2$ does
not disturb $p_3$ and that $p_1$ does not prevent the application of
$p_2$.  Notice that both of them are covered in our previous
explanation (in the two productions case), and thus we just need to
ensure that $p_1$ does not exclude $p_3$, taking into account that
$p_2$ is applied in between:
\begin{enumerate}
\item $p_1$ does not delete ($e^E_1$) any edge used ($L^E_3$) by $p_3$
  and not added ($\overline{r^E_2}$) by $p_2$:
  \begin{equation}
    e^E_1 L^E_3 \overline{r^E_2} = 0.
  \end{equation}
\item Production $p_3$ does not add -- $r^E_3$ -- any edge stemming
  from $p_1$ -- $R^E_1$ -- and not deleted by $p_2$ -- $e^E_2$ --:
  \begin{equation}
    r^E_3 R^E_1 \overline{e^E_2} = 0.
  \end{equation}
\end{enumerate}

Again, the last condition is needed in order to obtain a simple
digraph.  Performing similar manipulations to those carried out for
$s_2$ we get the full condition for $s_3$, given by the equation:

\begin{equation}\label{eq:ThreeProductions}
  L^E_2 e^E_1 \vee L^E_3 \left( e^E_1 \, \overline{r^E_2} \vee
    e^E_2 \right) \vee  R^E_1 \left( \overline{e^E_2} r^E_3 \vee r^E_2
  \right) \vee R^E_2 r^E_3 = 0.
\end{equation}

\begin{figure}[htbp]
  \centering
  \includegraphics[scale = 0.64]{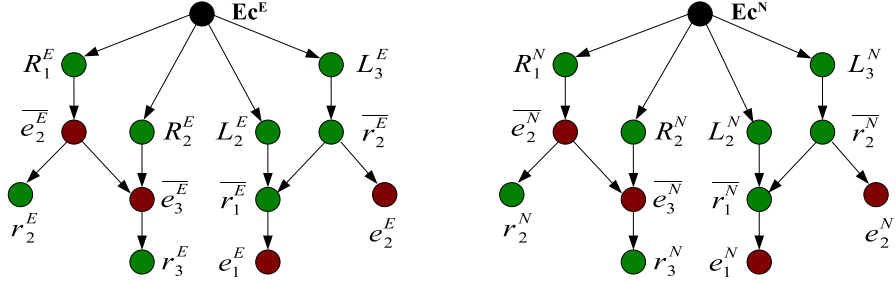}
  \caption{Coherence Conditions for Three Productions}
  \label{fig:CoherenceThreeProds}
\end{figure}

\noindent Proceeding as before, identity~(\ref{eq:ThreeProductions})
is completed:
\begin{eqnarray}\label{eq:ThreeProdTransformed}
  L^E_2 e^E_1 \, \overline{r^E_1} & \vee &
  L^E_3 \overline{r^E_2} \left( e^E_1 \, \overline{r^E_1} \vee  e^E_2
  \right)  \vee \nonumber\\
  & \vee & R^E_1 \overline{e^E_2} \left( r^E_2 \vee \overline{e^E_3}
    r^E_3 \right) \vee R^E_2 \overline{e^E_3} r^E_3 = 0.
\end{eqnarray}
Its representation is shown in Fig.~\ref{fig:CoherenceThreeProds} for
both nodes and edges.

Lemma~\ref{SeqCoherenceProp} can be extended slightly to include three
productions in an obvious way, but we will not discuss this further
because the generalization to cover \emph{n} productions is Theorem~\ref{th:SeqCoherenceTheorem}. 

\noindent\textbf{Example}.$\square$Recall productions $q_1$, $q_2$ and $q_3$
introduced in
Figs.~\ref{fig:FirstProduction}~and~\ref{fig:SecThirdProd} (on
pp.~\pageref{fig:FirstProduction} and~\pageref{fig:SecThirdProd},
respectively).  Sequences $q_3;q_2;q_1$ and $q_1;q_3;q_2$ are
coherent, while $q_3;q_1;q_2$ is not.  The latter is due to the fact
that edge $(5,5)$ is deleted (D) by $q_2$, used (U) by $q_1$ and added
(A) by $q_3$, being two pairs of forbidden actions.  For the former
sequences we have to check all actions performed on all edges and
nodes by the productions in the order specified by the concatenation,
verifying that they do not exclude each other.\proofend

\newtheorem{CoherenceOperators}[matrixproduct]{Definition}
\begin{CoherenceOperators}
  \index{operator!delta}\index{operator!nabla}Let $F(x,y)$ and
  $G(x,y)$ be two Boolean functions dependent on parameters $x, y \in
  I$ in some index set $I$. Operators delta $\bigtriangleup$ and nabla
  $\bigtriangledown$ are defined through the equations:
  \begin{equation}\label{eq:TriangleUp}
    \bigtriangleup_{t_0} ^{t_1} \left( F(x,y) \right) = \bigvee
    _{y=t_0} ^{t_1} \left( \bigwedge_{x=y} ^{t_1} \left( F(x,y)
      \right) \right)
  \end{equation}
  \begin{equation}\label{eq:TriangleDown}
    \bigtriangledown_{t_0} ^{t_1} \left( G(x,y) \right) = \bigvee
    _{y=t_0} ^{t_1} \left( \bigwedge_{x=t_0} ^{y} \left( G(x,y)
      \right) \right).
  \end{equation}
\end{CoherenceOperators}

These operators will be useful for the general case of $n$ productions
with coherence, initial digraphs, G-congruence and other concepts.  A
simple interpretation for both operators will be given at the end of
the section.

\noindent\textbf{Example}.$\square$Let $F(x,y) = G(x,y) = \overline{r}_x e_y $,
then we have:
\begin{eqnarray}
  \bigtriangleup_1^3 \left( \overline{r}_x e_y \right) & = & \bigvee
  _{y=1}^3 \left( \bigwedge_{x=y}^3 \left( \overline{r}_x e_y \right) \right) =
  \overline{r}_3 e_3 \vee \overline{r}_3 \overline{r}_2 e_2 \vee
  \overline{r}_3 \overline{r}_2 \overline{r}_1 e_1 \nonumber = e_3
  \vee \overline{r}_3 e_2 \vee \overline{r}_3 \overline{r}_2 e_1. \\
  \bigtriangledown_3^5 \left( \overline{r}_x e_y \right) & = &
  \bigvee_{y=3}^5 \left( \bigwedge_{x=3}^{x=y} \left( \overline{r}_x
      e_y \right) \right) = \overline{r}_3 e_3 \vee \overline{r}_3
  \overline{r}_4 e_4 \vee \overline{r}_3 \overline{r}_4 \overline{r}_5
  e_5 = e_3 \vee \overline{r}_3 e_4 \vee \overline{r}_3 \overline{r}_4
  e_5. \nonumber
\end{eqnarray}

Expressions have been simplified applying
Proposition~\ref{prop:simpleEqualities}. \proofend

Now we are ready to characterize coherent sequences of arbitrary
finite length.

\newtheorem{CoherenceTheorem}[matrixproduct]{Theorem}
\begin{CoherenceTheorem}\label{th:SeqCoherenceTheorem}
  The dangling-free concatenation $s_n = p_n;p_{n-1};\ldots;p_2;p_1$
  is coherent if for edges and nodes we have:
  \begin{equation}\label{eq:CoherenceFormula}
    \bigvee_{i=1}^n \left( R^E_i \bigtriangledown_{i+1}^n \left(
        \overline{e^E_x} \, r^E_y \right) \vee L^E_i
      \bigtriangleup_1^{i-1} \left( e^E_y \, \overline{r^E_x} \right)
    \right) = 0
  \end{equation}
  \begin{equation}
    \bigvee_{i=1}^n \left( R^N_i \bigtriangledown_{i+1}^n \left(
        \overline{e^N_x} \, r^N_y \right) \vee L^N_i
      \bigtriangleup_1^{i-1} \left( e^N_y \, \overline{r^N_x} \right)
    \right) = 0.
  \end{equation}
\end{CoherenceTheorem}

\noindent \emph{Proof}\\*
$\square$Induction on the number of productions (see cases $s_2$ and
$s_3$ studied above).\proofend

\begin{figure}[htbp]
  \centering
  \includegraphics[scale = 0.53]{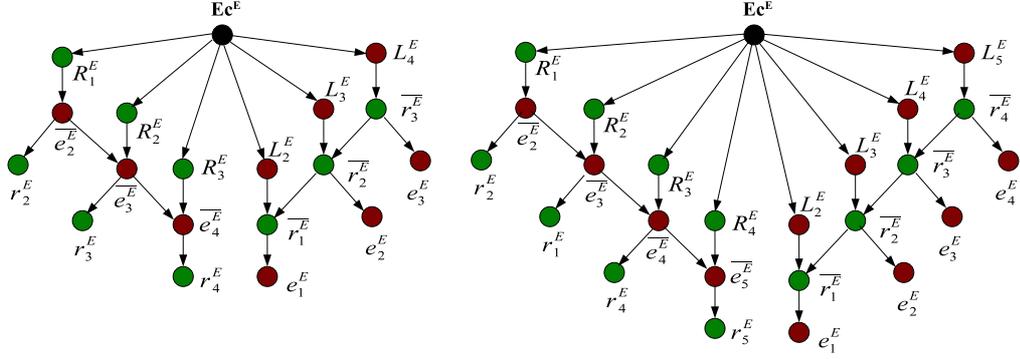}
  \caption{Coherence. Four and Five Productions}
  \label{fig:CoherenceFourFiveProds}
\end{figure}

Figure~\ref{fig:CoherenceFourFiveProds} includes the graph
representation of the formulas for coherence for $s_4 =
p_4;p_3;p_2;p_1$ and $s_5 = p_5;p_4;p_3;p_2;p_1$.

\begin{figure}[htbp]
  \centering
  \includegraphics[scale = 0.57]{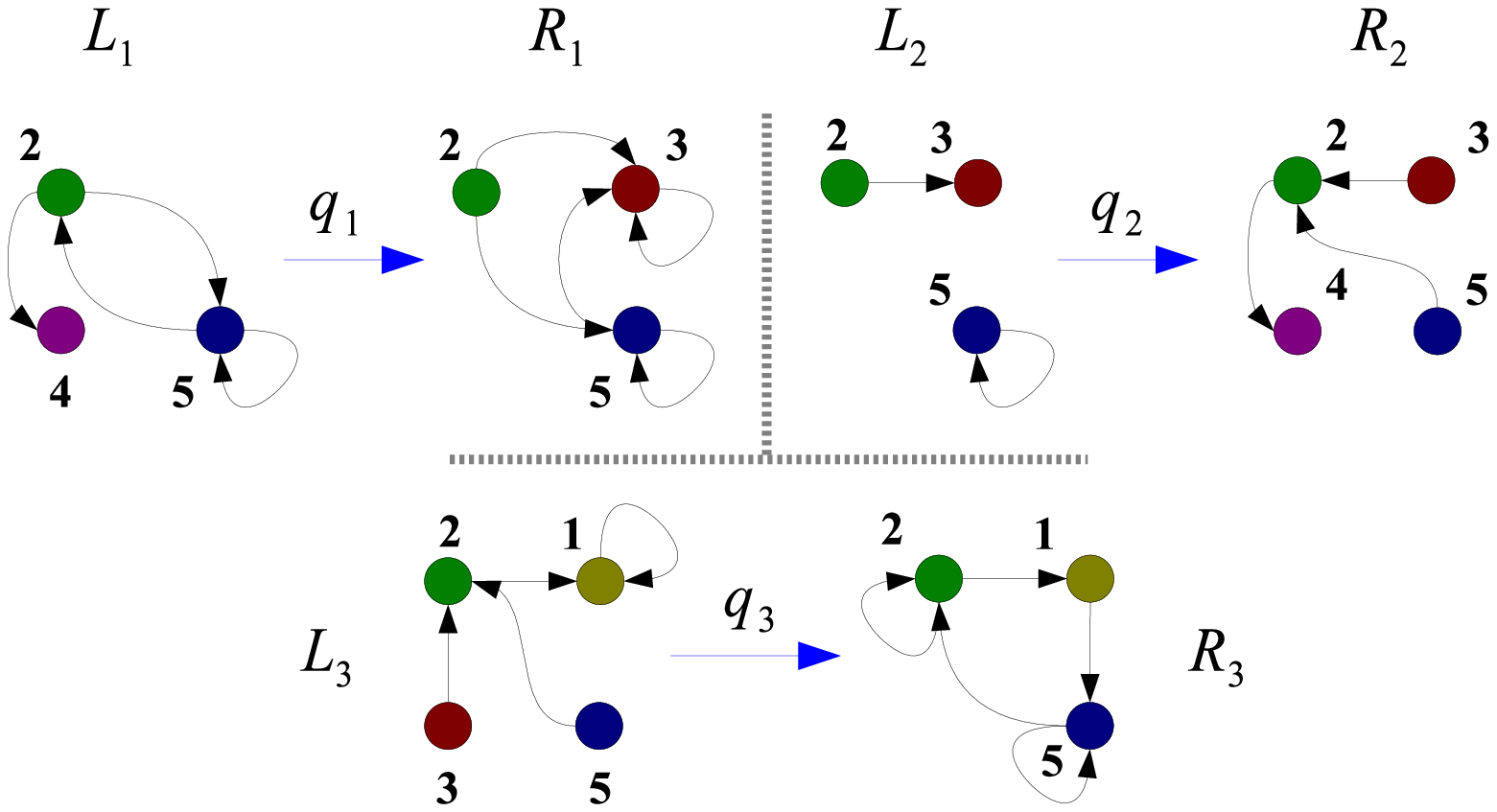}
  \caption{Productions $q_1$, $q_2$ and $q_3$ (Rep.)}
  \label{fig:prodsAgain}
\end{figure}

\noindent\textbf{Example.}$\square$We are going to verify that $s_1 =
q_1;q_3;q_2$ is coherent (only for edges), where $q_i$ are the
productions introduced in previous examples.  Productions are drawn
again in Fig.~\ref{fig:prodsAgain} for the reader convenience.  We
start expanding formula~(\ref{eq:CoherenceFormula}) for $n=3$:
\begin{eqnarray}\label{eq:CoherenceFormulaCase_3}
  \bigvee_{i=1}^3 \left( \right. & \!\!R^E_i &
  \!\!\left. \bigtriangledown_{i+1}^3 \left( \overline{e^E_x} \, r^E_y
    \right) \vee L^E_i \bigtriangleup_1^{i-1} \left( e^E_y \,
      \overline{r^E_x} \right) \right) = R^E_1 \left( \overline{e^E_2}
    \, r^E_2 \vee \overline{e^E_2} \, \overline{e^E_3} \, r^E_3
  \right) \vee \nonumber \\
  & \!\!\vee & \!\!\! R^E_2 \overline{e^E_3} r^E_3 \vee L^E_2
  \overline{r^E_1} \, e^E_1 \vee L^E_3 \left( \overline{r^E_1} \,
    \overline{r^E_2} \, e^E_1 \vee \overline{r^E_2} \, e^E_2 \right) =
  \nonumber \\
  & \!\!= & \!\!\!R^E_1 \left( r^E_2 \vee \overline{e^E_2} \, r^E_3
  \right) \vee R^E_2 r^E_3 \vee L^E_2 e^E_1 \vee L^E_3 \left( e^E_1 \,
    \overline{r^E_2} \vee e^E_2 \right). \nonumber
\end{eqnarray}
which should be zero.

Note that this equation applies to concatenation $s = q_3;q_2;q_1$ and
thus we have to map $(1,2,3) \mapsto (2,3,1)$ to obtain
\begin{equation}\label{eq:CohThreeProds}
  \underbrace{R^E_2 \left( r^E_3 \vee \overline{e^E_3} \, r^E_1
    \right)}_{(*)} \vee  \underbrace{R^E_3 r^E_1 \vee L^E_3
    e^E_2}_{(**)} \vee  \underbrace{L^E_1 \left( e^E_2 \,
      \overline{r^E_3} \vee e^E_3 \right)}_{(***)} = 0.
\end{equation}

Before checking whether these expressions are zero or not, we have to
complete the involved matrices.  All calculations have been divided
into three steps and, as they are operated with \textbf{or}, the
result will not be null if one fails to be zero.

Only the second term \textbf{(**)} is expanded, with ordering of nodes
not specified for a matter of space.  Nodes are sorted $[2\; 3\; 5\;
1\; 4]$ both by columns and by rows, meaning for example that element
$(3,4)$ is an edge starting in node 5 and ending in node 1.
\begin{equation}
  \left[
    \begin{array}{ccccc}
      \vspace{-6pt}
      1 & 0 & 0 & 1 & 0 \\
      \vspace{-6pt}
      0 & 0 & 0 & 0 & 0 \\
      \vspace{-6pt}
      1 & 0 & 1 & 0 & 0 \\
      \vspace{-6pt}
      0 & 0 & 1 & 0 & 0 \\
      \vspace{-6pt}
      0 & 0 & 0 & 0 & 0 \\
      \vspace{-10pt}
    \end{array} \right]
  \left[
    \begin{array}{ccccc}
      \vspace{-6pt}
      0 & 1 & 0 & 0 & 0 \\
      \vspace{-6pt}
      0 & 1 & 0 & 0 & 0 \\
      \vspace{-6pt}
      0 & 1 & 0 & 0 & 0 \\
      \vspace{-6pt}
      0 & 0 & 0 & 0 & 0 \\
      \vspace{-6pt}
      0 & 0 & 0 & 0 & 0 \\
      \vspace{-10pt}
    \end{array} \right]
  \vee
  \left[
    \begin{array}{ccccc}
      \vspace{-6pt}
      0 & 0 & 0 & 1 & 0 \\
      \vspace{-6pt}
      1 & 0 & 0 & 0 & 0 \\
      \vspace{-6pt}
      1 & 0 & 0 & 0 & 0 \\
      \vspace{-6pt}
      0 & 0 & 0 & 1 & 0 \\
      \vspace{-6pt}
      0 & 0 & 0 & 0 & 0 \\
      \vspace{-10pt}
    \end{array} \right]
  \left[
    \begin{array}{ccccc}
      \vspace{-6pt}
      0 & 1 & 0 & 0 & 0 \\
      \vspace{-6pt}
      0 & 0 & 0 & 0 & 0 \\
      \vspace{-6pt}
      0 & 0 & 1 & 0 & 0 \\
      \vspace{-6pt}
      0 & 0 & 0 & 0 & 0 \\
      \vspace{-6pt}
      0 & 0 & 0 & 0 & 0 \\
      \vspace{-10pt}
    \end{array} \right]
  = 0, \nonumber
\end{equation}

\noindent so the sequence is coherent\footnote{It is also necessary to
  check that $(*) = (***) = 0$.} where, as usual, a matrix filled up
with zeros is represented by $0$.

Now consider sequence $s'_3 = q_2;q_3;q_1$ where $q_2$ and $q_3$ have
been swapped with respect to $s_3$.  The condition for its coherence
is:
\begin{equation}\label{eq:CoherenceCounterExample}
  \underbrace{R^E_1 \left( r^E_3 \vee \overline{e^E_3} \, r^E_2 \right)}_{(*)} \vee 
  \underbrace{R^E_3 r^E_2 \vee L^E_3 e^E_1}_{(**)} \vee 
  \underbrace{L^E_2 \left( e^E_1 \, \overline{r^E_3} \vee e^E_3 \right)}_{(***)}=0.
\end{equation}

If we focus just on the first term \textbf{(*)} in
equation~(\ref{eq:CoherenceCounterExample})
\begin{equation}
  \left[
    \begin{array}{ccccc}
      \vspace{-6pt}
      0 & 1 & 1 & 0 & 0 \\
      \vspace{-6pt}
      0 & 1 & 0 & 0 & 0 \\
      \vspace{-6pt}
      0 & 1 & 1 & 0 & 0 \\
      \vspace{-6pt}
      0 & 0 & 0 & 0 & 0 \\
      \vspace{-6pt}
      0 & 0 & 0 & 0 & 0 \\
      \vspace{-10pt}
    \end{array} \right]
  \left(
    \left[
      \begin{array}{ccccc}
        \vspace{-6pt}
        1 & 0 & 0 & 0 & 0 \\
        \vspace{-6pt}
        0 & 0 & 0 & 0 & 0 \\
        \vspace{-6pt}
        0 & 0 & 1 & 0 & 0 \\
        \vspace{-6pt}
        0 & 0 & 1 & 0 & 0 \\
        \vspace{-6pt}
        0 & 0 & 0 & 0 & 0 \\
        \vspace{-10pt}
      \end{array} \right]
    \vee
    \left[
      \begin{array}{ccccc}
        \vspace{-6pt}
        1 & 1 & 1 & 1 & 1 \\
        \vspace{-6pt}
        0 & 1 & 1 & 1 & 1 \\
        \vspace{-6pt}
        1 & 1 & 1 & 1 & 1 \\
        \vspace{-6pt}
        1 & 1 & 1 & 0 & 1 \\
        \vspace{-6pt}
        1 & 1 & 1 & 1 & 1 \\
        \vspace{-10pt}
      \end{array} \right]
    \left[
      \begin{array}{ccccc}
        \vspace{-6pt}
        0 & 0 & 0 & 0 & 1 \\
        \vspace{-6pt}
        1 & 0 & 0 & 0 & 0 \\
        \vspace{-6pt}
        1 & 0 & 0 & 0 & 0 \\
        \vspace{-6pt}
        0 & 0 & 0 & 0 & 0 \\
        \vspace{-6pt}
        0 & 0 & 0 & 0 & 0 \\
        \vspace{-10pt}
      \end{array} \right]
  \right) \nonumber
\end{equation}
we obtain a matrix filled up with zeros except in position (3,3) which
corresponds to an edge that starts and ends in node $5$.  Ordering of
nodes has been omitted again due to lack of space, but it is the same
as above: $[2\; 3\; 5\; 1\; 4]$.

We do not only realize that the sequence is not coherent, but in
addition information on which node or edge may present problems when
applied to an actual host graph is provided.\proofend

Note that a sequence not being coherent does not necessarily mean that
the grammar is not well defined, but that we have to be especially
careful when applying it to a host graph because it is mandatory for
the match to identify all problematic parts in different places.

This information could be used when actually finding the match; a
possible strategy, if parallel matching for different productions is
required, is to start with those elements which may present a
problem.\footnote{The same remark applies to \emph{G-congruence}, to
  be studied in Sec.~\ref{sec:gCongruence}.}

This section ends providing a simple interpretation of $\nabla$ and
$\bigtriangleup$, which in essence are a generalization of the
structure of a sequence of productions.  A sequence $p_2;p_1$ is a
complex operation: To some potential digraph, one should start by
deleting elements specified by $e_1$, then add elements in $r_1$,
afterwards delete elements in $e_2$ and finally add elements in $r_2$.
\emph{Generalization} means that this same structure can be applied
but not limited to matrices $e$ and $r$, i.e. there is an alternate
sequence of ``delete'' and ``add'' operations with general expressions
rather than just matrices $e$ and $r$.  For example, $\nabla_1^3
\left( \overline{e_x} R_x \vee L_y \vee r_y \right)$.

Operators $\nabla$ and $\bigtriangleup$ represent ascending and
descending sequences.  For example, $\nabla_1^3 \overline{e_x}r_y =
p_1p_2(r_3)$ and $\bigtriangleup_1^3 \overline{e_x} r_y =
p_3p_2(r_1)$.  In some detail:
\begin{eqnarray}
  \nabla_1^3 \overline{e}_x \, r_y & = & \overline{e}_1 r_1 \vee
  \overline{e}_1 \, \overline{e}_2 \, r_2 \vee \overline{e}_1 \,
  \overline{e}_2 \, \overline{e}_3 \, r_3 = \nonumber \\
  & = & r_1 \vee \overline{e}_1 r_2 \vee \overline{e}_1 \,
  \overline{e}_2 r_3 = r_1 \vee \overline{e}_1 \left( r_2 \vee
    \overline{e}_2 r_3 \right) = p_1\left( p_2 \left( r_3 \right)
  \right). \nonumber
\end{eqnarray}

We will make good use of this interpretation in
Chap.~\ref{ch:matching} to establish the equivalence between coherence
plus compatibility of a derivation and finding its minimal and
negative initial digraphs in the host graph and its negation,
respectively.

As commented above, we shall return to coherence in
Sec.~\ref{sec:coherenceRevisited}, which is further generalized
in~\cite{MGG_Combinatorics} through so-called \emph{Boolean
  complexes}.

\section{Coherence Revisited}
\label{sec:coherenceRevisited}

In this section we shall extend the results of
Sec.~\ref{sec:sequencesAndCoherence} taking into account potential
dangling edges. To this end we need to introduce the nihil matrix $K$,
which will be very useful in the rest of the book.

Our plan now is to first make explicit all elements that should not be
present in a potential match of the left hand side of a rule in a host
graph, and then characterize them for a finite sequence. This is
carried out defining something similar to the minimal initial digraph,
the \emph{negative initial digraph}. In order to keep our philosophy
of making our analysis as general as possible (independent of any
concrete host graph) only the elements appearing on the LHS of the
productions that make up the sequence plus their actions will be taken
into account.

We will refer to elements that should not be present as
\emph{forbidden elements}. There are two sets of elements that for
different reasons should not appear in a potential initial digraph:

\begin{enumerate}
\item Edges added by the production, as we are limited for now to
  simple digraphs.
\item Edges incident to some node deleted by the production (dangling
  edges).
\end{enumerate}

\index{nihilation matrix}To consider elements just described, the
notation to represent productions is extended with a new graph $K$
that we will call the \emph{nihilation matrix}.\footnote{It will be
  normally represented by $K$. Subscripts will be used to distinguish
  nihil matrices of different productions, e.g. $K_2$ for the nihil
  matrix of production $p_2$. When dealing with sequences, e.g.
  sequence $s_3$, we shall prefer the notation $K(s_3)$.} Note that
the concept of grammar rule remains unaltered because we are just
making explicit some implicit information.

To further justify the naturalness of this matrix let's oppose its
meaning to that of the LHS and its interpretation as a \emph{positive
  application condition} (the LHS must exist in the host graph in
order to apply the grammar rule). In effect, $K$ can be seen as a
\emph{negative application condition}: If it is found in the host
graph then the production can not be applied. We will dedicate a whole
chapter (Chap.~\ref{ch:restrictionsOnRules}) to develop these
ideas.\footnote{In a negative application condition we will be allowed
  to add information of what elements must not be present. Probably it
  is more precise to speak of $K$ as an \emph{implicit negative
    application condition}.}

The order in which matrices are derived is enlarged to cope with the
nihilation matrix $K$:
\begin{equation}
  \left( L, R \right) \longmapsto \left( e, r \right) \longmapsto K.
\end{equation}
Otherwise stated, a production is \emph{statically} determined by its
left and right hand sides $p = \left( L, R \right)$, from which it is
possible to give a \emph{dynamic} definition $p = \left(L, e, r
\right)$, to end up with a full specification including its
\emph{environmental}\footnote{Environmental because $K$ specifies some
  elements in the surroundings of $L$ that should not exist. If the
  LHS has been completed -- probably because it belongs to some
  sequence -- then the nihilation matrix will consider those nodes
  too.}  behaviour $p = \left(L, K, e, r \right)$.

\newtheorem{dynamicProd}[matrixproduct]{Definition}
\begin{dynamicProd}[Production - Dynamic
  Formulation]\label{def:dynamicProduction}
  \index{production!dynamic formulation}A production $p$ is a
  morphism\footnote{In fact, a partial function since some elements in
    $L$ do not have an image in $R$.} between two simple digraphs $L$
  and $R$, and can be specified by the tuple
  \begin{equation}\label{eq:dynProdDefinition}
    p = \left( L^E, K^E, e^E, r^E, L^N, K^N, e^N, r^N \right).
  \end{equation}
\end{dynamicProd}

Compare with Dfinition~\ref{def:prodDef}, the static formulation of
production.  As commented earlier in the book, it should be possible
to consider nodes and edges together using the tensorial construction
of Chap.~\ref{ch:reachability}.

Next lemma shows how to calculate $K$ using the production $p$, by
applying it to a certain matrix:

\newtheorem{nihilationMatrix}[matrixproduct]{Lemma}
\begin{nihilationMatrix}[Nihilation
  matrix]\label{lemma:nihilationMatrix}
  Using tensor notation (see Sec.~\ref{sec:tensorAlgebra}) let's
  define $D = \overline{e^N} \otimes \overline{\left( e^N
    \right)}^{\,t}$, where $t$ denotes transposition.  Then,
  \begin{equation}\label{eq:nihilMatrix}
    K^E = p \left( \overline{D} \right).
  \end{equation}
\end{nihilationMatrix}

\noindent \emph{Proof}\\*
$\square$The following matrix specifies potential dangling edges
incident to nodes appearing in the left hand side of $p$:

\begin{equation}
  \overline{D} = d^i_{\!j} = \left\{
    \begin{array}{ll}
      1 & \qquad if \; \left( e^i \right)^N = 1 \; or \; \left(e_j
      \right)^N = 1. \\
      0 & \qquad otherwise.
    \end{array} \right.
\end{equation}

Note that $D = \overline{e^N} \otimes \overline{\left( e^N
  \right)}^{\,t}$.  Every element incident to a node that is going to
be deleted becomes dangling except edges deleted by the production. In
addition, edges added by the rule can not be present, thus we have
$K^E = r^E \vee \overline{e^E} \left( \overline{D} \right) = p \left(
  \overline{D} \right)$.\proofend

\begin{figure}[htbp]
  \centering
  \includegraphics[scale = 0.62]{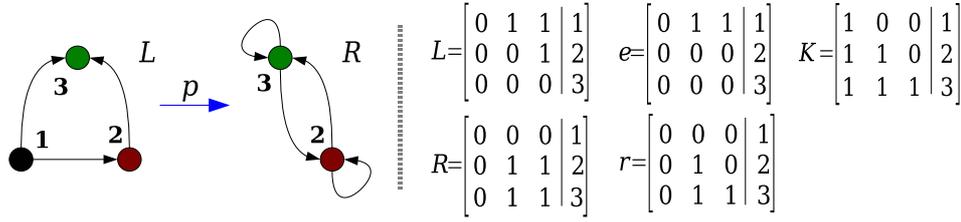}
  \caption{Example of Nihilation Matrix}
  \label{fig:nihilationMatrix}
\end{figure}

\noindent\textbf{Example}.$\square$We will calculate the elements appearing in
Lemma~\ref{lemma:nihilationMatrix} for the production of
Fig.~\ref{fig:nihilationMatrix}:
\begin{equation}
  \overline{\overline{e^N} \otimes \overline{\left( e^N \right)}^{\,
      t}} = 
  \overline{ \left[
      \begin{array}{c}
        \vspace{-6pt}
        0 \\
        \vspace{-6pt}
        1 \\
        \vspace{-6pt}
        1 \\
        \vspace{-10pt}
      \end{array}
    \right] \otimes
    \left[
      \begin{array}{c}
        \vspace{-6pt}
        0 \\
        \vspace{-6pt}
        1 \\
        \vspace{-6pt}
        1 \\
        \vspace{-10pt}
      \end{array}
    \right]^t} =
  \left[
    \begin{array}{ccc}
      \vspace{-6pt}
      1 & 1 & 1 \\
      \vspace{-6pt}
      1 & 0 & 0 \\
      \vspace{-6pt}
      1 & 0 & 0 \\
      \vspace{-10pt}
    \end{array}
  \right]
  \nonumber
\end{equation}
The nihilation matrix is given by equation~(\ref{eq:nihilMatrix}):
\begin{equation}
  K = r \vee \overline e \overline{D} =
  \left[
    \begin{array}{ccc}
      \vspace{-6pt}
      0 & 0 & 0 \\
      \vspace{-6pt}
      0 & 1 & 0 \\
      \vspace{-6pt}
      0 & 1 & 1 \\
      \vspace{-10pt}
    \end{array}
  \right] \vee
  \left[
    \begin{array}{ccc}
      \vspace{-6pt}
      1 & 0 & 0 \\
      \vspace{-6pt}
      1 & 1 & 0 \\
      \vspace{-6pt}
      1 & 1 & 1 \\
      \vspace{-10pt}
    \end{array} 
  \right]
  \left[
    \begin{array}{ccc}
      \vspace{-6pt}
      1 & 1 & 1 \\
      \vspace{-6pt}
      1 & 0 & 0 \\
      \vspace{-6pt}
      1 & 0 & 0 \\
      \vspace{-10pt}
    \end{array} 
  \right] =
  \left[
    \begin{array}{ccc}
      \vspace{-6pt}
      1 & 0 & 0 \\
      \vspace{-6pt}
      1 & 1 & 0 \\
      \vspace{-6pt}
      1 & 1 & 1 \\
      \vspace{-10pt}
    \end{array}
  \right]. \nonumber
\end{equation}

This matrix shows that node 1 can not have a self loop (it would
become a dangling edge as it is not deleted by the production) but
edges $(1,2)$ and $(1,3)$ may be present (in fact they must be present
as they belong to $L$). Edge $(2,1)$ must not exist for the same
reason. The self loop for node 2 can not be found because it is added
by the rule. A similar reasoning tells us that no edge starting in
node 3 can exist: The self loop and edge $(3,2)$ because they are
going to be added and $(3,1)$ because it would become a dangling
edge.\proofend

It is worth stressing that matrix $\overline{D}$ do not tell actions
of the production to be performed in the complement of the host graph,
$\overline{G}$. Actions of productions are specified exclusively by
matrices $e$ and $r$.

Some questions of importance remain unsolved regarding forbidden
elements and productions: How are the elements in 
the nihil matrix transformed by a production $p$? Otherwise stated,
if the forbidden elements in the LHS of the production are those given
by $K$, what are the forbidden elements in the RHS according to $p$?

Although this question will be studied in detail in
Sec.~\ref{sec:movingConditions} -- in particular in
Prop.~\ref{th:Nevolution} on p.~\pageref{th:Nevolution} -- we need to
advance the answer: for a production $p: L \rightarrow R$
with nihil part $K$, the forbidden elements (we shall use the letter
$Q$) are given by inverse of the grammar rule:
\begin{equation*}
  Q = p^{-1}(K).
\end{equation*}

Now we are in the position to extend the results of
Sec.~\ref{sec:sequencesAndCoherence} by considering potential dangling
edges. We shall prove that:

\newtheorem{CoherenceTheorem2}[matrixproduct]{Theorem}
\begin{CoherenceTheorem2}\label{th:SeqCoherenceTheorem2}
  The concatenation $s_n = p_n;\ldots;p_1$ is coherent if besides
  eq.~\eqref{eq:CoherenceFormula}, identity
  \begin{equation}
    \label{eq:CoherenceFormula2}
    \bigvee_{i=1}^n \left( Q_i \bigtriangledown_{i+1}^n \left(
        e_y \, \overline{r}_x \right) \vee K_i \bigtriangleup_1^{i-1}
      \left( r_y \, \overline{e}_x \right) \right).
  \end{equation}
  is also fulfilled.
\end{CoherenceTheorem2}

\noindent \emph{Proof}\\*
$\square$We proceed as for
Theorem~\ref{th:SeqCoherenceTheorem}. First, let's consider a sequence
of two productions $s_2 = p_2;p_1$. In order to decide whether the
application of $p_1$ does not exclude $p_2$ (regarding elements that
appear in the nihil parts) the following conditions must be demanded:
\begin{enumerate}
\item No common element is deleted by both productions:
  \begin{equation}
    \label{eq:20}
    e_1 e_2 = 0.
  \end{equation}
\item Production $p_2$ does not delete any element that the production
  $p_1$ demands not to be present and that besides is not added by
  $p_1$:
  \begin{equation}
    \label{eq:21}
    e_2 K_1 \overline{r}_1 = 0.
  \end{equation}
\item The first production does not add any element that is demanded
  not to exist by the second production:
  \begin{equation}
    \label{eq:23}
    r_1 K_2 = 0.
  \end{equation}
\end{enumerate}
Altogether we can write
\begin{equation}
  \label{eq:47}
  e_1 e_2 \vee \overline{r}_1 e_2 K_1 \vee r_1 K_2 = e_2 (e_1 \vee
  \overline{r}_1 K_1) \vee r_1 K_2 = e_2 Q_1 \vee r_1 K_2 = 0,
\end{equation}
which is equivalent to
\begin{equation}
  \label{eq:44}
  e_2 \overline{r}_2 Q_1 \vee \overline{e}_1 r_1 K_2 = 0
\end{equation}
due to basic properties of MGG productions (see
Prop.~\ref{prop:simpleEqualities}).

In the case of a sequence that consists of three productions, $s_3 =
p_3;p_2;p_1$, the procedure is to apply the same reasoning to
subsequences $p_2;p_1$ (restrictions on $p_2$ actions due to $p_1$)
and $p_3;p_2$ (restrictions on $p_3$ actions due to $p_1$) and
\textbf{or} them. Finally, we have to deduce which conditions have to
be imposed on the actions of $p_3$ due to $p_1$, but this time taking
into account that $p_2$ is applied in between. Again, we can put all
conditions in a single expression:
\begin{equation}
  \label{eq:36}
  Q_1 \left( e_2 \vee \overline{r}_2 e_3 \right) \vee Q_2 e_3 \vee
  K_2 r_1 \vee K_3 \left( r_1 \overline{e}_2 \vee r_2 \right) = 0.
\end{equation}

\begin{table}[htbp]
  \centering
  \begin{tabular}{||c|c||c|c||c|c||}
    \hline
    \phantom{HH} $D_2;D_1$ \phantom{HH} & \phantom{H} \eqref{eq:23}
    \phantom{H} & \phantom{HH} $D_2;P_1$ \phantom{HH} & \phantom{H}
    $\surd$ \phantom{H} & \phantom{HH} $D_2;A_1$ \phantom{HH} &
    \phantom{H} $\surd$ \phantom{H} \\
    \hline
    $P_2;D_1$ & \eqref{eq:23} & $P_2;P_1$ & $\surd$ & $P_2;A_1$ &
    $\surd$ \\ 
    \hline
    $A_2;D_1$ & $\surd$ & $A_2;P_1$ & \eqref{eq:21} & $A_2;A_1$ &
    \eqref{eq:20} \\
    \hline
  \end{tabular}
  \caption{Possible Actions (Two Productions Incl. Dangling Edges)}
  \label{tab:actionsForTwoProductionsNeg}
\end{table}

We now check that eqs.~\eqref{eq:44}~and~\eqref{eq:36} do imply
coherence. To see that eq.~\eqref{eq:44} implies coherence we only
need to enumerate all possible actions on the nihil parts. It might be
easier if we think in terms of the negation of a potential host graph
to which both productions would be applied $\left( \overline{G}
\right)$ and check that any problematic situation is ruled out. See
table~\ref{tab:actionsForTwoProductionsNeg} where \textbf{D} is
deletion of one element from $\overline{G}$ (i.e., the element is
added to $G$), \textbf{A} is addition to $G$ and \textbf{P} is
preservation. Notice that these definitions of \textbf{D}, \textbf{A}
and \textbf{P} are opposite to those given for the certainty case
above.\footnote{Preservation means that the element is demanded to be
  in $\overline{G}$ because it is demanded not to exist by the
  production (it appears in $K_1$) and it remains as non-existent
  after the application of the production (it appears also in $Q_1$).}
For example, action $A_2;A_1$ tells that in first place $p_1$ adds one
element $\varepsilon$ to $\overline{G}$.  To do so this element has to
be in $e_1$, or incident to a node that is going to be deleted. After
that, $p_2$ adds the same element, deriving a conflict between the
rules.

So far we have checked coherence for the case $n = 2$. When the
sequence has three productions, $s = p_3;p_2;p_1$, there are $27$
possible combinations of actions. However, some of them are considered
in the subsequences $p_2;p_1$ and $p_3;p_2$. Table
\ref{tab:actionsForThreeProductions} summarizes them.

\begin{table}[htbp]
  \centering
  \begin{tabular}{||c|c||c|c||c|c||}
    \hline
    \phantom{H} $D_3;D_2;D_1$ \phantom{H} & \phantom{H}
    \eqref{eq:23} \phantom{H} & \phantom{H} $D_3;D_2;P_1$
    \phantom{H} & \phantom{H} \eqref{eq:23} \phantom{H} &
    \phantom{H} $D_3;D_2;A_1$ \phantom{H} & \phantom{H}
    \eqref{eq:23} \phantom{H} \\
    \hline
    \phantom{H} $P_3;D_2;D_1$ \phantom{H} & \phantom{H}
    \eqref{eq:23} \phantom{H} & \phantom{H} $P_3;D_2;P_1$
    \phantom{H} & \phantom{H} \eqref{eq:23} \phantom{H} &
    \phantom{H} $P_3;D_2;A_1$ \phantom{H} & \phantom{H}
    \eqref{eq:23} \phantom{H} \\
    \hline
    \phantom{H} $A_3;D_2;D_1$ \phantom{H} & \phantom{H}
    \eqref{eq:23} \phantom{H} & \phantom{H} $A_3;D_2;P_1$
    \phantom{H} & \phantom{H} $\surd$ \phantom{H} & \phantom{H}
    $A_3;D_2;A_1$ \phantom{H} & \phantom{H} $\surd$ \phantom{H} \\
    \hline
    \phantom{H} $D_3;P_2;D_1$ \phantom{H} & \phantom{H}
    \eqref{eq:23} \phantom{H} & \phantom{H} $D_3;P_2;P_1$
    \phantom{H} & \phantom{H} $\surd$ \phantom{H} & \phantom{H}
    $D_3;P_2;A_1$ \phantom{H} & \phantom{H} $\surd$ \phantom{H} \\
    \hline
    \phantom{H} $P_3;P_2;D_1$ \phantom{H} & \phantom{H}
    \eqref{eq:23} \phantom{H} & \phantom{H} $P_3;P_2;P_1$
    \phantom{H} & \phantom{H} $\surd$ \phantom{H} & \phantom{H}
    $P_3;P_2;A_1$ \phantom{H} & \phantom{H} $\surd$ \phantom{H} \\
    \hline
    \phantom{H} $A_3;P_2;D_1$ \phantom{H} & \phantom{H}
    \eqref{eq:23}/\eqref{eq:21} \phantom{H} & \phantom{H} $A_3;P_2;P_1$
    \phantom{H} & \phantom{H} \eqref{eq:21} \phantom{H} & \phantom{H}
    $A_3;P_2;A_1$ \phantom{H} & \phantom{H} \eqref{eq:21} \phantom{H} \\
    \hline
    \phantom{H} $D_3;A_2;D_1$ \phantom{H} & \phantom{H}
    $\surd$ \phantom{H} & \phantom{H} $D_3;A_2;P_1$
    \phantom{H} & \phantom{H} \eqref{eq:21} \phantom{H} & \phantom{H}
    $D_3;A_2;A_1$ \phantom{H} & \phantom{H} \eqref{eq:20} \phantom{H} \\
    \hline
    \phantom{H} $P_3;A_2;D_1$ \phantom{H} & \phantom{H}
    $\surd$ \phantom{H} & \phantom{H} $P_3;A_2;P_1$
    \phantom{H} & \phantom{H} \eqref{eq:21} \phantom{H} & \phantom{H}
    $P_3;A_2;A_1$ \phantom{H} & \phantom{H} \eqref{eq:20} \phantom{H} \\
    \hline
    \phantom{H} $A_3;A_2;D_1$ \phantom{H} & \phantom{H}
    \eqref{eq:20} \phantom{H} & \phantom{H} $A_3;A_2;P_1$
    \phantom{H} & \phantom{H} \eqref{eq:20} \phantom{H} & \phantom{H}
    $A_3;A_2;A_1$ \phantom{H} & \phantom{H} \eqref{eq:20} \phantom{H} \\
    \hline
  \end{tabular}
  \caption{Possible Actions (Three Productions Incl. Dangling Edges)}
  \label{tab:actionsForThreeProductions}
\end{table}

There are four forbidden actions:\footnote{Those actions appearing in
  table~\ref{tab:actionsForTwoProductions} updated for $p_3$.}
$D_3;D_1$, $A_3;P_1$, $P_3;D_1$ and $A_3;A_1$. Let's consider the
first one, which corresponds to $r_1 r_3$ (the first production adds
the element -- it is erased from $\overline{G}$ -- and the same for
$p_3$). In Table~\ref{tab:actionsForThreeProductions} we see that
related conditions appear in positions $(1,1)$, $(4,1)$ and $(7,1)$.
The first two are ruled out by conflicts detected in $p_2;p_1$ and
$p_3;p_2$, respectively. We are left with the third case which is in
fact allowed. The condition $r_3 r_1$ taking into account the presence
of $p_2$ in the middle in eq.~\eqref{eq:36} is contained in $K_3 r_1
\overline{e}_2$, which includes $r_1 \overline{e}_2 r_3$. This must be
zero, i.e. it is not possible for $p_1$ and $p_3$ to remove from
$\overline{G}$ one element if it is not added to $\overline{G}$ by
$p_2$. The other three forbidden actions can be checked similarly.

The proof can be finished by induction on the number of productions.
The induction hypothesis leaves again four cases: $D_n;D_1$,
$A_n;P_1$, $P_n;D_1$ and $A_n;A_1$. The corresponding table changes
but it is not difficult to fill in the details. \proofend

There are some duplicated conditions, so it could be possible to
``optimize''
equations~\eqref{eq:CoherenceFormula}~and~\eqref{eq:CoherenceFormula2}. The
form considered in
Theorems~\ref{th:SeqCoherenceTheorem}~and~\ref{th:SeqCoherenceTheorem2}
is preferred because we may use $\bigtriangleup$ and
$\bigtriangledown$ to synthesize the expressions. Some comments on
previous proof follow:
\begin{enumerate}
\item Notice that eq.~\eqref{eq:20} is already considered in
  Theorem~\ref{th:SeqCoherenceTheorem} because
  eq.~\eqref{eq:FirstCondTwoProd} which demands $e_1 L_2 = 0$ (as $e_2
  \subset L_2$ we have that $e_1 L_2 = 0 \Rightarrow e_1 e_2 = 0$).
\item Condition~\eqref{eq:21} is $e_2 K_1 \overline{r}_1 = e_2
  \overline{r}_1 r_1 \vee e_2 \overline{r}_1 \overline{e}_1
  \overline{D}_1 = e_2 \overline{e}_1 \overline{D}_1$, where we have
  used that $K_1 = p \left( \overline{D}_1 \right)$. Note that those
  $\overline{e}_1 \overline{D}_1 \neq 0$ are the dangling edges not
  deleted by $p_1$.
\item Equation~\eqref{eq:23} is $r_1 K_2 = r_1 p_2 \left(
    \overline{D}_2 \right) = r_1 \left( r_2 \vee \overline{e}_2
    \overline{D}_2 \right) = r_1 r_2 \vee r_1 \overline{e}_2
  \overline{D}_2$. The first term $(r_1 r_2)$ is already included in
  Theorem~\ref{th:SeqCoherenceTheorem} and the second term is again
  related to dangling edges.
\end{enumerate}

Potential dangling edges appear in coherence which might indicate a
possible link between coherence and compatibility. Compatibility for
sequences is characterized in
Sec.~\ref{sec:compositionAndCompatibility}). Coherence takes into
account dangling edges, but only those that appear in the ``actions''
of the productions (in matrices $e$ and $r$).

\section{Summary and Conclusions}
\label{sec:summaryAndConclusions4_1}

In this chapter we have introduced two equivalent definitions of
production, one emphasizing the static part of grammar rules and the
other stressing its dynamics.

Also, completion has been addressed. To some extent it allows us to
study productions, forgetting about the state to which the rule is to
be applied. It provides us with a means to relate elements in different
graphs, a kind of horizontal identification of elements among the
rules in a sequence.

Sequences of productions have been introduced together with
compatibility and coherence. The first ensures that the underlying
structure (simple digraph) is kept, i.e. it is closed under the
operations defined in the sequence. Coherence guarantees that actions
specified by one production do not disturb productions following it.

Coherence can be compared with \emph{critical pairs}, used in the
categorical approach to graph grammars to detect conflicts between
grammar rules. There are differences, though. The main one is that
coherence in our approach covers any finite sequence of productions
while critical pairs are limited to two productions. Among other
things, coherence would be able to detect if a potential problem
between two productions is actually fixed by some intermediate rule.

In this and the next chapter (devoted to initial digraphs and
composition) we develop some analytical techniques independent to
some extent of the initial state of the system to which the grammar
rules will be applied. This allows us to obtain information about
grammar rules themselves, for example at design time. This information
may be useful during runtime.  We will return to this point in future
chapters.
\chapter{Initial Digraphs and Composition}
\label{ch:mggFundamentals2}

In this chapter, which builds in Chapter~\ref{ch:mggFundamentals1}, we
will mainly deal with initial digraphs and composition, providing
more analysis techniques independent to some extent of the initial
state of the grammar.

Initial digraphs (minimal and negative) are simple digraphs with
enough elements to permit the application of a given sequence. They
can be thought of as a proxy of a real initial state. The advantage is
that they allow us to study a grammar without considering a concrete
initial state.

Composition is an operation that defines a single production out of a
given sequence of productions. In some sense, composition and
concatenation (sequentialization, studied in
Chapter~\ref{ch:mggFundamentals1}) are opposite operations.

These analysis techniques (initial digraphs and composition) will be
of importance in addressing the problems posed in
Chapter~\ref{ch:introduction}. In particular they will be used to
tackle applicability (problem~\ref{prob:applicability}), sequential
independence (problem~\ref{prob:sequentialIndependence}) and
reachability (problem~\ref{prob:reachability}).

This chapter is organized as follows. The problem of finding those
elements that must be present (\emph{minimal initial digraph}) or must
not appear (\emph{negative initial digraph}) are addressed in
Secs.~\ref{sec:MID}~and~\ref{sec:NID}.  At times it is of
interest to build a rule that performs the same actions than a given
coherent sequence but is applied in a single step, i.e. no
intermediate states are generated.  This is \emph{composition}, as
normally defined in mathematics.  As they are related, the definition
of compatibility for a sequence of productions is also introduced and
characterized in Sec.~\ref{sec:compositionAndCompatibility}.  Finally,
as in every chapter, there is a section with a summary and some
conclusions.

\section{Minimal Initial Digraph}
\label{sec:MID}

Compatibility and composition plus matching in MGG are our main
motivations for introducing the concepts and results in this and the
next sections (minimal and negative initial digraphs). Next few
paragraphs clarify these points.

\begin{figure}[htbp]
  \centering \makebox{ \xymatrix{
      L_1 \ar[d]^{m_1} \ar@[blue][r]^{p_1} & R_1 \ar[d]_{m_1^*} & L_2
      \ar[dl]_{m_2} \ar@[blue][r]^{p_2} & R_2 \ar[d]_{m_2^*}& L_3
      \ar[dl]_{m_3} \ar@[blue][r]^{p_3} & R_3
      \ar[d]^{m_3^*} & \ar@{.>}[dl] \\
      G_0 \ar@[red][r]^{p^*_1} & G_1 \ar@[red][rr]^{p_2^*} && G_2
      \ar@[red][rr]^{p_3^*} && G_3 \ar@{.>}@[red][r] &
    } }
  \caption{Example of Sequence and Derivation}
  \label{fig:sequence}
\end{figure}
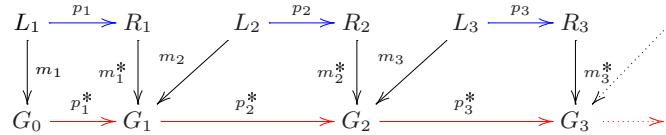

Matches find the left hand side of the production in the host graph
(see Chap.~\ref{ch:matching}) and, as side effect, relate and unrelate
elements among productions. We may think of matching as a
\emph{vertical identification} of nodes -- and hence edges -- relating
as a side effect elements, so to speak, \emph{horizontally} (see
Fig.~\ref{fig:sequence}). For example, if $L_1$ and $L_2$ have each
one a node of type 3 and $m_1:L_1 \rightarrow G_0$ and $m_2:L_2
\rightarrow G_1$ match this node in the same place of $G_0$ and $G_1$
(suppose it is not deleted by $p_1$) then this node is
\emph{horizontally} related. In Sec.~\ref{sec:MID} we will study
in detail this sort of relations.

Compatibility is determined by the result of applying a production to
an initial graph and checking nodes and edges of the result. If we try
to define compatibility for a concatenation or its composition, we
have to decide which is the initial graph (see the next example) but
we would prefer not to begin our analysis of matches yet.

\begin{figure}[htbp]
  \centering
  \includegraphics[scale = 0.57]{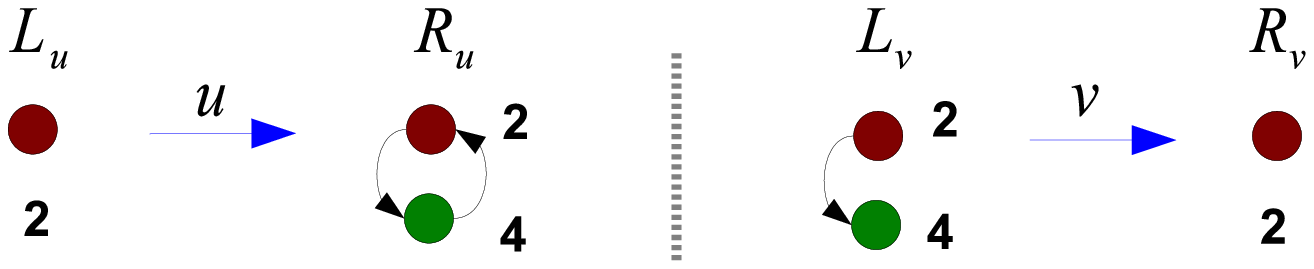}
  \caption{Non-Compatible Productions}
  \label{fig:NonComp}
\end{figure}

\noindent\textbf{Example.}$\square$Consider productions $u$ and $v$ defined in
Fig.~\ref{fig:NonComp}.  It is easy to see that $v;u$ is coherent but
not compatible.  It seems a bit more difficult to define their
composition $v \circ u$, as if they were applied to the same nodes, a
dangling edge would be obtained.  Although coherence itself does not
guarantee applicability of a sequence, we will see that compatibility
is sufficient (generalized to consider concatenations, not only graphs
or single productions as in Defs.~\ref{def:compatibilityDefinition}
and~\ref{def:prodCompatibility}).

Two possibilities are found in the literature (for the categorical
approach) in order to define a match, depending whether DPO or SPO is
followed (see
Secs.~\ref{sec:DPO}~and~\ref{sec:otherCategoricalApproaches}~or~\cite{handbook}).
In the latter, deletion prevails so in the present example production
$v$ would delete edge $(4,2)$. Our approximation to the match of a
production is slightly different, considering it as an operator that
acts on a space whose elements are productions (see
Chap.~\ref{ch:matching}).\footnote{In the SPO approach -- see
  Sec.~\ref{sec:otherCategoricalApproaches} -- rewriting has as side
  effect the deletion of dangling edges. One important difference is
  that in our approach it is defined as an operator that enlarges the
  production or the sequence of productions by adding new ones.}
\proofend

The example shows a problem that led us to consider not only
productions, but also the context in which they are to be applied. In
fact, the minimal context in which they can be applied. This situation
might be overcome if we were able to define a minimal and
unique\footnote{Unique once the concatenation has been completed.
  Minimal initial digraph makes \emph{horizontal identification} of
  elements explicit.} ``host graph'' with enough elements to permit
all operations of a given concatenation or composition of productions,
we would avoid to some extent considering matches and would remain
within the realm of productions alone.

In fact, as we shall see, it is possible to define such graphs. We
name it \emph{minimal initial digraph}.  Note that we were able to
give a definition of compatibility in
Def.~\ref{def:compatibilityDefinition} for a single production because
it is clear (so obvious that we did not mention it) which one is the
minimal initial digraph: Its left hand side.

Any production demands elements to exist in the host graph in order to
be applied. Also, some elements must not be present.  We will touch on
``forbidden'' elements in Sec.~\ref{sec:NID}.  Both are quite
useful concepts because they allow us to ignore matching if staying at
a grammar definition level is desired (to study its potential
behaviour or to define concepts independently of the host graph), and
also the applicability problem (see problem~\ref{prob:applicability})
can be characterized through them.  We will return to these concepts
once matching is introduced and characterized, in
Sec.~\ref{sec:initialDigraphSet} and also in
Chap.~\ref{ch:restrictionsOnRules} when we define graph constraints
and application conditions.

Let's turn to define and characterize minimal initial digraphs. One
graph is known which fulfills all demands of the coherent sequence
$s_n = p_n; \ldots; p_1$ -- namely $\mathscr{L} = \bigvee_{i=1}^n L_i$
-- in the sense that it has enough elements to carry out all
operations specified in the sequence. Graph $\mathscr{L}$ is not
completed (each $L_i$ with respect to the rest). If there are
coherence issues among all grammar rules, then probably all nodes in
all LHS of the rules will be unrelated giving rise to the disjoint
union of $L_i$. If, on the contrary, there are no coherence problems
at all, then we can identify across productions as many nodes of the
same type in $L_i$ as desired.

\newtheorem{minimalInitialDigraph}[matrixproduct]{Definition}
\begin{minimalInitialDigraph}[Minimal Initial
  Digraph]\label{minimalInitialDigraph}
  \index{minimal initial digraph}Let $s_n = p_n ; \ldots ; p_1$ be a
  completed sequence, a \emph{minimal initial digraph} is a simple
  digraph which permits all operations of $s_n$ and does not contain
  any proper subgraph with the same property.
\end{minimalInitialDigraph}

This concept will be slightly generalized in
Sec.~\ref{sec:initialDigraphSet},
Definition~\ref{def:initialDigraphSet}, in which we consider the set
of all potential minimal initial digraphs for a given (non-completed)
sequence and analyze its structure. In fact, $\mathscr{L}$ is not a
digraph but this initial digraph set.  Through completion one actual
digraph can be fixed.

\newtheorem{MinimumProd}[matrixproduct]{Theorem}
\begin{MinimumProd}\label{th:minProdTh}
  Given a completed coherent sequence of productions $s_n = p_n;
  \ldots ;p_1$, the minimal initial digraph is defined by the
  equation:
  \begin{equation}\label{eq:FirstMinDigraph}
    M_n = \nabla_1^n \left( \overline{r_x} L_y \right).
  \end{equation}
\end{MinimumProd}

\begin{figure}[htbp]
  \centering
  \includegraphics[scale = 0.6]{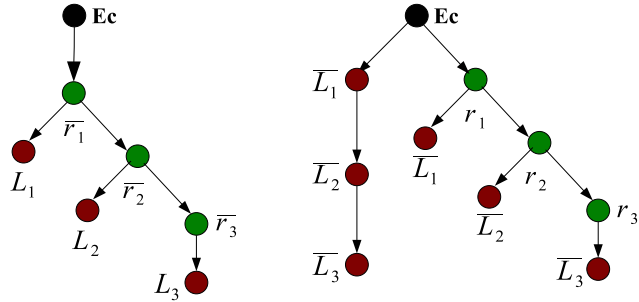}
  \caption{Formulas~(\ref{eq:FirstMinDigraph})~and~(\ref{eq:NegationMinInitDigraph}) for Three Productions}
  \label{fig:MinimumFourProds_1}
\end{figure}

Superscripts are omitted to make formulas easier to read (i.e. they
apply to both nodes and edges). In Fig.~\ref{fig:MinimumFourProds_1}
on p.~\pageref{fig:MinimumFourProds_1},
formula~(\ref{eq:FirstMinDigraph}) and its
negation~(\ref{eq:NegationMinInitDigraph}) are expanded for three
productions.

\noindent \emph{Proof}\\*
$\square$To properly prove this theorem we have to check that $M_n$
has enough edges and nodes to apply all productions in the specified
order, that it is minimal and finally that it is unique (up to
isomorphisms). We will proceed by induction on the number of
productions.

By hypothesis we know that the concatenation is coherent and thus the
application of one production does not exclude the ones coming after
it. In order to see that there are sufficient nodes and edges, it is
enough to check that $s_n \left( \bigvee_{i=1}^n L_i \right) = s_n
\left( M_n \right)$, as the most complete digraph to start with is
$\mathscr{L} = \bigvee_{i=1}^n L_i$, which has enough elements due to
coherence.\footnote{Recall that $\mathscr{L}$ is not completed so it
  somehow represents some digraph with enough elements to apply $s_n$
  to. This is not necessarily the \emph{maximal initial digraph} as
  introduced in Sec.~\ref{sec:initialDigraphSet}.}

If we had a sequence consisting of only one production $s_1 = p_1$,
then it should be obvious that the minimal digraph needed to apply the
concatenation is $L_1$.

In the case of a sequence of two productions, say $s_2 = p_2;p_1$,
what $p_1$ uses $\left( L_1 \right)$ is again needed.  All edges that
$p_2$ uses ($L_2$), except those added ($\overline{r}_1$) by the first
production, are also mandatory.  Note that the elements added ($r_1$)
by $p_1$ are not considered in the minimal initial digraph.  If an
element is preserved (used and not erased, $\overline{e_1} \, L_1$) by
$p_1$, then it should not be taken into account:

\begin{equation}
  \label{eq:minInitDig_tmp1}
  L_1 \vee L_2 \overline{r_1} \, \overline{\left( \overline{e_1} L_1
    \right)} = L_1 \vee L_2 \overline{r_1} \left( e_1 \vee
    \overline{L_1} \right) = L_1 \vee L_2 \overline{R_1}.
\end{equation}

This formula can be paraphrased as ``elements used by $p_1$ plus those
needed by $p_2$'s left hand side, except the ones resulting from
$p_1$'s application''.  It provides enough elements to $s_2$:
\begin{eqnarray}
  p_2;p_1 \left( L_1 \vee L_2 \overline{R_1} \right) & = & r_2
  \vee \overline{e_2} \left( r_1 \vee \overline{e_1} \left( L_1
      \vee L_2 \overline{R_1} \right) \right) = \nonumber \\
  & = & r_2 \vee \overline{e_2} \left( R_1 \vee r_1
    \overline{R_1} L_2 \vee \overline{e_1} \overline{R_1} L_2
  \right) = \nonumber \\
  & = & r_2 \vee \overline{e_2} \left( R_1 \vee r_1 L_2 \vee
    \overline{e_1} L_2 \right) = \nonumber \\
  & = & r_2 \vee \overline{e_2} \left( r_1 \vee \overline{e_1}
    \left( L_1 \vee L_2 \right) \right) = p_2;p_1 \left( L_1
    \vee L_2 \right). \nonumber
\end{eqnarray}

Let's move one step forward with the sequence of three productions
$s_3 = p_3;p_2;p_1$.  The minimal digraph needs what $s_2$ needed
($L_1 \vee L_2 \overline{R_1}$), but even more so.  We have to add
what the third production uses ($L_3$), except what comes out from
$p_1$ and is not deleted by production $p_2$ (this is, $R_1 \,
\overline{e_2}$), and finally remove what comes out ($R_2$) from
$p_2$:
\begin{equation}\label{eq:minInitDig_tmp2}
  M_3 = L_1 \vee L_2 \overline{R_1} \vee L_3 \overline{ \left(
      \overline{e_2} \, R_1 \right)} \, \overline {R_2} = L_1
  \vee L_2 \overline{R_1} \vee L_3 \overline{R_2} \left( e_2
    \vee \overline{R_1} \right).
\end{equation}

Similarly to what has already been done for $s_2$, we check that the
minimal initial digraph has enough elements so it is possible
to apply $p_1$, $p_2$ and $p_3$:
\begin{align*}
  p_3;p_2;p_1 \left( M_3 \right) &= r_3 \vee \overline{e_3} \left( r_2
    \vee \overline{e_2} \left( r_1 \vee \overline{e_1} \left( L_1 \vee
        L_2 \overline{R_1} \vee L_3 \overline{R_2} \left( e_2 \vee
          \overline{R_1} \right) \right) \right) \right) = \\
  &= r_3 \vee \overline{e_3} \left( r_2 \vee \overline{e_2} \left(
      \overline{e_1} L_2 \vee \overline{e_1} e_2 L_3 \overline{R_2}
      \vee \underbrace{R_1 \vee L_3 \overline{e_1} \overline{R_1}
        R_2}_{=R_1 \vee L_3 \overline{e_1} R_2} \right) \right) = \\
  &= r_3 \vee \overline{e_3} \left( \underbrace{\overline{e_2} r_1
      \vee \overline{e_2} \, \overline{e_1} L_1}_{=\overline{e}_2 R_1}
    \vee \overline{e_2} \, \overline{e_1} L_2 \vee \underbrace{ r_2
      \vee L_3 \overline{e_1} \, \overline{e_2} \, \overline{r_2}
      \overline{L_2}}_{=r_2 \vee L_3 \overline{e_1} \, \overline{e_2}
      \overline{L_2}} \right) = \\
  &= r_3 \vee \overline{e_3} \left( r_2 \vee \overline{e_2} \left( r_1
      \vee \overline{e_1} \left( L_1 \vee L_2 \vee L_3 \right) \right)
  \right) = \\
  &= p_3;p_2;p_1 \left( L_1 \vee L_2 \vee L_3 \right).
\end{align*}

The same reasoning applied to the case of four productions yields:
\begin{equation}
  M_4 = L_1 \vee L_2 \overline{R_1} \vee L_3 \overline{ \left(
      \overline{e_2} \, R_1 \right)} \, \overline {R_2} \vee L_4
  \overline{\left( \overline{e_3}\,\overline{e_2} R_1 \right)}\;
  \overline{\left( \overline{e_3} R_2 \right)}\; \overline{R_3}.
\end{equation}

Minimality is inferred by construction, because for each $L_i$ all
elements added by a previous production and not deleted by any
production $p_j$, $j<i$, are removed.  If any other element is erased
from the minimal initial digraph, then some production in $s_n$ would
miss some element.

Now we want to express previous formulas using operators $\nabla$ and
$\bigtriangledown$.  The expression
\begin{equation}
  L_1^E \vee \bigvee_{i=2}^n \left[ L^E_i \bigtriangleup_1^{i-1}
    \left( \overline{R^E_x} \, e^E_y \right) \right]
\end{equation}
is close but we would be adding terms that include $\overline{R_1^E}
e_1^E$, and clearly $\overline{R_1^E} e_1^E \neq \overline{R_1^E}$,
which is what we have in the minimal initial digraph.\footnote{Not in
  formula~(\ref{eq:FirstMinDigraph}) but in expressions derived up to
  now for minimal initial digraph:
  formulas~(\ref{eq:minInitDig_tmp1})~and~(\ref{eq:minInitDig_tmp2}).}
Thus, considering the fact that $\overline{a}b \vee \overline{a} \,
\overline{b} = \overline{a}$ (see Sec.~\ref{sec:logics}) we eliminate
them by performing \textbf{or} operations:

\begin{equation}
  \overline{e_1^E}\bigtriangledown_1^{n-1} \left( \overline{R_x^E}
    L_{y+1} \right).
\end{equation}

we have arrived at a formula for the minimal initial digraph which is
slightly different from that in the theorem:
\begin{equation}
  M_n = L_1 \vee \overline{e_1}\bigtriangledown_1^{n-1} \left(
    \overline{R_x} L_{y+1} \right) \vee \bigvee_{i=2}^n \left[ L_i
    \bigtriangleup_1^{i-1} \left( \overline{R_x} \, e_y \right)
  \right].
  \label{eq:firstFormulaMinInitialDigraph}
\end{equation}

\begin{figure}[htbp]
  \centering
  \includegraphics[scale = 0.6]{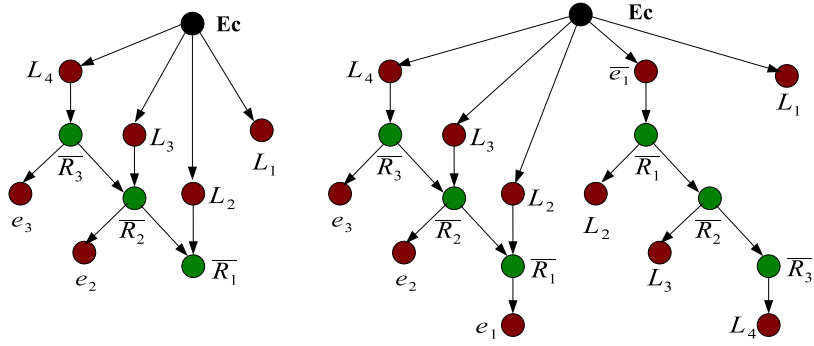}
  \caption{Minimal Initial Digraph (Intermediate Expression). Four
    Productions}
  \label{fig:MinimumFourProds}
\end{figure}

Please refer to Fig.~\ref{fig:MinimumFourProds} where, to the right,
expression~(\ref{eq:firstFormulaMinInitialDigraph}) is represented
while to the left the same equation, but simplified, is depicted for
$n=4$.

Our next step is to show that previous identity is equivalent to

\begin{equation}
  M_n = L_1 \vee \overline{e_1}\bigtriangledown_1^{n-1} \left(
    \overline{r_x} L_{y+1} \right) \vee \bigvee_{i=2}^n \left[ L_i
    \bigtriangleup_1^{i-1} \left( \overline{r_x} \, e_y \right)
  \right],
  \label{eq:secondFormulaMinInitialDigraph}
\end{equation}

\noindent illustrating the way to proceed for $n=3$.  To this end,
equation~(\ref{eq:L_And_Not_e}) is used as well as the fact that $a
\vee \overline{a}b = a \vee b$ (see Sec.~\ref{sec:logics}):

$M_3 = L_1 \vee L_2 \overline{R_1} \vee L_3 \overline{R_2} \left( e_2
  \vee \overline{R_1} \right) = $ \newline $\phantom{111111} = L_1
\vee L_2 \overline{r_1} \left( e_1 \vee \overline{L_1} \right) \vee
\left( L_3 \overline{r_2} e_2 \vee L_3 \overline{r_2} \overline{L_2}
\right) \left( e_2 \vee \overline{r_1} e_1 \overline{r_1}
  \overline{L_1} \right) = $ \newline $\phantom{111111} = L_1 \vee L_2
\overline{r_1} \overline{L_1} \vee L_2 e_1 \vee L_3 e_2 \vee
\underbrace{L_3 e_2 e_1 \vee L_3 e_2 \overline{r_1} \overline{L_1}
  \vee L_3 e_2 \overline{L_2}}_{\textrm{disappears due to }L_3 e_2}
\vee $ \newline $\phantom{11111111111} \vee L_3 \overline{r_2}
\overline{L_2} \overline{r_1} \overline{L_1} \vee L_3 \overline{r_2}
\overline{L_2} e_1 = $ \newline $\phantom{111111} = L_1 \vee L_2
\left( \overline{r_1} \vee e_1 \right) \vee L_3 \overline{L_2}
\overline{r_2} \, \overline{r_1} \vee L_3 e_2 \vee L_3 \overline{L_2}
\overline{r_2} e_1 = $ \newline $\phantom{111111} = L_1 \vee L_2
\overline{r_1} \vee L_3 \overline{r_2} \left( e_2 \vee \overline{r_1}
\right).$

But~(\ref{eq:secondFormulaMinInitialDigraph}) is what we have in the
theorem, because as the concatenation is coherent, the third term
in~(\ref{eq:secondFormulaMinInitialDigraph}) is zero:\footnote{This is
  precisely the second term in~(\ref{eq:CoherenceFormula}), the
  equation that characterizes coherence.}

\begin{equation}
  \bigvee_{i=2}^n \left[ L_i \bigtriangleup_1^{i-1} \left(
      \overline{r_x} \, e_y \right) \right] = 0.
\end{equation}

Finally, as $L_1 = L_1 \vee e_1$, it is possible to omit
$\overline{e_1}$ and obtain~(\ref{eq:FirstMinDigraph}), recalling that
$\overline r L = L$ (by Prop.~\ref{prop:simpleEqualities}).

Uniqueness can be proved by contradiction. Use
equation~(\ref{eq:FirstMinDigraph}) and induction on the number of
productions.\proofend

\begin{figure}[htbp]
  \centering
  \includegraphics[scale = 0.6]{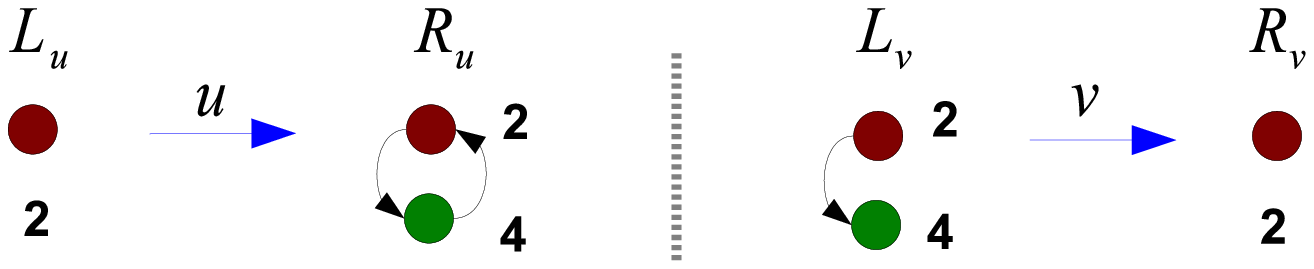}
  \caption{Non-Compatible Productions (Rep.)}
  \label{fig:nonCompAgain}
\end{figure}

\begin{figure}[htbp]
  \centering
  \includegraphics[scale =
  0.7]{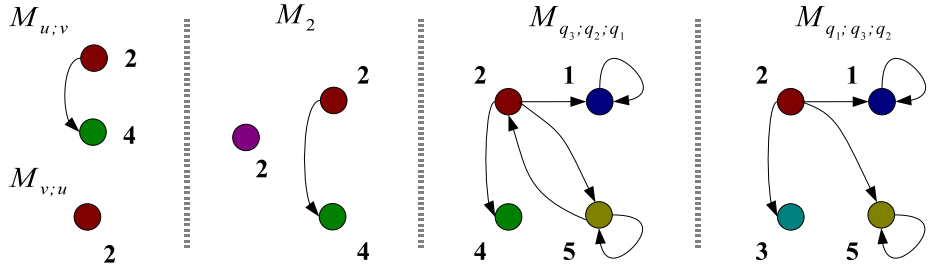}
  \caption{Minimal Initial Digraph. Examples and Counterexample}
  \label{fig:MinDigraphFirstSecThirdProd}
\end{figure}

\noindent\textbf{Example.}\label{ex:thrid}$\square$Let $s_2 = u;v$ and $s'_2 =
v;u$ (first introduced in Fig.~\ref{fig:NonComp} on
p.~\pageref{fig:NonComp} and reproduced in Fig.~\ref{fig:nonCompAgain}
for the reader convenience).  Minimal initial digraphs for these
productions are represented in
Fig.~\ref{fig:MinDigraphFirstSecThirdProd}.

The way we have introduced the concept of minimal initial digraph,
$M_2$ cannot be considered as such because either for sequence $u;v$
or $v;u$ there are subgraphs that permit their application. In the
same figure the minimal initial digraphs for productions $q_3;q_2;q_1$
and $q_1;q_3;q_2$ are also represented. Productions $q_i$ can be found
in Fig.~\ref{fig:prodsAgain} on p.~\pageref{fig:prodsAgain}.

We will explicitly compute the minimal initial digraph for the
concatenation $q_3;q_2;q_1$.  In this example, and in order to
illustrate some of the steps used to prove the previous theorem,
formula~(\ref{eq:firstFormulaMinInitialDigraph}) is used.  Once
simplified, it lays the equation:

\begin{equation}
  \underbrace{L_1^E \vee L_2^E \overline{R_1^E}}_{(*)} \vee \underbrace{L_3^E \overline{R_2^E} \left( e_2^E \vee \overline{R_1^E} \right)}_{(**)}. \nonumber
\end{equation}

The ordering of nodes is $[2 \; 3 \; 5 \; 1 \; 4]$.  We will only
display the computation for \textbf{(*)}, being \textbf{(**)} very
similar:

\begin{eqnarray}
  \left[
    \begin{array}{ccccc}
      \vspace{-6pt}
      0 & 0 & 1 & 0 & 1 \\
      \vspace{-6pt}
      0 & 0 & 0 & 0 & 0 \\
      \vspace{-6pt}
      1 & 0 & 1 & 0 & 0 \\
      \vspace{-6pt}
      0 & 0 & 0 & 0 & 0 \\
      \vspace{-6pt}
      0 & 0 & 0 & 0 & 0 \\
      \vspace{-10pt}
    \end{array} \right]
  \vee \left[
    \begin{array}{ccccc}
      \vspace{-6pt}
      0 & 1 & 0 & 0 & 0 \\
      \vspace{-6pt}
      0 & 0 & 0 & 0 & 0 \\
      \vspace{-6pt}
      0 & 0 & 1 & 0 & 0 \\
      \vspace{-6pt}
      0 & 0 & 0 & 0 & 0 \\
      \vspace{-6pt}
      0 & 0 & 0 & 0 & 0 \\
      \vspace{-10pt}
    \end{array} \right]
  \left[
    \begin{array}{ccccc}
      \vspace{-6pt}
      1 & 0 & 0 & 1 & 1 \\
      \vspace{-6pt}
      1 & 0 & 1 & 1 & 1 \\
      \vspace{-6pt}
      1 & 0 & 0 & 1 & 1 \\
      \vspace{-6pt}
      1 & 1 & 1 & 1 & 1 \\
      \vspace{-6pt}
      1 & 1 & 1 & 1 & 1 \\
      \vspace{-10pt}
    \end{array} \right]
  & = &
  \left[
    \begin{array}{ccccc}
      \vspace{-6pt}
      0 & 0 & 1 & 0 & 1 \\
      \vspace{-6pt}
      0 & 0 & 0 & 0 & 0 \\
      \vspace{-6pt}
      1 & 0 & 1 & 0 & 0 \\
      \vspace{-6pt}
      0 & 0 & 0 & 0 & 0 \\
      \vspace{-6pt}
      0 & 0 & 0 & 0 & 0 \\
      \vspace{-10pt}
    \end{array} \right] \nonumber \\
  (*) \vee (**) 
  = 
  \left[
    \begin{array}{cccccc}
      \vspace{-6pt}
      0 & 0 & 1 & 0 & 1 & \vert\; 2 \\
      \vspace{-6pt}
      0 & 0 & 0 & 0 & 0 & \vert\; 3 \\
      \vspace{-6pt}
      1 & 0 & 1 & 0 & 0 & \vert\; 5 \\
      \vspace{-6pt}
      0 & 0 & 0 & 0 & 0 & \vert\; 1 \\
      \vspace{-6pt}
      0 & 0 & 0 & 0 & 0 & \vert\; 4 \\
      \vspace{-10pt}
    \end{array} 
  \right]
  \vee
  \left[
    \begin{array}{cccccc}
      \vspace{-6pt}
      0 & 0 & 0 & 1 & 0 & \vert\; 2 \\
      \vspace{-6pt}
      0 & 0 & 0 & 0 & 0 & \vert\; 3 \\
      \vspace{-6pt}
      1 & 0 & 0 & 0 & 0 & \vert\; 5 \\
      \vspace{-6pt}
      0 & 0 & 0 & 1 & 0 & \vert\; 1 \\
      \vspace{-6pt}
      0 & 0 & 0 & 0 & 0 & \vert\; 4 \\
      \vspace{-10pt}
    \end{array} 
  \right] 
  & = &
  \left[
    \begin{array}{cccccc}
      \vspace{-6pt}
      0 & 0 & 1 & 1 & 1 & \vert\; 2 \\
      \vspace{-6pt}
      0 & 0 & 0 & 0 & 0 & \vert\; 3 \\
      \vspace{-6pt}
      1 & 0 & 1 & 0 & 0 & \vert\; 5 \\
      \vspace{-6pt}
      0 & 0 & 0 & 1 & 0 & \vert\; 1 \\
      \vspace{-6pt}
      0 & 0 & 0 & 0 & 0 & \vert\; 4 \\
      \vspace{-10pt}
    \end{array} 
  \right] \nonumber 
\end{eqnarray}

\noindent Depicted to the center of
Fig.~\ref{fig:MinDigraphFirstSecThirdProd}. \proofend

A closed formula for the effect of the application of a coherent
concatenation can be useful if we want to operate in the general case.
This is where next corollary comes in.

\newtheorem{CompositionCor}[matrixproduct]{Corollary}
\begin{CompositionCor} \label{cor:CompLemma} Let $s_n =
  p_n;\ldots;p_1$ be a coherent concatenation of completed
  productions, and $M_n$ its minimal initial digraph as defined in
  ~(\ref{eq:FirstMinDigraph}). Then,
  \begin{equation}\label{eq:EdgeConc}
    s_n \left( M^E_n \right) = \bigwedge_{i=1}^n \left(
      \overline{e^E_i} M^E_n \right) \vee
    \bigtriangleup_1^n \left( \overline{e^E_x} \, r^E_y
    \right)
  \end{equation}
  \begin{equation}\label{eq:EdgeConcNegation}
    \overline{s_n \left( M^E_n \right)} = \bigwedge_{i=1}^n \left(
      \overline{r^E_i} \, \overline{M^E_n} \right) \vee
    \bigtriangleup_1^n \left( \overline{r^E_x} \, e^E_y \right)
  \end{equation}
\end{CompositionCor}

\noindent \emph{Proof}\newline $\square$Theorem~\ref{th:minProdTh}
proves that $s_n \left( M_n^E \right) = s_n \left( \bigvee_{i=1}^{n}
  L_i \right)$.  To derive the formulas apply induction on the number
of productions and eq.~(\ref{eq:r_And_e}).\proofend

\noindent\textbf{Remark}.$\square$Equation~(\ref{eq:EdgeConcNegation}) will be
useful in Sec.~\ref{sec:compositionAndCompatibility} to calculate the
compatibility of a sequence. More interestingly, note that
equation~(\ref{eq:EdgeConc}) has the same shape as a single production
$p = r \vee \overline{e} \, L$, where:
\begin{eqnarray}
  \overline{e} & = & \bigwedge_{i=1}^n \left( \overline{e^E_i} \right) \nonumber \\
  r & = & \bigtriangleup_1^n \left( \overline{e^E_x} \, r^E_y \right). \nonumber
\end{eqnarray}

However, in contrast to what happens with a single production, the
order of application does matter, being necessary to carry out
deletion first and addition afterwards. The first equation are those
elements not deleted by any production and the second is what a
grammar rule adds and no previous production deletes (\emph{previous}
with respect to the order of application).

Equation~(\ref{eq:EdgeConc}) is closely related to composition of a
sequence of productions as defined in
Sec.~\ref{sec:compositionAndCompatibility},
Prop.~\ref{prop:CoherenceImpliesComposition}. This explains why it is
possible to interpret a coherent sequence of productions as a single
production. Recall that any sequence is coherent if the appropriate
\emph{horizontal identifications} are performed.\proofend

The negation of the minimal initial digraph that appears in
equation~(\ref{eq:EdgeConcNegation}) -- seen in
Fig.~\ref{fig:MinimumFourProds_1} -- can be explicitly calculated in
terms of operator nabla:

\begin{equation}\label{eq:NegationMinInitDigraph}
  \overline{M_n} = \nabla_{1}^{n-1} \left( \overline{L_x} \, r_y \right) \vee \bigwedge_{i=1}^n \overline{L_i}.
\end{equation}

\begin{figure}[htbp]
  \centering
  \includegraphics[scale = 0.6]{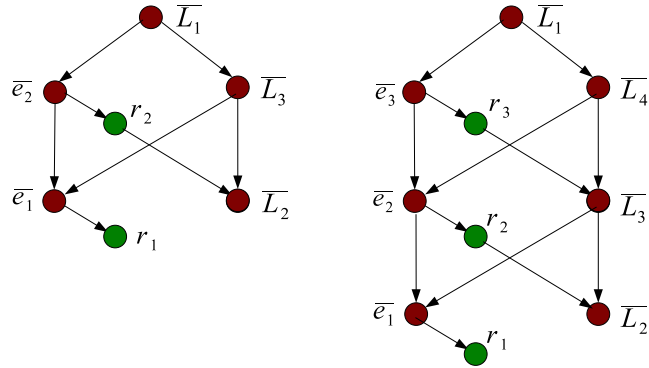}
  \caption{Equation~(\ref{eq:secondFormulaMinInitialDigraph}) for 3
    and 4 Productions (Negation of MID)}
  \label{fig:NegationMinDigraph_3_4}
\end{figure}

For the sake of curiosity, if we used
formula~(\ref{eq:secondFormulaMinInitialDigraph}) to calculate the
minimal initial digraph, the representation of its negation is
included in Fig.~\ref{fig:NegationMinDigraph_3_4} for $n=3$ and $n=4$.
It might be useful to find an expression using operators
$\bigtriangledown$ and $\nabla$ for these digraphs.

\section{Negative Initial Digraph}
\label{sec:NID}

We will make use in this section of forbidden elements and the nihil
matrix $K$ as introduced in Sec.~\ref{sec:coherenceRevisited}.

The \emph{negative initial digraph} $K(s_n)$ for a coherent sequence
$s_n=p_n;\ldots ;p_1$ is the smallest simple digraph whose elements
can not be found in the host graph to guarantee the applicability of
$s_n$.\footnote{It is not possible to speak of applicability because
  we are not considering matches yet.  This is just a way to
  intuitively introduce the concept.}  It is the symmetric concept to
minimal initial digraph, but for nihilation matrices.

\newtheorem{negativeInitialDigraph}[matrixproduct]{Definition}
\begin{negativeInitialDigraph}[Negative Initial
  Digraph]\label{negativeInitialDigraph}
  \index{negative!initial digraph}Let $s_n = p_n ; \ldots ; p_1$ be a
  completed sequence, a negative initial digraph is a simple digraph
  containing all elements that can spoil any of the operations of
  $s_n$.
\end{negativeInitialDigraph}

Negative initial digraphs depend on the way productions are completed
(minimal initial digraphs too).  In fact, as minimal and negative
initial digraphs are normally calculated at the same time for a given
sequence, there is a close relationship between them (in the sense
that one conditions the other).  This concept will be addressed again
in Sec.~\ref{sec:initialDigraphSet}, together with minimal initial
digraphs and \emph{initial sets}.

Let's introduce the dual notion to that of negative initial digraph:
\begin{equation}
  \label{eq:42}
  T = \left( \overline{ \overline{r} \otimes \overline{r}^t} \right)
  \wedge \left( \overline{e} \otimes \overline{e}^t \right).
\end{equation}

$T$ are the newly available edges after the application of a
production due to the addition of nodes.\footnote{This is why $T$
  does not appear in the calculation of the coherence of a sequence:
  coherence takes care of real actions $(e,r)$ and not of potential
  elements that may or may not be available $\left( \overline{D}, T
  \right)$.} The first term, $\overline{ \overline{r} \otimes
  \overline{r}^t}$, has a one in all edges incident to a vertex that
is added by the production. We have to remove those edges that are
incident to some node deleted by the production, which is what
$\overline{e} \otimes \overline{e}^t$ does.

\begin{figure}[htbp]
  \centering
  \includegraphics[scale = 0.67]{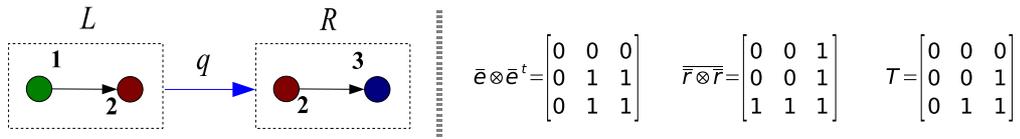}
  \caption{Available and Unavailable Edges After the Application of a
    Production}
  \label{fig:matrixT}
\end{figure}

\noindent \textbf{Example}.$\square$Figure \ref{fig:matrixT} depicts
to the left a production $q$ that deletes node $1$ and adds node $3$.
Its nihil term and its image are
\begin{equation}
  \label{eq:17}
  K = q \left( \overline{D} \right) = r \vee \overline{e} \overline{D}
  = \left[
    \begin{array}{ccc}
      \vspace{-6pt}
      1 & 0 & 1 \\ 
      \vspace{-6pt}
      1 & 0 & 1 \\
      \vspace{-6pt}
      1 & 0 & 0 \\
      \vspace{-10pt}
    \end{array}
  \right] \qquad Q = q^{-1}(K) = e \vee \overline{r}K = \left[
    \begin{array}{ccc}
      \vspace{-6pt}
      1 & 1 & 1 \\ 
      \vspace{-6pt}
      1 & 0 & 0 \\
      \vspace{-6pt}
      1 & 0 & 0 \\
      \vspace{-10pt}
    \end{array}
  \right] \nonumber
\end{equation}

To the right of Fig. \ref{fig:matrixT}, matrix $T$ is included. It
specifies those elements that are not forbidden once production $q$
has been applied. We will prove how the nihil matrix evolves according
to the production in Sec.~\ref{sec:movingConditions} -- in particular in
Prop.~\ref{th:Nevolution} on p.~\pageref{th:Nevolution}.~\proofend

As commented in Sec.\ref{sec:coherenceRevisited} for the matrix
$\overline{D}$, notice that $T$ do not tell actions of the production
to be performed in the complement of the host graph,
$\overline{G}$. Actions of productions are specified exclusively by
matrices $e$ and $r$.

\newtheorem{negInitialDigraph}[matrixproduct]{Theorem}
\begin{negInitialDigraph}\label{th:negInitialDigraph}
  Given a completed coherent sequence of productions $s_n = p_n;
  \ldots ;p_1$, the negative initial digraph is given by the equation:
  \begin{equation}\label{eq:NID}
    K(s_n) = \nabla_1^n \left( \overline{e}_x \overline{T}_x K_y
    \right).
  \end{equation}
\end{negInitialDigraph}

\noindent \emph{Proof (Sketch)}\\*
$\square$We can prove the result taking into account elements added by
productions in the sequence but not dangling edges for now. The proof
is similar to that of Theorem~\ref{th:minProdTh}, so it can be used to
fill in the gaps. A more detailed proof can be found
in~\cite{MGG_Combinatorics}.

Let's concentrate on what should not be found in the host graph
assuming that what a production adds is not interfered by actions of
previous productions. Note that this is coherence, assumed by
hypothesis. Consider for example sequence $s_2 = p_2;p_1$. Coherence
detects those elements added by both productions ($r_1 r_2 = 0$) and
also if $p_2$ adds what $p_1$ uses but does not delete ($r_2
\overline{e}_1 L_2 = 0$).\footnote{This is precisely the part of
  coherence (equation~\ref{eq:CoherenceFormula}) not used in the proof
  of Theorem~\ref{th:minProdTh}, the one for minimal initial digraphs:
  $\bigvee_{i=1}^n \left[ R^E_i \bigtriangledown_{i+1}^n \left(
      \overline{e^E_x} \, r^E_y \right)\right]$. Another reason for
  the naturalness of $K$.} Hence, we may not care about them. In the
proof of Theorem~\ref{th:minProdTh}, the final part precisely
addresses this point.

Now we proceed by induction. The case for one production $p_1$
considers elements added by $p_1$, i.e. $r_1$. For two productions
$s_2 = p_2;p_1$, besides what $p_1$ rejects, what $p_2$ is going to
add can not be found, except if $p_1$ deleted it: $r_1 \vee r_2
\overline{e}_1$. Three productions $s_3 = p_3;p_2;p_1$ should reject
what $s_2$ rejects and also what $p_3$ adds and no previous production
deletes: $r_1 \vee r_2 \overline{e}_1 \vee r_3 \overline{e}_2
\overline{e}_1$. We are using coherence here because the case in which
$p_1$ deletes edge $\epsilon$ and $p_2$ adds edge $\epsilon$ (we
should have a problem if $p_3$ also added $\epsilon$) is ruled out. By
induction we finally obtain:
\begin{equation}
  \nabla_{i=1}^n \left( \overline{e}_x r_y \right).
\end{equation}

Now, instead of considering as forbidden only those elements to be
appended by a production (not deleted by previous ones), any potential
dangling edge\footnote{Of course edges incident to nodes considered in
  the productions. There is no information at this point on edges
  provided by other nodes that might be in the host graph (to distance
  one to a node that is going to be deleted).} is also taken into
account, i.e. $r_y$ can be substituted by $K_y$ (note that
$\overline{e}_\alpha K_\alpha = K_\alpha$). To derive
eq.~\eqref{eq:NID} just put $\overline{T_x}$ for those edges that are
available again.\proofend

\begin{figure}[htbp]
  \centering
  \includegraphics[scale = 0.57]{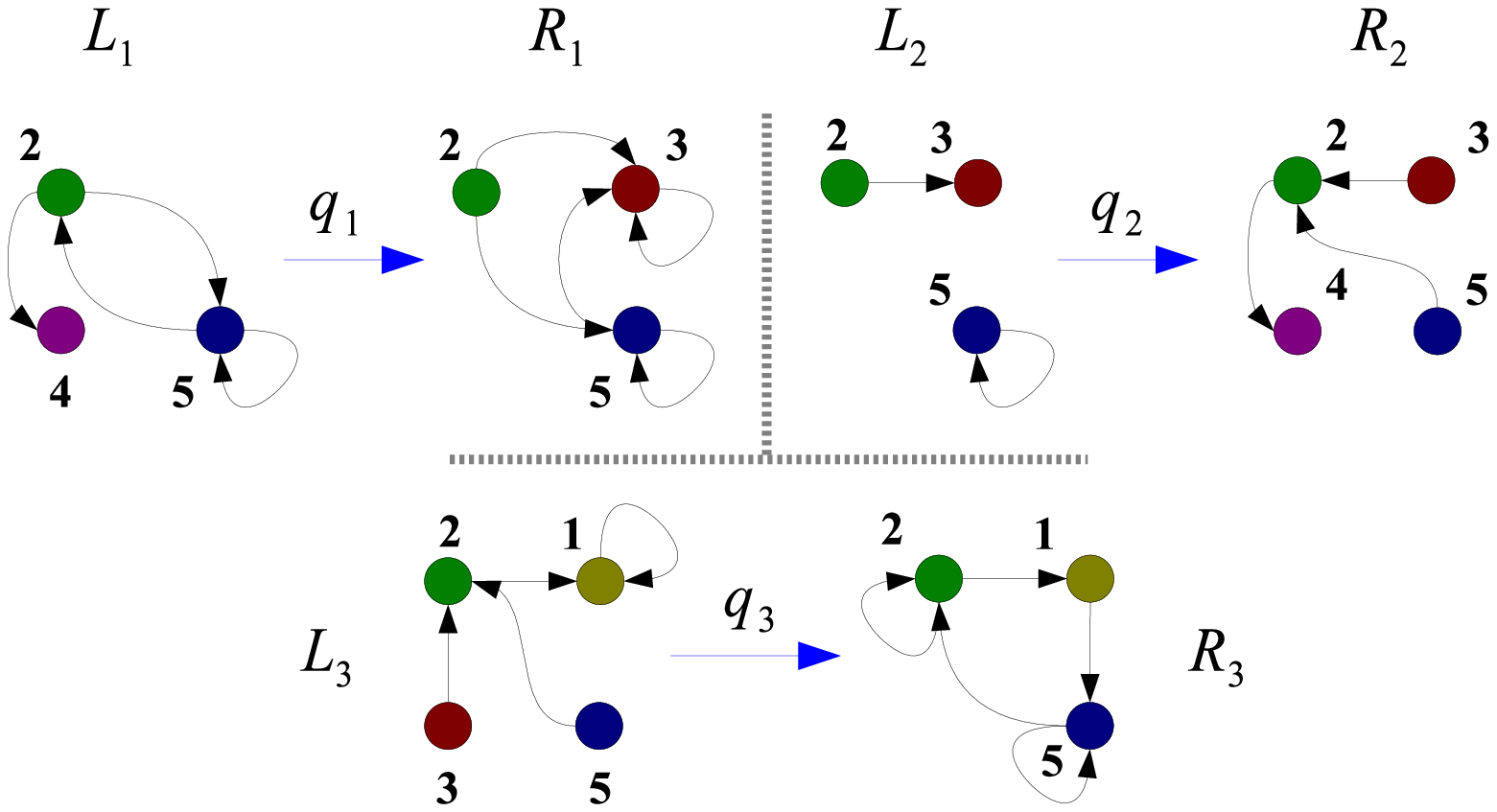}
  \caption{Productions $q_1$, $q_2$ and $q_3$ (Rep.)}
  \label{fig:prodsRepAgain}
\end{figure}

\noindent\textbf{Example}.$\square$Recall productions $q_1$
(Fig.~\ref{fig:FirstProduction} on p.~\pageref{fig:FirstProduction}),
$q_2$ and $q_3$ (Fig.~\ref{fig:SecThirdProd} on
p.~\pageref{fig:SecThirdProd}), reproduced in
Fig.~\ref{fig:prodsRepAgain} for the reader convenience.  We will
calculate the negative initial digraph for sequence $s_3 =
q_3;q_2;q_1$.  Its minimal initial digraph can be found in
Fig.~\ref{fig:MinDigraphFirstSecThirdProd}, on
p.~\pageref{fig:MinDigraphFirstSecThirdProd}. Expanding
equation~(\ref{eq:NID}) for $s_3$ we get:

\begin{equation}\label{eq:exampleNID}
  K(s_3) = K_1 \vee \overline{e}_1 K_2 \vee \overline{e}_1
  \overline{e}_2 K_3.
\end{equation}

\begin{figure}[htbp]
  \centering
  \includegraphics[scale = 0.6]{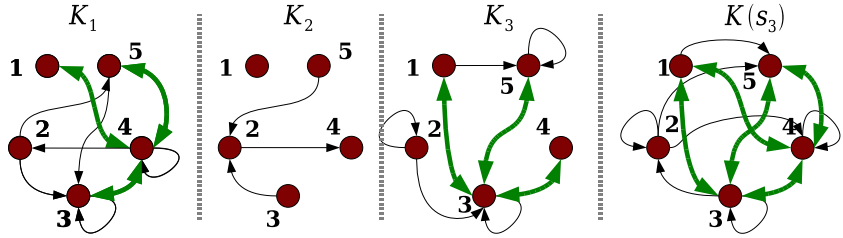}
  \caption{NID for $s_3 = q_3;q_2;q_1$ (Bold = Two Arrows)}
  \label{fig:exampleNID}
\end{figure}

In Fig.~\ref{fig:exampleNID} we have represented negative graphs for
the productions ($K_i$) and graph $K$ for $s_3$.  As there are quite a
lot of arrows, if two nodes are connected in both directions then a
single bold arrow is used.  Adjacency matrices (ordered $\left[ 2\;
  4\; 5\; 3\; 1 \right]$) for first three graphs are:
\begin{equation}
  K_1 = \left[
    \begin{array}{ccccc}
      \vspace{-6pt}
      0 & 0 & 0 & 1 & 0 \\
      \vspace{-6pt}
      1 & 1 & 1 & 1 & 1 \\
      \vspace{-6pt}
      0 & 1 & 0 & 1 & 0 \\
      \vspace{-6pt}
      0 & 1 & 0 & 1 & 0 \\
      \vspace{-6pt}
      0 & 1 & 0 & 0 & 0 \\
      \vspace{-10pt}
    \end{array}
  \right]; \;
  K_2 = r_2 = \left[
    \begin{array}{ccccc}
      \vspace{-6pt}
      0 & 1 & 0 & 0 & 0 \\
      \vspace{-6pt}
      0 & 0 & 0 & 0 & 0 \\
      \vspace{-6pt}
      1 & 0 & 0 & 0 & 0 \\
      \vspace{-6pt}
      1 & 0 & 0 & 0 & 0 \\
      \vspace{-6pt}
      0 & 0 & 0 & 0 & 0 \\
      \vspace{-10pt}
    \end{array}
  \right]; \;
  K_3 = \left[
    \begin{array}{ccccc}
      \vspace{-6pt}
      0 & 0 & 0 & 1 & 0 \\
      \vspace{-6pt}
      1 & 1 & 1 & 1 & 1 \\
      \vspace{-6pt}
      0 & 1 & 0 & 1 & 0 \\
      \vspace{-6pt}
      0 & 1 & 0 & 1 & 0 \\
      \vspace{-6pt}
      0 & 1 & 0 & 0 & 0 \\
      \vspace{-6pt}
    \end{array}
  \right] \nonumber
\end{equation}
The rest of matrices and calculations are omitted for space
considerations.\proofend

Matrix $K$ provides information on what will be called \emph{internal
  $\varepsilon$-productions} in
Sec.~\ref{sec:internalAndExternalProductions}.  These
$\varepsilon$-productions are grammar rules automatically generated to
deal with dangling edges.  We will distinguish between \emph{internal}
and \emph{external}, being internal (to the sequence) those that deal
with edges added by a previous production.

As above, think of $G$ as an ``ambient graph'' in which operations
take place. A final remark is that $\overline{T}$ makes the number of
edges in $\overline{G}$ as small as possible. For example, in
$\overline{e_1} \, \overline{e_2} \overline{T_1} \, \overline{T_2}
K_2$ we are in particular demanding $\overline{e_1} \overline{T_1}
\overline{T_2} r_2$ (because $K_2 = r_2 \vee \overline{e_2}
\overline{D_2}$). If we start with a compatible host graph, it is not
necessary to ask for the absence of edges incident to nodes that are
added by a production (\emph{potentially available}). Notice that
these edges could not be in the host graph as they would be dangling
edges or we would be adding an already existent
node. \label{rem:forgetT}Summarizing, if compatibility is assumed or
demanded by hypothesis, we may safely ignore $\overline{T}_x$ in the
formula for the initial digraph. This remark will be used in the proof
of the G-congruence characterization theorem in
Sec.~\ref{sec:gCongruence}.

\section{Composition and Compatibility}
\label{sec:compositionAndCompatibility}

Next we are going to introduce compatibility for sequences (extending
Definition~\ref{def:prodCompatibility}) and also composition.
Composition defines a unique production that to a certain
extent\footnote{If a production inside a sequence deletes a node and
  afterwards another production adds that same node, the overall
  effect is that the node is not touched. This may affect the deletion
  of dangling edges in an actual host graph (those incident to some
  node not appearing in the productions).} performs the same actions
than its corresponding sequence (the one that defines it).

Recall that compatibility is a means to deal with dangling edges,
equivalent to the dangling condition in DPO.  When a concatenation of
productions is considered, we are not only concerned with the final
result but also with intermediate states -- partial results -- of the
sequence.  Compatibility should take this into account and thus a
concatenation is said to be compatible if the overall effect on its
minimal initial digraph gives as result a compatible digraph starting
from the first production and increasing the sequence until we get the
full concatenation. We should then check compatibility for the growing
sequence of concatenations $S = \{ s_1, s_2, \ldots ,s_n \}$ where
$s_m = q_m;q_{m-1};\ldots ;q_1$, $1 \leq m \leq n$.

\newtheorem{CompConc}[matrixproduct]{Definition}
\begin{CompConc}
  \index{compatibility!sequence}A coherent sequence $s_n = q_n; \ldots
  ; q_1$ is said to be \emph{compatible} if the following identity is
  verified:
  \begin{equation}\label{eq:SequenceCompatibility}
    \bigvee_{m=1}^n\left\| \left[ s_m \left( M_m^E \right) \vee
        \left( s_m \left( M_m^E \right)\right)^t \right] \odot
      \overline{s_m \left( M_m^N \right)} \right\|_1 = 0.
  \end{equation}
\end{CompConc}

Corollary~\ref{cor:CompLemma} --
equations~(\ref{eq:EdgeConc})~and~(\ref{eq:EdgeConcNegation}) -- give
closed form formulas for the terms
in~(\ref{eq:SequenceCompatibility}).

Of course this definition coincides with
Def.~\ref{def:prodCompatibility} for one production and with
Def.~\ref{def:compatibilityDefinition} for the case of a single graph
(consider the identity production, for example).

Coherence examines whether actions specified by a sequence of
productions are feasible.  It warns us if one production adds or
deletes an element that it should not, as some later production might
need that element to carry out an operation that becomes impossible.
Compatibility is a more \emph{basic} concept because it examines if
the result is a digraph, that is, if the class of all digraphs is
closed under the operations specified by the sequence.

\begin{figure}[htbp]
  \centering
  \includegraphics[scale = 0.63]{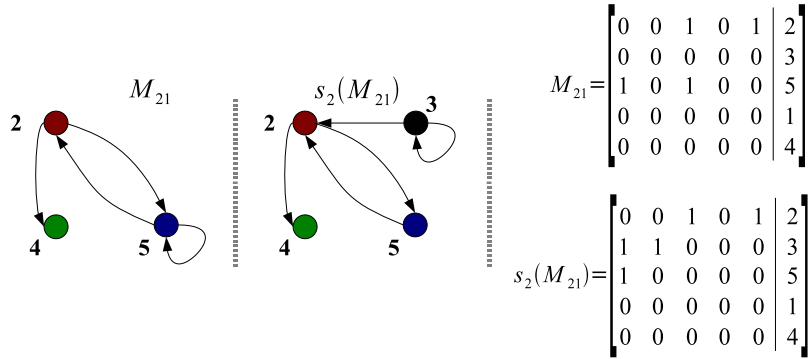}
  \caption{Minimal Initial Digraphs for $s_2 = q_2; q_1$}
  \label{fig:seqCompatibility}
\end{figure}

\begin{figure}[htbp]
  \centering
  \includegraphics[scale = 0.63]{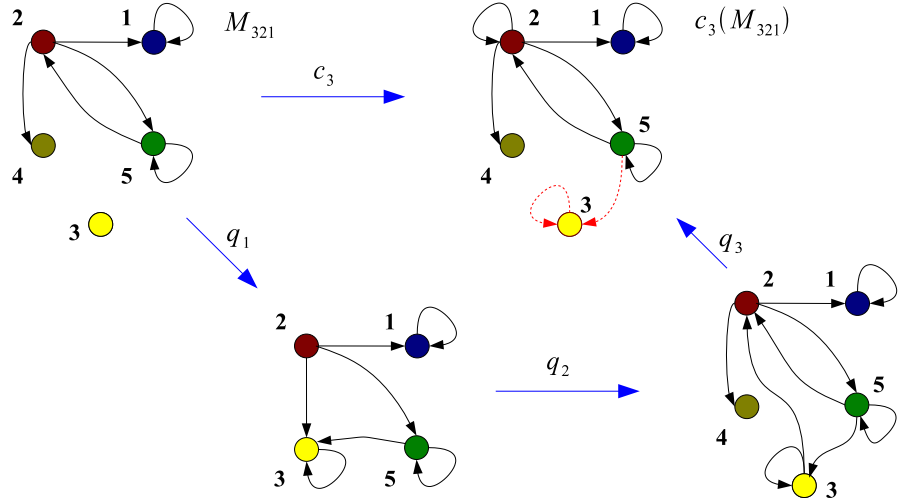}
  \caption{Composition and Concatenation of a non-Compatible Sequence}
  \label{fig:CompExample}
\end{figure}

\noindent\textbf{Example.}$\square$Consider sequence $s_3 = q_3;q_2;q_1$, with
$q_i$ as defined in
Figs.~\ref{fig:FirstProduction}~and~\ref{fig:SecThirdProd} on
pp.~\pageref{fig:FirstProduction}~and~\pageref{fig:SecThirdProd},
respectively.  In order to check
equation~(\ref{eq:SequenceCompatibility}) we need the minimal initial
digraphs $M_1$ (the LHS of $q_1$), $M_{21}$ (coincides with the LHS of
$q_1$) and $M_{321}$, that can be found in
Figs.~\ref{fig:seqCompatibility}~and~\ref{fig:CompExample} on
p.~\pageref{fig:CompExample}.

Equation~(\ref{eq:SequenceCompatibility}) for $m=1$ is compatibility
of production $q_1$ which has been calculated in the example of
p.~\pageref{ex:first}.  For $m=2$ we have
\begin{equation}\label{eq:twoProds}
  \left\| \left[ s_2 \left( M_{21}^E \right) \vee \left( s_2 \left(
          M_{21}^E \right)\right)^t \right] \odot \overline{s_2 \left(
        M_{21}^N \right)} \right\|_1
\end{equation}
which should be zero with nodes ordered as before, $[2\; 3\; 5\; 1\;
4]$.  The evolution of the vector of nodes is $\left[ 1 \; 0 \; 1 \; 0
  \; 1\right] \stackrel{q_1}{\longmapsto} \left[1 \; 1 \; 1 \; 0 \;
  0\right] \stackrel{q_2}{\longmapsto} \left[ 1 \; 1 \; 1 \; 0 \;
  1\right]$. Making all substitutions according to values displayed in
Fig.~\ref{fig:seqCompatibility} we obtain:
\begin{equation}
  (\ref{eq:twoProds}) = \left(
    \left[
      \begin{array}{ccccc}
        \vspace{-6pt}
        0 & 0 & 1 & 0 & 1 \\
        \vspace{-6pt}
        1 & 1 & 0 & 0 & 0 \\
        \vspace{-6pt}
        1 & 0 & 0 & 0 & 0 \\
        \vspace{-6pt}
        0 & 0 & 0 & 0 & 0 \\
        \vspace{-6pt}
        0 & 0 & 0 & 0 & 0 \\
        \vspace{-10pt}
      \end{array} \right]
    \vee
    \left[
      \begin{array}{ccccc}
        \vspace{-6pt}
        0 & 1 & 1 & 0 & 0 \\
        \vspace{-6pt}
        0 & 1 & 0 & 0 & 0 \\
        \vspace{-6pt}
        1 & 0 & 0 & 0 & 0 \\
        \vspace{-6pt}
        0 & 0 & 0 & 0 & 0 \\
        \vspace{-6pt}
        1 & 0 & 0 & 0 & 0 \\
        \vspace{-10pt}
      \end{array} \right] \right)
  \odot
  \left[
    \begin{array}{c}
      \vspace{-6pt}
      0 \\
      \vspace{-6pt}
      0 \\
      \vspace{-6pt}
      0 \\
      \vspace{-6pt}
      1 \\
      \vspace{-6pt}
      0 \\
      \vspace{-10pt}
    \end{array} \right] 
  = 
  \left[
    \begin{array}{cc}
      \vspace{-6pt}
      0 & \vert\; 2 \\
      \vspace{-6pt}
      0 & \vert\; 3 \\
      \vspace{-6pt}
      0 & \vert\; 5 \\
      \vspace{-6pt}
      0 & \vert\; 1 \\
      \vspace{-6pt}
      0 & \vert\; 4 \\
      \vspace{-10pt}
    \end{array} \right] \nonumber
\end{equation}

As commented above, we can make use of
identities~(\ref{eq:EdgeConc})~and~(\ref{eq:EdgeConcNegation}). The
case $m = 3$ is very similar to $m = 2$. There is another example
below (on p.~\pageref{ex:fourth}) with the graphical evolution of the
states of the system.\proofend

Once we have seen compatibility for a sequence the following corollary
to Theorems~\ref{th:minProdTh}~and~\ref{th:negInitialDigraph} can be
stated:

\newtheorem{eqNIDMID}[matrixproduct]{Corollary}
\begin{eqNIDMID}
  \label{cor:eqNIDMID}
  Let $M$ be a minimal initial digraph and $K$ the corresponding
  negative initial digraph for a coherent and compatible sequence,
  then $M \wedge K = 0$.
\end{eqNIDMID}

\noindent \emph{Proof} \\*
$\square$Just compare equations $M = \nabla_1^n \left( \overline{r_x}
  L_y \right)$ and $K = \nabla_1^n \left( \overline{e}_x
  \overline{T}_x K_y \right)$.
We know that elements added and deleted by a production are disjoint.
This implies that the negation of the corresponding adjacency matrices
have no common elements.\proofend

Intuitively, if we interpret matrices $M$ and $K$ as elements that
must be and must not be present in a potential host graph in order to
apply the sequence, then it should be clear that $L_i$ and $K_i$ must
also be disjoint.  This point will be addressed in
Chap.~\ref{ch:restrictionsOnRules}. The next proposition is a sort of
converse to Corollary~\ref{cor:eqNIDMID}.

\newtheorem{eqNIDMIDconv}[matrixproduct]{Proposition}
\begin{eqNIDMIDconv}
  \label{prop:seqComp}
  Let $s = p_n; \ldots; p_1$ be a sequence consisting of compatible
  productions. If
  \begin{equation}
    \label{eq:39}
    \bigtriangledown_1^n \left( \overline{e}_x \overline{r}_x M(s_y)
      K(s_y)\right) = 0
  \end{equation}
  then $s$ is compatible, where $M(s_m)$ and $K(s_m)$ are the
  minimal and negative initial digraphs of $s_m = p_m; \ldots; p_1$,
  $m \in \{1, \ldots, n\}$.
\end{eqNIDMIDconv}

\noindent \emph{Proof (Sketch)}\\
$\square$Equation \eqref{eq:39} is a restatement of the definition of
compatibility for a sequence of productions. The condition appears
when the certainty and nihil parts are demanded to have no common
elements. Compatibility of each production is used to simplify terms
of the form $L_i K_i$. \proofend

As happened with coherence -- and will happen with graph congruence in
Sec.~\ref{sec:gCongruence} -- eq.~\eqref{eq:39} for compatibility
provides information on which elements may prevent it. Compatibility
and coherence are related notions but only to some extent. Coherence
deals with actions of productions, while compatibility with potential
presence or absence of elements.

So far we have presented compatibility and will end this section
studying composition and the circumstances under which it is possible
to define a \emph{single production} if a coherent concatenation is
given.

When we introduced the notion of production, we first defined its LHS
and RHS and then we associated some matrices ($e$ and $r$) to them.
The situation for defining composition is similar, but this time we
first observe the overall effect (its dynamics, i.e. matrices $e$ and
$r$) of the production and then decide its left and right hand sides.

Assume $s_n = p_n; \ldots ; p_1$ is coherent, then the composition of
its productions is again a production defined by the rule $c = p_n
\circ p_{n-1} \circ \ldots \circ p_1$.\footnote{The concept and
  notation are those commonly used in mathematics.}  The description
of its erasing and its addition matrices $e$ and $r$ are given by
equations:
 
\begin{equation}\label{eq:e_r_compositionEdges}
  S^E = \sum_{i=1}^n \left( r^E_i - e^E_i \right)
\end{equation}
\begin{equation}\label{eq:e_r_compositionNodes}
  S^N = \sum_{i=1}^n \left( r^N_i - e^N_i \right).
\end{equation}

Due to coherence we know that elements of $S^E$ and $S^N$ are either
$+1$, $0$ or $-1$, so they can be split into their positive and
negative parts,
\begin{equation}
  S^E = r^E_{+} - e^E_{-}, \qquad S^N = r^N_{+} - e^N_{-},
\end{equation}
where all $r_+$ and $e_-$ elements are either zero or one.  We have:
\newtheorem{composition}[matrixproduct]{Proposition}
\begin{composition}\label{prop:CoherenceImpliesComposition}
  \index{composition}Let $s_n = p_n; \ldots ;p_1$ be a coherent and
  compatible concatenation of productions.  Then, the composition $c =
  p_n \circ p_{n-1} \circ \ldots \circ p_1$ defines a production with
  matrices $r^E = r^E_+$, $r^N = r^N_+$ and $e^E = - \, e^E_-$, $e^N =
  - \, e^N_-$.
\end{composition}

\noindent \emph{Proof}\newline $\square$Follows from comments
above.\proofend

The LHS is the minimal digraph necessary to carry out all operations
specified by the composition (plus those preserved by the
productions). As it is only one production, its LHS equals its erasing
matrix plus preserved elements and its right hand side is just the
image. The concept of composition is closely related to the formula
which outputs the image of a compatible and coherent sequence. Refer
to Corollary~\ref{cor:CompLemma}.

Note that preserved elements do depend on the order of productions in
the sequence. For example, sequence $s_3 = p_3;p_2;p_1$ first
preserves (appears in $L_1$ and $R_1$) then deletes ($p_2$) and
finally adds ($p_3$) element $\alpha$. This element is necessary in
order to apply $s_3$. However, the permutation $p_3' = p_2;p_1;p_3$
first adds $\alpha$, then preserves it and finally deletes it. It
cannot be applied if the element is present.

\newtheorem{CompEqConc}[matrixproduct]{Corollary}
\begin{CompEqConc}
  With the notation as above, $c \left( M_n \right) = s_n \left( M_n
  \right)$.
  \label{CompEqualsConc}
\end{CompEqConc}

Composition is helpful when we have a coherent concatenation and
intermediate states are useless or undesired.  It will be utilized in
sequential independence and explicit parallelism
(Secs.~\ref{sec:sequentializationGrammarRules}~and~\ref{sec:explicitParallelism}).

\noindent\textbf{Example.}\label{ex:fourth}$\square$We finish this section
considering sequence $s_3 = q_3;q_2;q_1$ again, calculating its
composition $c_3$ and comparing its result with that of $s_3$. Recall
that $S^E \left( s_3 \right) = \sum_{i=1}^3 \left( r_i^E - e_i^E
\right) = r_+^E - e_-^E$.

\begin{equation}
  \sum_{i=1}^3 r_i^E =
  \left[
    \begin{array}{cccccc}
      \vspace{-6pt}
      1 & 1 & 0 & 0 & 1 & \vert\; 2 \\
      \vspace{-6pt}
      1 & 1 & 0 & 0 & 0 & \vert\; 3 \\
      \vspace{-6pt}
      1 & 1 & 1 & 0 & 0 & \vert\; 5 \\
      \vspace{-6pt}
      0 & 0 & 0 & 0 & 0 & \vert\; 1 \\
      \vspace{-6pt}
      0 & 0 & 0 & 0 & 0 & \vert\; 4 \\
      \vspace{-10pt}
    \end{array} \right] 
  \quad \sum_{i=1}^3 e_i^E =
  \left[
    \begin{array}{cccccc}
      \vspace{-6pt}
      0 & 1 & 0 & 0 & 1 & \vert\; 2 \\
      \vspace{-6pt}
      1 & 0 & 0 & 0 & 0 & \vert\; 3 \\
      \vspace{-6pt}
      1 & 0 & 1 & 0 & 0 & \vert\; 5 \\
      \vspace{-6pt}
      0 & 0 & 0 & 1 & 0 & \vert\; 1 \\
      \vspace{-6pt}
      0 & 0 & 0 & 0 & 0 & \vert\; 4 \\
      \vspace{-10pt}
    \end{array} \right] \nonumber
\end{equation}

\begin{equation}
  S^E \left( s_3 \right) \! = \!
  \left[ \begin{array}{ccccc}
      \vspace{-6pt}
      1 & 0 & 0 & 0 & 0 \\
      \vspace{-6pt}
      0 & 1 & 0 & 0 & 0 \\
      \vspace{-6pt}
      0 & 1 & 0 & 0 & 0 \\
      \vspace{-6pt}
      0 & 0 & 0 & -1 & 0 \\
      \vspace{-6pt}
      0 & 0 & 0 & 0 & 0 \\
      \vspace{-10pt}
    \end{array} \right] \! = \!
  \left[ \begin{array}{ccccc}
      \vspace{-6pt}
      1 & 0 & 0 & 0 & 0 \\
      \vspace{-6pt}
      0 & 1 & 0 & 0 & 0 \\
      \vspace{-6pt}
      0 & 1 & 0 & 0 & 0 \\
      \vspace{-6pt}
      0 & 0 & 0 & 0 & 0 \\
      \vspace{-6pt}
      0 & 0 & 0 & 0 & 0 \\
      \vspace{-10pt}
    \end{array} \right] \!\! - \!\!
  \left[ \begin{array}{ccccc}
      \vspace{-6pt}
      0 & 0 & 0 & 0 & 0 \\
      \vspace{-6pt}
      0 & 0 & 0 & 0 & 0 \\
      \vspace{-6pt}
      0 & 0 & 0 & 0 & 0 \\
      \vspace{-6pt}
      0 & 0 & 0 & 1 & 0 \\
      \vspace{-6pt}
      0 & 0 & 0 & 0 & 0 \\
      \vspace{-10pt}
    \end{array} \right] \!\! = r_+^E \! - \! e_-^E.\nonumber
\end{equation}

Sequence $s_3$ has been chosen not only to illustrate composition, but
also compatibility and the sort of problems that may arise if it is
not fulfilled.  In this case, $q_3$ deletes node 3 and edge (3,2) but
does not specify anything about edges (3,3) and (3,5) -- the red
dotted elements in Fig.~\ref{fig:CompExample} --.  In order to apply
the composition, either the composed production is changed by
considering these elements or elements have to be related in other way
(in this case, unrelated).\proofend

Previous example provides us with some clues on how the match could be
defined.  The basic idea is to introduce an operator over the set of
productions, so once a match identifies a place in the host graph
where the rule might be applied, the operator modifies the rule
enlarging the deletion matrix. This way no dangling edge appears (it
should enlarge the grammar rule to include the context of the original
rule in the graph, adding all elements on both LHS and RHS).  In
essence, a match should be an injective morphism (in Matrix Graph
Grammars) plus an operator.  Pre-calculated information for coherence,
sequentialization, and the like, should help and hopefully reduce the
amount of calculations during runtime.  We will study this in
Chap.~\ref{ch:matching}.

This section ends noting that, in Matrix Graph Grammars, one
production is a morphism between two simple digraphs and thus it may
carry out just one action on each element.  When the composition of a
concatenation is performed we get a single production. Suppose one
production specifies the deletion of an element and another its
addition, the overall mathematical result of the composition should
leave the element unaltered.  When a match is considered, depending on
the chosen approach, all dangling edges incident to those erased nodes
should be removed, establishing an important difference between a
sequence and its composition.

\section{Summary and Conclusions}
\label{sec:summaryAndConclusions4_2}

Minimal and negative initial digraphs are of fundamental importance,
demanding the minimal (maximal) set of elements that must be found
(must not be found) in order to apply the sequence under
consideration. In particular they will be used to give one
characterization of the applicability problem
(problem~\ref{prob:applicability}).

Also, composition and the main differences between this and
concatenation have been addressed. Composition can be a useful tool to
study concurrency. Recall from
Sec.~\ref{sec:compositionAndCompatibility} that differences in the
image of the composition are not due to the order in which operations
are performed but in those elements needed by the productions, i.e. in
the initial digraph. This also gives information on initial digraphs
and its calculation. This topic -- which we call \emph{G-congruence}
-- will be addressed in deeper detail in Sec.~\ref{sec:gCongruence}.

So far we have developed some analytical techniques independent (to
some extent) of the initial state of the system to which the grammar
rules will be applied. This allows us to obtain information about
grammar rules themselves, for example at design time. This information
may be useful during runtime.  We will return to this point in future
chapters.

Chapter~\ref{ch:matching} starts with the semantics of a grammar rule
application, so a host graph or initial state will be considered.
Among other things the fundamental concept of direct derivation is
introduced. We will see what can be recovered of what we have
developed so far and how it can be used.
\chapter{Matching}
\label{ch:matching}

There are two fundamental parts in a grammar: Actions to be performed
in every single step (grammar rules) and where these actions are to be
performed in a system (\emph{matching}).  Previous chapter deals with
the former and this chapter with the latter. Also, restrictions on the
applicability of rules and their embedding in the host graph need to be
addressed. This topic is studied in Chap.~\ref{ch:restrictionsOnRules}.

If a rule is applied we automatically have the pair
\textbf{(production, match}) -- normally called \emph{direct
  derivation} -- which in essence specifies \emph{what} to do and
\emph{where} to do it.  If instead of a single rule we consider a
sequence with their corresponding matches then we will speak of
\emph{derivation}.  These initial definitions, together with the
matching are studied in Sec.~\ref{sec:matchAndExtendedMatch} in which
we will make use of some functional analysis notation (see
Sec.~\ref{sec:functionalAnalysis}).  When a match is considered, there
is the possibility that a new production (so called
$\varepsilon$-production) is concatenated to the original
one.\footnote{$\varepsilon$-productions take care of those edges --
  dangling edges -- not specified by the production and incident to
  some node that is going to be deleted.}  Both productions must be
applied (matched) to the same nodes.  The mechanism to obtain this
effect can be found in Sec.~\ref{sec:marking} (marking).  An important
issue is to study to what extent the notions introduced at
specification time (coherence, composition, etc) can be recovered
when a host graph is considered. They will be revisited considering
minimal and negative initial digraphs (see
Secs.~\ref{sec:MID}~and~\ref{sec:NID}) in a wider context in
Sec.~\ref{sec:initialDigraphSet}.  A classification of
$\varepsilon$-productions -- helpful in Chap.~\ref{ch:reachability} --
is accomplished in Sec.~\ref{sec:internalAndExternalProductions}.  The
chapter ends with a summary in Sec.~\ref{sec:summaryAndConclusions4}.

\section{Match and Extended Match}
\label{sec:matchAndExtendedMatch}

Matching is the operation of identifying the LHS of a rule inside a
host graph.  This identification is not necessarily unique, becoming
one source of non determinism.\footnote{In fact there are two sources
  of non-determinism. Apart from the one already mentioned, the rule
  to be applied is also chosen non-deterministically.}  The match can
be considered as one of the ways of completing $L$ with respect to
$G$.

\newtheorem{Match}[matrixproduct]{Definition}
\begin{Match}[Match]\label{def:match}
  \index{match!MGG}Given a production $p:L \rightarrow R$ and a simple
  digraph $G$, any tuple $m = \left( m_L, m_K \right)$ is called a
  match (for $p$ in $G$), with $m_L : L \rightarrow G$ and $m_K: K^E
  \rightarrow \overline{G^E}$ total injective morphisms.  Besides,
  \begin{equation}
    m_L(n) = m_K(n), \forall n \in L^N.
  \end{equation}
\end{Match}

The two main differences with respect to matches as defined in the
literature is that Def.~\ref{def:match} demands the non-existence of
potential problematic elements and that $m$ must be injective.

It is useful to consider the structure defined by the negation of the
host graph, $\overline G=(\overline {G^E}, \overline{G^N})$.  It is
made up of the graph $\overline {G^E}$ and the vector of nodes
$\overline{G^N}$.  Note that the negation of a graph (both, the
adjacency matrix and the node vector) is not a graph because in general
compatibility will fail.  Of course, the adjacency matrix alone
$\left( \, \overline{G^E} \, \right)$ does define a graph.

The negation of a graph is equivalent to taking its complement.  In
general this complement will be taken inside some ``bigger graph'',
normally constructed by performing the completion with respect to
other graphs involved in the operations.  For example, when checking
if graph $A$ is in $\overline{G^E}$ (suppose that $A$ has a node that
is not in $G$) we obtain that $A$ cannot be found in $\overline{G^E}$,
unless $\overline{G^E}$ is previously completed with that node and all
its incident edges.

Notice that the negation of a graph $G$ coincides with its complement.
Probably it should be more appropriate to keep the negation symbol
(the overline) when there is no completion (in other words, complement
is taken with respect to the graph itself) and use ${}^c$ when other
graphs are involved. From now on the overline will be used in all
cases. This abuse of notation should not be confusing.

Next, a notion of direct derivation that covers not only elements that
must be present ($L$) but also those that should not appear ($K$) is
presented. This extends the concept of derivation found in the
literature, which only considers explicitly positive\emph{}
information.

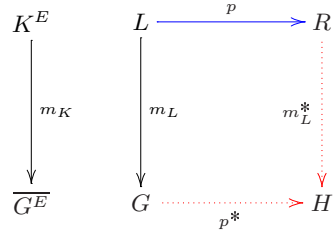
\begin{figure}[htbp]
  \begin{displaymath}
    \xymatrix{
      K^E \ar[dd]^{m_K} & L \ar[dd]^{m_L} \ar@[blue][rr]^p && R
      \ar@{.>}@[red][dd]_{m^*_L} \\ 
      &&& \\
      \overline{G^E} & G \ar@{.>}@[red][rr]_{p^*} && H
    }
  \end{displaymath}
  \caption{Production Plus Match (Direct Derivation)}
  \label{fig:productionPlusMatch}
\end{figure}

\newtheorem{directDerivationDef}[matrixproduct]{Definition}
\begin{directDerivationDef}[Direct
  Derivation]\label{def:directDerivationDef}
  \index{direct derivation!MGG}Given a production $p:L \rightarrow R$
  as in Fig.~\ref{fig:productionPlusMatch} and a match $m = \left(
    m_L, m_K \right)$, $d = \left( p, m \right)$ is called a
  \emph{direct derivation} with result $H = p^* \left( G \right)$ if
  the square is a pushout:
  \begin{equation}
    m^*_L \circ p \left( L \right) = p^* \circ m_L \left( L \right).
  \end{equation}
  The standard notation in this case is $G
  \stackrel{(p,m)}{\Longrightarrow} H$, or even $G \Longrightarrow H$
  if $p$, $m$ or both are not relevant.
\end{directDerivationDef}

We will see below that it is not necessary to rely on category theory
to define direct derivations in Matrix Graph Grammars. It is included
to ease comparison with DPO and SPO approaches.

Figure~\ref{fig:productionPlusMatch} displays a production $p$ and a
match $m$ for $p$ in G.  It is possible to close the diagram making it
commutative $\left( m^* \circ p = p^* \circ m \right)$, using the
pushout construction (see~\cite{Fundamentals}) on category
\textbf{Graph$^\mathbf{P}$} of simple digraphs and partial functions.
This categorical construction for relational graph rewriting is carried
out in~\cite{mizoguchi}.  See Sec.~\ref{sec:relationAlgebraicApproach}
for a quick overview on the relational approach.\footnote{There is a
  slight difference, though, as we have a simpler case. We demand
  matchings to be injective which, by Prop. 2.6 in~\cite{mizoguchi},
  implies that comatches are injective.}
 
If a concatenation $s = p_n;\ldots;p_1$ is considered together with
the set of matchings $m = \{ m_1, \ldots, m_n\}$, then $d = \left( s,
  m \right)$ is a \emph{derivation}.  In this case the notation $G
\Longrightarrow^* H$ is used.

When applying a rule to a host graph, the main problem to concentrate
on is that of so-called \emph{dangling edges}, which is differently
addressed in DPO and SPO (see
Secs.~\ref{sec:DPO}~and~\ref{sec:otherCategoricalApproaches}).  In
DPO, if one edge comes to be dangling then the rule is not applicable
for that match.  SPO allows the production to be applied by deleting
any dangling edge.

For Matrix Graph Grammars we propose an SPO-like behaviour as in our
case a DPO-like behaviour\footnote{In future sections we will speak of
  \emph{fixed} and \emph{floating} grammars, respectively.} would be a
particular case if compatibility is considered as an application
condition (see Chap.~\ref{ch:restrictionsOnRules}).\footnote{If
  $\varepsilon$-productions are not allowed and a rule can be applied
  if the output is again a simple digraph (compatibility) then we
  obtain a DPO-like behaviour.}


\begin{figure}[htbp]
  \setlength{\unitlength}{1.12cm}
  \begin{picture}(10.0, 4.5) \put(0.5, 4){ \xymatrix{
        L \ar[d]^{c} \ar[rr]^p && R \ar @{.>}[d]^{c^*} \\
        L \ar@/_15pt/@[green][dd]_{m_G} \ar[d]^{m_L} \ar@[blue][rr]^p
        && R \ar @{.>}@[red][d]^{m^*_{L}} \\
        G \ar[d]^{m_\varepsilon}\ar @{.>}@[red][rr]_{p*} && H
        \ar@{.>}@[red][d]^{m_\varepsilon^*}\\
        G \ar @{.>}@[red][rr]_{\widehat p ^*} && H } } \put(5, 3.7){
      \xymatrix@R=0.7cm@C=0.7cm{ L \ar[dd]_{i_L}
        \ar@[green][dddr]_(0.4){m_G} \ar[dr]^{m_L} \ar@[blue][rr]^p &&
        R \ar@{.>}[dd]^(0.3){i_R}
        \ar@[red][dr]^{m^*_L} & \\
        & G \ar[dl]_(0.2){i_G} \ar[dd]^(0.35){m_\varepsilon}
        \ar@[red][rr]^(0.35){p^*} && H \ar@{.>}[dl]^{i_H}
        \ar@[red][dd]_{m^*_\varepsilon}\\
        L + G \ar[dr] _{\widehat{m}}\ar @{.>}[rr]_{\qquad \widehat p}
        && R + H \ar@{.>}[dr]^{\widehat{m}^*} & \\
        & G \ar@[red][rr]^{\widehat p^*} && H } }
  \end{picture}
  \caption{(a) Neighborhood. (b) Extended Match}
  \label{fig:matches}
\end{figure}
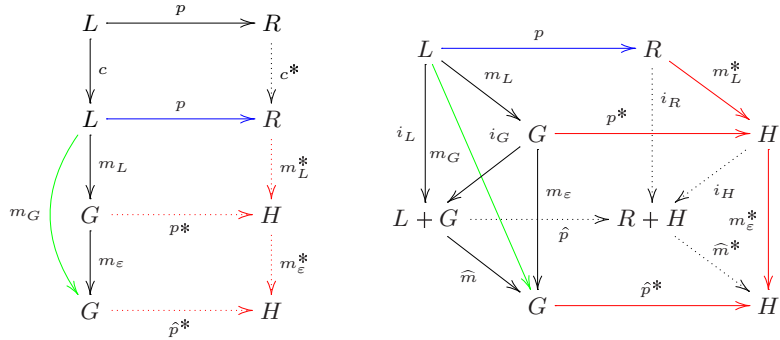

Figure~\ref{fig:matches} shows our strategy to handle dangling edges:
\begin{enumerate}
\item \label{firstStep}Complete $L$ with respect to $G$ ($c$ and $c^*$
  to the left of Fig.~\ref{fig:matches}). It is necessary to match $L$
  in $G$ to this end.\footnote{Abusing a little of the notation,
    graphs before completion and after completion are represented with
    the same letter, $L$ and $R$.}
\item Morphism $m_L$ will identify rule's left hand side (after
  completion) in the host graph.
\item A neighborhood of $m(L)\subseteq G$ covering all relevant extra
  elements is selected taking into account all dangling edges not
  considered by match $m_L$ with their corresponding source and target
  nodes.  This is performed by a morphism to be studied later,
  represented by $m_{\varepsilon}$.
\item Finally, $p$ is enlarged erasing any potential dangling edge.
  This is carried out by an operator that we will write as
  $T_\varepsilon$.  See definition below on p.~\pageref{pr:Tepsilon}.
\end{enumerate}

The order of previous steps is important as potential dangling
elements must be identified and erased before any node is deleted by
the original rule.


The coproduct in Fig.~\ref{fig:matches} should be understood as a
means to couple $L$ and $G$.  The existence of a morphism $p^*$ that
closes the top square on the right of Fig.~\ref{fig:matches} is not
guaranteed.  This is where $m_\varepsilon$ comes in.  This mapping, as
explained in point 2 above, extends the production to consider any
edge to distance 1 from nodes appearing in the left hand side of
$p$.\footnote{The idea may resemble analytical continuation in complex
  variable, when a function defined in a smaller domain is uniquely
  extended to a larger one.}

Note that if it is possible to define $p^*$ (to close the square) then
$m_\varepsilon$ would be the identity, and vice versa.  In other words,
if there are no dangling edges then it is possible to make the top
square in Fig.~\ref{fig:productionPlusMatch} commute and, hence, it is
not necessary to carry out any production ``continuation''.  The
converse is also true.

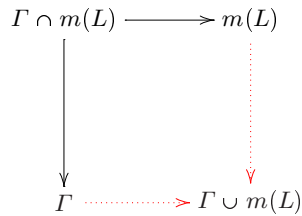
\begin{figure}[htbp]
  \begin{displaymath}
    \xymatrix{
      \Gamma \cap m(L) \ar[dd] \ar[r] & m(L) \ar@{.>}@[red][dd] \\
      & \\
      \Gamma \ar@{.>}@[red][r] & \Gamma \cup m(L)
    }
  \end{displaymath}
  \caption{Match Plus Potential Dangling Edges}
  \label{fig:poConstruction}
\end{figure}

Let be given a production $p:L \rightarrow R$, a host graph $G$ and a
match $m:L \rightarrow G$.  The graph $\Gamma$ is the set of dangling
edges together with their source and target nodes.  Abusing a little
bit of the notation (justified by the pushout construction in
Fig.~\ref{fig:poConstruction}) we will write $\Gamma \cup m(L)$ for
the graph consisting of the image of $L$ by the match plus its
potential dangling edges (and any incident node).  Recall nihilation
matrix definition, especially Lemma~\ref{lemma:nihilationMatrix}.

\newtheorem{extendedMatch}[matrixproduct]{Definition}
\begin{extendedMatch}[Extended Match]\label{def:extendedMatchDef}
  \index{match!extended}With notation as above (refer also to
  Fig.~\ref{fig:matches}), the extended match $\widehat m : L + G
  \rightarrow G$ is a morphism with image $\Gamma \cup m \left( L
  \right)$.
\end{extendedMatch}

As commented above, coproduct in Fig.~\ref{fig:matches} is used just
for coupling $L$ and $G$, being the first embedded into the second by
morphism $m_L$.  We will use the notation
\begin{equation}
  \underline L \stackrel{def}{=} m_G \left( L \right) \stackrel{def}{=} \left( m_\varepsilon \circ m \right)\left( L \right)
\end{equation}
when the image of the LHS is extended with its potential dangling
edges, i.e. extended digraphs are underlined and defined by composing
$m$ and $m_\varepsilon$.\footnote{There is a notational trick here,
  where ``continuation'' is represented as composition of morphisms
  $\left( m_L \circ m_\varepsilon \right)$.  This is not correct
  unless, as explained in Sec.~\ref{sec:completion}, matrices are
  completed.  Recall that completion extends the domain of morphisms
  (interpreting matrices as morphisms between digraphs). This is
  precisely step 1 on p.~\pageref{firstStep}.}

\begin{figure}[htbp]
  \centering
  \includegraphics[scale = 0.4]{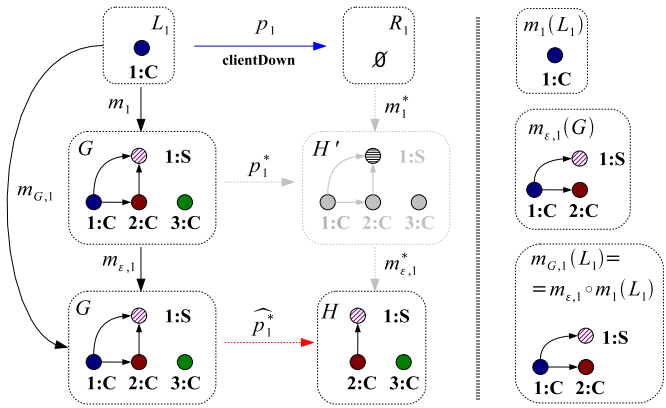}
  \caption{Matching and Extended Match}
  \label{fig:extended_match}
\end{figure}

\noindent\textbf{Example.}$\square$Consider the digraph $L_1$, the host graph
$G$ and the morphism match depicted to the left of
Fig.~\ref{fig:extended_match}.  On the top right side in the same
figure $m_1(L_1)$ is drawn and $m_G \left( L_1 \right)$ on the bottom
right side.  Nodes 2 and 3 and edges $\left( 2, 1 \right)$ and
$\left(2, 3 \right)$ have been added to $m_G \left( L \right)$ which
would become dangling in the image ``graph'' of $G$ by $p_1$ (as it
can not be defined it has been drawn shadowed).  This is why $p^*_1$
can not be defined: node $(1:C)$ would be deleted but not edges
$(1:C, 2:C)$ nor $(1:C, 1:S)$, so $H'$ would not be a digraph.

As commented above, the composition is performed because $m_1$ and
$m_{\varepsilon,1}$ are functions between Boolean matrices that have
been completed.\proofend


Actually it is not necessary to rely on category theory to define
direct derivations.  The basic idea is given precisely by that of
analytical continuation.  What morphism $m_\varepsilon$ does is to
extend the left hand side of the production, i.e. it adds elements to
$L$.  As matches are total functions, they can not delete elements
(nodes or edges) in contrast to productions.

Hence, a match can be seen as a particular type of production with
left hand side $L$ and right hand side $G$.  The LHS of the production
is enlarged with any potential dangling edge and the same for the RHS
except for edges incident to nodes deleted by the production (as they
are not added to its RHS, these edges will be deleted).  This way, a
direct derivation would be
\begin{equation}
  H = \widehat p (m(L)).
\end{equation}

Advancing some material from the next section, $m$ is essentially used
to mark nodes in which $p$ acts.  Production $p$ is the identity in
almost all elements except in some nodes (edges) marked by
$m$.\footnote{Note that $p$'s erasing and addition matrices, although
  as big as the entire system state -- probably huge -- would be zero
  almost everywhere. 
}

The rest of the section is devoted to the interpretation of this
``continuation technique'' as a production, in particular that of
$m_\varepsilon$.  

\label{pr:Tepsilon}Once we are able to complete the rule's LHS we have
to do the same with the rest of the rule.  To this end we define an
operator $T_\varepsilon:\mathfrak{G} \rightarrow \mathfrak{G}'$, where
$\mathfrak{G}$ is the original grammar and $\mathfrak{G}'$ is the
grammar transformed once $T_\varepsilon$ has modified the production.
In words, $T_\varepsilon$ extends production $p$ such that
$T_\varepsilon (p)$ has the same effect than $p$ but also deletes any
dangling edge.

\index{functional representation!production}The notation that we use
from now on is borrowed from functional analysis (see
Sec.~\ref{sec:functionalAnalysis}).  Bringing this notation to graph
grammar rules, a rule is written as $R=\left\langle L, p
\right\rangle$ (separating the static and dynamic parts of the
production) while the grammar rule transformation including matching
is:
\begin{equation}
  \underline R = \left\langle m_G \left( L \right), T_{\varepsilon} p \right\rangle.
\end{equation}


\newtheorem{prodExtension}[matrixproduct]{Proposition}
\begin{prodExtension}\label{prop:prodExtension}
  With notation as above, production $p$ can be extended to consider
  any dangling edge, $\underline R = \left\langle m_G \left( L
    \right), T_{\varepsilon} p \right\rangle$.
\end{prodExtension}
\emph{Proof}\\*
$\square$What we do is to split the identity operator in such a way
that any problematic element is taken into account (erased) by the
production.  In some sense, we first add elements to $p$'s LHS and
afterwards enlarge $p$ to delete them.  Otherwise stated, $m_G^* =
T_\varepsilon^{-1}$ and $T_\varepsilon^* = m_G^{-1}$, so we have:
\begin{equation}
  R = \left\langle L, p \right\rangle = \left\langle L, \left(
      T_\varepsilon ^{-1} \circ T_\varepsilon \right) p \right\rangle
  \nonumber = \left\langle m_G \left( L \right), T_\varepsilon \left(
      p \right) \right\rangle = \underline R.
\end{equation}
The equality $\underline R = R$ is valid only for edges as
$\underline{R}^N$ has the source and target nodes of the dangling
edges.\proofend

The effect of a match can be interpreted as a new production
concatenated to the original production.  Let $p_\varepsilon
\stackrel{def}{=} T^*_\varepsilon$,
\begin{eqnarray}
  \underline{R} = \left\langle m_G \left( L \right), T_\varepsilon
    \left( p \right) \right\rangle & = & \left\langle T_\varepsilon^*
    \left( m_G \left( L \right) \right), p
  \right\rangle = \\ \nonumber
  & = & p \left( T_\varepsilon^* \left( m_G \left( L \right) \right)
  \right) = p \, ; \, p_\varepsilon \, ; \, m_G \left( L \right) = p
  \, ; \, p_\varepsilon \left( \underline L \right).
\end{eqnarray}

\index{p@$\varepsilon$-production}\index{production!$\varepsilon$}Production
$p_\varepsilon$ is the $\varepsilon$-production associated to
production $p$.  Its aim is to delete potential dangling edges.  The
dynamic definition of $p_\varepsilon$ is given
in~(\ref{eq:epsilonProd1})~and~(\ref{eq:epsilonProd2}).

The fact of taking the match into account can be interpreted as a
temporary modification of the grammar, so it can be said that the
grammar modifies the host graph and the host graph interacts with the
grammar (altering it temporarily).

If we think of $m_G$ and $T_\varepsilon^*$ as productions respectively
applied to $L$ and $m_G \left( L \right)$, it is necessary to specify
their erasing and addition matrices.  To this end, recall matrix
$\overline{D}$ defined in Lemma~\ref{lemma:nihilationMatrix}, with
elements in row $i$ and column $i$ equal to one if node $i$ is to be
erased by $p$ and zero otherwise, which considers any potential
dangling edge.

For $m_G$ we have that $\underline e^N = \underline e^E = 0$, and
$\underline{r} = \underline L \, \overline{L}$ (for both nodes and
edges), as the production has to add the elements in $\underline L$
that are not present in $L$.  Let $p_\varepsilon = \left(
  e^E_{T_\varepsilon} , r^E_{T_\varepsilon}, e^N_{T_\varepsilon},
  r^N_{T_\varepsilon} \right)$, then
\begin{eqnarray}
  e^N_{T_\varepsilon} & = & r^E_{T_\varepsilon} = r^N_{T_\varepsilon} = 0 \label{eq:epsilonProd1} \\
  e^E_ {T_\varepsilon} & = & \overline{D} \wedge \underline{L}^E.\label{eq:epsilonProd2}
\end{eqnarray}

\begin{figure}[htbp]
  \centering
  \includegraphics[scale = 0.45]{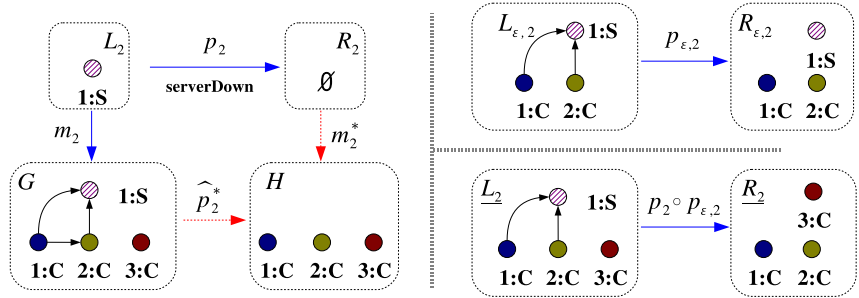}
  \caption{Full Production and Application}
  \label{fig:match_comatch_example}
\end{figure}

\noindent\textbf{Example}.$\square$Consider rules depicted in
Fig.~\ref{fig:match_comatch_example}, in which \texttt{serverDown} is
applied to model a server failure.  We have:

\begin{equation}
  e^E =	r^E = L^E =	\left[
    \begin{array}{c|c}
      0 & \, 1 \\
    \end{array}	\right]; \quad	\nonumber
  e^N =	\left[
    \begin{array}{c|c}
      1 & \, 1 \\
    \end{array}	\right] \nonumber
\end{equation}
\begin{equation}
  r^N =	\left[
    \begin{array}{c|c}
      0 & \, 1 \\
    \end{array}	\right]; \quad
  L^N =	\left[
    \begin{array}{c|c}
      1 & \, 1 \\
    \end{array}	\right]; \quad
  R^E = R^N =	\emptyset. \nonumber
\end{equation}

Once $m_G = \left(L^E, L^N, \underline{r}^E, 0, 0, 0 \right)$ and
operator $T_\varepsilon$ have been applied, giving rise to
$p_\varepsilon = \left( \underline{L}^E, \underline{L}^N, 0, 0,
  e^E_{T_\varepsilon}, 0 \right)$, the resulting matrices are:

\begin{equation}
  \underline{r}^E =	\left[
    \begin{array}{ccc}
      \vspace{-6pt}
      0 & 0 & 0 \\
      \vspace{-6pt}
      1 & 0 & 0 \\
      \vspace{-6pt}
      1 & 0 & 0 \\
      \vspace{-10pt}
    \end{array}	\right], \;
  \underline{L}^E =	\left[
    \begin{array}{ccc}
      \vspace{-6pt}
      0 & 0 & 0 \\
      \vspace{-6pt}
      1 & 0 & 0 \\
      \vspace{-6pt}
      1 & 0 & 0 \\
      \vspace{-10pt}
    \end{array}	\right], \;
  \underline{R}^E =	\left[
    \begin{array}{cc}
      \vspace{-6pt}
      0 & 0 \\
      \vspace{-6pt}
      0 & 0 \\
      \vspace{-10pt}
    \end{array}	\right], \;
  e_{T_\varepsilon}^E =	\left[
    \begin{array}{ccc}
      \vspace{-6pt}
      0 & 0 & 0 \\
      \vspace{-6pt}
      1 & 0 & 0 \\
      \vspace{-6pt}
      1 & 0 & 0 \\
      \vspace{-10pt}
    \end{array}	\right],	\nonumber
\end{equation}
where ordering of nodes is $[1:S, 1:C, 2:C]$ for matrices
$\underline{r}^E$, $\underline{L}^E$ and $e_{T_\varepsilon}^E$ and
$[1:C,2:C]$ for $\underline{R}^E$.  Matrix $\underline{r}^E$, besides
edges added by the production, specifies those to be added by $m_G$ to
the LHS in order to consider any potential dangling edge (in this case
$\left( 1:C, 1:S \right)$ and $\left( 2:C, 1:S \right)$).  As neither
$m_G$ nor production \texttt{serverDown} delete any element,
$\underline{e}^E = 0$.  Finally, $p_{\varepsilon}$ removes all
potential dangling edges (check out matrix $e_{T_\varepsilon}^E$) but
it does not add any, so $r_{T_\varepsilon}^E = 0$.  Vectors for nodes
have been omitted.\proofend

\index{e@$\varepsilon$-production!adjoint operator}Let
$T^*_\varepsilon = \left( {T^*_\varepsilon}^N, {T^*_\varepsilon}^E
\right)$ be the adjoint operator of $T_\varepsilon$.  We will end this
section giving an explicit formula for $T^*_\varepsilon$.  Define
$e^E_{\varepsilon}$ and $r^E_{\varepsilon}$ respectively as the
erasing and addition matrices of $T_{\varepsilon} \left( p \right)$.
It is clear that $r^E_{\varepsilon} = \underline{r}^E = r^E$ and
$e^E_{\varepsilon} = e^E \vee \overline{D} \, \underline{L}^E$, so
\begin{eqnarray}
  \underline{R}^E & = & \left\langle \, \underline{L}^E, T_\varepsilon
    \left( p \right) \, \right\rangle = r_\varepsilon^E \vee
  \overline{e_\varepsilon^E} \, \underline{L}^E = r^E \vee
  \overline{\left( e^E \vee \overline{D} \, \underline{L}^E \right)}
  \, \underline{L}^E = \nonumber \\ 
  & = & r^E \vee \left( D \vee \overline{ \underline{L}^E} \right)
  \overline{e^E} \underline{L}^E = r^E \vee \overline{e^E} D
  \underline{L}^E.
\end{eqnarray}

Previous identities show that $\underline{R}^E = \left\langle
  \underline{L}^E, T_{\varepsilon}^E \left( p^E \right) \right\rangle
= \left\langle D \, \underline{L}^E, p^E \right\rangle$, which proves
the identity:

\begin{equation}
  T^*_\varepsilon = \left( {T^*_\varepsilon}^N, {T^*_\varepsilon}^E \right) = \left( id, D \right).
\end{equation}

Summarizing, when a match $m$ is considered for a production $p$, the
production itself is first modified in order to consider all potential
dangling edges. Morphism $m$ is automatically transformed into a match
which is free from any dangling element and, in a second step, a
pre-production $p_\varepsilon$ is appended to form the
concatenation\footnote{It is also possible to define it as the
  composition: $\widehat{p}^* = p^* \circ p^*_\varepsilon$.}
\begin{equation}\label{eq:compProd}
  \widehat{p}^* = p^* \,; p^*_\varepsilon.
\end{equation}

Note that as injectiveness of matches is demanded, there is no problem
such as elements identified by matches that are both kept and deleted.

Depending on the operator $T_\varepsilon$, side effects are permitted
(SPO-like behaviour) or not (DPO-like behaviour).\index{fixed grammar}
A \emph{fixed grammar} or fixed Matrix Graph Grammar is one in which
(mandatory) the operator $T_\varepsilon$ is the identity. If the
operator is not forced to be the identity, we will speak of a
\index{floating grammar}\emph{floating grammar} or floating Matrix
Graph Grammar. Notice that the existence of side effects is equivalent
to transforming a production into a sequence. This will also be the
case when we deal with graph constraints and application conditions
(Chap.~\ref{ch:restrictionsOnRules}).

\section{Marking}
\label{sec:marking}

In previous section the problem of dangling edges has been addressed
by adding an $\varepsilon$-production which deletes any problematic
edge, so the original rule can be applied as it is.  However there is
no way to guarantee that both productions will use the same elements
(recall that in general matches are non-deterministic).  The same
problem exists with application conditions
(Sec.~\ref{sec:functionalRepresentation}) or whenever a rule is split
into subrules and applying them to the same elements in the host graph
is desired.

This topic is studied in~\cite{Schurr} (for a different reason) and
the solution proposed there is to ``pass'' the match from one
production to the other.  \index{marking!operator}We will tackle this
problem in a different way that consists in defining an operator
$T_{\mu,\alpha}$ for a label $\alpha$ acting on production $p$ as
follows:

\begin{itemize}
\item If no node is typed $\alpha$ in $p$ then a new node labeled
  $\alpha$ is added and connected to every already existing node.
\item If, on the contrary, there exists a node of that type then it is
  deleted.
\end{itemize}

The basic idea is to mark nodes and related productions with a node of
type $\alpha$.  The operator behaves differently depending on whether
it is marking the state (it adds node $\alpha$) or it is extending the
productions ($\alpha$-typed nodes are removed).

\begin{figure}[htbp]
  \centering
  \includegraphics[scale = 0.54]{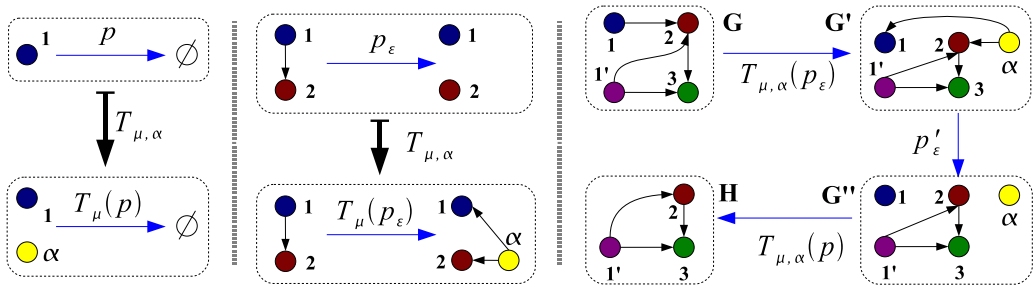}
  \caption{Example of Marking and Sequence $s = p;p_\varepsilon$}
  \label{fig:marking}
\end{figure}

For an example of a short sequence of two productions, please refer to
Fig.~\ref{fig:marking}.  Using functional analysis notation:
\begin{equation}
  R = \left\langle L, p\right\rangle \longmapsto \underline{R} =
  \left\langle m_\varepsilon (L), T_{\varepsilon} (p)\right\rangle
  \longmapsto \underline{\underline{R}} = \left\langle m_\varepsilon (L), T_\mu
    \circ T_{\varepsilon} (p)\right\rangle
\end{equation}
where, as in Sec.~\ref{sec:matchAndExtendedMatch}, $\underline R$ is
the extended rule's RHS that considers any dangling edge.

If a production is split into two subproductions, say $p \longmapsto
T_\varepsilon(p) = p\,; p_\varepsilon$ and we want them to be applied
in the same nodes of the host graph, we may proceed as follows:
\begin{itemize}
\item Enlarge $p_{\varepsilon}$ to add one node of some non-existent
  type ($\alpha$) together with edges starting in this node and ending
  in nodes used by $p_\varepsilon$.
\item Enlarge $p$ to delete $\alpha$ nodes of previous step.
\end{itemize}

It is important to note that $p$ must be enlarged to delete only the
previously added node ($\alpha$) and not the edges starting in
$\alpha$ appended by $T_\mu$ to $p_\varepsilon$.  The reason is that
in case of a sequence in which the $\varepsilon$-production is
advanced several positions, there exists the possibility to create
unreal dependencies between $p$ and some production applied before $p$
but after $p_\varepsilon$ (the example below illustrates this point in
particular).

Marking will normally create new $\varepsilon$-productions related to
$p$.  Note however that no recursive process should arise as there
shouldn't be any interest in permuting (advancing) this new
$\varepsilon$-productions.

For $\varepsilon$-productions all this makes sense just in case we do
not compose $p \circ p_\varepsilon$ (no marking would be needed).  Two
different operators, one for $\alpha$ nodes addition and another for
$\alpha$ nodes deletion (instead of just one) are not defined because
marking always acts on different productions.  This should not cause
any confusion.

\noindent\textbf{Example.}$\square$Figure~\ref{fig:marking} illustrates the
process for a simple production $p$ that deletes node 1 and is applied
to a host graph in which one or two dangling edges (depending on the
match, $1$ or $1'$) would be generated, $(1,2)$ or $(1',2)$ and
$(1',3)$.

We have chosen node $1$ for the match so there should be one dangling
edge $(1,2)$.  In order to avoid it, an $\varepsilon$-production
$p_\varepsilon$ which deletes $(1,2)$ is appended to $p$.

The marking process modifies $p_\varepsilon$ and $p$ becoming
$p_\varepsilon \mapsto T_\mu (p_\varepsilon)$ and $p \mapsto T_\mu
(p)$, respectively.  Note that $T_\mu (p)$ generates two dangling
edges -- $(\alpha,1)$ and $(\alpha,2)$ -- so a new
$\varepsilon$-production $p'_\varepsilon$ ought to be added.

When the production is applied, a sequence is generated as operators
act on the production -- $p \mapsto T_\varepsilon (p) \mapsto T_\mu
\circ T_\varepsilon (p) \mapsto T_\varepsilon \circ T_\mu \circ
T_\varepsilon (p)$ -- giving rise to the following sequence of
productions:
\begin{equation}
  p \longmapsto p\,; p_\varepsilon \longmapsto T_\mu(p);
  p'_{\varepsilon}; T_\mu (p_\varepsilon).
\end{equation}

The reason why it is important to specify only the new node deletion
($\alpha$) and not the edges starting in this node is not difficult but
might be a bit subtle.  It has been mentioned above.  The rest of the
example is devoted to explaining it.

If we specified the edges also, say $(\alpha,1)$ and $(\alpha,2)$ as
above, then the transformed production $T_\mu(p)$ would use node 2 as
it should appear in its LHS and RHS (remember that $p$ did not act on
node $2$).

Now imagine that we are interested in advancing the
$\varepsilon$-production three positions, for example because we know
that it is external (see
Sec.~\ref{sec:internalAndExternalProductions}) and independent: $p\,;
p_\varepsilon;p_2;p_1 \mapsto p\,;p_2;p_1;p_\varepsilon$.  Suppose
that production $p_1$ (placed between $p$ and the new allocation of
$p_\varepsilon$) deletes node 2 and production $p_2$ adds it. If $p$
was sequential independent with respect to $p_1$ and $p_2$ then it
would not be anymore due to the edge ending in node $2$ because now
$p$ would use node $2$ (appears in its left and right hand
sides).\proofend

Note that as the marking process can be easily automated, we can
safely ignore it and assume that it is somehow being performed, by
some runtime environment for example.

\section{Initial Digraph Set and Negative Digraph Set}
\label{sec:initialDigraphSet}


Concerning minimal and negative initial digraphs there may be
different ways to complete rule matrices, depending on the matches.
Therefore, we no longer have a unique initial digraph but a set (if we
assume any possible match). In fact two sets, one for elements that
must be found in the host graph and another for those that must be
found in its complement.  This section is closely related to
Secs.~\ref{sec:MID}~and~\ref{sec:NID} and extends results
therein proved.

The \emph{initial digraph set} contains all graphs that can be
potentially identified by matches in concrete host graphs.

\newtheorem{InitDigraphSet}[matrixproduct]{Definition}
\begin{InitDigraphSet}[Initial Digraph
  Set]\label{def:initialDigraphSet}
  \index{initial digraph!set}Given sequence $s_n$, its associated
  \emph{initial digraph set} $\mathfrak{M} \left( s_n \right)$ is the
  set of simple digraphs $M_i$ such that $\forall M_i \in \mathfrak{M}
  \left( s_n \right)$:
  \begin{enumerate}
  \item $M_i$ has enough nodes and edges for every production of the
    concatenation to be applied in the specified order.
  \item $M_i$ has no proper subgraph with previous property (keeping
    identifications).
  \end{enumerate}
\end{InitDigraphSet}

Every element $M_i \in \mathfrak{M} \left( s_n \right)$ is said to be
an \emph{initial digraph} for $s_n$.  It is easy to see that $\forall
s_n$ finite sequence of productions we have $\mathfrak{M} \left( s_n
\right) \neq \emptyset$.

In Sec.~\ref{sec:sequencesAndCoherence} coherence was used in a more
or less absolute way when dealing with sequences, assuming some
horizontal identification of elements.  Now we see that, due to
matching, coherence is a property that may depend on the given initial
digraph so, depending on the context, it might be appropriate to say
that $s_n$ is coherent with respect to initial digraph $M_i$ (just in
case direct derivations are considered).  Note that what we fix by
choosing an initial digraph is the relative matching of nodes across
productions (one of the actions of completion).

For the initial digraph set we can define the \emph{maximal initial
  digraph} as the element $M_n \in \mathfrak{M} \left( s_n \right)$
that considers all nodes in $p_i$ to be different.  This element is
unique up to isomorphism, and corresponds to considering the parallel
application of every production in the sequence, i.e. the LHS of every
production in the sequence is matched in disjoint parts of the host
graph.

This concept has already been used although it was not explicitly
mentioned: In the proof of Theorem~\ref{th:minProdTh} we started with
$\bigvee_{i=1}^n L_i$, a digraph that had enough nodes to perform all
actions specified by the sequence.

In a similar way, $M_i \in \mathfrak{M} \left( s_n \right)$ in which
all possible identifications are performed are known as \emph{minimal
  initial digraphs}.  Contrary to the maximal initial digraph, minimal
initial digraphs need not be unique as the following example shows.

\begin{figure}[htbp]
  \centering
  \includegraphics[scale = 0.49]{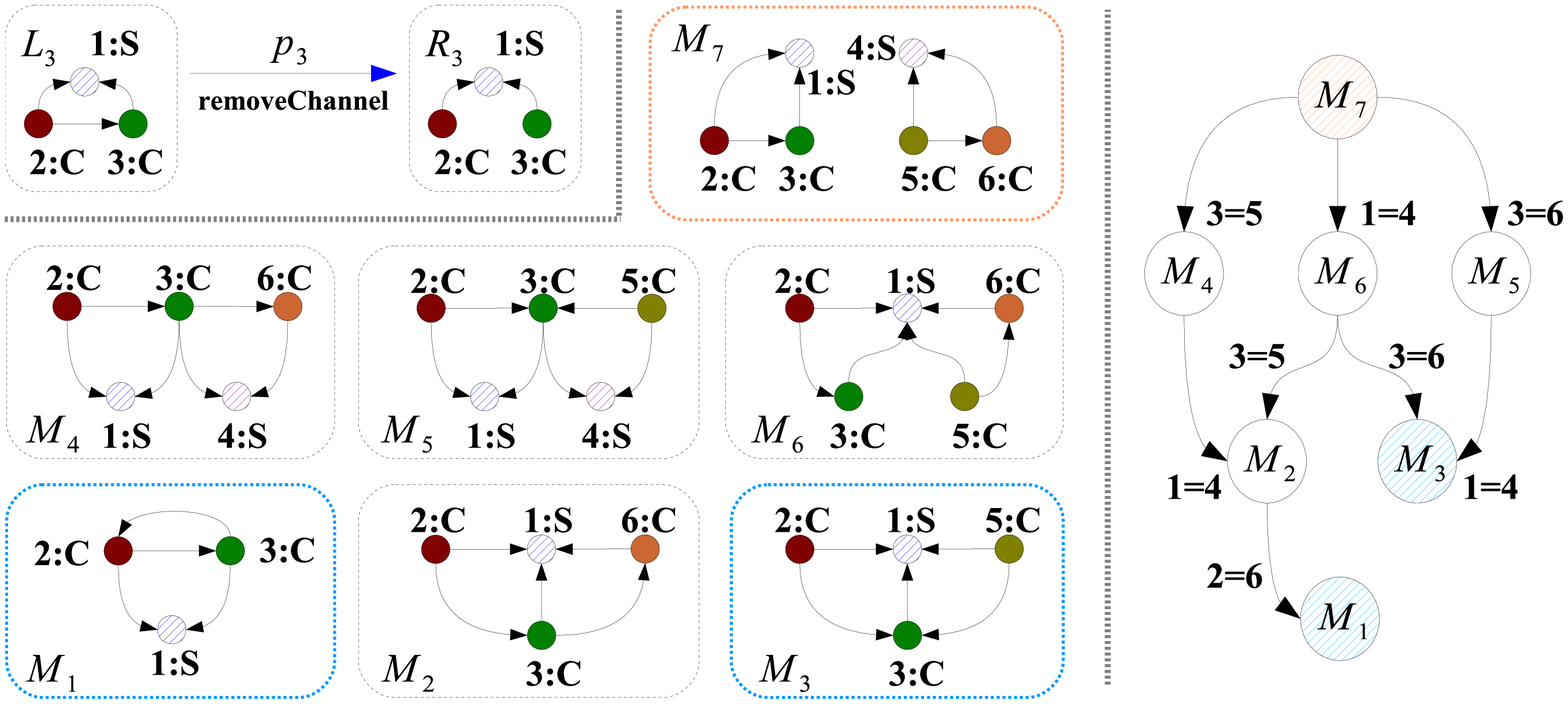}
  \caption{Initial Digraph Set for
    \texttt{s=remove\_channel;remove\_channel}}
  \label{fig:digraphSet}
\end{figure}

\label{ex:removeChannel}\noindent\textbf{Example}.$\square$In
Figure~\ref{fig:digraphSet} we have represented the minimal digraph set
for the sequence \texttt{s = removeChannel;removeChannel}.  The
production is also depicted in the figure where \textbf{S} stands for
\emph{server} and \textbf{C} for \emph{client}.  Note that it is not
coherent if all nodes in $L_3$ are identified because the link between
two clients is deleted twice.  Therefore, the initial digraphs should
provide at least (in fact, at most) two different links between
clients.

In the figure, the maximal initial digraph is $M_7$ and $M_1$ and
$M_3$ are the two minimal initial digraphs. Identifications are
written as $i=j$ meaning that nodes $i$ and $j$ become one and the
same.  A top-bottom procedure has been followed, starting out with the
biggest digraph $M_7$ and ending in the smallest.  Numbers on labels
are all different to ease identifications on the initial digraph tree
to the right of Fig.~\ref{fig:digraphSet}.\proofend

We can provide $\mathfrak{M} \left( s_n \right)$ with some structure
$\mathfrak{T} \left( s_n \right)$. See the right side of
Fig.~\ref{fig:digraphSet}.  Every node in $\mathfrak{T}$ represents
an element of $\mathfrak{M}$.  A directed edge from one node to
another stands for one operation of identification between
corresponding nodes in the LHS and RHS of productions of the sequence
$s_n$.

Following with the example above, node $M_7$ is the maximal initial
digraph, as it only has outgoing edges.  Nodes $M_1$ and $M_3$ are
minimal as they only have ingoing edges. The structure $\mathfrak T$
is an acyclic digraph with single root node (recall that there is just
one maximal initial digraph), known as \emph{graph-structured stack}.

It is possible to make a similar construction for negative initial
digraphs that we will call \emph{negative initial set}.  It will be
represented by $\mathfrak{N}(s_n)$ where $s_n$ is the sequence under
study.

\begin{figure}[htbp]
  \centering
  \includegraphics[scale = 0.49]{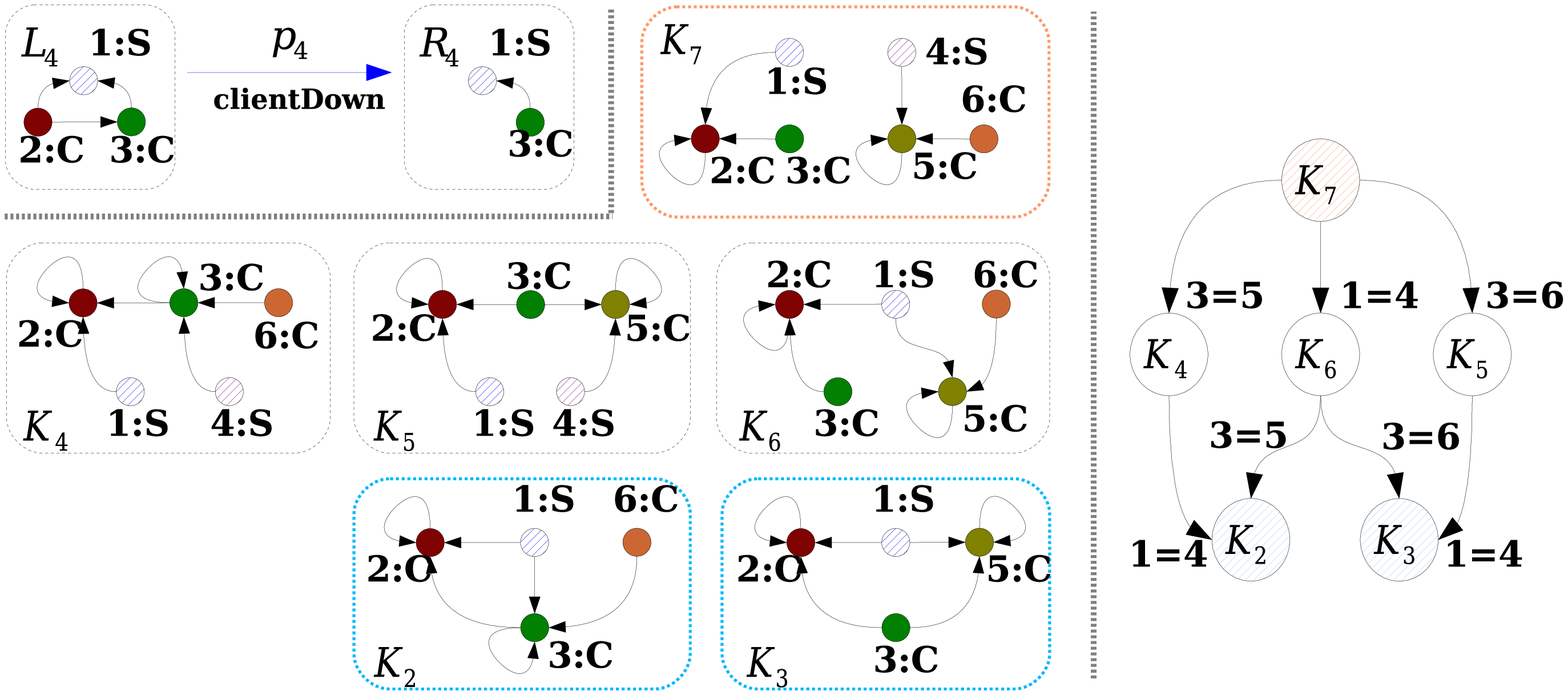}
  \caption{Negative Digraph Set for \texttt{s=clientDown;clientDown}}
  \label{fig:negativeSet}
\end{figure}

\newtheorem{NegInitialSet}[matrixproduct]{Definition}
\begin{NegInitialSet}[Negative Initial Set]\label{def:negInitialSet}
  \index{negative!initial set}Given sequence $s_n$, its associated
  \emph{negative initial set} $\mathfrak{N} \left( s_n \right)$ is the
  set of simple digraphs $K_i$ such that $\forall K_i \in \mathfrak{N}
  \left( s_n \right)$:
  \begin{enumerate}
  \item $K_i$ specifies all edges that can potentially prevent the
    application of some production of $s_n$.
  \item $K_i$ has no proper subgraph with previous property (keeping
    identifications).
  \end{enumerate}
\end{NegInitialSet}

\noindent\textbf{Example}.$\square$We study the sequence
\texttt{s=clientDown;clientDown} very similar to that in the example
of p.~\pageref{ex:removeChannel} but deleting one node and two edges.
It is depicted in Fig.~\ref{fig:negativeSet} and represents the
failure of a client connected to a server and to another client.

\begin{figure}[htbp]
  \centering
  \includegraphics[scale = 0.54]{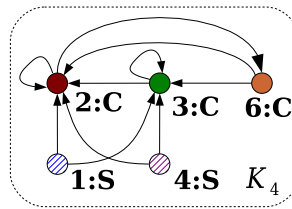}
  \caption{Complete Negative Initial Digraph $K_4$}
  \label{fig:negativeSetComplete}
\end{figure}

The same labeling criteria has been followed to ease comparison.
Minimal digraphs are very similar to those in
Fig.~\ref{fig:digraphSet} and in fact identifications have been
performed such that $K_i$ corresponds to $M_i$.  Graphs do not include
all edges that should not appear because there would be many edges,
probably being a confusing instead of a clarifying example.  For
instance, in $K_4$ there can not be any edge incident to node $(6:C)$
(except those coming from $(1:S)$ and $(4:S)$), in particular edge
$(2:C, 6:C)$ which is not represented.  Complete graph $K_4$ can be
found in Fig.~\ref{fig:negativeSetComplete}.  Note that for $K_4$ the
order of deletion is important, first node $(2:C)$ and then node
$(3:C)$.\proofend


The relationship between elements in $\mathfrak{M}$ and $\mathfrak{N}$
is compiled in Corollary~\ref{cor:eqNIDMID}.  Note that the
cardinality of both sets do not necessarily coincide.  In the example
of p.~\pageref{ex:removeChannel}, production $s$ does not add any edge
nor deletes any node (hence, no forbidden element) so its negative
digraph set is
empty.

Although in this book we are staying at a more theoretical level, we
will make a small digression on application of these concepts and
possible implementations.

Let's take as an example the calculation of $M_0$ in
Proposition~\ref{prop:SeqIndProp}, which states that two derivations
$d$ and $d'$ are sequential independent if they have a common initial
digraph for some identification of nodes, i.e. if $\mathfrak{M}(d)
\cap \mathfrak{M}(d') \neq \emptyset$.  We see that it is possible to
follow two complementary approaches:
\begin{itemize}
\item \textbf{Top-bottom}. Begin with the \emph{maximal initial
    digraph} and start identifying elements until we get the desired
  initial digraph or eventually get a contradiction.
\item \textbf{Bottom-up}. Start with different initial digraphs and
  unrelate nodes until an answer is reached.
\end{itemize}

In Fig.~\ref{fig:digraphSet} on p.~\pageref{fig:digraphSet} either we
begin with $M_7$ and start identifying nodes, eventually getting any
element of the minimal initial set, or we start with $M_1$ -- which is
not necessarily unique -- and build up the whole set, or stop as soon
as we get the desired minimal initial digraph.

Let the matrix filled up with 1's in all positions be represented by
\textbf{1}.  For the first case the following identity may be of some
help:
\begin{equation}\label{eq:satSolver}
  M_d = M_{d'} \Leftrightarrow M_d M_{d'} \vee \overline{M}_d \overline{M}_{d'} = 1.
\end{equation}

A SAT solver can be used on~(\ref{eq:satSolver}) to obtain conditions,
setting all elements in $M$ as variables except those already known.
In order to store $M$, binary decision diagrams -- BDD -- can be
employed. Refer to~\cite{CLR90}.

The same alternative processes might be applied to the negative
initial set to eventually reach any of its elements.

\section{Internal and External $\varepsilon$-productions}
\label{sec:internalAndExternalProductions}

Dangling edges can be classified into two disjoint sets according to
the \emph{place} where they appear, whether they have been added by a
previous production or not.

For example, given the sequence $p_2;p_1$, suppose that rule $p_1$
uses but does not delete edge $\left( 4, 1 \right)$, that rule $p_2$
specifies the deletion of node $1$ and that we have identified both
nodes $1$.  It is mandatory to add one $\varepsilon$-production
$p_{\varepsilon, 2}$ to the grammar with the disadvantage that there
is an unavoidable problem of coherence between $p_1$ and
$p_{\varepsilon, 2}$ if we wanted to advance the application of
$p_{\varepsilon, 2}$ to $p_1$, i.e. they are sequentially dependent.

Hence, edges of $\varepsilon$-productions are of two different types:
\begin{itemize}
  \index{edge!external}\index{initial digraph!actual}\item
  \textbf{External}: Any edge not appearing explicitly in the grammar
  rules, i.e. edges of the host graph ``in the surroundings'' of the
  actual initial digraph.\footnote{Among all possible initial digraphs
    in the initial digraph set for a given concatenation, if one is
    already fixed (matches have already been chosen), it will be known
    as \emph{actual initial digraph}.}  Examples are edges $\left(
    1:C, 1:S \right)$ and $\left( 2:C, 1:S \right)$ in
  Fig.~\ref{fig:match_comatch_example} on
  p.~\pageref{fig:match_comatch_example}.  \index{edge!internal}\item
  \textbf{Internal}: Any edge used or appended by a previous
  production in the concatenation.  One example is edge $\left( 4, 1
  \right)$ mentioned above.
\end{itemize}

\index{p@$\varepsilon$-production!internal}\index{p@$\varepsilon$-production!external}$\varepsilon$-productions
can be classified in \textbf{internal} $\varepsilon$-productions if
any of its edges is internal and \textbf{external}
$\varepsilon$-productions otherwise.

The ``advantage'' of internal over external $\varepsilon$-productions
is that the former can be considered (are known) during rule
specification while external remain unknown until the production is
applied.  This, in turn, may spoil coherence, compatibility and other
calculations performed during grammar definition.

On the other hand, external $\varepsilon$-productions do not interfere
with grammar rules so they can be advanced to the beginning and they
can even be composed to get a single production if so desired (these
are called \emph{exact derivations}, defined below).

\begin{figure}[htbp]
  \centering
  \includegraphics[scale = 0.45]{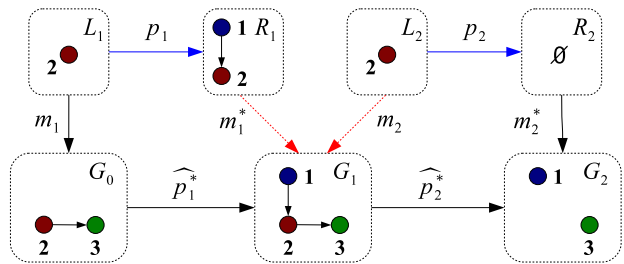}
  \caption{Example of Internal and External Edges}
  \label{fig:internalExternalExample}
\end{figure}

\noindent\textbf{Example}.$\square$Let's consider the derivation $d_2 =
p_2;p_1$ (see Fig.~\ref{fig:internalExternalExample}).  Edge $(1,2)$
in graph $G_1$ is internal (it has been added by production $p_1$)
while edge $(2,3)$ in the same graph is external (it already existed
in $G_0$).\proofend

Given a host graph $G$ in which $s_n$ -- coherent and compatible -- is
to be applied, and assuming a match which identifies $s_n$'s actual
initial digraph ($M_n$) in $G$ (defining a derivation $d_n$ out of
$s_n$), we check whether for some $\widehat m$ and
$\widehat{T_{\varepsilon}}$, which respectively represent all changes
to be done to $M_n$ and all modifications to $s_n$, it is correct to
write
\begin{equation}\label{eq:exactDer}
  H_n = d_n \left(M_n \right) = \left\langle \widehat{m} \left( M_n
    \right), \widehat{T_{\varepsilon}} \left( s_n \right)
  \right\rangle,
\end{equation}
where $H_n$ is the subgraph of the final state $H$ corresponding to
the image of $M_n$.

Equation~(\ref{eq:EdgeConc}) allows us to consider a concatenation
almost as a production, justifying operators
$\widehat{T_{\varepsilon}}$ and $\widehat{m}$ in
eq.~\eqref{eq:exactDer} and our abuse of notation (recall that bra and
kets apply to productions and not to sequences).

All previous considerations together with the following example are
compiled into the definition of \emph{exact sequence}.

\noindent\textbf{Example.}$\square$Let $s_2 = p_2;p_1$ be a coherent and
compatible concatenation.  Using operators we can write
\begin{equation}
  H=\left\langle m_{G,2} \left( \left\langle m_{G,1} \left( M_2
        \right), T_{\varepsilon , 1}\left( p_1 \right) \right\rangle
    \right), T_{\varepsilon , 2}\left( p_2 \right) \right\rangle,
\end{equation}
which is equivalent to $H=p_2; \, p_{\varepsilon, 2} ;p_1; \,
p_{\varepsilon, 1} \left(\underline{M_2}\right)$, with actual initial
digraph twice modified $\underline{M_2} = m_{G,2} \left( m_{G,1}
  \left( M_2 \right)\right) = \left(m_{G,2} \circ m_{G,1} \right)
\left( M_2 \right)$.  \proofend

\newtheorem{exactDerivation}[matrixproduct]{Definition}
\begin{exactDerivation}[Exact Derivation]\label{def:exactDerivation}
  \index{derivation!exact}Let $d_n = \left( s_n, m_n \right)$ be a
  derivation with actual initial digraph $M_n$, sequence $s_n =
  p_n;\ldots;p_1$, matches $m_n = \{ m_{G,1}, \ldots, m_{G,n}\}$ and
  $\varepsilon$-productions $\{ p_{\varepsilon, 1}, \ldots,
  p_{\varepsilon, n} \}$.  It is an \emph{exact} derivation if there
  exist $\widehat{m}$ and $\widehat{T}_{\varepsilon}$ such that
  equation~(\ref{eq:exactDer}) is fulfilled.
\end{exactDerivation}

Equation~(\ref{eq:exactDer}) is satisfied if once all matches are
calculated, the following identity holds:
\begin{equation}
  p_n; p_{\varepsilon, n}; \ldots; p_1; p_{\varepsilon, 1} = p_n;
  \ldots; p_1; p_{\varepsilon, n}; \ldots; p_{\varepsilon, 1}.
\end{equation}

\newtheorem{ExactnessProp1}[matrixproduct]{Proposition}
\begin{ExactnessProp1}\label{prop:ExactnessProp1}
  With notation as in Def.~\ref{def:exactDerivation}, if
  $p_{\varepsilon, j} \bot \left( p_{j-1}; \ldots ;p_1 \right)$,
  $\forall j$, then $d_n$ is exact.
\end{ExactnessProp1}

\noindent \emph{Proof}\\*
$\square$Operator $\widehat{T_\varepsilon}$ modifies the sequence
adding a unique $\varepsilon$-production, the composition of all
$\varepsilon$-productions $p_{\varepsilon, i}$.  To see this, if one
edge is to dangle, it should be eliminated by the corresponding
$\varepsilon$-production so no other $\varepsilon$-production deletes
it unless it is added by a subsequent production.  But by hypothesis
there is sequential independence of every $p_{\varepsilon, j}$ with
respect to all preceding productions and hence $p_{\varepsilon, j}$
does not delete any edge used by $p_{j-1}, \ldots, p_1$.  In
particular no edge added by any of these productions is erased.

In Def.~\ref{def:exactDerivation}, $\widehat{m}$ is the extension of
the match $m$ which identifies the actual initial digraph in the host
graph, so it adds to $m \left( M_n \right)$ all nodes and edges to
distance one to nodes that are going to be erased.  A symmetrical
reasoning to that of $\widehat{T_\varepsilon}$ shows that
$\widehat{m}$ is the composition of all $m_{G,i}$.  \proofend

With Def.~\ref{def:exactDerivation} and
Prop.~\ref{prop:ExactnessProp1} it is feasible to obtain a
concatenation where all $\varepsilon$-productions are applied first,
and all grammar rules afterwards, recovering the original
concatenation.  Despite some obvious advantages, all dangling edges
are deleted at the beginning which may be counterintuitive or even
undesired if, for example, the deletion of a particular edge is used
for synchronization purposes.

The following corollary states that exactness can only be ruined by
internal $\varepsilon$-productions.

\newtheorem{ExactnessProp2}[matrixproduct]{Corollary}
\begin{ExactnessProp2}\label{cor:ExactnessProp2}
  Let $s_n$ be a sequence to be applied to a host graph $G$ and $M_k
  \in \mathfrak{M} \left( s_n \right)$.  Assume there exists at least
  one match in $G$ for $M_k$ that does not add any internal
  $\varepsilon$-production.  Then, $d_n$ is exact.
\end{ExactnessProp2}

\noindent \emph{Proof (sketch)}\\*
$\square$All potential dangling elements are edges surrounding the
actual initial digraph.  It is thus possible to adapt the part of the
host graph modified by the sequence at the beginning, so applying
Prop.~\ref{prop:ExactnessProp1} we get exactness.\proofend

We are now in the position to characterize applicability,
problem~\ref{prob:applicability} stated on
p.~\pageref{prob:applicability}. In essence, applicability
characterizes when a sequence is a derivation with respect to a given
initial graph. 

\newtheorem{applicabilityCharac_2}[matrixproduct]{Theorem}
\begin{applicabilityCharac_2}[Applicability
  Characterization]\label{th:applicabilityCharac_2}
  A sequence $s_n$ is applicable to $G$ if there are matches for every
  production (define the derivation $d_n$ as the sequence $s_n$ plus
  these matches) such that any of the two following equivalent
  conditions is fulfilled:
  \begin{itemize}
  \item Derivation $d_n$ is coherent and compatible.
  \item $d_n$'s minimal initial digraph is in $G$ and $d_n$'s negative
    initial digraph is in $\overline{G}$.
  \end{itemize}
\end{applicabilityCharac_2}
\emph{Proof}\\*
$\square$$\blacksquare$
%

%
%

\section{Summary and Conclusions}
\label{sec:summaryAndConclusions4}

In this chapter we have seen how it is possible to match the left hand
side of a production in a given graph. We have not given a matching
algorithm, but the construction of derivations out of
productions.  

There are two properties that we would like to highlight.  The
expressive power of Matrix Graph Grammars lies in between that of
other approaches such as DPO and SPO:
\begin{itemize}
\item We find it more intuitive and convenient to demand injectiveness
  on matches.  This can be seen as a limitation on the semantics of
  the grammar but, on the other hand, not asking for injectiveness
  might present a serious problem. For example, when injectivity is
  necessary for some rules or non-injectivity is not allowed in some
  parts of the host graph. In a limit situation, it can be the case
  that several nodes and edges collapse to a single node and a single
  edge.
\item Rules can be applied even if they do not consider every edge
  that can appear in some given state.  The grammar designer can
  concentrate on the algorithm at a more abstract level, without
  worrying about every single case in which a concrete rule needs to
  be applied.\footnote{In cases of hundreds of rules, when every rule
    adds and deletes nodes and edges, it can be very difficult to keep
    track if some actions are still available. The canonical example
    would be a rule $p$ that deletes some special node but can not be
    applied because some other production eventually added one
    incident edge that is not considered in the left hand side of
    $p$.}
\end{itemize}

An advantage of $\varepsilon$-productions over previous approaches to
dangling edges is that they are erased by productions. This increases
our analysis abilities as there are no side effects.

We have also introduced marking, useful in many situations in which it
is necessary to guarantee that some parts of two or more rules will be
matched in the same area of the host graph.
It will be used throughout the rest of the book.

Initial and negative digraph sets are a generalization of minimal and
negative initial digraphs in which some or all possible
identifications are considered.  Actually, these concepts could have
been introduced in Chap.~\ref{ch:mggFundamentals2}, but we have
postponed their study because we find it more natural to consider them
once matching has been introduced.

We have classified the productions generated at runtime in internal
and external. In fact, it would be more appropriate to speak of
internal and external edges, but this classification suffices for our
purposes.

Applicability (problem~\ref{prob:applicability} stated on
p.~\pageref{prob:applicability}) will be used in
Chap.~\ref{ch:restrictionsOnRules} to characterize \emph{consistency}
of \emph{application conditions} and graph constraints.

In the next chapter sequentialization and parallelism are studied in
detail. Problem~\ref{prob:sequentialIndependence}, sequential
independence (stated on p.~\pageref{prob:sequentialIndependence}),
will be addressed and, in doing so, we will touch on parallelism and
related topics.

Chapter~\ref{ch:restrictionsOnRules} generalizes graph constraints and
application conditions and adapts them to Matrix Graph Grammars.  This
step is not necessary but convenient to study reachability,
problem~\ref{prob:reachability} stated on
p.~\pageref{prob:reachability}, which will be carried out in
Chap.~\ref{ch:reachability}.
\chapter{Sequentialization and Parallelism}
\label{ch:sequentializationAndParallelism}

In this chapter we will study in some detail
problem~\ref{prob:sequentialIndependence} (sequential independence,
p.~\pageref{prob:sequentialIndependence}) which is a particular case
of problem~\ref{prob:independence} (independence,
p.~\pageref{prob:independence}). Recall from
Chap.~\ref{ch:introduction} that two derivations $d$ and $d'$ are
\emph{independent} for a given state $G$ if $d(G) = H \cong H' =
d'(G)$.  We call them \emph{sequential independent} if, besides,
$\exists \; \sigma$ permutation such that $d' = \sigma (d)$.

Applicability (problem~\ref{prob:applicability}) is one of the
premises of independence, establishing an obvious connection between
them.  In Chap.~\ref{ch:reachability} we will sketch the relationship
with \emph{reachability} (problem~\ref{prob:reachability}) and
conjecture one with \emph{confluence} (problem~\ref{prob:confluence})
in Chap.~\ref{ch:conclusionsAndFurtherResearch}.

In Sec.~\ref{sec:gCongruence} \emph{G-congruence} is presented, which
in essence poses conditions for two derivations (one permutation of
the other) to have the same minimal and negative initial digraphs.
The idea behind \emph{sequential independence} is that changes of
order in the position of productions inside a sequence do not alter
the result of their application.  This is addressed in
Sec.~\ref{sec:sequentializationGrammarRules} for sequences and in
Sec.~\ref{sec:sequentialIndependenceDerivations} for derivations.  If
a quick review of permutation groups notation is needed, please see
Sec.~\ref{sec:graphTheory}.  In Sec.~\ref{sec:explicitParallelism} we
will see that there is a close link between sequential independence
and parallelization (see Church-Rosser theorems in,
e.g.~\cite{DPO:handbook}).  As in every chapter, we will close with a
summary (Sec.~\ref{sec:summaryAndConclusions5}).

\section{Graph Congruence}
\label{sec:gCongruence}

Sameness of minimal and negative initial digraphs for two sequences --
one a permutation of the other -- or for two derivations if some
matches have been given, will be known as graph congruence or
\emph{G-congruence}.  This concept helps in characterizing sequential
independence (see
Theorems~\ref{th:CoherenceImpliesSeqInd}~and~\ref{th:advanceDelayProdTheor}).

\newtheorem{congruenceDef}[matrixproduct]{Definition}
\begin{congruenceDef}[G-congruence]\label{congruenceDefinition}
  \index{G-congruence}Two coherent sequences $s_n$ and $\sigma \left(
    s_n \right)$, where $\sigma$ is a permutation, are called
  \emph{G-congruent} if they have the same minimal and negative
  initial digraphs, $M(s_n) = M(\sigma \left( s_n \right))$ and
  $K(s_n) = K(\sigma \left( s_n \right))$.
\end{congruenceDef}

We will identify the conditions that must be fulfilled in order to
guarantee equality of initial digraphs, first for productions
advancement and then for delaying, starting with two productions,
continuing with three and four to end up setting the theorem for the
general case.

The basic remark that justifies the way we tackle G-congruence is that
a sequence and a permutation of it perform the same actions but in
different order.  Initial digraphs depend on actions and the order in
which they are performed.  The idea is to concentrate on how a change
in the order of actions may affect initial
digraphs.

Suppose we have a coherent sequence made up of two productions $s_2 =
p_2;p_1$ with minimal initial digraph $M_2$ and, applying the (only
possible) permutation $\sigma_2$, get another coherent concatenation
$s_2' = p_1;p_2$ with minimal initial digraph $M_2'$.  Production
$p_1$ does not delete any element added by $p_2$ because, otherwise,
if $p_1$ in $s_2$ deleted something, it would mean that it already
existed (as $p_1$ is applied first in $s_2$) while $p_2$ adding that
same element in $s_2'$ would mean that this element was not present
(because $p_2$ is applied first in $s_2'$).  This condition can be
written:
\begin{equation}\label{eq:firstPosCond}
  e_1 r_2 = 0.
\end{equation}
A similar reasoning states that $p_1$ can not add any element that
$p_2$ is going to use:
\begin{equation}\label{eq:secondCondApplicability}
  r_1 L_2 = 0.
\end{equation}
Analogously for $p_2$ against $p_1$, i.e. for $s'_2=p_1;p_2$, we have:
\begin{eqnarray}
  e_2 r_1 & = & 0 \label{eq:3rd}\\
  r_2 L_1 & = & 0.
\end{eqnarray}

As a matter of fact two equations are redundant --
\eqref{eq:firstPosCond} and \eqref{eq:3rd} -- because they are already
contained in the other two.  Note that $e_i L_i = e_i$, i.e.  in some
sense $e_i \subset L_i$, so it is enough to ask for:
\begin{equation}\label{eq:2prodsPos}
  r_1 L_2 \vee r_2 L_1 = 0.
\end{equation}

It is easy to check that these conditions make minimal initial
digraphs coincide, $M_2 = M_2'$.  In detail:
\begin{eqnarray}
  M_2 = M_2 \vee r_1 L_2 = L_1 \vee \overline{r}_1 L_2 \vee r_1 L_2 & = & L_1 \vee L_2 \nonumber \\
  M'_2 = M'_2 \vee r_2 L_1 = L_2 \vee \overline{r}_2 L_1 \vee r_2 L_1 & = & L_2 \vee L_1. \nonumber
\end{eqnarray}

We will very briefly compare conditions for two productions with those
of the SPO approach.  In references~\cite{handbook, handbook3},
sequential independence is defined and categorically
characterized. See also
Secs.~\ref{sec:DPO}~and~\ref{sec:otherCategoricalApproaches}, in particular
equations~\eqref{eq:spoSequentialIndependence1}~and~\eqref{eq:spoSequentialIndependence2}). It
is not difficult to translate those conditions to our matrix language:
\begin{eqnarray}
  r_1 L_2 & = & 0 \\
  e_2 R_1 \equiv e_2 r_1 \vee e_2 \, \overline{e}_1 \, L_1 & = &
  0.
  \label{eq:firstGermanCondition}
\end{eqnarray}

First condition is eq.~\eqref{eq:secondCondApplicability} and, as
mentioned above, first part of second condition ($e_2 r_1 = 0$) is
already considered in eq.~\eqref{eq:secondCondApplicability}.  Second
part of second equation ($e_2 \, \overline{e}_1 \, L_1$ = 0) is
demanded for coherence, in fact something a bit stronger: $e_2 L_1 =
0$.  Hence G-congruence plus coherence imply sequential independence
in the SPO case, at least for a sequence of two productions.  The
converse does not hold in general.  Our conditions are more demanding
because we consider simple
digraphs. 

Let's now turn to the negative initial digraph, for which the first
production should not delete any element forbidden for $p_2$. In such
a case these elements would be in $\overline{G}$ for $p_1;p_2$ and in
$G$ for $p_2;p_1$:
\begin{equation}
  \label{eq:firstCondNeg}
  0 = e_1 K_2 = e_1 r_2 \vee e_1 \overline{e}_2 \overline{D}_2.
\end{equation}

Note that we already had $e_1 r_2 = 0$ in
equation~\eqref{eq:firstPosCond}.  A symmetrical reasoning yields $e_2
\overline{e}_1 \overline{D}_1 = 0$, and altogether:
\begin{equation}
  \label{eq:2prodsNeg}
  e_1 \overline{e}_2 \overline{D}_2 \vee e_2 \overline{e}_1 \overline{D}_1 = 0.
\end{equation}

\begin{figure}[htbp]
  \centering
  \includegraphics[scale = 0.7]{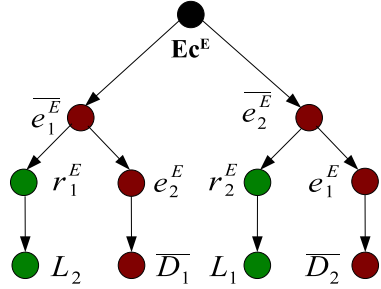}
  \caption{G-congruence for $s_2 = p_2;p_1$}
  \label{fig:twoProdsG_congruence}
\end{figure}

First monomial in eq.~\eqref{eq:2prodsNeg} simply states that no
potential dangling edge for $p_2$ (not deleted by $p_2$) can be
deleted by $p_1$.
Equations~\eqref{eq:2prodsPos}~and~\eqref{eq:2prodsNeg} are
schematically represented in Fig.~\ref{fig:twoProdsG_congruence}.

It is straightforward to show that equation~\eqref{eq:2prodsNeg}
guarantees the same negative initial digraph.  In $p_2;p_1$ the
negative initial digraph is given by $K_1 \vee \overline{e}_1 K_2$.
Condition~\eqref{eq:firstCondNeg} demands $e_1 K_2 = 0$ so we can
\textbf{or} them to get:
\begin{equation}
  K_1 \vee \overline{e}_1 K_2 \vee e_1 K_2 = K_1 \vee K_2.
\end{equation}
A similar reasoning applies to $p_1;p_2$, obtaining the same result.

We will proceed with three productions so, following a consistent
notation, we set $s_3 = p_3;p_2;p_1$, $s_3' = p_2;p_1;p_3$ with
permutation $\sigma_3 = [\, 1 \; 3 \; 2 \,]$ and their corresponding
minimal initial digraphs $M_3 = L_1 \vee \overline{r}_1 \, L_2 \vee
\overline{r}_1 \, \overline{r}_2 \, L_3$ and $M_3' = \overline{r}_3 \,
L_1 \vee \overline{r}_3 \,\overline{r}_2 \, L_2 \vee L_3$.  Conditions
are deduced similarly to the two productions case:\footnote{As far as
  we know, there is no rule of thumb to deduce the conditions for
  G-congruence. They depend on the operations that productions define
  and their relative order.}
\begin{equation}
  r_3 L_1 = 0 \qquad r_3 L_2 \overline{r_1} = 0 \qquad r_1 L_3 = 0
  \qquad r_2 L_3 \overline{e_1} = 0.
\end{equation}

Let's interpret them all.  $r_3 L_1 = 0$ says that $p_3$ cannot add an
edge that $p_1$ uses.  This is because this would mean (by $s_3$) that
the edge is in the host graph (it is used by $p_1$) but $s'_3$ says
that it is not there (it is going to be added by $p_3$).  The second
condition is almost equal but with $p_2$ in the role of $p_1$, which
is why we demand $p_1$ not to add the element $\left( \overline{r_1}
\right)$.  Third equation is symmetrical with respect to the first.
The fourth equation states that we would derive a contradiction if the
second production adds something $(r_2)$ that production $p_3$ uses
$(L_3)$ and $p_1$ does not delete $\left( \overline{e_1} \right)$.
This is because by $s_3$ the element was not in the host graph.  Note
that $s'_3$ says the opposite, as $p_3$ (to be applied first) uses it.
All can be put together in a single expression:
\begin{equation}\label{eq:threeProdsApplicability}
  L_3 \left( r_1 \vee \overline{e_1}\, r_2 \right) \vee r_3 \left( L_1 \vee \overline{r_1} \, L_2 \right) = 0.
\end{equation}

For the sake of completeness let's point out that there are other four
conditions but they are already considered
in~\eqref{eq:threeProdsApplicability}:
\begin{equation}\label{eq:posCondThree}
  e_1 r_3 = 0 \qquad r_3 e_2 \overline{r_1} = 0 \qquad e_3 r_1 = 0
  \qquad r_2 e_3 \overline{e_1} = 0.
\end{equation}

Now we deal with those elements that must not be present.  Four
conditions similar to those for two productions -- compare
with equations in~\eqref{eq:firstCondNeg} -- are needed:
\begin{eqnarray}\label{eq:realNegCC}
  e_1 K_3 = e_1 r_3 \vee e_1 \overline{e}_3 \overline{D}_3 & = & 0
  \nonumber \\
  e_3 K_1 = e_3 r_1 \vee e_3 \overline{e}_1 \overline{D}_1 & = & 0
  \nonumber \\
  e_3 K_2 \overline{e}_1 = e_3 r_2 \overline{e}_1 \vee e_3
  \overline{e}_1 \overline{e}_2 \overline{D}_2 & = & 0 \nonumber \\
  e_2 K_3 \overline{r}_1 = e_2 r_3 \overline{r}_1 \vee e_2
  \overline{r}_1 \overline{e}_3 \overline{D}_3 & = & 0.
\end{eqnarray}

Note that the first monomial in every equation can be discarded as
they are already considered in~\eqref{eq:threeProdsApplicability}. We
put them altogether to get:
\begin{eqnarray}
  e_1 \overline{e}_3 \overline{D}_3 & \vee & e_3 \overline{e}_2
  \overline{e}_1 \overline{D}_2 \vee e_3 \overline{e}_1 \overline{D}_1
  \vee e_2 \overline{e}_3 \overline{r}_1 \overline{D}_3 = \nonumber \\
  & = & e_3 \left( \overline{e}_1 \overline{D}_1 \vee \overline{e}_1
    \overline{e}_2 \overline{D}_2 \right) \vee \overline{e}_3
  \overline{D}_3 \left( e_1 \vee \overline{r}_1 e_2 \right).
\end{eqnarray}

\begin{figure}[htbp]
  \centering
  \includegraphics[scale = 0.7]{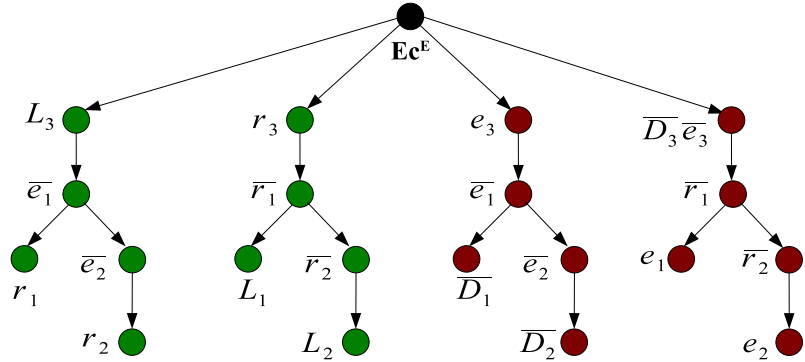}
  \caption{G-congruence for Sequences $s_3 = p_3;p_2;p_1$ and $s'_3 =
    p_2;p_1;p_3$}
  \label{fig:threeProdsG_congruence}
\end{figure}

In Fig.~\ref{fig:threeProdsG_congruence} there is a schematic
representation of all G-congruence conditions for sequences $s_3 =
p_3;p_2;p_1$ and $s'_3 = p_2;p_1;p_3$.  These conditions guarantee
sameness of the minimal and negative initial digraphs, which will be
proved below, in Theorem~\ref{th:GCongruence}.\footnote{Notice that by
  Prop.~\ref{prop:simpleEqualities}, equations \eqref{eq:r_And_e} and
  \eqref{eq:L_And_Not_e} in particular, we can put $\overline{r_i}
  L_i$ instead of just $L_i$ and $\overline{e_i} r_i$ instead of just
  $r_i$. It will be useful in order to find a closed formula in terms
  of $\nabla$.}

Moving one production three positions forward in a sequence of four
productions, i.e. $p_4;p_3;p_2;p_1 \mapsto p_3;p_2;p_1;p_4$, while
maintaining the minimal initial digraph has as associated conditions
those given by the equation:
\begin{equation}\label{eq:fourProdsApplicability} L_4 \left( r_1 \vee
    \overline{e}_1\, r_2 \vee \overline{e}_1 \, \overline{e}_2 \, r_3
  \right) \vee r_4 \left( L_1 \vee \overline{r}_1 \, L_2 \vee
    \overline{r}_1 \, \overline{r}_2 \, L_3 \right) = 0.
\end{equation}
and for the negative initial digraph we have:
\begin{equation}\label{eq:fourProdsNegIni}
  e_4 \left( \overline{e}_1 \, \overline{D}_1 \vee \overline{e}_1 \,
    \overline{e}_2 \overline{D}_2 \vee \overline{e}_1 \,
    \overline{e}_2 \, \overline{e}_3 \overline{D}_3 \right) \vee
  \overline{e}_4 \, \overline{D}_4 \left( e_1 \vee \overline{r}_1 \,
    e_2 \vee \overline{r}_1 \, \overline{r}_2 \, e_3 \right) = 0.
\end{equation}

\begin{figure}[htbp]
  \centering
  \includegraphics[scale = 0.7]{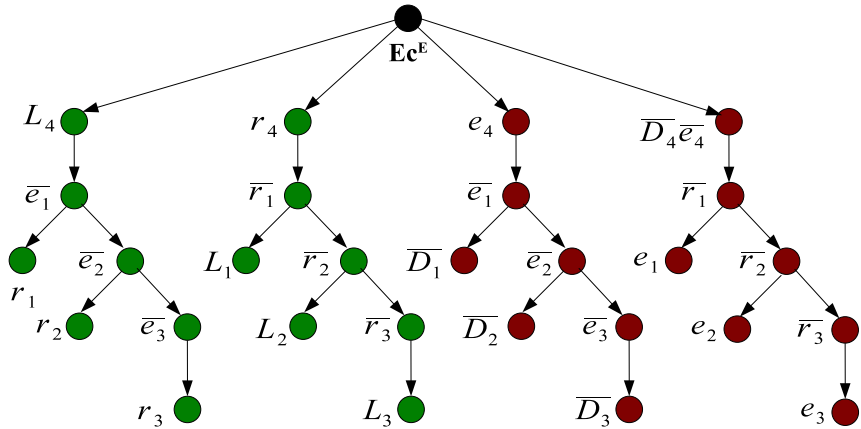}
  \caption{G-congruence for $s_4 = p_4;p_3;p_2;p_1$ and $s'_4 =
    p_3;p_2;p_1;p_4$}
  \label{fig:fourProdsG_congruence}
\end{figure}

Equations~\eqref{eq:fourProdsApplicability}~and~\eqref{eq:fourProdsNegIni}
together give G-congruence for $s_4$ and $s'_4$ are depicted on
Fig.~\ref{fig:fourProdsG_congruence}.

Before moving to the general case, let's briefly introduce and put an
example of a simple notation for cycles moving forward and backward a
single production:
\begin{enumerate}
\item Advance production $n-1$ positions: $\phi_n = [\, 1 \quad n \,
  \quad \, n-1 \quad \ldots \quad 3 \quad 2 \,]$.
\item Delay production $n-1$ positions: $\delta_n = [\, 1 \quad 2
  \quad \ldots \quad n-1 \quad n \, ]$.
\end{enumerate}

\noindent\textbf{Example.}$\square$Consider advancing three positions the
production $p_5$ inside the sequence $s_5 = p_5;p_4;p_3;p_2;p_1$ to
get $\phi_4 \left(s_5\right) = p_4;p_3;p_2;p_5;p_1$, where $\phi_4 = [
\, 1 \; 4 \; 3 \; 2 \, ]$.

To illustrate the way in which we represent delaying a production,
moving backwards production $p_2$ two places $p_5; p_4; p_3; p_2; p_1
\longmapsto p_5; p_2; p_4; p_3; p_1$ has as associated cycle $\delta_4
= [ \, 2 \; 3 \; 4 \, ]$.  Note that the numbers in the permutation
refer to the place the production occupies in the sequence, numbering
from left to right, and not to its subindex.\proofend

\index{congruence condition} \index{congruence condition!positive}
\index{congruence condition!negative} Conditions that must be
fulfilled in order to maintain the minimal and negative initial
digraphs will be called \emph{congruence conditions} and will be
abbreviated as \textbf{CC}, \emph{positive CC} if they refer to
minimal initial digraph and \emph{negative CC} for the negative
initial digraph.

By induction it can be proved that for advancement of one production
$n-1$ positions inside the sequence of $n$ productions $s_n = p_n;
\ldots; p_1$, the equation which contains all \emph{positive CC} can
be expressed in terms of operator $\nabla$ and has the form:
\begin{equation}\label{eq:posCC}
  CC_n^+ \left( \phi _{n}, s_n \right) = L_n \nabla_{1}^{n-1} \left( \overline{e_x} \, r_y \right) \vee r_n \nabla_{1}^{n-1}\left( \overline{r_x} \, L_y \right) = 0.
\end{equation}
and for the \emph{negative CC}:
\begin{equation}\label{eq:negCC}
  CC_n^- \left( \phi _{n}, s_n \right) = \overline{D}_n \,
  \overline{e}_n \nabla_{1}^{n-1} \left( \overline{r_x} \, e_y \right)
  \vee e_n \nabla_{1}^{n-1}\left( \overline{e_x} \, \overline{D}_y
  \right) = 0.
\end{equation}

\noindent\textbf{Remark}.$\square$Some monomials were discarded in
eq.~\eqref{eq:realNegCC} because they were already considered in
eq.~\eqref{eq:threeProdsApplicability}.  If~\eqref{eq:negCC} is not
used in conjunction with~\ref{eq:posCC}, then the more complete form
\begin{equation}
  CC_n^- \left( \phi _{n}, s_n \right) = K_n \nabla_{1}^{n-1} \left(
    \overline{r}_x e_y \right) \vee e_n \nabla_1^{n-1} \left(
    \overline{e}_x K_y \right)
\end{equation}
should be preferred.  Recall that $K_h = r_h \vee \overline{e}_h
\overline{D}_h$.  The point is that $\overline{e}_h \overline{D}_h$
considers potential dangling edges while $K_h$ also includes those to
be added.\proofend

\begin{figure}[htbp]
  \centering
  \includegraphics[scale = 0.7]{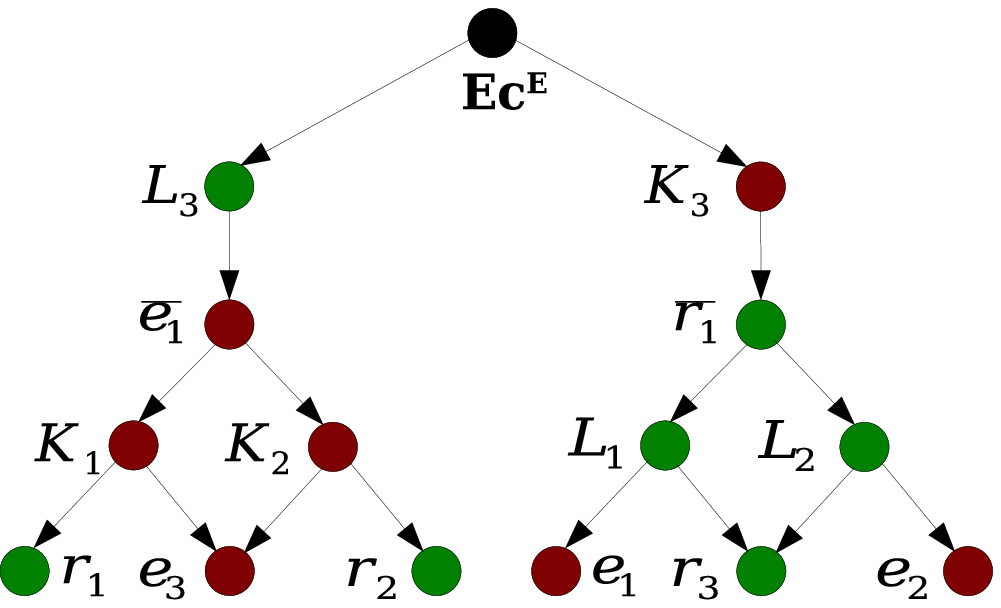}
  \caption{G-congruence (Alternate Form) for $s_3$ and $s'_3$}
  \label{fig:threeGcongruenceAlt}
\end{figure}

It is possible to put eqs.~\eqref{eq:posCC}~and~\eqref{eq:negCC} in
terms of $L_i$ and $K_i$.  We will do it for sequences $s_3$ and
$s'_3$ to obtain an equivalent form of
Fig.~\ref{fig:threeProdsG_congruence} (represented in
Fig.~\ref{fig:threeGcongruenceAlt}).

What we do is to merge the first branch in
Fig.~\ref{fig:threeProdsG_congruence} with the third branch and the
second branch with the fourth.  One illustrating example should
suffice:\footnote{The term $\overline{r}_1$ can be omitted.}

\begin{eqnarray}
  r_3 \overline{r}_1 L_1 \vee \overline{D}_3 \overline{e}_3
  \overline{r}_1 e_1 & = & \overline{r}_1 L_1 \left( r_3 \vee e_1
    \overline{e}_3 \overline{D}_3 \right) = \nonumber \\
  & = & \overline{r}_1 L_1 \left( r_3 e_1 \vee r_3 \overline{e}_1 \vee
    e_1 \overline{e}_3 \overline{D}_3 \right) = \nonumber \\
  & = & \overline{r}_1 L_1 \left( e_1 K_3 \vee r_3 \overline{e}_1
  \right) = \overline{r}_1 L_1 K_3 \left( e_1 \vee r_3 \right).
\end{eqnarray}

\begin{figure}[htbp]
  \centering
  \includegraphics[scale = 0.7]{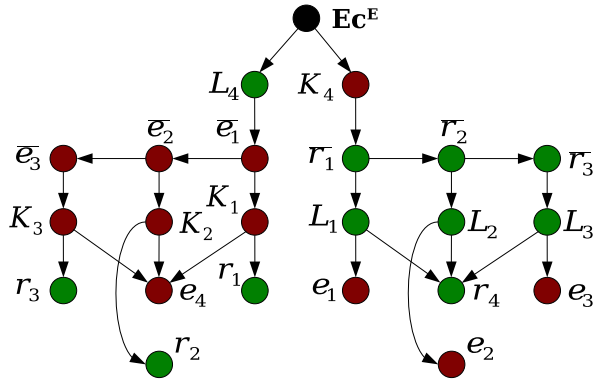}
  \caption{G-congruence (Alternate Form) for $s_4$ and $s'_4$}
  \label{fig:fourGcongruenceAlt}
\end{figure}

Last equality holds because $K_i r_i = r_i \vee r_i \overline{D}_i =
r_i$ and $a \vee \overline{a} b = a \vee b$.  We have also used that
$K_i \overline{e}_i = \overline{e}_i \left( r_i \vee \overline{e}_i
  \overline{D}_i \right) = K_i$.  The same sort of calculations for
$s_4$ and $s'_4$ are summarized in Fig.~\ref{fig:fourGcongruenceAlt}.

A formula considering the positive~\eqref{eq:posCC} and the
negative~\eqref{eq:negCC} parts can be derived by induction.  It is
presented as a proposition:

\newtheorem{ccPosMerged}[matrixproduct]{Proposition}
\begin{ccPosMerged}\label{prop:GCongruencePos}
  Positive and negative congruence conditions for sequences $s_n$ and
  $s'_n = \phi_n(s_n)$ are given by:
  \begin{equation}\label{eq:advCC}
    CC_n \left( \phi_n, s_n \right) = L_n \nabla_1^{n-1}
    \overline{e}_x K_y \left(r_y \vee e_n \right) \vee K_n
    \nabla_1^{n-1} \overline{r}_x L_y \left( e_y \vee r_n \right).
  \end{equation}
\end{ccPosMerged}

\noindent \emph{Proof}\\*
$\square$$\blacksquare$

$G$-congruence is obtained when $CC_n \left( \phi_n, s_n \right) =
0$. An equivalent reasoning does it for a production delayed $n-1$
positions, giving very similar formulas.  Suppose that production
$p_1$ is moved backwards in concatenation $s_n$ to get $s_n'' =
p_1;p_n;\ldots;p_2$, i.e. $\delta_{n}$ is applied.  The positive part
of the condition is:
\begin{equation}\label{eq:posCCDelay}
  CC^+_n \left( \delta _{n}, s_n \right) = L_1 \nabla_{2}^{n} \left(
    \overline{e_x} \, r_y \right) \vee r_1 \nabla_{2}^{n}\left(
    \overline{r_x} \, L_y \right) = 0
\end{equation}
and the negative part:
\begin{equation}\label{eq:negCCDelay}
  CC^-_n \left( \delta _{n}, s_n \right) = \overline{D}_1 \,
  \overline{e}_1 \nabla_{2}^{n} \left( \overline{r_x} \, e_y \right)
  \vee e_1 \nabla_{2}^{n}\left( \overline{e_x} \, \overline{D}_y
  \right) = 0.
\end{equation}

As in the positive case it is possible to merge
equations~\eqref{eq:posCCDelay}~and~\eqref{eq:negCCDelay} to get a
single expression:
\newtheorem{ccNegMerged}[matrixproduct]{Proposition}
\begin{ccNegMerged}\label{prop:GCongruenceNeg}
  Positive and negative congruence conditions for sequences $s_n$ and
  $s''_n = \delta_n(s_n)$ are given by:
  \begin{equation}
    CC_n \left( \delta_n, s_n \right) = L_1 \nabla_2^{n}
    \overline{e}_x K_y \left(r_y \vee e_1 \right) \vee K_1
    \nabla_2^{n} \overline{r}_x L_y \left( e_y \vee r_1 \right).
  \end{equation}
\end{ccNegMerged}

\noindent \emph{Proof}\\*
$\square$$\blacksquare$

It is necessary to show that these conditions guarantee sameness of
minimal and negative initial digraphs, but first we need a technical
lemma that provides us with some identities used to transform the
minimal initial digraphs.
Advancement and delaying are very similar so only advancement is
considered in the rest of the section.

\newtheorem{applicabilityAdvanceLemma}[matrixproduct]{Lemma}
\begin{applicabilityAdvanceLemma}
  \label{lemma:applicabilityAdvanceLemmaReference}
  Suppose $s_n = p_n; \ldots ; p_1$ and $s_n' = \sigma \left( s_n
  \right) = p_{n-1}; \ldots ; p_1; p_n$ and that $CC^+_n \left( \phi_n
  \right)$ is satisfied.  Then the following identity may be
  \textbf{or}ed to $s_n$'s minimal initial digraph $M_n$ without
  changing it:
  \begin{equation}\label{eq:appAdvanceLemma}
    DC^+_n (\phi_n, s_n) = L_n \nabla_1^{n-2} \left(\overline{r_x} \,
      e_y \right).
  \end{equation}
\end{applicabilityAdvanceLemma}

\noindent \emph{Proof}\\*
$\square$Let's start with three productions.  Recall that $M_3 = L_1
\,\vee $ \emph{other\_terms} and that $L_1 = L_1 \vee e_1 = L_1 \vee
e_1 \vee e_1 L_3$ (last equality holds in propositional logics $a \vee
a b = a$).  Note that $e_1 L_3$ is eq.~\eqref{eq:appAdvanceLemma} for
$n=3$.

For $n=4$, apart from $e_1 L_4$, we need to get $e_2 \overline{r_1}
L_4$ (because the full condition is $DC^+_4 = L_4 \left( e_1 \vee
  \overline{r}_1 e_2 \right)$).  Recall again the minimal initial
digraph for four productions whose first two terms are $M_4 = L_1 \vee
\overline{r_1} L_2$.  It is not necessary to consider all terms in
$M_4$ to get $DC^+_4$:
\begin{eqnarray}
  M_4 & = & \left( L_1 \vee e_1 \right) \vee \left( \overline{r_1} L_2 \vee \overline{r_1} e_2 \right) \vee \ldots \nonumber = \\
  & = & \left( L_1 \vee e_1 \vee e_1 L_4 \right) \vee \left( \overline{r}_1 L_2 \vee \overline{r}_1 e_2 \vee \overline{r}_1 e_2 L_4 \right) \vee \ldots = \nonumber \\
  & = & \left( L_1 \vee e_1 L_4 \right) \vee \left( \overline{r}_1 L_2 \vee \overline{r}_1 e_2 L_4 \right) \vee \ldots = \nonumber \\
  & = & M_4 \vee DC^+_4. \nonumber
\end{eqnarray}
The proof can be finished by induction.\proofend

Next lemma states a similar result for negative initial digraphs.  We
will need it to prove invariance of the negative initial digraph.

\newtheorem{appAdvNIDLemma}[matrixproduct]{Lemma}
\begin{appAdvNIDLemma}\label{lemma:appAdvNIDLemma}
  With notation as above and assuming that $CC^-_n \left( \phi_n
  \right)$ is satisfied, the following identity may be \textbf{or}ed
  to the negative initial digraph $K$ without changing it:
  \begin{equation}\label{eq:appAdvNIDLemma}
    DC^-_n (\phi_n, s_n) = \overline{e}_n \overline{D}_n
    \nabla_1^{n-2} \left(\overline{e_x} \, r_y \right).
  \end{equation}
\end{appAdvNIDLemma}

\noindent \emph{Proof}\\*
$\square$We follow the same scheme as in the proof of
Lemma~\ref{lemma:applicabilityAdvanceLemmaReference}.  Let's start
with three productions.  Recall that $K_3 = K_1 \,\vee $
\emph{other\_terms} and that $K_1 = K_1 \vee r_1 = K_1 \vee r_1 \vee
r_1 \overline{e}_3 \overline{D}_3$.  Note that $r_1 \overline{e}_3
\overline{D}_3$ is eq.~\eqref{eq:appAdvNIDLemma} for $n=3$.

For $n=4$, besides the term $r_1 \overline{e}_4 \overline{D}_4$ we
need to get $\overline{e}_1 r_2 \overline{e}_4 \overline{D}_4$
(because $DC^-_4 = \overline{e}_4 \overline{D}_4 \left( r_1 \vee
  \overline{e}_1 r_2 \right)$).  The first two terms of the negative
initial digraph for four productions are $K_4 = K_1 \vee
\overline{e_1} K_2$.  Again, it is not necessary to consider the whole
formula for $K_4$:
\begin{eqnarray}
  K_4 & = & \left( K_1 \vee r_1 \right) \vee \left( \overline{e}_1 K_2
    \vee r_2 \overline{e}_1 \right) \vee \ldots = \nonumber \\
  & = & \left( K_1 \vee r_1 \vee r_1 \overline{e}_4 \overline{D}_4
  \right) \vee \left( \overline{e}_1 K_2 \vee \overline{e}_1 r_2 \vee
    \overline{e}_1 r_2 \overline{e}_4 \overline{D}_4 \right) \vee
  \ldots = \nonumber \\
  & = & \left( K_1 \vee r_1 \overline{e}_4 \overline{D}_4 \right) \vee
  \left( \overline{e}_1 K_2 \vee \overline{e}_1 r_2 \overline{e}_4
    \overline{D}_4 \right) \vee \ldots = \nonumber \\
  & = & K_4 \vee DC^-_4. \nonumber
\end{eqnarray}
The proof can be finished by induction.\proofend

\begin{figure}[htbp]
  \centering
  \includegraphics[scale = 0.7]{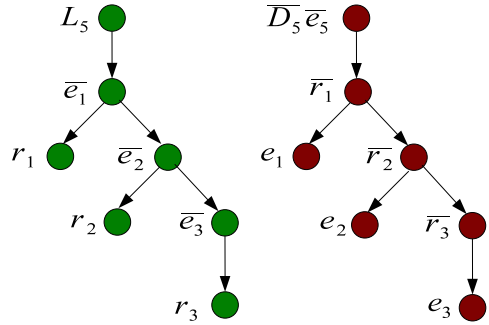}
  \caption{Positive and Negative DC Conditions, $DC^+_5$ and $DC^-_5$}
  \label{fig:DC_Conditions}
\end{figure}

Both, $DC^+_5$ and $DC^-_5$ are depicted in
Fig.~\ref{fig:DC_Conditions} for advancement of a single production
$s_5 = p_5;p_4;p_3;p_2;p_1 \longmapsto
s'_3=p_4;p_3;p_2;p_1;p_5$. Notice the similarities with first and
fourth branches of Fig.~\ref{fig:fourProdsG_congruence}.

\noindent\textbf{Remark}.$\square$If $CC^-_n$ and $DC^-_n$ are applied
independently of $CC^+_n$ and $DC^+_n$ then the expression
\begin{equation}
  DC^-_n (\phi_n, s_n) = K_n \nabla_1^{n-2} \left(\overline{e_x} \, r_y \right)
\end{equation}
should be used instead of the definition given by
equation~\eqref{eq:appAdvNIDLemma}.\proofend

We are ready to formally state a characterization of G-congruence in
terms of congruence conditions $CC$:

\newtheorem{applicabilityAdvance}[matrixproduct]{Theorem}
\begin{applicabilityAdvance}\label{th:GCongruence}
  With notation as above, if $s_n$ and $s_n' = \phi_n \left( s_n
  \right)$ are coherent and condition $CC \left( \phi _{n}, s_n
  \right)$ is satisfied then they are G-congruent.
\end{applicabilityAdvance}

\noindent \emph{Proof}\\*
$\square$First, using $CC^+_i$ and $DC^+_i$, we will prove $M_i =
M'_i$ for three and five productions.  Identities $a \vee \overline{a}
\, b = a \vee b$ and $\overline{a} \vee a \, b = \overline{a} \vee b$
will be used:
\begin{eqnarray}
  M_3 \vee CC^+_3 \vee DC^+_3 & = & \left[ L_1 \vee \overline{r}_1 \,
    L_2 \vee \overline{r}_1 \, \overline{r}_2 \, L_3 \right] \vee
  \left[ r_1 L_3 \vee \overline{e}_1 \, r_2 L_3 \vee r_3 L_1 \vee
  \right. \nonumber \\
  & \vee & \left. \overline{r}_1 \, r_3 L_2 \right] \vee \left[ e_1
    L_3 \right] = L_1 \vee \overline{r}_1 \, L_2 \vee
  \not\!{\overline{r}_1} \, \overline{r}_2 \, L_3 \vee r_1 L_3 \vee
  \nonumber \\
  & \vee & \not\!{\overline{e}_1} \, r_2 L_3 \vee e_1 L_3 = L_1 \vee
  \overline{r}_1 \, L_2 \vee \not\!{\overline{r}_2} \, L_3 \vee r_2
  L_3 \vee \nonumber \\
  & \vee & L_3 \left( r_1 \vee e_1 \right) = L_1 \vee \overline{r}_1
  \, L_2 \vee \, L_3. \nonumber
\end{eqnarray}

In our first step, as neither $r_3 L_1$ nor $\overline{r}_1 \, r_3
L_2$ are applied to $M_3$, they have been omitted (for example, $L_1
\vee r_3 L_1 = L_1$).  Once $r_1 L_3$, $e_1 L_3$ and $r_2 L_3$ have
been used, they are omitted as well.

Let's check out $M_3'$, where in the second equality $r_1 L_3$ and
$r_2 \, \overline{e}_1 \, L_3$ are ruled out since they are not used:
\begin{eqnarray}
  M_3' \vee CC^+_3 & = & \left[ \overline{r}_3 \, L_1 \vee
    \overline{r}_1 \, \overline{r}_3 \, L_2 \vee L_3 \right] \vee
  \left[ r_1 L_3 \vee r_2 \, \overline{e}_1 \, L_3 \vee r_3 L_1 \vee
    \overline{r}_1 \, r_3 L_2 \right] = \nonumber \\ \nonumber
  & = & \not\!{\overline{r}_3} \, L_1 \vee \overline{r}_1 \not
  \!{\overline{r}_3} \, L_2 \vee L_3  \vee r_3 L_1 \vee \overline{r}_1
  \, r_3 L_2 = \\ \nonumber
  & = & L_1 \vee \overline{r}_1 \, L_2 \vee L_3. \nonumber
\end{eqnarray}

The case for five productions is almost equal to that of three
productions but it is useful to illustrate in detail how $CC^+_5$ and
$DC^+_5$ are used to prove that $M_5 = M_5'$ in a more complex
situation. The key point is the transformation $\overline{r}_1 \,
\overline{r}_2 \, \overline{r}_3 \, \overline{r}_4 \, L_5 \longmapsto
L_5$ and the following identities show the way to proceed:
\begin{eqnarray}
  \quad \not\!{\overline{r}_1} \, \overline{r}_2 \, \overline{r}_3 \,
  \overline{r}_4 \, L_5 \vee r_1 L_5 & = & \overline{r}_2 \,
  \overline{r}_3 \, \overline{r}_4 \, L_5 \nonumber \\
  \quad \not\!{\overline{r}_2} \, \overline{r}_3 \, \overline{r}_4 \,
  L_5 \vee \not\!{\overline{e}_1} \, r_2 L_5 \vee e_1 L_5 & = &
  \overline{r}_3 \, \overline{r}_4 \, L_5 \nonumber \\
  \quad \not\!{\overline{r}_3} \, \overline{r}_4 \, L_5 \vee
  \not\!{\overline{e}_1} \not\!{\overline{e}_2} r_3 L_5 \vee e_1 L_5
  \vee \not\!{\overline{r}_1} e_2 L_5 \vee r_1 L_5 & = &
  \overline{r}_4 \, L_5 \nonumber \\
  \quad \not\!{\overline{r}_4} L_5 \vee \not\!{\overline{e}_1}
  \not\!{\overline{e}_2} \not\!{\overline{e}_3} r_4 L_5 \vee e_1 L_5
  \vee \not\!{\overline{r}_1} e_2 L_5 \vee r_1 L_5 & {} & {} \nonumber
  \\
  \vee \not\!{\overline{r}_1} \not\!{\overline{r}_2} e_3 L_5 \vee
  \not\!{\overline{e}_1} r_2 L_5 & = & L_5. \nonumber
\end{eqnarray}

Note that we are in a kind of iterative process: What we get on the
right of the equality is inserted and simplified on the left of the
following one, until we get $L_5$.  For $L_4$ the process is similar.

Now one example for the negative initial digraph is studied, $K(s_3)
\vee CC^-_3 \vee DC^-_3 = K'(s_3) \vee CC^-_3$:
\begin{align*}
  K'(s_3) \vee CC^-_3 & = \left[ \overline{e}_3 \, K_1 \vee
    \overline{e}_1 \, \overline{e}_3 \, K_2 \vee K_3 \right] \vee
  \left[ e_1 K_3 \vee e_2 \, \overline{r}_1 \, K_3 \vee e_3 K_1 \vee
    \overline{e}_1 \, e_3 K_2 \right] =  \\
  & = \not\!{\overline{e}_3} \, K_1 \vee \overline{e}_1
  \not\!{\overline{e}_3} \, K_2 \vee K_3  \vee e_3 K_1 \vee
  \overline{e}_1 \, e_3 K_2 = \\
  & = K_1 \vee \overline{e}_1 \, K_2 \vee K_3.
\end{align*}
\begin{align*}
  K'(s_3) \vee CC^-_3 & = \left[ \overline{e}_3 \, K_1 \vee
    \overline{e}_1 \, \overline{e}_3 \, K_2 \vee K_3 \right] \vee
  \left[ e_1 K_3 \vee e_2 \, \overline{r}_1 \, K_3 \vee e_3 K_1 \vee
    \overline{e}_1 \, e_3 K_2 \right] = \\
  & = \not\!{\overline{e}_3} \, K_1 \vee \overline{e}_1
  \not\!{\overline{e}_3} \, K_2 \vee K_3  \vee e_3 K_1 \vee
  \overline{e}_1 \, e_3 K_2 = \\
  & = K_1 \vee \overline{e}_1 \, K_2 \vee K_3.
\end{align*}
The procedure followed to show $K(s_3) = K'(s_3)$ is completely
analogous to that of $M_3 = M'_3$.\proofend

\begin{figure}[htbp]
  \centering
  \includegraphics[scale = 0.57]{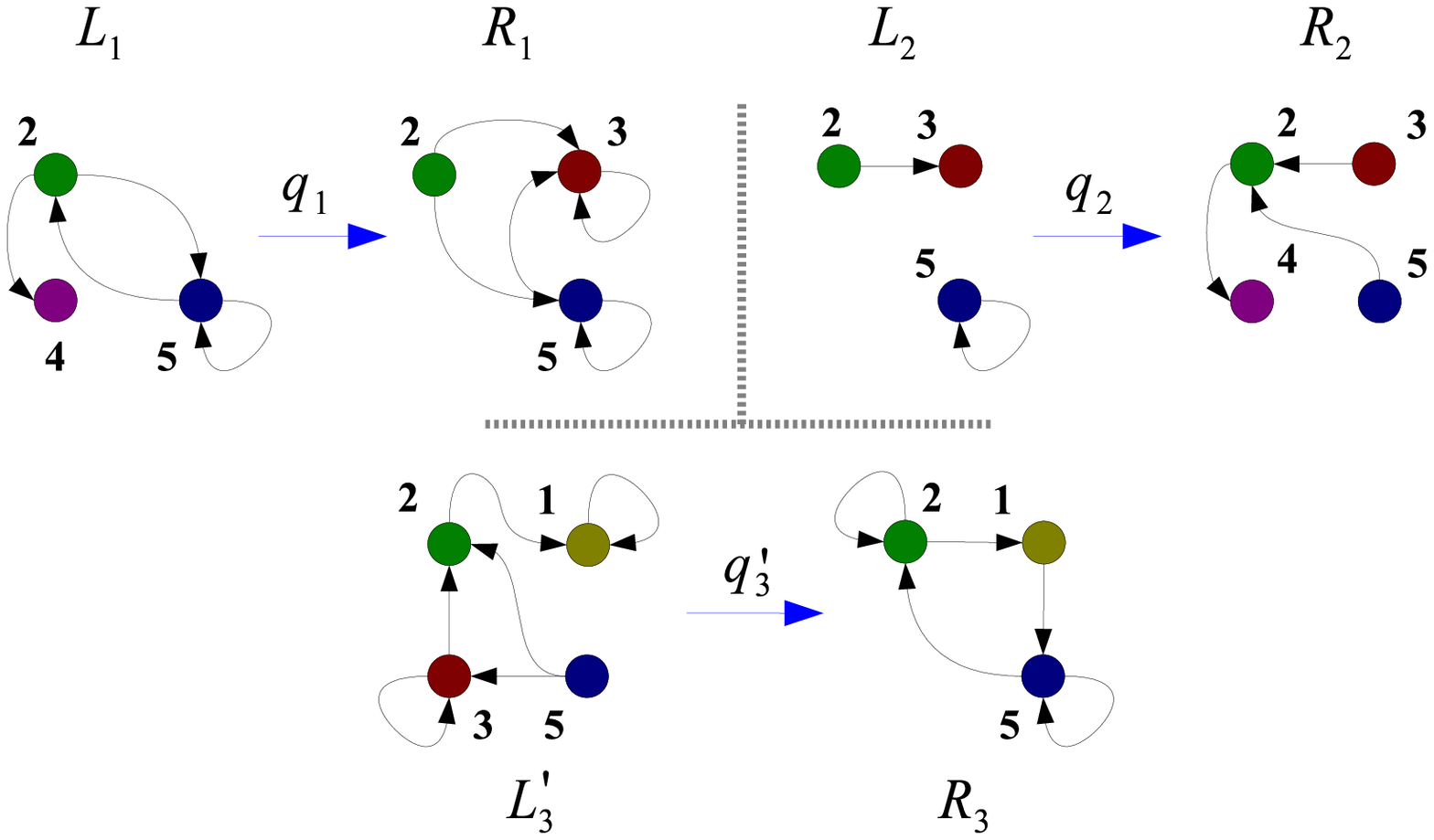}
  \caption{Altered Production $q'_3$ Plus Productions $q_1$ and $q_2$}
  \label{fig:ThirdProdAltered_1}
\end{figure}

\noindent\textbf{Remark}$\square$Congruence conditions report what elements
prevent graph congruence. In this way not only information of sameness
of minimal and negative initial digraphs is available but also what
elements prevent G-congruence.  For example, another way to see
congruence conditions is as the difference of the minimal initial
digraphs in the positive case.\proofend

\noindent\textbf{Example.}$\square$Reusing productions introduced so far
($q_1$, $q_2$ and $q_3$),\footnote{In examples on
  pp.~\pageref{ex:first},~\pageref{ex:second},~\pageref{ex:thrid}~and~\pageref{ex:fourth}.}
we are going to check G-congruence for a sequence of three productions
in which one is directly delayed two positions, i.e. it is not delayed
in two steps but just in one. As commented before, it is mandatory to
change $q_3$ in order to keep compatibility, so a new production
$q_3'$ is introduced, depicted in Fig.~\ref{fig:ThirdProdAltered_1}.

\begin{figure}[htbp]
  \centering
  \includegraphics[scale = 0.65]{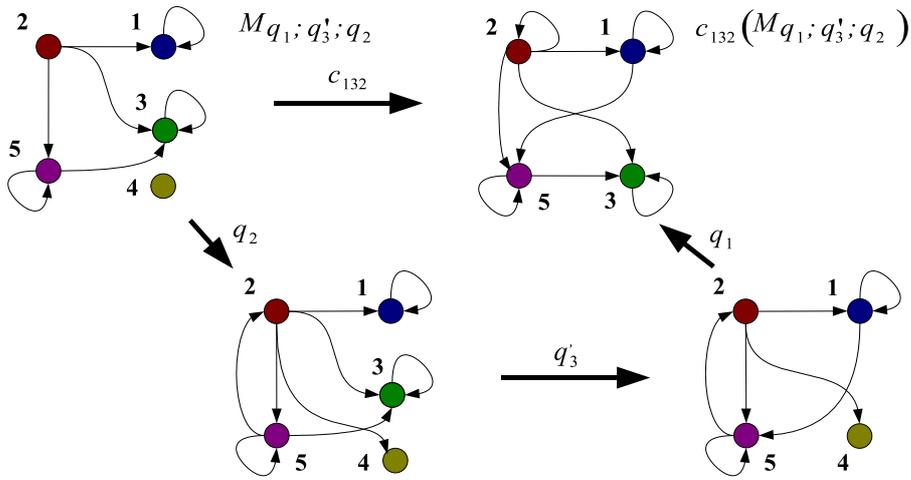}
  \caption{Composition and Concatenation. Three Productions}
  \label{fig:SeqExampleThreeProds}
\end{figure}

The minimal initial digraph for the sequence $q_3';q_2;q_1$ remains
unaltered, i.e. $M_{q_3';q_2;q_1} = M_{q_3;q_2;q_1}$ (compare with
Fig.~\ref{fig:CompExample} on p.~\pageref{fig:CompExample}), but the
one for $q_1;q_3';q_2$ is slightly different and can be found in
Fig.~\ref{fig:SeqExampleThreeProds} along with the concatenation
$s_{123}' = q_1;q_3';q_2$ and its intermediate states.

In this example, production $q_1$ is delayed two positions inside $s_3
= q_3';q_2;q_1$ to obtain $\delta_3\left( s_3 \right) = q_1;q_3';q_2$.
Such permutation can be expressed as $\delta_3 = [1 \; 2 \;\,
3]$.\footnote{Numbers 1, 2 and 3 in the permutation mean position
  inside the sequence, not production subindex.}  Only the
\emph{positive} case $CC^+_3\left(\delta_3,s_3\right)$ is illustrated.
Formula~\eqref{eq:posCCDelay} expanded and simplified is:
\begin{equation}\label{eq:exPosDelay}
  \underbrace{L_1 \left( r_2 \vee \overline{e_2} r_3 \right)}_{(*)} \vee \underbrace{r_1 \left( L_2 \vee \overline{r_2} L'_3 \right)}_{(**)}.
\end{equation}

If the minimal initial digraphs are equal, then
equation~\eqref{eq:exPosDelay} should be zero.  Node ordering is $[2
\; 3 \; 5 \; 1 \; 4]$, not included due to lack of space.

\begin{equation}
  \left[
    \begin{array}{ccccc}
      \vspace{-6pt}
      0 & 0 & 1 & 0 & 1 \\
      \vspace{-6pt}
      0 & 0 & 0 & 0 & 0 \\
      \vspace{-6pt}
      1 & 0 & 1 & 0 & 0 \\
      \vspace{-6pt}
      0 & 0 & 0 & 0 & 0 \\
      \vspace{-6pt}
      0 & 0 & 0 & 0 & 0 \\
      \vspace{-10pt}
    \end{array} \right] \left(
    \left[
      \begin{array}{ccccc}
        \vspace{-6pt}
        0 & 0 & 0 & 0 & 1 \\
        \vspace{-6pt}
        1 & 0 & 0 & 0 & 0 \\
        \vspace{-6pt}
        1 & 0 & 0 & 0 & 0 \\
        \vspace{-6pt}
        0 & 0 & 0 & 0 & 0 \\
        \vspace{-6pt}
        0 & 0 & 0 & 0 & 0 \\
        \vspace{-10pt}
      \end{array} \right]  
    \vee \left[
      \begin{array}{ccccc}
        \vspace{-6pt}
        1 & 0 & 1 & 1 & 1 \\
        \vspace{-6pt}
        1 & 1 & 1 & 1 & 1 \\
        \vspace{-6pt}
        1 & 1 & 0 & 1 & 1 \\
        \vspace{-6pt}
        1 & 1 & 1 & 1 & 1 \\
        \vspace{-6pt}
        1 & 1 & 1 & 1 & 1 \\
        \vspace{-10pt}
      \end{array} \right] 
    \left[
      \begin{array}{ccccc}
        \vspace{-6pt}
        1 & 0 & 0 & 0 & 0 \\
        \vspace{-6pt}
        0 & 0 & 0 & 0 & 0 \\
        \vspace{-6pt}
        0 & 0 & 1 & 0 & 0 \\
        \vspace{-6pt}
        0 & 0 & 1 & 0 & 0 \\
        \vspace{-6pt}
        0 & 0 & 0 & 0 & 0 \\
        \vspace{-10pt}
      \end{array} \right]   \nonumber
  \right)
\end{equation}
similarly for $(**)$:
\begin{equation}
  \left[
    \begin{array}{ccccc}
      \vspace{-6pt}
      0 & 1 & 0 & 0 & 0 \\
      \vspace{-6pt}
      0 & 1 & 0 & 0 & 0 \\
      \vspace{-6pt}
      0 & 1 & 0 & 0 & 0 \\
      \vspace{-6pt}
      0 & 0 & 0 & 0 & 0 \\
      \vspace{-6pt}
      0 & 0 & 0 & 0 & 0 \\
      \vspace{-10pt}
    \end{array} \right] \left(
    \left[
      \begin{array}{ccccc}
        \vspace{-6pt}
        0 & 1 & 0 & 0 & 0 \\
        \vspace{-6pt}
        0 & 0 & 0 & 0 & 0 \\
        \vspace{-6pt}
        0 & 0 & 1 & 0 & 0 \\
        \vspace{-6pt}
        0 & 0 & 0 & 0 & 0 \\
        \vspace{-6pt}
        0 & 0 & 0 & 0 & 0 \\
        \vspace{-10pt}
      \end{array} \right]  
    \vee \left[
      \begin{array}{ccccc}
        \vspace{-6pt}
        1 & 1 & 1 & 1 & 0 \\
        \vspace{-6pt}
        0 & 1 & 1 & 1 & 1 \\
        \vspace{-6pt}
        0 & 1 & 1 & 1 & 1 \\
        \vspace{-6pt}
        1 & 1 & 1 & 1 & 1 \\
        \vspace{-6pt}
        1 & 1 & 1 & 1 & 1 \\
        \vspace{-10pt}
      \end{array} \right] 
    \left[
      \begin{array}{ccccc}
        \vspace{-6pt}
        0 & 0 & 0 & 1 & 0 \\
        \vspace{-6pt}
        1 & 1 & 0 & 0 & 0 \\
        \vspace{-6pt}
        1 & 1 & 0 & 0 & 0 \\
        \vspace{-6pt}
        0 & 0 & 0 & 1 & 0 \\
        \vspace{-6pt}
        0 & 0 & 0 & 0 & 0 \\
        \vspace{-10pt}
      \end{array} \right]   \nonumber
  \right)
\end{equation}

\begin{figure}[htbp]
  \centering
  \includegraphics[scale = 0.6]{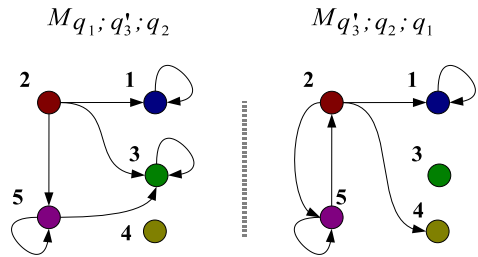}
  \caption{Example of Minimal Initial Digraphs}
  \label{fig:twoMID}
\end{figure}

We detect nonzero elements $\left( 1,5 \right)$ and $\left( 3,1
\right)$ in (*) and $\left( 1,2 \right)$, $\left( 2,3 \right)$ and
$\left( 3,2 \right)$ in (**).  They correspond to edges $\left( 2,4
\right)$, $\left( 5,2 \right)$, $\left( 2,3 \right)$, $\left( 3,3
\right)$ and $\left( 5,3 \right)$, respectively.  Both minimal initial
digraphs are depicted together in Fig.~\ref{fig:twoMID} to ease
comparison.\proofend

Previous results not only detect if the application of a permutation
(limited to advancing or delaying a single element) leaves minimal
initial digraphs unaltered, but also what elements are changed.



\section{Sequentialization -- Grammar Rules}
\label{sec:sequentializationGrammarRules}

In this section we will deal with position interchange inside a
sequence of productions.  For example, let $s_3 = p_3;p_2;p_1$ be a
coherent sequence made up of three productions and suppose we wanted
to move $p_3$ forward one position to obtain $\sigma \left( s_3
\right) = p_2;p_3;p_1$.  This can be seen as a permutation $\sigma$
acting on $s_3$'s indexes.\footnote{Notation of permutation groups is
  summarized in Sec.~\ref{sec:groupTheory}}

Although we are not considering matches in this section, there is a
close relationship between position interchange and
problem~\ref{prob:sequentialIndependence} that we will explore in this
and next sections.

This section first introduces sequential independence for productions
and a characterization through G-congruence, compatibility and
coherence.  G-congruence and related conditions have been studied in
Sec.~\ref{sec:gCongruence}.  Similar results for coherence
(advancement and delaying of a single production) are also derived.

\newtheorem{PotSeqInd}[matrixproduct]{Definition}
\begin{PotSeqInd}[Sequential Independence]
  \index{sequential independence!generalization}Let $s_n = p_n ;
  \ldots ; p_1$ be a sequence and $\sigma$ a permutation.  Then, $s_n$
  and $\sigma \left( s_n \right)$ are said to be sequential
  independent if both add and remove the same elements and have the
  same minimal and negative initial digraphs.
\end{PotSeqInd}

Compatibility and coherence imply sequential independence provided
$s_n$ and $\sigma(s_n)$ have the same minimal and initial digraphs.

\newtheorem{CoherenceImpliesSeqInd}[matrixproduct]{Theorem}
\begin{CoherenceImpliesSeqInd}
  \label{th:CoherenceImpliesSeqInd}
  With notation as above, if $s_n$ is compatible and coherent and
  $\sigma \left( s_n \right)$ is compatible and coherent and both are
  G-congruent, then they are sequential independent.
\end{CoherenceImpliesSeqInd}

\noindent \emph{Proof}\\*
$\square$By hypothesis we can define two productions $c_s$, $c_{\sigma
  (s)}$ which are respectively the compositions coming from $s_n$ and
$\sigma (s_n)$.  Using commutativity of sum in
formulas~\eqref{eq:e_r_compositionEdges}
and~\ref{eq:e_r_compositionNodes}) -- i.e. the order in which elements
are added does not matter -- we directly see that $s_n$ and $\sigma
(s_n)$ add and remove the same elements.  G-congruence guarantees
sameness of minimal and negative initial digraphs.\proofend

Note that, even though the final result is the same when moving
sequential independent productions inside a given concatenation,
intermediate states can be very different.

In the rest of this section we will discuss permutations that move one
production forward or backward a certain number of positions, yielding
the same result. This means, using
Theorem~\ref{th:CoherenceImpliesSeqInd} and assuming compatibility and
G-congruence, finding out the conditions to be satisfied such that
starting with a coherent sequence we again obtain a coherent sequence
after applying the permutation.

\newtheorem{advanceDelayProd}[matrixproduct]{Theorem}
\begin{advanceDelayProd}
  \label{th:advanceDelayProdTheor}
  Consider coherent sequences $t_n = p_\alpha ;p_n;p_{n-1}; \ldots
  ;p_2;p_1$ and $s_n = p_n;p_{n-1}; \ldots ;p_2;p_1;p_\beta$ and
  permutations $\phi_{n+1}$ and $\delta_{n+1}$.
  \begin{enumerate}
  \item $\phi_{n+1} \left( t_n \right)$ -- advances $p_\alpha$
    application -- is coherent if
    \begin{equation}\label{eq:advProd}
      e^E_\alpha \bigtriangledown_1^n \left( \overline{r^E_x} \, L^E_y \right) \vee R^E_\alpha \bigtriangledown_1^n \left( \overline{e^E_x} \, r^E_y \right) = 0.
    \end{equation} 
  \item $\delta_{n+1} \left( s_n \right)$ -- delays $p_\beta$
    application -- is coherent if
    \begin{equation}\label{eq:delayProd}
      L^E_\beta \bigtriangleup_1^n \left( \overline{r^E_x} \, e^E_y \right) \vee r^E_\beta \bigtriangleup_1^n \left( \overline{e^E_x} \, R^E_y \right) = 0.
    \end{equation}
  \end{enumerate}
\end{advanceDelayProd}


\noindent \emph{Proof}\\*
$\square$Both cases have a very similar proof so only production
advancement is included.  The way to proceed is to check differences
between the original sequence $t_n$ and the swapped one, $\phi_{n+1}
\left( t_n \right)$, discarding conditions already imposed by $t_n$.

We start with $t_2 = p_\alpha ;p_2;p_1 \longmapsto \phi_3 \left( t_2
\right) = p_2;p_1;p_\alpha $, where $\phi_3 = [1 \; 3 \; 2]$.
Coherence of both sequences specify several conditions to be
fulfilled, included in Table~\ref{tab:coherenceAdvTwoProds}.  Note
that conditions (t.1.7) and (t.1.10) can be found in the original
sequence -- (t.1.2) and (t.1.5) -- so they can be disregarded.

\begin{table*}[hbtp]
  \centering
  \begin{tabular}{|r|r|r|}
    \hline
    \begin{Large}\phantom{I}\end{Large}Coherence of
    $p_{\alpha};p_2;p_1$ \begin{Large}\phantom{I}\end{Large} &
    Coherence of $p_2;p_1;p_{\alpha}$
    \begin{Large}\phantom{I}\end{Large}\\
    \hline
    \hline
    \begin{Large}\phantom{I}\end{Large}$e^E_2 \, L^E_\alpha = 0 \qquad
    (t.1.1)$ \begin{Large}\phantom{I}\end{Large} & $e^E_1 \, L^E_2 = 0
    \qquad (t.1.7)$ \begin{Large}\phantom{I}\end{Large} \\
    \begin{Large}\phantom{I}\end{Large}$e^E_1 \, L^E_2 = 0 \qquad
    (t.1.2)$ \begin{Large}\phantom{I}\end{Large} & $e^E_\alpha \,
    L^E_1 = 0 \qquad (t.1.8)$ \begin{Large}\phantom{I}\end{Large} \\
    \begin{Large}\phantom{I}\end{Large}$e^E_1 \, L^E_\alpha \,
    \overline{r^E_2} = 0 \qquad (t.1.3)$
    \begin{Large}\phantom{I}\end{Large} &
    \begin{Large}\phantom{I}\end{Large} $e^E_\alpha \, L^E_2 \, 
    \overline{r^E_1} = 0 \qquad (t.1.9)$
    \begin{Large}\phantom{I}\end{Large} \\
    \hline
    \hline
    \begin{Large}\phantom{I}\end{Large}$r^E_\alpha R^E_2 = 0 \qquad
    (t.1.4)$ \begin{Large}\phantom{I}\end{Large} &
    \begin{Large}\phantom{I}\end{Large} $r^E_2 R^E_1 = 0 \qquad
    (t.1.10)$ \begin{Large}\phantom{I}\end{Large} \\
    \begin{Large}\phantom{I}\end{Large}$r^E_2 R^E_1 = 0 \qquad
    (t.1.5)$ \begin{Large}\phantom{I}\end{Large} &
    \begin{Large}\phantom{I}\end{Large} $r^E_1 R^E_\alpha = 0 \qquad
    (t.1.11)$ \begin{Large}\phantom{I}\end{Large} \\
    \begin{Large}\phantom{I}\end{Large}$r^E_\alpha R^E_1
    \overline{e^E_2} = 0 \qquad (t.1.6)$
    \begin{Large}\phantom{I}\end{Large} &
    \begin{Large}\phantom{I}\end{Large} $r^E_2 R^E_\alpha
    \overline{e^E_1} = 0 \qquad (t.1.12)$
    \begin{Large}\phantom{I}\end{Large} \\
    \hline
  \end{tabular}
  \caption{Coherence for Advancement of Two Productions}
  \label{tab:coherenceAdvTwoProds}
\end{table*}

We would like to express all previous identities using operators
delta~\eqref{eq:TriangleUp} and nabla~\eqref{eq:TriangleDown} for
which equation~\ref{eq:L_And_Not_e} is used on (t.1.8) and (t.1.9):
\begin{eqnarray} \label{eq:Cond_2_Realtered} e^E_\alpha \, L^E_1 \,
  \overline{r^E_1} & = & 0\\
  \label{eq:Cond_3_Realtered} e^E_\alpha \, L^E_2 \, \overline{r^E_2}
  \, \overline{r^E_1} & = & 0.
\end{eqnarray}

For the same reason, applying~\eqref{eq:r_And_e} to conditions
(t.1.11) and (t.1.12):
\begin{eqnarray} \label{eq:Cond_5_Realtered} r^E_1 \overline{e^E_1}
  R^E_\alpha & = & 0 \\
  \label{eq:Cond_6_Realtered} r^E_2 \overline{e^E_2} R^E_\alpha
  \overline{e^E_1} & = & 0.
\end{eqnarray}

Condition (t.1.4) can be split into two parts --
recall~\eqref{eq:SecondCondTwoProd}~and~\ref{eq:ThirdCondTwoProd}) --
being $r^E_2 r^E_3 = 0$ one of them.  Doing the same operation on
(t.1.12), $r^E_2 r^E_3 \, \overline{e^E_1} = 0$ is obtained, which is
automatically verified and therefore should not be considered.  It is
not ruled out since, as stated above, we want to get formulas
expressible using operators delta and nabla. Finally we obtain the
equation:
\begin{equation}
  \label{eq:AdvanceThreeProd}
  R^E_\alpha \, \overline{e^E_1} \left( r^E_1 \vee \overline{e^E_2}
    r^E_2 \right) \vee e^E_\alpha \, \overline{r^E_1} \left( L^E_1
    \vee \overline{r^E_2} \, L^E_2 \right) = 0.
\end{equation}

\begin{figure}[htbp]
  \centering
  \includegraphics[scale =
  0.68]{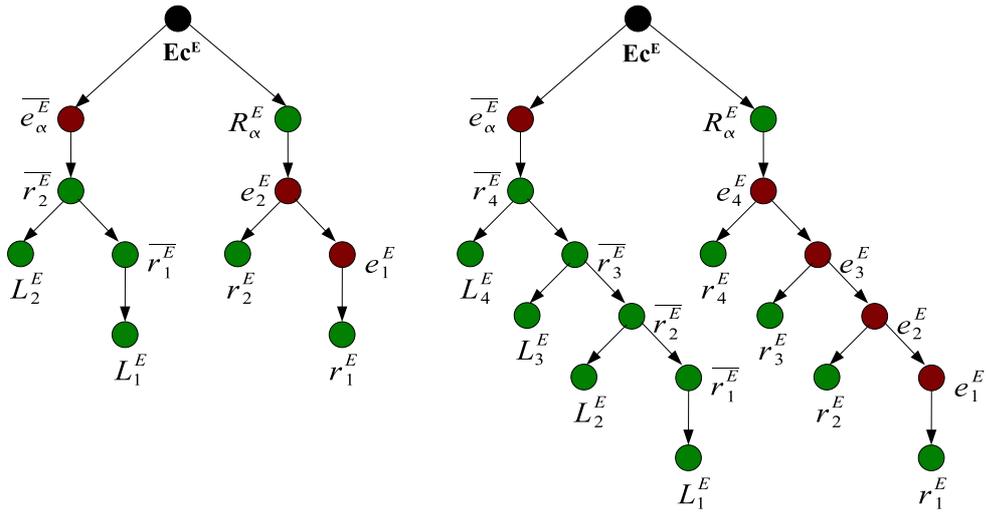}
  \caption{Advancement. Three and Five Productions}
  \label{fig:AdvancingThreeFiveProds}
\end{figure}

Performing similar manipulations on the sequence $t_3 = p_\alpha
;p_3;p_2;p_1$ we get $\phi_4 \left( t_3 \right) =
p_3;p_2;p_1;p_\alpha$ (with $\phi_4 = [ \, 1 \,\, 4 \,\, 3 \,\, 2 \,
]$); we find out that the condition to be satisfied is:
\begin{eqnarray}
  \label{eq:AdvanceFourProd}
  R^E_\alpha \overline{e^E_1} \Bigg( r^E_1 & \vee & \overline{e^E_2}
  \Big[ r^E_2 \vee \overline{e^E_3} r^E_3 \Big] \Bigg) \vee
  \nonumber\\
  & \vee & e^E_\alpha \, \overline{r^E_1} \Bigg( L^E_1 \vee
  \overline{r^E_2} \Big[ L^E_2 \vee \, \overline{r^E_3} \, L^E_3 \Big]
  \Bigg) = 0.
\end{eqnarray}

Figure~\ref{fig:AdvancingThreeFiveProds} includes the associated
graphs to previous example and to $n = 4$.  The proof can be finished
by induction.\proofend

Previous theorems foster the following notation: If
eq.~\eqref{eq:advProd} is satisfied and we have sequential
independence, we will write $p_\alpha \, \bot \, \left( p_n; \ldots ;
  p_1 \right)$ whereas if equation~\eqref{eq:delayProd} is true and
again they are sequential independent, it will be represented by
$\left( p_n; \ldots ; p_1 \right) \, \bot \, p_\beta$.  Note that if
we have the coherent sequence made up of two productions $p_2;p_1$ and
we have that $p_1;p_2$ is coherent we can write $p_2 \bot p_1$ to mean
that either $p_2$ may be moved to the front or $p_1$ to the back.

\noindent\textbf{Example.}$\square$It is not difficult to put an example of
three productions $t_3 = w_3; w_2; w_1$ where the advancement of the
third production two positions to get $t_3' = w_2;w_1;w_3$ has the
following properties: Their associated minimal initial digraphs -- $M$
and $M'$, respectively -- coincide, they are both coherent (and thus
sequential independent) but $t_3'' = w_2;w_3;w_1$ can not be
performed, so it is not possible to advance $w_3$ one position and,
right afterwards, another one, i.e. the advancement of two places must
be carried out in a single step.

\begin{figure}[htbp]
  \centering
  \includegraphics[scale = 0.52]{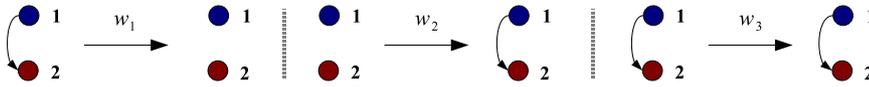}
  \caption{Three Simple Productions}
  \label{fig:threeVerySimpleProds}
\end{figure}

As drawn in Fig.~\ref{fig:threeVerySimpleProds}, $w_1$ deletes edge
$\left( 1,2 \right)$, $w_2$ adds it while it is preserved by $w_3$
(appears on its left hand side but it is not deleted).

Using previous notation, this is an example where $w_3 \bot \left(
  w_2;w_1 \right)$ but $w_3 \not \bot w_2$. As far as we know, in SPO
or DPO approaches, testing whether $w_3 \bot \left( w_2;w_1 \right)$
or not has to be performed in two steps: $w_3 \bot w_2$, that would
allow for $w_3;w_2;w_1 \mapsto w_2;w_3;w_1$, and $w_3 \bot w_1$ to get
the desired result: $w_2;w_1;w_3$.\proofend

\begin{figure}[htbp]
  \centering
  \includegraphics[scale = 0.57]{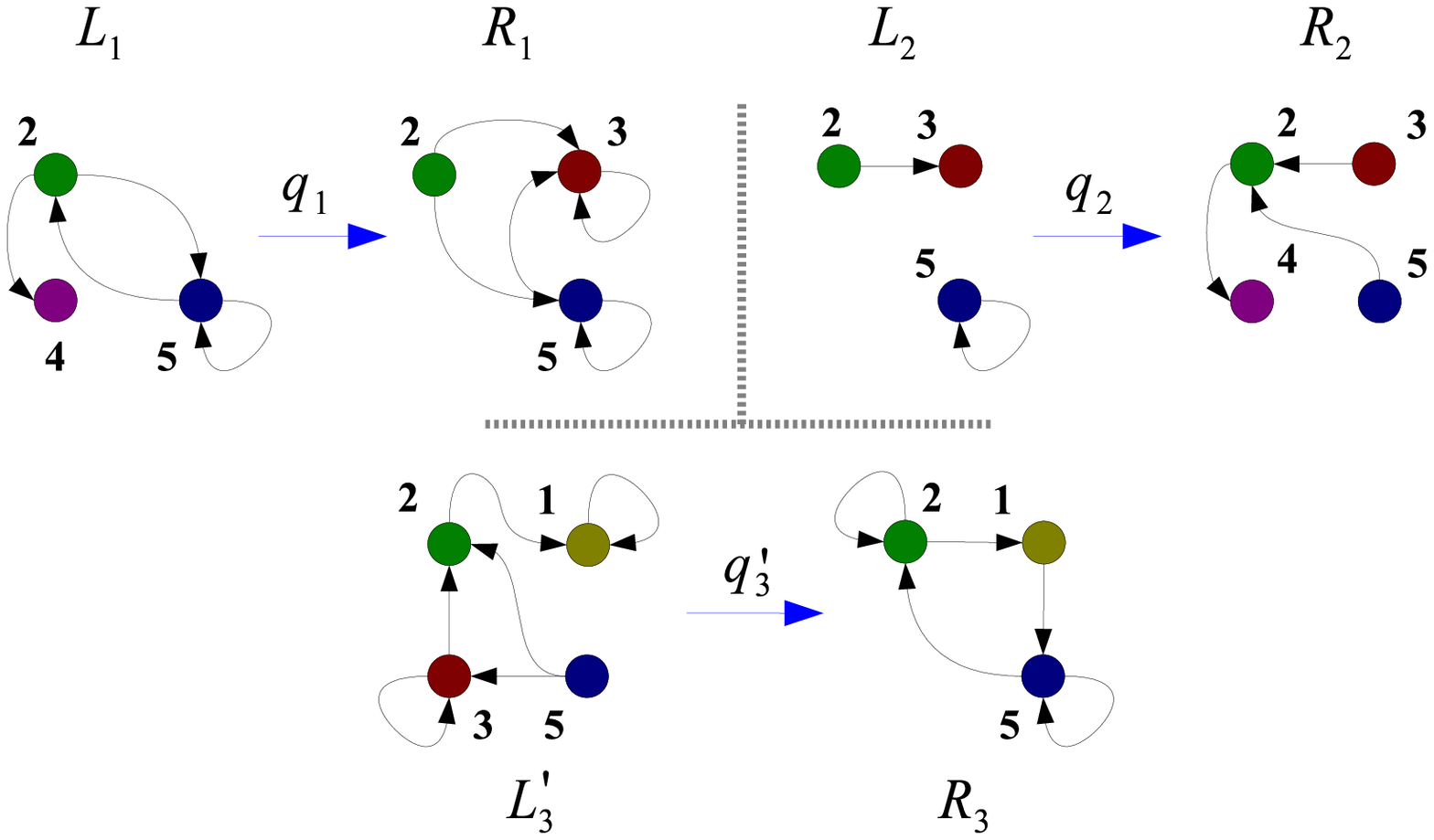}
  \caption{Altered Production $q'_3$ Plus Productions $q_1$ and $q_2$
    (Rep.)}
  \label{fig:ThirdProdAltered}
\end{figure}

\noindent\textbf{Example.}$\square$We will use productions $q_1$, $q_2$ and
$q'_3$ (reproduced again in Fig.~\ref{fig:ThirdProdAltered}).
Production $q'_3$ is advanced two positions inside $q_3';q_2;q_1$ to
obtain $q_2;q_1;q_3'$.  Such permutation can be expressed as $\phi_3 =
[1 \; 3 \; 2]$.\footnote{Numbers 1, 2 and 3 in the permutation mean
  position inside the sequence, not production subindex.}
Formula~\eqref{eq:advProd} expanded, simplified and adapted for this
case is:
\begin{equation}
  \underbrace{e_3 \left( L_1 \vee \overline{r_1} L_2 \right)}_{(*)} \vee \underbrace{R_3 \left( r_1 \vee \overline{e_1} r_2 \right)}_{(**)}.
\end{equation}

Finally, all elements are substituted and the operations are
performed, checking that the result is the null matrix.  Node ordering
is $[2 \; 3 \; 5 \; 1 \; 4]$, not included due to lack of space.  The
first part $(*)$ is zero:
\begin{equation}
  \left[
    \begin{array}{ccccc}
      \vspace{-6pt}
      0 & 0 & 0 & 0 & 0 \\
      \vspace{-6pt}
      1 & 1 & 0 & 0 & 0 \\
      \vspace{-6pt}
      0 & 1 & 0 & 0 & 0 \\
      \vspace{-6pt}
      0 & 0 & 0 & 1 & 0 \\
      \vspace{-6pt}
      0 & 0 & 0 & 0 & 0 \\
      \vspace{-10pt}
    \end{array} \right] \left(
    \left[
      \begin{array}{ccccc}
        \vspace{-6pt}
        0 & 0 & 1 & 0 & 1 \\
        \vspace{-6pt}
        0 & 0 & 0 & 0 & 0 \\
        \vspace{-6pt}
        1 & 0 & 1 & 0 & 0 \\
        \vspace{-6pt}
        0 & 0 & 0 & 0 & 0 \\
        \vspace{-6pt}
        0 & 0 & 0 & 0 & 0 \\
        \vspace{-10pt}
      \end{array} \right]  
    \vee \left[
      \begin{array}{ccccc}
        \vspace{-6pt}
        1 & 0 & 1 & 1 & 1 \\
        \vspace{-6pt}
        1 & 0 & 1 & 1 & 1 \\
        \vspace{-6pt}
        1 & 0 & 1 & 1 & 1 \\
        \vspace{-6pt}
        1 & 1 & 1 & 1 & 1 \\
        \vspace{-6pt}
        1 & 1 & 1 & 1 & 1 \\
        \vspace{-10pt}
      \end{array} \right] 
    \left[
      \begin{array}{ccccc}
        \vspace{-6pt}
        0 & 1 & 0 & 0 & 0 \\
        \vspace{-6pt}
        0 & 0 & 0 & 0 & 0 \\
        \vspace{-6pt}
        0 & 0 & 1 & 0 & 0 \\
        \vspace{-6pt}
        0 & 0 & 0 & 0 & 0 \\
        \vspace{-6pt}
        0 & 0 & 0 & 0 & 0 \\
        \vspace{-10pt}
      \end{array} \right]   \nonumber
  \right)
\end{equation}
and the same for $(**)$:
\begin{equation}
  \left[
    \begin{array}{ccccc}
      \vspace{-6pt}
      1 & 0 & 0 & 1 & 0 \\
      \vspace{-6pt}
      0 & 0 & 0 & 0 & 0 \\
      \vspace{-6pt}
      1 & 0 & 1 & 0 & 0 \\
      \vspace{-6pt}
      0 & 0 & 1 & 0 & 0 \\
      \vspace{-6pt}
      0 & 0 & 0 & 0 & 0 \\
      \vspace{-10pt}
    \end{array} \right] \left(
    \left[
      \begin{array}{ccccc}
        \vspace{-6pt}
        0 & 1 & 0 & 0 & 0 \\
        \vspace{-6pt}
        0 & 1 & 0 & 0 & 0 \\
        \vspace{-6pt}
        0 & 1 & 0 & 0 & 0 \\
        \vspace{-6pt}
        0 & 0 & 0 & 0 & 0 \\
        \vspace{-6pt}
        0 & 0 & 0 & 0 & 0 \\
        \vspace{-10pt}
      \end{array} \right]  
    \vee \left[
      \begin{array}{ccccc}
        \vspace{-6pt}
        1 & 1 & 1 & 1 & 0 \\
        \vspace{-6pt}
        1 & 1 & 1 & 1 & 1 \\
        \vspace{-6pt}
        0 & 1 & 1 & 1 & 1 \\
        \vspace{-6pt}
        1 & 1 & 1 & 1 & 1 \\
        \vspace{-6pt}
        1 & 1 & 1 & 1 & 1 \\
        \vspace{-10pt}
      \end{array} \right] 
    \left[
      \begin{array}{ccccc}
        \vspace{-6pt}
        0 & 0 & 0 & 0 & 1 \\
        \vspace{-6pt}
        1 & 0 & 0 & 0 & 0 \\
        \vspace{-6pt}
        0 & 0 & 0 & 0 & 1 \\
        \vspace{-6pt}
        0 & 0 & 0 & 0 & 0 \\
        \vspace{-6pt}
        0 & 0 & 0 & 0 & 0 \\
        \vspace{-10pt}
      \end{array} \right]   \nonumber
  \right)
\end{equation}
and hence the permutation is also coherent.\proofend

\section{Sequential Independence -- Derivations}
\label{sec:sequentialIndependenceDerivations}

Sequential independence for derivations is very similar to sequences
studied in previous section, the main difference being that there is a
state now to be taken into account.

Here $\sigma$ will represent an element of the group of permutations
and derivation $d_n$ will have associated sequence $s_n$.  Note that
two sequences $s_n$ and $s'_n = \sigma (s_n)$ carry out the same
operations but in different order.

\newtheorem{SeqIndDef}[matrixproduct]{Definition}
\begin{SeqIndDef}\label{def:SeqIndDef}
  \index{sequential independence!generalization}Two derivations $d_n$
  and $d'_n = \sigma \left( d_n \right)$ are sequential independent
  with respect to $G$ if $d_n \left( G \right) = H_n \cong H'_n = d'_n
  \left( G \right)$.
\end{SeqIndDef}

Compare with problem~\ref{prob:sequentialIndependence} on
p.~\pageref{prob:sequentialIndependence}.  Even though $s'_n = \sigma
(s_n)$, if $\varepsilon$-productions appear because the same
productions are matched to different places in the host graph, then it
might not be true that $d'_n = \sigma (d_n)$. A restatement of
Def.~\ref{def:SeqIndDef} is the following proposition.

\newtheorem{SeqIndProp}[matrixproduct]{Proposition}
\begin{SeqIndProp}\label{prop:SeqIndProp}
  If for two applicable derivations $d_n$ and $d'_n = \sigma (d_n)$
  \begin{enumerate}
  \item $\exists M_0 \subset G$ such that $\emptyset \ne M_0 \in
    \mathfrak{M} \left( s_n \right) \cap \; \mathfrak{M} \left( s'_n
    \right)$ and
  \item the corresponding negative initial digraph $K_0 \in
    \mathfrak{N} \left( s_n \right) \cap \; \mathfrak{N} \left( s'_n
    \right)$,
  \end{enumerate}
  then $d_n (M_0)$ and $d'_n (M_0)$ are sequential
  independent.
\end{SeqIndProp}

\noindent \emph{Proof}\\*
$\square$Existence of a minimal initial digraph and its corresponding
negative initial digraph guarantees coherence and compatibility.  As
it is the same in both cases, they are G-congruent.  A derivation and
any of its permutations carry out the same actions, but in different
order.  Hence, their result must be isomorphic.\proofend

If two derivations (with underlying permuted sequences) are not a
permutation of each other due to $\varepsilon$-productions but are
confluent (their image graphs are isomorphic), then in fact it is
possible to write them as a permutation of each other:

\newtheorem{SeqIndProp1}[matrixproduct]{Proposition}
\begin{SeqIndProp1}\label{prop:SeqIndProp1}
  If $d_n$ and $d'_n$ are sequential independent and $s'_n = \sigma
  (s_n)$, then $\exists \,\hat \sigma \; \vert \;d'_n = \hat \sigma
  (d_n)$ for some appropriate composition of
  $\varepsilon$-productions.
\end{SeqIndProp1}

\noindent \emph{Proof}\\*
$\square$Let $\widehat T:p_\varepsilon \mapsto \widehat
T(p_\varepsilon)$ be an operator acting on $\varepsilon$-productions,
which splits them into a sequence of $n$ productions each with one
edge.\footnote{More on operator $\widehat T$ in
  Chap.~\ref{ch:restrictionsOnRules}. It is used in
  Sec.~\ref{sec:functionalRepresentation} for application conditions.}

If $\widehat T$ is applied to $d_n$ and $d'_n$ we must get the same
number of $\varepsilon$-productions.  Moreover, the number must be the
same for every type of edge or a contradiction can be derived as
$\varepsilon$-productions only delete elements.\proofend

\begin{figure}[htbp]
  \centering
  \includegraphics[scale = 0.57]{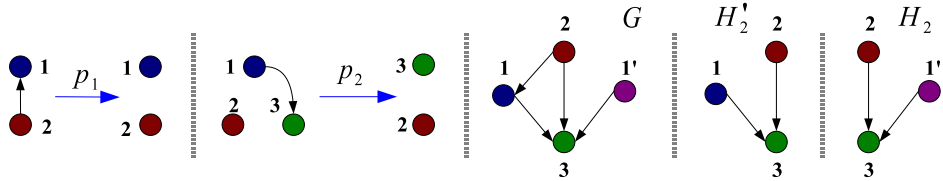}
  \caption{Sequential Independence with \emph{Free} Matching}
  \label{fig:SeqIndExample}
\end{figure}

\noindent\textbf{Example.}$\square$Define two productions $p_1$ and $p_2$,
where $p_1$ deletes edge $\left( 2, 1 \right)$ and $p_2$ deletes node
$1$ and edge $\left(1, 3 \right)$.  Define sequences $s_2 = p_2; p_1$
and $s'_2 = p_1; p_2$ and apply them to graph $G$ depicted in
Fig.~\ref{fig:SeqIndExample} to get $H_n$ and $H'_n$, respectively.
Note that $p_1$ and $p_2$ are not sequential independent in the sense
of Sec.~\ref{sec:sequentializationGrammarRules} with this
identification.

\begin{figure}[htbp]
  \centering
  \includegraphics[scale = 0.51]{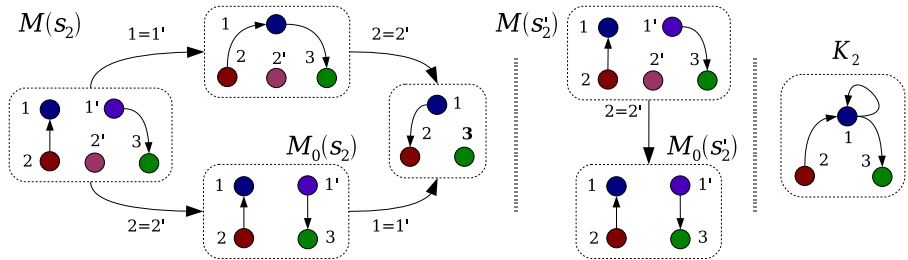}
  \caption{Associated Minimal and Negative Initial Digraphs}
  \label{fig:SeqIndExample_MNID}
\end{figure}

Suppose that in $s'_2$ the match $m_2$ for production $p_2$ identifies
node 1.  In this case an $\varepsilon$-production $p_{\varepsilon, 2}$
should appear deleting edge $\left( 2, 1 \right)$, transforming the
concatenation to $s'_2 = p_1; p_2; p_{\varepsilon, 2}$ and making
$p_1$ inapplicable.  If $m_2$ identifies node $1'$ instead of $1$,
then we have $H_n \cong H'_n$ with the obvious isomorphism $\left( 1,
  2, 3 \right) \mapsto \left( 1', 2, 3 \right)$, getting in this case
$p_2 \bot p_1$.  Note that $M_0 \left( s'_2 \right) \in \mathfrak{M}
\left( s_2 \right) \cap \mathfrak{M} \left( s'_2 \right)$ (see
Fig.~\ref{fig:SeqIndExample_MNID}).

Neither sequence $s_2$ nor $s'_2$ add any edge and only $p_2$ deletes
one node. The negative digraph set has just one element that has been
called $K_2$, also depicted in
Fig.~\ref{fig:SeqIndExample_MNID}.\proofend

The theory developed so far fits well here.  Results for sequential
independence such as Theorem~\ref{th:CoherenceImpliesSeqInd}, for
coherence (Theorems~\ref{th:SeqCoherenceTheorem},
\ref{th:advanceDelayProdTheor} and~\ref{th:advanceDelayProdTheor}) and
for minimal and negative initial digraphs are recovered.

Marking (see Sec.~\ref{sec:marking}) can be used to freeze the place
in which productions are applied.  For example, if a production is
advanced and we already know that there is sequential independence,
any node identification across productions should be kept because if
the production was applied at a different match sequential
independence could be ruined.


\section{Explicit Parallelism}
\label{sec:explicitParallelism}

This chapter finishes analyzing which productions or group of
productions can be computed in parallel and what conditions guarantee
this operation.  Firstly we will take into account productions only,
without initial state.

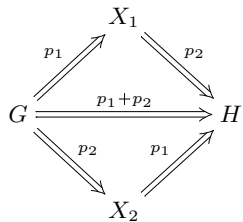
\begin{figure}[htb]
  \centering \makebox{ \xymatrix{
      & X_1\ar@{=>}[dr]^{p_2} & \\
      G\ar@{=>}[ur]^{p_1}\ar@{=>}[dr]^{p_2}\ar@{=>}[rr]^{p_1+p_2} && H \\
      & X_2\ar@{=>}[ur]^{p_1} & \\
    } }
  \caption{Parallel Execution}
  \label{fig:catSeqParallel}
\end{figure}

In the categorical approach the definition for two productions is
settled considering the two alternative sequential ways in which they
can be composed, looking for equality in their final state.
Intermediate states are disregarded using categorical coproduct of the
involved productions (see Sec.~\ref{sec:DPO}).  Then, the main
difference between sequential and parallel execution is the existence
of intermediate states in the former, as seen in
Fig.~\ref{fig:catSeqParallel}.  We follow the same approach saying
that it is possible to execute two productions in parallel if the
result does not depend on generated intermediate states.

\newtheorem{PotParInd}[matrixproduct]{Definition}
\begin{PotParInd}
  \label{ParallelismTwoProd}
  \index{true concurrency}Two productions $p_1$ and $p_2$ are said to
  be truly concurrent if it is possible to define their composition
  and it does not depend on the order:
  \begin{equation}
    p_2 \circ p_1 = p_1 \circ p_2.
  \end{equation}
\end{PotParInd}

We use the notation $p_1 \parallel p_2$ to denote true concurrency.
True concurrency defines a symmetric relation so it does not matter
whether $p_1 \parallel p_2$ or $p_2 \parallel p_1$ is written.

Next proposition compares true concurrency and sequential independence
for two productions, in the style of the \emph{parallelism theorem} --
see~\cite{DPO:handbook} --.\footnote{However, in DPO it is possible to
  identify elements once the coproduct has been performed through
  non-injective matches.}  The proof is straightforward in our case
and is not included.

\newtheorem {ParSeqTwoProd}[matrixproduct]{Proposition}
\begin{ParSeqTwoProd}
  Let $s_2=p_2;p_1$ be a coherent and compatible concatenation, then:
  \begin{equation}
    p_1 \parallel p_2 \Longleftrightarrow p_2 \bot p_1.
  \end{equation}
  \label{prop:parVsSeqInd}
\end{ParSeqTwoProd}

\noindent \emph{Proof}\\*
$\square$Assuming compatibility frees us from
$\varepsilon$-productions.\proofend

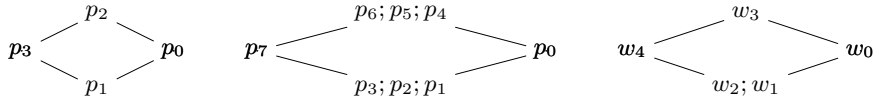
\begin{figure}[htbp]
  \setlength{\unitlength}{1cm} \centering
  \begin{picture}(10, 2.0) \put(-0.5, 1.0){ \xy \PATH ~={**\dir{-}}
      '(0,0)*+{p_3} '(10,5)*+{p_2} '(20,0)*+{p_0} \PATH ~={**\dir{-}}
      '(0,0)*+{p_3} '(10,-5)*+{p_1} '(20,0)*+{p_0}
      \endxy
    }
    \put(2.6, 1.0){
      \xy \PATH ~={**\dir{-}}
      '(0,0)*+{p_7} '(19,5)*+{p_6;p_5;p_4} '(38,0)*+{p_0}
      \PATH ~={**\dir{-}}
      '(0,0)*+{p_7} '(19,-5)*+{p_3;p_2;p_1} '(38,0)*+{p_0}
      \endxy
    }
    \put(7.5, 1.0){
      \xy \PATH ~={**\dir{-}}
      '(0,0)*+{w_4} '(15,5)*+{w_3} '(30,0)*+{w_0}
      \PATH ~={**\dir{-}}
      '(0,0)*+{w_4} '(15,-5)*+{w_2;w_1} '(30,0)*+{w_0}
      \endxy
    }
  \end{picture}
  \caption{Examples of Parallel Execution}
  \label{fig:ParallelExample}
\end{figure}

So far we have just considered one production per branch when
parallelizing, as represented to the left of
Fig.~\ref{fig:ParallelExample}.  One way to deal with more general
schemes -- center and right of the same figure -- is to test
parallelism for each element in one branch against every element in
the other.

Consider the scheme in the middle of Fig.~\ref{fig:ParallelExample}.
Sequences $s_1 = p_6;p_5;p_4$ and $s_2 =p_3;p_2;p_1$ can be computed
in parallel if there is sequential independence for every
interleaving.  This is true if $p_i \parallel p_j$, $\forall i \in \{
4, 5, 6\}$, $\forall j \in \{ 1, 2, 3 \}$.  There are many
combinations that keep the relative order of $s_1$ and $s_2$, for
example $p_6;p_3;p_2;p_5;p_1;p_4$ or $p_3;p_6;p_2;p_5;p_1;p_4$.  In
order to apply these two sequences in parallel, all interleavings that
maintain the relative order should have the same result.

Although it is not true in general, in many cases it is not necessary
to check true concurrency for every two productions.  The following
example illustrates the idea that is developed afterwards.

\noindent\textbf{Example.}$\square$Let be given the concatenation
$w_4;w_3;w_2;w_1;w_0$. See Fig.~\ref{fig:ParallelExample}
(right). Some of its productions are depicted in
Fig.~\ref{fig:threeVerySimpleProds} on
p.~\pageref{fig:threeVerySimpleProds}.  Rule $w_1$ deletes one edge,
$w_2$ adds the same edge while $w_3$ preserves it.

We already know that $w_3;w_2;w_1$ is compatible and coherent and that
$w_3 \bot \left( w_2;w_1 \right)$.  Both have the same minimal initial
digraph.  Following our previous study for two productions we would
like to put $w_3$ and $w_2;w_1$ in parallel, as depicted to the right
of Fig.~\ref{fig:ParallelExample}.

From a sequential point of view this diagram can be interpreted in
different ways, depending on how they are computed.  There are three
dissimilar interleavings: (1) $w_3;w_2;w_1$, (2) $w_2;w_1;w_3$ and (3)
$w_2;w_3;w_1$.

Any problem involving the first two possibilities is ruled out by
coherence.  As a matter of fact, $w_3$ and $w_2;w_1$ can not be
parallelized because it could be the case that $w_3$ is using edge
$(1,2)$ when $w_1$ has just deleted it and before $w_2$ adds it, which
is what the third case expresses, leaving the system in an
inconsistent state.  Thus, we do not have $w_3 \parallel w_2$ nor $w_3
\parallel w_1$ -- we do not have sequential independence -- but both
$w_3; w_2; w_1$ and $w_2; w_1; w_3$ are coherent.\proofend

One possibility to proceed is to use the fact that although it could
be the case that $p_3 \not \perp p_2$, it still might be possible to
advance the production with the help of another production, i.e. $p_3
\perp \left( p_2;p_1 \right)$ as seen in
Secs.~\ref{sec:sequentializationGrammarRules}~and~\ref{sec:sequentialIndependenceDerivations}.

Although there are some similarities between this concept and the
theorem of concurrency,\footnote{See
  Sec.~\ref{sec:DPO}~or~\cite{Fundamentals}.} here we rely on the
possibility to characterize production advancement or delaying inside
sequences more than just one position, hence, being more general.

\newtheorem{AdvanceDelayPar}[matrixproduct]{Theorem}
\begin{AdvanceDelayPar}
  Let $s_n = p_n; \ldots ; p_1$ and $t_m = q_m; \ldots ;q_1$ be two
  compatible and coherent sequences with the same minimal initial
  digraph, where either $n=1$ or $m=1$.  Suppose $r_{m+n} = t_m; s_n$
  is compatible and coherent and either $t_m \bot s_n$ or $s_n \bot
  t_m$.  Then, $t_m \parallel s_n$ through composition.
  \label{th:advanceDelayParallelism}
\end{AdvanceDelayPar}

\noindent \emph{Proof}\\*
$\square$Using Proposition~\ref{prop:parVsSeqInd}.\proofend

\emph{Through composition} means that the concatenation with length
greater than one must be transformed into a single production using
composition.  This is possible because it is coherent and compatible
-- refer to Prop.~\ref{prop:CoherenceImpliesComposition} --.  In fact
it should not be necessary to transform the whole concatenation using
composition, but only the parts that present a problem.

Setting $n=1$ corresponds to advancing a production in sequential
independence, while $m=1$ to moving a production backwards inside a
concatenation.  In addition, in the hypothesis we ask for coherence of
$r_n$ and either $t_m \bot s_n$ or $s_m \bot t_n$.  In fact, if
$r_{m+n}$ is coherent and $t_m \bot s_n$, then $s_n \bot t_m$.  It is
also true that if $r_{m+n}$ is coherent and $s_n \bot t_m$, then $t_m
\bot s_n$ (it could be proved by contradiction).

The idea behind Theorem~\ref{th:advanceDelayParallelism} is to erase
intermediate states through composition but, in a real system, this is
not always possible or desirable if for example these states were used
for synchronization of productions or states.  All this section can be
extended easily to consider derivations.

\section{Summary and Conclusions}
\label{sec:summaryAndConclusions5}

In this chapter we have studied in more detail sequences and
derivations, paying special attention to sequential independence.  We
remark once more that certain properties of sequences can be gathered
during grammar specification.  This information can be used for an
a-priori analysis of the graph transformation system (grammar if an
initial state is also provided) or, if properly stored, during
runtime.

In essence, sequential independence corresponds to the concept of
commutativity $\left( a\,;b = b\,;a \right)$ or a generalization of
it, because commutativity is defined for two elements and here we
allow $a$ or $b$ to be sequences.  It can be used to reduce the size
of the state space associated to the grammar.  From a theoretical or
practical-theoretical point of view, sequential independence helps by
reducing the amount of productions combinatorics in sequences or
derivations.  This is of interest, for example, for confluence
(problem~\ref{prob:confluence} on p.~\pageref{prob:confluence}).

Besides sequential independence for concatenations and derivations, we
have also studied G-congruence, which guarantees sameness of the
minimal and negative initial digraphs, and explicit parallelism,
useful for parallel computation.

%

One of the objectives of the present book is to tackle
problems~\ref{prob:independence}~and~\ref{prob:sequentialIndependence}, 
independence and sequential independence, respectively, defined in
Sec.~\ref{sec:motivation}.  The whole chapter is directed to this end,
but with success in the restricted case of advancing or delaying a
single production an arbitrary number of positions in a sequence.
This is achieved in
Theorems~\ref{th:CoherenceImpliesSeqInd}~and~\ref{th:advanceDelayProdTheor},
which rely on Theorem~\ref{th:GCongruence} (G-congruence), and also in
Props.~\ref{prop:SeqIndProp}~and~\ref{prop:SeqIndProp1}.

These results can be generalized by addressing other types of
permutations such as advancing or delaying blocks of productions.
Another possibility is to study the swap of two productions inside a
sequence. It can be addressed following the same sort of development
along this chapter. Swaps of two productions are 2-cycles and it is
well known that any permutation is the product of 2-cycles.

In order to link this chapter with the next one and
Chapter~\ref{ch:transformationOfRestrictions}, which deal with
application conditions and restrictions on graphs, let's note that 
conditions that need to be fulfilled in order to obtain sequential
independence can be interpreted as graph constraints and application
conditions.  Graph constraints and application conditions are
important both from the theoretical and from the practical points of
view.
\chapter{Restrictions on Rules}
\label{ch:restrictionsOnRules}

In this chapter graph constraints and application conditions -- that
we call \emph{restrictions} -- for Matrix Graph Grammars will be
studied, generalizing previous approaches to this topic.  For us, a
restriction is just a condition to be fulfilled by some graph. This
study will be completed in the following chapter.

In the literature there are two kinds of restrictions:
\emph{Application conditions} and \emph{graph constraints}.  Graph
constraints express a global restriction on a graph while application
conditions are normally thought of as local properties, namely in the
area where the match identifies the LHS of the grammar rule.  By
generalizing graph constraints and application conditions we will see
that they can express both local and global properties and, moreover,
that application conditions are a particular case of graph
constraints.

It is at times advisable to speak of \emph{properties} rather than
\emph{restrictions}.  For a given grammar, restrictions can be set
either during rule application (application conditions, to be checked
before the rule is applied or after it is applied) or on the shape of
the state (graph constraints, which can be set on the input state or
on the output state).

Application conditions are important from both the practical and the
theoretical points of view. On the practical side, they are convenient
to concisely express properties or to synthesize productions. They
also open the possibility to partially act on the nihilation matrix.
On the theoretical side, application conditions put into a new
perspective the left and right hand sides of a production. They also
enlarge the scope of Matrix Graph Grammars, including multidigraphs
(though this will be addressed in
Chap.~\ref{ch:transformationOfRestrictions}).

This book extends previous approaches using monadic second
order logic (MSOL, see Sec.~\ref{sec:logics} for a quick overview).
Section~\ref{sec:graphConstraintsAndApplicationConditions} sets the
basics for graph constraints and application conditions by introducing
\emph{diagrams} and their semantics.  In
Sec.~\ref{sec:extendingDerivations} derivations and diagrams are put
together, showing that diagrams are a natural generalization of graphs
$L$ and $K$ (in the precondition case).
Section~\ref{sec:functionalRepresentation} expresses all these results
using the functional notation introduced in
Sec.~\ref{sec:matchAndExtendedMatch} (see also
Sec.~\ref{sec:functionalAnalysis}). We prove that any application
condition is equivalent to some (set of) sequence(s) of
productions. Section~\ref{sec:summaryAndConclusions6} closes the
chapter with a summary and some more comments.

\section{Graph Constraints and Application Conditions}
\label{sec:graphConstraintsAndApplicationConditions}

\index{diagram}A graph constraint (GC) in Matrix Graph Grammars is
defined as a \emph{diagram} (a set of graphs and partial injective
morphisms) plus a MSOL formula.\footnote{MSOL corresponds to regular
  languages~\cite{Courcelle}, which are appropriate to express
  patterns.}  The diagram is made of a set of graphs and morphisms
(partial injective functions) which specify the relationship between
elements of the graphs. The formula specifies the conditions to be
fulfilled in order to make the host graph $G$ satisfy the GC, i.e. we
check whether $G$ is a model for the diagram and the formula.

The \emph{domain of discourse} are simple digraphs, and the diagram is
a means to represent the \emph{interpretation function}
\textbf{I}. Recall that in essence the domain of discourse is a set of
individual elements which can be quantified over. The interpretation
function assigns meanings (semantics) to symbols. See
Sec.~\ref{sec:logics} and references therein for more details.

\begin{figure}[htb]
  \centering \makebox{ \xymatrix{ A_0
      \ar@{.>}@[blue]@/_/[rrrrdd]^{m_{A_0}} && A_1 \ar[ll]_{d_{10}}
      \ar@{.>}@[blue]@/_/[ddrr]^{m_{A_1}} && L
      \ar@/_17pt/[llll]_{d_{L0}} \ar[ll]_{d_{L1}}
      \ar@{.>}@[blue][rr]^p \ar@{.>}@[blue][dd]^{m_L} && R
      \ar@{.>}@[red][dd] \\ \\
      &&&& G \ar@{.>}@[red][rr] && H } }
  \caption{Application Condition on a Rule's Left Hand Side}
  \label{fig:firstExAC}
\end{figure}
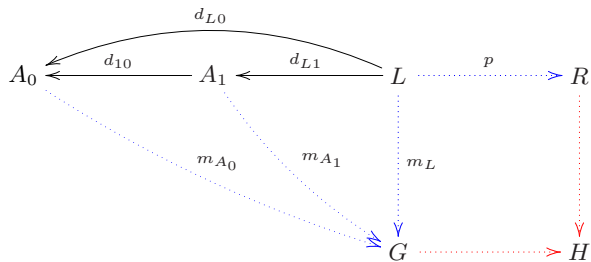

\noindent\textbf{Example}.$\square$Figure~\ref{fig:firstExAC} shows a diagram
associated to the left hand side of a production $p: L \rightarrow R$
matched to a host graph $G$ by $m_L$.  An example of associated
formula can be $\mathfrak{f} = \exists L \forall A_0 \exists A_1
\left[ L \left( A_0 \Rightarrow A_1 \right) \right]$.\proofend

We will focus on logical expressions encoding that one simple
digraph is contained in another, because this is in essence what
matching does. To this end, the following two predicates are
introduced:
\begin{align}
  P(X_1, X_2) &= \forall m [F(m, X_1) \Rightarrow F(m, X_2)] \\
  Q(X_1, X_2) & = \exists e [F(e, X_1) \wedge F(e, X_2)],
\end{align}
which rely on predicate $F(m, X)$, ``node or edge $m$ is in digraph
$X$'', or on $F(e, X)$, ``edge $e$ is in digraph $X$''. Predicate
$P(X_1, X_2)$ holds if and only if $X_1 \subset X_2$ and $Q(X_1, X_2)$
is true if and only if $X_1 \cap X_2 \neq \emptyset$.  Formula $P$
will deal with total morphisms and $Q$ with non-empty partial
morphisms (see graph constraint satisfaction,
Def.~\ref{def:GCSatisfied}).

\begin{figure}[htbp]
  \centering
  \includegraphics[scale = 0.6]{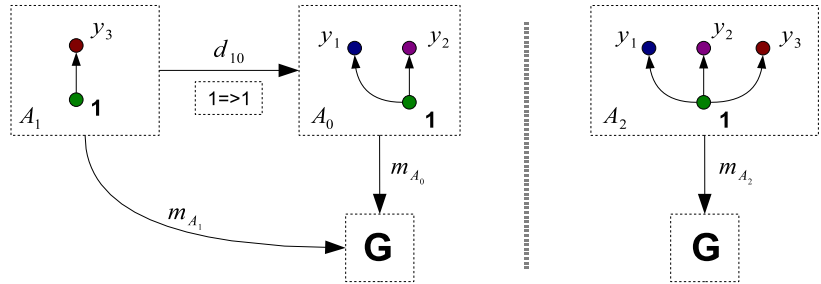}
  \caption{At Most Two Outgoing Edges}
  \label{fig:AtMostTwoOutgoingEdges}
\end{figure}

\noindent\textbf{Remark}.$\square$$P^E(X_1, X_2)$ says that every
edge\footnote{Mind the superscript $E$ in $P^E$. As in previous
  chapters, an $E$ superscript means \emph{edge} and an $N$ superindex
  stands for \emph{node}.} in graph $X_1$ should also be present in
$X_2$, so a morphism $d_{12}:X_1 \rightarrow X_2$ is demanded.  The
diagram may already include one such morphism (which can be seen as
restrictions imposed on function \textbf{I}) and we can either allow
extensions of $d_{12}$ (relate more nodes if necessary) or keep it as
defined in the diagram.  This latter possibility will be represented
appending the subscript $U$ to $P^E \mapsto P^E_U$.  Predicate $P^E_U$
can be expressed\footnote{Non-extensible existence of $d_{10}$ for a
  graph constraint is $\forall x \in A_0, \forall y \in A_1,
  m_{A_0}\!\!\left( x \right) = m_{A_1}\!\!\left( y \right)
  \Leftrightarrow y = d_{10}\!\!\left( x \right)$, with notation as in
  Fig.~\ref{fig:AtMostTwoOutgoingEdges}. In words: When elements are
  matched in the host graph (or in other graphs through different
  $d_{ij}$) elements unrelated by $d_{10}$ remain unrelated.} using
$P^E$:
\begin{equation}
  \label{eq:7}
  P^E_U(X_1,X_2) = \forall a [\neg \left( F(a,D) + F(a,coD) \right)] =
  P^E(D,coD) \wedge P^E(D^c,coD^c)  
\end{equation}
where $D = Dom(d_{12})$, $coD = coDom(d_{12})$, ${}^c$ stands for the
complement ($D^c$ is the complement of $Dom(d_{12})$ w.r.t. $X_1$) and
$+$ is the \textbf{xor} operation. For example, following the notation
in Fig.~\ref{fig:AtMostTwoOutgoingEdges}, $P_U\left( A_1, A_0 \right)$
would mean that it is not possible to further relate another element
apart from 1 between $A_0$ and $A_1$.  This could only happen when
$A_0$ and $A_1$ are matched in the host graph.

$P_U$ will be used as a means to indicate that elements not related by
their morphisms in the diagram must remain unrelated. These
relationships (forbidden according to $P_U$) could be specified either
by other morphisms in the diagram or by matches in the host graph. For
example, two unrelated nodes of the same type in different graphs of
the diagram can be identified as the same node by the corresponding
matches in the host graph. Hence, even though not explicitly
specified, there would exist a morphism relating these nodes in the
diagram. $P_U$ prevents this side effect of matches. The same can
happen if there is a chain of morphisms in the diagram such as $A_0
\rightarrow A_1 \rightarrow A_2$. There might exist an implicit
unspecified morphism $A_0 \rightarrow A_2$.\proofend

\begin{figure}[htbp]
  \centering
  \includegraphics[scale = 0.55]{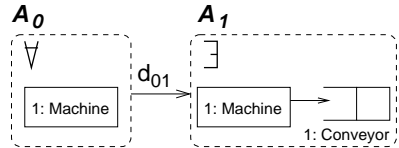}
  \caption{Example of Diagram} \label{fig:DiagExample}
\end{figure}

\noindent\textbf{Example}.$\square$Before starting with formal definitions, we
give an intuition of GCs.  The following GC is satisfied if for every
$A_0$ in $G$ it is possible to find a related $A_1$ in $G$: $\forall
A_0 \exists A_1 \left[ A_0 \Rightarrow A_1\right]$, equivalent by
definition to $\forall A_0 \exists A_1 \left[ P\!\!\left(A_0, G
  \right) \Rightarrow P\!\! \left(A_1, G\right) \right]$. Nodes and
edges in $A_0$ and $A_1$ are related through the diagram shown in
Fig.~\ref{fig:DiagExample}, which relates elements with the same
number and type. As a notational convenience, to enhance readability,
each graph in the diagram has been marked with the quantifier given in
the formula. The graph constraint in Fig.~\ref{fig:DiagExample}
expresses that each machine should have an output conveyor. \proofend

It is interesting for restrictions to be able to express negative
conditions, that is, to express that some elements should not be
present in the host graph. By elements we mean nodes, edges or both.
When some elements are requested not to exist in $G$, one possibility
is to find them in the complementary graph.

To this end we will define a structure $\overline{G} = \left(
  \overline{G^E}, \overline{G^N} \right)$ that in first instance
consists of the negation of the adjacency matrix of $G$ and the
negation of its vector of nodes.  We speak of structure because the
negation of a digraph is not a digraph.  In general, compatibility
fails for $\overline{G}$.\footnote{In Chap.~\ref{ch:mggFundamentals1}
  a matrix for edges and a vector for nodes were introduced to
  differentiate one from the other, mainly because operations could be
  performed on nodes or on edges. Recall that compatibility related
  both of them and completion permitted operations on matrices of
  different size (with a different number of nodes).}

Although it has been commented already, we will insist in the
difference between completion and negation of the adjacency matrix.
The complement of a graph coincides with the negation of the adjacency
matrix, but while negation is just the logical operation, taking the
complement means that a completion operation has been performed
before. Hence, taking the complement of a matrix $G$ is the negation
with respect to some appropriate completion of $G$.  As long as no
confusion arises negation and complements will not be syntactically
distinguished.
Graph with respect to which the completion (if any) is performed will
not be explicitly written from now on.

\begin{figure}[htbp]
  \centering
  \includegraphics[scale = 0.6]{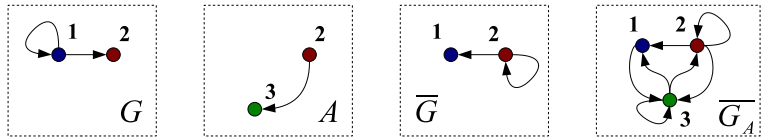}
  \caption{Finding Complement and Negation}
  \label{fig:firstNAC}
\end{figure}

\noindent\textbf{Example}.$\square$Suppose we have two graphs $A$ and $G$ as
those depicted in Fig.~\ref{fig:firstNAC} and that we want to check
that $A$ is not in $G$.  Note that $A$ is not contained in $\overline
G$ (node $3$ does not even appear) but it does appear in the negation
of the completion with respect to $A$ of $G$ (graph $\overline{G}_{A}$
in the same figure).\proofend

The notation (syntax) will be alleviated a bit more by making the host
graph $G$ the default second argument for predicates $P$ and $Q$.
Besides, it will be assumed that by default total morphisms are
demanded. That is, predicate $P$ will be assumed unless otherwise
stated. Our proposal to simplify the notation is to omit $G$ and $P$
in these cases. Also, it is not necessary to repeat quantifiers that
are together, e.g. $\forall A_0 \exists A_1 \exists A_2\forall A_3$
can be abbreviated as $\forall A_0 \exists A_1 A_2 \forall A_3$.

\noindent\textbf{Example}.$\square$A sophisticated way of demanding the
existence of one graph $\exists A [A]$ is:
\begin{displaymath}
  \exists A^N \exists A^E \left[ P\left( A^N, A^E \right) \wedge A^N
    \wedge A^E \right]
\end{displaymath}
that reads \emph{it is possible to find in $G$ the set of nodes of $A$
  and its set of edges in the same place -- $P\left( A^N, A^E \right)$
  --}.  In this case it is possible to use the universal quantifier
instead, as there is a single occurrence of $A^N$ in $A^E$ up to
isomorphisms:
\begin{displaymath}
  \forall A^N \exists A^E \left[ P\left( A^N, A^E \right) \wedge A^N
    \wedge A^E \right].
\end{displaymath}

As another example, the following graph constraint is fulfilled if for
every $A_0$ in $G$ it is possible to find a related $A_1$ in $G$:
\begin{equation}\label{eq:exNotation}
  \forall A_0 \exists A_1 \left[ A_0 \Rightarrow A_1\right],
\end{equation}
which by definition is equivalent to
\begin{equation}
  \forall A_0 \exists A_1 \left[ P\!\!\left(A_0, G \right) \Rightarrow
    P\!\! \left(A_1, G\right) \right].
\end{equation}
These syntax simplifications just try to simplify most commonly used
rules.\proofend

Negations inside abbreviations must be applied to the corresponding
predicate, e.g. $\exists A \left[ \overline{A} \right] \equiv \exists
A \left[ \overline{P} \left( A, G \right) \right]$ is not the negation
of $A$'s adjacency matrix. For the case of edges, the following
identity is fulfilled:
\begin{equation}\label{eq:relPQ}
  \overline{P^E}(A,G) = Q(A,\overline{G^E}).
\end{equation}

The part that takes care of the nodes is easier, so from now on we
will mainly concentrate on edges and adjacency
matrices.\footnote{Using the tensor product it is possible to embed
  the node vector into the adjacency matrix. This is not used in this
  book except in Chap.~\ref{ch:reachability}. See the definition of
  the \emph{incidence tensor} in Sec.~\ref{sec:dPOLikeMGG}.}

A bit more formally, the syntax of well-formed formulas is inductively
defined as in monadic second-order logic, which is first-order logic
plus variables for the subset of the domain of discourse. Across this
chapter, formulas will normally have one variable term $G$ which
represents the host graph. Usually, the rest of the terms will be
given (they will be constant terms). Predicates will consist of $P$
and $Q$ and combinations of them through negation and binary
connectives. Next definition formally presents the notion of {\em
  diagram}.

\newtheorem{diagramDef}[matrixproduct]{Definition}
\begin{diagramDef}[Diagram]\label{def:diagram}
  \index{diagram}A \emph{diagram} $\mathfrak{d}$ is a set of simple
  digraphs $\{A_i \}_{i \in I}$ and a set of partial injective
  morphisms $\{d_k \}_{k \in K}, d_k: A_i \rightarrow A_j$. We will
  say that a diagram is \emph{well defined} if every cycle of
  morphisms commute.
\end{diagramDef}

\begin{figure}[htbp]
  \centering
  \includegraphics[scale = 0.7]{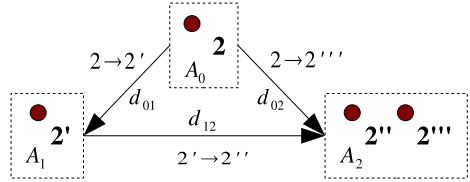}
  \caption{non-Injective Morphisms in Application Condition}
  \label{fig:wellDefinedness}
\end{figure}

\index{well-definedness}To illustrate well-definedness consider the
diagram of Fig.~\ref{fig:wellDefinedness}. Node typed $2$ has two
different images, $2''$ and $2'''$, depending if morphism $d_{12}
\circ d_{01}$ is considered or $d_{02}$. There would be an
inconsistency if $d_{01}(2) = 2'$, $d_{02}(2) = 2'''$ and $d_{12}(2')
= 2''$ because $d_{12} \circ d_{01}(2) = 2''$ while. Notice that node
$2$ would have two different images and we have imposed by hypothesis
that all morphisms must be injective.

\index{ground formula}The term \emph{ground formula} will mean a MSO
closed formula which uses $P$ and $Q$ with constant nodes (i.e. nodes
of a concrete type which can be matched with nodes of the same type).

The formulae in the constraints use variables in the set $\{A_i\}_{i
  \in I}$, and predicates $P$ and $Q$. Formulae are restricted to have
no free variables except for the default second argument of predicates
$P$ and $Q$, which is the host graph $G$ in which we evaluate the
GC. Next definition presents the notion of GC.

\newtheorem{graphConstraint}[matrixproduct]{Definition}
\begin{graphConstraint}[Graph Constraint]\label{def:graphConstraint}
  \index{graph constraint} $GC = ( \mathfrak{d} = ( \{A_i \}_{i \in
    I}, \{ d_j \}_{j \in J} ), \mathfrak{f} ) $ is a graph constraint,
  where $\mathfrak{d}$ is a well defined diagram and $\mathfrak{f}$ a
  sentence with variables in $\{A_i\}_{i \in I}$. A constraint is
  called {\em basic} if $|I|=2$ (with one bound variable and one free
  variable) and $J=\emptyset$.
\end{graphConstraint}

In general, there will be an outstanding variable among the $A_i$
representing the host graph, being the only free variable in
$\mathfrak{f}$. In previous paragraphs it has been denoted by $G$, the
default second argument for predicates $P$ and $Q$. We sometimes speak
of a ``GC defined over G''. A basic GC will be one made of just one
graph and no morphisms in the diagram (recall that the host graph is
not represented by default in the diagram nor included in the
formulas). For now we will limit to ground formulas and it will not be
until Sec.~\ref{sec:fromSimpleDigraphsToMultidigraphs} that
\emph{variable nodes} are considered. A variable node is one whose
type is not specified.

How graph constraints can be expressed using diagrams and logic
formulas will be illustrated with some examples\footnote{Examples ``at
  most two outgoing edges'' below and ``3-vertex colorable graph'' on
  p.~\pageref{ex:3VertexColorGraph} have been adapted
  from~\cite{Courcelle}.} throughout this section, comparing with the
way they should be written using FOL and
MSOL.

\label{ex:twoOutgoingEdges}\noindent\textbf{Example (at most two outgoing
  edges).}$\square$Let's characterize graphs in which every node of
type $1$ has at most two outgoing edges. Using FOL:
\begin{eqnarray}
  \mathfrak{f}_1 = \forall y_1, y_2, y_3 [ edg \left( 1, y_1 \right) &
  \wedge & edg \left( 1, y_2 \right) \wedge \nonumber \\
  & \wedge & edg \left( 1, y_3 \right) \Rightarrow y_1 = y_2 \vee y_1
  = y_3 \vee y_2 = y_3 ],
\end{eqnarray}
where function $edg \left( x, y \right)$ is \textbf{true} if there
exists an edge starting in node $x$ and ending in node $y$.  In our
case, we consider the diagram to the left of
Fig.~\ref{fig:AtMostTwoOutgoingEdges} together with the formula:
\begin{equation}\label{eq:f_1_OutEdges}
  \mathfrak{f}_1 = \forall A_0 \not{\exists} A_1 \left[A_0 \Rightarrow
    \left(A_1 \wedge P_U(D,coD)\right)\right]
\end{equation}
where $D = Dom(d_{10})$ and $coD = coDom(d_{10})$.

There must be two total injective morphisms $m_{A_0}:A_0 \rightarrow
G$, $m_{A_1}:A_1 \rightarrow G$ and a partial injective morphism
$m_{A_1A_0}:A_1 \rightarrow A_0$ which does not extend $d_{10}$
($m_{A_1A_0} = d_{10}$), i.e. elements of type $1$ are related and
variables $y_1$ and $y_2$ remain unrelated with $y_3$.  Hence, two
outgoing edges are allowed but not three.

In this case it is also possible to consider the diagram to the right
of Fig.~\ref{fig:AtMostTwoOutgoingEdges} together with the much
simpler formula $\mathfrak{f}_2 = \not{\exists} A_2[A_2]$.  This form
will be used when the theory is extended to cope with multidigraphs in
Sec.~\ref{sec:fromSimpleDigraphsToMultidigraphs}.\proofend

A graph constraint is a limitation on the shape of a graph, i.e. what
elements it is made up of.  This is something that can always be
demanded on any graph, irrespective of the existence of a grammar or
rule.  This is not the case for application conditions which need the
presence of productions.

In the following few paragraphs, application conditions will be
introduced.  Out of the definition it is not difficult to see
application conditions as a particular case of graph constraints in
this framework: one of the graphs in the diagram is the rule's LHS
(existentially quantified over the host graph) and another one is the
graph induced by the nihilation matrix (existentially quantified over
the negation of the host graph).

\newtheorem{WeakPrecondDef}[matrixproduct]{Definition}
\begin{WeakPrecondDef}[Weak Precondition]\label{def:weakPrecond}
  \index{precondition!weak}Given a production $p:L \rightarrow R$ with
  nihilation matrix $K$, a weak precondition is a graph constraint
  over $G$ satisfying:
  \begin{enumerate}
  \item $\exists ! i,j$ such that $A_i = L$ and $A_j = K$.
  \item $\exists ! k$ such that $A_k = G$ is the only free variable.
  \item $\mathfrak{f}$ must demand the existence of $L$ in $G$ and the
    existence of $K$ in $\overline{G^E}$.
  \end{enumerate}
\end{WeakPrecondDef}

The simple graph $G$ can be thought of as a host graph to which some
grammar rules are to be applied. For simplicity, we usually do not
explicitly show the condition 3 in the formulae of ACs, nor the
nihilation matrix $K$ in the diagram.  However, if omitted, both $L$
and $K$ are existentially quantified before any other graph of the
AC. Thus, an AC has the form $\exists L \nexists K ... [L \wedge
P(K, \overline G) \wedge ...]$.

For technical reasons to be clarified in
Sec.~\ref{sec:movingConditions}, it is better not to have morphisms
whose codomains are $L$ or $K$, for example $d_i: A_i \rightarrow L$
or $d_j: A_j \rightarrow K$.  This is not a big issue as we may always
use their inverses due to $d_i$'s injectiveness, i.e. one may consider
$d_i^{-1}:L \rightarrow A_i$ and $d_j^{-1}:K \rightarrow
A_j\label{eq:domain}$ instead.

Note the similarities between Def.~\ref{def:weakPrecond} and that of
derivation in Sec.~\ref{def:directDerivationDef}.  Actually, this
definition interprets the left hand side of a production and its
nihilation matrix as a weak precondition.  Hence, any well defined
production has a natural associated weak precondition.

\index{postcondition!weak}Starting with the definition of weak
precondition we define \emph{weak postconditions} similarly but using
the comatch $m_R:R \rightarrow H$, $H = p\left( G \right)$.
\index{precondition!MGG}\index{postcondition!MGG}A \emph{precondition}
is a weak precondition plus a match $m_L:L \rightarrow G$ and,
symmetrically, a \emph{postcondition} is a weak postcondition plus a
comatch $m_R:R \rightarrow H$.

Every production naturally specifies a weak postcondition.  Elements
that must be present are those found at $R$, while $e \vee
\overline{D}$ should not be found by the comatch.

Weak application conditions, weak preconditions and weak
postconditions permit the specification of restrictions at a grammar
definition stage with no need for matches, as in
Chaps.~\ref{ch:mggFundamentals1}~and~\ref{ch:mggFundamentals2}.

\newtheorem{applCond}[matrixproduct]{Definition}
\begin{applCond}[(Weak) Application Condition]\label{def:applCond}
  \index{application condition!weak} \index{application condition!in
    MGG} For a production $p$, a \emph{(weak) application condition}
  is a (weak) precondition plus a (weak) postcondition, $AC = \left(
    AC_L, AC_R \right)$.
\end{applCond}

\begin{figure}[htbp]
  \centering
  \includegraphics[scale = 0.7]{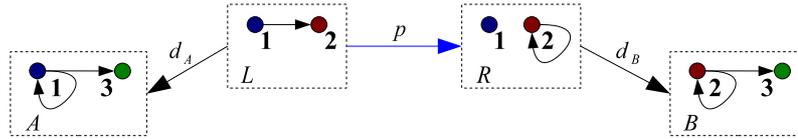}
  \caption{Example of Precondition Plus Postcondition}
  \label{fig:prePostCondsEx}
\end{figure}

\noindent\textbf{Example}.$\square$Figure~\ref{fig:prePostCondsEx} depicts a
production with diagram $\mathfrak{d}_{LHS} = \{A\}$ for its LHS and
diagram $\mathfrak{d}_{RHS} = \{B\}$ for its RHS. If the associated
formula for $\mathfrak{d}_{LHS}$ is $\mathfrak{f}_{LHS} = \exists L
\exists A \left[ L \, \overline{A} \right]$ then there are two
different possibilities depending on how morphism $d_{A}$ is defined:
\begin{enumerate}
\item $d_A$ identifies node 1 in $L$ and $A$.  Whenever $L$ is matched
  in a host graph there can not be at least one $A$, i.e. at least for
  one matching of $A$ -- with node 1 in common with $L$ -- in the host
  graph either edge $(1,1)$ or edge $(1,3)$ are missing.
\item $d_A$ does not identify node 1 in $L$ and $A$.  This does not
  necessarily mean that they must be different when matched in an
  actual host graph.  Now, it is sufficient not to find one $A$ which
  would be fine for any match of $L$ in the host graph.
\end{enumerate}

Recall that the interpretation of the quantified parts $\exists L$ and
$\exists A$ are, respectively, to find nodes $1$ and $2$ and $1$ and
$3$ (edges too). In the first bullet above, both nodes 1 must coincide
while in the second case they may coincide or they may be different.

The story varies if formula $\mathfrak{f}_{LHS} = \exists L \forall A
\left[ L \, \overline{A} \right]$ is considered.  There are again two
cases, but now:
\begin{enumerate}
\item $d_A$ identifies node 1 in $L$ and $A$.  No other node $3$ can
  be linked to node $1$ if it has a self loop.
\item $d_A$ does not identify node 1 in $L$ and $A$.  The same as
  above, but now both nodes $1$ need not be the same.
\end{enumerate}

A similar interpretation can be given to the postcondition
$\mathfrak{d}_{RHS}$ together with formula $\mathfrak{f}_{RHS} =
\exists R \exists A \left[ R \, \overline{A} \right]$ and
$\mathfrak{f}_{RHS} = \exists R \forall A \left[ R \, \overline{A}
\right]$.\proofend

\noindent\textbf{Remark (local vs. global properties)}.$\square$As commented
in the introduction of this chapter, graph constraints are normally
thought of as global conditions on the entire graph while application
conditions are local properties, defined in the neighborhood of the
match (and usually not beyond).

In our setting, the use of quantifiers on restrictions permit
``local'' graph constraints and ``global'' application conditions.
The first by using existential quantifiers (so as soon as the
restriction is fulfilled in one piece of the host graph, the graph
constraint is fulfilled) and the latter through universal quantifiers
(for every potential match of the application condition it must be
fulfilled).\proofend

\noindent\textbf{Remark (semantics of quantification)}.$\square$In GCs or ACs,
graphs are quantified either existentially or universally. We now give
the intuition of the semantics of such quantification applied to basic
formulae. Thus, we consider four cases: (i) $\exists A[A]$, (ii)
$\forall A[A]$, (iii) $\nexists A[A]$, (iv) $\slash\!\!\forall A [A]$.

\begin{figure}[htbp]
  \centering
  \includegraphics[scale = 0.5]{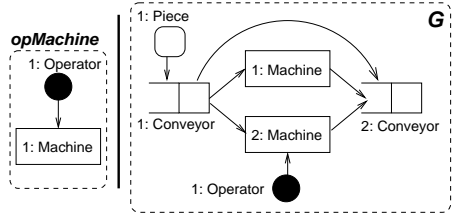}
  \caption{Quantification Example}
  \label{fig:quantifier_semantics}
\end{figure}

Case (i) states that a graph $A$ should be found in $G$.  For example,
in Fig.~\ref{fig:quantifier_semantics}, the GC $\exists
opMachine[opMachine]$ demands an occurrence of $opMachine$ in $G$
(which exists).

Case (ii) demands that, for all {\em potential occurrences} of $A$ in
$G$, the shape of graph $A$ is actually found. The term potential
occurrences means all distinct maximal partial matches\footnote{A
  match is partial if it does not identify all nodes or edges of the
  source graph.  The domain of a partial match should be a graph.}
(which are total on nodes) of $A$ in $G$. A non-empty partial match in
$G$ is maximal if it is not strictly included in another partial or
total match.  For example, consider the GC $\forall
opMachine[opMachine]$ in the context of
Fig.~\ref{fig:quantifier_semantics}.  There are two possible
instantiations of $opMachine$ (as there are two machines and one
operator), and these are the two input elements to the formula. As
only one of them satisfies $P(opMachine, G)$ (the expanded form of
$[opMachine]$) the GC is not satisfied by $G$.

Case (iii) demands that, for all potential occurrences of $A$, none of
them should have the shape of $A$.  The term potential occurrence have
the same meaning as in case (ii). In
Fig.~\ref{fig:quantifier_semantics}, there are two potential
instantiations of the GC $\nexists opMachine[opMachine]$. As one of
them actually satisfies $P(opMachine, G)$, the formula is not
satisfied by $G$.

Finally, case (iv) is equivalent to $\exists A[\overline A]$, where by
definition $\overline A \equiv \overline P(A, G)$.  This GC states
that for all possible instantiations of $A$, one of them does not have
the shape of $A$. This means that a non-empty partial morphism should
be found from $A$ to $\overline G$.  In
Fig.~\ref{fig:quantifier_semantics}, the GC $\exists
opMachine[\overline{opMachine}]$ is satisfied by $G$, because again
there are two possible instantiations, and one of them actually does
not have an edge between the operator and the machine.\proofend

Some notation for the set of morphisms and isomorphisms between two
graphs is needed in order to interpret basic constraints satisfaction.
\begin{eqnarray}
  par^{\max}(A_i, A_j) & = &  \left\{ f:A_i \rightarrow A_j \; \vert
    \, f \textrm{ maximal non-empty partial morphism}\right. \nonumber
  \\ 
  & & \left. \textrm{$\qquad\qquad\qquad$ with $Dom(f)^N = A^N$}
  \right\} \nonumber \\
  tot(A_i, A_j) & = & \left\{ f : A_i \rightarrow A_j \; \vert \, f
    \textrm{ is a total morphism} \right\} \subseteq par^{max}(A, G)
  \nonumber \\
  iso(A_i, A_j) & = & \left\{ f:A_i \rightarrow A_j \; \vert \, f
    \textrm{ is an isomorphism} \right\} \subseteq tot(A, G) \nonumber
\end{eqnarray}
where $Dom(f)^N$ are the nodes of the graph in the domain of $f$.
Thus, $par^{max}(A, G)$ denotes the set of all potential occurrences
of a given constraint graph $A$ in $G$ (where we require all nodes in
$A$ to be present in the domain of $f$). Note that each $f \in
par^{max}$ may be empty in edges.

\newtheorem{BGCSatisfied}[matrixproduct]{Definition}
\begin{BGCSatisfied}[Basic Constraint Satisfaction]
  \label{def:BGCSatisfied} 
  The four most basic graph constraint satisfactions are: 
  \begin{itemize}
  \item Graph $G$ satisfies $\exists A [A]$ iff $\exists f \in
    par^{\max}(A,G) \; \vert \, f \in tot(A,G)$.
  \item Graph $G$ satisfies $\forall A [A]$ iff $\forall f \in
    par^{\max}(A,G) \; \vert \, f \in tot(A,G)$.
  \item Graph $G$ satisfies $\not \exists A [A]$ iff $\forall f \in
    par^{\max}(A,G) \; \vert \, f \not\in tot(A,G)$.
  \item Graph $G$ satisfies $\not\!\forall A [A]$ iff $\exists f \in
    par^{\max}(A,G) \; \vert \, f \not\in tot(A,G)$.
  \end{itemize}
\end{BGCSatisfied}

The diagrams associated to the formulas in previous definition have
been omitted for simplicity as they consist of a single element:
$A$. Recall that by default predicate $P$ is assumed as well as $G$ as
second argument, e.g. the first formula in previous definition
$\exists A[A]$ is actually $\exists A[P(A,G)]$. In fact, only the
first two cases are needed because one has $\nexists A[P(A,G)] \equiv
\forall A[\overline P(A, G)]$ and $\slash\!\!\forall A[P(A,G)] \equiv
\exists A[\overline P(A, G)] $.

Given a graph $G$ and a graph constraint $GC$, the next step is to
state when $G$ satisfies $GC$.  This definition also applies to
application conditions.

\newtheorem{GCSatisfied}[matrixproduct]{Definition}
\begin{GCSatisfied}[Graph Constraint Satisfaction]
  \label{def:GCSatisfied}
  \index{graph constraint!fulfillment}We say that $\mathfrak{d}_0 =
  (\{A_i\}, \{d_j\})$ satisfies the graph constraint $GC =
  (\mathfrak{d} = (\{X_i\},$$\{d_j\}), \mathfrak{f})$ under the
  interpretation function $I$, written $(I, \mathfrak{d}_0) \models
  \mathfrak{f}$, if $\mathfrak{d}_0$ is a model for $\mathfrak f$ that
  satisfies the element relations\footnote{As any mapping, $d_j$
    assigns elements in the domain to elements in the codomain.
    Elements so related should be mapped to the same element. For
    example, Let $a \in X_1$ and $d_{1i}:X_1 \rightarrow X_i$ with $b
    = d_{12}(a)$ and $c = d_{13}(a)$. Further, assume $d_{23}:X_2
    \rightarrow X_3$, then $d_{23}(b) = c$.} specified by the diagram
  $\mathfrak d$, and the following interpretation for the predicates
  in $\mathfrak{f}$:
  \begin{enumerate}
  \item $I\left(P \left( X_i, X_j \right)\right) = m^T: X_i
    \rightarrow X_j$ total injective morphism.
  \item $I\left(Q \left( X_i, X_j \right)\right) = m^P: X_i
    \rightarrow X_j$ partial injective morphism, non-empty in edges.
  \end{enumerate}
  where $m^T \vert_D = d_k = m^P \vert_D$ with\footnote{It can be the
    case that $Dom\left(m^P \right) \cap Dom \left(d_k \right) =
    \emptyset$.} $d_k:X_i \rightarrow X_j$ and $D=Dom\left(d_k
  \right)$. The interpretation of quantification is as in
  Def.~\ref{def:BGCSatisfied} but setting $X_i$ and $X_j$ instead
  of $A$ and $G$, respectively.
\end{GCSatisfied}

Recall that we say that a morphism is \emph{total} if its domain
coincides with the initial set and \emph{partial} if it is a proper
subset.

\noindent\textbf{Remark}.$\square$There can not exist a model if there is any
contradiction in the definition of the graph constraint.  A
contradiction is to ask for an element to appear in $G$ and also to be
in $\overline{G}$.  In the case of an application condition, some
contradictions are avoidable while others are not.  We will return to
this point in Sec.~\ref{sec:extendingDerivations} with an example and
appropriate definitions.\proofend

The four basic constraint satisfactions of Def.~\ref{def:BGCSatisfied}
can be written $G \models \exists A [A]$, $G \models \forall A [A]$, $G
\models \not \exists A [A]$ and $G \models \not\!\forall A [A]$. The
notation deserves the following comments:
\begin{enumerate}
\item The notation $(I, \mathfrak{d}_0) \models \mathfrak{f}$ means
  that the formula $\mathfrak{f}$ is satisfied under interpretation
  given by $I$, assignments given by morphisms specified in
  $\mathfrak{d}_0$ and substituting the variables in $\mathfrak{f}$
  with the graphs in $\mathfrak{d}_0$.
\item As commented after Def.~\ref{def:graphConstraint}, in many cases
  the formula $\mathfrak{f}$ will have a single variable (the one
  representing the host graph $G$) and always the interpretation
  function will be that given in Def.~\ref{def:GCSatisfied}. We may
  thus write $G \models \mathfrak{f}$. The notation $G \models GC$ may
  also be used.
\item Similarly, as an AC is just a GC where $L$, $K$ and $G$ are
  present, we may write $G \models AC$. For practical purposes, we are
  interested in checking whether, given a host graph $G$, a certain
  match $\abb{m_L}{L}{G}$ satisfies the AC. In this case we write $(G,
  m_L) \models AC$. In this way, the satisfaction of an AC by a match
  and a host graph is like the satisfaction of a GC by a graph $G$,
  where a morphism $m_L$ is already specified in the diagram of the
  GC.
\end{enumerate}

\begin{figure}[htbp]
  \centering
  \includegraphics[scale = 0.6]{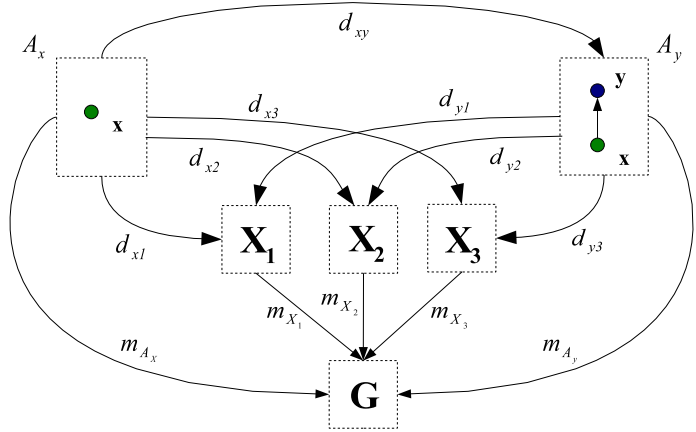}
  \caption{Diagram for Three Vertex Colorable Graph Constraint}
  \label{fig:ThreeVertexColorableGraph}
\end{figure}

\label{ex:3VertexColorGraph}\noindent\textbf{Example (3-vertex colorable
  graph).}$\square$In order to express that a graph $G$ is 3-vertex
colorable we need to state two basic facts: First, every single node
belongs to one of three disjoint sets, called $X_1$, $X_2$ and $X_3$:
Check first three lines in formula~\eqref{eq:MSform_2}.  Second, every
two nodes joined by one edge must belong to different $X_i$, $i=1, 2,
3$, which is stated in the last two lines of~\eqref{eq:MSform_2}.
Using MSOL:
\begin{eqnarray}
  \mathfrak{f}_2 = \exists X_1, X_2, X_3 [ & \!\! \forall & \!\! x \;
  ( x \in X_1 \vee x \in X_2 \vee x \in X_3  ) \wedge \nonumber \\
  & \!\! \forall & \!\! x \; ( \psi \left( x, X_1, X_2, X_3 \right)
  \wedge \psi \left( x, X_2, X_1, X_3 \right) \wedge  \nonumber \\
  && \quad \; \psi \left( x, X_3, X_2, X_1 \right)) \wedge \nonumber \\
  & \!\! \forall & \!\! x, y \; ( edg \left( x, y \right) \wedge
  \left(x \ne y \right) \Rightarrow \phi \left( x, y, X_1 \right)
  \wedge \nonumber \\
  && \qquad \qquad \qquad \qquad \phi \left( x, y, X_2 \right) \wedge
  \phi \left( x, y, X_3 \right))]
  \label{eq:MSform_2}
\end{eqnarray}
where,
\begin{eqnarray}
  \psi \left( x, X, Y, Z \right) & = & \left[ x \in X \Rightarrow x
    \notin Y \wedge x \notin Z \right] \nonumber \\
  \phi \left( x, y, X \right) & = & \left[ \neg \left( x \in X \wedge
      y \in X \right) \right] = \left[ x \notin X \vee y \notin X
  \right]. \nonumber
\end{eqnarray}

In our case, we consider the diagram of
Fig.~\ref{fig:ThreeVertexColorableGraph} and formula
\begin{equation}\label{eq:f_2_OutEdges}
  \mathfrak{f}_2 = \exists X_1 \exists X_2 \exists X_3 \forall A_x \!
  \not{\exists} A_y \left[ \left( \bigwedge_{i=1}^3 X_i \right)
    \Rightarrow \left[ A \wedge A_y \right] \right]
\end{equation}
where $A = (P(A_x,X_1) + P(A_x,X_2) + P(A_x, X_3))$. Digraphs $X_i$
split $G$ into three disjoint subsets (the three colors) through
predicate $A$, which states the disjointness of $X_i$ and, with the
rest of the clause, the coverability of $G$, $G = X_1 \bigcup X_2
\bigcup X_3$.\proofend

\begin{figure}[htbp]
 \centering
 \includegraphics[scale = 0.5]{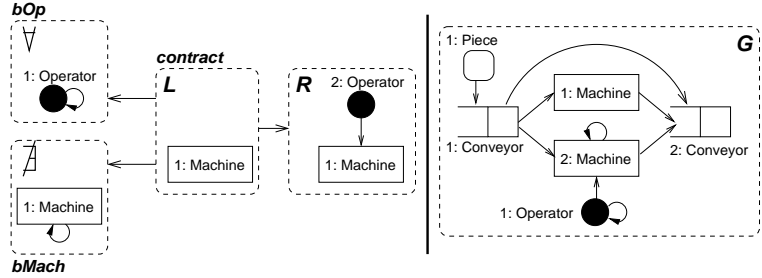}
 \caption{Satisfaction of Application Condition.}
 \label{fig:example_fulfill1}
\end{figure}

\noindent\textbf{Example}$\square$Figure~\ref{fig:example_fulfill1} shows rule
{\em contract}, with an AC given by the diagram in the figure (where
morphisms identify elements with the same type and number, this
convention is followed throughout the paper), together with formula
$\exists L \: \nexists bMach \: \forall bOp [L \wedge bMach \wedge
bOp]$. The rule creates a new operator, and assigns it to a machine.
The rule can be applied if there is a match of the LHS (a machine is
found), the machine is not busy ($\nexists bMach[bMach]$), and all
operators are busy ($\forall bOp[bOp]$). Graph $G$ to the right
satisfies the AC, with the match that identifies the machine in the
LHS with the machine in $G$ with the same number.

Using the terminology of ACs in the algebraic
approach~\cite{Fundamentals}, $\nexists bMach[bMach]$ is a negative
application condition (NAC). On the other hand, there is nothing
equivalent to $\forall bOp[bOp]$ in the algebraic approach, but in
this case it could be emulated by a diagram made of two graphs stating
that if an operator exists then it does not have a self-loop. However,
this is not possible in all cases as next example shows.\proofend

\begin{figure}[htbp]
  \centering
  \includegraphics[scale = 0.35]{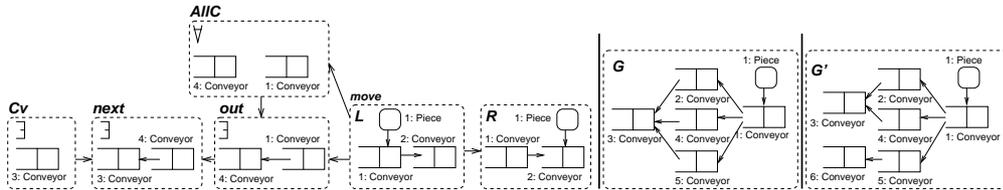}
  \caption{Example of Application Condition.} \label{fig:example_AC}
\end{figure}

\noindent\textbf{Example}.$\square$Figure~\ref{fig:example_AC} shows rule {\em
  move}, which has an application condition with formula: $\exists Cv
\: \forall AllC \: \exists out \: \exists next [ (AllC \wedge out)
\Rightarrow $ $( next \wedge Cv )]$. As previously stated, in this
example and the followings, the rule's LHS and the nihilation matrix
are omitted in the AC's formula.  The example AC checks whether all
conveyors connected to conveyor 1 in the LHS reach a common target
conveyor in one step. We can use ``global'' information, as graph $Cv$
has to be found in $G$ and then all output conveyors are checked to be
connected to it ($Cv$ is existentially quantified in the formula
before the universal). Note that we first obtain all possible
conveyors ($\forall AllC$).  As the identifications of the morphism
$L\rightarrow AllC$ have to be preserved, we consider only those
potential instances of $AllC$ with $1: Conveyor$ equal to $1:
Conveyor$ in $L$. From these, we take those that are connected
($\exists out$), and which therefore have to be connected with the
conveyor identified by the LHS.  Graph $G$ satisfies the AC, while
graph $G'$ does not, as the target conveyor connected to $5$ is not
the same as the one connected to $2$ and $4$.  To the best of our
efforts it is not possible to express this condition using the
standard ACs in the DPO approach given
in~\cite{Fundamentals}. \proofend

\section{Embedding Application Conditions into Rules}
\label{sec:extendingDerivations}

The question of whether our definition of direct derivation is
powerful enough to deal with application conditions (from a semantical
point of view) will be proved in Theorem~\ref{th:embeddingGC} and
Corollary~\ref{cor:embedding} in this section.  It is necessary to
check that direct derivations can be the codomain of the
interpretation function, i.e. ``MGG + AC = MGG'' and ``MGG + GC =
MGG''.

Note that a direct derivation in essence corresponds to the formula:
\begin{equation}
  \exists L \exists K \left[ L \wedge P\left(K, \overline{G^E} \right)
  \right]
\end{equation}
but additional application conditions (AC) may represent much more
general properties, due to universal quantifiers and partial
morphisms.  Normally, for different reasons, other approaches to graph
transformation do not care about elements that can not be present at a
rule specification level.  If so, a direct derivation would be as
simple as:
\begin{equation}
  \exists L [L].
\end{equation}
Thus, one way to embed ACs into grammar rules is to seek for a means
to translate universal quantifiers and partial morphisms into
existential quantifiers and total morphisms. To this end, we introduce
two operations on basic diagrams: \emph{Closure} ($\mathfrak{C}$) and
\emph{Decomposition} ($\mathfrak{D}$).  The first deals with universal
quantifiers and the second with partial morphisms.  In some sense they
are complementary (compare
equations~\eqref{eq:closure}~and~\eqref{eq:decomp}). 

The closure operator converts a universal quantification into a number
of existentials, as many as maximal partial matches there are in the
host graph (see Definition~\ref{def:BGCSatisfied}).  Thus, given a
host graph $G$, demanding the universal appearance of graph $A$ in $G$
is equivalent to asking for the existence of as many replicas of $A$
as partial matches of $A$ are in $G$.

\newtheorem{closureDef}[matrixproduct]{Definition}
\begin{closureDef}[Closure]\label{def:closureDef}
  \index{closure}Given the $GC = \left( \mathfrak{d}, \mathfrak{f}
  \right)$ with diagram $\mathfrak{d} = \{ A \}$, ground formula
  $\mathfrak{f} = \forall A [A]$ and a host graph $G$, the result of
  applying $\mathfrak{C}$ to $GC$ is calculated as follows:
  \begin{eqnarray}\label{eq:closure}
    \mathfrak{d} & \longmapsto & \mathfrak{d}' = \left(\{ A^1, \ldots,
      A^n \}, d_{ij}:A^i \rightarrow A^j \right) \nonumber \\
    \mathfrak{f} & \longmapsto & \mathfrak{f}' = \exists A^1 \ldots
    \exists A^n \left[ \bigwedge_{i=1}^n A^i \bigwedge_{i,j = 1,
        \,j>i} P_U \left( A_i, A_j \right)\right]
  \end{eqnarray} 
  with $A^i \cong A$, $d_{ij} \not \in iso(A^i, A^j)$, $\mathfrak{C}
  \left( GC \right) = GC' = \left( \mathfrak{d}', \mathfrak{f}'
  \right)$ and $n = \left \lvert par^{\max}(A,G) \right \rvert$.
\end{closureDef}

The condition that morphism $d_{ij}$ must not be an isomorphism means
that at least one element of $A^i$ and $A^j$ will be identified in
different places of $G$. This is accomplished by means of predicate
$P_U$ (see its definition in equation~\eqref{eq:7}) which ensures
that the elements not related by $\abb{d_{ij}}{A^i}{A^j}$, are not
related in $G$.

\begin{figure}[htbp]
  \centering
  \includegraphics[scale = 0.45]{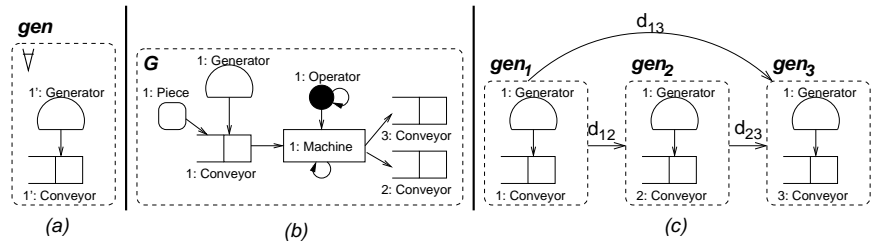}
  \caption{(a) GC diagram (b) Graph to which GC applies (c)
    Closure of GC}
  \label{fig:example_closure1}
\end{figure}

\noindent\textbf{Example}.$\square$Assume the diagram to the left of
Fig.~\ref{fig:example_closure1}, made of just graph $gen$, together
with formula $\forall gen[gen]$, and graph $G$, where such GC is to be
evaluated. The GC asks $G$ for the existence of all potential
connections between each generator and each conveyor.  Performing
closure we obtain $\mathfrak{C}((gen, \forall
gen[gen]))=(\mathfrak{d}_C, \exists gen_1 \exists gen_2 \exists gen_3
[gen_1 \wedge gen_2 \wedge gen_3 \wedge P_U(gen_1, gen_2) \wedge
P_U(gen_1, gen_3) \wedge P_U(gen_2, gen_3) ])$, where diagram
$\mathfrak{d}_C$ is shown to the right of
Fig.~\ref{fig:example_closure1}, and each $d_{ij}$ identifies elements
with the same number and type. The closure operator makes explicit
that three potential occurrences must be found (as $|par^{max}(gen,
G)|=3$), thus, taking information from the graph where the GC is
evaluated and placing it in the GC itself. There is another example
right after the definition of the \emph{decomposition} operator, on
p.~\pageref{ex:closureDecomposition}. \proofend

The interpretation of the closure operator is that demanding the
universal appearance of a graph is equivalent to the existence of all
of its potential instances in the specified digraph ($G$,
$\overline{G}$ or whatever).  Whenever nodes in $A$ are identified in
$G$, edges of $A$ must also be found. Therefore, each $A^i$ contains
the image of a possible match of $A$ in $G$ (there are $n$ possible
occurrences of $A$ in $G$) and $d_{ij}$ identifies elements considered
equal.

Now we turn to \emph{decomposition}. The idea behind it is to split a
graph into its components to transform partial morphisms into total
morphisms of one of its parts. If nodes are considered as the building
blocks of graphs for this purpose, then if two graphs share a node of
the same type there would be a partial match between them,
irrespective of the links established by the edges of the graphs.
Also, as stated above, we are more interested in the behavior of
edges (which to some extent comprises nodes as source and target
elements of the edges, except for isolated nodes) than on nodes alone
as they define the \emph{topology} of the graph.\footnote{This is why
  predicate $Q$ was defined to be true in the presence of a partial
  morphism non-empty in edges.} These are the reasons why
decomposition operator $\mathfrak{D}$ is defined to split a digraph
$A$ into its edges, generating as many digraphs as edges in $A$.

If so desired, in order to consider isolated nodes, it is possible to
define two decomposition operators, one for nodes and one for edges.
Note however that decomposition for nodes makes sense mostly for
graphs made up of isolated nodes, or for parts of graphs consisting of
isolated nodes only. In this case, we would be dealing with sets more
than with graphs.

\newtheorem{decompDef}[matrixproduct]{Definition}
\begin{decompDef}[Decomposition]\label{def:decompDef}
  \index{decomposition}Given a $GC = \left( \mathfrak{d},
    \mathfrak{f} \right)$ with ground formula $\mathfrak{f} = \exists
  A [Q(A)]$, diagram $\mathfrak{d} = \{ A \}$ and host graph $G$,
  $\mathfrak{D}$ acts on $GC$ -- $\mathfrak{D} \left( GC \right) = GC'
  = \left( \mathfrak{d}', \mathfrak{f}' \right)$ -- in the following
  way:
  \begin{eqnarray}
    \mathfrak{d} & \longmapsto & \mathfrak{d}' = \left(\{ A^1, \ldots,
      A^n \}, d_{ij}:A^i \rightarrow A^j \right) \nonumber \\
    \mathfrak{f} & \longmapsto & \mathfrak{f}' = \exists A^1 \ldots
    \exists A^n \left[ \bigvee_{i=1}^n A^i \right] \label{eq:decomp}
  \end{eqnarray}
  where $n = \#\{edg(A)\}$, the number of edges of $A$.  So $A^i
  \subset A$, containing a single edge of digraph $A$.
\end{decompDef}

In words: Demanding a partial morphism is equivalent to asking for the
existence of a total morphism of some of its edges, i.e. each $A^i$
contains one and only one of the edges of $A$. It does not seem to be
relevant whether $A^i$ includes all nodes of $A$ or just the source
and target nodes. Notice that decomposition is not affected by the
host graph.

\begin{figure}[htbp]
  \centering
  \includegraphics[scale = 0.73]{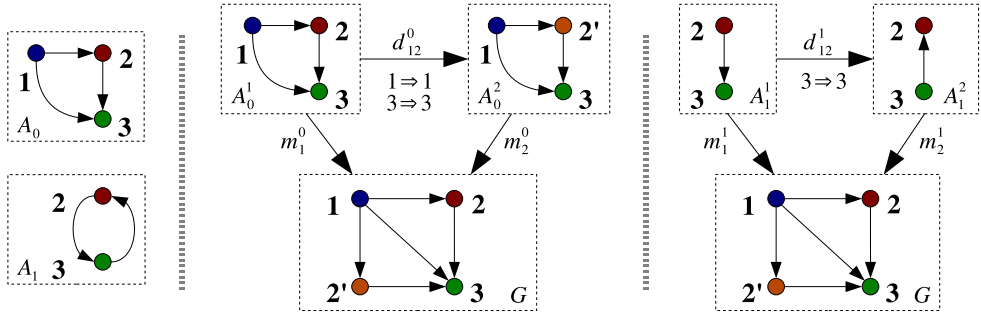}
  \caption{Closure and Decomposition}
  \label{fig:closureDecompEx}
\end{figure}

\noindent\textbf{Example}.$\square$We will consider conditions represented in
Fig.~\ref{fig:closureDecompEx}, $A_0$ for closure and $A_1$ for
decomposition, to illustrate Defs.~\ref{def:closureDef} (again)
and~\ref{def:decompDef}.

Recall that the formula associated to closure is $\mathfrak{f} =
\forall A [A]$.  Closure applied to $A_0$ outputs two digraphs,
$A^1_0$ and $A^2_0$, and a morphism $d^0_{12}$ that identifies nodes
$1$ and $3$.  Any further match of $A_0$ in $G$ would imply an
isomorphism.  Equation~\eqref{eq:closure} for $A_0$ is
\begin{equation}\label{ex:closureDecomposition}
  \mathfrak{f}' = \exists A^1_0 \exists A^2_0 \left[ A_0^1 \wedge A_0^2\right]
\end{equation}
with associated diagram
\begin{equation}
  \mathfrak{d}' = \left(\{ A_0^1, A_0^2 \}, d^0_{12}:A_0^1 \rightarrow A_0^2 \right)
\end{equation}
depicted to the center of Fig.~\ref{fig:closureDecompEx}.  Note that
the maximum number of non-empty partial morphisms not being
isomorphisms is 2. 

Formula associated to $\mathfrak{D}$ is $\mathfrak{f} = \exists A
[Q(A,G)]$.  Decomposition can be found to the right of the same
figure, in this case with associated formulas:
\begin{eqnarray}
  \mathfrak{d}' & = & \left(\{ A_1^1, A_1^2 \}, d^1_{12}:A_1^1
    \rightarrow A_1^2 \right) \nonumber \\
  \mathfrak{f}' & = & \exists A_1^1 \exists A_1^2 \left[ A_1^1 \vee
    A_1^2 \right].
\end{eqnarray}

The number of edges that make up the graph is 2, which is the number
of different graphs $A_1^i$.\proofend

Now we get to the main result of this section.  The following theorem
states that it is possible to reduce any formula in a graph constraint
(or application condition) to one using existential quantifiers and
total morphisms. Recall that, in Matrix Graph Grammars, matches are
total morphisms.\footnote{In fact in any approach to graph
  transformation, to the best of our knowledge.}

\newtheorem{embeddingGC}[matrixproduct]{Theorem}
\begin{embeddingGC}\label{th:embeddingGC}
  Let $GC = \left( \mathfrak{d}, \mathfrak{f} \right)$ be a graph
  constraint such that $\mathfrak{f} = \mathfrak{f} \left( P, Q
  \right)$ is a ground function.  Then, $\mathfrak{f}$ can be
  transformed into a logically equivalent $\mathfrak{f'} =
  \mathfrak{f}'(P)$ with existential quantifiers only.
\end{embeddingGC}

\noindent \emph{Proof}\\*
$\square$Define the depth of a graph for a fixed node $n_0$ to be the
maximum over the shortest path (to avoid cycles) starting in any node
different from $n_0$ and ending in $n_0$.  The diagram $\mathfrak{d}$
is a graph\footnote{Where nodes are digraphs $A_i$ and edges are
  morphisms $d_{ij}$.} with a special node $G$.  We will use the
notation $depth\left( GC \right) = depth\left(\mathfrak{d}\right)$,
the depth of the diagram.

In order to prove the theorem we apply induction on the depth,
checking out every case.  There are sixteen possibilities for
$depth\left(\mathfrak{d}\right) = 1$ and a single element $A$,
summarized in Table~\ref{tab:possibilitiesSingleCase}.

\begin{table*}[hbtp]
  \centering
  \begin{tabular}{|rl|rl||rl|rl|}
    \hline
    \begin{Large}\phantom{I}\end{Large}(1) & $\exists A [ A
    ]$\begin{Large}\phantom{I}\end{Large} &
    \begin{Large}\phantom{I}\end{Large}(5) & $\not\!\forall A
    [\overline{A}]$\begin{Large}\phantom{I}\end{Large} &
    \begin{Large}\phantom{I}\end{Large} (9) & $\exists A
    [\overline{Q}(A)]$\begin{Large}\phantom{I}\end{Large} &
    \begin{Large}\phantom{I}\end{Large} (13) & $\not\!\forall A
    [Q(A)]$\begin{Large}\phantom{I}\end{Large}\\
    \hline
    \begin{Large}\phantom{I}\end{Large}(2) & $\exists A
    [\overline{A}]$ & (6) & $\not\!\forall A [A]$ & (10) & $\exists A
    [Q(A)]$ & (14) & $\not\!\forall A [\overline{Q}(A)]$\\
    \hline
    \begin{Large}\phantom{I}\end{Large}(3) & $\not{\exists} A
    [\overline{A}]$ & (7) & $\forall A [A]$ & (11) & $\not{\exists} A
    [Q(A)]$ & (15) & $\forall A [\overline{Q}(A)]$\\
    \hline
    \begin{Large}\phantom{I}\end{Large}(4) & $\not{\exists} A [ A ]$ &
    (8) & $\forall A [\overline{A}]$ & (12) & $\not{\exists} A
    [\overline{Q}(A)]$ & (16) & $\forall A [Q(A)]$\\
    \hline
  \end{tabular}
  \caption{All Possible Diagrams for a Single Element}
  \label{tab:possibilitiesSingleCase}
\end{table*}

Elements in the same row for each pair of columns are related using
equalities $\not{\exists} A[A] = \forall A[\overline{A}]$ and $\not
\!\!\forall A[A] = \exists A[\overline{A}]$, so it is possible to
reduce the study to cases (1)--(4) and (9)--(12).\footnote{Notice that
  $\not\! \forall$ should be read ``not for all\ldots'' and not
  ``there isn't any\ldots''.} Identities $\overline{Q}(A) = P(A,
\overline{G})$ and $Q(A) = \overline{P}(A, \overline{G})$ (see also
equation~\eqref{eq:relPQ}) reduce (9)--(12) to formulas (1)--(4):
\begin{eqnarray}
  \exists A [\overline{Q}(A)] & = & \exists A\left[P(A,
    \overline{G})\right] \nonumber \\
  \exists A [Q(A)] & = & \exists A\left[\overline{P}(A,
    \overline{G})\right] \nonumber \\
  \not{\exists} A [Q(A)] & = & \not{\exists} A\left[\overline{P}(A,
    \overline{G})\right] \nonumber \\
  \not{\exists} A [\overline{Q}(A)] & = & \not{\exists} A\left[P(A,
    \overline{G})\right]. \nonumber
\end{eqnarray}

What we mean with this is that it is enough to study the first four
cases, although it will be necessary to specify if $A$ must be found
in $G$ or in $\overline{G}$.  Finally, every case in the first column
can be reduced to (1):
\begin{itemize}
\item (1) is the definition of match in Sec.~\ref{sec:matchAndExtendedMatch}.
\item (2) can be transformed into total morphisms (case 1) using
  operator $\mathfrak{D}$:
  \begin{equation}\label{eq:second}
    \exists A\left[\overline{A}\right] =\exists A
    \left[Q(A,\overline{G})\right] = \exists A^1\ldots\exists A^n
    \left[\bigvee_{i=1}^n P\left(A^i, \overline{G}\right)\right].
  \end{equation}
\item (3) can be transformed into total morphisms (case 1) using
  operator $\mathfrak{C}$:
  \begin{equation}\label{eq:third}
    \not{\exists} A\left[\overline{A}\right] = \forall A [A] = \exists
    A^1\ldots\exists A^n \left[\bigwedge_{i=1}^n A^i\right].
  \end{equation}
  The conditions on $P_U$ are supposed to be satisfied and thus have
  not been included.
\item (4) combines (2) and (3), where operators $\mathfrak{C}$ and
  $\mathfrak{D}$ are applied in order $\mathfrak{D}\circ\mathfrak{C}$
  (see remark after the end of this proof). Again, conditions on $P_U$
  are supposed to be fulfilled and thus have been omitted:
  \begin{equation}\label{eq:fourth}
    \not{\exists} A[A] = \forall A \left[\overline{A}\right] = \exists
    A^{11}\ldots\exists A^{mn} \left[\bigwedge_{i=1}^m \bigvee_{j=1}^n
      P\left(A^{ij}, \overline{G}\right)\right].
  \end{equation}
\end{itemize}

If there is more than one element at depth 1, this same procedure can
be applied mechanically. Note that if depth is 1, graphs on the
diagram are unrelated (otherwise, depth $>$ 1). Well-definedness
guarantees independence with respect to the order in which elements
are selected.

For the induction step, when there is a universal quantifier $\forall
A$, according to eq.~\eqref{eq:closure}, elements of $A$ are
replicated as many times as potential instances of this graph can be
found in the host graph.  Suppose the connected graph is called
$B$. There are two possibilities: Either $B$ is existentially
quantified $\forall A \exists B$ or universally quantified $\forall A
\forall B$.

If $B$ is existentially quantified then it is replicated as many times
as $A$. There is no problem as morphisms $d_{ij}: B_i \rightarrow B_j$
can be isomorphisms.\footnote{If for example there are three instances
  of $A$ in the host graph but only one of $B$, then the three
  replicas of $B$ are matched to the same part of $G$.} Mind the
importance of the order: $\forall A \exists B \neq \exists B \forall
A$.

If $B$ is universally quantified, again it is replicated as many times
as $A$. Afterwards, $B$ itself needs be replicated due to its
universality. Note that the order in which these replications are
performed is not relevant, $\forall A \forall B = \forall B \forall
A$. The order in the general case is given by the formula
$\mathfrak{f}$.  More in detail, when closure is applied to $A$, we
iterate on all graphs $B_j$ in the diagram:
\begin{itemize}
\item If $B_j$ is existentially quantified after $A$ ($\forall A ...
  \exists B_j$) then it is replicated as many times as $A$.
  Appropriate morphisms are created between each $A^i$ and $B^i_j$ if
  a morphism $d:A \rightarrow B$ existed. The new morphisms identify
  elements in $A^i$ and $B^i_j$ according to $d$.  This allows finding
  different matches of $B_j$ for each $A^i$, some of which can be
  equal.\footnote{If for example there are three instances of $A$ in
    the host graph but only one of $B_j$, then the three replicas of
    $B_j$ are matched to the same part of $G$.}
\item If $B_j$ is existentially quantified before $A$ ($\exists B_j
  ... \forall A$) then it is not replicated, but just connected to
  each replica of $A$ if necessary. This ensures that a unique $B_j$
  has to be found for each $A^i$. Moreover, the replication of $A$ has
  to preserve the shape of the original diagram. That is, if there is
  a morphism $d:B \rightarrow A$, then each $d_i : B \rightarrow A^i$
  has to preserve the identifications of $d$ (this means that we take
  only those $A^i$ which preserve the structure of the diagram).
\item If $B_j$ is universally quantified (no matter if it is
  quantified before or after $A$), again it is replicated as many
  times as $A$. Afterwards, $B_j$ itself needs to be replicated
  due to its universality. The order in which these replications are
  performed is not relevant as $\forall A \forall B_j = \forall B_j
  \forall A$.\proofend
\end{itemize}

\noindent\textbf{Remark}.\label{remark:nonCommutatutivity}$\square$It is not
difficult to see that $\mathfrak{C}$ and $\mathfrak{D}$ commute, i.e.
$\mathfrak{C} \circ \mathfrak{D} = \mathfrak{D} \circ \mathfrak{C}$.
In fact in equation~\eqref{eq:fourth} it does not matter whether
$\mathfrak{D} \circ \mathfrak{C}$ or $\mathfrak{D} \circ \mathfrak{C}$
is considered.

Composition $\mathfrak{D} \circ \mathfrak{C}$ is a direct translation
of $\forall A [\overline{A}]$ which, in first instance, considers all
appearances of nodes in $A$ and then splits these occurrences into
separate digraphs.  This is the same as considering every pair of
single nodes connected in $A$ by one edge and take their closure, i.e.
$\mathfrak{C} \circ \mathfrak{D}$.\proofend

\begin{figure}[htbp]
  \centering
  \includegraphics[scale = 0.6]{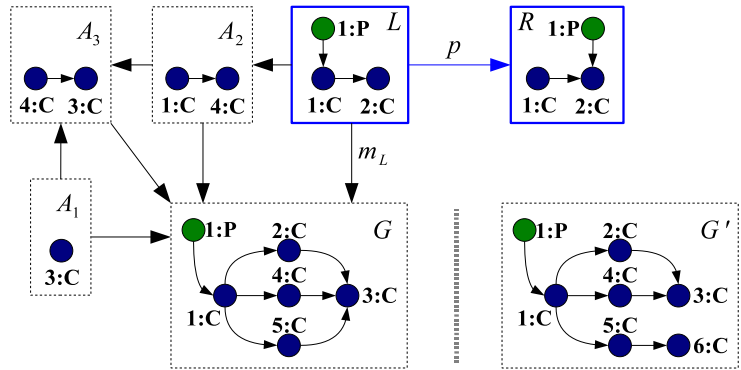}
  \caption{Application Condition Example}
  \label{fig:AC_example}
\end{figure}

\noindent\textbf{Examples}.$\square$Let be given a diagram like the
one that appears in Figure~\ref{fig:AC_example} with formula
$\mathfrak{f} = \exists A_1 \forall A_2 \exists A_3 \left[ A_2
  \Rightarrow \left( A_1 \wedge A_3 \right) \right]$.  Say \emph{C}
stands for conveyor.\footnote{Taken from the study case in
  App.~\ref{app:caseStudy}.}  If a conveyor is connected to three
conveyors, then they are eventually joint into a single
conveyor. Graph $G$ in the same figure satisfies the application
condition as elements $(2:C)$, $(4:C)$ and $(5:C)$ are connected to a
single node $(3:C)$.  Graph $G'$ does not satisfy the application
condition.\label{ex:exampleDiff2Repre} Note that:
\begin{equation}\label{eq:exSimplification}
  \mathfrak{f} = \exists A_1 \forall A_2 \exists A_3 \left[ A_2
    \Rightarrow \left( A_1 \wedge A_3 \right) \right] = \exists A_1
  \forall A_2 \exists A_3 \left[ \overline{A_2} \vee \left( A_1 \wedge
      A_3 \right) \right].
\end{equation}

Suppose that the second form of $\mathfrak{f}$
in~\eqref{eq:exSimplification} is used. Closure applies to $A_2$, so
it is copied three times with the additional property of mandatory
being identified in different parts of the host graph.  As $A_3$ is
connected to $A_2$ it is also replicated. $A_1$ has no common element
with $A_2$ so it needs not be replicated. Hence, a single $A_1$
appears when the closure operator is applied. Note however that there
is no difference if $A_1$ is also replicated because all different
copies can be identified in the same part of the host graph.

\begin{figure}[htbp]
  \centering
  \includegraphics[scale = 0.47]{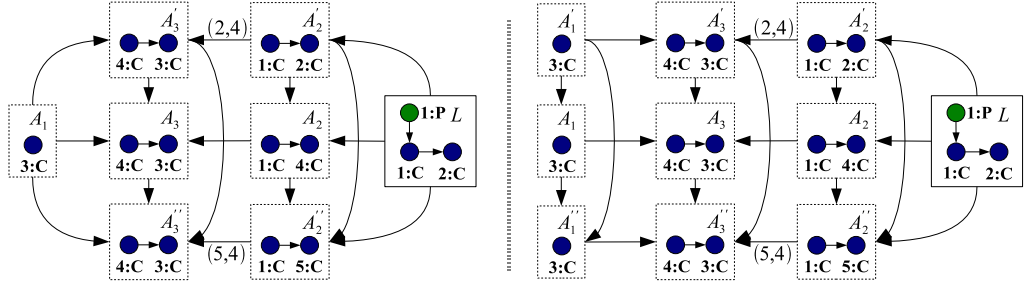}
  \caption{Closure Example}
  \label{fig:AC_closure_Example}
\end{figure}

The key point is that $A_2$ must be matched in different places of the
host graph (otherwise there should be some isomorphism) and the same
may apply to $A_3$ (as long as node $(4:C)$ in $A_3$ is different for
$A_3$, $A_3'$ and $A_3''$) but $A_1$, $A_1'$ and $A_1''$ can be
matched in the same place. Here there is no difference in asking for
three matches of $A_1$ or a single match, as long as they can be
matched in the same place. $A_1$, $A_1'$ and $A_1''$ are depicted to
the right of Fig.~\ref{fig:AC_closure_Example}.

\begin{figure}[htbp]
  \centering
  \includegraphics[scale = 0.6]{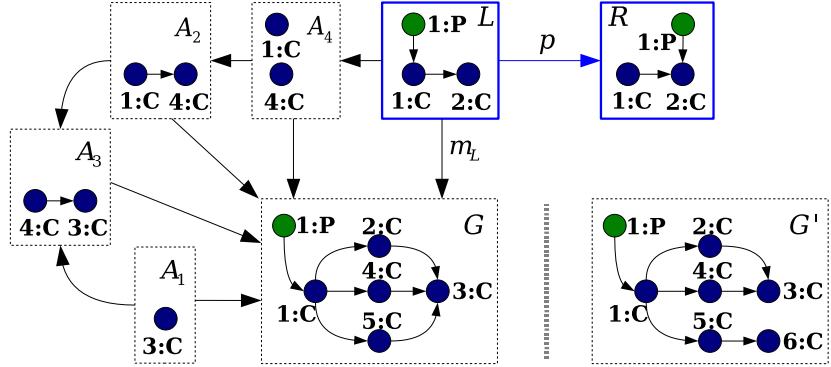}
  \caption{Application Condition Example Corrected}
  \label{fig:AC_example_modified}
\end{figure}

In fact, there is something wrong in our previous reasoning because
$\forall A_2$ demands all potential matches of $A_2$. This includes
the graph made up of nodes $(1:C)$ and $(3:C)$ and the edge joining
the first with the second. To obtain the behavior described in
previous paragraphs we need to add another graph $A_4$ that has only
nodes $(1:C)$ and $(4:C)$, modify the formula
\begin{equation}
  \label{eq:6}
  \mathfrak{f} = \exists A_1 \forall A_4 \exists A_2 \exists A_3
  \left[ \left( A_4 \wedge A_2 \right) \Rightarrow \left( A_1 \wedge
      A_3 \right) \right]
\end{equation}
and also the morphisms in the diagrams. It is all depicted in Fig.~\ref{fig:AC_example_modified}.\proofend

Theorem~\ref{th:embeddingGC} is of interest because derivations as
defined in Matrix Graph Grammars (the matching part) use only total
morphisms and existential quantifiers.  An application condition $AC =
\left( \mathfrak{d}_{AC} , \mathfrak{f}_{AC} \right)$ is a graph
constraint $GC = \left( \mathfrak{d}_{GC} , \mathfrak{f}_{GC} \right)$
with\footnote{Actually, it is not necessary to demand the existence of
  the nodes of $K$ because they are the same as those of $L$.}
\begin{equation}\label{eq:equivAC_GC}
  \mathfrak{f}_{AC} = \exists L \exists K \left[L \wedge P\left( K,
      \overline{G} \right) \wedge \mathfrak{f}_{GC} \right],
\end{equation}
so Theorem~\ref{th:embeddingGC} can be applied to application
conditions.

\newtheorem{embeddingAC}[matrixproduct]{Corollary}
\begin{embeddingAC}\label{cor:embedding}
  Any application condition $AC = \left( \mathfrak{d}, \mathfrak{f}
  \right)$ such that $\mathfrak{f} = \mathfrak{f} \left( P, Q \right)$
  is a ground function can be embedded into its corresponding direct
  derivation.
\end{embeddingAC}

This corollary asserts that any application condition can be expressed
in terms of Matrix Graph Grammars rules. So we have proved the
informal equations MGG + AC = MGG + GC = MGG. Examples illustrating
formulas~\eqref{eq:second},~\eqref{eq:third}~and~\eqref{eq:fourth} and
Corollary~\ref{cor:embedding} can be found in
Sec.~\ref{sec:functionalRepresentation}.

\section{Sequentialization of Application Conditions}
\label{sec:functionalRepresentation}

In this section, operators $\mathfrak{C}$ and $\mathfrak{D}$ are
translated into the functional notation of previous chapters (see
Sec.~\ref{sec:functionalAnalysis} for a quick introduction), inspired
by the Dirac or bra-ket notation, where productions can be written as
$R=\left\langle L, p\right\rangle$.  This notation is very convenient
for several reasons, for example, it splits the static part (initial
state, $L$) from the dynamics (element addition and deletion, $p$).
Besides, this will permit us to interpret application conditions as
sequences or sets of sequences to e.g. study their consistency through
applicability (Sec.~\ref{sec:consistencyAndCompatibility}).

Operators $\mathfrak{C}$ and $\mathfrak{D}$ will be formally
represented as $\widecheck T$ and $\widehat T$, respectively.  Recall
that $\widehat T$ has been used in the proof of
Prop.~\ref{prop:SeqIndProp1}.

Let $p:L \rightarrow R$ be a production with application condition
$AC=\left(\mathfrak{d}, \mathfrak{f}\right)$.  We will follow a case
by case study of the proof of Theorem~\ref{th:embeddingGC} to
structure this section.

The first case addressed in the proof of Theorem~\ref{th:embeddingGC}
is the most simple: If the nodes of $A$ are found in $G$ then its
edges must also be matched.
\begin{equation}\label{eq:existence}
  \mathfrak{d} = \left(A, d:L \rightarrow A \right), \quad
  \mathfrak{f} = \exists A [A].
\end{equation}

Let $id_{\!A}$ be the production that does nothing on $A$ -- $id_A(A)
= A$ -- and also the operator that demands\footnote{Operator
  $id_{A}(p)$ could be thought of as a ``production'' that in a single
  step deletes and adds the elements of $A$.} the existence of
$A$. The set of identities
\begin{equation}\label{eq:firstSetIds}
  \left\langle L \vee A, p \right\rangle = \left\langle L, id_{\!A}(p)
  \right\rangle = \left \langle L, p \circ id_{\!A} \right \rangle
\end{equation}
proves that
\begin{equation}\label{eq:firstAdj}
  id^*_{\!A}\!\left( L \right) = L \vee A,
\end{equation}
which is the adjoint operator of $id_{\!A}$. Here, \textbf{or} is
carried out according to identifications specified by $d$. Production
$id_{\!A}$ can be seen as an operator (adjoints are defined only for
operators).  As a matter of fact, it is easy to prove that any
production is in particular an operator.\footnote{Just define its
  action.}

So if $AC$ asks for the existence of a graph like in
eq.~\eqref{eq:existence}, it is possible to enlarge the production $p
\mapsto p \circ id_{\!A}$.  The marking operator $T_\mu$
(Sec.~\ref{sec:marking}) enables us to use concatenation instead of
composition as in equation~\eqref{eq:firstSetIds}:
\begin{equation}\label{eq:matchTransPre}
  \left\langle L \vee A, p \right\rangle = p ; id_{\!A},
\end{equation}
to be understood in the sense of applicability. The following lemma
has just been proved:

\newtheorem{firstCase}[matrixproduct]{Lemma}
\begin{firstCase}[Match]\label{lemma:firstCase}
  \index{functional representation!match}Let $p:L\rightarrow R$ be a
  production together with an application condition as in
  eq.~\eqref{eq:existence}.  Its applicability is equivalent to the 
  applicability of the sequence $p;id_{A}$, as in
  equation~\eqref{eq:matchTransPre}.
\end{firstCase}

\begin{figure}[htbp]
  \centering
  \includegraphics[scale = 0.65]{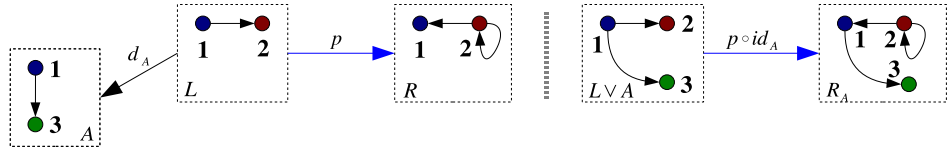}
  \caption{Production Transformation According to Lemma~\ref{lemma:firstCase}}
  \label{fig:matchTransformation}
\end{figure}

\noindent\textbf{Examples}.$\square$To the left of Fig.~\ref{fig:matchTransformation} a production and the diagram of its weak
application condition is depicted. Let its formula be $\exists A [A]$.
To the right, its transformation according to~\eqref{eq:matchTransPre}
is represented, but using composition instead of
concatenation.

\begin{figure}[htbp]
  \centering
  \includegraphics[scale = 0.37]{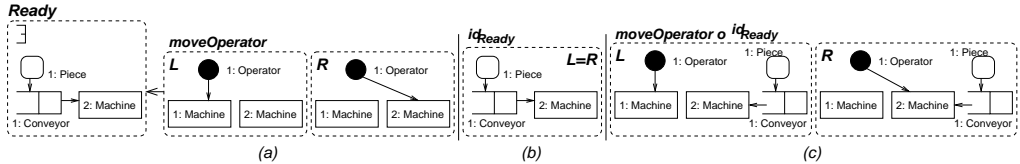}
  \caption{Transforming $\exists Ready[Ready]$ into a
    Sequence.}\label{fig:example_Exists}
\end{figure}

The AC of rule $moveOperator$ in Fig.~\ref{fig:example_Exists} (a) has
associated formula $\exists Ready [Ready]$ (i.e. the operator may move
to a machine with an incoming piece).  Using previous construction, we
obtain that the rule is equivalent to sequence $moveOperator^{\flat};
id_{Ready}$, where $moveOperator^{\flat}$ is the original rule without
the AC. Rule $id_{Ready}$ is shown in Fig.~\ref{fig:example_Exists}
(b).  Alternatively, we could use composition to obtain
$moveOperator^{\flat} \circ id_{Ready}$ as shown in
Fig.~\ref{fig:example_Exists} (c). \proofend

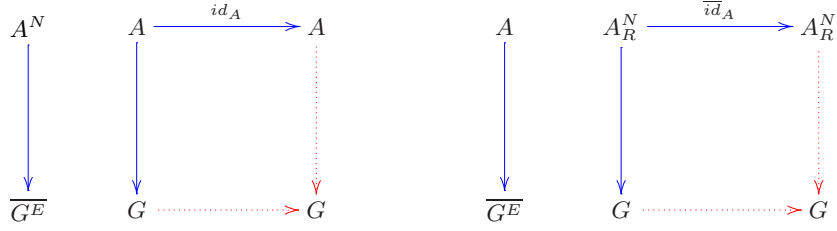
\begin{figure}[htb]
  \centering \makebox{ \xymatrix{ A^N \ar@[blue][dd] & A
      \ar@[blue][dd] \ar@[blue][rr]^{id_A} && A \ar@{.>}@[red][dd] &&
      A \ar@[blue][dd] & A^N_R \ar@[blue][dd]
      \ar@[blue][rr]^{\overline{id}_A} && A^N_R \ar@{.>}@[red][dd] \\
      \\
      \overline{G^E} & G \ar@{.>}@[red][rr] && G && \overline{G^E} & G
      \ar@{.>}@[red][rr] && G } }
  \caption{Identity $id_A$ and Conjugate $\overline{id}_A$ for Edges}
  \label{fig:idAndConj}
\end{figure}

\index{identity conjugate}We will introduce a kind of conjugate of
production $id_{A}$, to be written $\overline{id}_{A}$.  To the left
of Fig.~\ref{fig:idAndConj} there is a representation of $id_A$.  It
simply preserves (uses but does not delete) all elements of $A$, which
is equivalent to demand their existence. To the right we have its
conjugate, $\overline{id}_A$, which asks for nothing to the host graph
except the existence of $A$ in the complement of $G$.

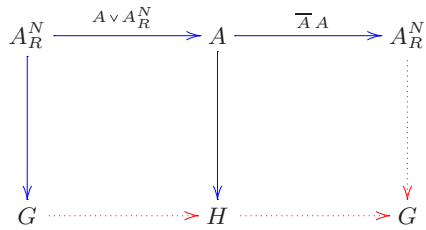
\begin{figure}[htb]
  \centering \makebox{ \xymatrix{ A^N_R \ar@[blue][dd]
      \ar@[blue][rr]^{A \vee A^N_R} && A \ar@[blue][dd]
      \ar@[blue][rr]^{\overline{A} \, A} && A^N_R
      \ar@{.>}@[red][dd] \\ \\
      G \ar@{.>}@[red][rr] && H \ar@{.>}@[red][rr] && G } }
  \caption{$\overline{id}_A$ as Sequence for Edges}
  \label{fig:idAsSeqForEdges}
\end{figure}

If instead of introducing $\overline{id}_A$ directly, a definition on
the basis of already known concepts is preferred we may proceed as
follows.  Recall that $K = r \vee \overline{e} \, \overline{D}$, so
our only chance to define $\overline{id}_A$ is to act on the elements
that some production adds.  Let
\begin{equation}
  p^e;p^r
\end{equation}
be a sequence such that the first production $\left( p^r \right)$ adds
elements whose presence is to be avoided and the second $\left( p^e
\right)$ deletes them (see Fig.~\ref{fig:idAsSeqForEdges}).  The
overall effect is the identity (no effect) but the sequence can be
applied if and only if elements of $A$ are in $\overline{G^E}$.

Note that a similar construction does not work for nodes because if a
node is already present in the host graph, a new one can be added
without any problem (adding and deleting a node does not guarantee
that the node is not in the host graph).

The way to proceed is to care only about nodes that are present in the
host graph as the others, together with their edges, will be present
in the completion of the complement of $G$. This is represented by
$A^N_R$, where $R$ stands for \emph{restriction}. Restriction and
completion are in some sense complementary operations.

Our analysis continues with the second case in the proof of Theorem~\ref{th:embeddingGC}, which states that some edges of $A$ can not be
found in $G$ for some identification of nodes in $G$, i.e.
$\not\!\forall A \left[ A \right] = \exists A \left[ \overline{A}
\right]$. This corresponds to operator $\widehat T_{\!A}$
(decomposition), defined by:
\begin{equation}
  \widehat T_{\!A}\left(p \right) = \left\{p_1, \ldots, p_n \right\}.
\end{equation}
Here, $p_i = p \circ \overline{id}_{\!A^i}$ with $A^i$ a graph
consisting of one edge of $A$ (together with its source and target
nodes) and $n = \#\{edg(A)\}$, the number of edges of $A$.
Equivalently, the formula is transformed into:
\begin{equation}\label{eq:nonExistence}
  \mathfrak{f} = \exists A [\overline{A}] \longmapsto \mathfrak{f}' =
  \exists \widehat{A^1} \ldots \exists \widehat{A^n} \left[
    \bigvee^n_{i=1} P \left( \widehat{A^i}, \overline{G}
    \right)\right],
\end{equation}
i.e. the matrix of edges that must not appear in order to apply the
production is enlarged $K_i = K \vee A^i$ (being $K_i$ the nihilation
matrix of $p_i$).

If composition is chosen, the grammar is modified by removing rule $p$
and adding the set of productions $\left\{p_1, \ldots, p_n \right\}$.
If the production is part of the sequence $q_2;p;q_1$ then we are
allowing variability on production $p$ as it can be substituted by any
$p_i$, $i \in \{1, \ldots, n\}$, i.e. $q_2;p;q_1 \longmapsto
q_2;p_i;q_1$.

A similar reasoning applies if we use concatenation instead of
composition but it is not necessary to eliminate production $p$ from
the grammar: $q_2;p;q_1 \mapsto q_2;p;\overline{id}_{A^i};q_1$.
Production $p$ and sequence $\overline{id}_{A^i}$ are related through
marking.

\newtheorem{secondCase}[matrixproduct]{Lemma}
\begin{secondCase}[Decomposition]\label{lemma:secondCase}
  \index{functional representation!decomposition}With notation as
  above, let $p:L\rightarrow R$ be a production together with an
  application condition as in eq.~\eqref{eq:nonExistence}.  Its
  applicability is equivalent to the applicability of any of the
  sequences
  \begin{equation}\label{eq:decompTransPre}
    s_i = p;\overline{id}_{\widehat{A^i}}
  \end{equation}
  where $\widehat{A^i}$ is defined as in
  equations~\eqref{eq:second}~or~\eqref{eq:nonExistence}.
\end{secondCase}

Before moving on to the third case in the proof of
Theorem~\ref{th:embeddingGC}, previous results will be clarified with
a simple example with similar conditions to those of
Fig.~\ref{fig:closureDecompEx}.

\begin{figure}[htbp]
  \centering
  \includegraphics[scale = 0.7]{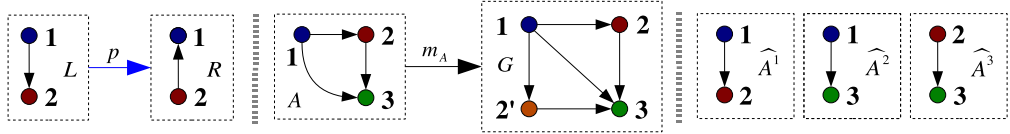}
  \caption{Decomposition Operator}
  \label{fig:closureDecompThEx}
\end{figure}

\noindent\textbf{Examples}.$\square$Consider production $p$ to the left of
Fig.~\ref{fig:closureDecompThEx} and application condition $A$ to the 
center of the same figure.  If the associated formula for $A$ is
$\mathfrak{f} = \exists A \left[\overline{A}\right]$ then three
sequences are derived ($p_i, i \in \{1,2,3\}$) with $p_i = p;
\overline{id}_{\widehat{A^i}}$, being $\widehat{A^i}$ those depicted
to the right of Fig.~\ref{fig:closureDecompThEx}.

\begin{figure}[htbp]
  \centering
  \includegraphics[scale = 0.47]{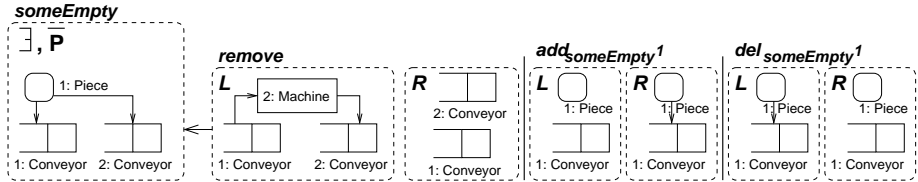}
  \caption{Transforming $\exists someEmpty[\overline {someEmpty}]$
    into a Sequence.}\label{fig:example_decomposition}
\end{figure}

The application condition of rule $remove$ in
Fig.~\ref{fig:example_decomposition} has as associated formula
$\exists someEmpty [\overline {someEmpty}]$. The formula states that
the machine can be removed if there is one piece that is not connected
to the input or output conveyor (as we must not find a total morphism
from $someEmpty$ to $G$).  Applying Lemma~\ref{lemma:secondCase}, rule
$remove$ is applicable if some of the sequences in the set $\{
remove^{\flat}; del_{someEmpty^i};$ $add_{someEmpty^i} \}_{i=\{1,
  2\}}$ is applicable, where productions $add_{someEmpty^2}$ and
$del_{someEmpty^2}$ are like the rules in the figure, but considering
conveyor $2$ instead. Thus $\overline {id}_{someEmpty^i}=
del_{someEmpty^i} \circ add_{someEmpty^i}$ \proofend

The third case in the proof of Theorem~\ref{th:embeddingGC} demands
that for any identification of nodes in the host graph every edge must
also be found.  Recall that $\not{\exists} A [ \overline{A} ] =
\forall A\left[A\right]$ which is associated to operator
$\widecheck{T}_{\!A}$ (closure). We will assume that all instances are
matched in their corresponding parts, so the $P_U$ part of equation~\eqref{eq:closure} is always fulfilled (is always true).\footnote{When
  dealing with morphisms $P_U$ was used. For operators, the marking
  operator $T_\mu$ acting on the host graph and on $A_i$ suffices.
  This remark applies to the rest of the chapter.} Hence,
\begin{equation}\label{eq:forall}
  \mathfrak{f} = \not{\exists} A [\overline{A}] \longmapsto \exists
  \widecheck{A^1} \ldots \exists \widecheck{A^n} \left[
    \bigwedge^n_{i=1} \widecheck{A^i} \right].
\end{equation}

This means that more edges must be present in order to apply the
production, $L \longmapsto \bigvee_{i=1}^n \left( L \vee A^i \right)$.
By a similar reasoning to that of the derivation of
eq.~\eqref{eq:firstAdj}:
\begin{equation}
  \left\langle \bigvee_{i=1}^n \left(\widecheck{A^i} \vee L \right), p
  \right\rangle = \left\langle L, \widecheck{T}_{\!A}\!\left(p\right)
  \right\rangle = \left \langle L, \left( id_{\widecheck{A^1}} \circ
      \ldots \circ id_{\widecheck{A^n}} \right) (p) \right \rangle =
  \left \langle L, p \circ id_{\widecheck{A}} \right \rangle,
\end{equation}
-- where $id_{\widecheck{A}} = id_{\widecheck{A^1}} \circ \ldots \circ
id_{\widecheck{A^n}}$ -- the adjoint operator can be calculated:
\begin{equation}
  \widecheck{T}^*_{A} \left( L \right) = L \vee \left( \bigvee_{i=1}^n
    \widecheck{A^i} \right).
\end{equation}

As commented above, the marking operator $T_\mu$ allows us to
substitute composition with concatenation:
\begin{equation}\label{eq:closureTransPre}
  \left\langle \bigvee_{i=1}^n \left(\widecheck{A^i} \vee L \right), p
  \right\rangle = p ; id_{\widecheck{A^1}} ; \ldots ;
  id_{\widecheck{A^n}} = p ; id_{\widecheck{A}}
\end{equation}
to be understood in the sense of applicability. We have proved the
following lemma:

\newtheorem{thirdCase}[matrixproduct]{Lemma}
\begin{thirdCase}[Closure]\label{lemma:thirdCase}
  \index{functional representation!closure}With notation as above, let
  $p:L\rightarrow R$ be a production together with an application
  condition as in eq.~\eqref{eq:forall}.  Its applicability is
  equivalent to the applicability of the sequence $p;id_{\widecheck
    A}$.
\end{thirdCase}

\begin{figure}[htbp]
  \centering
  \includegraphics[scale = 0.7]{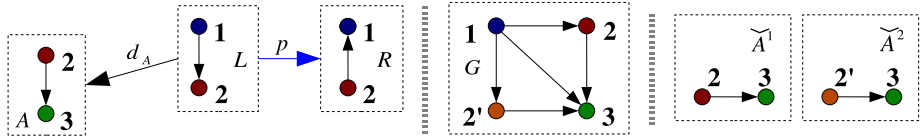}
  \caption{Closure Operator}
  \label{fig:closureTransformation}
\end{figure}

\noindent\textbf{Example}.$\square$Consider production $p$ to the left of
Fig.~\ref{fig:closureTransformation} and application condition $A$ to
the center of the same figure. If the associated formula for $A$ is
$\mathfrak{f} = \forall A \left[ A \right]$ then two sequences are
derived ($p_i, i \in \{1,2\}$) with $p_i = p;
\overline{id}_{\widecheck{A^i}}$, being $\widecheck{A^i}$ those
depicted to the right of Fig.~\ref{fig:closureTransformation}.\proofend

The fourth case is equivalent to that known in the literature as
\emph{negative application condition}, NAC, which is a mixture of
cases (2) and (3), in which the order of composition does not matter
due to the fact that $\widecheck{T}$ and $\widehat{T}$
commute.\footnote{See remark on
  p.~\pageref{remark:nonCommutatutivity}.}  It says that there does 
not exist an identification of nodes of $A$ for which all edges in $A$
can also be found, $\not{\exists}A[A]$, i.e. for every identification
of nodes there is at least one edge in $\overline{G}$.  If we define
\begin{equation}
  \widetilde{T}_{A}\! \left(p\right) = \left(\widehat{T}_{A} \circ
    \widecheck{T}_{A} \right)\! (p) = \left(\widecheck{T}_{A} \circ
    \widehat{T}_{A} \right)\! (p),
\end{equation}
then
\begin{equation}\label{eq:nacsDecomp}
  \mathfrak{f} = \forall A [\overline{A}] \longmapsto \exists
  \widetilde{A^{11}} \ldots \exists \widetilde{A^{mn}} \left[
    \bigwedge_{i=1}^m\bigvee_{\!j=1}^n \widetilde{A^{ij}} \right].
\end{equation}

In more detail, if we first apply closure to $A$ then we obtain a
sequence of $m+1$ productions, $p \longmapsto p ; id_{\widecheck{A^1}}
; \ldots ; id_{\widecheck{A^m}}$, assuming $m$ different matches of
$A$ in the host graph $G$.  Right afterwards, decomposition splits
every $\widecheck{A^i}$ into its components (in this case there are
$n$ edges in $A$).  So every match of $A$ in $G$ is transformed to
look for at least one missing edge, $id_{\widecheck{A^1}} \longmapsto
\overline{id}_{\widetilde{A^{11}}} \vee \ldots \vee
\overline{id}_{\widetilde{A^{1n}}}$.

Operator $\widetilde{T}_A$ acting on a production $p$ with a weak
precondition $A$ results in a set of productions
\begin{equation}
  \widetilde{T}_{\!A}\left(p \right) = \left\{p_1, \ldots, p_r
  \right\} \nonumber
\end{equation}
where $r = m^n$.  Each $p_k$ is the composition of $m+1$ productions,
defined as $p_k = p \circ \overline{id}_{\widetilde{A^{u_0v_0}}} \circ
\ldots \circ \overline{id}_{\widetilde{A^{u_mv_m}}}$.  Marking
operator $T_\mu$ of Sec.~\ref{sec:marking} permits concatenation
instead of composition:
\begin{equation}\label{eq:nacsImage}
  \widetilde{T}_A (p) = \left\{ p_k \; \vert \; p_k = p;
    \overline{id}_{\widetilde{A^{u_0v_0}}} ; \ldots ;
    \overline{id}_{\widetilde{A^{u_mv_m}}}\right\}_{k \in \{1, \ldots,
    m^n\}}.
\end{equation}


\newtheorem{nacs}[matrixproduct]{Lemma}
\begin{nacs}[Negative Application Conditions]\label{lemma:nacs}
  \index{functional representation!negative application
    condition}Keeping notation as above, let $p:L\rightarrow R$ be a
  production together with an application condition as in
  eq.~\eqref{eq:nacsDecomp}, then its applicability is equivalent to
  the applicability of some of the sequences derived from
  equation~\eqref{eq:nacsImage}.
\end{nacs}

\noindent\textbf{Example}.$\square$If there are two matches and $A$ has three
edges, $i=3$ and $j=2$, then equation~\eqref{eq:nacsDecomp} becomes:
\begin{eqnarray}
  \bigwedge_{i=1}^3 \bigvee_{j=1}^2 \widetilde{A^{ij}} & = &
  \left(\widetilde{A^{11}} \vee
    \widetilde{A^{12}}\right)\left(\widetilde{A^{21}} \vee
    \widetilde{A^{22}}\right)\left(\widetilde{A^{31}} \vee
    \widetilde{A^{32}}\right) \nonumber \\
  & = & \widetilde{A^{11}}\widetilde{A^{21}}\widetilde{A^{31}} \vee
  \widetilde{A^{11}}\widetilde{A^{21}}\widetilde{A^{32}} \vee \ldots
  \vee \widetilde{A^{12}}\widetilde{A^{22}}\widetilde{A^{31}} \vee
  \widetilde{A^{12}}\widetilde{A^{22}}\widetilde{A^{32}}. \nonumber
\end{eqnarray}
For example, the first monomial
$\widetilde{A^{11}}\widetilde{A^{21}}\widetilde{A^{31}}$ is the
sequence
\begin{equation}
  p;\overline{id}_{\widetilde{A^{11}}};\overline{id}_{\widetilde{A^{21}}};
  \overline{id}_{\widetilde{A^{31}}}
  \nonumber
\end{equation} \proofend


Summarizing in a sort of rule of thumb, there are two operations --
\textbf{and} and \textbf{or} -- that might be combined using the rules
of monadic second order logics. These operations are transformed in
the following way:
\begin{itemize}
\item Operation \textbf{and} in the $\mathfrak{f}$ of an application
  condition becomes an \textbf{or} when calculating an equivalent
  production.
\item Operation \textbf{or} enlarges the grammar with new productions,
  removing the original rule if composition instead of concatenation
  is chosen.
\end{itemize}

\begin{figure}[htb]
  \centering \makebox{ \xymatrix{ A_0 \ar@/_/[rrrrdd]^{m_{A_0}} && A_1
      \ar[ll]_{d_{10}} \ar@/_/[ddrr]^{m_{A_1}} && L
      \ar@/_17pt/[llll]_{d_{L0}} \ar[ll]_{d_{L1}}
      \ar@{.>}@[blue][rr]^p \ar@{.>}@[blue][dd]^{m_L}
      && R \ar@{.>}@[red][dd] \\ \\
      &&&& G \ar@{.>}@[red][rr] && H } }
  \caption{Example of Diagram with Two Graphs}
  \label{fig:exDiagramSec}
\end{figure}
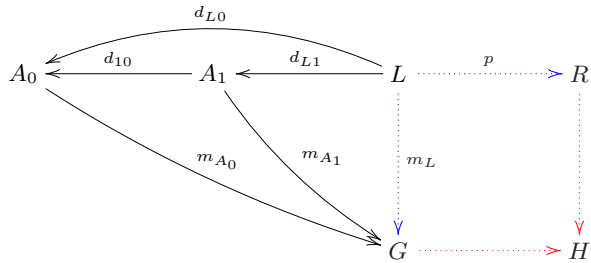

\begin{figure}[htbp]
  \centering
  \includegraphics[scale =
  0.52]{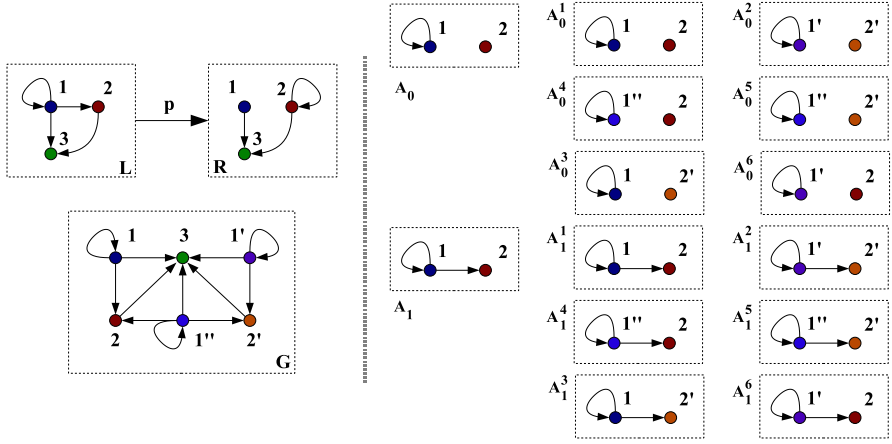}
  \caption{Precondition and Postcondition}
  \label{fig:exSectionFiveConstraints}
\end{figure}

\label{ex:twoDiagrams}\noindent\textbf{Example}.$\square$Let
$AC=\left(\mathfrak{d},\mathfrak{f}\right)$ be a graph constraint with
diagram $\mathfrak{d}$ depicted in Fig.~\ref{fig:exDiagramSec} (graphs
shown in Fig.~\ref{fig:exSectionFiveConstraints}) and associated
formula $\mathfrak{f}=\exists L \forall A_0 \exists A_1 \left[ L
  \left( A_0 \Rightarrow A_1 \right)\right]$,
$d_{L0}\left(\{1\}\right)=\{1\}$.  Let morphisms be defined as
follows: $d_{L1}\left(\{1\}\right)=\{1\}$,
$d_{10}\left(\{1\}\right)=\{1\}$ and $d_{10}\left(\{2\}\right)=\{2\}$.

The interpretation of $\mathfrak{f}$ is that $L$ must be found in $G$
(for simplicity $K$ is omitted) and whenever nodes of $A_0$ are found
then there must exist a match for the nodes of $A_1$ such that there
is an edge joining both nodes.

Note that matching of nodes of $A_0$ and $A_1$ must coincide (this is
what $d_{10}$ is for) and that node $1$ has to be the same as that
matched by $m_L$ for $L$ in $G$ (morphisms $d_{L0}$ and $d_{L1}$).

Application of operator $\widecheck{T}$ for the universal quantifier
yields six digraphs for $A_0$ and another six for $A_1$, represented
in Fig.~\ref{fig:exSectionFiveConstraints}.  Note that in this case we
have $\overline{A_0^i} = P^E\left(A_0^i, \overline{G}\right)$ because
$A_0^i$ has only one edge.  Suppose that
$m_L(\{1,2,3\})=\{1'',2',3\}$, then $\mathfrak{f}$ becomes
\begin{equation}\label{eq:exUniversalAC}
  \mathfrak{f}_1 = \exists L \exists A_0^4 \exists A_0^5 \exists A_1^4
  \exists A_1^5 \left[L \left(\overline{A_0^4} \vee A_1^4
    \right)\left(\overline{A_0^5} \vee A_1^5 \right)\right]. 
\end{equation}

Different matches and relations among components of the application
condition derive different formulas $\mathfrak{f}$.  For example, we
could fix only node $1$ in $d_{10}$, allowing node $2$ to be
differently matched in $G$.  Notice that neither $A^3_1$ nor $A^6_1$
exist in $G$ so the condition would not be fulfilled for $A^3_0$ or
$A^6_0$ because terms $\overline{A^3_0} \vee A^6_0$ and
$\overline{A^3_1} \vee A^6_1$ would be false ($A^3_0$ and $A^6_0$ are
in $G$ for any identification of nodes).\proofend

Previous lemmas prove that weak preconditions can be reduced to
studying sequences of productions.  If instead of weak preconditions
we have preconditions then we should study derivations (or sets of
derivations) instead of sequences.

\newtheorem{reductionPre}[matrixproduct]{Theorem}
\begin{reductionPre}\label{th:reductionPre}
  Any weak precondition can be reduced to the study of the
  corresponding set of sequences.
\end{reductionPre}

\noindent \emph{Proof} \\*
$\square$This result is the sequential version of Theorem~\ref{th:embeddingGC}.  The four cases of its proof correspond to
Lemmas~\ref{lemma:firstCase} through~\ref{lemma:nacs}.\proofend

\noindent\textbf{Example}.$\square$Continuing example on p.~\pageref{ex:twoDiagrams}, equation~\eqref{eq:exUniversalAC} put in
normal disjunctive form reads
\begin{equation}\label{eq:exUniversalAC_2}
  \mathfrak{f}_1 = \exists L \exists A_0^4 \exists A_0^5 \exists A_1^4
  \exists A_1^5 \left[L\overline{A_0^4}\,\overline{A_0^5} \vee
    L\overline{A_0^4}A_1^5  \vee L A_1^4 \overline{A_0^5} \vee L A_1^4
    A_1^5 \right]
\end{equation}
which is equivalent to
\begin{equation}
  \mathfrak{f}_1 = \exists L \exists A_0^4 \exists A_0^5 \exists A_1^4
  \exists A_1^5 \left[L A_1^4 A_1^5 \right] \nonumber
\end{equation}
because $A^4_0$ and $A^5_0$ can be found in $G$.  This is the same as
applying the sequence $p;id_{A^4_1};id_{A^5_1}$ or
$p;id_{A^5_1};id_{A^4_1}$ (because $id_{A^4_1} \bot id_{A^5_1}$).

So the satisfaction of an $AC$, once match $m_L$ has been
fixed,\footnote{In this example. In general it is not necessary to fix
  the match in advance.} is equivalent to the applicability of the
sequence to which equation~\eqref{eq:exUniversalAC_2} gives
rise.\proofend

\section{Summary and Conclusions}
\label{sec:summaryAndConclusions6}

In this chapter, graph constraints and application conditions have been
introduced and studied in detail for the Matrix Graph Grammar
approach.  Our proposal considerably generalizes previous efforts in
other approaches such as SPO or DPO.

Generalization is not necessarily good in itself, but in our opinion
it is interesting in this case.  We have been able to ``reduce'' graph
constraints and application conditions one to each other (which will
be useful in Sec.~\ref{sec:fromSimpleDigraphsToMultidigraphs}).
Besides, the left hand side, right hand side and nihilation matrices
appear as particular cases of this more general framework, giving the
impression of being a very natural extension of the theory.  Also, it
is always possible to embed application conditions in Matrix Graph
Grammars direct derivations (Theorem~\ref{th:embeddingGC} and
Corollary~\ref{cor:embedding}).  We have managed to study
preconditions, postconditions and their weak counterparts,
independently to some extent of any match.

Other interesting points are that application conditions seem to be a
good way to synthesize closely related grammar rules.  Besides, they
allow us to partially act on the nihilation matrices $K$ and $Q$
(recall that the nihilation matrix was directly derived out of $L$,
$e$ and $r$).

Representing application conditions using the functional notation
introduced for productions and direct derivations allowed us to prove
a very useful fact: Any application condition is equivalent to some
sequence of productions (or a set of them).  See
Theorem~\ref{th:reductionPre} (and also Theorem~\ref{th:reductionPost}
in the next chapter). It is worth stressing the importance of the
relationship between application conditions and sequences of
productions and will be used extensively in
Chap.~\ref{ch:transformationOfRestrictions}.

Chapter~\ref{ch:transformationOfRestrictions} continues our study of
restrictions with concepts such as consistency, the transformation of
preconditions into postconditions and vice versa and a
practical--theoretical application: the extension of Matrix Graph
Grammars to cope with multidigraphs with no major modification of the
theory.

Chapter~\ref{ch:reachability} addresses one fundamental topic in
grammars: Reachability.  This topic has been stated as
problem~\ref{prob:reachability} and is widely addressed in the
literature, specially in the theory of Petri nets.
\chapter{Transformation of Restrictions}
\label{ch:transformationOfRestrictions}

In this chapter we continue the study of graph constraints and
application conditions -- restrictions -- started in
Chap.~\ref{ch:restrictionsOnRules}.

Section~\ref{sec:consistencyAndCompatibility} introduces consistency,
compatibility and coherence of application
conditions. Section~\ref{sec:movingConditions} tackles the
transformation of application conditions imposed to a rule's LHS into
one equivalent application condition but on the rule's RHS. The
converse, more natural from a practical point of view, is also
addressed. Besides, we shall outline how to move application
conditions from one production to another inside the same sequence. As
an application of restrictions to Matrix Graph Grammars,
Sec.~\ref{sec:fromSimpleDigraphsToMultidigraphs} shows how to make MGG
deal with multidigraphs instead of just simple digraphs without major
modifications to the theory. Section~\ref{sec:summaryAndConclusions7}
closes the chapter with a summary and some more comments.

\section{Consistency and Compatibility}
\label{sec:consistencyAndCompatibility}

We shall start by defining some (desirable) properties of application
conditions. As pointed out above, any application condition is
equivalent to some sequence or set of sequences so we will be able to
characterize these properties using the theory developed so far.

\newtheorem{consistentAC}[matrixproduct]{Definition}
\begin{consistentAC}[Consistency, Coherence,
  Compatibility]\label{def:consistentAC}
  Let $AC = \left( \mathfrak{d}, \mathfrak{f} \right)$ be a weak
  application condition on the grammar rule $p:L \rightarrow R$. We
  say that the $AC$ is:
  \begin{itemize}
  \item \index{application condition!coherent}\emph{coherent}
    if it is not a fallacy (i.e., false in all scenarios).
  \item \index{application condition!compatible} \emph{compatible} if,
    together with the rule's actions, produces a simple
    digraph.
  \item \index{application condition!consistent}\emph{consistent} if
    $\exists G$ host graph such that $G \models AC$ to which the
    production is applicable.
  \end{itemize}
\end{consistentAC}

The definitions for application conditions instead of their weak
counterparts are almost the same, except that consistency does not ask
for the existence of some host graph but takes into account the one
already considered.

Coherence of ACs studies whether there are contradictions in it
preventing its application in any scenario. Typically, coherence is
not satisfied if the condition simultaneously asks for the existence
and non-existence of some element. Compatibility of ACs checks whether
there are conflicts between the AC and the rule's actions. Here we
have to check for example that if a graph of the $AC$ demands the
existence of some edge, then it can not be incident to a node that is
deleted by production $p$. Consistency is a kind of well-formedness of
the AC when a production is taken into account. Next, we show some
examples of non-consistent, non-compatible and non-coherent ACs.

\begin{figure}[htbp]
  \centering
  \includegraphics[scale = 0.5]{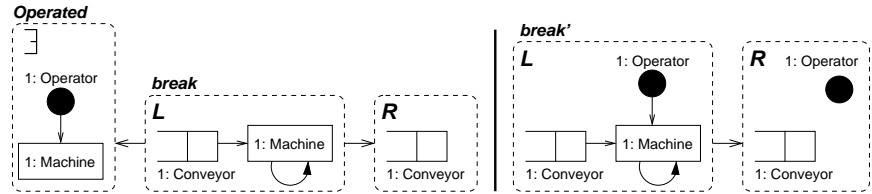}
  \caption{Non-Compatible Application
    Condition}\label{fig:example_non_compat}
\end{figure}

\noindent\textbf{Examples}.$\square$Non-compatibility can be avoided at times
just rephrasing the AC and the rule.  Consider the example to the left
of Fig.~\ref{fig:example_non_compat}. The rule models the breakdown of
a machine by deleting it. The AC states that the machine can be broken
if it is being operated. The AC has associated diagram $\mathfrak{d} =
\{Operated\}$ and formula $\mathfrak{f} = \exists Operated
[Operated]$. As the production deletes the machine and the AC asks for
the existence of an edge connecting the operator with the machine, it
is for sure that if the rule is applied we will obtain at least one
dangling edge.

The key point is that the AC asks for the existence of the edge but
the production demands its non-existence as it is included in the
nihilation matrix $K$.  In this case, the rule $break'$ depicted to
the right of the same figure is equivalent to $p$ but with no
potential compatibility issues.

Notice that coherence is fulfilled in the example to the left of
Fig.~\ref{fig:example_non_compat} (the AC alone does not encode any
contradiction) but not consistency as no host graph can satisfy it.

\begin{figure}[htbp]
  \centering
  \includegraphics[scale = 0.5]{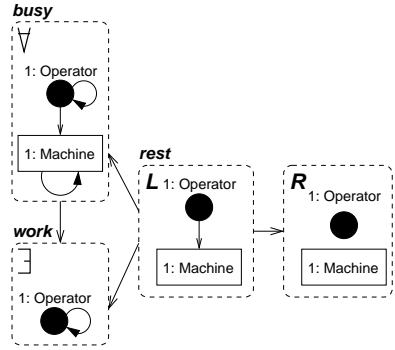}
  \caption{Non-Coherent Application
    Condition}\label{fig:example_non_coherent}
\end{figure}

An example of non-coherent application condition can be found in
Fig.~\ref{fig:example_non_coherent}.  The AC has associated formula
$\mathfrak{f} = \forall busy \exists work [ busy \wedge$ $P ( work,
\overline{G}) ]$.  There is no problem with the edge deleted by the
rule, but with the self-loop of the operator.  Note that due to
$busy$, it must appear in any potential host graph but $work$ says
that it should not be present. \proofend

Just to clarify the terminology, we will see that an application
condition is coherent if and only if its associated sequence is
coherent, and the same for compatibility (this is why these concepts
have been named this way).  We will also see that an application
condition is consistent if its associated sequence is applicable.
Here, morphisms play a similar role in the graphs that make up the
application condition to completion in sequences of rules. Another
example follows.

\begin{figure}[htbp]
  \centering
  \includegraphics[scale = 0.65]{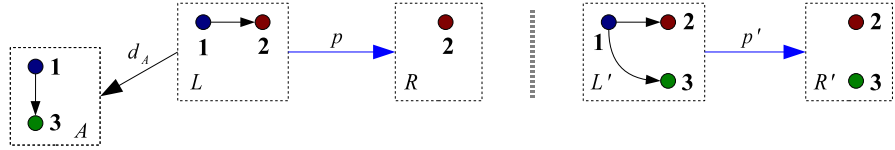}
  \caption{Avoidable non-Compatible Application Condition}
  \label{fig:fallacyEx}
\end{figure}

\noindent\textbf{Example}.$\square$As commented above, non-compatibility can be
avoided at times just rephrasing the condition and the rule. Consider
the weak precondition $A$ as represented to the left of
Fig.~\ref{fig:fallacyEx}.  There is a diagram $\mathfrak{d} = \{A\}$
with associated formula $\mathfrak{f} = \exists A [A]$, being morphism
$d_A(1) = 1$.  As the production deletes node 1 and the application
condition asks for the existence of edge $(1,3)$, it is for sure that
if the rule is applied we will obtain at least one dangling edge.

The key point is that the condition asks for the existence of edge
$(1,3)$ but the production demands its non-existence as it is included
in the nihilation matrix $K$. In this case, the rule $p'$ depicted to
the right of the same figure is completely equivalent to $p$ but with
no potential compatibility issues.

\begin{figure}[htbp]
  \centering
  \includegraphics[scale = 0.65]{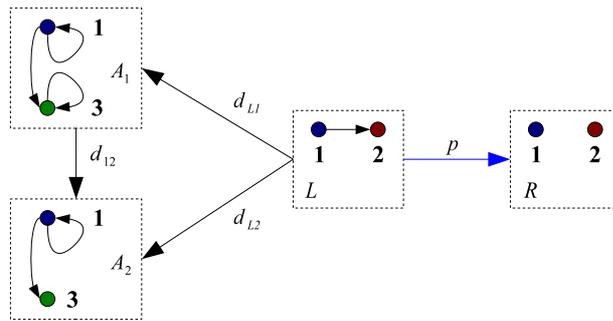}
  \caption{non-Coherent Application Condition}
  \label{fig:nonCoherent_AC}
\end{figure}

A non-coherent application condition can be found in
Fig.~\ref{fig:nonCoherent_AC}.  Morphisms identify all nodes:
$d_{Li}(\{1\}) = \{1\} = d_{12}(\{1\})$, $d_{Li}(\{2\}) = \{2\}$,
$d_{12}(\{3\}) = \{3\}$ with formula $\mathfrak{f} = \exists L \forall
A_1 \exists A_2 \left[ L \Rightarrow A_1 \wedge P\left( A_2,
    \overline{G} \right) \right]$.  There is no problem with edge
$\left( 1,2 \right)$ but with $\left( 1,1 \right)$ there is one.  Note
that due to $A_1$, it must appear in any potential host graph but
$A_2$ says that it should not be present. \proofend

A direct application of Theorem~\ref{th:reductionPre} allows us to
test if a weak precondition specifies a tautology or a fallacy. It
will also be used in the next section to study how to construct weak
postconditions equivalent to given weak preconditions. It is also
useful to proceed in the opposite way, i.e. to transform
postconditions into equivalent preconditions.

\newtheorem{equivPreAC_seqs}[matrixproduct]{Corollary}
\begin{equivPreAC_seqs}\label{cor:equivPreAC_seqs}
  A weak precondition is coherent if and only if its associated
  sequence (set of sequences) is coherent.  Also, it is compatible if
  and only if its sequence (set of sequences) is compatible and it is
  consistent if and only if its sequence (set of sequences) is
  applicable.
\end{equivPreAC_seqs}

\noindent\textbf{Example}.$\square$For coherence we will change the formula of
previous example (Fig.~\ref{fig:nonCoherent_AC}) a little.  Consider
$\mathfrak{f}_2 = \exists L \forall A_0 \exists A_1 \left[ L \left(
    A_1 \Rightarrow \overline{A_0} \right)\right]$.  Note that
$\mathfrak{f}_2$ cannot be fulfilled because on the one hand edges
$(1,1)$ and $(1,2)$ must be found in $G$ and on the other edge $(1,1)$
must be in $\overline{G}$.

To simplify the example, suppose that some match is already given.
The sequence to study is $p;id_{A_1};\overline{id}_{A_0}$, which is
not coherent because in its equivalent form $p;id_{A_1};p^e_0;p^r_0$
production $p^e_0$ deletes edge $(1,1)$ used by
$id_{A_1}$.\proofend

\newtheorem{precondConsCohComp}[matrixproduct]{Corollary}
\begin{precondConsCohComp}\label{cor:precondConsCohComp}
  A weak precondition is consistent if and only if it is coherent and
  compatible.
\end{precondConsCohComp}


\noindent\textbf{Examples}.$\square$Compatibility for ACs tells us whether
there is a conflict between an AC and the rule's action. As stated in
Corollary~\ref{cor:equivPreAC_seqs}, this property is studied by
analyzing the compatibility of the resulting sequence.  Rule {\em
  break} in Fig.~\ref{fig:example_non_compat} has an AC with formula
$\exists Operated[Operated]$.  This results in sequence:
$break^{\flat}; id_{Operated}$, where the machine in both rules is
identified (i.e. has to be the same).  Our analysis technique for
compatibility~\cite{JuanPP_1} outputs a matrix with a $1$ in the
position corresponding to edge $(1:Operator, 1: Machine)$, thus
signaling the dangling edge.

Coherence detects conflicts between the graphs of the AC (which
includes $L$ and $K$) and we can study it by analyzing coherence of
the resulting sequence.  For the case of rule ``rest'' in
Fig.~\ref{fig:example_non_coherent}, we would obtain a number of
sequences, each testing that ``busy'' is found, but the self-loop of
``work'' is not. This is not possible, because this self-loop is also
part of ``busy''. Coherence detects such conflict and the problematic
element. \proofend

In addition, we can also use the MGG techniques of previous chapters to
analyze application conditions and gather more information. This is
reviewed in the rest of the section.

\begin{itemize}
\item \textbf{Sequential Independence}. We can use MGG results for
  sequential independence of sequences to investigate if, once several
  rules with ACs are translated into sequences, we can for example
  delay all the rules checking the AC constraints to the end of the
  sequence. Note that usually, when transforming an AC into a
  sequence, the original flat rule should be applied last. Sequential
  independence allows us to choose some other order.  Moreover, for a
  given sequence of productions, ACs are to some extent delocalized
  inside the sequence. In particular it could be possible to pass
  conditions from one production to others inside a sequence (paying
  due attention to compatibility and coherence). For example, a
  post-condition for $p_1$ in the sequence $p_2; p_1$ might be
  translated into a pre-condition for $p_2$, and vice versa.
\end{itemize}

\noindent\textbf{Example}.$\square$The sequence resulting from the rule in
Fig.~\ref{fig:example_Exists} is $moveOperator^{\flat}; id_{Ready}$.
In this case, both rules are independent and can be applied in any
order. This is due to the fact that the rule effects do not affect the
AC. \proofend

\begin{itemize}
\item \textbf{Minimal and Negative Initial Digraphs}. The concepts of
  MID and NID allow us to obtain the (set of) minimal graph(s) able to
  satisfy a given GC (or AC), or to obtain the (set of) minimal
  graph(s) which cannot be found in $G$ for the GC (or AC) to be
  applicable. In case the AC results in a single sequence, we can
  obtain a minimal graph; if we obtain a set of sequences, we get a
  set of minimal graphs. In case universal quantifiers are present, we
  have to complete all existing partial matches so it might be useful
  to limit the number of nodes in the host graph under
  study.\footnote{This, in many cases, arises naturally. For example
    in~\cite{MGG_computation} MGG is studied as a model of computation
    and a formal grammar, and also it is compared to Turing machines
    and Boolean Circuits. Recall that Boolean Circuits have fixed
    input variables, giving rise to MGGs with a fixed number of
    nodes. In fact, something similar happens when modeling Turing
    machines, giving rise to the so-called (MGG) nodeless model of
    computation.}  A direct application of the MID/NID technique
  allows us to solve the problem of finding a graph that satisfies a
  given AC. The technique can be extended to cope with more general
  GCs.
\end{itemize}


\begin{figure}[htbp]
 \centering
 \includegraphics[scale = 0.44]{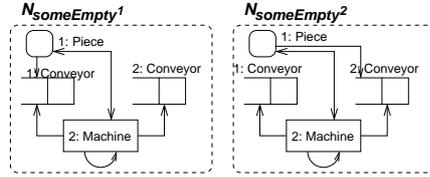}
 \caption{Negative Graphs Disabling the Sequences in
   Fig.~\ref{fig:example_decomposition}}
 \label{fig:example_NID}
\end{figure}

\noindent\textbf{Examples}.$\square$Rule $remove$ in
Figure~\ref{fig:example_decomposition} results in two sequences. In
this case, the minimal initial digraph enabling the applicability for
both is equal to the LHS of the rule.  The two negative initial
digraphs are shown in Fig.~\ref{fig:example_NID} (and both assume a
single piece in $G$). This means that the rule is not applicable if
$G$ has any edge stemming from the machine, or two edges stemming from
the piece to the two conveyors.

\begin{figure}[htbp]
  \centering
  \includegraphics[scale = 0.42]{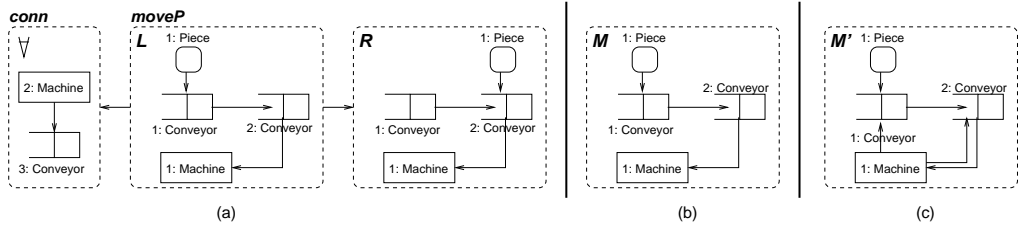}
  \caption{(a) Example rule (b) MID without AC (c) Completed MID}
  \label{fig:MID_Completion}
\end{figure}

Figure~\ref{fig:MID_Completion} shows the minimal initial digraph for
executing rule $moveP$. As the rule has a universally quantified
condition ($\forall conn[conn]$), we have to complete the two partial
matches of the initial digraph so as to enable the execution of the
rule. \proofend

\begin{itemize}
\item \textbf{G-congruence}.Graph congruence characterizes sequences
  with the same initial digraph. Therefore, it can be used to study
  when two GCs/ACs are equivalent for all morphisms or for some of
  them. See Section~7 in~\cite{MGG_Combinatorics} or
  Section~\ref{sec:gCongruence}.
\end{itemize}

The current approach to restrictions allows us to analyze properties
which up to now have been analyzed either without ACs or with NACs,
but not with arbitrary ACs:

\begin{itemize}
\item \textbf{Critical Pairs}. A critical pair is a minimal graph in
  which two rules are applicable, and applying one disables the
  other~\cite{Heckel:ICGT02}. Critical pairs have been studied for
  rules without ACs~\cite{Heckel:ICGT02} or for rules with
  NACs~\cite{Lambers}. The techniques in MGG however enable the study
  of critical pairs with any kind of AC. This can be done by
  converting the rules into sequences, calculating the graphs which
  enable the application of both sequences, and then checking whether
  the application of a sequence disables the other.

  In order to calculate the graphs enabling both sequences, we derive
  the minimal digraph set for each sequence as described in previous
  item. Then, we calculate the graphs enabling both sequences (which
  now do not have to be minimal, but we should have jointly surjective
  matches from the LHS of both rules) by identifying the nodes in each
  minimal graph of each set in every possible way. Due to universals,
  some of the obtained graphs may not enable the application of some
  sequence. The way to proceed is to complete the partial matches of
  the universally quantified graphs, so as to make the sequence
  applicable.

  Once we have the set of starting graphs, we take each one of them
  and apply one sequence. Then, the sequence for the second rule is
  recomputed -- as the graph has changed -- and applied to the
  graph. If it can be applied, there are no conflicts for the given
  initial graph, otherwise there is a conflict.  Besides the conflicts
  known for rules without ACs or with NACs (delete-use and
  produce-forbid~\cite{Fundamentals}, our ACs may produce additional
  kinds of conflicts. For example, a rule can create elements which
  produce a partial match for a universally quantified constraint in
  another AC, thus making the latter sequence inapplicable.
\end{itemize}

\begin{figure}[htbp]
  \centering
  \includegraphics[scale = 0.44]{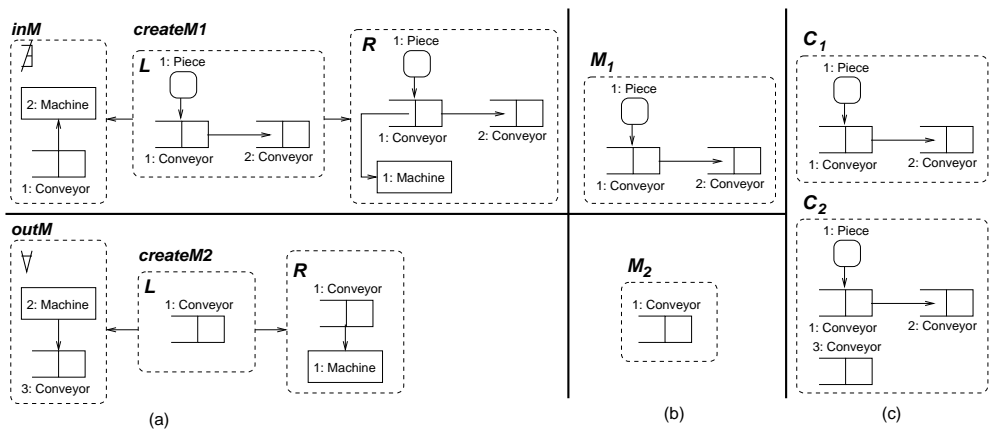}
  \caption{(a) Example Rules (b) MIDs (c) Starting Graphs for
    Analyzing Conflicts}
  \label{fig:Critical_Pairs}
\end{figure}

\noindent\textbf{Example}. Figure~\ref{fig:Critical_Pairs}(a) shows two rules,
$createM1$ and $createM2$, with ACs $\nexists inM[inM]$ and $\forall
outM[outM]$, respectively. The center of the same figure depicts the
minimal digraphs $M_1$ and $M_2$, enabling the execution of the
sequences derived from $createM1$ and $createM2$, respectively. In
this case, both are equal to the LHS of each rule.  The right of the
figure shows the two resulting graphs once we identify the nodes in
$M_1$ and $M_2$ in each possible way. These are the starting graphs
that are used to analyze the conflicts.  The rules present several
conflicts. First, rule $createM1$ disables the execution of
$createM2$, as the former creates a new machine, which is not
connected to all conveyors, thus disabling the $\forall outM[outM]$
condition of $createM2$.  The conflict is detected by executing the
sequence associated to $createM1$ (starting from either $C_1$ or
$C_2$), and then recomputing the sequence for $createM2$, taking the
modified graph as the starting one. Similarly, executing rule
$createM2$ may disable $createM1$ if the new machine is created in the
conveyor with the piece (this is a conflict of type
produce-forbid~\cite{Lambers}). \proofend

\begin{itemize}
\item \textbf{Rule Independence}. In Matrix Graph Grammars, we convert
  the rules into sets of sequences and then check each combination of
  sequences of the two rules.
\end{itemize}

\section{Moving Conditions}
\label{sec:movingConditions}

Roughly speaking, there have been two basic ideas in previous sections
that allowed us to check consistency of the definition of direct
derivations with weak preconditions, and also provided us with some
means to use the theory developed so far in order to continue the
study of application conditions:
\begin{itemize}
\item Embed application conditions into the production or derivation.
  The left hand side $L$ of a production receives elements that must
  be found -- $P(A,G)$ -- and $K$ those whose presence is forbidden --
  $\overline{P}(A,G)$ --.
\item Find a sequence or a set of sequences whose behavior is
  equivalent to that of the production plus the application condition.
\end{itemize}

In this section we will care about how (weak) preconditions can be
transformed into (weak) postconditions and vice versa: Given a weak
precondition $A$, what is the equivalent weak postcondition (if any)
and how can one be transformed into the other? Before this, it is
necessary to state the main results of previous sections for
postconditions.

The notation needs to be further enlarged so we will append a left
arrow on top of conditions to indicate that they are (weak)
preconditions and an upper right arrow for (weak) postconditions.
Examples are $\stackrel{\leftarrow}{A}$ for a weak precondition and
$\stackrel{\rightarrow}{A}$ for a weak postcondition.  If it is clear
from the context, we will omit arrows.

There is a direct translation of Theorem~\ref{th:embeddingGC} for
postconditions. Operators $\widehat{T}_{\stackrel{\rightarrow}{A}}$
and $\widecheck{T}_{\stackrel{\rightarrow}{A}}$ are defined similarly
for weak postconditions.  Again, if it is clear from context, it will
not be necessary to over-elaborate the
notation.

Equivalent results to lemmas in
Sec.~\ref{sec:functionalRepresentation}, in particular to
equations~\eqref{eq:matchTransPre}, \eqref{eq:decompTransPre},
\eqref{eq:closureTransPre}~and~\eqref{eq:nacsImage} are given in the
following proposition:

\newtheorem{postConds}[matrixproduct]{Proposition}
\begin{postConds}\label{prop:postConds}
  Let $\stackrel{\rightarrow}{A} = \left( \mathfrak{f}, \mathfrak{d}
  \right) = \left( \mathfrak{f}, \left(\{A\}, d:R \rightarrow A
    \right)\right)$ be a weak postcondition. Then we can obtain a set
  of equivalent sequences to given basic formulae as follows:
  \begin{eqnarray}
    \index{functional representation!match}\label{eq:postMatch}
    \textrm{(Match)} \quad \mathfrak{f} & = & \exists A [A] \quad
    \longmapsto \quad T_A\left(p\right) = id_A ; p. \\
    \index{functional representation!decomposition}
    \textrm{(Decomposition)} \quad 
    \mathfrak{f} & = & \exists A [\overline{A}] \quad \longmapsto
    \quad \widehat{T}_A\left(p\right) = \overline{id}_A ; p. \\
    \index{functional representation!closure}\textrm{(Closure)} \quad
    \mathfrak{f} & = & \not{\exists} A [\overline{A}] \quad
    \longmapsto \quad \widecheck{T}_A\left(p\right) = id_{A^1} ;
    \ldots ; id_{A^m}; p. \\
    \index{functional representation!negative application
      condition}\label{eq:postNAC} \textrm{(NAC)} \quad \mathfrak{f} &
    = & \not{\exists} A [A] \quad \longmapsto \quad
    \widetilde{T}_A\left(p\right) = \overline{id}_{A^{u_0v_0}} ;
    \ldots ; \overline{id}_{A^{u_mv_m}}; p.
  \end{eqnarray}
\end{postConds}

\noindent \emph{Proof} \\*
$\square$$\blacksquare$

There is a symmetric result to Theorem~\ref{th:reductionPre} for weak
postconditions that directly stems from Prop.~\ref{prop:postConds}.
The development and ideas are the same, so we will not repeat them
here.

\newtheorem{reductionPost}[matrixproduct]{Theorem}
\begin{reductionPost}\label{th:reductionPost}
  Any weak postcondition can be reduced to the study of the
  corresponding set of sequences.
\end{reductionPost}

\noindent \emph{Proof} \\*
$\square$$\blacksquare$

Corollaries~\ref{cor:equivPreAC_seqs} and~\ref{cor:precondConsCohComp}
have their versions for postconditions which are explicitly stated
for further reference.

\newtheorem{equivPostAC_seqs}[matrixproduct]{Corollary}
\begin{equivPostAC_seqs}\label{cor:equivPostAC_seqs}
  A weak postcondition is coherent if and only if its associated
  sequence (set of sequences) is coherent.  Also, it is compatible if
  and only if its sequence (set of sequences) is compatible and it is
  consistent if and only if its sequence (set of sequences) is
  applicable.
\end{equivPostAC_seqs}

\newtheorem{postcondConsCohComp}[matrixproduct]{Corollary}
\begin{postcondConsCohComp}\label{cor:postcondConsCohComp}
  A weak postcondition is consistent if and only if it is coherent and
  compatible.
\end{postcondConsCohComp}

Let $p:L \rightarrow R$ be a production applied to graph $G$ such that
$p(G) = H$.  Elements to be found in $G$ are those that appear in $L$.
Similarly, elements that are mandatory in the ``post'' side are those
in $R$. The evolution of the positive part (to be added to $L$) of a
weak application condition is given by the grammar rule itself.

The evolution of the negative part $K$ has not been addressed up to
now as it has not been needed. Recall that $K$ represents the negative
elements of the LHS of the production and let's represent by $Q$ those
elements that must not be present in the RHS.\footnote{Note that $K$
  and $Q$ precede $L$ and $R$ in the alphabet.}

\newtheorem{Nevolution}[matrixproduct]{Proposition}
\begin{Nevolution}\label{th:Nevolution}
  Let $p:L \rightarrow R$ be a compatible production with negative
  left hand side $K$ and negative right hand side $Q$. Then,
  \begin{equation}
    \label{eq:1}
    Q = p^{-1} \left( K \right).
  \end{equation}
\end{Nevolution}

\noindent \emph{Proof} \\*
$\square$First suppose that $K$ is the one naturally defined by the
production, i.e. the one found in Lemma~\ref{lemma:nihilationMatrix}.
The only elements that should not appear in the RHS are potential
dangling edges and those deleted by the production: $e \vee
\overline{D}$. It coincides with \eqref{eq:1} as shown by the
following set of identities:
\begin{equation}
  \label{eq:3}
  p^{-1} \left( K \right) = e \vee \overline{r} \, K = e \vee
  \overline{r} \left( r \vee \overline{e} \, \overline{D} \right) = e
  \vee \overline{e} \, \overline{r} \, \overline{D} = e \vee
  \overline{r} \, \overline{D} = e \vee \overline{D}.
\end{equation}
In the last equality of \eqref{eq:3} compatibility has been used,
$\overline{r} \, \overline{D} = \overline{D}$. Now suppose that $K$
has been modified, adding some elements that should not be found in
the host graph (Theorem~\ref{th:reductionPre}). There are three
possibilities:
\begin{itemize}
\item The element is erased by the production. This case is ruled out
  by Corollary~\ref{cor:equivPreAC_seqs} as the weak precondition
  could not be coherent.
\item The element is added by the production. Then, in fact, the
  condition is superfluous as it is already considered in $K$ without
  modifications, i.e. \eqref{eq:3} can be applied.
\item None of the above. Then equation \eqref{eq:1} is trivially
  fulfilled because the production does not affect this element.
\end{itemize}
Just a single element has been considered to ease
exposition.\proofend

\noindent\textbf{Remark}.$\square$Though strange at a first glance, a dual
behavior of the negative part of a production with respect to the
positive part should be expected. The fact that $K$ uses $p^{-1}$
rather than $p$ for its evolution is quite natural. When a production
$p$ erases one element, it asks its LHS to include it, so it demands
its presence. The opposite happens when $p$ adds some element. For $K$
things happen quite in the opposite direction. If the production asks
for the addition of some element, then the size of $K$ is increased
while if some element is deleted, $K$ \emph{shrinks}.\proofend

Now we can proceed to prove that it is possible to transform
preconditions into postconditions and back again. Proposition~\ref{th:Nevolution} allows us to consider the positive part only. The
negative part would follow using the inverse of the productions.

There is a restricted case that can be directly addressed using
equations~\eqref{eq:postMatch}~--~\eqref{eq:postNAC}, Theorems~\ref{th:reductionPre}~and~\ref{th:reductionPost} and Corollaries~\ref{cor:equivPreAC_seqs}~and~\ref{cor:equivPostAC_seqs}. It is the
case in which the transformed postcondition for a given precondition
does not change.\footnote{Note that this is not so unrealistic. For
  example, if the production preserves all elements appearing in the
  precondition.} The question of whether it is always possible to
transform a precondition into a postcondition -- and back again --
would be equivalent to asking for sequential independence of the
production and identities, i.e. whether $id_{ A^i} \perp p$ or not.

In general the production may act on elements that appear on the
definition of the graphs of the precondition. Recall that one demand
on precondition specification is that $L$ and $K$ are always the
domain of their respective morphisms $d_{L}$ and $d_{K}$ (refer to
comments on p.~\pageref{eq:domain}). The reason for doing so will be
clarified shortly.

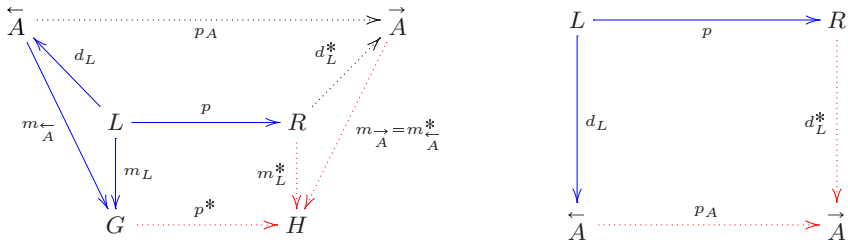
\begin{figure}[htb]
  \centering \makebox{ \xymatrix{ \stackrel{\leftarrow}{A}
      \ar@[blue][ddr]_{m_{\stackrel{\leftarrow}{A}}}
      \ar@{.>}[rrrr]_{p_A} &&&& \stackrel{\rightarrow}{A}
      \ar@{.>}@[red][ddl]^{m_{\stackrel{\rightarrow}{A}} =
        m^*_{\stackrel{\leftarrow}{A}}} && L \ar@[blue][rrr]_{p}
      \ar@[blue][dd]^{d_L} &&& R
      \ar@{.>}@[red][dd]_{d^{*}_L} \\
      & L \ar@[blue][rr]^p \ar@[blue][d]^{m_L} \ar@[blue][ul]_{d_L} &&
      R \ar@{.>}@[red][d]_{m^*_L} \ar@{.>}[ur]^{d^*_L} \\
      & G \ar@{.>}@[red][rr]^{p^*} && H &&& \stackrel{\leftarrow}{A}
      \ar@{.>}@[red][rrr]^{p_A} &&& \stackrel{\rightarrow}{A} } }
  \caption{(Weak) Precondition to (Weak) Postcondition Transformation}
  \label{fig:prePostTrans}
\end{figure}

Theorems on this and previous sections make it possible to interpret
preconditions and postconditions as sequences. The only difference is
that preconditions are represented by productions to be applied before
$p$ while postconditions need to be applied after $p$. Hence, the only
thing we have to do to transform a precondition into a postcondition
(or vice versa) is to pass productions from one part to the other. The
case in which we have sequential independence has been studied above.
If there is no sequential independence the transformation can be
reduced to a pushout construction\footnote{The square made up of $L$,
  $R$, $\stackrel{\leftarrow}{A}$ and $\stackrel{\rightarrow}{A}$ is a
  pushout where $p$, $L$, $d_L$, $R$ and $\stackrel{\leftarrow}{A}$
  are known and $\stackrel{\rightarrow}{A}$, $p_A$ and $d_L$ need to
  be calculated. Recall from Sec.~\ref{sec:matchAndExtendedMatch} that
  production composition can be used instead of pushout constructions.
  The same applies here, but we will not enter this topic for now.} --
as for direct derivation definition -- except for one detail: In
direct derivations matches are total morphisms while here $d_{L}$ and
$d_{K}$ need not be (see Fig.~\ref{fig:prePostTrans}).

\begin{figure}[htbp]
  \centering
  \includegraphics[scale = 0.66]{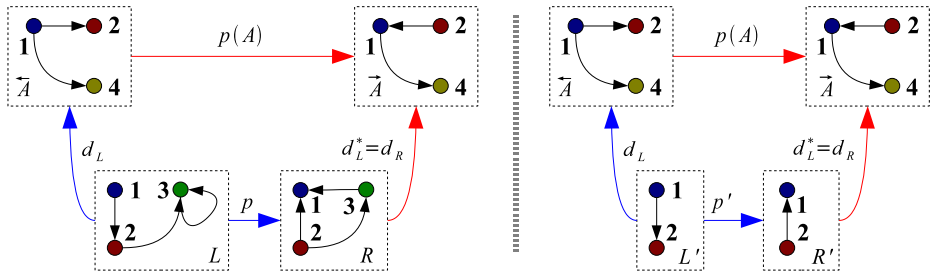}
  \caption{Restriction to Common Parts: Total Morphism}
  \label{fig:pre2PostSimpleEx}
\end{figure}

The way to proceed is to restrict to the part in which the morphisms
are defined (they are trivially total in that part). For example, the
transformation for the weak application condition depicted to the left
of Fig.~\ref{fig:pre2PostSimpleEx} is a pushout. It is again
represented to the right of the same figure.

The notation is extended to represent this transformation of
preconditions into postconditions as follows:
\begin{equation}\label{eq:prePostTrans_1}
  \stackrel{\rightarrow}{A} = p \left( \stackrel{\leftarrow}{A} \right).
\end{equation}

To see that precondition satisfaction is equivalent to postcondition
satisfaction all we have to do is to use their representation as
sequences of productions
(Theorems~\ref{th:reductionPre}~and~\ref{th:reductionPost}).  Note
that applying $p$ delays the application of the result (the $id_A$ or
$\overline{id}_A$ productions) in the sequence, i.e. we have a kind of
sequential independence except that productions can be different
($id_{\stackrel{\leftarrow}{A}} \neq id_{\stackrel{\rightarrow}{A}}$)
because they may be modified by the production:
\begin{equation}\label{eq:prePostTrans_2}
  p ; id_{\stackrel{\leftarrow}{A}} \; \longmapsto \;
  id_{\stackrel{\rightarrow}{A}} ; p.
\end{equation}

If the weak precondition is consistent so must the weak postcondition
be. There can not be any compatibility issue and coherence is
maintained (again, $id_A$ and $\overline{id}_A$ may be modified by the
production). Production $p$ deals with the positive part of the
precondition and, by Proposition~\ref{th:Nevolution}, $p^{-1}$ will
manage the part associated to $K$. For the post-to-pre transformation
roles of $p$ and $p^{-1}$ are interchanged.

Pre-to-post or post-to-pre transformations do not affect the shape of
the formula associated to a diagram except in the case where redundant
graphs are discarded. There are two clear examples of this:
\begin{itemize}
\item The application condition requires the graph to appear and the
  production deletes all its elements.
\item The application condition requires the graph not to appear and
  the production adds all its elements.
\end{itemize}

Recalling that there can not be any compatibility nor coherence
problem due to precondition consistency, consistency permits the
transformation, proving the main result of this section:

\newtheorem{prePostPre}[matrixproduct]{Theorem}
\begin{prePostPre}\label{th:prePostPre}
  Any consistent (weak) precondition is equivalent to some consistent
  (weak) postcondition and vice versa.
\end{prePostPre}

\noindent \emph{Proof (Sketch)} \\*
$\square$What has been addressed in previous pages is the equivalent
to the first case in the proof of Theorem~\ref{th:embeddingGC} or to
Lemma~\ref{lemma:firstCase}. Hence, a similar procedure using closure,
decomposition or both proves the result. Notice that it is necessary
to consider the host graph in order to calculate the
equivalence.\proofend

This result allows us to extend the notation to consider the
transformation of a precondition.  A postcondition is the image of
some precondition, and vice versa:
\begin{equation}\label{eq:prePostTransFuncNotation}
  \stackrel{\rightarrow}{A} = \left\langle\
    \!\!\stackrel{\leftarrow}{A}, p \right\rangle.
\end{equation}

As commented above, for a given application condition AC it is not
necessarily true that $A = p^{-1};p(A)$ because some new elements may
be added and some obsolete elements can be discarded.  What we will
get is an equivalent condition adapted to $p$ that holds whenever $A$
holds and fails to be true whenever $A$ is false.

\begin{figure}[htbp]
  \centering
  \includegraphics[scale = 0.7]{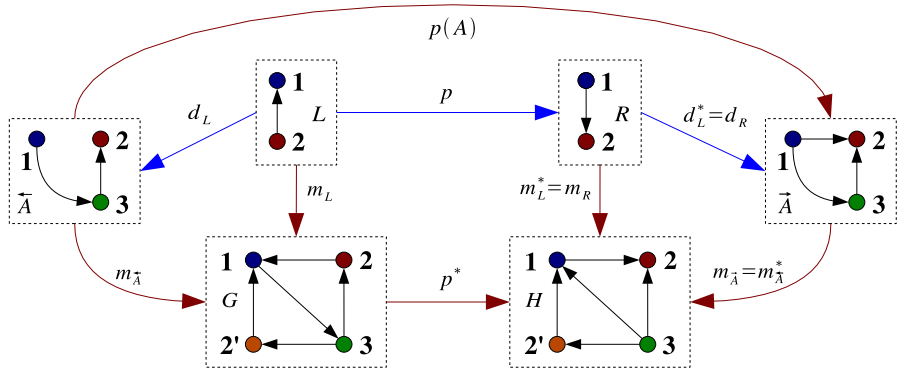}
  \caption{Precondition to Postcondition Example}
  \label{fig:pre2PostEx}
\end{figure}

\noindent\textbf{Example}.$\square$In Fig.~\ref{fig:pre2PostEx} there is a very
simple transformation of a precondition into a postcondition through
morphism $p(A)$. The production deletes one arrow and adds a new one.
The overall effect is reverting the direction of the edge between
nodes $1$ and $2$.

The opposite transformation, from postcondition to precondition, can
be obtained by reverting the arrow, i.e. through $p^{-1}(A)$.  More
general schemes can be studied applying the same principles, although
diagrams will be a bit cumbersome with only a few application
conditions.

Let $\mathcal{A} = p^{-1} \circ p \left( \stackrel{\leftarrow}{A}
\right)$. If a pre-post-pre transformation is carried out, we will
have $\stackrel{\leftarrow}{A} \,\neq \mathcal{A}$ because edge (2,1)
would be added to $\stackrel{\leftarrow}{A}$. However, it is true that
$\mathcal{A} = p^{-1}\circ p \left( \mathcal{A} \right)$.

Note that in fact $id_{\stackrel{\leftarrow}{A}} \bot p$ if we limit
ourselves to edges, so it would be possible to simply move the
precondition to a postcondition as it is. Nonetheless, we have to
consider nodes 1 and 2 as the common parts between $L$ and
$\stackrel{\leftarrow}{A}$. This is the same kind of restriction than
the one illustrated in Fig.~\ref{fig:pre2PostSimpleEx}.\proofend

If the pre-post-pre transformation is thought of as an operator $T_p$
acting on application conditions, then it fulfills
\begin{equation}
  \label{eq:2}
  T_p^2 = id,
\end{equation}
where $id$ is the identity. The same would also be true for a
post-pre-post transformation.

Theorem~\ref{th:prePostPre} can be generalized at least in two ways.
We will just sketch how to proceed as it is not difficult with the
theory developed so far.

Firstly, an application condition has been transformed into an
equivalent sequence of productions (or set of sequences) but no
$\varepsilon$-productions have been introduced to help with
compatibility of the application condition. Think of a production that
deletes one node and that some graph of the application condition has
an edge incident to that node (and that edge is not deleted by the
production). So to speak, we have a fixed grammar pre to post
transformation theorem. It should not be very difficult to proceed as
in Chap.~\ref{ch:matching} to define a floating grammar behavior.


Secondly, application conditions can now be thought of as properties
of the production, and not necessarily as part of its left or right
hand sides.  It is not difficult to see that, for a given sequence of
productions, application conditions are to some extent
\emph{delocalized} in the sequence.  In particular it would be
possible to pass conditions from one production to others inside a
sequence (paying due attention to compatibility and coherence).  Note
that a postcondition for $p_1$ in the sequence $p_2;p_1$ might be
translated into a precondition for $p_2$, and vice versa.\footnote{This
  transformation can be carried out under appropriate circumstances,
  but we are not limited to sequential independence. Recall that
  productions specifying constraints can be advanced or delayed even
  though they are not sequential independent with respect to the
  productions that define the sequence.}

When defining diagrams some ``practical problems'' may turn up.  For
example, if the diagram $\mathfrak{d} = \left( L
  \stackrel{d_{L0}}{\rightarrow} A_0 \stackrel{d_{10}}{\leftarrow} A_1
\right)$ is considered then there are two potential problems:
\begin{enumerate}
\item The direction in the arrow $A_0 \leftarrow A_1$ is not the
  natural one.  Nevertheless, injectiveness allows us to safely revert
  the arrow, $d_{01} = d^{-1}_{10}$.
\item Even though we only formally state $d_{L0}$ and $d_{10}$, other
  morphisms naturally appear and need to be checked out, e.g. $d_{L1}:
  R \rightarrow A_1$.  New morphisms should be considered if they
  relate at least one element.\footnote{Otherwise stated: Any
    condition made up of $n$ graphs $A_i$ can be identified as the
    complete graph $K_n$, in which nodes are $A_i$ and morphisms are
    $d_{ij}$.  Whether this is a directed graph or not is a matter of
    taste (morphisms are injective).}
\end{enumerate}

A possible interpretation of eq.~\eqref{eq:2} is that the definition
of the application condition can vary from the \emph{natural} one,
according to the production under consideration. Pre-post-pre or
post-pre-post transformations adjust application conditions to the
corresponding production.

Let's end this section relating graph constraints and moving
conditions. Recall equation \eqref{eq:equivAC_GC} in which a first
relationship between application conditions and graph constraints is
established. That equation states how to enlarge the requirements
already imposed by a graph constraint to a given host graph if,
besides, a given production is to be applied.

Another different though related point is how to make productions
respect some properties of a graph. This topic is addressed in the
literature, for example in~\cite{Fundamentals}. The proposed way to
proceed is to transform a graph constraint into a postcondition first
and a precondition right afterwards. The equivalent condition to
\eqref{eq:equivAC_GC} would be
\begin{equation}
  \label{eq:4}
  \mathfrak{f}_{PC} = \exists R \exists Q \left[ R \wedge P\left( Q,
      \overline{G} \right) \wedge \mathfrak{f}_{GC}\right],
\end{equation}
being $\mathfrak{f}_{GC}$ the graph constraint to be kept by the
production.


\section{From Simple Digraphs to Multidigraphs}
\label{sec:fromSimpleDigraphsToMultidigraphs}

In this section we show how it is possible to consider multidigraphs
(directed graphs allowing multiple parallel edges) without changing
the theory developed so far.  At first sight this might seem a hard
task as Matrix Graph Grammars heavily depend on adjacency matrices.
Adjacency matrices are well suited for simple digraphs but can not
deal with parallel edges. This section is a \emph{theoretical
  application} of graph constraints and application conditions to
Matrix Graph Grammars.

Before addressing multidigraphs, variable nodes are introduced as one
depends on the other.  We will follow reference~\cite{Hof05} to which
the reader is referred for further details.

\index{graph pattern}If instead of nodes of fixed type variable types
are allowed, we get a so called \emph{graph pattern}.  \index{rule
  scheme}A \emph{rule scheme} is just a production in which graphs are
graph patterns.  \index{substitution function}A \emph{substitution
  function} $\iota$ specifies how variable names taking place in a
production are substituted.  A rule scheme $p$ is instantiated via
substitution functions producing a particular production.  For
example, for substitution function $\iota$ we get $p^{\,\iota}$.  The
set of production instances for $p$ is defined as the set
$\mathcal{I}(p) = \left\{ p^{\,\iota} \; \vert \right.$ $\iota$ is a
substitution$\left.\right\}$.

\index{kernel (graph)}The kernel of a graph $G$, $ker(G)$, is defined
as the graph resulting when all variable nodes are removed.  It might
be the case that $ker(G) = \emptyset$.

The basic idea is to reduce any rule scheme to a set of rule
instances.  Note that it is not possible in general to generate
$\mathcal{I}(p)$ because this set can be infinite.  The way to proceed
is simple:
\begin{enumerate}
\item Find a match for the kernel of $L$.
\item Induce a substitution $\iota$ such that the match for the kernel
  becomes a full match $m:L^{\iota} \rightarrow G$.
\item Construct the instance $R^{\,\iota}$ and apply $p^{\,\iota}$ to
  get the direct derivation $G \stackrel{p^{\,\iota}}{\Longrightarrow}
  H$.
\end{enumerate}

Mind the non-determinism of step (2), which is matching.  Rule schemes
are required to satisfy two conditions:
\begin{enumerate}
\item Any variable name occurs at most once in $L$.
\item Rule schemes do not add variable nodes.
\end{enumerate}

These two conditions greatly simplify rule application when there are
variable nodes, specially for the DPO approach.  In our case they are
not that important because, among other things, matches in Matrix
Graph Grammars are injective.


Let's start with multidigraphs and how it is possible to extend Matrix
Graph Grammars to cope with them without any major modification.  The
idea is not difficult: \index{multinode}A special kind of node (call
it \emph{multinode}) associated to every edge in the graph is
introduced.  Graphically, they will be represented by a filled square.

Now two or more edges can join the same nodes, as in fact there are
multinodes in the middle that convert them into simple digraphs.
\index{simple!node}The term multinode is just a means to distinguish
them from the rest of ``normal'' nodes that we will call \emph{simple
  nodes} and will be represented as usual with colored circles.  They
are not of a different kind as for example hyperedges with respect to
edges (see Sec.~\ref{sec:hyperedgeReplacement}).  In our case, simple
nodes and multinodes are defined similarly and obey the same rules,
although their semantics differ.

There are some restrictions to be imposed on the actions that can be
performed on multinodes (application conditions) or, more precisely,
the shape or topology of permitted graphs (graph constraints).

Operations previously specified on edges now act on multinodes.  Edges
are managed through multinodes: Adding an edge is transformed into a
multinode addition and edge deletion becomes multinode deletion.
Still, there are edges in the ``old'' sense, to link multinodes to
their source and target simple nodes.  We will touch on
$\varepsilon$-productions later in this section.

\begin{figure}[htbp]
  \centering
  \includegraphics[scale = 0.73]{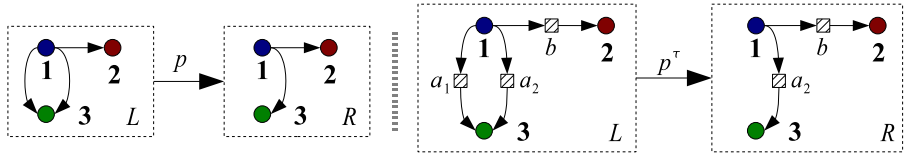}
  \caption{Multidigraph with Two Outgoing Edges}
  \label{fig:simpleMultiGraph}
\end{figure}

\noindent\textbf{Example}.$\square$Consider the simple production in Fig.~\ref{fig:simpleMultiGraph} with two edges between nodes $1$ and $3$.
Multinodes are represented by square nodes while normal nodes are left
unchanged.  When $p$ deletes an edge, $p^\tau$ deletes a multinode.
Adjacency matrices for $p^\tau$ are:

\begin{equation}
  L = \left[
    \begin{array}{ccccccc}
      \vspace{-6pt}
      0 & 0 & 0 & 1 & 1 & 1 & \vert\; 1_{\phantom{2}} \\
      \vspace{-6pt}
      0 & 0 & 0 & 0 & 0 & 0 & \vert\; 2_{\phantom{2}} \\
      \vspace{-6pt}
      0 & 0 & 0 & 0 & 0 & 0 & \vert\; 3_{\phantom{2}} \\
      \vspace{-6pt}
      0 & 0 & 1 & 0 & 0 & 0 & \vert\; a_1 \\
      \vspace{-6pt}
      0 & 0 & 1 & 0 & 0 & 0 & \vert\; a_2 \\
      \vspace{-6pt}
      0 & 1 & 0 & 0 & 0 & 0 & \vert\; b_{\phantom{2}} \\
      \vspace{-10pt}
    \end{array} \right] \qquad
  R = \left[
    \begin{array}{cccccc}
      \vspace{-6pt}
      0 & 0 & 0 & 1 & 1 & \vert\; 1_{\phantom{2}} \\
      \vspace{-6pt}
      0 & 0 & 0 & 0 & 0 & \vert\; 2_{\phantom{2}} \\
      \vspace{-6pt}
      0 & 0 & 0 & 0 & 0 & \vert\; 3_{\phantom{2}} \\
      \vspace{-6pt}
      0 & 0 & 1 & 0 & 0 & \vert\; a_2 \\
      \vspace{-6pt}
      0 & 1 & 0 & 0 & 0 & \vert\; b_{\phantom{2}} \\
      \vspace{-10pt}
    \end{array} \right] \nonumber
\end{equation}
\begin{equation}
  K = \left[
    \begin{array}{ccccccc}
      \vspace{-6pt}
      0 & 0 & 0 & 0 & 0 & 0 & \vert\; 1_{\phantom{2}} \\
      \vspace{-6pt}
      0 & 0 & 0 & 1 & 0 & 0 & \vert\; 2_{\phantom{2}} \\
      \vspace{-6pt}
      0 & 0 & 0 & 1 & 0 & 0 & \vert\; 3_{\phantom{2}} \\
      \vspace{-6pt}
      1 & 1 & 0 & 1 & 1 & 1 & \vert\; a_1 \\
      \vspace{-6pt}
      0 & 0 & 0 & 1 & 0 & 0 & \vert\; a_2 \\
      \vspace{-6pt}
      0 & 0 & 0 & 1 & 0 & 0 & \vert\; b_{\phantom{2}} \\
      \vspace{-10pt}
    \end{array} \right] \qquad
  e = \left[
    \begin{array}{ccccccc}
      \vspace{-6pt}
      0 & 0 & 0 & 1 & 0 & 0 & \vert\; 1_{\phantom{2}} \\
      \vspace{-6pt}
      0 & 0 & 0 & 0 & 0 & 0 & \vert\; 2_{\phantom{2}} \\
      \vspace{-6pt}
      0 & 0 & 0 & 0 & 0 & 0 & \vert\; 3_{\phantom{2}} \\
      \vspace{-6pt}
      0 & 0 & 1 & 0 & 0 & 0 & \vert\; a_1 \\
      \vspace{-6pt}
      0 & 0 & 0 & 0 & 0 & 0 & \vert\; a_2 \\
      \vspace{-6pt}
      0 & 0 & 0 & 0 & 0 & 0 & \vert\; b_{\phantom{2}} \\
      \vspace{-10pt}
    \end{array} \right]  \nonumber
\end{equation}

\proofend

Adjacency matrices are more sparse because simple nodes are not
directly connected by edges anymore.  Note that the number of edges
must be even.

In a real situation, a development tool such as AToM$^3$ should take
care of all these representation issues.  A user would see what
appears to the left of Fig.~\ref{fig:simpleMultiGraph} and not what is
depicted to the right of the same figure.  From a representation point
of view we can safely draw $p$ instead of $p^\tau$.  In fact,
according to Theorem~\ref{th:multiGraph}, it does not matter which one
is used.

Some restrictions on what a production can do to a multidigraph are
necessary in order to obtain a multidigraph again. Think for example
the case in which after applying some productions we get a graph in
which there is an isolated multinode (which would stand for an edge
with no source nor target nodes).

\begin{figure}[htbp]
  \centering
  \includegraphics[scale = 0.65]{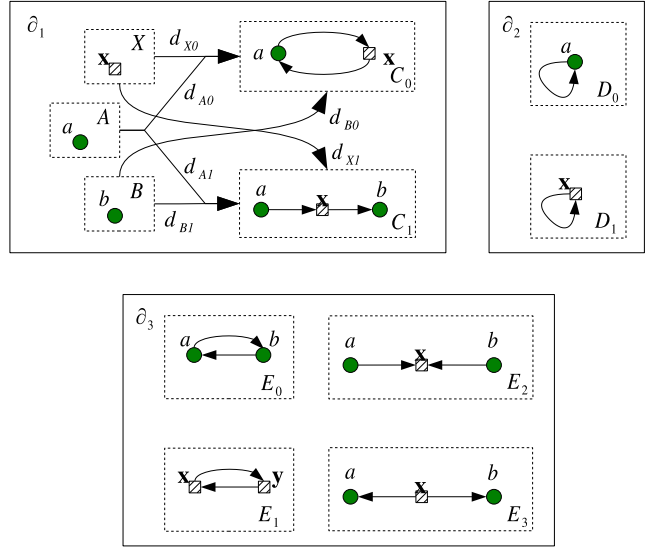}
  \caption{Multidigraph Constraints}
  \label{fig:multidigraphGC}
\end{figure}

The question is to find the properties that define one edge and impose
them on multinodes as graph constraints.  This way, multinodes will
behave as edges.  In the bullets that follow, graphs between brackets
can be found in Fig.~\ref{fig:multidigraphGC}:
\begin{itemize}
\item One edge always connects two nodes (diagram $\mathfrak{d}_1$,
  digraphs $C_0$ and $C_1$).
\item Simple nodes can not be directly connected by one edge ($D_0$
  and $E_0$).  Now edges start in a simple node and end in a multinode
  or vice versa, linking simple nodes with multinodes but not simple
  nodes between them.
\item A multinode can not be directly connected to another multinode
  ($D_1$ and $E_1$).  The contrary would mean that an edge in the
  simple digraph case is incident to another edge, which is not
  possible.
\item Edges always have a single simple node as source ($E_2$) and a
  single simple node as target ($E_3$).\footnote{This condition can be
    relaxed in case hyperedges were considered. See Sec.~\ref{sec:hyperedgeReplacement}.}
\end{itemize}

The graph constraint consists of three parts: First diagram
$\mathfrak{d}_1$ is closely related to compatibility of the
multidigraph\footnote{Note that now there are ``two levels'' when
  talking about a graph.  For example, if we say compatibility we may
  mean compatibility of the multidigraph (left side in Fig.~\ref{fig:simpleMultiGraph}) or of the underlying simple digraph
  (right side in Fig.~\ref{fig:simpleMultiGraph}) which are quite
  different.  In the first case we talk about edges connecting nodes
  while in the second we speak of edges connecting some node with some
  multinode.}  and has associated formula:
\begin{equation}\label{eq:multidigraph_1}
  \mathfrak{f}_1 = \forall X \exists C_0 \exists C_1 \exists A \exists
  B \left[ X A \left(C_0 \vee B \, C_1 \right)\right].
\end{equation}
Diagram $\mathfrak{d}_2$ and formula
\begin{equation}\label{eq:multidigraph_2}
  \mathfrak{f}_2 = \forall D_0 \forall D_1 \left[ \overline{D}_0
    \overline{D}_1 \right]
\end{equation}
prevents that a simple node or a multinode could be linked by an edge
to itself.  Self loops should be represented as in
$C_0$.

Finally, when considering two or more simple nodes or multinodes,
configurations in diagram $\mathfrak{d}_3$ are not allowed.  Its
associated formula is:
\begin{equation}\label{eq:multidigraph_3}
  \mathfrak{f}_3 = \forall E_0 \forall E_1\forall E_2\forall E_3
  \left[ \overline{Q} \left( E_0 \right) \overline{Q} \left( E_1
    \right) \overline{E}_2 \, \overline{E}_3 \right].
\end{equation}

\index{multidigraph constraints}This set of constraints will be known
as \emph{multidigraph constrains}, and the abbreviation $MC = \left(
  \mathfrak{d}_1 \cup \mathfrak{d}_2 \cup \mathfrak{d}_3,
  \mathfrak{f}_1 \wedge \mathfrak{f}_2 \wedge \mathfrak{f}_3 \right)$
will be used. Refer to Fig.~\ref{fig:multidigraphGC}.

\begin{figure}[htbp]
  \centering
  \includegraphics[scale = 0.66]{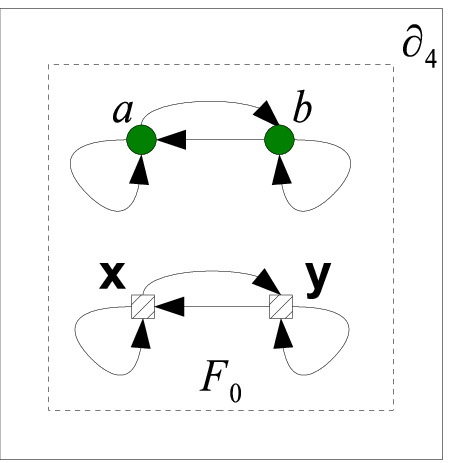}
  \caption{Simplified Diagram for Multidigraph Constraint}
  \label{fig:simplifiedDiagram}
\end{figure}

Some of these diagrams could be merged, also unifying (and simplifying
a little bit) their corresponding formulas.  For example, instead of
$D_0$, $D_1$, $E_0$ and $E_1$ we could have considered the diagram in
Fig.~\ref{fig:simplifiedDiagram}.  Its associated formula would have
been $\mathfrak{f}_4 = \forall F_0 \left[ \overline{Q}(F_0) \right]$.
However, a new constraint needs to consider the case in which a single
simple node or a single multinode is found in the host graph (as these
two cases are not taken into account by $\left( \mathfrak{d_4},
  \mathfrak{f}_4 \right)$).

\newtheorem{multiGraph}[matrixproduct]{Theorem}
\begin{multiGraph}[Multidigraphs]\label{th:multiGraph}
  Any multidigraph is isomorphic to some simple digraph $G$ together
  with multidigraph constraint $MC = \left( \mathfrak{f}, \mathfrak{d}
  \right)$, with $\mathfrak{d}$ as defined in
  Fig.~\ref{fig:multidigraphGC} and $\mathfrak{f}$ as
  in
  eqs.~\eqref{eq:multidigraph_1},~\eqref{eq:multidigraph_2}~and~\eqref{eq:multidigraph_3}.
\end{multiGraph}

\noindent \emph{Proof (sketch)}\\*
$\square$A graph with multiple edges $M=\left( V,E,s,t \right)$
consists of disjoint finite sets $V$ of nodes and $E$ of edges and
source and target functions $s:E \rightarrow V$ and $t:E \rightarrow
V$, respectively.  Function $v = s(e)$, $v \in V$, $e \in E$ returns
the node source $v$ for edge $e$.  We are considering multidigraphs
because the pair function $(s,t):E\rightarrow V\times V$ need not be
injective, i.e. several different edges may have the same source and
target nodes.  We have digraphs because there is a distinction between
source and target nodes.  This is the standard definition found in any
textbook.

It is clear that any $M$ can be represented as a multidigraph $G$
satisfying $MC$.  The converse also holds.  To see it, just consider
all possible combinations of two nodes and two multinodes and check
that any problematic situation is ruled out by $MC$. Induction
finishes the proof.\proofend

The multidigraph constraint $MC = \left( \mathfrak{f}, \mathfrak{d}
\right)$ must be fulfilled by any host graph.  If there is a
production $p:L \rightarrow R$ involved, $MC$ has to be transformed
into an application condition over $p$.  In fact, the multidigraph
constraint should be demanded both as precondition and postcondition
(recall that we can transform preconditions into postconditions and
vice versa). In Sec.~\ref{sec:graphConstraintsAndApplicationConditions}
we saw that this is an easy task in Matrix Graph Grammars: See
equations~\eqref{eq:equivAC_GC}~and~\eqref{eq:4}. This is a clear
advantage of being able to relate graph constraints and application
conditions.


This section is closed analyzing what behavior we have for
multidigraphs with respect to dangling edges.  With the theory as
developed so far, if a production specifies the deletion of a simple
node then an $\varepsilon$-production would delete any edge incident
to this simple node, connecting it to any surrounding multinode.  But
restrictions imposed by the multidigraph constraint do not allow this
so any production with potential dangling edges can not be applied.
Thus, we have a DPO-like behavior with respect to dangling edges for
multidigraphs.

In order to have a SPO-like behavior $\varepsilon$-productions need
to be restated, defining them at a multidigraph level, i.e.
$\varepsilon$-productions have to delete any potential ``dangling
multinode''.  \index{x@$\Xi$-production}A new type of productions
($\Xi$-productions) are introduced to get rid of annoying
edges\footnote{Edges connect simple nodes and multinodes.} that would
dangle when multinodes are also deleted by $\varepsilon$-productions.

We will not develop it in detail and will limit to describe the
concepts.  The way to proceed is very similar to what has been studied
in Sec.~\ref{sec:matchAndExtendedMatch}, by defining the appropriate
operator $T_\Xi$ and redefining $T_\varepsilon$.

\begin{figure}[htbp]
  \centering
  \includegraphics[scale = 0.61]{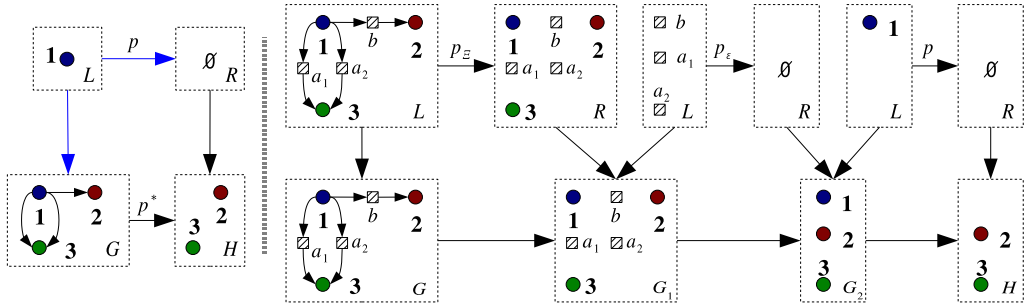}
  \caption{$\varepsilon$-production and $\Xi$-production}
  \label{fig:spoMultidigraph}
\end{figure}

A production $p:L \rightarrow R$ between multidigraphs that deletes
one simple node may give rise to one $\varepsilon$-production that
deletes one or more multinodes.  This $\varepsilon$-production can in
turn be applied only if any edge has already been erased, hence
possibly provoking the appearance of one $\Xi$-production.

This process is depicted in Fig.~\ref{fig:spoMultidigraph} where, in
order to apply production $p$, productions $p_\varepsilon$ and $p_\Xi$
need to be applied before
\begin{equation}
  p \longrightarrow p; p_\varepsilon; p_\Xi
\end{equation}

Eventually, one could simply compose the $\Xi$-production with its
$\varepsilon$-production, renaming it to $\varepsilon$-production and
defining it as the way to deal with dangling edges in case of multiple
edges, fully recovering a SPO-like behavior.  As commented above, a
potential user of a development tool such as AToM$^3$ would still see
things as in the simple digraph case, with no need to worry about
$\Xi$-productions.


Another theoretical use of application conditions and graph
constraints is the encoding of Turing Machines and Boolean
Circuits using Matrix Graph Grammars. See~\cite{MGG_computation}. In
Sec.~\ref{sec:mggTechniquesForPetriNets} we will see how to encode
Petri nets using Matrix Graph Grammars.

\section{Summary and Conclusions}
\label{sec:summaryAndConclusions7}

This chapter is a continuation of Chap.~\ref{ch:restrictionsOnRules}
in the study of graph constraints and application conditions. Besides,
we have seen how the nihilation matrix evolves with grammar rules. The
usefulness of the transformation of application conditions into
sequences is apparent in this chapter:

\begin{itemize}
\item to characterize properties such as consistency of application
  conditions and graph constraints in
  Sec.~\ref{sec:consistencyAndCompatibility};
\item to transform preconditions into postconditions and vice versa in
  Sec.~\ref{sec:movingConditions};
\item to extend MGG to deal with multidigraphs in
  Sec.~\ref{sec:fromSimpleDigraphsToMultidigraphs}.
\end{itemize}
 
We have also seen that to some extent application conditions are
delocalized inside sequences of productions. Besides, we have sketched
the usefulness of the analysis techniques of previous chapters to
study application conditions.

The next chapter addresses one fundamental topic in grammars:
Reachability.  This topic has been stated as
problem~\ref{prob:reachability} and is widely addressed in the
literature, specially in the theory of Petri nets. We will prove that
Petri nets can be interpreted as a proper subset of MGG, thus all
techniques developed so far can be used to study them. MGG will
benefit also from this relationship and algebraic techniques for
reachability in Petri nets will be generalized to cope with more
general grammars.

Chapter~\ref{ch:conclusionsAndFurtherResearch} closes the theory in
this book with a general summary, some more conclusions and proposals
for further research.  Appendix~\ref{app:caseStudy} presents a worked
out example to illustrate all the theory developed in this book,
focusing more on the \emph{practical side} of the theory.
\chapter{Reachability}
\label{ch:reachability}

In this chapter we will brush over reachability, presented as
problem~\ref{prob:reachability} in Sec.~\ref{sec:motivation}.  It is
an important concept for both, practice and theory.  Given a grammar
$\mathfrak{G}$ recall that, for some fixed initial $S_0$ and final
$S_T$ states, reachability solves the question of whether it is
possible to go from $S_0$ to $S_T$ with productions in $\mathfrak{G}$.
It should be of capital importance to provide one or more sequences
that carry this out, or identify that $S_T$ is unreachable.  At least,
it should be very valuable to gather some information of what
productions would be involved and the number of times that they
appear.

So far, this problem is easily related to (in the sense that it
depends on) problem~\ref{prob:applicability}, applicability, because
we look for a sequence applicable to $S_0$.  Also
problem~\ref{prob:sequentialIndependence} contributes because if it is
not possible to give a concrete sequence but a set of productions (the
order is unknown) together with the number of times that production
appears in the sequence, problem~\ref{prob:sequentialIndependence} may
reduce the size of the search space (to find out one concrete sequence
that transforms $S_0$ into $S_T$).

The chapter is organized as follows.
Section~\ref{sec:crashCourseInPetriNets} introduces Petri nets and
explains why in our opinion the state equation is a necessary but not
a sufficient condition.  In Sec.~\ref{sec:mggTechniquesForPetriNets}
Petri nets are interpreted as a proper subset of Matrix Graph
Grammars.  Also, the concept of \emph{initial marking} (minimal
initial digraph) is defined and the main concepts of Matrix Graph
Grammars are revisited for Petri nets.  The rest of the chapter
enlarges the state equation to cope with more general graph grammars.
We will make use of the tensor notation introduced in
Sec.~\ref{sec:tensorAlgebra}.  First, in Sec.~\ref{sec:dPOLikeMGG} for
fixed Matrix Graph Grammars (grammars with no dangling edges) and in
Sec.~\ref{sec:sPOLikeMGG} for general Matrix Graph Grammars (floating
grammars). As in every chapter, we finish with a summary in
Sec.~\ref{sec:summaryAndConclusions8} with some further comments, in
particular on other problems that can be addressed similarly to what
is done here for reachability.

\section{Crash Course in Petri Nets}
\label{sec:crashCourseInPetriNets}

\index{Petri net}A Petri net (also a Place/Transition net or P/T net)
is a mathematical representation of a discrete distributed system,~\cite{Murata}.  The structure of the distributed system is depicted as
a bipartite digraph.  \index{place}\index{transition}There are place
nodes, transition nodes and arcs connecting places with transitions.
\index{token}A place may contain any number of tokens.
\index{marking}A distribution of tokens over the places is called a
\emph{marking}.  \index{transition!enabled}A transition is enabled if
it can fire.  \index{transition!firing}When a transition fires
consumes tokens from its input places and puts a number of tokens in
its output places. The execution of Petri nets is non-deterministic,
so they are appropriate to model concurrent behaviour of distributed
systems. More formally,

\newtheorem{PNDef}[matrixproduct]{Definition}
\begin{PNDef}[Petri Net]
  \label{def:PNDef}
  \index{Petri net!definition}A Petri net is a 5-tuple $PN = \left( P,
    T, F, W, M_0 \right)$ where
  \begin{itemize}
  \item $P = \{p_1, \ldots, p_m\}$ is a finite set of places.
  \item $T = \{t_1, \ldots, t_n\}$ is a finite set of transitions.
  \item $F \subseteq (P \times T) \cup (T \times P)$ is a set of arcs.
  \item $W:F \rightarrow \mathbb{N}>1$ is a weight function.
  \item $M_0:P \rightarrow \mathbb{N}$ is the initial marking.
  \item $P \cap T = \emptyset$ and $P \cup T \neq \emptyset$.
  \end{itemize}
\end{PNDef}

The set of arcs establishes the flow direction. A \emph{Petri net
  structure} is the 4-tuple $N = \left( P, T, F, W \right)$ in which
the initial marking is not specified. Normally, a Petri net with a
initial marking is written $PN = \left( N, M_0 \right)$.

Algebraic techniques for Petri nets are based on the representation of
the net with an incidence matrix $A$ in which columns are transitions.
Element $A^i_j$ is the number of tokens that transition $i$ removes --
negative -- or adds -- positive -- to place $j$.

\index{reachability}One of the problems that can be analyzed using
algebraic techniques is \emph{reachability}.  Given an initial marking
$M_0$ and a final marking $M_d$, a necessary condition to reach $M_d$
from $M_0$ is to find a solution $x$ to the equation $M_d = M_0 + A
x$, which can be rewritten as the linear system
\begin{equation}\label{eq:necCond}
  M = A x.
\end{equation}

\index{Parikh vector}Solution $x$ -- known as \emph{Parikh vector} --
specifies the number of times that each transition should be fired,
but not the order.  \index{state equation}Identity~\eqref{eq:necCond}
is the \emph{state equation}.  Refer to~\cite{Murata} for a more
detailed explanation.

The ideas presented up to the end of the section are interpretations
of the author and should not be considered as standard in the theory
of Petri nets.

\begin{figure}[htbp]
  \centering
  \includegraphics[scale = 0.4]{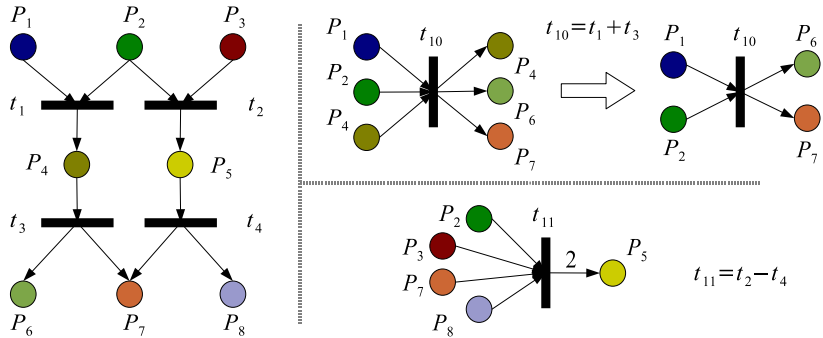}
  \caption{Linear Combinations in the Context of Petri Nets}
  \label{fig:ReprGeomOperaciones}
\end{figure}

The state equation introduces a matrix, which conceptually can be
thought of as associating a vector space to the dynamic behaviour of
the Petri net.  It is interesting to graphically interpret the
operations involved in linear combinations: Addition and
multiplication by scalars, as depicted in
Fig.~\ref{fig:ReprGeomOperaciones}.

The addition of two transitions is again a transition $t_k = t_i +
t_j$ for which input places are the addition of input places of every
transition and the same for output places.  If a place appears as
input and output place in $t_k$, then it can be removed.

Multiplication by $-1$ inverts the transition, i.e. input places
become output places and vice versa, which in some sense is equivalent
to disapplying the transition.

One important issue is that of notation.  Linear algebra uses an
additive notation (addition and subtraction) which is normally
employed when an Abelian structure is under consideration.  For
non-commutative structures, such as permutation groups, the
multiplicative notation (composition and inverses) is preferred.  The
basic operation with productions is the definition of sequences
(concatenation) for which historically a multiplicative notation has
been chosen, but substituting composition ``$\circ$'' by the
concatenation ``$;$'' operation.\footnote{This is the reason why
  Chap.~\ref{ch:mggFundamentals1} introduces ``;'' to be read from
  right to left, contrary to the Graph Transformation Systems
  literature.}

From a conceptual point of view, we are interested in relating linear
combinations and sequences of productions.\footnote{Linear
  combinations are the building blocks of vector spaces, and the
  structure to be kept by matrix application.}  Note that, due to
commutativity, linear combinations do not have an associated notion of
ordering, e.g. linear combination $PV_1 = p_1 + 2p_2 + p_3$ coming
from Parikh vector $\left[ 1, 2, 1 \right]$ can represent sequences
$p_1; p_2; p_3; p_2$ or $p_2; p_2; p_3; p_1$, which can be quite
different.  The fundamental concept that deals with commutativity is
precisely sequential independence.

Following this reasoning, we can find the problem that makes the state
equation a necessary but not a sufficient condition: Some transitions
can temporarily owe some tokens to the net.  The Parikh vector
specifies a linear combination of transitions and thus, negatives are
temporarily allowed (subtraction).

\newtheorem{StateEqSufCond}{Proposition}
\begin{StateEqSufCond}\label{prop:StateEqSufCond}
  Sufficiency of the state equation can only be ruined by transitions
  temporarily borrowing tokens from the Petri net.
\end{StateEqSufCond}

\noindent\emph{Proof}\\*
$\square$If there are enough tokens in every place then the
transitions can be fired (equiv., productions can be applied).  In
this case the state equation guarantees reachability.  A negative
number of tokens in one place (temporarily) represents a coherence
problem in the sequence.  Note that due to the way in which Petri nets
are defined there can not be compatibility issues, hence reachability
depends exclusively on coherence.\proofend

In the proof we have used Matrix Graph Grammars concepts such as
sequences and coherence. Notice that we have not stated how a Petri
net is coded in Matrix Graph Grammars. This point is addressed in
Sec.~\ref{sec:mggTechniquesForPetriNets}.

Proposition~\ref{prop:StateEqSufCond} does not provide any criteria
based on the topology of the Petri net, as Theorems 16, 17, 18 and
Corollaries 2 and 3 in~\cite{Murata}, but contains the essential idea
in their proofs: The hypothesis in previously mentioned theorems
guarantee that cycles in the Petri net will not ruin coherence.

\section{MGG Techniques for Petri Nets}
\label{sec:mggTechniquesForPetriNets}

In this section we will brush over some of the concepts developed so
far for Matrix Graph Grammars and see how they can be applied to Petri
nets.  Given a Petri net, we will consider it as the initial host
graph in our Matrix Graph Grammar.

\begin{figure}[htbp]
  \centering
  \includegraphics[scale = 0.73]{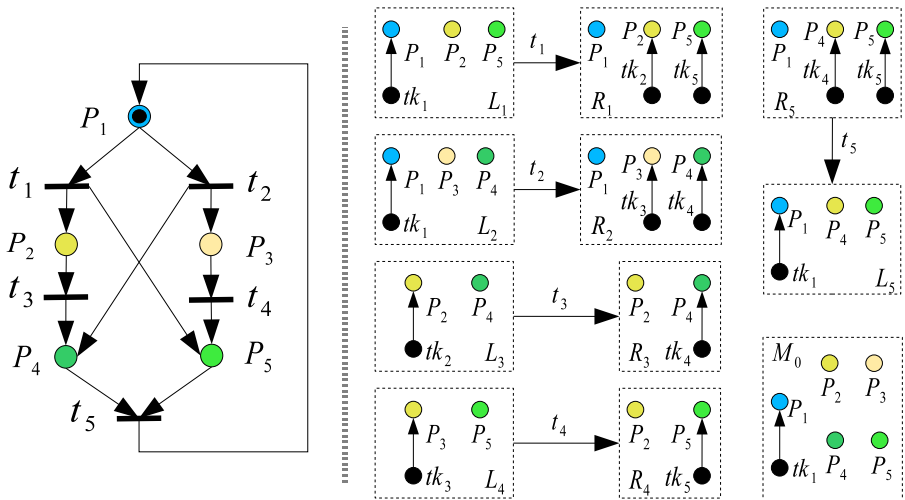}
  \caption{Petri Net with Related Production Set}
  \label{fig:PN_andAssociatedProds}
\end{figure}

One production is associated to every transition in which places and
tokens are nodes and there is an arrow joining each token to its
place.  In fact, we represent places for illustrative purposes only as
they are not strictly necessary (including tokens alone is enough).
Figure~\ref{fig:PN_andAssociatedProds} shows an example, where
production $p_i$ corresponds to transition $t_i$.  The firing of a
transition corresponds to the application of a rule.

Thus, Petri nets can be considered as a proper subset of Matrix Graph
Grammars with two important properties:
\begin{enumerate}
\item \label{prop:firstPN} There are no dangling edges when applying
  productions (firing transitions).
\item \label{prop:secondPN} Every production can only be applied in
  one part of the host graph.
\end{enumerate}

Properties (1) and (2) somehow allow us to safely ``ignore'' matchings
as introduced in Chap.~\ref{ch:matching}. In~\cite{MGG_computation}
nodeless MGGs are introduced. The main property of this submodel of
computation is to avoid dangling edges, as property (1)
above. Property (2) prevents one of the two types of non-determinism
associated to MGGs: where a production should be applied in case there
were more than one matching. Permitting non-determinism in which
production to apply is one of the characteristics of Petri nets,
useful to describe concurrence.

\index{Petri net!pure}We shall consider Petri nets with no
self-loops.\footnote{Petri nets without self-loops are called
  \emph{pure Petri nets}. A place $p$ and a transition $t$ are on a
  self-loop if $p$ is both an input and an output place of $t$.}
Translating to Matrix Graph Grammars, this means that one production
either adds or deletes nodes of a concrete type, but there is never a
simultaneous addition and deletion of nodes of the same type.  This
agrees with the expected behaviour of Matrix Graph Grammars
productions with respect to nodes (which is the behaviour of edges as
well, see Sec.~\ref{sec:characterizationAndBasicConcepts}) and will be
kept throughout the present chapter, mainly because rules in floating
grammars are adapted depending on whether a given production deletes
nodes or not (refer to Sec.~\ref{sec:sPOLikeMGG}).

\noindent\textbf{Remark}.$\square$It is advisable that elements are not
relative integers.  A number four must mean that production $p$ adds
four nodes of type $x$ and not that $p$ adds four nodes more than it
deletes of type $x$.  If we had one such production $p$, a possible
way to proceed is to split $p$ into two rules, one performing the
addition actions, $p_r$, and another for the deletion ones, $p_e$.
Sequentially, $p$ should be decomposed as $p = p_r; p_e$. \proofend

\begin{figure}[htbp]
  \centering
  \includegraphics[scale =
  0.61]{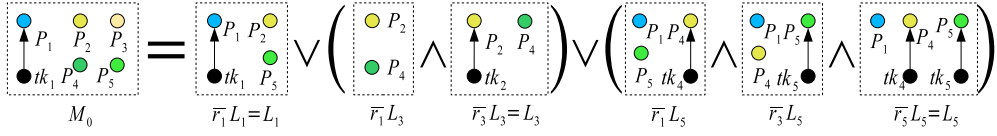}
  \caption{Minimal Marking Firing Sequence $t_5; t_3; t_1$}
  \label{fig:PN_MID}
\end{figure}

\textbf{Minimal Marking}.  \index{marking!minimal}The concept of
minimal initial digraph can be used to find the minimum marking able
to fire a given transition sequence.  For example,
Fig.~\ref{fig:PN_MID} shows the calculation of the minimal marking
able to fire transition sequence $t_5 ; t_3 ; t_1$ (from right to
left).  Notice that $(\overline r_1 L_1) \vee (\overline r_1
L_2)(\overline r_2 L_2) \vee \cdots \vee (\overline r_1 L_n) \cdots
(\overline r_n L_n)$ is the expanded form of
equation~\eqref{eq:FirstMinDigraph}.  The formula is transformed
according to $[1 \; 2 \; 3] \longmapsto [1 \; 3 \; 5]$.

\textbf{Reachability.} The reachability problem can also be expressed
using Matrix Graph Grammar concepts, as the following definition
shows.

\newtheorem{reachDef}[matrixproduct]{Definition}
\begin{reachDef}[Reachability]
  \label{def:reachability}
  \index{reachability}For a grammar $\mathfrak{G} = \left( M_0, \{
    p_1, \ldots, p_n\} \right)$, a state $M_d$ is called
  \emph{reachable} starting in state $M_0$, if there exists a coherent
  concatenation made up of productions $p_i \in \mathfrak{G}$ with
  minimal initial digraph contained in $M_0$ and image in $M_d$.
\end{reachDef}

This definition will be used to extend the state equation from Petri
nets to Matrix Graph Grammars.

\textbf{Compatibility and Coherence.}  \index{compatibility}As pointed
out in the proof of Prop.~\ref{prop:StateEqSufCond}, there can not be
compatibility issues for Petri nets as no dangling edge may ever
happen.  \index{coherence}Coherence of the sequence of transition
firing implies applicability (problem~\ref{prob:applicability}).  It
will be possible to unrelate problematic nodes (make the sequence
coherent) if there are enough nodes in the current state, which
eventually depends on the initial marking.


\section{Fixed Matrix Graph Grammars}
\label{sec:dPOLikeMGG}

In this and next sections we will be concerned with the generalization
of the state equation to wider types of grammars.

Recall from Sec.~\ref{sec:matchAndExtendedMatch} that by a fixed
Matrix Graph Grammar we understand a grammar as introduced in Chap.~\ref{ch:mggFundamentals1}, but in which rule application is not allowed
to generate dangling edges, i.e. any production $p$ that deletes a
node but not all of its incoming and outgoing edges can not be
applied. In other words, operator $T_\varepsilon$ is forced to be the
identity. Property~\ref{prop:secondPN} of Petri nets (see
Sec.~\ref{sec:mggTechniquesForPetriNets},
p.~\pageref{sec:mggTechniquesForPetriNets}) is relaxed because now a
single production may eventually be applied in several different
places of the host graph.

The approach of this section can be used with classical DPO graph
grammars~\cite{Fundamentals}.  However, following the discussion after
Prop.~\ref{prop:simpleEqualities} on
p.~\pageref{prop:simpleEqualities}, we restrict to DPO rules in which
nodes (or edges) of the same type are not rewritten (deleted and
created) in the same rule.

In order to perform an \emph{a priori} analysis it is mandatory to get
rid of matches. To this end, either an approach as proposed in
Chaps.~\ref{ch:mggFundamentals1},~\ref{ch:mggFundamentals2}~and~\ref{ch:matching}
is followed (as we did in Sec.~\ref{sec:mggTechniquesForPetriNets}) or
types of nodes are taken into account instead of nodes themselves.
The second alternative is chosen\footnote{Notice that this abstraction
  provokes information loss unless there is a single node per type.
  The problem here is that of non-determinism inside the host graph
  (where the production is to be applied).}  so productions, initial
state and final state are transformed such that types of elements are
considered, obtaining matrices with elements in $\mathbb{Z}$.

Tensor notation will be used in the rest of the chapter to extend the
state equation.  Although it will be avoided whenever possible, five
indexes may be used simultaneously, $\;{}^{E}_0\!\textrm{A}
^{i}_{jk}$.  Top left index indicates whether we are working with
nodes (N) or with edges (E).  Bottom left index specifies the position
inside a sequence, if any.  Top right and bottom right are
contravariant and covariant indexes, respectively, where $k = k_0$ is
the adjacency matrix (with types of elements, as commented above)
corresponding to production $p_{k_0}$.

\newtheorem{IncidenceMatrixNodes}[matrixproduct]{Definition}
\begin{IncidenceMatrixNodes} \label{def:IncidenceMatrixNodes} Let
  $\mathfrak{G} = \left( {}_0 M , \{p_1, \ldots, p_n\} \right)$ be a
  fixed graph grammar and $m$ the number of different types of nodes
  in $\mathfrak{G}$.  \index{incidence matrix}The incidence matrix for
  nodes ${}^{N}\!\mathrm{A} = \left(\mathrm{A}^i_k \right)$ where $i
  \in \{ 1, \ldots, n \}$ and $k \in \{ 1, \ldots, m\}$ is defined by
  the identity
  \begin{equation}
    \mathrm{A}^i_k = 
    \left\{ \begin{array}{ll}
        +r & \qquad\textrm{if production \emph{k} adds \emph{r} nodes of type \emph{i}} \\
        -r & \qquad\textrm{if production \emph{k} deletes \emph{r} nodes of type \emph{i}} \\
      \end{array} \right.
  \end{equation}
\end{IncidenceMatrixNodes}

It is straightforward to deduce for nodes an equation similar
to~\eqref{eq:necCond}:
\begin{equation}\label{eq:nodesEqComps}
  {}_d^N\!\mathrm{M}^i = {}_0^N\!\mathrm{M}^i + \sum_{k=1}^n
  {}^N\!\mathrm{A}_{k}^i \mathrm{x}^k.
\end{equation}

\index{incidence tensor!matrices}The case for edges is similar, with
the peculiarity that edges are represented by matrices instead of
vectors and thus the incidence matrix becomes the \emph{incidence
  tensor} ${}^E\!\mathrm{A}^i_{jk}$.  Again, only types of edges, and
not edges themselves, are taken into account.  Two edges $e_1$ and
$e_2$ are of the same type if their starting nodes are of the same
type and their terminal nodes are of the same type.

\begin{figure}[htbp]
  \centering
  \includegraphics[scale =
  0.55]{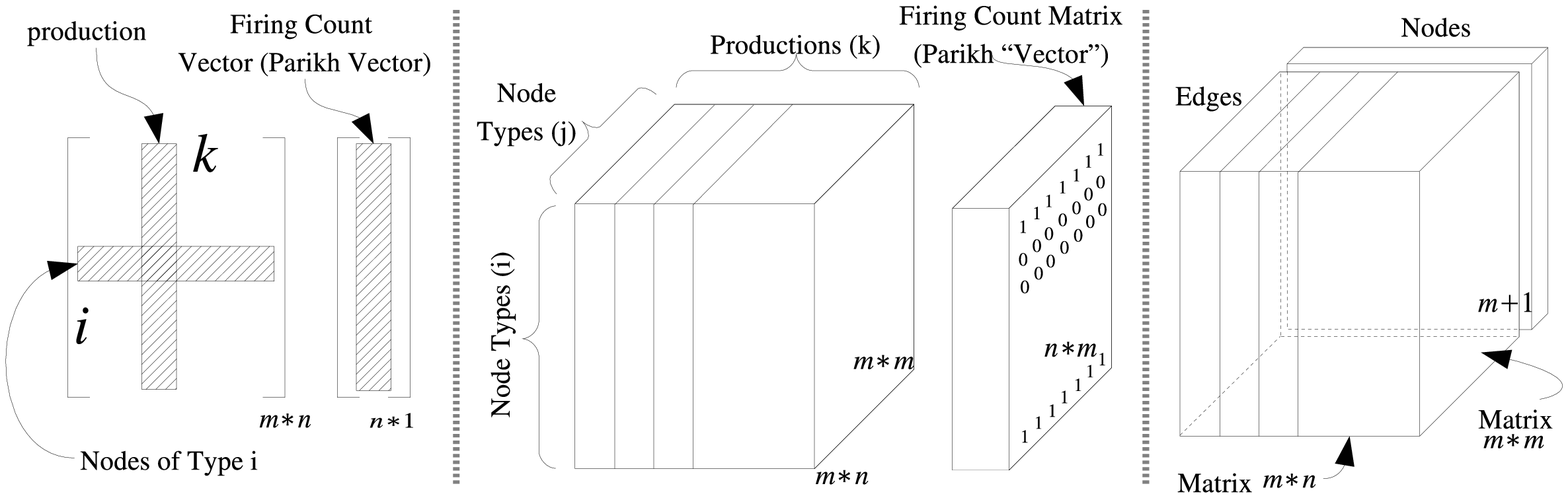}
  \caption{Matrix Representation for Nodes, Tensor for Edges and Their
    Coupling}
  \label{fig:MatricesAndTensorsExplanation}
\end{figure}

Source nodes will be assumed to have a contravariant behaviour (index
on top, $i$) while target nodes (first index, $j$) and productions
(second index, $k$) will behave covariantly (index on bottom).  See
diagram to the center of Fig.~\ref{fig:MatricesAndTensorsExplanation}.

\begin{figure}[htbp]
  \centering
  \includegraphics[scale = 0.45]{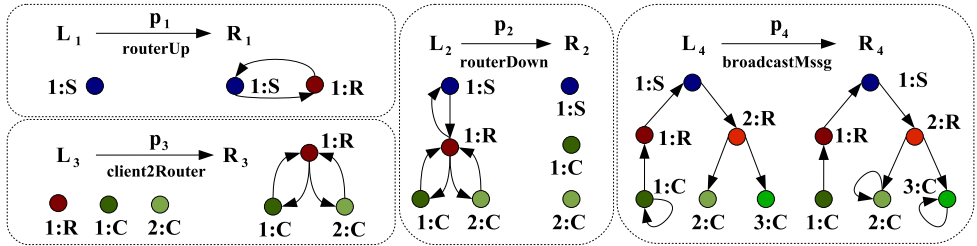}
  \caption{Rules for a Client-Server Broadcast-Limited System}
  \label{fig:ExampleClientServer}
\end{figure}

\noindent\textbf{Example}.$\square$Some rules for a simple client-server system
are defined in Fig.~\ref{fig:ExampleClientServer}.  There are three
types of nodes: Clients (C), servers (S) and routers (R), and messages
(self-loops in clients) can only be broadcasted.

In the Matrix Graph Grammar approach, this transformation system will
behave as a fixed or floating grammar depending on the initial state.
Note that production $p_4$ adds and deletes edges of the same type
$\left( C, C \right)$.  For now, the rule will not be split into its
addition and deletion components as suggested in
Sec.~\ref{sec:mggTechniquesForPetriNets}.  See
Subsec.~\ref{subsec:external} for an example of this splitting.

Incidence tensor (edges) for these rules can be represented
componentwise, each component being the matrix associated to the
corresponding production.

\begin{displaymath}
  {}^E\!\!A^i_{j1} = \left[
    \begin{array}{rrrl}
      \vspace{-6pt}
      0 & 0 & 0 & \vert\; C \\
      \vspace{-6pt}
      0 & 0 & 1 & \vert\; R \\
      \vspace{-6pt}
      0 & 1 & 0 & \vert\; S \\
      \vspace{-10pt}
    \end{array}
  \right] ; \quad
  {}^E\!\!A^i_{j2} = \left[
    \begin{array}{rrrl}
      \vspace{-6pt}
      0 & -2 & 0 & \vert\; C \\
      \vspace{-6pt}
      -2 & 0 & -1 & \vert\; R \\
      \vspace{-6pt}
      0 & -1 & 0 & \vert\; S \\
      \vspace{-10pt}
    \end{array}
  \right]
\end{displaymath}

\begin{displaymath}
  {}^E\!\!A^i_{j3} = \left[
    \begin{array}{cccl}
      \vspace{-6pt}
      0 & 2 & 0 & \vert\; C \\
      \vspace{-6pt}
      2 & 0 & 0 & \vert\; R \\
      \vspace{-6pt}
      0 & 0 & 0 & \vert\; S \\
      \vspace{-10pt}
    \end{array}
  \right] ; \qquad \quad \;
  {}^E\!\!A^i_{j4} = \left[
    \begin{array}{cccl}
      \vspace{-6pt}
      1 & 0 & 0 & \vert\; C \\
      \vspace{-6pt}
      0 & 0 & 0 & \vert\; R \\
      \vspace{-6pt}
      0 & 0 & 0 & \vert\; S \\
      \vspace{-10pt}
    \end{array}
  \right]
\end{displaymath}
Columns follow the same ordering $[C \; R \; S]$.\proofend

\newtheorem{reachDPOEdges}[matrixproduct]{Lemma}
\begin{reachDPOEdges}\label{lemma:ReachDPOEdges}
  With notation as above, a necessary condition for state
  ${}_d\mathrm{M}$ to be reachable from state ${}_0\mathrm{M}$ is
  \begin{equation}\label{eq:stateEqEdgesDPO}
    {}_d \mathrm{M} - {}_0 \mathrm{M}={}^E\!\mathrm{M} =
    {}^E\!\mathrm{M}_j^i = \sum_{k=1}^n {}^E\!\mathrm{A}_{\!jk}^i
    x^k_j = \sum_{\substack{k=1\\p=k}}^n\left( {}^E\!\mathrm{A}
      \otimes x \right)^{ip}_{jk},
  \end{equation}
  where $i, j \in \{1, \ldots , m \}$.
\end{reachDPOEdges}

Last equality in equation~\eqref{eq:stateEqEdgesDPO} is the definition
of and inner product -- see Sec.~\ref{sec:tensorAlgebra} -- so we
further have:
\begin{equation}
  {}_d \mathrm{M} - {}_0 \mathrm{M} = \left\langle {}^E\!A, x
  \right\rangle.
\end{equation}

\noindent \emph{Proof}\\*
$\square$Consider the construction depicted to the center of
Fig.~\ref{fig:MatricesAndTensorsExplanation} in which tensor
$A^i_{\!jk}$ is represented as a cube.  Setting $k=k_0$ fixes
production $p_{k_0}$.  A product for this object is defined in the
following way: Every vector in the cube perpendicular to matrix $x$
acts on the corresponding row of the matrix in the usual way, i.e. for
every fixed $i = i_0$ and $j = j_0$ in eq.~\eqref{eq:stateEqEdgesDPO},
\begin{equation}
  {}_d^E\!\mathrm{M}^{i_0}_{\!j_0} = {}_0^E\!\mathrm{M}^{i_0}_{\!j_0}
  + \sum_{k=1}^n {}^E\!\mathrm{A}^{i_0}_{\!j_0 k}
  \mathrm{x}^k_{\!j_0}.
  \label{eq:stateEqComps}
\end{equation}

Every column in matrix $x$ is a Parikh vector as defined for Petri
nets.  Its elements specify the amount of times that every production
must be applied, so all rows must be equal and hence
equation~\eqref{eq:stateEqComps} needs to be enlarged with some
additional identities:
\begin{equation}
  \left\{
    \begin{array}{rll}
      \mathrm{M}^i_{\!j} & = & \displaystyle \sum_{k=1}^{n}
      {}\!\mathrm{A}^i_{\!jk} \mathrm{x}^k_{\!j} \\
      x^k_p & = & x^k_q \\
    \end{array}
  \right.
\end{equation}
with $p,q \in \{ 1, \ldots, m\}$.  This uniqueness together with
previous equations provide the intuition to
raise~\eqref{eq:stateEqEdgesDPO}.

Informally, we are enlarging the space of possible solutions and then
projecting according to some restrictions.  To see that it is a
necessary condition suppose that there exists a sequence $s_n$ such
that $s_n \left( {}_0 \mathrm{M} \right) = {}_d \mathrm{M}$ and that
equation~\eqref{eq:stateEqComps} does not provide any solution.
Without loss of generality we may assume that the first column fails
(the one corresponding to nodes emerging from the first node), which
produces an equation completely analogous to the state equation for
Petri nets, deriving a contradiction.\proofend


\begin{figure}[htbp]
  \centering
  \includegraphics[scale = 0.5]{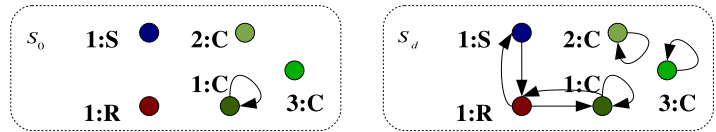}
  \caption{Initial and Final States for Productions in
    Fig.~\ref{fig:ExampleClientServer}}
  \label{fig:ExampleClientServerStates}
\end{figure}

\noindent\textbf{Example (Cont'd)}.$\square$Let's check whether it is possible
to move from state $S_0$ to state $S_d$ (see
Fig.~\ref{fig:ExampleClientServerStates}) with the productions defined
in Fig.~\ref{fig:ExampleClientServer} on
p.~\pageref{fig:ExampleClientServer}.  Matrices for the states (edges
only) and their difference are:
\begin{displaymath}
  {}^E\!\!S_0 = \left[
    \begin{array}{rrrl}
      \vspace{-6pt}
      1 & 0 & 0 & \vert\; C \\
      \vspace{-6pt}
      0 & 0 & 0 & \vert\; R \\
      \vspace{-6pt}
      0 & 0 & 0 & \vert\; S \\
      \vspace{-10pt}
    \end{array}
  \right] ; \;
  {}^E\!\!S_d = \left[
    \begin{array}{rrrl}
      \vspace{-6pt}
      3 & 1 & 0 & \vert\; C \\
      \vspace{-6pt}
      1 & 0 & 1 & \vert\; R \\
      \vspace{-6pt}
      0 & 1 & 0 & \vert\; S \\
      \vspace{-10pt}
    \end{array}
  \right] ; \;
  {}^E\!\!S = {}^E\!\!S_d - {}^E\!\!S_0 = \left[
    \begin{array}{rrrl}
      \vspace{-6pt}
      2 & 1 & 0 & \vert\; C \\
      \vspace{-6pt}
      1 & 0 & 1 & \vert\; R \\
      \vspace{-6pt}
      0 & 1 & 0 & \vert\; S \\
      \vspace{-10pt}
    \end{array}
  \right]
\end{displaymath}

The proof of Prop.~\ref{prop:ReachabilityDPO} poses the following
matrices, where the ordering on rows and columns is $[C \; R \; S]$:
\begin{displaymath}
  {}^E\!\!A^i_{\!1k} = \left[
    \begin{array}{rrrr}
      \vspace{-6pt}
      0 & 0 & 0 & 1  \\
      \vspace{-6pt}
      0 & -2 & 2 & 0 \\
      \vspace{-6pt}
      0 & 0 & 0 & 0 \\
      \vspace{-10pt}
    \end{array}
  \right] ; \quad 
  {}^E\!\!A^i_{\!2k} = \left[
    \begin{array}{rrrr}
      \vspace{-6pt}
      0 & -2 & 2 & 0 \\
      \vspace{-6pt}
      0 & 0 & 0 & 0  \\
      \vspace{-6pt}
      1 & -1 & 0 & 0 \\
      \vspace{-10pt}
    \end{array}
  \right] ; \quad 
  {}^E\!\!A^i_{\!3k} = \left[
    \begin{array}{rrrr}
      \vspace{-6pt}
      0 & 0 & 0 & 0  \\
      \vspace{-6pt}
      1 & -1 & 0 & 0 \\
      \vspace{-6pt}
      0 & 0 & 0 & 0 \\
      \vspace{-10pt}
    \end{array}
  \right].
\end{displaymath}

These matrices act on matrix $x = \left( x^p_q \right)$, $p \in \{ 1,
2, 3, 4 \}$, $q \in \{ 1, 2, 3 \}$ to obtain:
\begin{equation}\label{eq:exStEq}
  \begin{array}{rcl}
    {}^E\!S_1 & = & \displaystyle \sum_{k=1}^4 {}^E\!A_{1k} x^k_1 = \left[
      \begin{array}{c}
        x^4_1 \\ -2x^2_1 + 2x^3_1 \\ 0
      \end{array} 
    \right] \\
    && \\
    {}^E\!S_2 & = & \displaystyle \sum_{k=1}^4 {}^E\!A_{2k} x^k_2 = \left[
      \begin{array}{c}
        -2x^2_2 + 2x^3_2 \\ 0 \\ x^1_2 - x^2_2
      \end{array}
    \right] \\
    && \\
    {}^E\!S_3 & = & \displaystyle \sum_{k=1}^4 {}^E\!A_{3k} x^k_3 = \left[
      \begin{array}{c}
        0 \\ x^2_3 - x^3_3 \\ 0
      \end{array}
    \right]
  \end{array}
\end{equation}

Recall that $x$ must satisfy:
\begin{equation}
  x^1_1=x^1_2=x^1_3;\quad x^2_1=x^2_2=x^2_3;\quad
  x^3_1=x^3_2=x^3_3;\quad x^4_1=x^4_2=x^4_3. \nonumber
\end{equation}

A contradiction is derived for example with equations $x^2_3 = x^2_2$,
$1 = x^2_3 - x^3_3$, $x^3_2 = x^3_3$ and $1 = -2 x^2_2 +
2x^3_2$.\proofend

\noindent\textbf{Remark}.$\square$If there is no development tool handy and you
need to write equations~\eqref{eq:exStEq} it is useful to remember the
following rules of thumb:
\begin{itemize}
\item The subscript of $S$ coincides with the subscripts of all $x$
  and it is the terminal node for edges.  Hence, there will be as many
  equations in $S_i$ as types of terminal nodes to which modified
  edges arrive.  The first thing to do is a list of these nodes.
\item For a fixed $S_j$, there will be as many equations in the vector
  of variables as initial nodes for modified edges.  The terminal node
  is $j$ in this case.
\item The superscript of $x$ is the production.  To derive each
  equation just count how many edges of the same type are added and
  deleted and sum up.
\end{itemize}
For a larger example see Sec.~\ref{sec:reachabilityAndConfluence}.
\proofend

It is straightforward to derive a unique equation for reachability
which considers both nodes and edges,
i.e.
equations~\eqref{eq:nodesEqComps}~plus~\eqref{eq:stateEqEdgesDPO}.
This is accomplished extending the incidence matrix $M$ from $M: E
\rightarrow E$ to $M:E \times N \rightarrow E$ (from $M_{m \times m}$
to $M_{m \times \left( m+1 \right)}$), where column $m+1$ corresponds
to nodes.

\newtheorem{IncidenceTensor}[matrixproduct]{Definition}
\begin{IncidenceTensor}[Incidence Tensor]\label{def:IncidenceTensor}
  \index{incidence tensor}Let $\mathfrak{G} = \left( {}_0\!M, \{ p_1,
    \ldots, p_n\} \right)$ be a Matrix Graph Grammar.  The incidence
  tensor $A^i_{\!jk}$ with $i \in \{ 1, \ldots, m\}$ and $j \in \{ 1,
  \ldots , m+1 \}$ is defined by eq.~\eqref{eq:stateEqEdgesDPO} if $1
  \leq j \leq m$ and by eq.~\eqref{eq:nodesEqComps} if $j = m+1$.
\end{IncidenceTensor}

Top left index in our notation works as follows: ${}^N\!\!A$ refers to
nodes, ${}^E\!\!A$ to edges and $A$ to their coupling.  Note that a
similar construction can be carried out for productions if it was
desired to consider nodes and edges in a single expression.  Almost
all the theory as developed so far would remain without major
notational changes.  The exception would probably be compatibility
which would need to be rephrased.

An immediate extension of Lemma~\ref{lemma:ReachDPOEdges} is:

\newtheorem{reachDPOGeneral}[matrixproduct]{Proposition}
\begin{reachDPOGeneral}[State Equation for Fixed
  MGG]\label{prop:ReachabilityDPO}
  Let notation be as above.  A necessary condition for state
  ${}_d\mathrm{M}$ to be reachable (from state ${}_0\mathrm{M}$) is:
  \begin{equation}\label{eq:stateEqGeneralDPO}
    M_j^{\!i} = \sum_{k=1}^n A_{\!jk}^i x^k.
  \end{equation}
\end{reachDPOGeneral}

\noindent \emph{Proof}\\*
$\square$$\blacksquare$

Equation~\eqref{eq:stateEqGeneralDPO} is a generalization
of eq.~\eqref{eq:necCond} for Petri nets.  If there is just one place
of application for each production then the state equation as stated
for Petri nets is recovered.

\section{Floating Matrix Graph Grammars}
\label{sec:sPOLikeMGG}

Our intention now is to relax the first property of Petri nets
(Sec.~\ref{sec:mggTechniquesForPetriNets}, p.~\pageref{prop:firstPN})
and allow production application even though some dangling edge might
appear (see Chap.~\ref{ch:matching}).  The plan is carried out in two
stages which correspond to the subsections that follow, according to
the classification of $\varepsilon$-productions in
Sec.~\ref{sec:internalAndExternalProductions}.

In Matrix Graph Grammars, if applying a production $p_0$ causes
dangling edges then the production can be applied but a new production
(a so-called $\varepsilon$-production) is created and applied first.
In this way a sequence $p_0; p_{\varepsilon 0}$ is obtained with the
restriction that $p_{\varepsilon 0}$ is applied at a match that
includes all nodes deleted by $p_0$.  See Chap.~\ref{ch:matching} for
details.

Inside a sequence, a production $p_0$ that deletes an edge or node can
have an \emph{external} or \emph{internal} behaviour, depending on the
identifications carried out by the match.  Following
Chap.~\ref{ch:matching}, if the deleted element was added or used by a
previous production the production is labeled as \emph{internal}
(according to the sequence).  On the other hand, if the deleted
element is provided by the host graph and it is not used until $p_0$'s
turn, then $p_0$ is an external production.

Their properties are (somewhat) complementary: While external
$\varepsilon$-productions can be advanced and composed to eventually
get a single initial production which adapts the host graph to the
sequence, internal $\varepsilon$-productions are more
\emph{static}\footnote{Maybe it is possible to advance their
  application but, for sure, not to the beginning of the sequence.} in
nature.  On the other hand, internal $\varepsilon$-productions depend
on productions themselves and are somewhat independent of the host
graph, in contrast to external $\varepsilon$-productions.  Note
however that internal nodes can be unrelated if, for example,
matchings identified them in different parts of the host graph, thus
becoming external.

\subsection{External $\varepsilon$-production}
\label{subsec:external}

The main property of external $\varepsilon$-productions, compared to
internal ones, is that they act only on edges that appear in the
initial state, so their application can be advanced to the beginning
of the sequence.  In this situation, the first thing to know for a
given Matrix Graph Grammar $\mathfrak{G} = \left( {}_0\!M, \{ p_1,
  \ldots , p_n \} \right)$ -- with at most external
$\varepsilon$-productions -- when applied to ${}_0\!M$ is the maximum
number of edges that can be erased from its initial state.

The potential dangling edges (those with any incident node to be
erased) are given by
\begin{equation}
  \mathfrak{e} = \bigvee_{k=1}^n \left( \overline{\overline{{}^N_k\!e}
      \otimes \overline{{}^N_k\!e}} \right),
\end{equation}
which is closely related to the nihilation matrix introduced in
Sec.~\ref{sec:coherenceRevisited}, in particular in
Lemma~\ref{lemma:nihilationMatrix}.

\newtheorem{reachSPOexternal}[matrixproduct]{Proposition}
\begin{reachSPOexternal}\label{prop:reachSPOexternal}
  A necessary condition for state ${}_d\!M$ to be reachable (from
  state ${}_0\!M$) is:
  \begin{equation}\label{eq:ReachSPOexternal}
    M^i_{\!j} = \sum_{k=1}^n \left( A^i_{\!jk} x^k \right) + b^i_{\!j},
  \end{equation}
  with the restriction ${}_0\!M \mathfrak{e} \leq b^i_{\!j} \leq 0$.
\end{reachSPOexternal}

\noindent \emph{Proof (Sketch)}\\*
$\square$According to Sec.~\ref{sec:internalAndExternalProductions},
all $\varepsilon$-productions can be advanced to the beginning of the
sequence and can be composed to obtain a single production, adapting
the initial digraph before applying the sequence, which in some sense
interprets matrix $b$ as \emph{the} production number $n+1$ in the
sequence (the first to be applied).  Because it is not possible to
know in advance the order of application of productions, all we can do
is to provide bounds for the number of edges to be erased. This is in
essence what $b$ does.\proofend

Note that equation~\eqref{eq:stateEqGeneralDPO} in
Prop.~\ref{prop:ReachabilityDPO} is recovered
from~\eqref{eq:ReachSPOexternal} if there are no external
$\varepsilon$-productions.

\begin{figure}[htbp]
  \centering
  \includegraphics[scale = 0.5]{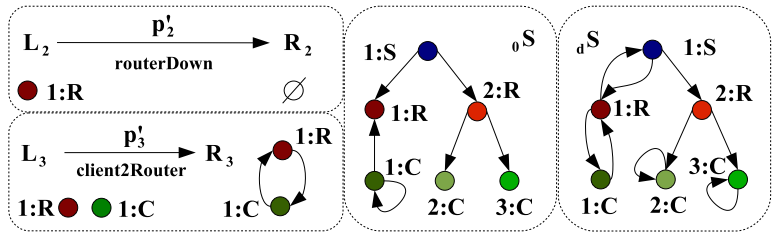}
  \caption{Initial and Final States (Based on Productions of Fig.
   ~\ref{fig:ExampleClientServer})}
  \label{fig:ExClientServerExternal}
\end{figure}

\label{ex:reachSPOexternal}\noindent\textbf{Example.}$\square$Consider the
initial and final states shown in
Fig.~\ref{fig:ExClientServerExternal}. Productions of previous
examples are used, but two of them are modified ($p_2$ and $p_3$).

In this case there are sequences that transform state ${}_0 S$ in
${}_d S$, for example, $s_4 = p_4; p'_3; p_1; p'_2$.  Note that the
problems are in edges $\left( 1:S, 1:R\right)$ and $\left( 1:C,
  1:R\right)$ of the initial state: Router $1$ is able to receive
packets from server $1$ and client $1$, but not to send them.

Next, matrices for the states and their difference are calculated.
The first three columns correspond to edges (first to clients, second
to routers and third to servers) and fourth to nodes which has been
split by a vertical line for illustrative purposes only.  The ordering
of nodes is $[C \; R \; S]$ both by columns and by rows.
\begin{displaymath}
  {}_0\!S = \left[
    \begin{array}{rrrc}
      \vspace{-6pt}
      1 & 1 & 0 & \vert\; 3 \\
      \vspace{-6pt}
      2 & 0 & 0 & \vert\; 2 \\
      \vspace{-6pt}
      0 & 2 & 0 & \vert\; 1 \\
      \vspace{-10pt}
    \end{array}
  \right] ; \quad
  {}_d\!S = \left[
    \begin{array}{rrrc}
      \vspace{-6pt}
      2 & 1 & 0 & \vert\; 3 \\
      \vspace{-6pt}
      3 & 0 & 1 & \vert\; 2 \\
      \vspace{-6pt}
      0 & 2 & 0 & \vert\; 1 \\
      \vspace{-10pt}
    \end{array}
  \right] ; \quad
  S = {}_d\!S - {}_0\!S = \left[
    \begin{array}{rrrc}
      \vspace{-6pt}
      1 & 0 & 0 & \vert\; 0 \\
      \vspace{-6pt}
      1 & 0 & 1 & \vert\; 0 \\
      \vspace{-6pt}
      0 & 0 & 0 & \vert\; 0 \\
      \vspace{-10pt}
    \end{array}
  \right]
\end{displaymath}

The incidence tensors for every production (recall that $p_2$ and
$p_3$ are as in Fig.~\ref{fig:ExClientServerExternal}) have the form
\begin{displaymath}
  A^i_{j1} = \left[
    \begin{array}{rrrcc}
      \vspace{-6pt}
      0 & 0 & 0 & \vert\; 0 & \vert\; C \\
      \vspace{-6pt}
      0 & 0 & 1 & \vert\; 1 & \vert\; R \\
      \vspace{-6pt}
      0 & 1 & 0 & \vert\; 0 & \vert\; S \\
      \vspace{-10pt}
    \end{array}
  \right] \qquad
  A^i_{j2} = \left[
    \begin{array}{rrrcc}
      \vspace{-6pt}
      0 & 0 & 0 & \vert\; \phantom{-}\phantom{,}0 & \vert\; C \\
      \vspace{-6pt}
      0 & 0 & 0 & \vert\, -1 & \vert\; R \\
      \vspace{-6pt}
      0 & 0 & 0 & \vert\; \phantom{-}\phantom{,}0 & \vert\; S \\
      \vspace{-10pt}
    \end{array}
  \right]
\end{displaymath}
\begin{displaymath}
  A^i_{j3} = \left[
    \begin{array}{ccccc}
      \vspace{-6pt}
      0 & 1 & 0 & \vert\; 0 & \vert\; C \\
      \vspace{-6pt}
      1 & 0 & 0 & \vert\; 0 & \vert\; R \\
      \vspace{-6pt}
      0 & 0 & 0 & \vert\; 0 & \vert\; S \\
      \vspace{-10pt}
    \end{array}
  \right] \quad \;\;\;
  A^i_{j4} = \left[
    \begin{array}{ccccc}
      \vspace{-6pt}
      1 & 0 & 0 & \vert\; 0 & \vert\; C \\
      \vspace{-6pt}
      0 & 0 & 0 & \vert\; 0 & \vert\; R \\
      \vspace{-6pt}
      0 & 0 & 0 & \vert\; 0 & \vert\; S \\
      \vspace{-10pt}
    \end{array}
  \right]
\end{displaymath}

Although it does not seem to be strictly necessary here, more
information is kept and calculations are more flexible if production
$p_4$ is split into the part that deletes messages and the part that
adds them, $p_4 = p_4^+;p_4^-$.  Refer to comments in
Sec.~\ref{sec:mggTechniquesForPetriNets}.

\begin{displaymath}
  A^{i-}_{j4} = \left[
    \begin{array}{ccccc}
      \vspace{-6pt}
      -1 & 0 & 0 & \vert\; 0 & \vert\; C \\
      \vspace{-6pt}
      0 & 0 & 0 & \vert\; 0 & \vert\; R \\
      \vspace{-6pt}
      0 & 0 & 0 & \vert\; 0 & \vert\; S \\
      \vspace{-10pt}
    \end{array}
  \right] \qquad
  A^{i+}_{j4} = \left[
    \begin{array}{ccccc}
      \vspace{-6pt}
      2 & 0 & 0 & \vert\; 0 & \vert\; C \\
      \vspace{-6pt}
      0 & 0 & 0 & \vert\; 0 & \vert\; R \\
      \vspace{-6pt}
      0 & 0 & 0 & \vert\; 0 & \vert\; S \\
      \vspace{-10pt}
    \end{array}
  \right]
\end{displaymath}

As in the example of Sec.~\ref{sec:dPOLikeMGG}, the following matrices
are more appropriate for calculations:
\begin{displaymath}
  A^i_{\!1k} \! = \! \left[
    \begin{array}{ccccc}
      \vspace{-6pt}
      0 & 0 & 0 & -1 & 2 \\
      \vspace{-6pt}
      0 & 0 & 1 & 0 & 0 \\
      \vspace{-6pt}
      0 & 0 & 0 & 0 & 0 \\
      \vspace{-10pt}
    \end{array}
  \right] \quad
  A^i_{\!2k}\! =\! \left[
    \begin{array}{ccccc}
      \vspace{-6pt}
      0 & 0 & 1 & 0 & 0 \\
      \vspace{-6pt}
      0 & 0 & 0 & 0 & 0 \\
      \vspace{-6pt}
      1 & 0 & 0 & 0 & 0 \\
      \vspace{-10pt}
    \end{array}
  \right]
\end{displaymath}
\begin{displaymath}
  A^i_{\!3k}\! =\! \left[
    \begin{array}{ccccc}
      \vspace{-6pt}
      0 & 0 & 0 & 0 & 0 \\
      \vspace{-6pt}
      1 & 0 & 0 & 0 & 0 \\
      \vspace{-6pt}
      0 & 0 & 0 & 0 & 0 \\
      \vspace{-10pt}
    \end{array}
  \right] \quad
  A^i_{\!4k}\! =\! \left[
    \begin{array}{ccccc}
      \vspace{-6pt}
      0 & 0 & 0 & 0 & 0 \\
      \vspace{-6pt}
      1 & -1 & 0 & 0 & 0 \\
      \vspace{-6pt}
      0 & 0 & 0 & 0 & 0 \\
      \vspace{-10pt}
    \end{array}
  \right]
\end{displaymath}

If equation~\eqref{eq:stateEqGeneralDPO} was directly applied, we
would get $x^1 = 0$ and $x^1 = 1$ (third row of $A^i_{\!2k}$ and
second of $A^i_{\!3k}$) deriving a contradiction.  The variations
permitted for the initial state are given by the matrix
\begin{equation}\label{eq:exLimit}
  {}_0\!M \mathfrak{e} = \left[
    \begin{array}{cccc}
      0 & \alpha^1_2 & 0 & 0 \\
      \alpha^2_1 & 0 & 0 & 0 \\
      0 & \alpha^3_2 & 0 & 0 \\
    \end{array}
  \right]
\end{equation}
with $\alpha^1_2 \in \{0,-1\}, \; \alpha^2_1, \alpha^3_2 \in
\{0,-1,-2\}$.  Setting $b^1_2 = -1$ and $b^3_2 = -1$ (one edge $\left(
  S, R \right)$ and one edge $\left( C, R \right)$ removed) the system
to be solved is
\begin{displaymath}
  \left[ \begin{array}{cccc}
      1 & 1 & 0 & 0 \\
      1 & 0 & 1 & 0 \\
      0 & 1 & 0 & 0 
    \end{array}	\right]
  =
  \left[ \begin{array}{cccc}
      -x^4+2x^4 \;& \; x^3 \; & 0 \; 			& 0 \\
      x^3 			\;& 0 				& \; x^1 \; & x^1-x^2 \\
      0 					& \; x^1 		& 0 				& 0 
    \end{array}	\right]
\end{displaymath}
with solution $x^1=x^2=x^3=x^4=1$, $s_4$ being one of its associated
sequences.  Notice that the restriction in
Prop.~\ref{prop:reachSPOexternal} is fulfilled, see
equation~\eqref{eq:exLimit}.\proofend

In previous example, as we knew a sequence ($s_4$) answer to the
reachability problem, we have fixed matrix $b$ directly to show how
Prop.~\ref{prop:reachSPOexternal} works.  Although this will not be
normally the case, the way to proceed is very similar: Relax matrix
$M$ by subtracting $b$, find a set of solutions $\{ x, b \}$ and
check whether the restriction for matrix $b$ is fulfilled or not.

\subsection{Internal $\varepsilon$-production}
\label{subsec:internal}

Internal $\varepsilon$-productions delete edges appended or used by
productions preceding it in the sequence.  In this subsection we first
limit to sequences which may have only internal
$\varepsilon$-productions and, by the end of the section, we will put
together Prop.~\ref{prop:reachSPOexternal} from
Subsec.~\ref{subsec:external} with results derived here to state
Theorem~\ref{th:reachGeneral} for floating Matrix Graph Grammars.

The proposed way to proceed is analogous to that of external
$\varepsilon$-productions.  The idea is to allow some variation in the
amount of edges erased by every production, but this variation is
constrained depending on the behaviour (definition) of the rest of the
rules.  Unfortunately, not so much information is gathered in this
case and what we are basically doing is ignoring this part of the
state equation.

Define $h^i_{\!jk} = \left[ A^i_{\!jk} \left( \mathfrak{e} \otimes
    \mathbb{I}_k \right)\right]^+ = \max \left( \left\{ A
    (\mathfrak{e}\otimes \mathbb{I}), 0 \right\} \right)$, where
vector $\mathbb{I}_k = [1, \ldots, 1]_{(1,k)}$.\footnote{$\mathfrak{e}
  \otimes \mathbb{I}(k)$ defines a tensor of type (1,2) which
  ``repeats'' matrix $\mathfrak{e}$ ``k'' times.}

\newtheorem{reachSPOinternal}[matrixproduct]{Proposition}
\begin{reachSPOinternal}\label{prop:reachSPOinternal}
  A necessary condition for state ${}_d\mathrm{M}$ to be reachable
  (from state ${}_0\mathrm{M}$) is:
  \begin{equation}\label{eq:reachSPOinternal}
    M^i_j = \sum_{k=1}^n \left( A^i_{\!jk} + V \right) x^k
  \end{equation}
  with the restriction $h^{i}_{\!jk} \leq V^i_{\!jk} \leq 0$.
\end{reachSPOinternal}

\noindent \emph{Proof}\\*
$\square$$\blacksquare$

In some sense, external $\varepsilon$-productions are the limiting
case of internal $\varepsilon$-productions and can be seen almost as a
particular case: As $\varepsilon$-productions do not interfere with
previous productions they have to act exclusively on the host graph.

The full generalization of the state equation for non-restricted
Matrix Graph Grammars is given in the next theorem.

\newtheorem{reachGeneral}[matrixproduct]{Theorem}
\begin{reachGeneral}[State Equation]\label{th:reachGeneral}
  \index{state equation}With notation as above, a necessary condition
  for state ${}_d\mathrm{M}$ to be reachable (from state
  ${}_0\mathrm{M}$) is
  \begin{equation}\label{eq:reachGeneral}
    M^i_{\!j} = \sum_{k=1}^n \left( A^i_{jk} + V \right) x^k + b^i_{\!j},
  \end{equation}
  with $b^i_{\!j}$ satisfying restrictions specified in
  Prop.~\ref{prop:reachSPOexternal} and $V$ satisfying those in
  Prop.~\ref{prop:reachSPOinternal}.
\end{reachGeneral}

\noindent \emph{Proof}\\*
$\square$$\blacksquare$

One interesting possibility of eq.~\eqref{eq:reachGeneral} is that we
can specify if productions acting on some edges must have a fixed or
floating behaviour, depending whether variances are permitted or not.

Strengthening hypothesis, formula~\eqref{eq:reachGeneral} becomes
those already studied for floating grammars with internal
$\varepsilon$-productions ($b=0$), with external
$\varepsilon$-productions ($V=0$), fixed grammars (from multilinear to
linear transformations) or Petri nets, fully recovering the original
form of the state equation.

\section{Summary and Conclusions}
\label{sec:summaryAndConclusions8}

The starting point of the present chapter is the study of Petri nets
as a particular case of Matrix Graph Grammars.  We have adapted
concepts of Matrix Graph Grammars to Petri nets, such as initial
marking.  Next, reachability and the state equation have been
reformulated and extended with the language of this approach, trying
to provide tools for grammars as general as possible.

Matrix Graph Grammars have also benefited from the theory developed
for Petri nets: Through the generalized state
equation~\eqref{eq:reachGeneral} it is possible to tackle
problem~\ref{prob:reachability}.

Despite the fact that the more general the grammar is, the less
information the state equation provides, Theorem~\ref{th:reachGeneral}
can be considered as a full generalization of the state equation.

Equation~\eqref{eq:reachGeneral} is more accurate as long as the rate
of the amount of types of nodes with respect to the amount of nodes
approaches one.  Hence, in general, it will be of little practical use
if there are many nodes but few types.

Although the use of vector spaces (as in Petri nets) and multilinear
algebra is almost straightforward, many other algebraic structures are
available to improve the results herein presented.  For example, Lie
algebras seem a good candidate if we think of the Lie bracket as a
measure of commutativity (recall Subsec.~\ref{sec:crashCourseInPetriNets} in which we saw that this is one of
the main problems of using linear combinations).

It should be possible to extend a little the Lie bracket to consider
two sequences instead of just two productions.\footnote{If sequences
  are coherent, composition can be used to recover a single
  production.}  With the theory of Chap.~\ref{ch:sequentializationAndParallelism} the case of one production
and one sequence can be directly addressed.

Other Petri nets concepts have algebraic characterizations and can be
studied with Matrix Graph Grammars.  Also, it is possible to extend
their definition from Petri nets to Matrix Graph Grammars.  A short
summary of some of them follows:
\begin{itemize}
\item \index{Petri net!conservative}\emph{Conservative} Petri nets are
  those for which the sum of the tokens remains constant.  For
  example, think of tokens as resources of the problem under
  consideration.
\item \index{invariants!place}\index{invariants!transition}An
  \emph{invariant} is some quantity that does not change during
  runtime.  They are divided in two main families: \emph{Place
    invariants} and \emph{transition invariants}.
\item \index{liveness}\emph{Liveness} studies whether transitions in a
  Petri net can be fired.  There are five levels ($L0$ to $L4$) with
  algebraic characterizations of necessary conditions.
\item \index{boundedness}\emph{Boundedness} of a Petri net studies the
  number of tokens in places (in particular if this number remains
  bounded).  Sufficient conditions are known.
\end{itemize}

Note that reachability can be directly used to study invariance under
sequences of initial states. If the initial state must not change, set
the initial and the final states as one and the same. This way, the
state equation must be equalized to zero. This is related to
termination because if there are sequences that leave some state
invariant, then there are cycles in the execution of the grammar,
preventing termination.

The book finishes in Chap.~\ref{ch:conclusionsAndFurtherResearch}, a
summary with further research proposals. Appendix~\ref{app:caseStudy}
presents a full worked out example that illustrates all relevant
concepts presented in this dissertation in a more or less realistic
case.  Its main objective is to show the use and practical utility of
compatibility, coherence, minimal and negative initial digraphs,
applicability, sequential independence and reachability.  In
particular properties of the system related to
problems~\ref{prob:applicability},~\ref{prob:sequentialIndependence}
and~\ref{prob:reachability} are addressed.
\chapter{Conclusions and Further Research}
\label{ch:conclusionsAndFurtherResearch}

This chapter closes the main body of the book. There is still Appendix
A. It includes a detailed real world case study in which much of the
theory developed so far is applied.

This chapter is organized in two sections. In
Sec.~\ref{sec:summaryAndShortTermResearch} we summarize the theory and 
highlight some topics that can be further investigated with Matrix
Graph Grammars as developed so
far. Sec.~\ref{sec:longTermResearchProgram} exposes a long term
program to address termination, confluence and complexity from the
point of view of Matrix Graph Grammars.

\section{Summary and Short Term Research}
\label{sec:summaryAndShortTermResearch}

In this book we have presented a new theory to study graph dynamics.
Also, increasingly difficult problems of graph grammars have been
addressed: Applicability, sequential independence and reachability.

First, two characterizations of \emph{action} over graphs (known as
productions or grammar rules) are defined, one emphasizing its static
part and one its dynamics.  To some extent it is possible to study
these actions without taking into account the initial state of the
system.  Hence, information on the grammar can be gathered at design
time, being potentially useful during runtime.  Nodes and edges are
considered independently, although related by compatibility.  It
should be possible, using the tensorial construction of Chap.~\ref{ch:reachability}, to define a single (algebraic) structure and
set compatibility as one of its axioms (a property to be fulfilled).

Sequences of productions are studied in great detail as they are
responsible for the dynamics of any grammar.  Composition, parallelism
and true concurrency have also been addressed.

The effect of a rule on a graph depends on where the rule is applied
(matching).  In Matrix Graph Grammars, matches are injective
morphisms.  As different productions in a sequence can be applied at
different places non-deterministically, \emph{marking} links parts of
productions guaranteeing their applicability on the same elements. It
is possible to define both matching and marking as operators acting on
productions.

Production application may have side effects, e.g. the deletion of
dangling edges. A special type of productions, known as
$\varepsilon$-productions, appear to keep compatibility.  It is shown
that they are the output of some operator acting on productions as
well as matching and marking.\footnote{Compatibility is a must. The
  operator may act appending new $\varepsilon$-productions, recovering
  a floating behavior or it can be ``deactivated'' getting a fixed
  behavior. Throughout this book we have focused on floating
  grammars, which are more general.} Operators for compatibility,
matching and marking can be translated into productions of a
sequence. This new perspective eases their analysis.

Minimal and negative initial digraphs are respectively generalized to
initial and negative digraph sets.  Two characterizations for
applicability are given.  One depends on coherence and compatibility
and the other on minimal and negative initial digraphs.

Sequential independence is closely related to commutativity, but with
the possibility to consider more than two elements at a time.  This
has been studied in the case of one production being advanced or
delayed an arbitrary (but finite) number of positions.

One interesting question is whether two sequences need the same
initial elements or not, especially when one is a permutation of the
other.  G-congruence and congruence conditions tackle this point again
for one production being advanced or delayed a finite number of
positions inside a sequence.  An interesting topic for further study
is to obtain similar results but considering moving blocks of
productions instead of a single production.

Graph constraints and particularly application conditions are of great
interest, mainly for two reasons: First, the left hand side and the
nihilation matrix are particular cases, and second it is possible to
deal with multidigraphs without any major modifications of the theory.
We have seen that application conditions are a particular case of
graph constraints and that a graph constraint can be reduced to an
application condition in the presence of a production.  Application
conditions can again be seen as operators acting on productions. This,
once more, means that they are equivalent to sequences of a certain
type. Among other things, this reduces the study of consistency of
application conditions to that of applicability.

As it is possible to transform preconditions into postconditions and
back again, they are in some sense \emph{delocalized} in a production.
Although this is sketched in some detail in
Chap.~\ref{ch:transformationOfRestrictions}, no concrete theorem is
established concerning the possibility to move application conditions
among productions inside a sequence.  We do not foresee, to the best
of our knowledge, any special difficulty in addressing this topic with
the theory developed so far.  This would be one application of
sequential independence -- problem~\ref{prob:sequentialIndependence}
-- to application conditions.

Finally, in order to consider reachability -- problem~\ref{prob:reachability} -- Petri nets are presented as a particular
case of Matrix Graph Grammars.  From this perspective, notions of
Matrix Graphs Grammars like the minimal initial digraph are directly
applied to Petri nets.  Also it is interesting that concepts and
results from Petri nets can be generalized to be included in Matrix
Graph Grammars.  Precisely, one example of this is reachability. Some
other concepts can also be investigated such as liveness, boundedness,
etc., and are left for future work.

For our research in reachability we have almost directly generalized
previous approaches (vector spaces) to reachability by using tensor
algebra.  It is worth studying other algebraic structures such as Lie
algebras.  Also, our study of reachability has not taken into account
the nihilation matrix nor application conditions, other two possible
directions for further research.

In our opinion, the main contribution of this book is the novelty of
the graph grammar representation, simple and powerful.  It naturally
brings in several branches of mathematics that can be applied to
Matrix Graph Grammars, allowing a potential use of advanced results to
solve old and new problems: First and second order logics, group
theory, tensor algebra, graph theory, category theory and functional
analysis.

\section{Long Term Research Program}
\label{sec:longTermResearchProgram}

On the practical side, as Appendix A shows, some tasks need to be
automated to ease further research. Manipulations can get rather
tedious and error prone. The development or improvement of a tool such
as AToM$^3$ would be very valuable. Besides, a good behavior of an
implementation of Matrix Graph Grammars is expected.

\begin{figure}[htbp]
  \centering
  \includegraphics[scale =
  0.38]{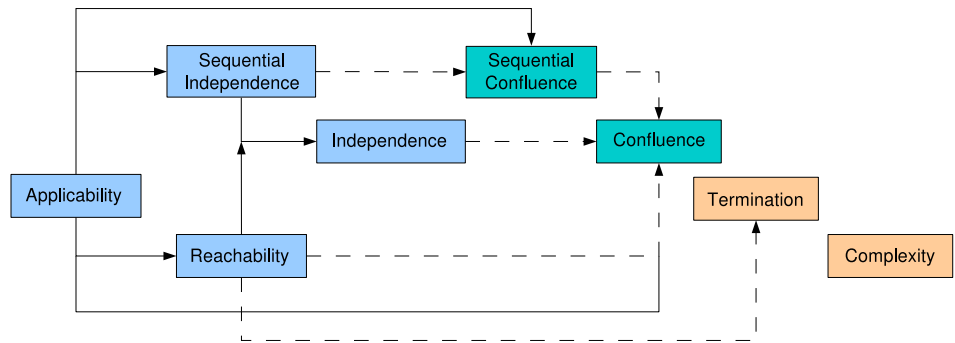}
  \caption{Diagram of Problem Dependencies.}
  \label{fig:problemsDepDiagram}
\end{figure}

At a more theoretical level we propose to study other three
increasingly difficult problems: Termination, confluence and
complexity. We think that the theory developed in this book can be
useful. See Fig.~\ref{fig:problemsDepDiagram}.

Termination, in essence, asks whether there is a solution for a given
problem (if some state is reached). In other branches of mathematics
this is the well-known concept of \emph{existence}. Reachability with
some improvements can be of help in two directions. Starting in some
initial state, if for some sequence of productions some invariant
state is reached, then the grammar can not be terminating (as it
enters a cycle as soon as it is reached). Second, to check the
invariance for a given state (if there exists some sequence that
leaves the graph unaltered), the state equation can also be used by
equaling the initial and final states.

If we have a terminating grammar we may wonder whether there is a
single final state or more than one: Confluence. In other branches of
mathematics this is the well-known concept of \emph{uniqueness}.
Sequential independence can be used in this case.

If a grammar is terminating and confluent, the next natural question
seems to be how much it takes to get to its final state. This is
complexity, which can also be addressed using Matrix Graph Grammars.
It is not difficult to interpret Matrix Graph Grammars as a new model
of computation, just as Boolean Circuits~\cite{Vollmer} or Turing
machines~\cite{Papadimitriou}. This is currently our main direction of
research. See~\cite{MGG_computation} for some initial results. The
main concept addressed in this book is sequentialization, whose
complexity is encoded the classes \textbf{P}, \textbf{NP} and more
generally in the Polynomial Hierarchy,
\textbf{PH}. See~\cite{Papadimitriou} for a comprehensive introduction
to this topic.

Notice that there are two properties that make Matrix Graph Grammars
differ from standard Turing machines: Its potential non-uniformity
(shared with Boolean Circuits) and the use of an oracle, in its
strongest version, whose associated decision problem is
\textbf{NP}-complete.

Non-uniformity is widely addressed in the theory of Boolean Circuits.
The same ideas possibly with some changes can be applied to Matrix
Graph Grammars.

The strongest version of Matrix Graph Grammars as introduced here use
an oracle whose associated decision problem is \textbf{NP}-complete:
The subgraph isomorphism problem, SI, to match the left hand side of a
production in the host graph. If problems that need to distinguish
lower level complexity classes (assuming \textbf{P}$\neq$\textbf{NP})
such as \textbf{P} are considered, it is possible to restrict
ourselves to some proper submodel of computation. For example, the
match operation can be forced to use GI instead.\footnote{GI, Graph
  Isomorphism, is widely believed not to be \textbf{NP}-complete,
  though this is still a conjecture. Problems that can be reduced to
  GI define the complexity class \textbf{GI}.}

Limitations on matching are not the only natural submodels of Matrix
Graph Grammars. The permitted operations can be constrained, for
example forbidding the addition and deletion of nodes (this would be
closely related to non-uniformity and the use of a
\textbf{GI}-complete problem rather than SI). Also, we can act on the
set of permitted graphs to derive submodels of computation. For
example, consider only those graphs with a single incoming and a
single outgoing edge in every node.\footnote{By the way, what standard
  and very well known mathematical structure is isomorphic to these
  graphs?.}
\appendix
\chapter{Case Study}
\label{app:caseStudy}

This Appendix presents a full worked out example that illustrates many
of the concepts and results of this book (more conceptual aspects such
as functional representations, adjoints and the like are omitted in
this appendix).  Although the aim of Matrix Graph Grammars is to be a
theoretical tool for the study of graph grammars and graph
transformation systems, we will see that it is also of practical
interest.

The case study herein presented tries to be simple enough to be
approached with paper and pencil but complex enough to look realistic.

As will be noticed throughout this appendix, Matrix Graph Grammars (as
well as any approach to graph transformation) encourages the
definition of a particular language to solve a particular problem.
\index{DSL, Domain-Specific Languages}These are known as
Domain-Specific languages (DSL). See~\cite{JVLC}.

Section~\ref{sec:PresentationOfTheScenario} presents an assembly line
with four types of machines (assembler, disassembler, quality and
packaging), one or more operators and some items to process.
Section~\ref{sec:sequences} presents some sequences and derivations,
together with possible states of the system.  Section~\ref{sec:initialDigraphSetsAndGcongruence} tackles minimal and
negative initial digraphs and G-congruence. As we progress, the
example will be enlarged to be more detailed. Section~\ref{sec:reachabilityAndConfluence} deals with
applicability, sequential independence, reachability and confluence.
Graph constraints and application conditions are exemplified in Sec.~\ref{sec:GraphConstraintsAndApplicationConditions}. Section~\ref{sec:derivations} returns to derivations, adding and modifying
productions. Dangling edges
and their treatment with $\varepsilon$-productions will show up
throughout the case study. 

\section{Presentation of the Scenario}
\label{sec:PresentationOfTheScenario}

In this section our sample scenario is set up. Some basic concepts
will be illustrated: Matrix representation of graphs and productions
(Sec.~\ref{sec:characterizationAndBasicConcepts}), compatibility
(Secs.~\ref{sec:graphTheory}, \ref{sec:characterizationAndBasicConcepts}~and~\ref{sec:compositionAndCompatibility}),
completion (Sec.~\ref{sec:completion}) and the nihilation matrix (Sec.~\ref{sec:coherenceRevisited}).

Our initial assembly line will consist of four machines that take as
input one or more items and output one or more items.  Depending on
the machine, items are processed transforming them into other items or
some decision is taken (reject, accept items) with no modification.

\begin{figure}[htbp]
  \centering
  \includegraphics[scale = 0.42]{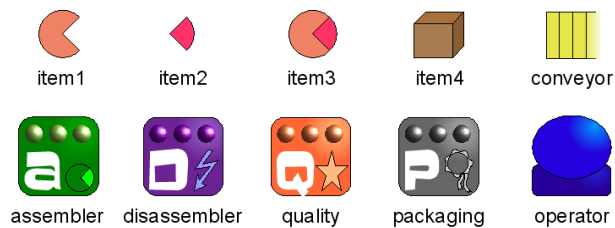}
  \caption{Graphical Representation of System Actors}
  \label{fig:exFigures}
\end{figure}

There are four types of items, \texttt{item1} -- \texttt{item4}.  One
assembly machine (named \texttt{assembler}, connected to two input
conveyors) processes one piece of \texttt{item1} and one piece of
\texttt{item2} to output in another conveyor one piece of type
\texttt{item3}.  There is a quality assurance machine --
\texttt{quality} -- that checks if \texttt{item3} fulfills certain
quality standards.  If it does, then \texttt{item3} is accepted and
packed to further produce \texttt{item4} through a \texttt{packaging} machine.
On the contrary, it is rejected and recycled through machine
\texttt{disassembler}. Machines need the presence of an
\texttt{operator} in order to work properly. Elements are graphically
represented in Fig.~\ref{fig:exFigures}.

\begin{figure}[htbp]
  \centering
  \includegraphics[scale = 0.4]{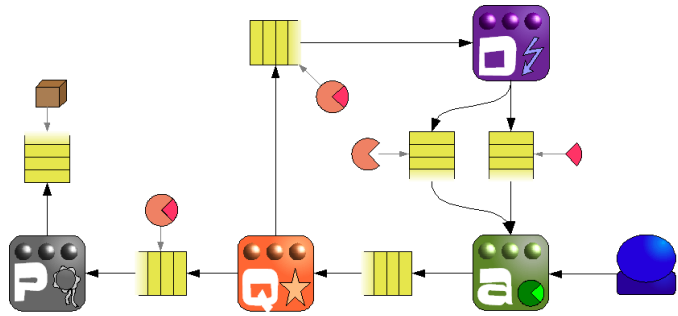}
  \caption{Snapshot of the Assembly Line}
  \label{fig:snapshot}
\end{figure}

In this case study types are those in Fig.~\ref{fig:exFigures}.  There
can be more than one element of each type, e.g there are six elements
of type conveyor in Fig.~\ref{fig:snapshot}, which shows a snapshot of
the state of an example of assembly line. For typing conventions refer
to comments on the example in p.~\pageref{ex:typing}.

Note that for now conveyors have infinite load capacity, elements in a
conveyor are not ordered and one operator can simultaneously manage
two or more machines.  It should be desirable that one operator
might look after different machines but only one at a time.  This can
be guaranteed only with graph constraints although if the initial
state fulfills this condition and productions observe this fact, there
should be no problem.  We will return to this point in
Sec.~\ref{sec:GraphConstraintsAndApplicationConditions}.

\begin{figure}[htbp]
  \centering
  \includegraphics[scale = 0.5]{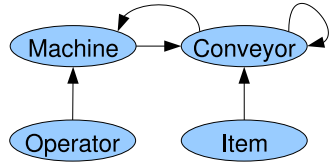}
  \caption{DSL Syntax Specification}
  \label{fig:DSLsyntax}
\end{figure}

When dealing with DSLs, it is customary to specify its syntax through
a meta-model. We will restrict connections among the different actors
of the system:
\begin{itemize}
\item Operators can only be connected to machines (by the end of Sec.~\ref{sec:sequences} this will be relaxed).
\item Items can only be connected to conveyors (until Sec.~\ref{sec:GraphConstraintsAndApplicationConditions} in which they
  will be allowed to be connected to other items).
\item Conveyors can only be connected to machines or to other
  conveyors.
\item Machines can be connected only to conveyors (by the end of Sec.~\ref{sec:sequences} this will be relaxed).
\end{itemize}
These restrictions have a natural graph representation (see
Fig.~\ref{fig:DSLsyntax}), which is sometimes referred to as typed
graphs,~\cite{Corradini}. Notice that for simplicity all actual types
have been omitted. For example, there should be four nodes for the 
different types of items ($\mathtt{item1}, \ldots, \mathtt{item4}$)
and the same for the machines.

\begin{figure}[htbp]
  \centering
  \includegraphics[scale = 0.28]{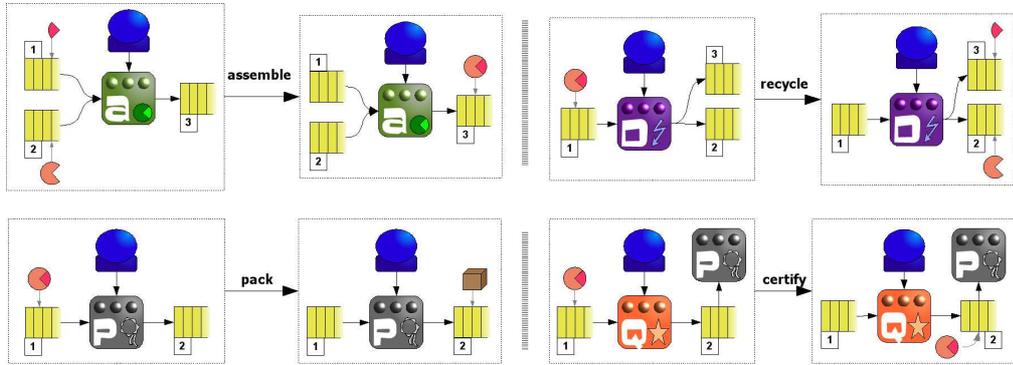}
  \caption{Basic Productions of the Assembly Line}
  \label{fig:basicProds}
\end{figure}

Now we describe the actions that can be performed.  These are the
grammar rules.  The state machine will evolve according to them.  See
Fig.~\ref{fig:basicProds} for the basic productions.  We will enlarge
or amend them and add some others in future sections.

Machines are not fully automatic so in this four productions one
operator is needed.  The four basic actions are assemble, disassemble,
certify and pack.  They correspond to productions \texttt{assemble},
\texttt{recycle}, \texttt{certify} and \texttt{pack}. Identifications
are obvious so they have not been made explicit (numbers between
different productions need not be related, i.e. \texttt{1:conv} in
production \texttt{assem} and \texttt{1:conv} in \texttt{certify} can
be differently identified in a host
graph). 

\begin{figure}[htbp]
  \centering
  \includegraphics[scale =
  0.32]{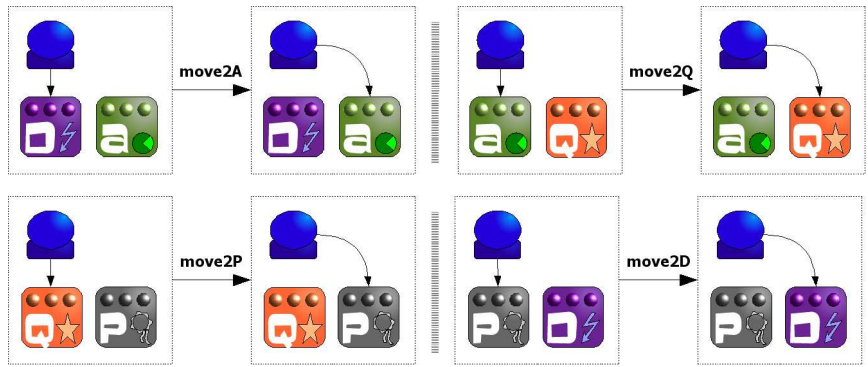}
  \caption{Productions for Operator Movement}
  \label{fig:moveOperator}
\end{figure}

There are four rules that permit operators to change from one machine
to another.  This movement is cyclic (to make the grammar a little bit
more interesting).  A practical justification could be that the
manager of the department obliges every operator passing near a
machine to check if there is any task pending, attending it just in
case.
We will start with a single operator to avoid collapses.  See grammar
rules \texttt{move2A}, \texttt{move2Q}, \texttt{move2P} and
\texttt{move2D} in Fig.~\ref{fig:moveOperator}.

\begin{figure}[htbp]
  \centering
  \includegraphics[scale = 0.36]{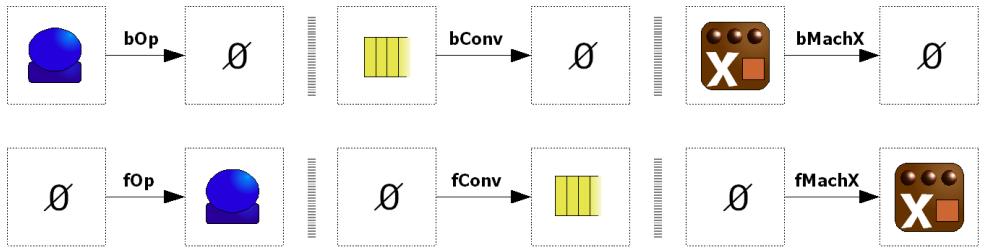}
  \caption{Break-Down and Fixing of Assembly Line Elements}
  \label{fig:breakDown}
\end{figure}

The last set of productions specify machines and operators break-down
(the 'b' in front of the productions).  Fortunately for the company
they can be fixed or replaced (the 'f' in front of the productions).
See Fig.~\ref{fig:breakDown} for the productions, where as usual
$\emptyset$ stands for the empty graph.  In order to save some space
we have summarized four rules (one per machine) substituting the name
of the machine by an $X$.  This is notationally convenient but we
should bear in mind that there are four rules for machines break down
(\texttt{bMachA}, \texttt{bMachQ}, \texttt{bMachP} and
\texttt{bMachD}) and another four for machines fixing
(\texttt{fMachA}, \texttt{fMachQ}, \texttt{fMachP} and
\texttt{fMachD}). Also, they can be thought of as abstract
rules\footnote{See reference~\cite{abstractRules}.} or variable nodes
as in Sec.~\ref{sec:fromSimpleDigraphsToMultidigraphs}.  The total
amount of grammar rules up to now is twenty.

Here we face the problem of $\varepsilon$-productions for the first
time.  If a conveyor with two items breaks (disappears) due to rule
\texttt{bConv}, there will be at least two dangling edges, one from
its input machine and another to its output machine.  These dangling
edges could be avoided defining one production per conveyor that takes
them into account.  If the conveyor had any item, then the
corresponding edge would also dangle.  Again this can be avoided if
there is a limit in the number of pieces that a conveyor can carry,
but a rule for each one is again needed.\footnote{A rule for the case
  in which a conveyor has one item, another for the case in which the
  conveyor has two items, etcetera.} Another possibility for DPO-like
graph transformation systems (what we have called fixed graph
grammars) is to define a sort of \emph{subgrammar} that takes care of
potential dangling edges. This subgrammar productions would be applied
iteratively until no edge can dangle.  This is a characteristic of
fixed graph transformation systems and in some situations can be a bit
annoying.  If there is no limit to the number of items (or the limit
is too high, e.g. a memory stack in a CPU RAM), it is possible to use
fixed graph grammars only to some extent.  Thus,
$\varepsilon$-productions are useful -- at times essential -- from a
practical point of view, among other things, to decrease the number of
productions in a grammar (this probably eases grammar definition and
maintenance and increases runtime efficiency).

Matrix representation of these rules is almost straightforward
according to Sec.~\ref{sec:characterizationAndBasicConcepts}.  We will
explicitly write the static (left and right hand sides) and dynamic
representations (deletion, addition and nihilation matrices) of
production \texttt{assemble}.

Elements are ordered \texttt{[1:item1 1:item2 1:conv 2:conv 3:conv
  1:macA 1:op]} for $L^E_{\textrm{assem}}$ and $L^N_{\textrm{assem}}$,
i.e. element $(1,3)$ of matrix $L^E_{\textrm{assem}}$ is the edge that
starts in node \texttt{(1:item1)} and ends in the first conveyor,
\texttt{(1:conv)}.  The ordering for productions
$R^E_{\textrm{assem}}$ and $R^N_{\textrm{assem}}$ is \texttt{[1:item3
  1:conv 2:conv 3:conv 1:macA 1:op]}.  Numbers in front of types are a
means to distinguish between elements of the same type in a given
graph (these are the numbers that appear in Fig.~\ref{fig:basicProds}).

\begin{displaymath}
  L^E_{assem} \! = \! \left[
    \begin{array}{ccccccc}
      \vspace{-6pt}
      0 & 0 & 1 & 0 & 0 & 0 & 0 \\
      \vspace{-6pt}
      0 & 0 & 0 & 1 & 0 & 0 & 0 \\
      \vspace{-6pt}
      0 & 0 & 0 & 0 & 0 & 1 & 0 \\
      \vspace{-6pt}
      0 & 0 & 0 & 0 & 0 & 1 & 0 \\
      \vspace{-6pt}
      0 & 0 & 0 & 0 & 0 & 0 & 0 \\
      \vspace{-6pt}
      0 & 0 & 0 & 0 & 1 & 0 & 0 \\
      \vspace{-6pt}
      0 & 0 & 0 & 0 & 0 & 1 & 0 \\
      \vspace{-10pt}
    \end{array}
  \right],
  R^E_{assem} \! = \! \left[
    \begin{array}{cccccc}
      \vspace{-6pt}
      0 & 0 & 0 & 1 & 0 & 0 \\
      \vspace{-6pt}
      0 & 0 & 0 & 0 & 1 & 0 \\
      \vspace{-6pt}
      0 & 0 & 0 & 0 & 1 & 0 \\
      \vspace{-6pt}
      0 & 0 & 0 & 0 & 0 & 0 \\
      \vspace{-6pt}
      0 & 0 & 0 & 1 & 0 & 0 \\
      \vspace{-6pt}
      0 & 0 & 0 & 0 & 1 & 0 \\
      \vspace{-10pt}
    \end{array}
  \right],
  L^N_{assem} \! = \! \left[
    \begin{array}{c}
      \vspace{-6pt}
      1 \\
      \vspace{-6pt}
      1 \\
      \vspace{-6pt}
      1 \\
      \vspace{-6pt}
      1 \\
      \vspace{-6pt}
      1 \\
      \vspace{-6pt}
      1 \\
      \vspace{-6pt}
      1 \\
      \vspace{-10pt}
    \end{array}
  \right], 
  R^N_{assem} \! = \! \left[
    \begin{array}{c}
      \vspace{-6pt}
      1 \\
      \vspace{-6pt}
      1 \\
      \vspace{-6pt}
      1 \\
      \vspace{-6pt}
      1 \\
      \vspace{-6pt}
      1 \\
      \vspace{-6pt}
      1 \\
      \vspace{-10pt}
    \end{array}
  \right].
\end{displaymath}

For $e^E$, $e^N$, $r^E$ and $r^N$ we have the same ordering of
elements.

\begin{displaymath}
  e^E_{assem} \! = \! \left[
    \begin{array}{ccccccc}
      \vspace{-6pt}
      0 & 0 & 1 & 0 & 0 & 0 & 0 \\
      \vspace{-6pt}
      0 & 0 & 0 & 1 & 0 & 0 & 0 \\
      \vspace{-6pt}
      0 & 0 & 0 & 0 & 0 & 0 & 0 \\
      \vspace{-6pt}
      0 & 0 & 0 & 0 & 0 & 0 & 0 \\
      \vspace{-6pt}
      0 & 0 & 0 & 0 & 0 & 0 & 0 \\
      \vspace{-6pt}
      0 & 0 & 0 & 0 & 0 & 0 & 0 \\
      \vspace{-6pt}
      0 & 0 & 0 & 0 & 0 & 0 & 0 \\
      \vspace{-10pt}
    \end{array}
  \right],
  r^E_{assem} \! = \! \left[
    \begin{array}{cccccc}
      \vspace{-6pt}
      0 & 0 & 0 & 1 & 0 & 0 \\
      \vspace{-6pt}
      0 & 0 & 0 & 0 & 0 & 0 \\
      \vspace{-6pt}
      0 & 0 & 0 & 0 & 0 & 0 \\
      \vspace{-6pt}
      0 & 0 & 0 & 0 & 0 & 0 \\
      \vspace{-6pt}
      0 & 0 & 0 & 0 & 0 & 0 \\
      \vspace{-6pt}
      0 & 0 & 0 & 0 & 0 & 0 \\
      \vspace{-10pt}
    \end{array}
  \right],
  e^N_{assem} \! = \! \left[
    \begin{array}{c}
      \vspace{-6pt}
      1 \\
      \vspace{-6pt}
      1 \\
      \vspace{-6pt}
      0 \\
      \vspace{-6pt}
      0 \\
      \vspace{-6pt}
      0 \\
      \vspace{-6pt}
      0 \\
      \vspace{-6pt}
      0 \\
      \vspace{-10pt}
    \end{array}
  \right],
  r^N_{assem} \! = \! \left[
    \begin{array}{c}
      \vspace{-6pt}
      1 \\
      \vspace{-6pt}
      0 \\
      \vspace{-6pt}
      0 \\
      \vspace{-6pt}
      0 \\
      \vspace{-6pt}
      0 \\
      \vspace{-6pt}
      0 \\
      \vspace{-10pt}
    \end{array}
  \right].
\end{displaymath}

The production is defined $R = p(L) = r \vee \overline{e} L$ both for
edges and for nodes.  To operate it is mandatory to complete the
matrices. See equation~\eqref{eq:nodesFirstEx} for the implicit
ordering of elements.

\begin{equation}\label{eq:assemMatrices1}
  \underbrace{\left[
      \begin{array}{cccccccc}
        \vspace{-6pt}
        0 & 0 & 0 & 0 & 0 & 0 & 0 & 0 \\
        \vspace{-6pt}
        0 & 0 & 0 & 0 & 0 & 0 & 0 & 0 \\
        \vspace{-6pt}
        0 & 0 & 0 & 0 & 0 & 1 & 0 & 0 \\
        \vspace{-6pt}
        0 & 0 & 0 & 0 & 0 & 0 & 1 & 0 \\
        \vspace{-6pt}
        0 & 0 & 0 & 0 & 0 & 0 & 1 & 0 \\
        \vspace{-6pt}
        0 & 0 & 0 & 0 & 0 & 0 & 0 & 0 \\
        \vspace{-6pt}
        0 & 0 & 0 & 0 & 0 & 1 & 0 & 0 \\
        \vspace{-6pt}
        0 & 0 & 0 & 0 & 0 & 0 & 1 & 0 \\
        \vspace{-10pt}
      \end{array}
    \right]}_{R^E_{assem}} = 
  \underbrace{\left[
      \begin{array}{cccccccc}
        \vspace{-6pt}
        0 & 0 & 0 & 0 & 0 & 0 & 0 & 0 \\
        \vspace{-6pt}
        0 & 0 & 0 & 0 & 0 & 0 & 0 & 0 \\
        \vspace{-6pt}
        0 & 0 & 0 & 0 & 0 & 1 & 0 & 0 \\
        \vspace{-6pt}
        0 & 0 & 0 & 0 & 0 & 0 & 0 & 0 \\
        \vspace{-6pt}
        0 & 0 & 0 & 0 & 0 & 0 & 0 & 0 \\
        \vspace{-6pt}
        0 & 0 & 0 & 0 & 0 & 0 & 0 & 0 \\
        \vspace{-6pt}
        0 & 0 & 0 & 0 & 0 & 0 & 0 & 0 \\
        \vspace{-6pt}
        0 & 0 & 0 & 0 & 0 & 0 & 0 & 0 \\
        \vspace{-10pt}
      \end{array}
    \right]}_{r^E_{assem}} \vee
  \underbrace{\overline{\left[
        \begin{array}{cccccccc}
          \vspace{-6pt}
          0 & 0 & 0 & 1 & 0 & 0 & 0 & 0 \\
          \vspace{-6pt}
          0 & 0 & 0 & 0 & 1 & 0 & 0 & 0 \\
          \vspace{-6pt}
          0 & 0 & 0 & 0 & 0 & 0 & 0 & 0 \\
          \vspace{-6pt}
          0 & 0 & 0 & 0 & 0 & 0 & 0 & 0 \\
          \vspace{-6pt}
          0 & 0 & 0 & 0 & 0 & 0 & 0 & 0 \\
          \vspace{-6pt}
          0 & 0 & 0 & 0 & 0 & 0 & 0 & 0 \\
          \vspace{-6pt}
          0 & 0 & 0 & 0 & 0 & 0 & 0 & 0 \\
          \vspace{-6pt}
          0 & 0 & 0 & 0 & 0 & 0 & 0 & 0 \\
          \vspace{-10pt}
        \end{array}
      \right]}}_{\overline{e^E_{assem}}}
  \underbrace{\left[
      \begin{array}{cccccccc}
        \vspace{-6pt}
        0 & 0 & 0 & 1 & 0 & 0 & 0 & 0 \\
        \vspace{-6pt}
        0 & 0 & 0 & 0 & 1 & 0 & 0 & 0 \\
        \vspace{-6pt}
        0 & 0 & 0 & 0 & 0 & 0 & 0 & 0 \\
        \vspace{-6pt}
        0 & 0 & 0 & 0 & 0 & 0 & 1 & 0 \\
        \vspace{-6pt}
        0 & 0 & 0 & 0 & 0 & 0 & 1 & 0 \\
        \vspace{-6pt}
        0 & 0 & 0 & 0 & 0 & 0 & 0 & 0 \\
        \vspace{-6pt}
        0 & 0 & 0 & 0 & 0 & 1 & 0 & 0 \\
        \vspace{-6pt}
        0 & 0 & 0 & 0 & 0 & 0 & 1 & 0 \\
        \vspace{-10pt}
      \end{array}
    \right]}_{L^E_{assem}}
\end{equation}

The expression for nodes is similar.  As pointed out in Sec.~\ref{sec:summaryAndConclusions7}, using a similar construction to that
of Sec.~\ref{sec:dPOLikeMGG} (in the definition of the incidence
tensor~\ref{def:IncidenceTensor}) it should be possible to get a
single expression for both nodes and edges instead of a formula for
edges and a formula for nodes.  This might be of interest for
implementations of Matrix Graph Grammars as more compact expressions
would be derived.

We shall mainly concentrate on edges because they define matrices
instead of just vectors and all problems such as inconsistencies
(dangling elements) come this way.

\begin{equation}\label{eq:nodesFirstEx}
  \underbrace{\left[
      \begin{array}{cl}
        \vspace{-6pt}
        0 & \vert \; \textrm{\texttt{1:item1}}\\
        \vspace{-6pt}
        0 & \vert \; \textrm{\texttt{1:item2}}\\
        \vspace{-6pt}
        1 & \vert \; \textrm{\texttt{1:item3}} \\
        \vspace{-6pt}
        1 & \vert \; \textrm{\texttt{1:conv}} \\
        \vspace{-6pt}
        1 & \vert \; \textrm{\texttt{2:conv}} \\
        \vspace{-6pt}
        1 & \vert \; \textrm{\texttt{3:conv}} \\
        \vspace{-6pt}
        1 & \vert \; \textrm{\texttt{1:machA}} \\
        \vspace{-6pt}
        1 & \vert \; \textrm{\texttt{1:op}} \\
        \vspace{-10pt}
      \end{array}
    \right]}_{R^N_{assem}} = 
  \underbrace{\left[
      \begin{array}{cl}
        \vspace{-6pt}
        0 & \vert \; \textrm{\texttt{1:item1}}\\
        \vspace{-6pt}
        0 & \vert \; \textrm{\texttt{1:item2}}\\
        \vspace{-6pt}
        1 & \vert \; \textrm{\texttt{1:item3}} \\
        \vspace{-6pt}
        0 & \vert \; \textrm{\texttt{1:conv}} \\
        \vspace{-6pt}
        0 & \vert \; \textrm{\texttt{2:conv}} \\
        \vspace{-6pt}
        0 & \vert \; \textrm{\texttt{3:conv}} \\
        \vspace{-6pt}
        0 & \vert \; \textrm{\texttt{1:machA}} \\
        \vspace{-6pt}
        0 & \vert \; \textrm{\texttt{1:op}} \\
        \vspace{-10pt}
      \end{array}
    \right]}_{r^N_{assem}} \vee
  \underbrace{\overline{\left[
        \begin{array}{cl}
          \vspace{-6pt}
          1 & \vert \; \textrm{\texttt{1:item1}}\\
          \vspace{-6pt}
          1 & \vert \; \textrm{\texttt{1:item2}}\\
          \vspace{-6pt}
          0 & \vert \; \textrm{\texttt{1:item3}} \\
          \vspace{-6pt}
          0 & \vert \; \textrm{\texttt{1:conv}} \\
          \vspace{-6pt}
          0 & \vert \; \textrm{\texttt{2:conv}} \\
          \vspace{-6pt}
          0 & \vert \; \textrm{\texttt{3:conv}} \\
          \vspace{-6pt}
          0 & \vert \; \textrm{\texttt{1:machA}} \\
          \vspace{-6pt}
          0 & \vert \; \textrm{\texttt{1:op}} \\
          \vspace{-10pt}
        \end{array}
      \right]}}_{\overline{e^N_{assem}}}
  \underbrace{\left[
      \begin{array}{cl}
        \vspace{-6pt}
        1 & \vert \; \textrm{\texttt{1:item1}}\\
        \vspace{-6pt}
        1 & \vert \; \textrm{\texttt{1:item2}}\\
        \vspace{-6pt}
        0 & \vert \; \textrm{\texttt{1:item3}} \\
        \vspace{-6pt}
        1 & \vert \; \textrm{\texttt{1:conv}} \\
        \vspace{-6pt}
        1 & \vert \; \textrm{\texttt{2:conv}} \\
        \vspace{-6pt}
        1 & \vert \; \textrm{\texttt{3:conv}} \\
        \vspace{-6pt}
        1 & \vert \; \textrm{\texttt{1:machA}} \\
        \vspace{-6pt}
        1 & \vert \; \textrm{\texttt{1:op}} \\
        \vspace{-10pt}
      \end{array}
    \right]}_{L^N_{assem}}
\end{equation}

Note that some elements in the node vectors are zero.  This means that
these nodes appear in the algebraic expressions but are not part of
the graphs.

The nihilation matrix in this case includes all edges incident to any
node that is deleted plus edges that are added by production
\texttt{assem}.  See Lemma~\ref{lemma:nihilationMatrix} for its
calculation formula:

\begin{equation}\label{eq:assemNihilMatrix}
  K_{\textrm{\texttt{assem}}} = \left[
    \begin{array}{ccccccccl}
      \vspace{-6pt}
      1 & 1 & 1 & 0 & 1 & 1 & 1 & 1 & \vert \; \textrm{\texttt{1:item1}} \\
      \vspace{-6pt}
      1 & 1 & 1 & 1 & 0 & 1 & 1 & 1 & \vert \; \textrm{\texttt{1:item2}} \\
      \vspace{-6pt}
      1 & 1 & 0 & 0 & 0 & 1 & 0 & 0 & \vert \; \textrm{\texttt{1:item3}} \\
      \vspace{-6pt}
      1 & 1 & 0 & 0 & 0 & 0 & 0 & 0 & \vert \; \textrm{\texttt{1:conv}} \\
      \vspace{-6pt}
      1 & 1 & 0 & 0 & 0 & 0 & 0 & 0 & \vert \; \textrm{\texttt{2:conv}} \\
      \vspace{-6pt}
      1 & 1 & 0 & 0 & 0 & 0 & 0 & 0 & \vert \; \textrm{\texttt{3:conv}} \\
      \vspace{-6pt}
      1 & 1 & 0 & 0 & 0 & 0 & 0 & 0 & \vert \; \textrm{\texttt{1:machA}} \\
      \vspace{-6pt}
      1 & 1 & 0 & 0 & 0 & 0 & 0 & 0 & \vert \; \textrm{\texttt{1:op}} \\
      \vspace{-10pt}
    \end{array}
  \right]
\end{equation}

Let's consider sequence \texttt{bOp;assem} to see how formula~\eqref{eq:compeq} works to check compatibility (Props.~\ref{prop:compatibilityUsingNorm}~and~\ref{prop:prodCompatibility}).
We can foresee a problem with edge \texttt{(1:op,1:machA)} because the
node disappears but not the edge.

According to eq.~\eqref{eq:SequenceCompatibility} we need to check
compatibility for the increasing set of sequences \texttt{s1 = assem}
and \texttt{s2 = bOp;assem}.  Note that the minimal initial digraph is
the same for both sequences and coincides with the left hand side of
\texttt{assem}.  Sequence \texttt{assem} is compatible, as the output
of production \texttt{assem} is a simple digraph again, i.e. rule
\texttt{assemble} is well defined:

\begin{eqnarray}
  \left\| \left[ s_1 \left( M_\textrm{\texttt{assem}}^E \right)
    \right.\right. & \! \vee & \!\left.\left.\left( s_1 \left(
          M_\textrm{\texttt{assem}}^E \right)\right)^t \right] \odot
    \overline{s_1 \left( M_\textrm{\texttt{assem}}^N \right)}
  \right\|_1 = \left\| \left[ R_\textrm{\texttt{assem}}^E \vee
      \left(R_\textrm{\texttt{assem}}^E\right)^t \right] \odot
    \overline{R_\textrm{\texttt{assem}}^N} \right\|_1 = \nonumber \\
  & = & \left\| \left(\left[
        \begin{array}{cccccccc}
          \vspace{-6pt}
          0 & 0 & 0 & 0 & 0 & 0 & 0 & 0 \\
          \vspace{-6pt}
          0 & 0 & 0 & 0 & 0 & 0 & 0 & 0 \\
          \vspace{-6pt}
          0 & 0 & 0 & 0 & 0 & 1 & 0 & 0 \\
          \vspace{-6pt}
          0 & 0 & 0 & 0 & 0 & 0 & 1 & 0 \\
          \vspace{-6pt}
          0 & 0 & 0 & 0 & 0 & 0 & 1 & 0 \\
          \vspace{-6pt}
          0 & 0 & 0 & 0 & 0 & 0 & 0 & 0 \\
          \vspace{-6pt}
          0 & 0 & 0 & 0 & 0 & 1 & 0 & 0 \\
          \vspace{-6pt}
          0 & 0 & 0 & 0 & 0 & 0 & 1 & 0 \\
          \vspace{-10pt}
        \end{array}
      \right] \vee \left[
        \begin{array}{cccccccc}
          \vspace{-6pt}
          0 & 0 & 0 & 0 & 0 & 0 & 0 & 0 \\
          \vspace{-6pt}
          0 & 0 & 0 & 0 & 0 & 0 & 0 & 0 \\
          \vspace{-6pt}
          0 & 0 & 0 & 0 & 0 & 0 & 0 & 0 \\
          \vspace{-6pt}
          0 & 0 & 0 & 0 & 0 & 0 & 0 & 0 \\
          \vspace{-6pt}
          0 & 0 & 0 & 0 & 0 & 0 & 0 & 0 \\
          \vspace{-6pt}
          0 & 0 & 1 & 0 & 0 & 0 & 1 & 0 \\
          \vspace{-6pt}
          0 & 0 & 0 & 1 & 1 & 0 & 0 & 1 \\
          \vspace{-6pt}
          0 & 0 & 0 & 0 & 0 & 0 & 0 & 0 \\
          \vspace{-10pt}
        \end{array}
      \right] \right) \odot \left[
      \begin{array}{c}
        \vspace{-6pt}
        1 \\
        \vspace{-6pt}
        1 \\
        \vspace{-6pt}
        0 \\
        \vspace{-6pt}
        0 \\
        \vspace{-6pt}
        0 \\
        \vspace{-6pt}
        0 \\
        \vspace{-6pt}
        0 \\
        \vspace{-6pt}
        0 \\
        \vspace{-10pt}
      \end{array}
    \right] \right\|_1 = 0 \nonumber
\end{eqnarray}

Thus, there is no problem with $s_1$.  Let's check $s_2$ out.
Operations are also easy for it.  Note that $r_{\texttt{bOp}} \vee
\overline{e}_{\texttt{bOp}} \left( R^E_{\texttt{assem}} \right) =
R^E_{\texttt{assem}}$, so:

\begin{eqnarray}
  \left\| \left[ s_2 \left( M^E \right) \right.\right. & \! \vee & \!
  \left.\left.\left( s_2 \left( M^E \right)\right)^t \right] \odot
    \overline{s_1 \left( M^N \right)} \right\|_1 = \left\| \left[
      \texttt{bOp} \left(R^E\right) \vee
      \left(\texttt{bOp}\left(R^E\right)\right)^t \right] \odot
    \overline{\textrm{\texttt{bOp}}\left(R^N\right)} \right\|_1
  \nonumber \\
  & = & \left\| \left[ R^E \vee \left( R^E \right)^t \right] \odot
    \overline{\texttt{bOp}\left(R^N\right)} \right\|_1 =
  \left\|\left[ \begin{array}{cccccccc}
        \vspace{-6pt}
        0 & 0 & 0 & 0 & 0 & 0 & 0 & 0 \\
        \vspace{-6pt}
        0 & 0 & 0 & 0 & 0 & 0 & 0 & 0 \\
        \vspace{-6pt}
        0 & 0 & 0 & 0 & 0 & 1 & 0 & 0 \\
        \vspace{-6pt}
        0 & 0 & 0 & 0 & 0 & 0 & 1 & 0 \\
        \vspace{-6pt}
        0 & 0 & 0 & 0 & 0 & 0 & 1 & 0 \\
        \vspace{-6pt}
        0 & 0 & 1 & 0 & 0 & 0 & 1 & 0 \\
        \vspace{-6pt}
        0 & 0 & 0 & 1 & 1 & 1 & 0 & \mathbf{1} \\
        \vspace{-6pt}
        0 & 0 & 0 & 0 & 0 & 0 & 1 & 0 \\
        \vspace{-10pt}
      \end{array}
    \right] \odot \left[
      \begin{array}{c}
        \vspace{-6pt}
        1 \\
        \vspace{-6pt}
        1 \\
        \vspace{-6pt}
        0 \\
        \vspace{-6pt}
        0 \\
        \vspace{-6pt}
        0 \\
        \vspace{-6pt}
        0 \\
        \vspace{-6pt}
        0 \\
        \vspace{-6pt}
        1 \\
        \vspace{-10pt}
      \end{array}
    \right] \right\|_1 = 1 \nonumber
\end{eqnarray}

This kind of formulas do not only assert compatibility for the
sequence, but also tells us which elements are problematic. In
previous equation we see that the final answer is 1 because of element
in position $(7,8)$ (bold).

In our case study as defined up to now, compatibility can only be
ruined by productions starting with a 'b' (\texttt{bOp}, etcetera).
Either an $\varepsilon$-production is appended or the result is not a
simple digraph (not a graph, actually).  Some information about
compatibility can be gathered at design time, on the basis of required
elements appearing on the left hand side of the productions, or
elements added.  For example, according to productions considered so
far any operator is connected to some machine so if production
\texttt{bOp} is applied it is very likely that some dangling edge will
appear.  Nihilation matrices can be automatically calculated as well
as completion of rules with respect to each other.



A typical snapshot of the evolution of our assembly line can be found
in Fig.~\ref{fig:snapshot}. It will be used in future sections as
initial state.

\section{Sequences}
\label{sec:sequences}

One topic not addressed in this book is how rules in a graph grammar
are selected for its application to an actual host graph.  There are
several possibilities.  To simplify the exposition rules will be
chosen randomly.  As commented in Secs.~\ref{sec:matchAndExtendedMatch}~and~\ref{sec:fromSimpleDigraphsToMultidigraphs}, this is the first -- out
of two -- source of non-determinism in graph transformation systems,
in particular in Matrix Graph Grammars.

\begin{figure}[htbp]
  \centering
  \includegraphics[scale = 0.4]{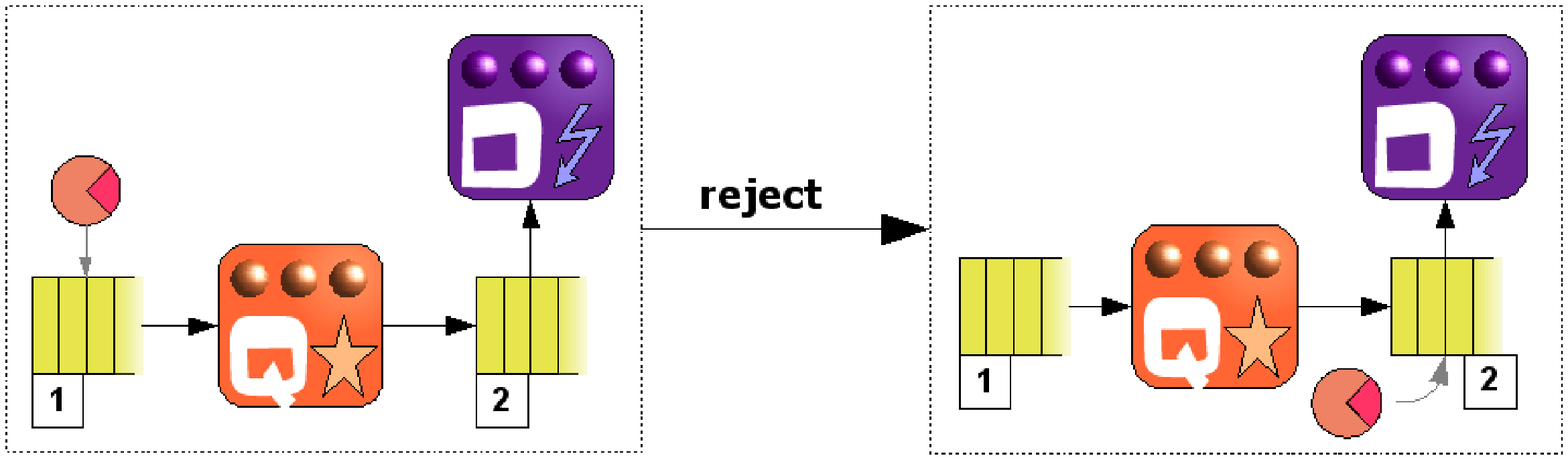}
  \caption{Graph Grammar Rule \texttt{reject}}
  \label{fig:ruleReject}
\end{figure}

We will add another rule -- \texttt{reject} -- that discards one
element once it has been assembled.  It is represented in Fig.~\ref{fig:ruleReject}.

We have two comments on this rule.  First, \texttt{reject} does not
need the presence of an operator to act, but it may also be applied if
an operator is on the machine.  Second, if grammar rules are applied
randomly following some probability distribution, elements will be
rejected according to the selected probability measure.


Let's begin with one sequence that starts with one piece of type
\texttt{item1} and one of type \texttt{item2} and produces one of type
\texttt{item4}:
\begin{equation}\label{eq:seqS0}
  \texttt{s}_\texttt{0} = \texttt{pack;certify;assem}
\end{equation}
which is compatible as no production generates any dangling edge.
Recall that compatibility also depends on the host graph: If
\texttt{item1} was connected to two different conveyors (should this
make any sense) then rule \texttt{assem} would produce one dangling
edge.

\begin{figure}[htbp]
  \centering
  \includegraphics[scale =
  0.36]{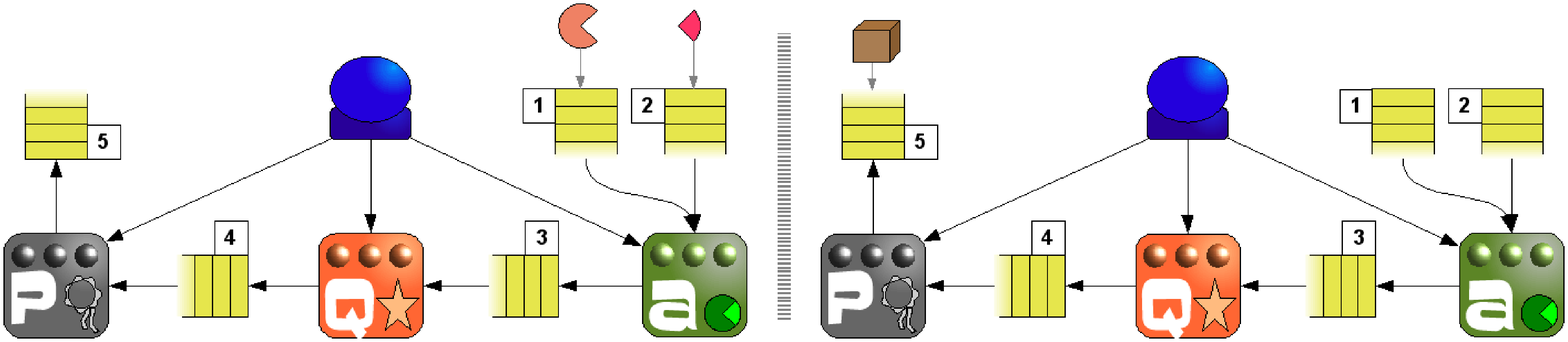}
  \caption{Minimal Initial Digraph and Image of Sequence
    $\texttt{s}_\texttt{0}$}
  \label{fig:minInitialDig_Ex}
\end{figure}

The minimal initial digraph of $\texttt{s}_\texttt{0}$ can be
calculated using eq.~\eqref{eq:FirstMinDigraph}),
$M_{\texttt{s}_\texttt{0}} = \nabla_1^3 \left( \overline{r_x} L_y
\right)$, where order of nodes is \texttt{[1:item1 1:item2 1:item3
  1:item4 1:conv 2:conv 3:conv 4:conv 5:conv 1:machA 1:machQ 1:machP
  1:op]}.  The completion we have performed identifies operators in
the productions as being the same.  Also, element \texttt{1:conv} in
rule \texttt{certify} (Fig.~\ref{fig:basicProds}) becomes
\texttt{3:conv} and \texttt{2:conv} is now \texttt{4:conv}.  Similar
manipulations have been performed for \texttt{pack}.  Theorem~\ref{th:minProdTh} demands coherence in order to apply eq.~(\ref{eq:FirstMinDigraph}), which is checked out in~\eqref{eq:s0_coherence}.  More attention will be paid to initial
digraphs in the next section.
\begin{equation}
  M_{\texttt{s}_\texttt{0}} = L_{\texttt{assem}} \vee \overline{r}_{\texttt{assem}} L_{\texttt{certify}} \vee \overline{r}_{\texttt{assem}} \overline{r}_{\texttt{certify}} L_{\texttt{pack}} = \left[
    \begin{array}{ccccccccccccc}
      \vspace{-6pt}
      0 & 0 & 0 & 0 & 1 & 0 & 0 & 0 & 0 & 0 & 0 & 0 & 0 \\
      \vspace{-6pt}
      0 & 0 & 0 & 0 & 0 & 1 & 0 & 0 & 0 & 0 & 0 & 0 & 0 \\
      \vspace{-6pt}
      0 & 0 & 0 & 0 & 0 & 0 & 0 & 0 & 0 & 0 & 0 & 0 & 0 \\
      \vspace{-6pt}
      0 & 0 & 0 & 0 & 0 & 0 & 0 & 0 & 0 & 0 & 0 & 1 & 0 \\
      \vspace{-6pt}
      0 & 0 & 0 & 0 & 0 & 0 & 0 & 0 & 0 & 1 & 0 & 0 & 0 \\
      \vspace{-6pt}
      0 & 0 & 0 & 0 & 0 & 0 & 0 & 0 & 0 & 1 & 0 & 0 & 0 \\
      \vspace{-6pt}
      0 & 0 & 0 & 0 & 0 & 0 & 0 & 0 & 0 & 0 & 0 & 0 & 0 \\
      \vspace{-6pt}
      0 & 0 & 0 & 0 & 0 & 0 & 0 & 0 & 0 & 0 & 0 & 1 & 0 \\
      \vspace{-6pt}
      0 & 0 & 0 & 0 & 0 & 0 & 0 & 0 & 0 & 0 & 0 & 0 & 0 \\
      \vspace{-6pt}
      0 & 0 & 0 & 0 & 0 & 0 & 0 & 1 & 0 & 0 & 0 & 0 & 0 \\
      \vspace{-6pt}
      0 & 0 & 0 & 0 & 0 & 0 & 0 & 0 & 0 & 0 & 0 & 0 & 0 \\
      \vspace{-6pt}
      0 & 0 & 0 & 0 & 0 & 0 & 0 & 1 & 1 & 0 & 0 & 0 & 0 \\
      \vspace{-6pt}
      0 & 0 & 0 & 0 & 0 & 0 & 0 & 0 & 0 & 1 & 1 & 1 & 0 \\
      \vspace{-10pt}
    \end{array}
  \right]
\end{equation}

The negative initial digraph is calculated using eq.~\eqref{eq:NID},
$K(\texttt{s}_\texttt{0}) = \nabla_1^3 \left( \overline{e}_x K_y
\right)$.  It is not shown in any figure because it has many edges.
In order to calculate $K(\texttt{s}_\texttt{0})$, the nihilation
matrices of productions \texttt{assem}~\eqref{eq:assemNihilMatrix},
\texttt{certify} and \texttt{pack} are needed.  Equation~\eqref{eq:nihilMatrix}, $K = p \left( \overline{D} \right)$, can be
used with the same ordering of nodes as for
$M_{\texttt{s}_\texttt{0}}$.

\begin{equation}
  K(\texttt{s}_\texttt{0}) = K_{\texttt{assem}} \vee
  \overline{e}_{\texttt{assem}} K_{\texttt{cert}} \vee
  \overline{e}_{\texttt{assem}} \overline{e}_{\texttt{cert}}
  K_{\texttt{pack}} = \left[
    \begin{array}{ccccccccccccc}
      \vspace{-6pt}
      1 & 1 & 1 & 1 & 0 & 1 & 1 & 1 & 1 & 1 & 1 & 1 & 1 \\
      \vspace{-6pt}
      1 & 1 & 1 & 1 & 1 & 0 & 1 & 1 & 1 & 1 & 1 & 1 & 1 \\
      \vspace{-6pt}
      1 & 1 & 1 & 1 & 1 & 1 & 0 & 1 & 1 & 1 & 1 & 1 & 1 \\
      \vspace{-6pt}
      0 & 0 & 0 & 0 & 0 & 0 & 0 & 1 & 0 & 0 & 0 & 0 & 0 \\
      \vspace{-6pt}
      1 & 1 & 1 & 0 & 0 & 0 & 0 & 0 & 0 & 0 & 0 & 0 & 0 \\
      \vspace{-6pt}
      1 & 1 & 1 & 0 & 0 & 0 & 0 & 0 & 0 & 0 & 0 & 0 & 0 \\
      \vspace{-6pt}
      1 & 1 & 1 & 0 & 0 & 0 & 0 & 0 & 0 & 0 & 0 & 0 & 0 \\
      \vspace{-6pt}
      1 & 1 & 1 & 0 & 0 & 0 & 0 & 0 & 0 & 0 & 0 & 0 & 0 \\
      \vspace{-6pt}
      1 & 1 & 1 & 0 & 0 & 0 & 0 & 0 & 0 & 0 & 0 & 0 & 0 \\
      \vspace{-6pt}
      1 & 1 & 1 & 0 & 0 & 0 & 0 & 0 & 0 & 0 & 0 & 0 & 0 \\
      \vspace{-6pt}
      1 & 1 & 1 & 0 & 0 & 0 & 0 & 0 & 0 & 0 & 0 & 0 & 0 \\
      \vspace{-6pt}
      1 & 1 & 1 & 0 & 0 & 0 & 0 & 0 & 0 & 0 & 0 & 0 & 0 \\
      \vspace{-6pt}
      1 & 1 & 1 & 0 & 0 & 0 & 0 & 0 & 0 & 0 & 0 & 0 & 0 \\
      \vspace{-10pt}
    \end{array}
  \right]
\end{equation}

The result of applying $\texttt{s}_\texttt{0}$ to
$M_{\texttt{s}_\texttt{0}}$ is given by eq.~\eqref{eq:EdgeConc},
$\texttt{s}_\texttt{0} \left( M^E_{\texttt{s}_\texttt{0}} \right) =
\bigwedge_{i=1}^3 \left( \overline{e^E_i} M^E_{\texttt{s}_\texttt{0}}
\right) \vee \bigtriangleup_1^3 \left( \overline{e^E_x} \, r^E_y
\right)$ and can be found to the right of
Fig.~\ref{fig:minInitialDig_Ex}.  For its calculation, it is possible
to interpret $\texttt{s}_\texttt{0}$ as a production according to the
remark that appears right after eq.~\eqref{eq:EdgeConc}.

Sequence $\texttt{s}_\texttt{0}$ is coherent with respect to the
identifications proposed in its minimal initial digraph
(Fig.~\ref{fig:minInitialDig_Ex}).  To see
this~\eqref{eq:CoherenceFormula} in
Theorem~\ref{th:SeqCoherenceTheorem} can be used, which once
simplified is eq.~\eqref{eq:ThreeProductions}:
\begin{eqnarray}
  L_{\texttt{cert}} e_{\texttt{assem}} & \vee & L_{\texttt{pack}}
  \left( e_{\texttt{assem}} \, \overline{r}_{\texttt{cert}} \vee
    e_{\texttt{cert}} \right) \vee \nonumber \\
  & \vee & R_{\texttt{assem}} \left( \overline{e}_{\texttt{cert}}
    r_{\texttt{pack}} \vee r_{\texttt{cert}} \right) \vee
  R_{\texttt{cert}} r_{\texttt{pack}} = 0.
  \label{eq:s0_coherence}
\end{eqnarray}

A very simple non-coherent sequence -- assuming that both rules act on
the same elements -- is $\texttt{t}_\texttt{0} = \texttt{reject} ;
\texttt{certify}$. It is obvious as both consume the same item. When
its coherence is calculated, not only will we be informed that
coherence fails but also what elements are responsible for this
failure.

\begin{figure}[htbp]
  \centering
  \includegraphics[scale = 0.35]{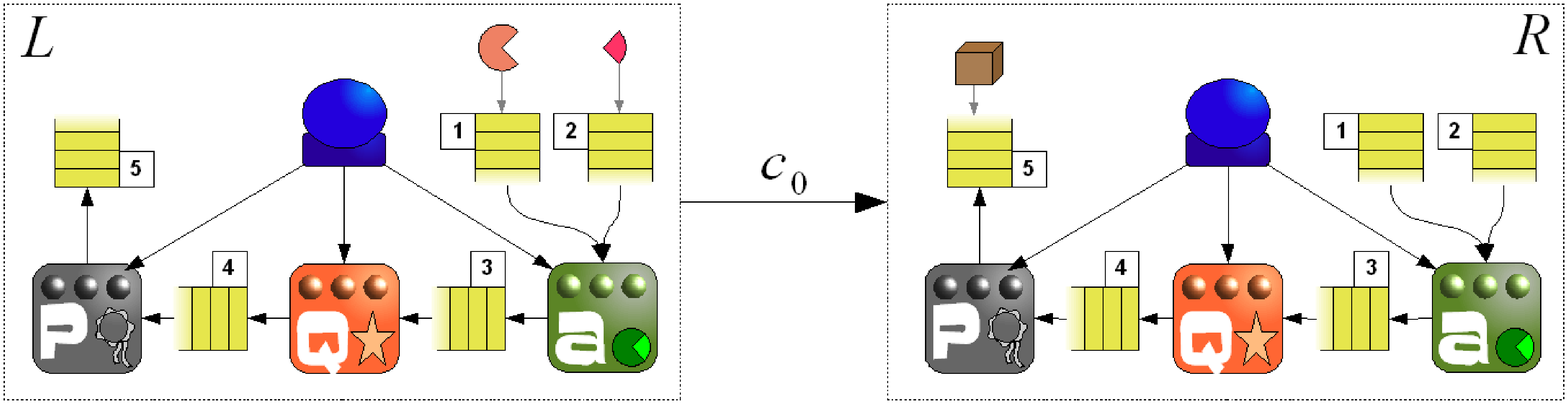}
  \caption{Composition of Sequence $\texttt{s}_\texttt{0}$}
  \label{fig:comp_s_0}
\end{figure}

Proposition~\ref{prop:CoherenceImpliesComposition} tells us that the
rules in $\texttt{s}_\texttt{0}$ can be composed if they are coherent
and compatible.  Let $c_0 = \left( L_c, e_c, r_c \right)$ be the rule
so defined.  Using
equations~\eqref{eq:e_r_compositionEdges}~and~\eqref{eq:e_r_compositionNodes}
its matrices can be found.  Also, taking advantage of previous
calculations for the image and using Corollary~\ref{cor:CompLemma}, we
can see that the composition is the one given in
Fig.~\ref{fig:comp_s_0}, closely related to
Fig.~\ref{fig:minInitialDig_Ex}.

Let $\texttt{mv}_\texttt{1} = \texttt{move2A};\texttt{move2D}$ and
$\texttt{mv}_\texttt{2} = \texttt{move2P};\texttt{move2Q}$ and define
the sequence $s_4 =
\texttt{pack};\texttt{mv}_\texttt{2};\texttt{assem};\texttt{mv}_\texttt{1}$.
Production \texttt{pack} is not sequentially independent of
$\texttt{mv}_\texttt{1}$ nor of
$\texttt{mv}_\texttt{2};\texttt{assem}$.  This is a simple example in
which it is possible to advance productions inside sequences only if
jumps of length strictly greater than one are allowed. To see that
$\texttt{pack}\,\bot \left(
  \texttt{mv}_\texttt{2};\texttt{assem}\,;\texttt{mv}_\texttt{1}
\right)$ it is necessary -- see
Theorem~\ref{th:CoherenceImpliesSeqInd} -- to check coherence of both
sequences and G-congruence.

Coherence for advancement of a single production inside a sequence is
given by eq.~\eqref{eq:advProd} in Theorem~\ref{th:advanceDelayProdTheor},
which should be zero.  It is straightforward to check that:
\begin{equation}
  e_{\texttt{pack}} \bigtriangledown_1^5 \left( \overline{r_x} \, L_y
  \right) \vee R_{\texttt{pack}} \bigtriangledown_1^5 \left(
    \overline{e_x} \, r_y \right) = 0.
\end{equation}

\begin{figure}[htbp]
  \centering
  \includegraphics[scale =
  0.5]{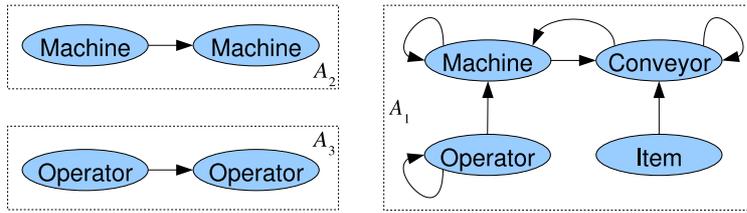}
  \caption{DSL Syntax Specification Extended}
  \label{fig:syntaxSpecExtended}
\end{figure}

By increasing the number of productions the system can be modelled in
greater detail.  For example, one operator can be busy or idle.  The
operator is busy if some action needs his attention.  This will be
represented by a self loop attached to the operator under
consideration. The same applies to a machine. The syntax as a DSL of
our grammar changes because there can exist self-loops for machines
and operators. This is not allowed in Fig.~\ref{fig:DSLsyntax}.
However, negative conditions are needed in the type graph (there can
be self-loops in machines or operators but not connections between two
operators or between two machines). See Fig.~\ref{fig:syntaxSpecExtended}. We need to demand $A_1$ for every single
edge (using the decomposition operator $\widehat{T}$ of Sec.~\ref{sec:functionalRepresentation}) and the nonexistence of matchings
with $A_2$ and $A_3$.

Up to now a single operator could be in charge of more than one
machine so if there are edges from the operator to several machines,
all machines may work simultaneously.  Besides, there can be more than
one operator working on the same machine.  In a probably more
realistic grammar, these two scenarios could not take place.  These
restrictions will be addressed in Sec.~\ref{sec:GraphConstraintsAndApplicationConditions}.

\begin{figure}[htbp]
  \centering
  \includegraphics[scale =
  0.34]{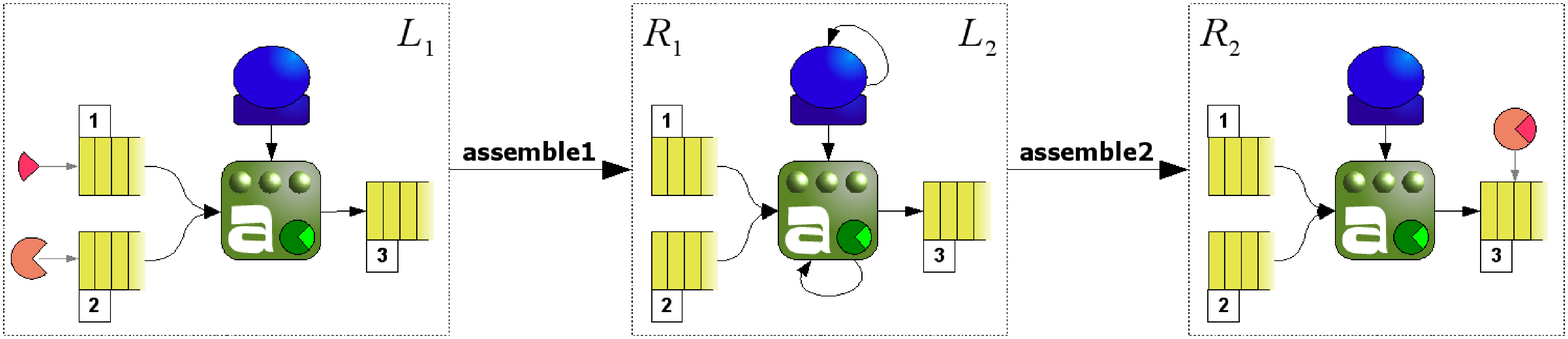}
  \caption{Production \texttt{assemble} in Greater Detail}
  \label{fig:assemExtended}
\end{figure}

The production process of any machine can be split into two phases: If
there are enough elements to start its job, then the input pieces
disappear and the machine and the operator become busy.  After that,
some output piece is produced and the machine and the operator become
idle again.  This is represented in the sequence of Fig.~\ref{fig:assemExtended}.  Note that $\texttt{assemble} =
\texttt{assemble}_\texttt{1} \circ \texttt{assemble}_\texttt{2}$.

If we limit our Matrix Graph Grammar to deal with simple digraphs we
have a built-in application condition ``for free''.  Even though one
operator can still be in charge of several machines simultaneously, he
will manage at most one machine at a time.  Otherwise, two self-loops
would be added violating compatibility.

Application conditions are needed if we want to set restrictions on
productions \texttt{move}. This can be permitted if the machine has a
kind of ``pause'', so the machine (which is busy as it has a self
loop) can resume as soon as an operator moves to it. It is not
necessary to specify a restriction to state that a machine can not
start a job when the operator is busy, as the rule would try to append
a second self-loop to the operator (something not allowed if we are
limited to simple digraphs).



Sequences can be generated at design time to debug the grammar or
during runtime to force a set of events.  They can also be
automatically generated by application conditions or can be associated
to other concepts, such as reachability.


\section{Initial Digraph Sets and G-Congruence}
\label{sec:initialDigraphSetsAndGcongruence}


To calculate the initial digraph set of sequence
$\texttt{s}_\texttt{0} = \texttt{pack}\, ; \texttt{certify} \,;
\texttt{assem}$ we start with the maximal initial digraph $M_0$, the
digraph that unrelates all elements for different productions.  It is
formed by the disjoint union of the left hand sides of the three
productions in sequence $\texttt{s}_\texttt{0}$. The rest of elements
$M_i$ of the initial digraph set $\mathfrak{M}\left( s_0 \right)$ are
derived by identifying nodes and edges in $M_0$.  These
identifications however can not be carried out arbitrarily because any
$M_i \in \mathfrak{M} \left( s_0 \right)$ must satisfy eq.~\eqref{eq:FirstMinDigraph}. Hence, there are identifications that
make some elements unnecessary.  For example, if the output conveyor
of production \texttt{certify} is identified with the input conveyor
of \texttt{pack}, then \texttt{item3} (mandatory for the application
of \texttt{pack}) is not needed anymore because it will be provided by
\texttt{certify}.

\begin{figure}[htbp]
  \centering
  \includegraphics[scale = 0.39]{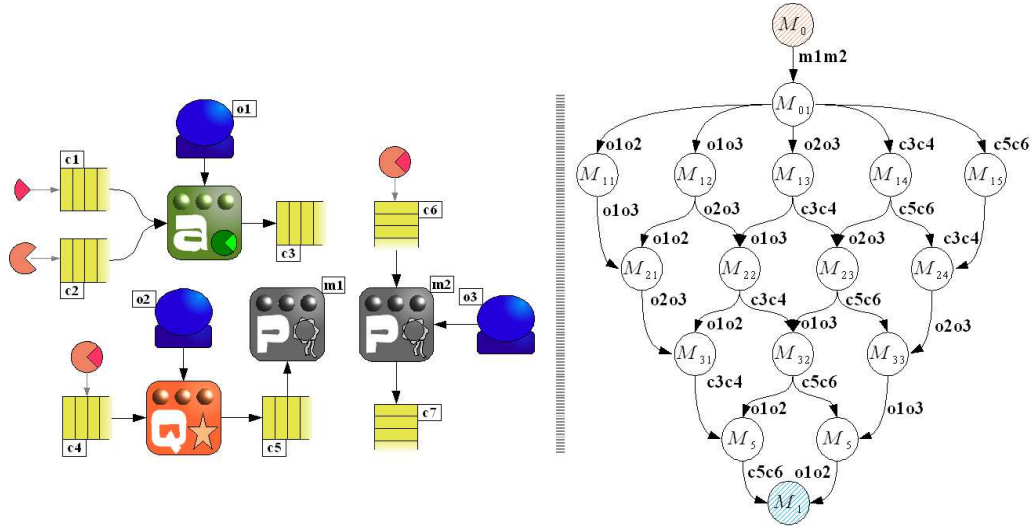}
  \caption{MID and Excerpt of the Initial Digraph Set of
    $\texttt{s}_\texttt{0} = \texttt{pack}\, ; \texttt{certify}\, ;
    \texttt{assem}$}
  \label{fig:excerptIDS}
\end{figure}

\begin{figure}[htbp]
  \centering
  \includegraphics[scale =
  0.35]{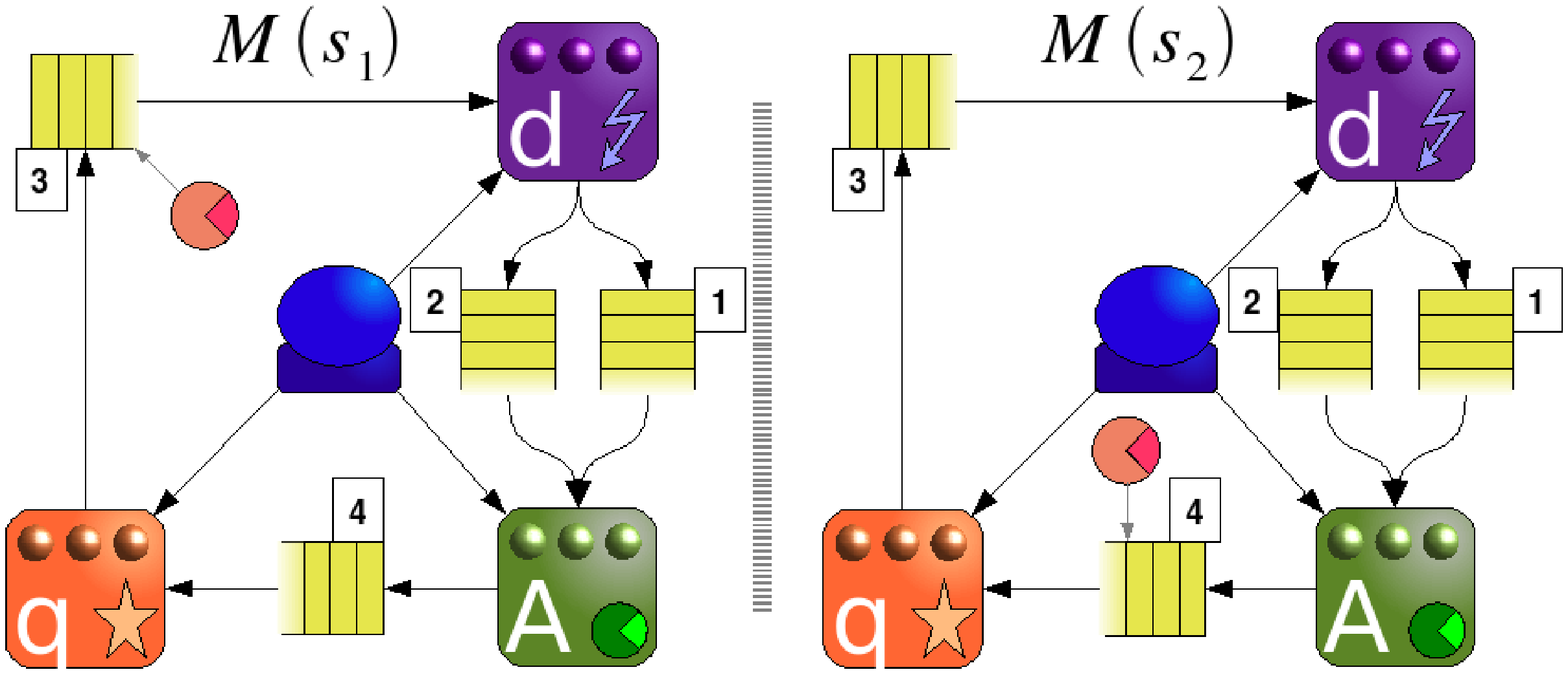}
  \caption{MID for Sequences $\texttt{s}_\texttt{1}$ and
    $\texttt{s}_\texttt{2}$}
  \label{fig:MIDrejectSample}
\end{figure}

For $s_0$ we will label $c1$ and $c2$ the input conveyors of
\texttt{assemble} and $c3$ its output conveyor.  Similarly, we have
$c4$ and $c5$ for \texttt{certify} and $c6$ and $c7$ for
\texttt{pack}.  Operators will be labelled accordingly so $o1$ is the
one in \texttt{assemble}, $o2$ in \texttt{certify} and $o3$ in
\texttt{pack}.  There are two machines for packing, $m1$ the one in
\texttt{certify} and $m_2$ in \texttt{pack}.  See the graph to the
left of Fig.~\ref{fig:MIDrejectSample}.  No identification prevents
any other\footnote{For an example in which not all identifications are
  permitted refer to Sec.~\ref{sec:initialDigraphSet},
  Fig.~\ref{fig:digraphSet}.} in $\mathfrak{M} \left(s_0\right)$, so
the number of elements in $\mathfrak{M}\left( s_0 \right)$ grows
factorially. In this case, since there are 6 possible identifications
we have 720 possibilities.  In Fig.~\ref{fig:excerptIDS} a part of the
initial digraph set can be found to the right.  The string that
appears close to each arrow specifies the identification (top-bottom)
performed to derive the corresponding initial digraph.

Initial digraph sets can be useful to debug a grammar.  By choosing
certain testing sequences it is possible to automatically select
``extreme'' cases in which as many elements as possible are identified
or unrelated.  For example, the development framework can tell that a
single operator may manage all machines with the grammar as defined so
far, but maybe this was not the intended behavior.

G-congruence and congruence conditions guarantee the sameness of the
minimal initial digraph.  They also provide information on what
elements are spoiling this property.  Consider the sequences
$\texttt{s}_\texttt{1} =
\texttt{reject}\,;\texttt{assemble}\,;\texttt{recycle}$ and
$\texttt{s}_\texttt{2} = \texttt{assemble}\,;\texttt{recycle}\,;$
$\texttt{reject}$, where in $\texttt{s}_\texttt{2}$ the application of
production $\texttt{reject}$ has been advanced two positions with
respect to $\texttt{s}_\texttt{1}$.  The minimal initial digraphs of
both sequences can be found in Fig.~\ref{fig:MIDrejectSample}. By the
way, notice that $M(s_i)$ are invariants for these transformations,
i.e.  $s_i\left( M\left( s_i \right)\right) = M\left( s_i \right)$.


G-congruence is characterized in terms of congruence conditions in
Theorem~\ref{th:GCongruence}.  Congruence conditions for the
advancement of a single production inside a sequence are stated in
Prop.~\ref{prop:GCongruencePos}, in particular in eq.~\eqref{eq:advCC}.
Simplified and adapted for this case with nodes ordered
\texttt{[1:item1 1:item2 1:item3 1:conv 2:conv 3:conv 4:conv 1:macA
  1:macQ 1:macD 1:op]}:\footnote{Where subscript 1 stands for rule
  \texttt{recycle}, subscript 2 is \texttt{assemble} and subscript 3
  is \texttt{reject}.}

\begin{eqnarray}
  CC & = & L_3 \nabla_1^2 \overline{e}_x K_y \left(r_y \vee e_3
  \right) \vee K_3 \nabla_1^2 \overline{r}_x L_y \left( e_y \vee r_3
  \right) = \nonumber \\
  & = & L_3 \left[K_1 \left( r_1 \vee e_3 \right) \vee \overline{e}_1
    K_2 \left(r_2 \vee e_3 \right)\right] \vee K_3 \left[ L_1 \left(
      e_1 \vee r_3 \right) \vee \overline{r}_1 L_2 \left( e_2 \vee r_3
    \right) \right] = \nonumber \\
  & = & \left[ \begin{array}{ccccccccccc} \vspace{-6pt}
      0 & 0 & 0 & 0 & 0 & 0 & 0 & 0 & 0 & 0 & 0 \\
      \vspace{-6pt}
      0 & 0 & 0 & 0 & 0 & 0 & 0 & 0 & 0 & 0 & 0 \\
      \vspace{-6pt}
      0 & 0 & 0 & 0 & 0 & 0 & 1 & 0 & 0 & 0 & 0 \\
      \vspace{-6pt}
      0 & 0 & 0 & 0 & 0 & 0 & 0 & 0 & 0 & 0 & 0 \\
      \vspace{-6pt}
      0 & 0 & 0 & 0 & 0 & 0 & 0 & 0 & 0 & 0 & 0 \\
      \vspace{-6pt}
      0 & 0 & 0 & 0 & 0 & 0 & 0 & 0 & 0 & 1 & 0 \\
      \vspace{-6pt}
      0 & 0 & 0 & 0 & 0 & 0 & 0 & 0 & 1 & 0 & 0 \\
      \vspace{-6pt}
      0 & 0 & 0 & 0 & 0 & 0 & 0 & 0 & 0 & 0 & 0 \\
      \vspace{-6pt}
      0 & 0 & 0 & 0 & 0 & 1 & 0 & 0 & 0 & 0 & 0 \\
      \vspace{-6pt}
      0 & 0 & 0 & 0 & 0 & 0 & 0 & 0 & 0 & 0 & 0 \\
      \vspace{-6pt}
      0 & 0 & 0 & 0 & 0 & 0 & 0 & 0 & 0 & 0 & 0 \\
      \vspace{-10pt}
    \end{array}	\right] \left[ \left[ \begin{array}{ccccccccccc}
        \vspace{-6pt}
        0 & 0 & 1 & 1 & 0 & 0 & 0 & 0 & 0 & 0 & 0 \\
        \vspace{-6pt}
        0 & 0 & 1 & 0 & 1 & 0 & 0 & 0 & 0 & 0 & 0 \\
        \vspace{-6pt}
        1 & 1 & 1 & 1 & 1 & 0 & 1 & 1 & 1 & 1 & 1 \\
        \vspace{-6pt}
        0 & 0 & 1 & 0 & 0 & 0 & 0 & 0 & 0 & 0 & 0 \\
        \vspace{-6pt}
        0 & 0 & 1 & 0 & 0 & 0 & 0 & 0 & 0 & 0 & 0 \\
        \vspace{-6pt}
        0 & 0 & 1 & 0 & 0 & 0 & 0 & 0 & 0 & 0 & 0 \\
        \vspace{-6pt}
        0 & 0 & 1 & 0 & 0 & 0 & 0 & 0 & 0 & 0 & 0 \\
        \vspace{-6pt}
        0 & 0 & 1 & 0 & 0 & 0 & 0 & 0 & 0 & 0 & 0 \\
        \vspace{-6pt}
        0 & 0 & 1 & 0 & 0 & 0 & 0 & 0 & 0 & 0 & 0 \\
        \vspace{-6pt}
        0 & 0 & 1 & 0 & 0 & 0 & 0 & 0 & 0 & 0 & 0 \\
        \vspace{-6pt}
        0 & 0 & 1 & 0 & 0 & 0 & 0 & 0 & 0 & 0 & 0 \\
        \vspace{-10pt}
      \end{array} \right] \left( \left[ \begin{array}{ccccccccccc}
          \vspace{-6pt}
          0 & 0 & 0 & 1 & 0 & 0 & 0 & 0 & 0 & 0 & 0 \\
          \vspace{-6pt}
          0 & 0 & 0 & 0 & 1 & 0 & 0 & 0 & 0 & 0 & 0 \\
          \vspace{-6pt}
          0 & 0 & 0 & 0 & 0 & 0 & 0 & 0 & 0 & 0 & 0 \\
          \vspace{-6pt}
          0 & 0 & 0 & 0 & 0 & 0 & 0 & 0 & 0 & 0 & 0 \\
          \vspace{-6pt}
          0 & 0 & 0 & 0 & 0 & 0 & 0 & 0 & 0 & 0 & 0 \\
          \vspace{-6pt}
          0 & 0 & 0 & 0 & 0 & 0 & 0 & 0 & 0 & 0 & 0 \\
          \vspace{-6pt}
          0 & 0 & 0 & 0 & 0 & 0 & 0 & 0 & 0 & 0 & 0 \\
          \vspace{-6pt}
          0 & 0 & 0 & 0 & 0 & 0 & 0 & 0 & 0 & 0 & 0 \\
          \vspace{-6pt}
          0 & 0 & 0 & 0 & 0 & 0 & 0 & 0 & 0 & 0 & 0 \\
          \vspace{-6pt}
          0 & 0 & 0 & 0 & 0 & 0 & 0 & 0 & 0 & 0 & 0 \\
          \vspace{-6pt}
          0 & 0 & 0 & 0 & 0 & 0 & 0 & 0 & 0 & 0 & 0 \\
          \vspace{-10pt}
        \end{array}	\right] \vee \right. \right. \nonumber
\end{eqnarray}
\begin{eqnarray}
  & \vee & \left. \left[ \begin{array}{ccccccccccc}
        \vspace{-6pt}
        0 & 0 & 0 & 0 & 0 & 0 & 0 & 0 & 0 & 0 & 0 \\
        \vspace{-6pt}
        0 & 0 & 0 & 0 & 0 & 0 & 0 & 0 & 0 & 0 & 0 \\
        \vspace{-6pt}
        0 & 0 & 0 & 0 & 0 & 0 & 1 & 0 & 0 & 0 & 0 \\
        \vspace{-6pt}
        0 & 0 & 0 & 0 & 0 & 0 & 0 & 0 & 0 & 0 & 0 \\
        \vspace{-6pt}
        0 & 0 & 0 & 0 & 0 & 0 & 0 & 0 & 0 & 0 & 0 \\
        \vspace{-6pt}
        0 & 0 & 0 & 0 & 0 & 0 & 0 & 0 & 0 & 0 & 0 \\
        \vspace{-6pt}
        0 & 0 & 0 & 0 & 0 & 0 & 0 & 0 & 0 & 0 & 0 \\
        \vspace{-6pt}
        0 & 0 & 0 & 0 & 0 & 0 & 0 & 0 & 0 & 0 & 0 \\
        \vspace{-6pt}
        0 & 0 & 0 & 0 & 0 & 0 & 0 & 0 & 0 & 0 & 0 \\
        \vspace{-6pt}
        0 & 0 & 0 & 0 & 0 & 0 & 0 & 0 & 0 & 0 & 0 \\
        \vspace{-6pt}
        0 & 0 & 0 & 0 & 0 & 0 & 0 & 0 & 0 & 0 & 0 \\
        \vspace{-10pt}
      \end{array} \right] \right) \vee \left[ \begin{array}{ccccccccccc}
      \vspace{-6pt}
      1 & 1 & 1 & 1 & 1 & 1 & 1 & 1 & 1 & 1 & 1 \\
      \vspace{-6pt}
      1 & 1 & 1 & 1 & 1 & 1 & 1 & 1 & 1 & 1 & 1 \\
      \vspace{-6pt}
      1 & 1 & 1 & 1 & 1 & 0 & 1 & 1 & 1 & 1 & 1 \\
      \vspace{-6pt}
      1 & 1 & 1 & 1 & 1 & 1 & 1 & 1 & 1 & 1 & 1 \\
      \vspace{-6pt}
      1 & 1 & 1 & 1 & 1 & 1 & 1 & 1 & 1 & 1 & 1 \\
      \vspace{-6pt}
      1 & 1 & 1 & 1 & 1 & 1 & 1 & 1 & 1 & 1 & 1 \\
      \vspace{-6pt}
      1 & 1 & 1 & 1 & 1 & 1 & 1 & 1 & 1 & 1 & 1 \\
      \vspace{-6pt}
      1 & 1 & 1 & 1 & 1 & 1 & 1 & 1 & 1 & 1 & 1 \\
      \vspace{-6pt}
      1 & 1 & 1 & 1 & 1 & 1 & 1 & 1 & 1 & 1 & 1 \\
      \vspace{-6pt}
      1 & 1 & 1 & 1 & 1 & 1 & 1 & 1 & 1 & 1 & 1 \\
      \vspace{-6pt}
      1 & 1 & 1 & 1 & 1 & 1 & 1 & 1 & 1 & 1 & 1 \\
      \vspace{-10pt}
    \end{array}	\right] \left[ \begin{array}{ccccccccccc}
      \vspace{-6pt}
      1 & 1 & 1 & 0 & 1 & 1 & 1 & 1 & 1 & 1 & 1 \\
      \vspace{-6pt}
      1 & 1 & 1 & 1 & 0 & 1 & 1 & 1 & 1 & 1 & 1 \\
      \vspace{-6pt}
      1 & 1 & 0 & 0 & 0 & 0 & 1 & 0 & 0 & 0 & 0 \\
      \vspace{-6pt}
      1 & 1 & 0 & 0 & 0 & 0 & 0 & 0 & 0 & 0 & 0 \\
      \vspace{-6pt}
      1 & 1 & 0 & 0 & 0 & 0 & 0 & 0 & 0 & 0 & 0 \\
      \vspace{-6pt}
      1 & 1 & 0 & 0 & 0 & 0 & 0 & 0 & 0 & 0 & 0 \\
      \vspace{-6pt}
      1 & 1 & 0 & 0 & 0 & 0 & 0 & 0 & 0 & 0 & 0 \\
      \vspace{-6pt}
      1 & 1 & 0 & 0 & 0 & 0 & 0 & 0 & 0 & 0 & 0 \\
      \vspace{-6pt}
      1 & 1 & 0 & 0 & 0 & 0 & 0 & 0 & 0 & 0 & 0 \\
      \vspace{-6pt}
      1 & 1 & 0 & 0 & 0 & 0 & 0 & 0 & 0 & 0 & 0 \\
      \vspace{-6pt}
      1 & 1 & 0 & 0 & 0 & 0 & 0 & 0 & 0 & 0 & 0 \\
      \vspace{-10pt}
    \end{array}	\right] \nonumber
\end{eqnarray}
\vspace{-0.6cm}
\begin{eqnarray}
  & & \left.\left(\left[ \begin{array}{ccccccccccc}
          \vspace{-6pt}
          0 & 0 & 0 & 0 & 0 & 0 & 0 & 0 & 0 & 0 & 0 \\
          \vspace{-6pt}
          0 & 0 & 0 & 0 & 0 & 0 & 0 & 0 & 0 & 0 & 0 \\
          \vspace{-6pt}
          0 & 0 & 0 & 0 & 0 & 0 & 1 & 0 & 0 & 0 & 0 \\
          \vspace{-6pt}
          0 & 0 & 0 & 0 & 0 & 0 & 0 & 0 & 0 & 0 & 0 \\
          \vspace{-6pt}
          0 & 0 & 0 & 0 & 0 & 0 & 0 & 0 & 0 & 0 & 0 \\
          \vspace{-6pt}
          0 & 0 & 0 & 0 & 0 & 0 & 0 & 0 & 0 & 0 & 0 \\
          \vspace{-6pt}
          0 & 0 & 0 & 0 & 0 & 0 & 0 & 0 & 0 & 0 & 0 \\
          \vspace{-6pt}
          0 & 0 & 0 & 0 & 0 & 0 & 0 & 0 & 0 & 0 & 0 \\
          \vspace{-6pt}
          0 & 0 & 0 & 0 & 0 & 0 & 0 & 0 & 0 & 0 & 0 \\
          \vspace{-6pt}
          0 & 0 & 0 & 0 & 0 & 0 & 0 & 0 & 0 & 0 & 0 \\
          \vspace{-6pt}
          0 & 0 & 0 & 0 & 0 & 0 & 0 & 0 & 0 & 0 & 0 \\
          \vspace{-10pt}
        \end{array}	\right] \vee \left[ \begin{array}{ccccccccccc}
          \vspace{-6pt}
          0 & 0 & 0 & 0 & 0 & 0 & 0 & 0 & 0 & 0 & 0 \\
          \vspace{-6pt}
          0 & 0 & 0 & 0 & 0 & 0 & 0 & 0 & 0 & 0 & 0 \\
          \vspace{-6pt}
          0 & 0 & 0 & 0 & 0 & 0 & 1 & 0 & 0 & 0 & 0 \\
          \vspace{-6pt}
          0 & 0 & 0 & 0 & 0 & 0 & 0 & 0 & 0 & 0 & 0 \\
          \vspace{-6pt}
          0 & 0 & 0 & 0 & 0 & 0 & 0 & 0 & 0 & 0 & 0 \\
          \vspace{-6pt}
          0 & 0 & 0 & 0 & 0 & 0 & 0 & 0 & 0 & 0 & 0 \\
          \vspace{-6pt}
          0 & 0 & 0 & 0 & 0 & 0 & 0 & 0 & 0 & 0 & 0 \\
          \vspace{-6pt}
          0 & 0 & 0 & 0 & 0 & 0 & 0 & 0 & 0 & 0 & 0 \\
          \vspace{-6pt}
          0 & 0 & 0 & 0 & 0 & 0 & 0 & 0 & 0 & 0 & 0 \\
          \vspace{-6pt}
          0 & 0 & 0 & 0 & 0 & 0 & 0 & 0 & 0 & 0 & 0 \\
          \vspace{-6pt}
          0 & 0 & 0 & 0 & 0 & 0 & 0 & 0 & 0 & 0 & 0 \\
          \vspace{-10pt}
        \end{array}	\right] \right) \right] \vee \left[ \begin{array}{ccccccccccc}
      \vspace{-6pt}
      0 & 0 & 0 & 0 & 0 & 0 & 0 & 0 & 0 & 0 & 0 \\
      \vspace{-6pt}
      0 & 0 & 0 & 0 & 0 & 0 & 0 & 0 & 0 & 0 & 0 \\
      \vspace{-6pt}
      0 & 0 & 0 & 0 & 0 & 1 & 0 & 0 & 0 & 0 & 0 \\
      \vspace{-6pt}
      0 & 0 & 0 & 0 & 0 & 0 & 0 & 0 & 0 & 0 & 0 \\
      \vspace{-6pt}
      0 & 0 & 0 & 0 & 0 & 0 & 0 & 0 & 0 & 0 & 0 \\
      \vspace{-6pt}
      0 & 0 & 0 & 0 & 0 & 0 & 0 & 0 & 0 & 0 & 0 \\
      \vspace{-6pt}
      0 & 0 & 0 & 0 & 0 & 0 & 0 & 0 & 0 & 0 & 0 \\
      \vspace{-6pt}
      0 & 0 & 0 & 0 & 0 & 0 & 0 & 0 & 0 & 0 & 0 \\
      \vspace{-6pt}
      0 & 0 & 0 & 0 & 0 & 0 & 0 & 0 & 0 & 0 & 0 \\
      \vspace{-6pt}
      0 & 0 & 0 & 0 & 0 & 0 & 0 & 0 & 0 & 0 & 0 \\
      \vspace{-6pt}
      0 & 0 & 0 & 0 & 0 & 0 & 0 & 0 & 0 & 0 & 0 \\
      \vspace{-10pt}
    \end{array} \right] \nonumber
\end{eqnarray}
\vspace{-0.6cm}
\begin{eqnarray}
  \left[ \left[ \begin{array}{ccccccccccc}
        \vspace{-6pt}
        0 & 0 & 0 & 0 & 0 & 0 & 0 & 0 & 0 & 0 & 0 \\
        \vspace{-6pt}
        0 & 0 & 0 & 0 & 0 & 0 & 0 & 0 & 0 & 0 & 0 \\
        \vspace{-6pt}
        0 & 0 & 0 & 0 & 0 & 1 & 0 & 0 & 0 & 0 & 0 \\
        \vspace{-6pt}
        0 & 0 & 0 & 0 & 0 & 0 & 0 & 0 & 0 & 0 & 0 \\
        \vspace{-6pt}
        0 & 0 & 0 & 0 & 0 & 0 & 0 & 0 & 0 & 0 & 0 \\
        \vspace{-6pt}
        0 & 0 & 0 & 0 & 0 & 0 & 0 & 0 & 0 & 1 & 0 \\
        \vspace{-6pt}
        0 & 0 & 0 & 0 & 0 & 0 & 0 & 0 & 0 & 0 & 0 \\
        \vspace{-6pt}
        0 & 0 & 0 & 0 & 0 & 0 & 0 & 0 & 0 & 0 & 0 \\
        \vspace{-6pt}
        0 & 0 & 0 & 0 & 0 & 0 & 0 & 0 & 0 & 0 & 0 \\
        \vspace{-6pt}
        0 & 0 & 0 & 1 & 1 & 0 & 0 & 0 & 0 & 0 & 0 \\
        \vspace{-6pt}
        0 & 0 & 0 & 0 & 0 & 0 & 0 & 0 & 0 & 1 & 0 \\
        \vspace{-10pt} \end{array} \right] 
    \left( \left[ \begin{array}{ccccccccccc}
          \vspace{-6pt}
          0 & 0 & 0 & 0 & 0 & 0 & 0 & 0 & 0 & 0 & 0 \\
          \vspace{-6pt}
          0 & 0 & 0 & 0 & 0 & 0 & 0 & 0 & 0 & 0 & 0 \\
          \vspace{-6pt}
          0 & 0 & 0 & 0 & 0 & 1 & 0 & 0 & 0 & 0 & 0 \\
          \vspace{-6pt}
          0 & 0 & 0 & 0 & 0 & 0 & 0 & 0 & 0 & 0 & 0 \\
          \vspace{-6pt}
          0 & 0 & 0 & 0 & 0 & 0 & 0 & 0 & 0 & 0 & 0 \\
          \vspace{-6pt}
          0 & 0 & 0 & 0 & 0 & 0 & 0 & 0 & 0 & 0 & 0 \\
          \vspace{-6pt}
          0 & 0 & 0 & 0 & 0 & 0 & 0 & 0 & 0 & 0 & 0 \\
          \vspace{-6pt}
          0 & 0 & 0 & 0 & 0 & 0 & 0 & 0 & 0 & 0 & 0 \\
          \vspace{-6pt}
          0 & 0 & 0 & 0 & 0 & 0 & 0 & 0 & 0 & 0 & 0 \\
          \vspace{-6pt}
          0 & 0 & 0 & 0 & 0 & 0 & 0 & 0 & 0 & 0 & 0 \\
          \vspace{-6pt}
          0 & 0 & 0 & 0 & 0 & 0 & 0 & 0 & 0 & 0 & 0 \\
          \vspace{-10pt}
	\end{array} \right] \vee \left[ \begin{array}{ccccccccccc}
          \vspace{-6pt}
          0 & 0 & 0 & 0 & 0 & 0 & 0 & 0 & 0 & 0 & 0 \\
          \vspace{-6pt}
          0 & 0 & 0 & 0 & 0 & 0 & 0 & 0 & 0 & 0 & 0 \\
          \vspace{-6pt}
          0 & 0 & 0 & 0 & 0 & 1 & 0 & 0 & 0 & 0 & 0 \\
          \vspace{-6pt}
          0 & 0 & 0 & 0 & 0 & 0 & 0 & 0 & 0 & 0 & 0 \\
          \vspace{-6pt}
          0 & 0 & 0 & 0 & 0 & 0 & 0 & 0 & 0 & 0 & 0 \\
          \vspace{-6pt}
          0 & 0 & 0 & 0 & 0 & 0 & 0 & 0 & 0 & 0 & 0 \\
          \vspace{-6pt}
          0 & 0 & 0 & 0 & 0 & 0 & 0 & 0 & 0 & 0 & 0 \\
          \vspace{-6pt}
          0 & 0 & 0 & 0 & 0 & 0 & 0 & 0 & 0 & 0 & 0 \\
          \vspace{-6pt}
          0 & 0 & 0 & 0 & 0 & 0 & 0 & 0 & 0 & 0 & 0 \\
          \vspace{-6pt}
          0 & 0 & 0 & 0 & 0 & 0 & 0 & 0 & 0 & 0 & 0 \\
          \vspace{-6pt}
          0 & 0 & 0 & 0 & 0 & 0 & 0 & 0 & 0 & 0 & 0 \\
          \vspace{-10pt}
        \end{array}	\right] \right) \right. \vee \nonumber
\end{eqnarray}
\vspace{-0.6cm}
\begin{eqnarray}
  \left[ \begin{array}{ccccccccccc}
      \vspace{-6pt}
      1 & 1 & 1 & 0 & 1 & 1 & 1 & 1 & 1 & 1 & 1 \\
      \vspace{-6pt}
      1 & 1 & 1 & 1 & 0 & 1 & 1 & 1 & 1 & 1 & 1 \\
      \vspace{-6pt}
      1 & 1 & 1 & 1 & 1 & 1 & 1 & 1 & 1 & 1 & 1 \\
      \vspace{-6pt}
      1 & 1 & 1 & 1 & 1 & 1 & 1 & 1 & 1 & 1 & 1 \\
      \vspace{-6pt}
      1 & 1 & 1 & 1 & 1 & 1 & 1 & 1 & 1 & 1 & 1 \\
      \vspace{-6pt}
      1 & 1 & 1 & 1 & 1 & 1 & 1 & 1 & 1 & 1 & 1 \\
      \vspace{-6pt}
      1 & 1 & 1 & 1 & 1 & 1 & 1 & 1 & 1 & 1 & 1 \\
      \vspace{-6pt}
      1 & 1 & 1 & 1 & 1 & 1 & 1 & 1 & 1 & 1 & 1 \\
      \vspace{-6pt}
      1 & 1 & 1 & 1 & 1 & 1 & 1 & 1 & 1 & 1 & 1 \\
      \vspace{-6pt}
      1 & 1 & 1 & 1 & 1 & 1 & 1 & 1 & 1 & 1 & 1 \\
      \vspace{-6pt}
      1 & 1 & 1 & 1 & 1 & 1 & 1 & 1 & 1 & 1 & 1 \\
      \vspace{-10pt}
    \end{array} \right] \left[ \begin{array}{ccccccccccc}
      \vspace{-6pt}
      0 & 0 & 0 & 1 & 0 & 0 & 0 & 0 & 0 & 0 & 0 \\
      \vspace{-6pt}
      0 & 0 & 0 & 0 & 1 & 0 & 0 & 0 & 0 & 0 & 0 \\
      \vspace{-6pt}
      0 & 0 & 0 & 0 & 0 & 0 & 0 & 0 & 0 & 0 & 0 \\
      \vspace{-6pt}
      0 & 0 & 0 & 0 & 0 & 0 & 0 & 1 & 0 & 0 & 0 \\
      \vspace{-6pt}
      0 & 0 & 0 & 0 & 0 & 0 & 0 & 1 & 0 & 0 & 0 \\
      \vspace{-6pt}
      0 & 0 & 0 & 0 & 0 & 0 & 0 & 0 & 0 & 0 & 0 \\
      \vspace{-6pt}
      0 & 0 & 0 & 0 & 0 & 0 & 0 & 0 & 0 & 0 & 0 \\
      \vspace{-6pt}
      0 & 0 & 0 & 0 & 0 & 0 & 1 & 0 & 0 & 0 & 0 \\
      \vspace{-6pt}
      0 & 0 & 0 & 0 & 0 & 0 & 0 & 0 & 0 & 0 & 0 \\
      \vspace{-6pt}
      0 & 0 & 0 & 0 & 0 & 0 & 0 & 0 & 0 & 0 & 0 \\
      \vspace{-6pt}
      0 & 0 & 0 & 0 & 0 & 0 & 0 & 1 & 0 & 0 & 0 \\
      \vspace{-10pt}
    \end{array}	\right] \left( \left[ \begin{array}{ccccccccccc}
        \vspace{-6pt}
        0 & 0 & 0 & 1 & 0 & 0 & 0 & 0 & 0 & 0 & 0 \\
        \vspace{-6pt}
        0 & 0 & 0 & 0 & 1 & 0 & 0 & 0 & 0 & 0 & 0 \\
        \vspace{-6pt}
        0 & 0 & 0 & 0 & 0 & 0 & 0 & 0 & 0 & 0 & 0 \\
        \vspace{-6pt}
        0 & 0 & 0 & 0 & 0 & 0 & 0 & 0 & 0 & 0 & 0 \\
        \vspace{-6pt}
        0 & 0 & 0 & 0 & 0 & 0 & 0 & 0 & 0 & 0 & 0 \\
        \vspace{-6pt}
        0 & 0 & 0 & 0 & 0 & 0 & 0 & 0 & 0 & 0 & 0 \\
        \vspace{-6pt}
        0 & 0 & 0 & 0 & 0 & 0 & 0 & 0 & 0 & 0 & 0 \\
        \vspace{-6pt}
        0 & 0 & 0 & 0 & 0 & 0 & 0 & 0 & 0 & 0 & 0 \\
        \vspace{-6pt}
        0 & 0 & 0 & 0 & 0 & 0 & 0 & 0 & 0 & 0 & 0 \\
        \vspace{-6pt}
        0 & 0 & 0 & 0 & 0 & 0 & 0 & 0 & 0 & 0 & 0 \\
        \vspace{-6pt}
        0 & 0 & 0 & 0 & 0 & 0 & 0 & 0 & 0 & 0 & 0 \\
        \vspace{-10pt}
      \end{array}	\right] \right. \vee \nonumber
\end{eqnarray}
\vspace{-0.6cm}
\begin{eqnarray}
  \left.\left.\left[ \begin{array}{ccccccccccc}
          \vspace{-6pt}
          0 & 0 & 0 & 0 & 0 & 0 & 0 & 0 & 0 & 0 & 0 \\
          \vspace{-6pt}
          0 & 0 & 0 & 0 & 0 & 0 & 0 & 0 & 0 & 0 & 0 \\
          \vspace{-6pt}
          0 & 0 & 0 & 0 & 0 & 1 & 0 & 0 & 0 & 0 & 0 \\
          \vspace{-6pt}
          0 & 0 & 0 & 0 & 0 & 0 & 0 & 0 & 0 & 0 & 0 \\
          \vspace{-6pt}
          0 & 0 & 0 & 0 & 0 & 0 & 0 & 0 & 0 & 0 & 0 \\
          \vspace{-6pt}
          0 & 0 & 0 & 0 & 0 & 0 & 0 & 0 & 0 & 0 & 0 \\
          \vspace{-6pt}
          0 & 0 & 0 & 0 & 0 & 0 & 0 & 0 & 0 & 0 & 0 \\
          \vspace{-6pt}
          0 & 0 & 0 & 0 & 0 & 0 & 0 & 0 & 0 & 0 & 0 \\
          \vspace{-6pt}
          0 & 0 & 0 & 0 & 0 & 0 & 0 & 0 & 0 & 0 & 0 \\
          \vspace{-6pt}
          0 & 0 & 0 & 0 & 0 & 0 & 0 & 0 & 0 & 0 & 0 \\
          \vspace{-6pt}
          0 & 0 & 0 & 0 & 0 & 0 & 0 & 0 & 0 & 0 & 0 \\
          \vspace{-10pt}
        \end{array}	\right] \right) \right] = \left[ \begin{array}{ccccccccccc|r}
      \vspace{-6pt}
      0 & 0 & 0 & 0 & 0 & 0 & 0 & 0 & 0 & 0 & 0 & i1\\
      \vspace{-6pt}
      0 & 0 & 0 & 0 & 0 & 0 & 0 & 0 & 0 & 0 & 0 & i2\\
      \vspace{-6pt}
      0 & 0 & 0 & 0 & 0 & 1 & 1 & 0 & 0 & 0 & 0 & i3\\
      \vspace{-6pt}
      0 & 0 & 0 & 0 & 0 & 0 & 0 & 0 & 0 & 0 & 0 & 1c\\
      \vspace{-6pt}
      0 & 0 & 0 & 0 & 0 & 0 & 0 & 0 & 0 & 0 & 0 & 2c\\
      \vspace{-6pt}
      0 & 0 & 0 & 0 & 0 & 0 & 0 & 0 & 0 & 0 & 0 & 3c\\
      \vspace{-6pt}
      0 & 0 & 0 & 0 & 0 & 0 & 0 & 0 & 0 & 0 & 0 & 4c\\
      \vspace{-6pt}
      0 & 0 & 0 & 0 & 0 & 0 & 0 & 0 & 0 & 0 & 0 & 1mA\\
      \vspace{-6pt}
      0 & 0 & 0 & 0 & 0 & 0 & 0 & 0 & 0 & 0 & 0 & 1mQ\\
      \vspace{-6pt}
      0 & 0 & 0 & 0 & 0 & 0 & 0 & 0 & 0 & 0 & 0 & 1mD\\
      \vspace{-6pt}
      0 & 0 & 0 & 0 & 0 & 0 & 0 & 0 & 0 & 0 & 0 & 1op\\
      \vspace{-10pt}
    \end{array}	\right] \nonumber
\end{eqnarray}

The congruence condition fails precisely in those elements that make
both minimal initial digraph different, $(i3,3c)$ and $(i3,4c)$.  See
Fig.~\ref{fig:MIDrejectSample}.

\begin{figure}[htbp]
  \centering
  \includegraphics[scale =
  0.27]{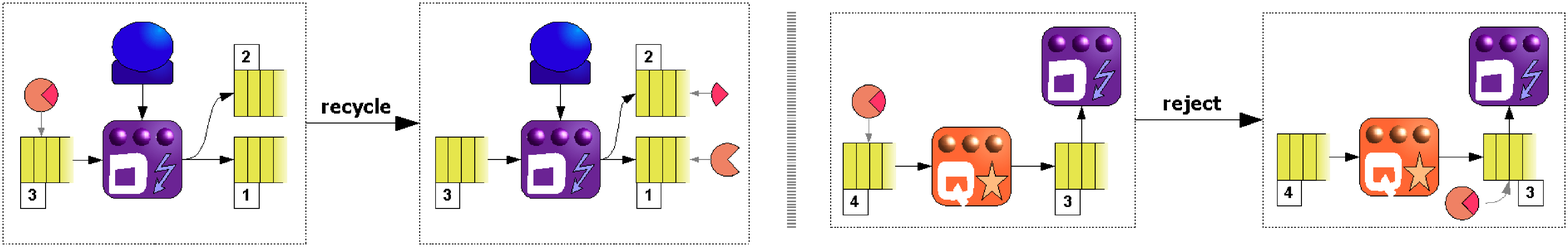}
  \caption{Ordered Items in Conveyors}
  \label{fig:recycleRejectMatrices}
\end{figure}

Relevant matrices in previous calculations can be found in eqs.~\eqref{eq:recycleMatrices}~and~\eqref{eq:rejectMatrices} for rules
\texttt{recycle} and \texttt{reject}, and in Sec.
\ref{sec:PresentationOfTheScenario} for \texttt{assemble}, in
particular equations~\eqref{eq:assemMatrices1}~and~\eqref{eq:assemNihilMatrix}. For identifications across productions
see Figs.~\ref{fig:MIDrejectSample}~and~\ref{fig:recycleRejectMatrices}.
\begin{eqnarray}\label{eq:recycleMatrices}
  K_{\textrm{\texttt{recycle}}} & = & \left[
    \begin{array}{ccccccccl}
      \vspace{-6pt}
      0 & 0 & 1 & 1 & 0 & 0 & 0 & 0 & \vert \; \textrm{\texttt{1:item1}} \\
      \vspace{-6pt}
      0 & 0 & 1 & 0 & 1 & 1 & 0 & 0 & \vert \; \textrm{\texttt{1:item2}} \\
      \vspace{-6pt}
      0 & 0 & 1 & 1 & 1 & 0 & 1 & 1 & \vert \; \textrm{\texttt{1:item3}} \\
      \vspace{-6pt}
      0 & 0 & 1 & 0 & 0 & 0 & 0 & 0 & \vert \; \textrm{\texttt{1:conv}} \\
      \vspace{-6pt}
      0 & 0 & 1 & 0 & 0 & 0 & 0 & 0 & \vert \; \textrm{\texttt{2:conv}} \\
      \vspace{-6pt}
      0 & 0 & 1 & 0 & 0 & 0 & 0 & 0 & \vert \; \textrm{\texttt{3:conv}} \\
      \vspace{-6pt}
      0 & 0 & 1 & 0 & 0 & 0 & 0 & 0 & \vert \; \textrm{\texttt{1:machD}} \\
      \vspace{-6pt}
      0 & 0 & 1 & 0 & 0 & 0 & 0 & 0 & \vert \; \textrm{\texttt{1:op}} \\
      \vspace{-10pt}
    \end{array}
  \right]
  L_{\texttt{recycle}} = \left[
    \begin{array}{cccccccc}
      \vspace{-6pt}
      0 & 0 & 0 & 0 & 0 & 0 & 0 & 0 \\
      \vspace{-6pt}
      0 & 0 & 0 & 0 & 0 & 0 & 0 & 0 \\
      \vspace{-6pt}
      0 & 0 & 0 & 0 & 0 & 1 & 0 & 0 \\
      \vspace{-6pt}
      0 & 0 & 0 & 0 & 0 & 0 & 0 & 0 \\
      \vspace{-6pt}
      0 & 0 & 0 & 0 & 0 & 0 & 0 & 0 \\
      \vspace{-6pt}
      0 & 0 & 0 & 0 & 0 & 0 & 1 & 0 \\
      \vspace{-6pt}
      0 & 0 & 0 & 1 & 1 & 0 & 0 & 0 \\
      \vspace{-6pt}
      0 & 0 & 0 & 0 & 0 & 0 & 1 & 0 \\
      \vspace{-10pt}
    \end{array}	\right] \\
  \label{eq:rejectMatrices} e_{\textrm{\texttt{reject}}} & = & \left[
    \begin{array}{cccccl}
      \vspace{-6pt}
      0 & 0 & 1 & 0 & 0 & \vert \; \textrm{\texttt{1:item3}} \\
      \vspace{-6pt}
      0 & 0 & 0 & 0 & 0 & \vert \; \textrm{\texttt{3:conv}} \\
      \vspace{-6pt}
      0 & 0 & 0 & 0 & 0 & \vert \; \textrm{\texttt{4:conv}} \\
      \vspace{-6pt}
      0 & 0 & 0 & 0 & 0 & \vert \; \textrm{\texttt{1:machD}} \\
      \vspace{-6pt}
      0 & 0 & 0 & 0 & 0 & \vert \; \textrm{\texttt{1:machQ}} \\
      \vspace{-10pt}
    \end{array}
  \right]
  \qquad\quad r_{\texttt{reject}} = \left[
    \begin{array}{ccccc}
      \vspace{-6pt}
      0 & 1 & 0 & 0 & 0 \\
      \vspace{-6pt}
      0 & 0 & 0 & 0 & 0 \\
      \vspace{-6pt}
      0 & 0 & 0 & 0 & 0 \\
      \vspace{-6pt}
      0 & 0 & 0 & 0 & 0 \\
      \vspace{-6pt}
      0 & 0 & 0 & 0 & 0 \\
      \vspace{-10pt}
    \end{array}
  \right]
\end{eqnarray}




\section{Reachability}
\label{sec:reachabilityAndConfluence}

In this section reachability is addressed together with some comments
on other problems such as confluence, termination and complexity (to
be addressed in a future contribution).

Throughout the book some techniques to deal with sequences have been
developed.  Sequences to be studied have to be supplied by the user.
Reachability is a more indirect source of sequences, because initial
and final states are specified and the system provides us with sets of
candidate sequences.

\begin{figure}[htbp]
  \centering
  \includegraphics[scale =
  0.34]{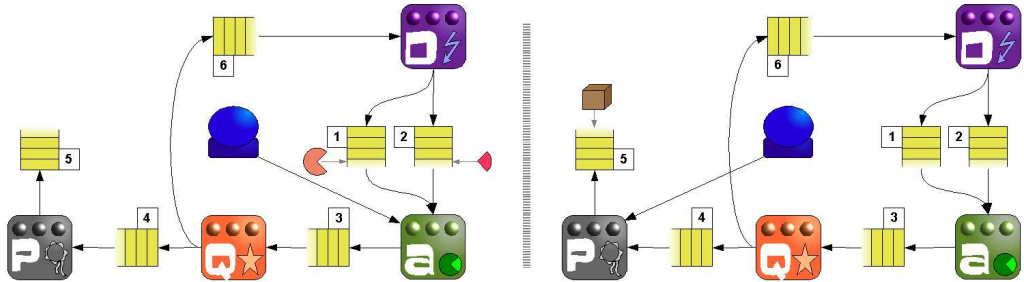}
  \caption{Initial and Final Digraphs for Reachability Example}
  \label{fig:reachExample}
\end{figure}

We shall use similar initial and final states as those in
Fig.~\ref{fig:minInitialDig_Ex} (see Fig.~\ref{fig:reachExample}). Our
grammar as defined so far has a fixed behavior, i.e. it is a fixed
graph grammar, whose state equation is given
by~\eqref{eq:stateEqGeneralDPO} in Prop.~\ref{prop:ReachabilityDPO}.

Let ${}_0S$ and ${}_dS$ be the initial and final states and the
ordering \texttt{[1:item1 1:item2 1:item3 1:item4 1:conv 2:conv 3:conv
  4:conv 5:conv 6:conv} \texttt{1:machA 1:machQ 1:machD 1:machP
  1:op]}.  Nodes appear in the last column.

\begin{equation}
  M_j^{i} = {}_dS - {}_0S = \sum_{k=1}^n A_{\!jk}^i x^k	=
  \left[ \begin{array}{cccccccccccccccc}
      \vspace{-6pt}
      0 & 0 & 0 & 0 & -1 & 0 & 0 & 0 & 0 & 0 & 0 & 0 & 0 & 0 & 0 & -1 \\
      \vspace{-6pt}
      0 & 0 & 0 & 0 & 0 & -1 & 0 & 0 & 0 & 0 & 0 & 0 & 0 & 0 & 0 & -1 \\
      \vspace{-6pt}
      0 & 0 & 0 & 0 & 0 & 0 & 0 & 0 & 0 & 0 & 0 & 0 & 0 & 0 & 0 & 0 \\
      \vspace{-6pt}
      0 & 0 & 0 & 0 & 0 & 0 & 0 & 0 & 1 & 0 & 0 & 0 & 0 & 0 & 0 & 1 \\
      \vspace{-6pt}
      0 & 0 & 0 & 0 & 0 & 0 & 0 & 0 & 0 & 0 & 0 & 0 & 0 & 0 & 0 & 0 \\
      \vspace{-6pt}
      0 & 0 & 0 & 0 & 0 & 0 & 0 & 0 & 0 & 0 & 0 & 0 & 0 & 0 & 0 & 0 \\
      \vspace{-6pt}
      0 & 0 & 0 & 0 & 0 & 0 & 0 & 0 & 0 & 0 & 0 & 0 & 0 & 0 & 0 & 0 \\
      \vspace{-6pt}
      0 & 0 & 0 & 0 & 0 & 0 & 0 & 0 & 0 & 0 & 0 & 0 & 0 & 0 & 0 & 0 \\
      \vspace{-6pt}
      0 & 0 & 0 & 0 & 0 & 0 & 0 & 0 & 0 & 0 & 0 & 0 & 0 & 0 & 0 & 0 \\
      \vspace{-6pt}
      0 & 0 & 0 & 0 & 0 & 0 & 0 & 0 & 0 & 0 & 0 & 0 & 0 & 0 & 0 & 0 \\
      \vspace{-6pt}
      0 & 0 & 0 & 0 & 0 & 0 & 0 & 0 & 0 & 0 & 0 & 0 & 0 & 0 & 0 & 0 \\
      \vspace{-6pt}
      0 & 0 & 0 & 0 & 0 & 0 & 0 & 0 & 0 & 0 & 0 & 0 & 0 & 0 & 0 & 0 \\
      \vspace{-6pt}
      0 & 0 & 0 & 0 & 0 & 0 & 0 & 0 & 0 & 0 & 0 & 0 & 0 & 0 & 0 & 0 \\
      \vspace{-6pt}
      0 & 0 & 0 & 0 & 0 & 0 & 0 & 0 & 0 & 0 & 0 & 0 & 0 & 0 & 0 & 0 \\
      \vspace{-6pt}
      0 & 0 & 0 & 0 & 0 & 0 & 0 & 0 & 0 & 0 & -1 & 0 & 0 & 1 & 0 & 0 \\
      \vspace{-10pt}
    \end{array}	\right] 
\end{equation}

For tensor $A_{\!jk}^i$ only the basic productions \texttt{assem},
\texttt{certify}, \texttt{reject}, \texttt{recycle} and \texttt{pack}
are considered plus those for operator movement \texttt{mov2*}.
Following Sec.~\ref{sec:mggTechniquesForPetriNets}, grammar rules that
add and delete elements of the same type are split in their
addition (+) and deletion (--) parts.  This includes only productions
\texttt{certify} and \texttt{reject}.\footnote{Note that neither
  \texttt{certify} nor \texttt{reject} add or delete the
  \texttt{item1} node. They only act on edges. These productions are
  split because the edge deleted and the edge added are of the same
  type, $(\mathtt{item1}, \mathtt{conv})$.}

The set of rules is $\{\texttt{assem}, \texttt{certify}^+,
\texttt{certify}^-, \texttt{reject}^+, \texttt{reject}^-,
\texttt{recycle}, \texttt{pack},$ $\texttt{mov2A}, \texttt{mov2Q},
\texttt{mov2D}, \texttt{mov2P}\}$, so $k \in \{1, \ldots, 11\}$.  This
ordering is kept in the equations from now on.

The following list summarizes all actions performed by the grammar
rules under consideration on nodes and edges.  A plus sign between
brackets means that the element is added and a minus sign that it is
deleted.
\begin{itemize}
\item $\left( \texttt{1:item1}, \texttt{1:conv} \right) \longmapsto
  \texttt{assem} \left( - \right), \texttt{recycle} \left( + \right)$
\item $\left( \texttt{1:item2}, \texttt{2:conv} \right) \longmapsto
  \texttt{assem} \left( - \right), \texttt{recycle} \left( + \right)$
\item $\left( \texttt{1:item3}, \texttt{3:conv} \right) \longmapsto
  \texttt{assem} \left( + \right), \texttt{certify}^- \left( -
  \right), \texttt{reject}^- \left( - \right)$
\item $\left( \texttt{1:item3}, \texttt{4:conv} \right) \longmapsto
  \texttt{certify}^+ \left( + \right), \texttt{pack} \left( - \right)$
\item $\left( \texttt{1:item3}, \texttt{6:conv} \right) \longmapsto
  \texttt{reject}^+ \left( + \right), \texttt{recycle} \left( -
  \right)$
\item $\left( \texttt{1:item4}, \texttt{5:conv} \right) \longmapsto
  \texttt{pack} \left( + \right)$
\item $\left( \texttt{1:op}, \texttt{1:machA} \right) \longmapsto
  \texttt{mov2A} \left( + \right), \texttt{mov2Q} \left( - \right)$
\item $\left( \texttt{1:op}, \texttt{1:machQ} \right) \longmapsto
  \texttt{mov2Q} \left( + \right), \texttt{mov2P} \left( - \right)$
\item $\left( \texttt{1:op}, \texttt{1:machD} \right) \longmapsto
  \texttt{mov2D} \left( + \right), \texttt{mov2A} \left( - \right)$
\item $\left( \texttt{1:op}, \texttt{1:machP} \right) \longmapsto
  \texttt{mov2P} \left( + \right), \texttt{mov2D} \left( - \right)$
\item $\left( \texttt{1:item1} \right) \longmapsto \texttt{assem}
  \left( - \right), \texttt{recycle} \left( + \right)$
\item $\left( \texttt{1:item2} \right) \longmapsto \texttt{assem}
  \left( - \right), \texttt{recycle} \left( + \right)$
\item $\left( \texttt{1:item3} \right) \longmapsto \texttt{assem}
  \left( + \right), \texttt{recycle} \left( - \right), \texttt{pack}
  \left( - \right)$
\item $\left( \texttt{1:item4} \right) \longmapsto \texttt{pack}
  \left( + \right)$
\end{itemize}

What is finally derived according to the methods proposed in
Chap.~\ref{ch:reachability} is a system of linear equations.  To those
arising from the tensor equations another thirteen must be appended:
\begin{eqnarray}
  \{x^k_p & = & x^k_q\}, \quad p,q \in \{1, \ldots, 11\} \nonumber \\
  x^2_p & = & x^3_q \nonumber \\
  x^4_p & = & x^5_q. \nonumber
\end{eqnarray}

The first set of equations guarantee that a rule is applied a concrete
number of times.  The second and the third equations do not allow
inconsistencies for rules \texttt{certify} and \texttt{reject}, that
have been split in their addition and deletion parts. They have to
be applied the same amount of times.

Only those columns of $M$ for which some ``activity'' is defined in
the productions are of interest, i.e. all except the first four.  Zero
elements are not included, but substituted by bold zeros:

\begin{equation}
  \begin{array}{rcl}
    \left[ \begin{array}{r} -1 \\ \textbf{0}	\end{array} \right] =
    M_5 & = & \displaystyle \sum_{k=1}^{11} A_{5k} x^k_5 = \left[
      \begin{array}{c} -x^1_5 + x^6_5 \\ \textbf{0} \end{array}
    \right] \\
    && \\
    \left[ \begin{array}{r}	0 \\ -1 \\ \textbf{0}	\end{array}
    \right] = M_6 & = & \displaystyle \sum_{k=1}^{11} A_{6k} x^k_6 =
    \left[	\begin{array}{c} 0 \\ -x^1_6 + x^6_6 \\ \textbf{0}
      \end{array} \right] \\
  \end{array} \nonumber
\end{equation}
\begin{equation}
  \begin{array}{rcl}
    \left[ \begin{array}{r} \textbf{0} \end{array} \right] =
    M_7 & = & \displaystyle \sum_{k=1}^{11} A_{7k} x^k_7 = \left[
      \begin{array}{c} 0 \\ 0 \\ x^1_7 - x^3_7 - x^5_7 \\ \textbf{0}
      \end{array} \right] \\
    && \\
    \left[ \begin{array}{r}	\textbf{0} \end{array} \right] = M_8 &
    = & \displaystyle \sum_{k=1}^{11} A_{8k} x^k_8 = \left[
      \begin{array}{c} 0 \\ 0 \\ x^2_8 - x^7_8 \\ \textbf{0}
      \end{array} \right] \\
    && \\
    \left[ \begin{array}{r} 0 \\ 0 \\ 0 \\ 1 \\ \textbf{0} \end{array}
    \right] = M_9 & = & \displaystyle \sum_{k=1}^{11} A_{9k} x^k_9 =
    \left[ \begin{array}{c} 0 \\ 0 \\ 0 \\ x^7_9 \\ \textbf{0}
      \end{array} \right] \\
    && \\
    \left[ \begin{array}{r}	\textbf{0} \end{array} \right] =
    M_{10} & = & \displaystyle \sum_{k=1}^{11} A_{10,k} x^k_{10} =
    \left[ \begin{array}{c} 0 \\ 0 \\ x^4_{10} - x^6_{10} \\
        \textbf{0} \end{array} \right] \\
    && \\
    \left[ \begin{array}{r} \textbf{0} \\ -1 \end{array} \right] =
    M_{11} & = & \displaystyle \sum_{k=1}^{11} A_{11,k} x^k_{11} =
    \left[ \begin{array}{c} \mathbf{0} \\ x^8_{11} - x^9_{11} \\ 0
      \end{array} \right] \\
    && \\
    \left[ \begin{array}{r}	\textbf{0} \end{array} \right] =
    M_{12} & = & \displaystyle \sum_{k=1}^{11} A_{12,k} x^k_{12} =
    \left[ \begin{array}{c} \mathbf{0} \\ x^9_{12} - x^{11}_{12} \\ 0
      \end{array} \right] \\
  \end{array} \nonumber
\end{equation}
\begin{equation}
  \begin{array}{rcl}
    \left[ \begin{array}{r} \textbf{0} \end{array} \right] =
    M_{13} & = & \displaystyle \sum_{k=1}^{11} A_{13,k} x^k_{13} =
    \left[ \begin{array}{c} \mathbf{0} \\ x^{10}_{13} - x^8_{13} \\ 0
      \end{array} \right] \\
    && \\
    \left[ \begin{array}{r}	\textbf{0} \\ 1 \end{array} \right] =
    M_{14} & = & \displaystyle \sum_{k=1}^{11} A_{14,k} x^k_{14} =
    \left[ \begin{array}{c} \textbf{0} \\ x^{11}_{14} - x^{10}_{14}
      \end{array} \right] \\
    && \\
    \left[ \begin{array}{r}	-1 \\ -1 \\ 0 \\ 1 \\ \textbf{0}
      \end{array} \right] = M_{16} & = & \displaystyle \sum_{k=1}^{11}
    A_{16,k} x^k_{16} = \left[ \begin{array}{c} x^{6}_{16} - x^1_{16}
        \\ x^6_{16} - x^1_{16} \\ x^1_{16} - x^6_{16} - x^7_{16} \\
        x^7_{16} \\ \textbf{0} \end{array} \right] \\
  \end{array} \nonumber
\end{equation}

$M_{16}$ corresponds to nodes.  Recall that $x$ must satisfy the
additional conditions $x^k_p = x^k_q$, $k \in \{1, \ldots , 11\}$.
The system has the solution:
\begin{equation}\label{eq:exSol}
  \left( x, 1, 1, x-1, x-1, x-1, 1, y-1, y, y-1, y \right) = \textbf{0}.
\end{equation}
being $s_0$ -- see equation \eqref{eq:seqS0} -- one of the sequences
for $x=1$, $y=1$. Note that solutions are uncoupled in two parts: The
one that rules operator movement $(y)$ and that of items processing
$(x)$.

%



This is a good example to study termination and confluence.  Any
evolution of the system having as initial state the one depicted to
the left of Fig.~\ref{fig:reachExample} will eventually get to the
state to the right of the same figure (termination).\footnote{In fact,
  it is not terminating because the productions that move the operator
  can still be applied. What we would need is another production that
  drives the system to a halting state.}\index{termination} The
grammar is confluent (there is a single \emph{solution}) although
there is no upper bound to the number of steps it will take to get to
its final state (complexity).\index{complexity} Depending on the
probability distribution there will be more chances to end up sooner
or later.  Independently of the distribution, larger sequences have
smaller probabilities, being their probability zero in the limit (if
the probability assigned to rejecting \texttt{item1} is different from
1).

%

\section{Graph Constraints and Application Conditions}
\label{sec:GraphConstraintsAndApplicationConditions}

Application conditions and graph constraints will make our case study
much more realistic. We will see two examples on how application
conditions can be used to limit the applicability of rules or to avoid
undesired behaviors.

\begin{figure}[htbp]
  \centering
  \includegraphics[scale =
  0.5]{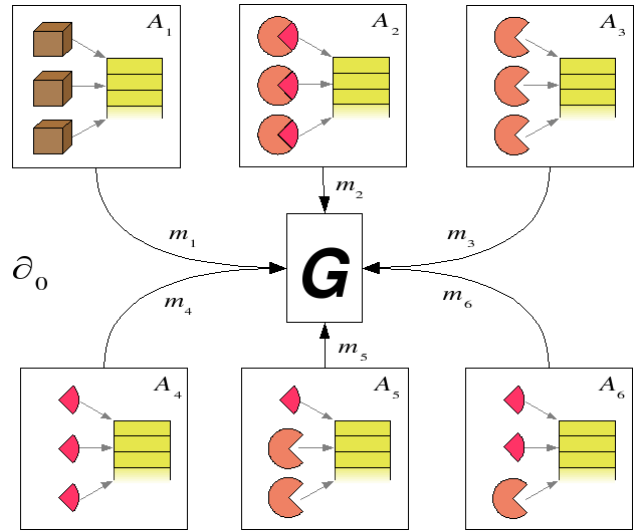}
  \caption{Graph Constraint on Conveyor Load}
  \label{fig:GC_convLoad_1_bmp}
\end{figure}

The first is based on the remark that conveyors as presented so far
have infinite capacity to load items.  Probably either due to a limit
of space or of load, conveyors can not transport more than, say, two
items. This is a constraint on the whole system, which can be modelled
as a graph constraint as introduced in Chap.~\ref{ch:restrictionsOnRules}. Figure~\ref{fig:GC_convLoad_1_bmp} shows
a diagram $\mathfrak{d}_0$ that sets this limit, with associated
formula:

\begin{equation}\label{eq:loadLimit}
  \mathfrak{f}_0 = \not \exists A_1 \ldots A_6 \left[ \bigvee_{i=1}^6
    A_i \right] = \forall A_1 \ldots A_6 \left[ \bigwedge_{i=1}^6
    \overline{A_i} \right].
\end{equation}

Recall that if the quantifier is not repeated it means that it applies
to every term, e.g. $\not \exists A_1A_2 \equiv \not \exists A_1 \not
\exists A_2$.

Graphs $A_5$ and $A_6$ are necessary because rule \texttt{recycle} may
mix elements of type \texttt{item1} and \texttt{item2} in the same
conveyor.  This graph constraint will be named $GC_0 = \left(
  \mathfrak{f}_0, \mathfrak{d}_0 \right)$.  By using variable nodes --
see Sec.~\ref{sec:fromSimpleDigraphsToMultidigraphs} -- the diagram
and the formula would be simpler, similar to the example on p.~\pageref{ex:twoOutgoingEdges}, in particular the right side of Fig.~\ref{fig:AtMostTwoOutgoingEdges}.  In the end, the diagram and the
formula would be instantiated to a graph constraint similar to what
appears on Fig.~\ref{fig:GC_convLoad_1_bmp} and equation~\eqref{eq:loadLimit}.

\begin{figure}[htbp]
  \centering
  \includegraphics[scale =
  0.5]{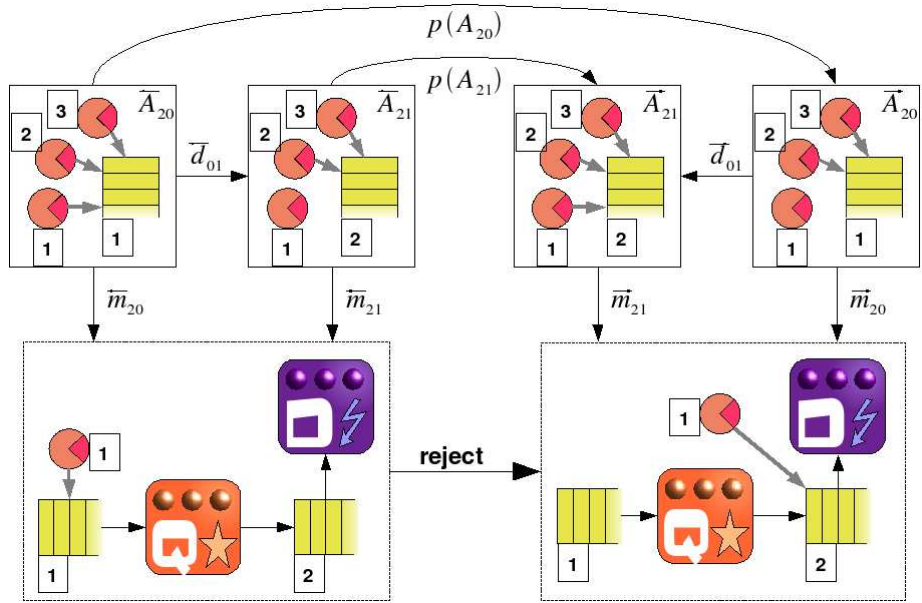}
  \caption{Graph Constraint as Precondition and Postcondition}
  \label{fig:GC_convLoad_2_bmp}
\end{figure}

The same graph constraint is depicted as precondition and
postcondition on Fig.~\ref{fig:GC_convLoad_2_bmp}.  The equations are
those adapted from~\eqref{eq:loadLimit}:
\begin{eqnarray}
  \stackrel{\leftarrow}{\mathfrak{f}_2} & = & \not\exists
  \stackrel{\leftarrow}{A_{20}} \stackrel{\leftarrow}{A_{21}} \left[
    \stackrel{\leftarrow}{A_{20}} \vee \stackrel{\leftarrow}{A_{21}}
  \right] \\
  \stackrel{\rightarrow}{\mathfrak{f}_2} & = & \not\exists
  \stackrel{\rightarrow}{A_{20}} \stackrel{\rightarrow}{A_{21}} \left[
    \stackrel{\rightarrow}{A_{20}} \vee \stackrel{\rightarrow}{A_{21}}
  \right].
\end{eqnarray}

Only the diagram in which elements of type \texttt{item3} appear has
been kept because we know that in conveyor labelled 1 there should not
be items of any other type (they would never be processed).  Actually,
with the definitions of rules given up to now, conveyors connecting
different machines are of the same kind.  Hence, all six diagrams
should appear on \texttt{reject}'s left hand side and their
transformation, according to Theorem~\ref{th:prePostPre}, on its right
hand side.


The precondition and the postcondition can be transformed into
equivalent sequences according to Theorems~\ref{th:reductionPre}~and~\ref{th:reductionPost}.  This is a negative application condition, see
Theorem~\ref{th:embeddingGC} and Lemma~\ref{lemma:nacs}. Hence, they
are split into two subconditions, each one demanding the nonexistence
of one element.  $\stackrel{\leftarrow}{A'_{20}}$ will ask for the
nonexistence of edge $\left( 2:\texttt{item3}, 1:\texttt{conv}
\right)$ and $\stackrel{\leftarrow}{A''_{20}}$ for $\left(
  3:\texttt{item3}, 1:\texttt{conv} \right)$.  Similarly we have
$\stackrel{\leftarrow}{A'_{21}}$ for $\left( 2:\texttt{item3},
  2:\texttt{conv} \right)$ and $\stackrel{\leftarrow}{A''_{21}}$ for
$\left( 3:\texttt{item3}, 2:\texttt{conv} \right)$.\footnote{To be
  precise, there would be other two conditions asking for the
  nonexistence of $\left( 1:\texttt{item3}, 1:\texttt{conv} \right)$,
  however this part of the application condition is inconsistent for
  the first conveyor (this edge is demanded because it has to be
  erased) and redundant for the second conveyor (it would be fulfilled
  always because this edge is going to be added, so it can not exist
  in the left hand side). This stems from the theory developed in
  Chap.~\ref{ch:restrictionsOnRules}.} At least one element in each
case must not be present, so there are four combinations:
\begin{eqnarray}
  \texttt{reject} & \longmapsto & \left\{\texttt{reject} ;
    \overline{id}_{\stackrel{\leftarrow}{A'_{21}}} ;
    \overline{id}_{\stackrel{\leftarrow}{A'_{20}}}, \quad \texttt{reject} ;
    \overline{id}_{\stackrel{\leftarrow}{A'_{21}}} ;
    \overline{id}_{\stackrel{\leftarrow}{A''_{20}}}, \right. \nonumber \\
  &  &  \left. \;\; \texttt{reject} ;
    \overline{id}_{\stackrel{\leftarrow}{A''_{21}}} ;
    \overline{id}_{\stackrel{\leftarrow}{A'_{20}}}, \quad \texttt{reject} ;
    \overline{id}_{\stackrel{\leftarrow}{A''_{21}}} ;
    \overline{id}_{\stackrel{\leftarrow}{A''_{20}}} \right\} 
\end{eqnarray}

The corresponding formula -- the left arrow on top is omitted -- can
be written:

\begin{equation}
  \label{eq:5}
  \exists A'_{20}A''_{20}A'_{210}A''_{21} \left[
    \left(\overline{A'_{20}} \vee \overline{A''_{20}}\right)\left(
      \overline{A'_{21}} \vee \overline{A''_{21}}\right)\right]
\end{equation}

Here postconditions and preconditions turn out to be the same because
$\texttt{reject} \, \bot \,
\overline{id}_{\stackrel{\leftarrow}{A'_{2x}}}$ and $\texttt{reject}
\, \bot \, \overline{id}_{\stackrel{\leftarrow}{A''_{2x}}}$, $x \in
\{0,1\}$.  For each sequence it is possible to compose all productions
and derive a unique rule. If so, as there are just elements that have
to be found in the complement of the host graph, they are appended to
the nihilation matrix of the composition.

For graph constraints, if something is to be forbidden, it is more
common to think in ``what should not be'', i.e. to think it as a
postcondition (graph constraint $GC_0$ is of this type).  On the
contrary, if something is to be demanded then it is normally easier to
describe it as a precondition.

Let's continue with another property of our system not addressed up to
now.  Note that conveyors clearly have a direction: Each one is the
output of one or more machines and input of one or more machines.  In
our example this is simplified so conveyors just join two different
machines.
What might be of interest is that items in conveyors are naturally
ordered.  Machines should pick the first ordered element.

\begin{figure}[htbp]
  \centering
  \includegraphics[scale =
  0.35]{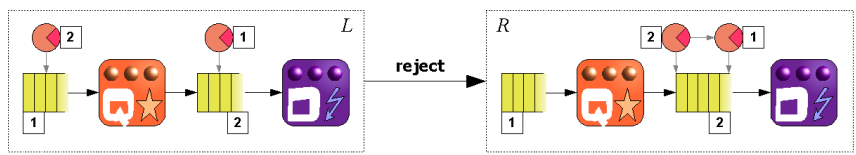}
  \caption{Ordered Items in Conveyors}
  \label{fig:orderedItems}
\end{figure}

To make our assembly line realize this feature, when the machine
processes a new item -- \texttt{2:item3} in Fig.~\ref{fig:orderedItems} -- and there is already an item in the output
conveyor -- \texttt{1:item3} in Fig.~\ref{fig:orderedItems} --, an
edge from \texttt{2:item3} to \texttt{1:item3} will be added.  A chain
is thus defined: The first element will have an incoming edge from
another item, but it will not be the source of any edge that ends in
other item.  The last item will not have any incoming edge but one
outgoing edge to another item.  It has been exemplified for rule
\texttt{reject} in Fig.~\ref{fig:orderedItems}.\footnote{We are not
  going to propose the modification of every single rule to handle
  ordering in conveyors. On the contrary, we are going to propose a
  method based on graph constraints and application conditions that
  automatically takes care of ordering.}

Again we have to change the allowable connections among types. The
diagram in Fig.~\ref{fig:syntaxSpecExtended} needs to be further
extended with a self-loop for items (there can be edges now) joining
two of them. However, concrete items can not have self-loops, so a new
graph constraint should take care of this.

This ordering convention poses two problems when the rule is applied:
\begin{enumerate}
\item If the input conveyor has two or more items, the first -- the
  one with incoming edges -- should be used.
\item If the output conveyor has one or more items, the new item must
  be linked to the last one.
\end{enumerate}

The first \emph{if} statement (pick the \emph{elder} item) can be
modelled by an application condition.  We have a precondition
$\stackrel{\leftarrow}{A} = \left( \mathfrak{f}_1, \mathfrak{d}_1
\right)$ with:
\begin{equation}\label{eq:firstItem}
  \mathfrak{f}_1 = \forall A_1 \exists A_2 \left[ \overline{A}_1
    \wedge \overline{A}_2 \right].
\end{equation}

\begin{figure}[htbp]
  \centering
  \includegraphics[scale =
  0.34]{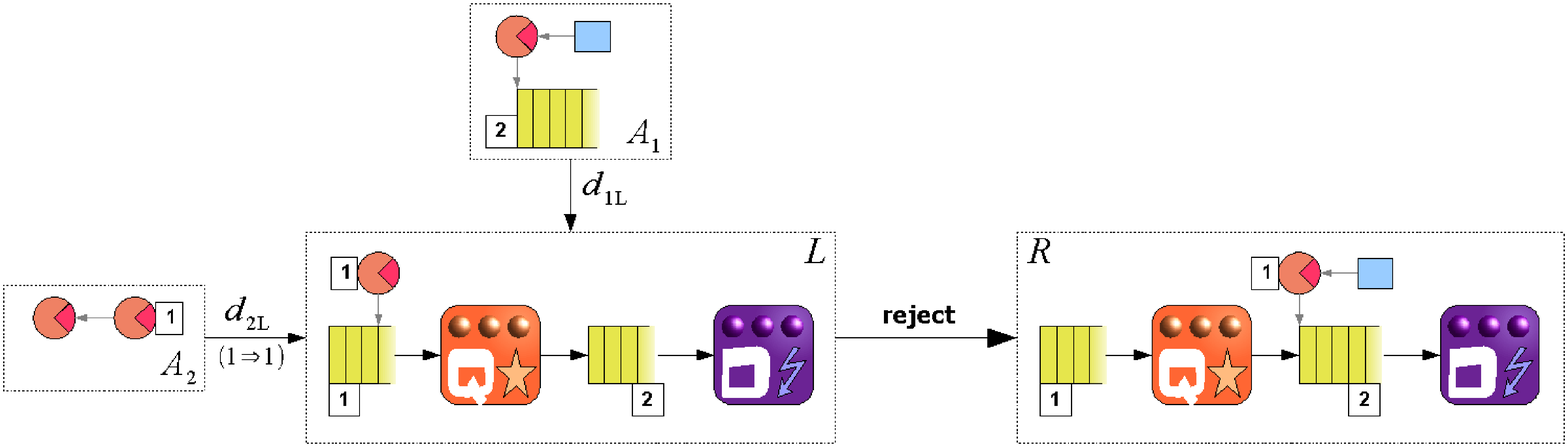}
  \caption{Expanded Rule \texttt{reject}}
  \label{fig:orderedRejectAC}
\end{figure}

The diagram is represented in Fig.~\ref{fig:orderedRejectAC}.
Numbered elements are related by the corresponding morphisms.  In
formula $\mathfrak{f}_1$ the term $\forall A_1 \ldots \left[
  \overline{A}_1 \ldots \right]$ prevents the application of the rule
if there is some marked item in the output conveyor (the blue square,
read below).  If the rule was applied then there would be two ``last''
items and it should become impossible to distinguish which one was
added first.  The term $\ldots \exists A_2 \left[ \ldots
  \overline{A}_2 \right]$ forces the rule to pick the first item in
the chain, just in case there was a chain.  Item \texttt{1:item3} will
be chosen either if it is the first in the chain or it is alone.  This
is equivalent to demand one item that has no outgoing edges to any
other item.

\begin{figure}[htbp]
  \centering
  \includegraphics[scale = 0.35]{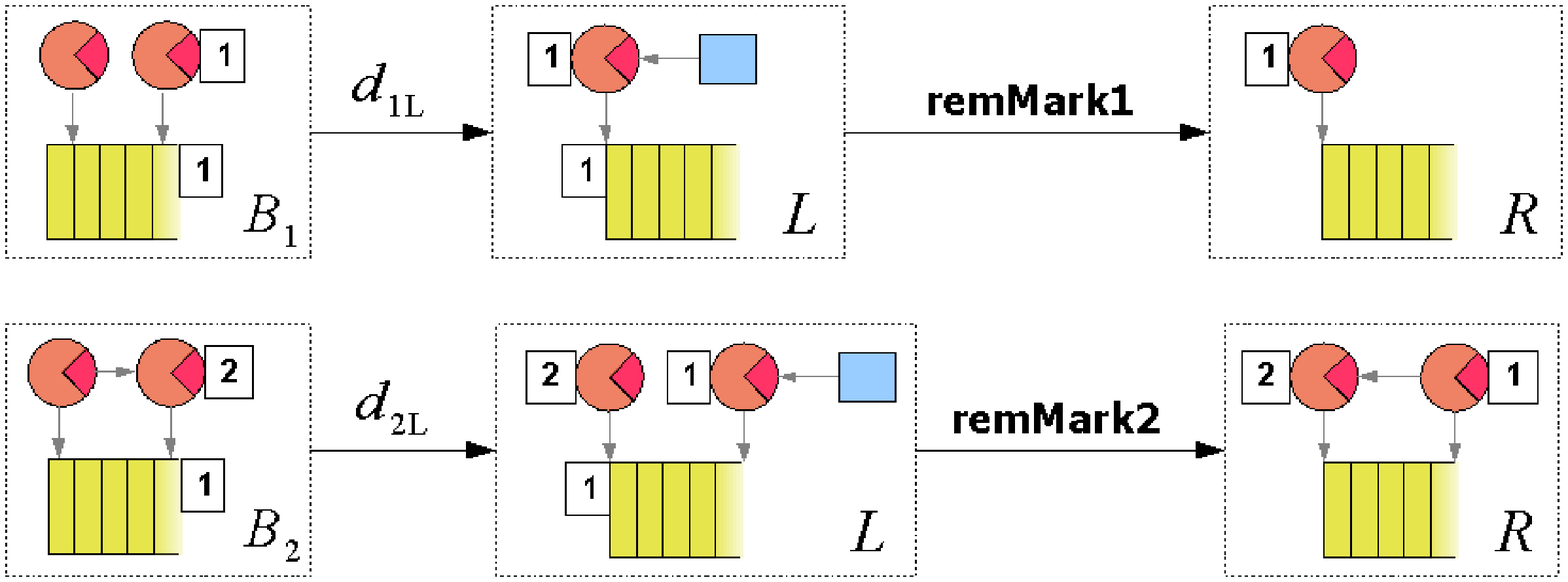}
  \caption{Rules to Remove Last Item Marks}
  \label{fig:removeMark}
\end{figure}

The second \emph{if} statement can not be modelled with an application
condition.  The reason is that we need to add one edge in case a
``last'' item exists in the output conveyor (if the output conveyor is
empty, then the rule should simply add the item).  Application
conditions are limited to checking whether (almost any arbitrary
combination of) elements are present or not, but they can not directly
modify the actions of the rules.  Anyway, the solution is not
difficult:
\begin{enumerate}
\item The newly added element needs to be marked so that the last item
  in the conveyor can be identified: The blue square of $A_1$ in
  Fig.~\ref{fig:orderedRejectAC} marks the last item added.
\item A precondition has to be imposed such that if there are marked
  items in the output conveyor, the rule can not be applied (this way
  at most one unlinked item will exist in each output conveyor).
  Again, see $A_1$ in Fig.~\ref{fig:orderedRejectAC} and the
  corresponding term in eq.~\eqref{eq:firstItem}.
\item The grammar is enlarged with a new rule that checks if there are
  unlinked items (linking them, \texttt{remMark2}) and another that
  unmarks them if they are alone in the conveyor, \texttt{remMark1}.
  See Fig.~\ref{fig:removeMark}
\end{enumerate}

Both productions \texttt{remMark1} and \texttt{remMark2} have
application conditions, $AC_1 = \left( \mathfrak{f}_1, \mathfrak{d}_1
  = \{B_1\} \right)$ and $AC_2 = \left( \mathfrak{f}_2, \mathfrak{d}_2
  = \{B_2\} \right)$, respectively.  The corresponding formulas are:
\begin{eqnarray}
  \mathfrak{f}_1 & = & \not \exists B_1 \left[ B_1 \right] \nonumber \\
  \mathfrak{f}_2 & = & \forall B_2 \left[ \overline{B}_2 \right] =
  \not \exists B_2 \left[ B_2 \right] \nonumber
\end{eqnarray}

Production \texttt{remMark1} can be applied only if there is just a
single item in the conveyor. \texttt{remMark2} applies when there is
more than one item. $B_2$ selects the last item: It is equivalent to
``the item with no incoming edges''.

There is no problem in transforming both preconditions of Fig.~\ref{fig:orderedRejectAC} into postconditions.  Note that there are no
dangling elements in $A_2$ because $\texttt{1:item3}$ is not erased
(which would mean removing and adding the same element, something
forbidden in Matrix Graph Grammars, see comments right after Prop.~\ref{prop:simpleEqualities}).

Notice that we have included ordering in conveyors with graph
constraints and application conditions (there exists the possibility
to transform one into the other) without really modifying existent
grammar rules. Ordering is a property of the system and not of the
productions, which should just take care of the actions to be
performed. We think that Matrix Graph Grammars clearly separate both
topics: It is feasible to specify grammar rules first and properties
of the system afterwards. With the theory developed in Chap.~\ref{ch:restrictionsOnRules} a framework -- such as AToM$^3$ -- can
relate one to the other more or less automatically.

Other examples of restrictions and limitations that can be imposed on
the case study are:
\begin{itemize}
\item Limitations on the number of operators, e.g. a maximum of four
  operators.
\item An operator can be in charge of at most one machine.
\item There should not be two operators working in the same machine,
  which is a restriction on rules of type \texttt{mov2*}.
\end{itemize}

More general constraints such as \emph{the number of operators can not
  exceed the number of machines} are also possible, although variable
nodes would be needed in this case.

The examples so far are simple and can be expressed with other
approaches to the topic. For other natural application conditions that
can only be addressed with Matrix Graph Grammar approaches (to the
best of our knowledge) please refer to the example on
p.~\pageref{ex:exampleDiff2Repre} or to~\cite{JuanPP_5}. The example
studied in this appendix is a extended version of the one that appears
there.

\section{Derivations}
\label{sec:derivations}

In this section a slight modification of the initial state depicted in
Fig.~\ref{fig:snapshot} together with a permutation of sequence $s_0$
will be used again, but enlarged with ordering of productions
(sequences) and restrictions of Sec.~\ref{sec:GraphConstraintsAndApplicationConditions}.  Internal and
external $\varepsilon$-productions will be addressed in passing.

\begin{figure}[htbp]
  \centering
  \includegraphics[scale =
  0.4]{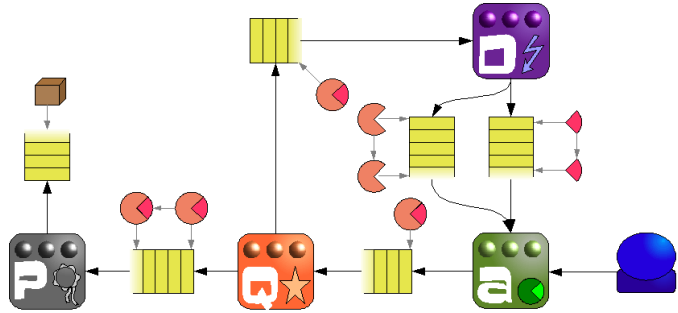}
  \caption{Grammar Initial State for $s'_5$}
  \label{fig:snapshotEnlarged}
\end{figure}

Let's consider as initial state the one depicted in Fig.~\ref{fig:snapshotEnlarged}.  Due to restrictions, sequence
$\texttt{s}_\texttt{0} = \texttt{pack} \, ; \texttt{certify} \, ;
\texttt{assem}$ is not applicable (three items would appear in the
input conveyor of \texttt{pack}).  However, productions are all
sequentially independent because they are applied to different items
(due to the amount of elements available in the initial state in Fig.~\ref{fig:snapshotEnlarged}) so sequence $s'_5 = \texttt{certify} \, ;
\texttt{pack} \, ;\texttt{assem}$ can be considered instead.

Sequence $s'_5$ can not be applied because the operator has to move to
the appropriate machine and ordering of items needs to be considered.
Let's suppose that the four basic rules have a higher probability --
or that they are in a higher layer, as e.g. in AGG\footnote{AToM$^3$
  has priorities.} -- so as soon as one of them is applicable it is in
fact applied.  According to the way an operator may move in our
assembly line, applying $s'_5$ would need at least the following
rules:
\begin{equation}
  s''_5 = \texttt{certify} \, ; \texttt{mov2Q} \, ; \texttt{mov2A} \,
  ; \texttt{recycle} \, ; \texttt{mov2D} \, ; \texttt{pack} \, ;
  \texttt{mov2P} \, ;  \texttt{mov2Q} \,; \texttt{assem}.
\end{equation}

Production \texttt{reject} could have been applied somewhere in the
sequence.
Again, as items are ordered and some dangling edges appear during the
process, this is not enough and some other productions need to be
appended:
\begin{eqnarray}
  s_5 & = & \left( \texttt{remMark2} \, ; \texttt{certify} \, ;
    \texttt{certify}_\varepsilon \right)\, ; \texttt{mov2Q} \, ;
  \texttt{mov2A} \, ; \texttt{recycle} \, ; \texttt{mov2D} \, ;
  \nonumber \\
  & & \left( \texttt{remMark2} \, ; \texttt{pack} \, ;
    \texttt{pack}_\varepsilon \right) \, ; \texttt{mov2P} \, ;
  \texttt{mov2Q} \,; \left( \texttt{remMark2} \, ; \texttt{assem} \, ;
    \texttt{assem}_\varepsilon \right) \nonumber
\end{eqnarray}

\begin{figure}[htbp]
  \centering
  \includegraphics[scale =
  0.4]{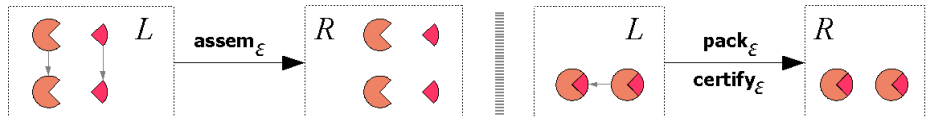}
  \caption{Production to Remove Dangling Edges (Ordering of Items in
    Conveyors)}
  \label{fig:exEpsilonProd}
\end{figure}

Parentheses are used to isolate subsequences that could probably be
composed to obtain more ``natural'' \emph{atomic} actions.  See
Fig.~\ref{fig:removeMark} for the definition of \texttt{remMark2} and
Fig.~\ref{fig:exEpsilonProd} for $\texttt{assem}_\varepsilon$,
$\texttt{pack}_\varepsilon$ and $\texttt{certify}_\varepsilon$.  In
this case, both $\texttt{assem}_\varepsilon$ and
$\texttt{pack}_\varepsilon$ are external while
$\texttt{certify}_\varepsilon$ is internal.  Productions between
brackets are related through a marking operator.  It is mandatory that
they act on the same nodes and edges.

\begin{figure}[htbp]
  \centering
  \includegraphics[scale =
  0.4]{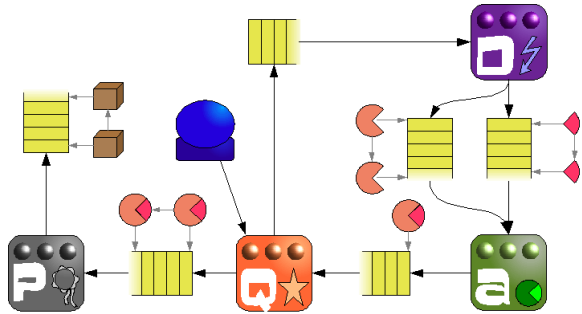}
  \caption{Grammar Final State for $s_5$}
  \label{fig:snapshotEnlargedFS}
\end{figure}

A user of a tool such as AToM$^3$ or AGG does not necessarily need to
know about $\varepsilon$-productions, even less about marking.
Probably in this case it should be better to compose productions that
include \texttt{remMark1} or \texttt{remMark2} and call them as the
original rule, e.g. $\texttt{remMark2} \, ; \texttt{assem} \longmapsto
\texttt{assem}$.  The final state for $s_5$ can be found in Fig.~\ref{fig:snapshotEnlargedFS}

A development framework should have facilities to ease visualization
of grammar rules, as diagrams can be quite cumbersome with only a few
constraints.  For example, it should be possible to keep graph
constraints apart from productions, calculating on demand how a
concrete constraint modifies a selected production, its left and right
hand sides and nihilation matrices.

\newpage

\backmatter




\renewcommand{\baselinestretch}{1.3} \printindex

\end{document}